\documentclass[prb,aps,amssymb,twocolumn,nofootinbib,superscriptaddress]{revtex4}
\usepackage{amsmath}
\usepackage{amssymb}
\usepackage{amsthm}
\usepackage{amsfonts}
\usepackage{listings}
\usepackage[colorlinks, citecolor=blue]{hyperref}
\usepackage{dsfont}
\usepackage{enumerate}
\usepackage{latexsym}
\usepackage{longtable}
\usepackage[normalem]{ulem}
\usepackage{float}
\usepackage{setspace}
\usepackage{multirow}

\usepackage{psfrag}

\usepackage{bm}
\usepackage{graphicx}
\usepackage[caption=false]{subfig}
\usepackage{xcolor}

\newlength{\bibitemsep}
\newlength{\bibparskip}
\let\oldthebibliography\thebibliography
\renewcommand\thebibliography[1]{%
  \oldthebibliography{#1}%
  \setlength{\parskip}{\bibitemsep}%
  \setlength{\itemsep}{\bibparskip}%
}

\newcommand{\beq}{\begin{equation}}
\newcommand{\eneq}{\end{equation}}

\newcommand{\ket}[1]{\left|#1\right\rangle}

\input{epsf}

\newcommand{\bb}{\boldsymbol{b}}

\newcommand{\mbf}{\mathbf}
\newcommand{\mbb}{\mathbb}

\newcommand{\mrm}{\mathrm}

\newcommand{\ovl}{\overline}

\def\beq#1\eeq{\begin{equation}#1\end{equation}}
\def\beqs#1\eeqs{\begin{align}#1\end{align}}

\def\pare#1{\left( #1 \right)}

\def\ket#1{| #1 \rangle}

\def\nono{\nonumber}

\def\pr{\prime}
\def\prpr{{\prime\prime}}

\def\ZZ{\mathbb{Z}}
\def\kk{\mathbf{k}}
\def\tt{\mathbf{t}}
\def\qq{\mathbf{q}}
\def\EBR{\mathcal{EBR}}
\def\CR{\mathcal{CR}}

\def\TRS{\mathcal{T}}
\def\INV{\mathcal{I}}

\newcommand{\up}{\uparrow}

\newcommand{\bpm}{\begin{pmatrix}}
\newcommand{\epm}{\end{pmatrix}}

\newcommand{\bal}{\begin{align}}

\DeclareMathOperator{\Tr}{Tr}

\DeclareMathOperator{\sgn}{sgn}

\begin{document}

\title{Magnetic Topological Quantum Chemistry}

\author{Luis Elcoro}
\thanks{These authors contributed equally to this work.}
\affiliation{Department of Condensed Matter Physics, 
University of the Basque Country UPV/EHU, 
Apartado 644, 48080 Bilbao, Spain}

\author{Benjamin J. Wieder$^\dag$}
\thanks{These authors contributed equally to this work.}
\affiliation{Department of Physics, Massachusetts Institute of Technology, Cambridge, MA 02139, USA}
\affiliation{Department of Physics, Northeastern University, Boston, MA 02115, USA}
\affiliation{Department of Physics,
Princeton University,
Princeton, NJ 08544, USA
	}

\author{Zhida Song}
\affiliation{Department of Physics,
Princeton University,
Princeton, NJ 08544, USA
	}

\author{Yuanfeng Xu}
\affiliation{Max Planck Institute of Microstructure Physics, 
06120 Halle, Germany
}

\author{Barry Bradlyn}
\affiliation{Department of Physics and Institute for Condensed Matter Theory, University of Illinois at Urbana-Champaign, Urbana, IL, 61801-3080, USA}

\author{B. Andrei Bernevig$^\dag$$^\ddag$}
\affiliation{Department of Physics,
Princeton University,
Princeton, NJ 08544, USA
	}
\affiliation{Donostia International Physics Center, P. Manuel de Lardizabal 4, 20018 Donostia-San Sebastian, Spain}
\affiliation{IKERBASQUE, Basque Foundation for Science, Bilbao, Spain}

\begin{abstract}
For over 100 years, the group-theoretic characterization of crystalline solids has provided the foundational language for diverse problems in physics and chemistry.  However, the group theory of crystals with commensurate magnetic order has remained incomplete for the past 70 years, due to the complicated symmetries of magnetic crystals.  In this work, we complete the 100-year-old problem of crystalline group theory by deriving the small corepresentations, momentum stars, compatibility relations, and magnetic elementary band corepresentations of the 1,421 magnetic space groups (MSGs), which we have made freely accessible through tools on the Bilbao Crystallographic Server.  We extend Topological Quantum Chemistry to the MSGs to form a complete, real-space theory of band topology in magnetic and nonmagnetic crystalline solids -- Magnetic Topological Quantum Chemistry (MTQC).  Using MTQC, we derive the complete set of symmetry-based indicators of electronic band topology, for which we identify symmetry-respecting bulk and anomalous surface and hinge states.
\end{abstract}

\maketitle

 {
\centerline{\bf Introduction}
\vspace{0.3in}
}

A crystal is defined by its discrete translation symmetry.  Over the past 140 years~\cite{sohncke1879book,fedorov1891symmetry}, a tremendous number of physical phenomena have been shown to arise from the complicated mathematical structures implied by this otherwise simple definition of a crystal.  For example, the symmetry and group theory of crystalline solids have been used to characterize phase transitions~\cite{LandauTheory}, identify biological structures like the DNA double helix~\cite{WatsonCrickDNA}, and, most recently, to elucidate the position-space origin of topological bands through the theories of Topological Quantum Chemistry (TQC)~\cite{QuantumChemistry,AndreiMaterials} and equivalent works~\cite{AshvinIndicators,AshvinMaterials,ChenMaterials}.

\begin{figure}[!t]
\centering
\includegraphics[width=\columnwidth]{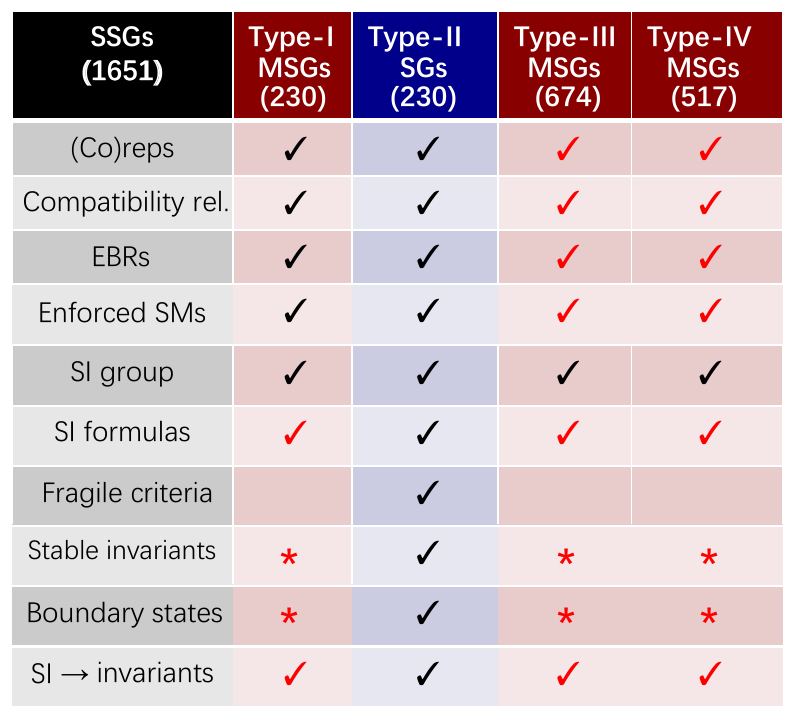}
\caption{Summary of results.  In this work, we have derived the complete sets of trivial bands [elementary band (co)representations (EBRs), see SA~\ref{sec:MEBRs}] and symmetry-indicated, spinful, stable topological bands in the 1,651 Shubnikov space groups [SSGs].  The EBRs subdivide into the physical EBRs of the 230 Type-II nonmagnetic space groups [SGs] and the magnetic EBRs [MEBRs] of the 1,421 Type-I, III, and IV magnetic SGs [MSGs, see SA~\ref{sec:MSGs}]~\cite{ShubnikovBook,BigBook,MagneticBook}.  We have additionally performed the first complete calculation of the small (co)representations [(co)reps] and compatibility relations [see SA~\ref{sec:momentumSpace}] for all 1,651 single and double SSGs, which we have made accessible through the tools listed in Table~\ref{tb:BCStoolsMain}.  These results comprise the theories of Magnetic Topological Quantum Chemistry (MTQC) and fermionic symmetry-based indicators (double SIs)~\cite{SlagerSymmetry,AshvinIndicators,ChenTCI,AshvinTCI,AshvinMagnetic}, which apply to all possible 3D magnetic and nonmagnetic crystals with mean-field Hamiltonians.  We have also determined the physical bases of all double (spinful) symmetry-based indicators (SIs), and symmetry-indicated topological bulk and anomalous boundary states for all 1,651 double SSGs (SA~\ref{sec:topologicalBands}).  Lastly, the MEBRs of the Type-III and Type-IV MSGs computed in this work also facilitate the complete enumeration of symmetry-enforced magnetic topological semimetals (SMs) -- examples are provided in Fig.~\ref{fig:mainFermion}(c) and in SA~\ref{sec:corepExampleYesAnti}.  In this figure, we have used red checks to indicate areas of magnetic topological band theory completed in this work, and we have used red stars to indicate areas in which we have solved complete subareas (such as the double SIs of the 1,651 double SSGs), but in which there remain topological features outside of the scope of this work, such as non-symmetry-indicated stable topological bands~\cite{HourglassInsulator,DiracInsulator,TMDHOTI,WiederAxion} and bosonic (spinless) topological crystalline insulators (TCIs).}
\label{fig:mainResults}
\end{figure}

In time-reversal- ($\mathcal{T}$-) symmetric, periodic systems -- which most familiarly include nonmagnetic crystalline solids -- the energy (Bloch) eigenstates respect the symmetries of the nonmagnetic (Type-II) Shubnikov space group (SSGs)~\cite{ShubnikovBook,BigBook,MagneticBook} [see Fig.~\ref{fig:mainResults} and Supplementary Appendix (SA)~\ref{sec:MSGs}].  Though there are 230 Type-II SSGs, including SSGs with complicated combinations of glide and screw symmetries, the group theory of nonmagnetic crystalline solids has been largely solved for over 40 years~\cite{BigBook}.  In particular, the enumeration of the irreducible momentum-space [small] corepresentations [coreps, see SA~\ref{sec:coreps}], and a partial enumeration of the space group (elementary band) coreps [EBRs, see SA~\ref{sec:MEBRs}] of the Type-II SSGs were completed prior to the advent of personal and distributed computing~\cite{BigBook,Wigner1932,wigner1959group,ZakBandrep1,EvarestovMEBR}.  In recent years, the group theory of Type-II SSGs has facilitated a revolution in the search for topological insulators (TIs)~\cite{CharlieTI,AndreiTI,FuKaneMele,FuKaneInversion,QHZ,HsiehDiracInsulator} and topological crystalline insulators (TCIs)~\cite{LiangTCIOriginal,HsiehTCI,HourglassInsulator,DiracInsulator}, including the recent discovery of higher-order TCIs (HOTIs)~\cite{WladTheory,HOTIBernevig,ChenRotation} through TQC and related methods~\cite{SlagerSymmetry,ChenTCI,AshvinTCI,HOTIBismuth,TMDHOTI}.

\begin{table}[t]
\begin{small}
\begin{tabular}{|c|c|c|}
\hline
\multicolumn{3}{|c|}{BCS Applications Implemented for MTQC} \\
\hline
Application & Contents & Description  \\
\hline
\hline
\href{http://www.cryst.ehu.es/cryst/mkvec}{MKVEC} & Momentum stars & SA~\ref{sec:MKVEC} \\
 & of the MSGs & \\
\hline
\href{http://www.cryst.ehu.es/cryst/corepresentations}{Corepresentations} & Small and full & SA~\ref{sec:coreps} \\
 & magnetic (co)reps & \\
\hline
\href{https://www.cryst.ehu.es/cryst/mcomprel}{MCOMPREL} & Compatibility relations & SA~\ref{sec:compatibilityRelations} \\
& in the MSGs & \\
\hline
\href{http://www.cryst.ehu.es/cryst/corepresentationsPG}{CorepresentationsPG} & Magnetic site-symmetry & SA~\ref{sec:magWannier} \\
 & group (co)reps & \\
\hline
\href{http://www.cryst.ehu.es/cryst/msitesym}{MSITESYM} & Magnetic small & SA~\ref{sec:induction} \\
 & (co)reps at one ${\bf k}$ point & \\ 
 & induced from a site ${\bf q}$ & \\
\hline
\href{http://www.cryst.ehu.es/cryst/mbandrep}{MBANDREP} & MEBRs of the MSGs & SA~\ref{sec:mbandrep} \\
\hline
\end{tabular}
\end{small}
\caption{Applications on the Bilbao Crystallographic Server implemented for MTQC.  For this work, we have implemented the Bilbao Crystallographic Server (BCS) programs listed in this table to access group-theoretic properties of the MSGs that we have computed to complete the theory of MTQC.  In order, this table contains the name of the program, the data accessible through the program, and the section of the SA in which the program is detailed.  In addition to the properties of the MSGs listed in this table, each tool contains the analogous properties of the 230 Type-II (nonmagnetic) SGs.  Therefore, as respectively detailed in each listed SA section, each program in this table subsumes the content of an existing program on the BCS.}
\label{tb:BCStoolsMain}
\end{table}

However, the 230 Type-II SSGs represent only a fraction of the 1,651 (magnetic and nonmagnetic) SSGs (MSGs and SGs, respectively, see Fig.~\ref{fig:mainResults} and SA~\ref{sec:MSGs}).  Specifically, while Type-II SGs contain unitary symmetries and $\mathcal{T}$ about any point ($\{\mathcal{T}|{\bf 0}\}$), there are also Type-I MSGs with only unitary symmetries, Type-III MSGs that contain combinations of $\mathcal{T}$ and rotation or reflection (\emph{e.g.} $\{C_{2z}\times\mathcal{T}|{\bf 0}\}$, in which $C_{ni}$ is a rotation by $2\pi/n$ about the $i$ axis), and Type-IV MSGs that contain the combination of $\mathcal{T}$ and fractional lattice translation ($\{\mathcal{T}|{\bf a}/2\}$, in which ${\bf a}$ is an odd-integer linear combination of lattice vectors).  The small (co)reps and magnetic EBRs [MEBRs] of the MSGs are necessary for a wide range of physical applications, including characterizing magnetic topological semimetals (SMs)~\cite{MagneticWeylZahid,MagneticWeylYulin,MagneticWeylHaim,WeylReview}, TIs~\cite{QAHScience1,QAHScience2}, and TCIs~\cite{AXIExp1,AXIExp2}.  Beyond topological materials, the magnetic small (co)reps are also required to construct theories of magnetic phase transitions with nonzero ${\bf q}$ vectors from magnetic structure data obtained through neutron diffraction experiments~\cite{HalpernNeutron,CracknellMagneticTrans}, and to characterize $\mathcal{T}$-breaking superconducting phases~\cite{ReadGreenSC} with nonzero Cooper-pair momenta, such as Fulde-Ferrell-Larkin-Ovchinnikov states~\cite{FuldeFerrellSC,LarkinNonuniformSC,NormanNonsymmorphicSC,YanaseNonsymmorphicSC}.  Nevertheless, due to the relative complexity of the MSGs, and despite a number of significant partial tabulations~\cite{millerTables,AshvinMagnetic}, progress towards completing the group theory of magnetic crystals has largely stalled for the past 70 years~\cite{ShubnikovBook,BigBook}.  

\begin{figure*}[!t]
\centering
\includegraphics[width=0.67\textwidth]{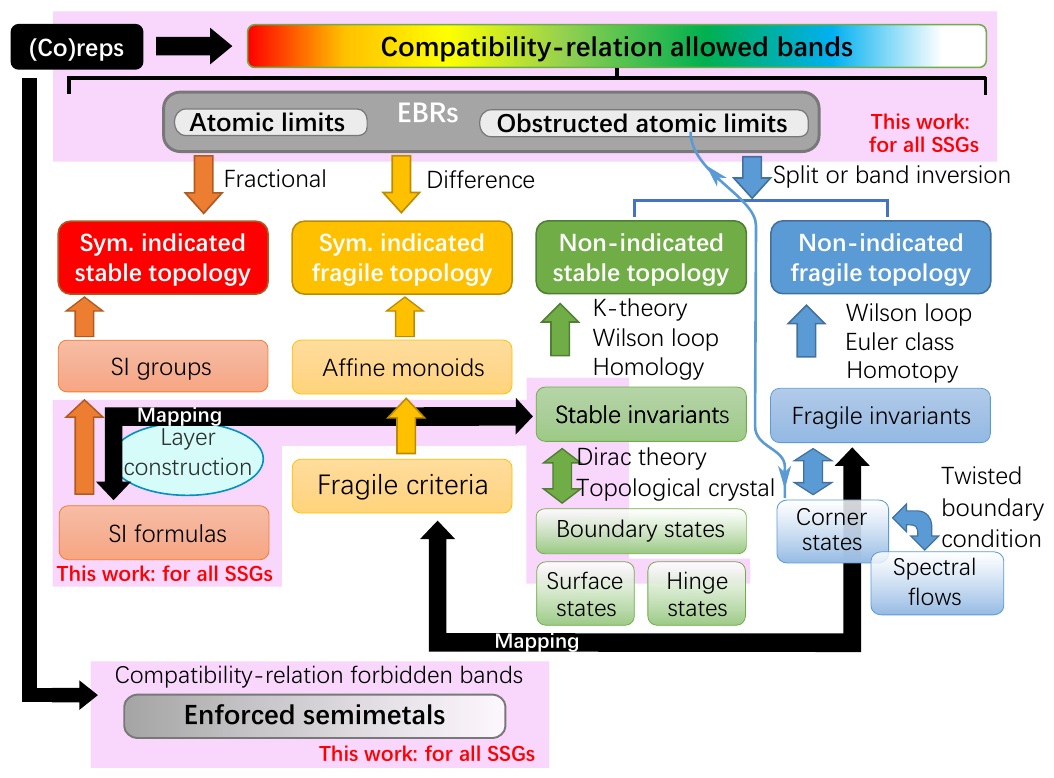}
\caption{Magnetic Topological Quantum Chemistry in the scheme of topological band theory.  The complete scheme of topological band theory for 3D crystals, following the framework and notation established in Refs.~\onlinecite{QuantumChemistry,ChenTCI,AndreiMaterials,ZhidaFragile,ZhidaFragile2}.  Through crystal symmetry eigenvalues [small (co)reps] in momentum space (SA~\ref{sec:coreps}), the compatibility relations (SA~\ref{sec:compatibilityRelations}) indicate whether a set of bands is allowed by symmetry to be energetically isolated from other bands in the energy spectrum.  If the bands are energetically isolated, then there exist a wide range of methods for diagnosing whether the bands exhibit the stable topology of topological insulators (TIs) and TCIs~\cite{CharlieTI,AndreiTI,FuKaneMele,FuKaneInversion,LiangTCIOriginal,HsiehTCI,HourglassInsulator,DiracInsulator,WladTheory,HOTIBernevig,ChenRotation,SlagerSymmetry,ChenTCI,AshvinTCI,HOTIBismuth,TMDHOTI,HingeSM}, fragile topology~\cite{AshvinFragile,JenFragile1,BarryFragile,ZhidaFragile,ZhidaFragile2,HingeSM,WiederAxion}, or the polarization-nontrivial topology of obstructed atomic limits~\cite{QuantumChemistry,multipole,HingeSM}.  For example, as detailed in Refs.~\onlinecite{SlagerSymmetry,AshvinIndicators,ChenTCI,AshvinTCI,AshvinMagnetic,ZhidaFragile,ZhidaFragile2}, the small (co)reps of a set of isolated bands comprise momentum-space symmetry data that can be mapped to position-space topology and boundary states through stable and fragile SIs and real-space invariants.  If the bands are instead required by symmetry to cross, then the bands characterize a topological SM, which may exhibit surface~\cite{WeylReview} or hinge~\cite{HingeSM,TMDHOTI} states.  In this figure, the pink boxes indicate areas of topological band theory completed in this work.}
\label{fig:mainScheme}
\end{figure*}

In this work, we use a combination of computational and analytic methods to derive the small (co)reps and MEBRs of the MSGs, completing the 100-year-old problem of crystalline group theory.  Using the small (co)reps and MEBRs, we construct a complete position-space theory of mean-field band topology in the 1,651 single (spinless) and double (spinful) SSGs -- Magnetic Topological Quantum Chemistry (MTQC) -- that subsumes the earlier theory of TQC~\cite{QuantumChemistry,AndreiMaterials} [see Fig.~\ref{fig:mainScheme}].  The completeness of MTQC stems from the completeness of our tabulation of the MEBRs.  Specifically, even in MSGs in which trivial and topological states cannot be distinguished by symmetry eigenvalue labels, the MEBRs provide a complete basis for constructing and analyzing all possible lattice models of trivial, gapless, and stable and fragile topological insulating phases (for specific examples of non-symmetry-indicated topological phases analyzed using EBRs, see Refs.~\onlinecite{BarryFragile,Bandrep3,HingeSM,WiederAxion,TMDHOTI}).  To access the data generated for this work, we have implemented several programs on the Bilbao Crystallographic Server (BCS)~\cite{BCS1,BCS2}, which are listed in Table~\ref{tb:BCStoolsMain}.  Each of the programs listed in Table~\ref{tb:BCStoolsMain} contains data for both the magnetic and nonmagnetic SSGs, and therefore replaces an existing tool on the BCS.  In the Results section below, we will first describe the underlying machinery of MTQC through which band (co)reps in momentum space are induced from magnetic atomic (Wannier) orbitals in position space.  Next, we will detail the topological information that can be inferred from the MEBRs, which include lattice models for magnetic exceptions to fermion doubling theorems~\cite{SteveMagnet,DiracInsulator}, and symmetry-based indicators (SIs)~\cite{SlagerSymmetry,AshvinIndicators,ChenTCI,AshvinTCI,AshvinMagnetic} for magnetic SMs, TIs, and TCIs (see SA~\ref{sec:topologicalBands}).  In particular in this work, going beyond the earlier tabulation of the magnetic SI groups in Ref.~\onlinecite{AshvinMagnetic}, we have for the first time generated the complete double SI formulas, as well as symmetry-respecting topological bulk and boundary states for all 1,651 double SSGs, which characterize spinful electronic states in solid-state materials.  Through this calculation, we have obtained the complete set of symmetry-indicated 3D spinful (fermionic) topological phases.

We find that many of the symmetry-indicated spinful magnetic topological phases consist of familiar Weyl SMs with surface Fermi arcs~\cite{AshvinWeyl,AndreiWeyl,SYWeyl}, 3D quantum anomalous Hall (QAH) phases constructed from layered integer quantum Hall states (2D Chern insulators)~\cite{HaldaneModel,QAHScience1}, and axion insulators (AXIs), which are equivalent to 3D TIs with magnetically gapped surface states on particular crystal facets~\cite{QHZ,VDBAxion,WiederAxion}.  However, we also in this work discover the existence of previously unidentified non-axionic magnetic HOTIs with mirror-protected helical hinge states (see SA~\ref{sec:newHOTIs}).  We conclude by briefly discussing future directions in magnetic group theory, including the prediction of spinless (bosonic) TCIs, and applications of magnetic crystal symmetry beyond mean-field theory.  We have also included an extensive set of Supplementary Appendices and Tables containing additional details of our methodology, historical commentary, references, documentation for the BCS programs introduced in this work, and data for the EBRs and double SIs (see SA~\ref{sec:appendixIntro} and~\ref{sec:supplementaryTables}).

 {
\vspace{0.06in}
\centerline{\bf Results}
\vspace{0.06in}
}

\begin{figure*}[!t]
\centering
\includegraphics[width=0.93\textwidth]{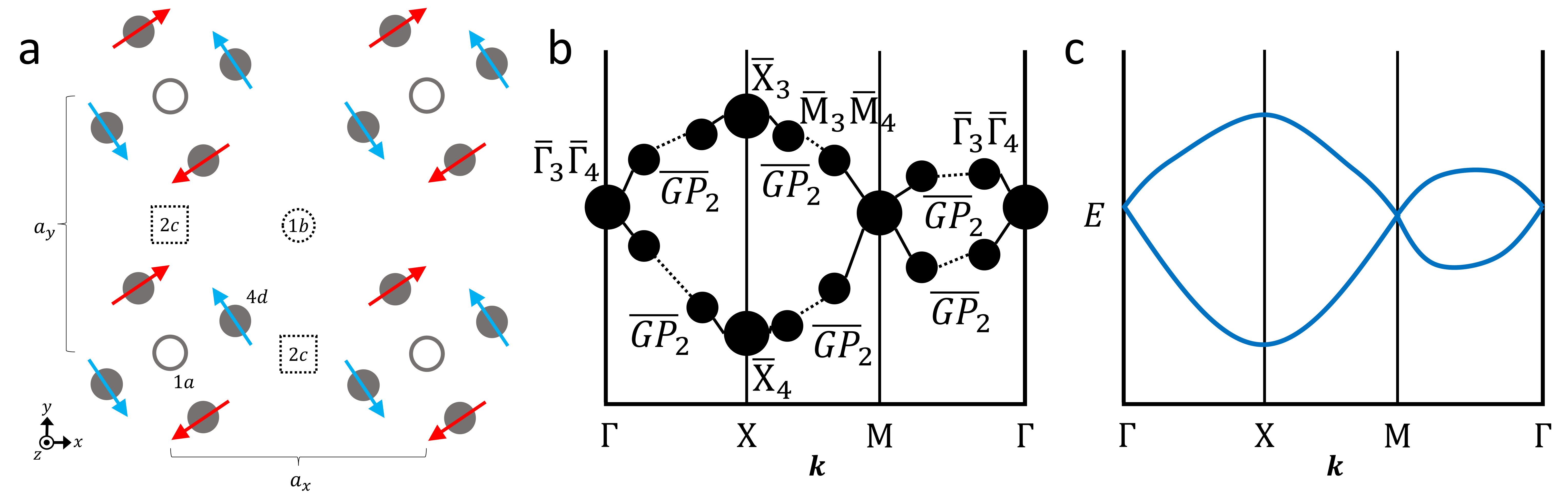}
\caption{Magnetic band (co)reps from magnetic atomic orbitals.  (a) A crystal with lattice-commensurate magnetic order.  In the mean-field, the basis states of the electronic Hamiltonian of the crystal in (a) are magnetic atomic orbitals (SA~\ref{sec:magWannier}).  When weakly coupled, the magnetic atomic orbitals in (a) continue to form a set of exponentially localized, symmetric Wannier orbitals~\cite{QuantumChemistry,Bandrep3,ZakBandrep1,EvarestovMEBR} that transform in the (co)reps of magnetic site-symmetry groups [SA~\ref{sec:SecFullWyckoff}].  (b) The magnetic site-symmetry (co)reps in (a) induce a band (co)rep in momentum [${\bf k}$] space.  (c) Correspondingly, the Bloch eigenstates of the Fourier-transformed electronic Hamiltonian of the magnetic crystal in (a) transform in the band (co)rep in (b) [see SA~\ref{sec:induction}].}
\label{fig:mainMTQC}
\end{figure*}

\textit{MEBRs from magnetic atomic orbitals} -- To construct the theory of MTQC, we first tabulate the EBRs of the 1,651 SSGs, which include the MEBRs of the MSGs [Fig.~\ref{fig:mainMTQC}(b) and SA~\ref{sec:MEBRs}].  In each SSG, the EBRs correspond to the independent topologically trivial bands.  Specifically, each EBR corresponds to a (set of) band(s) that can be inverse-Fourier-transformed into exponentially localized, symmetric Wannier orbitals, and the set of EBRs in each SSG forms the basis for all energetically isolated sets of trivial bands (\emph{i.e.} bands without stable or fragile topology)~\cite{QuantumChemistry,Bandrep3,AndreiMaterials,AshvinIndicators,AshvinMaterials,ChenMaterials,ZakBandrep1,EvarestovMEBR,SlagerSymmetry,ChenTCI,AshvinTCI,HOTIBismuth,TMDHOTI,AshvinFragile,JenFragile1,BarryFragile,ZhidaFragile,ZhidaFragile2,HingeSM,WiederAxion}.

We begin by considering a nonmagnetic crystal that is furnished with atomic orbitals that are sufficiently weakly coupled as to not invert bands at any ${\bf k}$ point in the Brillouin zone (BZ).  Each atomic orbital occupies a site in a Wyckoff position of a Type-II SG.  Crucially, the atomic orbitals on each site transform in direct sums of the irreducible coreps of the site-symmetry group (SA~\ref{sec:SecFullWyckoff} and~\ref{sec:magWannier}), which is necessarily isomorphic to one of the 32 nonmagnetic point groups (PGs, see SA~\ref{sec:siteSymmetry}).

We next consider the case in which the crystal undergoes a transition into a phase with lattice-commensurate magnetic order [Fig.~\ref{fig:mainMTQC}(a)].  The onset of magnetism lowers the crystal symmetry from a Type-II SG into either a Type-I, III, or IV MSG (see Refs.~\onlinecite{ShubnikovBook,BigBook,MagneticBook} and SA~\ref{sec:type1},~\ref{sec:type3}, and~\ref{sec:type4}, respectively).  Specifically, in the limit in which the magnetic moments are taken to be decoupled from the underlying lattice, the crystal of moments may appear to exhibit additional symmetries, such global and local spin rotation.  However, when coupling between the spins and the underlying lattice is not ignored, the magnetic phase transition strictly lowers the system symmetry to that of a magnetic Shubnikov subgroup $M$ of the Type-II SG $G$ of the parent nonmagnetic crystal~\cite{BigBook}.

Hence, the magnetic order also lowers the symmetry at each site in the crystal.  This can be seen by recognizing that $\{\mathcal{T}|{\bf 0}\}$ is an element of every site-symmetry group in a nonmagnetic crystal, but cannot be an element of any site-symmetry group in a magnetic crystal (SA~\ref{sec:Wyckoff}).  For example, in a solid-state material with magnetic atoms, the orbitals of nonmagnetic atoms elsewhere in the unit cell are necessarily subject to a background magnetic potential (see SA~\ref{sec:WyckoffExample}).  While the energy scale of the magnetic potential is detail-dependent, the magnetic potential on the atoms considered to be nonmagnetic is only exactly zero in a fine-tuned limit.  This statement remains valid whether individual atoms in the magnetic crystal are taken to host localized magnetic dipole moments, or whether the magnetic crystal is taken to consist of multi-atom clusters with higher magnetic multipole moments~\cite{SuzukiMagneticMultipoleCluster1,SuzukiMagneticMultipoleCluster2}.  Consequently, independent of the phenomenological microscopic treatment of the magnetic order, each site-symmetry group in the magnetic crystal is isomorphic to one of the 90 crystallographic magnetic point groups (MPGs, see SA~\ref{sec:siteSymmetry}).  In a solid-state material in which the effects of magnetism can be approximated through mean-field theory, the atomic orbitals of the original crystal [\emph{e.g.} $s$ and $p_{x,y}$] split into magnetic atomic orbitals [\emph{e.g.} $s$ and $p_{x}\pm i p_{y}$] that transform in (co)reps of the MPGs [see SA~\ref{sec:singleFirstSiteSym},~\ref{sec:singleSecondSiteSym}, and~\ref{sec:singleThirdSiteSym}].  For this work, we have implemented the~\href{http://www.cryst.ehu.es/cryst/corepresentationsPG}{CorepresentationsPG} tool (\url{http://www.cryst.ehu.es/cryst/corepresentationsPG}, detailed in SA~\ref{sec:magWannier}), through which users can access the (co)reps of all 122 single and double PGs and MPGs.

Next, the magnetic site-symmetry (co)reps in each Wyckoff position in the crystal induce a band (co)rep into $M$ [Fig.~\ref{fig:mainMTQC}(b)].  The set of all possible band (co)reps in each MSG is spanned by the MEBRs of $M$.  In this work, we have for the first time computed the 22,611 MEBRs of all 1,191 single and double Type-III and Type-IV MSGs, which -- along with the 5,641 MEBRs of the 230 Type-I MSGs and the 4,757 EBRs of the 230 Type-II SGs previously calculated for TQC~\cite{QuantumChemistry,Bandrep3,ZakBandrep1,EvarestovMEBR} [Fig.~\ref{fig:mainResults}] -- can be accessed through the~\href{http://www.cryst.ehu.es/cryst/mbandrep}{MBANDREP} tool on the BCS (\url{http://www.cryst.ehu.es/cryst/mbandrep}, further detailed in SA~\ref{sec:mbandrep}).  To enumerate the MEBRs of each MSG $M$, we begin by inducing band (co)reps from each irreducible (co)rep of one site-symmetry group within each of the highest-symmetry [\emph{i.e.} maximal, see SA~\ref{sec:Wyckoff}] Wyckoff positions in $M$.  We next exclude the exceptional cases in which the induced band (co)reps are equivalent to direct sums of other band (co)reps [SA~\ref{sec:exceptions} and~\ref{sec:exceptionalTables}].  The remaining band (co)reps are defined as elementary [\emph{i.e.} MEBRs]; statistics and further details for the MEBRs are provided in SA~\ref{sec:mebrStats} and~\ref{sec:EBCRdimension}.

Importantly, just as each MEBR is the Fourier-transformed description of a crystal of site-symmetry (co)reps, the Wannierizable bands that transform in each MEBR are the Bloch eigenstates of the Fourier-transformed electronic Hamiltonian of weakly coupled magnetic atomic orbitals [Fig.~\ref{fig:mainMTQC}(c) and SA~\ref{sec:induction}].  Consequently, in each momentum star of each MSG -- which are accessible through the~\href{http://www.cryst.ehu.es/cryst/mkvec}{MKVEC} tool (\url{http://www.cryst.ehu.es/cryst/mkvec}, see SA~\ref{sec:MKVEC}) -- each MEBR contains a set of full (co)reps that is specified by the Wyckoff position from which the MEBR is induced.  Each full (co)rep can be reduced through subduction to a set of irreducible small (co)reps at each ${\bf k}$ point that are known as the symmetry data [Fig.~\ref{fig:mainMTQC}(b)].  The complete set of small and full (co)reps of each MSG and direct dependencies between the site-symmetry (co)reps at ${\bf q}$ and the induced symmetry data at ${\bf k}$ are respectively accessible through the~\href{http://www.cryst.ehu.es/cryst/corepresentations}{Corepresentations}~(\url{http://www.cryst.ehu.es/cryst/corepresentations}, detailed in SA~\ref{sec:coreps}) and~\href{http://www.cryst.ehu.es/cryst/msitesym}{MSITESYM}~(\url{http://www.cryst.ehu.es/cryst/msitesym}, detailed in SA~\ref{sec:induction}) tools.  Lastly, to determine whether the bands that transform in the induced symmetry data are required by symmetry to be degenerate or cross along high-symmetry paths in the BZ, we have computed the magnetic small (co)rep compatibility relations, which are accessible through the~\href{https://www.cryst.ehu.es/cryst/mcomprel}{MCOMPREL} tool introduced in this work~(\url{https://www.cryst.ehu.es/cryst/mcomprel}, detailed in SA~\ref{sec:compatibilityRelations}).

Before discussing topological applications of the MEBRs and the small and full (co)reps of each MSG, we will first briefly discuss the advances made in this work in the context of previous studies of magnetic symmetry and group theory.  First, in the 1960's, Miller and Love in Ref.~\onlinecite{millerTables} performed the largest tabulation of magnetic small (co)reps prior to this work.  Specifically, in Ref.~\onlinecite{millerTables}, Miller and Love computed the single- and double-valued irreducible small (co)reps of the little groups of each MSG at high-symmetry points and along high-symmetry lines, but not along high-symmetry planes or in the BZ interior, which are required to complete the insulating compatibility relations for each MSG (SA~\ref{sec:compatibilityRelations}) and to compute the MEBRs (SA~\ref{sec:MEBRs}).  Additionally, the magnetic small (co)reps computed in Ref.~\onlinecite{millerTables} are displayed in difficult-to-read tables outputted directly from computer code, and are hence difficult to verify.  For this work, we have implemented the~\href{http://www.cryst.ehu.es/cryst/corepresentations}{Corepresentations} tool on the BCS [SA~\ref{sec:coreps}], which represents the first complete and publicly available online tabulation of the magnetic small (co)reps.  Through~\href{http://www.cryst.ehu.es/cryst/corepresentations}{Corepresentations}, users may obtain the matrix representatives in each magnetic small (co)rep of the generating symmetries of the magnetic little group at each ${\bf k}$ point in each MSG in an accessible format readily suited towards analyzing the output of tight-binding and first-principles calculations [see SA~\ref{sec:corepExampleNoAnti} and~\ref{sec:corepExampleYesAnti} for representative examples of the output of~\href{http://www.cryst.ehu.es/cryst/corepresentations}{Corepresentations}].  We additionally note that prior to this work, Evarestov Smirnov, and Egorov in Ref.~\onlinecite{EvarestovMEBR} introduced a method for obtaining the MEBRs of the MSGs and computed representative examples, but did not perform a large-scale tabulation of MEBRs or establish a connection to magnetic band topology.  In this work, we have employed a method equivalent to the procedure in Ref.~\onlinecite{EvarestovMEBR} to perform the first complete tabulation of the single- and double-valued MEBRs of the 1,421 MSGs  (see SA~\ref{sec:mbandrep}), which we have additionally made publicly accessible through the~\href{http://www.cryst.ehu.es/cryst/mbandrep}{MBANDREP} tool on the BCS.

Having computed the MEBRs of the single and double MSGs and established the theory of MTQC, we will next describe two applications of the MEBRs and MTQC to the discovery and characterization of novel topological phases of matter: elucidating the relationship between topological SMs and TCIs through symmetry-enhanced fermion doubling theorems, and extending the SIs of stable band topology~\cite{SlagerSymmetry,AshvinIndicators,ChenTCI,AshvinTCI,TMDHOTI} to the MSGs.

\textit{Symmetry-enhanced fermion doubling theorems} -- The surface states of each $d$-dimensional [$d$-D] TI and TCI are termed anomalous because the surface states cannot be stabilized in a $(d-1)$-D lattice model with the symmetries of the TI or TCI surface.  In 3D TIs, AXIs, and Chern (QAH) insulators, the boundary anomaly and bulk response can be understood from the perspective of well-known high-energy field theories~\cite{QHZ,VDBAxion,HaldaneModel}.   For example, the bulk of a 3D TI is characterized by a quantized axionic magnetoelectric response governed by a Lagrangian density $\mathcal{L}_{EM}\propto \theta{\bf E}\cdot {\bf B}$ in which the axion angle $\theta$ is pinned to the nontrivial value $\theta\text{ mod }2\pi=\pi$ by $\{\mathcal{T}|{\bf 0}\}$ symmetry~\cite{QHZ,VDBAxion}.  As a consequence of the bulk axionic topology, each surface of a 3D TI exhibits an odd number of twofold-degenerate Dirac cones, representing an exception to the 2D parity anomaly -- a fermion doubling theorem that mandates the existence of an even number of symmetry-stabilized twofold Dirac cones in any 2D system with a lattice (-regularized) description~\cite{FuKaneMele,FuKaneInversion,QHZ,VDBAxion,DiracInsulator}.  However in other gapped topological phases, such as 3D helical TCIs and HOTIs, the boundary anomalies and bulk response theories have not yet been elucidated in the language of high-energy field theory~\cite{DiracInsulator,HOTIBernevig,ChenRotation,AshvinTCI,TMDHOTI,WiederAxion}.  Nevertheless, as shown in Refs.~\onlinecite{DiracInsulator,ChenRotation,AshvinTCI}, the anomalous surface states of $d$-D TIs and TCIs may be classified through a comparison to the complete set of $(d-1)$-D lattice models of symmetry-stabilized topological SMs.

It is possible to evade a fermion doubling theorem by either stabilizing the anomalous nodal point[s] on the $(d-1)$-D boundary of a $d$-D topological [crystalline] insulator [\emph{i.e.} through spectral flow], or by modifying one of the system symmetries so that the symmetry is represented differently at low and high energies.  For example, the matrix representatives of $\{\mathcal{T}|{\bf 0}\}$ and $\{\mathcal{T}|{\bf a}/2\}$ are the same near ${\bf k}={\bf 0}$, but differ at larger ${\bf k}$ (see SA~\ref{sec:corepExampleYesAnti}).  In effect, systems with $\{\mathcal{T}|{\bf 0}\}$ symmetry and integer lattice translations are nonmagnetic (see SA~\ref{sec:type2}) and constrained by fermion doubling theorems that derive from $\{\mathcal{T}|{\bf 0}\}$ symmetry~\cite{DiracInsulator}, whereas systems generated by $\{\mathcal{T}|{\bf a}/2\}$ and integer lattice translations are antiferromagnetic (see SA~\ref{sec:type4}), and are not constrained by the same doubling theorems~\cite{SteveMagnet}.  As discussed in Ref.~\onlinecite{PotterNonLocalExceptions}, it is desirable to identify lattice-regularizable systems that circumvent fermion doubling theorems, because correlation effects in these systems can be modeled without also incorporating complicated and numerically intensive bulk degrees of freedom.  Many of the symmetry-enhanced fermion doubling theorems exceptions discovered to date rely on emergent unitary particle-hole symmetries that act nonlocally~\cite{PotterNonLocalExceptions,DamThanhSonCompositeDirac}, and relate to the anomalous surface states of particle-hole-symmetric TCIs in Class AIII in the nomenclature of Ref.~\onlinecite{KitaevClass}.  However, emergent unitary particle-hole is typically only a valid symmetry in a handful of solid-state materials, and only then at low energies.  As we will discuss below, by considering nodal degeneracies stabilized by MSG symmetries -- which are conversely valid in solid-state magnetic materials at all energies without fine tuning -- it is possible to systematically enumerate symmetry-enhanced, single-particle fermion doubling theorems, as well as materials-relevant models that circumvent symmetry-enhanced fermion doubling.

The elucidation of a (symmetry-enhanced) fermion doubling theorem and an example of its evasion has historically required a significant theoretical effort.  For example, in Ref.~\onlinecite{Steve2D}, it was shown that unpaired fourfold-degenerate Dirac fermions cannot be stabilized in lattice models of 2D, $\mathcal{T}$-symmetric SMs.  Through an exhaustive analysis of the symmetry-enforced spectral flow in 3D crystals, a 3D $\mathcal{T}$-symmetric TCI with an unpaired (anomalous), symmetry-stabilized, fourfold surface Dirac fermion was identified in Ref.~\onlinecite{DiracInsulator}.  Crucially, using the fourfold Dirac fermion doubling theorem established in Ref.~\onlinecite{Steve2D}, the authors of Ref.~\onlinecite{DiracInsulator} were able to diagnose the surface fourfold Dirac fermion as anomalous without establishing a bulk or boundary field theory.  Lastly, it was subsequently shown in Ref.~\onlinecite{SteveMagnet} that fourfold Dirac fermion doubling can also be evaded in lattice models of 2D magnetic SMs with the symmetry $\{\mathcal{T}|{\bf a}/2\}$ common to Type-IV 2D symmetry (wallpaper or layer) groups (see SA~\ref{sec:corepExampleYesAnti}).  Hence, one may infer the existence of novel quantized response effects and condensed-matter realizations of high-energy anomalies by exploiting the restrictions imposed by crystal symmetries on lattice models of SMs, TIs, and TCIs.

Because a complete tabulation of the magnetic small (co)reps was previously unavailable, then earlier theoretical searches for magnetic exceptions to fermion doubling theorems, such as Ref.~\onlinecite{SteveMagnet}, were performed \emph{ad hoc}.  However, the magnetic small (co)reps, the magnetic compatibility relations, and the MEBRs computed in this work allow, for the first time, the immediate enumeration of the complete set of lattice models of symmetry-stabilized magnetic SMs in three or fewer dimensions.  Below, we will outline the method for enumerating the complete set of stable magnetic SMs using the data generated in this work.  We will then detail the simplest possible magnetic fermion doubling exception that can be obtained by considering the set of lattice models of 1D magnetic SMs inferred from the 1D MEBRs.  Despite the simplicity of the example below, we find that it has not been addressed from the intuitive picture of mean-field magnetic band theory in previous literature.  In SA~\ref{sec:HOTIfermionDoubling}, we also introduce a doubling theorem for twofold Dirac fermions in magnetic 2D symmetry groups, which we find to be evaded on the surfaces of the non-axionic magnetic HOTIs discovered in this work (see SA~\ref{sec:HOTItbModel}).

To begin, by occupying the bands that transform in each connected branch of each MEBR with integer-valued numbers of electrons increasing from one to one less than the dimension of the MEBR (see Refs.~\onlinecite{QuantumChemistry,AndreiMaterials} and SA~\ref{sec:compatibilityRelations},~\ref{sec:mebrStats}, and~\ref{sec:EBCRdimension}), we have obtained the exhaustive list of connectivity-enforced 3D magnetic SMs.  The remaining stable 3D SMs can then be obtained through band inversion in lattice models constructed from sums of MEBRs (or branches of decomposable MEBRs, see SA~\ref{sec:mebrStats}) using the magnetic compatibility relations, as well as previously established topological invariants for nodal fermions at low-symmetry ${\bf k}$ points.  Specifically, in each MSG, the minimal multiplicity of stable nodal points may be obtained by considering the small (co)reps along all high-symmetry BZ lines and planes [which are accessible through~\href{http://www.cryst.ehu.es/cryst/corepresentations}{Corepresentations}, see SA~\ref{sec:coreps}], in addition to the nodal points stabilized by topological invariants evaluated along closed manifolds in the BZ (\emph{e.g.} Weyl points, see Refs.~\onlinecite{AshvinWeyl,AndreiWeyl,SYWeyl,WiederAxion,ChenRotation}).  Lastly, the complete set of 2D and 1D lattice models of magnetic SMs may be obtained by restricting the above procedure to MSGs that are isomorphic modulo integer lattice translations to layer and rod groups, respectively (see SA~\ref{sec:MSGs} and Refs.~\onlinecite{SteveMagnet,DiracInsulator,HingeSM}).

\begin{figure}[!t]
\centering
\includegraphics[width=\columnwidth]{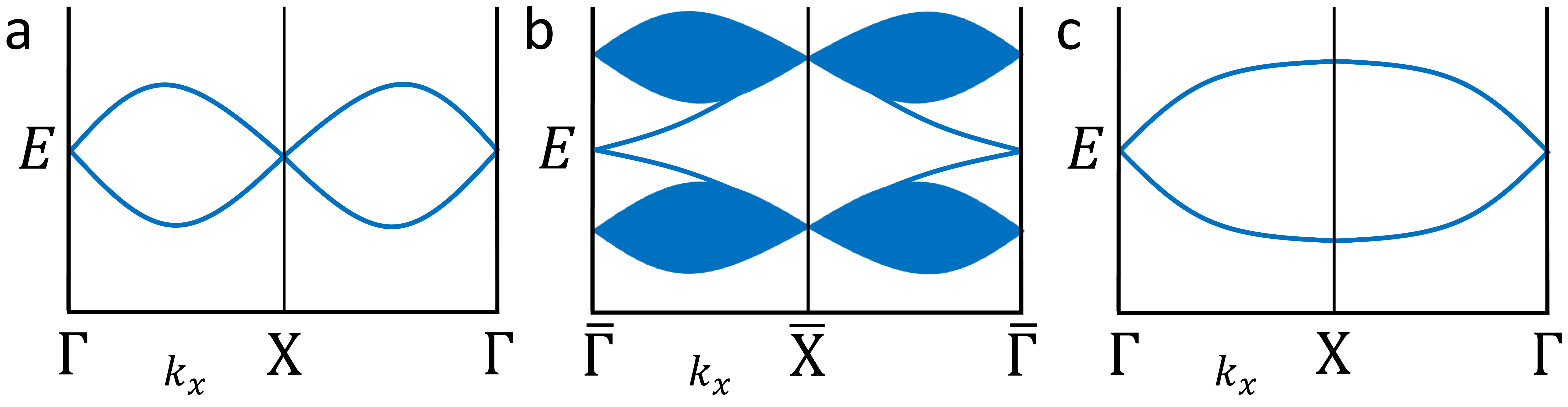}
\caption{Dirac fermion doubling from elementary band (co)representations.  (a) A pair of spinful bands that transform in the double-valued EBR of a Type-II line group generated by $\{\mathcal{T}|0\}$ and lattice translation [isomorphic to Type-II double SG 1.2 $P11'$ modulo lattice translations].  At half filling, there are two, twofold Dirac fermions in (a), representing an example of twofold Dirac fermion doubling in 1D.  (b) The edge spectrum of a 2D TI features an unpaired twofold Dirac fermion that circumvents the doubling theorem in (a)~\cite{CharlieTI,AndreiTI,QHZ}.  (c) A pair of spinful bands that transform in the double-valued MEBR of a Type-IV magnetic line group generated by $\{\mathcal{T}|1/2\}$ [isomorphic to Type-IV double MSG 1.3 $P_{S}1$ modulo lattice translations].  At half filling, the spectrum in (c) consists of an unpaired twofold Dirac fermion with the same $k\cdot p$ Hamiltonian as the Dirac points at $\Gamma$ and $X$ in (a) and the 2D TI edge in (b), representing a magnetic exception to twofold Dirac fermion doubling in 1D.}
\label{fig:mainFermion}
\end{figure}

\begin{figure*}[!t]
\centering
\includegraphics[width=\textwidth]{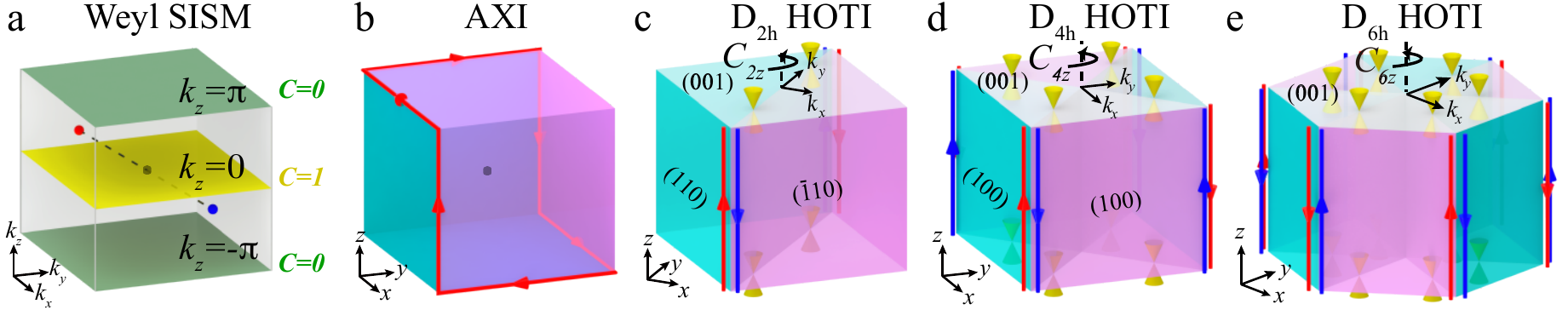}
\caption{The five families of 3D symmetry-indicated, spinful, strong topological phases.  In this work, we have computed the complete set of symmetry-indicated spinful topological phases of 3D magnetic and nonmagnetic crystalline solids (see SA~\ref{sec:topologicalBands}).  We find that, for spinful bands in 3D crystals that satisfy the insulating the compatibility relations along all high-symmetry lines and planes [see SA~\ref{sec:compatibilityRelations}], there are only five families of symmetry-indicated strong topological phases: (a) Smith-index Weyl SMs (Weyl SISMs), (b) axion insulators (AXIs) and 3D TIs defined by the nontrivial axion angle~\cite{QHZ,VDBAxion,WiederAxion,FuKaneMele,FuKaneInversion} $\theta=\pi$ [\emph{e.g.} MnBi$_2$Te$_4$~\cite{AXIExp1,AXIExp2}], (c) helical TCIs and higher-order TCIs (HOTIs) equivalent to two superposed AXIs with the same orbital hybridization and twofold rotation or rotoinversion symmetry [\emph{e.g.} bismuth~\cite{HOTIBismuth} and MoTe$_{2}$~\cite{TMDHOTI}], (d) helical TCIs and HOTIs equivalent to four superposed AXIs with the same orbital hybridization~\cite{HingeSM} and fourfold rotation or screw symmetry [\emph{e.g.} SnTe~\cite{HsiehTCI,HOTIBernevig}], and (e) helical TCIs and HOTIs equivalent to six superposed AXIs with the same orbital hybridization and sixfold rotation or screw symmetry.  Through the double SIs calculated for this work (Table~\ref{tb:minimalSIsMain} and SA~\ref{sec:34minimal} and~\ref{sec:summaryDoubleSIs}), we have discovered the existence of helical magnetic HOTIs with mirror-protected hinge states and bulk topology respectively enforced by the mirror and rotation symmetries of (c) double MPG 8.1.24 $mmm$ [\emph{i.e.} $D_{2h}$, see Ref.~\onlinecite{BigBook}], (d) double MPG 15.1.53 $4/mmm$ [$D_{4h}$], and (e) double MPG 27.1.100 $6/mmm$ [$D_{6h}$], where we have labeled MPGs using the notation of the~\href{http://www.cryst.ehu.es/cryst/corepresentationsPG}{CorepresentationsPG} tool (see SA~\ref{sec:magWannier}).  The magnetic HOTIs in (c-e) are respectively indicated by the minimal double SIs (c) $z_{4}=2$ in double MSG 47.249 $Pmmm$, (d) $z_{8}=4$ in double MSG 123.339 $P4/mmm$, and (e) $z_{12}=6$ in double MSG 191.233 $P6/mmm$ [as well as trivial values for all other independent minimal double SIs, see Table~\ref{tb:minimalSIsMain} and SA~\ref{sec:newHOTIs} for further details].}
\label{fig:mainHOTI}
\end{figure*}

In Fig.~\ref{fig:mainFermion}, we show the simplest example of a fermion doubling exception obtained using the MEBRs.  First, in Fig.~\ref{fig:mainFermion}(a), we show a pair of spinful bands in a nonmagnetic 1D crystal that transform in the double-valued EBR of the Type-II 1D double symmetry (line) group generated by $\{\mathcal{T}|0\}$ and lattice translation.  At half filling, the band structure in Fig.~\ref{fig:mainFermion}(a) exhibits two, twofold Dirac fermions per 1D BZ.  Additionally, in the absence of chiral symmetry -- which is not generically a symmetry of crystalline solids -- unpaired nodal points away from $\Gamma$ and $X$ in Fig.~\ref{fig:mainFermion}(a) cannot be stabilized.  Specifically, even if a nodal point stabilized by reflection or rotation symmetry is present at a point $k_{x}$, $\{\mathcal{T}|0\}$ symmetry mandates the existence of a second stable nodal point at $-k_{x}$.  By further investigating the symmetry-allowed band connectivities in all Type-II 1D (line and rod) supergroups of the line group in Fig.~\ref{fig:mainFermion}(a) (which can be inferred from the~\href{http://www.cryst.ehu.es/cryst/corepresentations}{Corepresentations},~\href{https://www.cryst.ehu.es/cryst/mcomprel}{MCOMPREL}, and~\href{http://www.cryst.ehu.es/cryst/mbandrep}{MBANDREP} tools in Table~\ref{tb:BCStoolsMain}), we conclude that an odd number of twofold Dirac fermions cannot be stabilized in 1D nonmagnetic, spinful lattice models.

However, it is well established that twofold Dirac fermion doubling in 1D is evaded on the edge of a 2D TI through spectral flow~\cite{CharlieTI,AndreiTI,QHZ} [Fig.~\ref{fig:mainFermion}(b)].  Recently, in Ref.~\onlinecite{MetlitskiAFMEdge}, the author performed an intensive, high-energy field-theory calculation demonstrating that a 1D lattice model with an unpaired twofold Dirac fermion could be formulated by invoking an exotic, non-on-site $\mathcal{T}$-like symmetry.  However, in this work, we recognize that a simpler, alternative interpretation of a non-on-site $\mathcal{T}$ symmetry is the antiferromagnetic (AFM) symmetry $\{\mathcal{T}|1/2\}$ common to all Type-IV magnetic line groups (SA~\ref{sec:type4}).  Correspondingly, in Fig.~\ref{fig:mainFermion}(c), we show a pair of spinful bands that transform in the double-valued MEBR of a Type-IV magnetic double line group generated by $\{\mathcal{T}|1/2\}$.  When the bands in Fig.~\ref{fig:mainFermion}(c) are half filled, the band structure features an unpaired twofold Dirac fermion with the same $k\cdot p$ Hamiltonian as the anomalous twofold Dirac fermion on the edge of a 2D TI [Fig.~\ref{fig:mainFermion}(b)].  Hence, the crystal in Fig.~\ref{fig:mainFermion}(c) represents a magnetic exception to twofold Dirac fermion doubling in 1D, analogous to the magnetic exception to fourfold Dirac fermion doubling in 2D demonstrated in Ref.~\onlinecite{SteveMagnet}.

\begin{table*}[t]
\centering
\begin{footnotesize}
\begin{tabular}{|c|c|c|c|c|c|c|}
\hline
\multicolumn{7}{|c|}{Minimal Double SIs of Spinful Band Topology the 1,651 Magnetic and Nonmagnetic Double SSGs} \\
\hline
SI & Minimal Double SSG(s) & Bulk Topology & $\ \ \ $ & SI & Minimal Double SSG(s) & Bulk Topology \\
\hline
\hline
$\eta_{4I}$ & 2.4 $P\bar{1}$ & WSM/QAH/AXI & & $z_{4m,\pi}^{\pm}$ & 83.43 $P4/m$ & weak TI/weak TCI \\
\hline
$z_{2I,i}$ & 2.4 $P\bar{1}$ & QAH & & $z_{4m,0}^{+}$ & 84.51 $P4_{2}/m$ & QAH/weak TI/weak TCI \\
\hline
$\eta_{2I}'$ & 2.4 $P\bar{1}$ & AXI & & $z_{8}$ & 83.44 $P4/m1'$, 123.339 $P4/mmm$ & AXI/TCI/HOTI \\
\hline
$z_{2R}$ & 3.1 $P2$, 41.215 $Ab'a'2$ & QAH & & $z_{3R}$ & 147.13 $P\bar{3}$ & QAH \\
\hline
$\delta_{2m}$ & 10.42 $P2/m$ & QAH/AXI/TCI & & $z_{6R}$ & 168.109 $P6$ & QAH \\
\hline
$z_{2m,\pi}^{\pm}$ & 10.42 $P2/m$ & QAH/weak TI/weak TCI & & $\delta_{3m}$ & 174.133 $P\bar{6}$ & QAH/AXI/TCI \\
\hline
$z_{4}$ & 2.5 $P\bar{1}1'$, 47.249 $Pmmm$, & AXI/TCI/HOTI & & $z_{3m,\pi}^\pm$ & 174.133 $P\bar{6}$ & weak TI/weak TCI \\
 & 83.45 $P4'/m$ & & & & & \\
\hline
$z_{4}'$ & 135.487 $P4_{2}'/mbc'$ & AXI/TCI & & $\delta_{6m}$ & 175.137 $P6/m$ & QAH/AXI/TCI  \\
\hline
$z_{2w,i}$ & 2.5 $P\bar{1}1'$, 47.249 $Pmmm$, & weak TI/weak TCI & & $z_{6m,\pi}^{\pm}$ & 175.137 $P6/m$ & weak TI/weak TCI  \\ 
 & 83.45 $P4'/m$ & & & & & \\
\hline
$z_{4R}$ & 75.1 $P4$ & QAH & & $z_{6m,0}^{+}$ & 176.143 $P6_{3}/m$ & QAH/weak TI/weak TCI \\
\hline
$z_{2R}'$, & 27.81 $Pc'c'2$, 54.342 $Pc'c'a$, & QAH & & $z_{12}$ & 175.138 $P6/m1'$, 191.233 $P6/mmm$ & AXI/TCI/HOTI \\
$z_{2R}''$ & 56.369 $Pc'c'n$, 60.424 $Pb'cn'$, & & & & & \\
& 77.13 $P4_{2}$, 110.249 $I4_{1}c'd'$ & & & & & \\
\hline
$z_{4S}$ & 81.33 $P\bar{4}$ & QAH & & $z_{12}'$ & 176.144 $P6_{3}/m1'$ & AXI/TCI/HOTI \\
\hline
$\delta_{2S}$ & 81.33 $P\bar{4}$ & WSM & & $z_{4R}'$ & 103.199 $P4c'c'$ & QAH \\
\hline
$z_{2}$ & 81.33 $P\bar{4}$ & AXI & & $z_{6R}'$ & 184.195 $P6c'c'$ & QAH \\
\hline
$\delta_{4m}$ & 83.43 $P4/m$ & QAH/AXI & & & & \\
\hline
\end{tabular}
\end{footnotesize}
\caption{The minimal double SIs of spinful band topology in all 1,651 double SSGs.  In order, this table contains the symbol of each double SI, the minimal double SSG(s) [\emph{i.e.} the lowest-symmetry SSG(s) in which the double SI predicts nontrivial band topology, see SA~\ref{sec:minimalSIProcedure} and~\ref{sec:minimalSSGTables}], and the bulk topological phase(s) associated to nontrivial values of the double SI.  All symmetry-indicated spinful SISM (specifically symmetry-indicated WSM), quantum anomalous Hall (QAH), TI, and TCI phases in magnetic and nonmagnetic crystalline solids necessarily exhibit nontrivial values of at least one of the double SIs listed in this table.  We note that, in this table, the symbol AXI refers to both magnetic AXIs and $\mathcal{T}$-symmetric 3D TIs, because AXI and 3D TI phases are both defined by the nontrivial bulk axion angle $\theta=\pi$ [Fig.~\ref{fig:mainHOTI}(b) and Refs.~\onlinecite{QHZ,VDBAxion,WiederAxion}]. Additionally, the symbols TCI and HOTI respectively indicate helical (\emph{i.e.} non-axionic) mirror Chern insulators~\cite{HsiehTCI} and HOTIs~\cite{HOTIBernevig,ChenRotation,DiracInsulator,ChenTCI,AshvinTCI}, which include the magnetic HOTIs in Fig.~\ref{fig:mainHOTI}(c-e) introduced in this work, as well as the nonmagnetic helical HOTI phases previously identified in bismuth~\cite{HOTIBismuth} and MoTe$_2$~\cite{TMDHOTI}.  Specific details of our SI calculations -- including explicit SI formulas, TI and TCI layer constructions, tight-binding models, and the minimal double SSG associated to each double SSG -- are provided in SA~\ref{sec:topologicalBands} and~\ref{sec:minimalSSGTables}.}
\label{tb:minimalSIsMain}
\end{table*}

\textit{Symmetry-based indicators of stable band topology in the 1,651 double SSGs} -- If a set of bands in a crystal is energetically isolated along all high-symmetry BZ lines and planes, then a subset of the topological properties of the bands may be inferred through the eigenvalues of unitary crystal symmetries.  Restricting focus to symmetry-indicated stable topological bands, which do not transform in integer-valued linear combinations of EBRs [see SA~\ref{sec:reviewTopologicalBands}], the crystal symmetry eigenvalues that indicate stable topology [encoded in the small (co)reps of the isolated bands, see SA~\ref{sec:coreps}] form the symmetry-based indicators (SIs) of stable band topology [see SA~\ref{sec:smithForm} and Refs.~\onlinecite{SlagerSymmetry,AshvinIndicators,ChenTCI,AshvinTCI,TMDHOTI}].  In each SSG, the SIs consist of an SI group (\emph{e.g.} $\mathbb{Z}_{4}\times\mathbb{Z}_{2}^{3}$) and an SI formula (\emph{e.g.} the Fu-Kane parity criterion for 3D TIs~\cite{FuKaneInversion}, see SA~\ref{sec:SIexP2} for an additional detailed example).  The complete SIs of spinful band topology in nonmagnetic 3D crystals -- which we term the double SIs of the 230 Type-II double SGs -- were previously computed in Refs.~\onlinecite{AshvinIndicators,ChenTCI,AshvinTCI}.  Following those works, the single and double SI groups in the 1,421 MSGs were computed in Ref.~\onlinecite{AshvinMagnetic}, but the authors of that work did not compute the SI formulas or determine the physical interpretation (\emph{i.e.} the bulk topology and anomalous boundary states) of the magnetic bands with nontrivial SIs [see Fig.~\ref{fig:mainResults}].

In this work, we have computed the complete set of double SI groups and formulas for spinful band topology in all 1,651 double SSGs.  We have further determined symmetry-respecting bulk and anomalous surface and hinge states for all nontrivial values of the double SIs.  The SI formulas introduced in this work (see SA~\ref{sec:34minimal} and~\ref{sec:summaryDoubleSIs}) have been unified into a consistent basis in which all previously identified nonmagnetic double SI formulas correspond to established nonmagnetic SM, TI, and TCI phases, and in which the SIs of symmetry-indicated TIs and TCIs with the same bulk topology (\emph{e.g.} 3D TIs and AXIs with the common nontrivial axion angle $\theta=\pi$) are related by intuitive SI subduction relations.  To summarize our calculation of the double SIs, we begin by considering a set of bands that is energetically isolated along all high-symmetry lines and planes, such that the Bloch states across all ${\bf k}$ points transform in small (co)reps that satisfy the insulating compatibility relations [see SA~\ref{sec:compatibilityRelations}].  If the bands exhibit nontrivial SIs, then the bands cannot be inverse-Fourier-transformed into exponentially localized, symmetric Wannier orbitals.  This can be seen by recognizing that the set of bands does not transform in an integer-valued linear combination of EBRs.  Consequently, the set of bands either forms a topological semimetal with nodal points in the BZ interior -- which we term a Smith-index SM (SISM), or corresponds to a stable TI or TCI phase with anomalous 2D surface or 1D hinge states~\cite{CharlieTI,AndreiTI,FuKaneMele,FuKaneInversion,LiangTCIOriginal,HsiehTCI,AshvinIndicators,HourglassInsulator,DiracInsulator,WladTheory,HOTIBernevig,ChenRotation,SlagerSymmetry,ChenTCI,AshvinTCI,HOTIBismuth,TMDHOTI,HingeSM}.

Because there are 1,651 double SSGs, then individually calculating the bulk and anomalous surface and hinge states and physical basis for each nontrivial SI in each double SSG is a practically intractable task.  However, in this work, we have reduced the size of the calculation by recognizing that the double SIs in each double SSG $G$ continue to exhibit unique, nontrivial values -- termed the minimal double SIs -- when the SI topological bands in $G$ are subduced onto a double SSG $M$ from the considerably smaller subset of 34 minimal double SSGs.  In SA~\ref{sec:minimalSIProcedure}, we rigorously detail the procedure for obtaining the minimal double SIs, and in SA~\ref{sec:minimalSSGTables}, we list the minimal double SSG associated to each double SSG.  Across all of the minimal double SIs, we have implemented a consistent physical basis for the SI formulas, determined symmetry-respecting topological bulk and boundary states, and formulated layer constructions of the stable TI and TCI phases -- the minimal double SIs are summarized in Table~\ref{tb:minimalSIsMain} and the details of our SI calculations are provided in SA~\ref{sec:topologicalBands}.

Using the subduction relations and layer constructions contained in SA~\ref{sec:34minimal}, we have determined by direct computation that, for spinful bands in 3D crystals, all symmetry-indicated topological phases are either strong topological Weyl SISMs, AXIs, 3D TIs, helical TCIs or HOTIs, or can be deformed into weak stacks of 2D TIs, mirror TCIs, or Chern insulators with nonzero net Chern numbers in each unit cell [termed QAH states].  Curiously, we find that there are no Type-IV minimal double SSGs (SA~\ref{sec:minimalSSGTables}).  This implies that symmetry-indicated spinful SISM, TI, and TCI phases in Type-IV MSGs are actually protected by the symmetries of Type-I or Type-III double MSGs, as opposed to the symmetry $\{\mathcal{T}|{\bf a}/2\}$ common to Type-IV MSGs [though, as shown in Fig.~\ref{fig:mainFermion}(c) and in Ref.~\onlinecite{SteveMagnet}, there exist topological SM phases unique to Type-IV MSGs].  For example, in Ref.~\onlinecite{AFMMooreTI}, the authors introduced $\mathcal{I}$-symmetric AFM TCIs in which $\theta=\pi$ was enforced by the symmetry $\{\mathcal{T}|{\bf a}/2\}$ common to all Type-IV MSGs.  However, we have shown that the spinful, symmetry-indicated TCI phases in Type-IV MSGs can be subduced onto Type-I or Type-III MSGs without closing a gap or changing the bulk topology.  Hence, the symmetry-indicated AFM TCIs introduced in Ref.~\onlinecite{AFMMooreTI} can more simply be understood as $\mathcal{I}$-symmetry-enforced AXIs that remain topological when subduced onto the minimal Type-I double MSG 2.4 $P\bar{1}$.  Through the layer constructions and double SI dependencies in SA~\ref{sec:34minimal} and~\ref{sec:minimalSSGTables}, we have also demonstrated that all of the 3D symmetry-indicated spinful magnetic TCIs with odd numbers of chiral modes on crystal hinges (edges) in the 1,421 double MSGs exhibit the nontrivial axion angle $\theta=\pi$, and are therefore AXIs~\cite{QHZ,VDBAxion,WiederAxion}.  Specifically, we find that all of the symmetry-indicated, spinful magnetic TCIs with chiral hinge states are AXIs in which $\theta=\pi$ is either quantized by $\mathcal{I}$, or by one of the rotoinversion symmetries $C_{4z}\times\mathcal{I}$ or $C_{6z}\times\mathcal{I}$ (see Table~\ref{tb:minimalSIsMain}).  This result is not necessarily intuitive -- for example, when cut into a rod with the same point group symmetry as the bulk MSG, an $\mathcal{I}$-symmetric AXI in Type-I double MSG 2.4 $P\bar{1}$ exhibits two chiral hinge states, whereas a $C_{4z}\times\mathcal{T}$-symmetric AXI in Type-III double MSG 83.45 $P4'/m$ exhibits four chiral hinge states; nevertheless, as shown in SA~\ref{sec:34minimal}, both AXI phases exhibit $\theta=\pi$.  We additionally note that there do not exist symmetry-indicated, spinful magnetic TCIs with even numbers of intrinsic copropagating chiral hinge states (though magnetic TCIs with mirror symmetry may in principle exhibit copropagating chiral hinge modes, depending on the bulk mirror Chern numbers and boundary termination details).

Overall, across the 1,651 double SSGs, we find that there are only five families of 3D symmetry-indicated, spinful, strong topological phases [Fig.~\ref{fig:mainHOTI}]: Weyl SISMs, AXIs and 3D TIs, and helical TCIs and HOTIs with twofold, fourfold, and sixfold symmetries.  We note that helical TCIs and HOTIs in particular exhibit trivial axion angles $\theta\text{ mod }2\pi=0$, and are therefore non-axionic.  In this work, we have discovered three novel variants of non-axionic magnetic HOTIs, which are shown in Fig.~\ref{fig:mainHOTI}(c-e).  Further details for the non-axionic HOTIs in Fig.~\ref{fig:mainHOTI}(c-e), including symmetry-enhanced fermion doubling theorems~\cite{DiracInsulator,ChenRotation} and tight-binding models, are provided in SA~\ref{sec:newHOTIs}.  When cut into the finite nanorod geometries shown in Fig.~\ref{fig:mainHOTI}(c-e), the non-axionic magnetic HOTIs exhibit helical, mirror-protected hinge states.  We note that, if the mirror-symmetric HOTI hinges in Fig.~\ref{fig:mainHOTI}(c-e) were sanded to expose mirror-symmetric 2D surfaces, each surface would exhibit two anomalous, mirror-protected, twofold Dirac cones, analogous to the mirror-protected helical hinge states of SnTe discussed in Ref.~\onlinecite{HOTIBernevig}.  Lastly, we emphasize that the magnetic HOTIs in Fig.~\ref{fig:mainHOTI}(c,e) exhibit the same nontrivial double SI $z_{4}=2$ as $\mathcal{T}$-symmetric helical HOTI phases in supergroups of Type-II double SG 2.5 $P\bar{1}1'$ (see Table~\ref{tb:minimalSIsMain} and Refs.~\onlinecite{HOTIBismuth,TMDHOTI,AndreiMaterials,ChenMaterials,AshvinMaterials}).  Unlike for AXIs and 3D TIs~\cite{QHZ,VDBAxion,FuKaneMele,FuKaneInversion}, the bulk response theories of helical HOTIs have not yet been elucidated.  In light of recent experiments demonstrating incipient signatures of helical higher-order topology in bismuth crystals~\cite{HOTIBismuth} and MoTe$_2$~\cite{PhuanOngMoTe2Hinge}, the absence of a response theory for helical HOTIs analogous to axion electrodynamics~\cite{QHZ,VDBAxion} has become an urgent issue.  The discovery in this work of helical magnetic HOTI phases whose bulk topology is solely enforced by the combination of unitary (spinful) mirror and rotation symmetries should provide crucial insight towards the elucidation of quantized response effects in helical HOTIs.

 {
\vspace{0.08in}
\centerline{\bf Discussion}
\vspace{0.08in}
}

The theory of MTQC can also be applied to a wide variety of problems beyond the topological applications highlighted in this work.  Most notably, while we have enumerated the spinful stable topological phases with nontrivial double SIs, the analogous enumeration of spinless magnetic SISMs and TCIs with nontrivial single SIs remains an open problem.  In particular, whereas bosonic, symmetry-indicated AXI phases protected by $\mathcal{I}$ and $\text{SU}(2)$ spin-rotation symmetry have been demonstrated in previous works~\cite{AshvinMagnetic,TMDHOTI}, it remains an open question whether there exist symmetry-indicated, non-axionic spinless (bosonic) TCIs.  Additionally, while we have restricted consideration to single-particle topological phases, the magnetic (co)reps computed in this work can also be used to characterize correlated systems, including spin (-orbital) liquids~\cite{DeconfinedQCPSenthil} and multipole tensor gauge theories~\cite{GromovFractonPRX}.  For example, if a correlated magnetic insulator admits a mean-field slave-rotor description~\cite{FlorensGeorgesSlaveRotor}, then the effective Hamiltonian of each quasiparticle species, such as spinon and chargeon degrees of freedom~\cite{PesinBalentsSlaveSpinonMott}, can separately be analyzed with MTQC.

 {
\vspace{0.08in}
\centerline{\bf Data Availability}
\vspace{0.08in}
}

The data supporting the findings of this study are available within the paper and through the BCS applications listed in Table~\ref{tb:BCStoolsMain}.  Additional information regarding the data generated for this study is available from the corresponding authors upon reasonable request.

 {
\vspace{0.08in}
\centerline{\bf Acknowledgments}
\vspace{0.08in}
}

$^\dag$Corresponding author: \url{bwieder@mit.edu} (B. J. W.), \url{bernevig@princeton.edu} (B. A. B.).  $^\ddag$Primary address.  We thank Mois I. Aroyo, Jennifer Cano, Claudia Felser, Nicolas Regnault, Maia G. Vergniory, and Zhijun Wang for crucial discussions during the early stages of this project.  B. J. W., B. B., and B. A. B. acknowledge the hospitality of the Donostia International Physics Center, where parts of this work were carried out.  The analytic calculations performed for this work were supported by the Department of Energy Grant No. DE-SC0016239.  B. J. W., Z. S., and B. A. B. were further supported by NSF EAGER Grant No. DMR 1643312, NSF-MRSEC Grant Nos. DMR-2011750 and DMR-142051, Simons Investigator Grant No. 404513, ONR Grant Nos. N00014-14-1-0330 and N00014-20-1-2303, the Packard Foundation, the Schmidt Fund for Innovative Research, the BSF Israel US Foundation Grant No. 2018226, the Gordon and Betty Moore Foundation through Grant No. GBMF8685 towards the Princeton theory program, and a Guggenheim Fellowship from the John Simon Guggenheim Memorial Foundation.  L. E. was supported by the Government of the Basque Country (Project IT1301-19) and the Spanish Ministry of Science and Innovation (PID2019-106644GB-I00).  L. E. and B. A. B. acknowledge additional support through the ERC Advanced Grant Superflat, and Y. X. and B. A. B. received additional support from the Max Planck Society.  B. B. acknowledges the support of the Alfred P. Sloan Foundation and the National Science Foundation Grant No. DMR-1945058.  Concurrently with the preparation of this work, the theory of MTQC was employed to perform a high-throughput search for magnetic topological materials~\cite{MTQCmaterials}.  Additionally, after the submission of this work, the authors of Ref.~\onlinecite{SpinSGMagTIs} used the group theory of magnetic spin space groups to analyze topological phases in crystals with commensurate magnetic order and negligible spin-orbit coupling.  Lastly, after the submission of this work, the authors of Ref.~\onlinecite{ChenMagneticCrystalSIs} computed the complete topological crystal constructions of all gapped spinful topological phases in all 1,421 double MSGs, related the resulting topological crystals to the magnetic SIs in each MSG, and deduced the spinful SISM phases in each MSG.  We have confirmed complete agreement between the calculations performed in Ref.~\onlinecite{ChenMagneticCrystalSIs} and the magnetic SIs and topological phases introduced in this work.

 {
\vspace{0.08in}
\centerline{\bf Author Contributions}
\vspace{0.08in}
}

All authors contributed equally to the intellectual content of this work.  The magnetic small coreps and MEBRs were computed by L. E.  The BCS tools for accessing the symmetry and group theory data generated in this study were implemented by L. E.  The topological analysis of the magnetic small coreps and MEBRs was performed by B. J. W. under the supervision of B. B. and B. A. B.  The double SIs of the 1,651 double SSGs were computed by L. E. and B. A. B. with help from B. B.  The physical interpretation of the double SIs was determined by B. J. W.  The double SIs were unified into a self-consistent, physical basis by Z. S. with help from B. J. W., L. E., and B. B.  Layer constructions for all symmetry-indicated magnetic TI, TCI, and HOTI phases were computed by Z. S.  Symmetry-enhanced fermion doubling theorems for the helical magnetic HOTIs were formulated by B. J. W. and Z. S.  Tight-binding calculations for the helical magnetic HOTIs were performed by Y. X. with help from Z. S. and B. J. W.  The manuscript was written by B. J. W. with help from all of the authors.  B. A. B. was responsible for the overall research direction.

 {
\vspace{0.08in}
\centerline{\bf Competing Interests}
\vspace{0.08in}
}

The authors declare no competing interests.

\clearpage
\onecolumngrid
\begin{appendix}

\begin{center}
{\bf Supplementary Appendices for ``Magnetic Topological Quantum Chemistry''}
\end{center}

\tableofcontents

\section{Introduction to Supplementary Appendices}
\label{sec:appendixIntro}

In this supplement, we provide proofs and tables that extend Topological Quantum Chemistry (TQC)~\cite{QuantumChemistry,Bandrep1,Bandrep2,Bandrep3,JenFragile1,BarryFragile} to the magnetic space groups (MSGs), to develop a complete theory of Magnetic Topological Quantum Chemistry (MTQC).  MTQC provides, for the first time, a predictive, position-space formulation of the characteristics of band structures -- including stable and fragile topology -- in all translationally invariant crystalline solids that are characterized by mean-field theory with a static background magnetic field.  Most relevant to the physical systems studied in this work, MTQC provides tools for characterizing the symmetry and topology of electronic states in solid-state materials with lattice-commensurate magnetism.  We begin in Appendix~\ref{sec:MSGs} by precisely defining the MSGs, drawing connection where possible to the more familiar nonmagnetic space groups (SGs).  We then discuss in Appendix~\ref{sec:siteSymmetry} the Wyckoff positions of the MSGs, whose sites are left invariant under the symmetries of site-symmetry groups that are necessarily isomorphic to crystallographic magnetic point groups (MPGs)~\cite{ShubnikovMagneticPoint,BilbaoPoint,PointGroupTables,MagneticBook,EvarestovBook,EvarestovMEBR,BCS1,BCS2,BCSMag1,BCSMag2,BCSMag3,BCSMag4,ArisLandauLevel}.  Next, in Appendix~\ref{sec:momentumSpace}, we introduce crystal momentum ${\bf k}$ in the MSGs, and discuss how spatial and magnetic symmetries are represented in momentum space.  To enumerate the set of symmetry-independent ${\bf k}$ points in each MSG and SG, we have implemented the~\href{http://www.cryst.ehu.es/cryst/mkvec}{MKVEC} tool (further detailed in Appendix~\ref{sec:MKVEC}), which is now available on the Bilbao Crystallographic Server (BCS)~\cite{BCS1,BCS2}.  In Appendix~\ref{sec:coreps}, we then describe how, in this work, we have for the first time derived the complete set of irreducible [small] little group and full [space group] (co)representations [(co)reps] of the MSGs, which can now be accessed on the BCS~\cite{BCS1,BCS2} through the~\href{http://www.cryst.ehu.es/cryst/corepresentations}{Corepresentations} tool [further detailed in Appendices~\ref{sec:corepExampleNoAnti} and~\ref{sec:corepExampleYesAnti}].  Lastly, by combining the results of Appendices~\ref{sec:MKVEC} and~\ref{sec:coreps}, we then in Appendix~\ref{sec:compatibilityRelations} derive the compatibility relations between small (co)reps in the MSGs, which we have made accessible through the~\href{https://www.cryst.ehu.es/cryst/mcomprel}{MCOMPREL} tool on the BCS.

Having established position- and momentum-space characterizations of the MSGs, we then in Appendix~\ref{sec:MEBRs} complete the theory of MTQC by enumerating the magnetic elementary band (co)representations [MEBRs]~\cite{ZakBandrep1,ZakBandrep2,EvarestovBook,EvarestovMEBR,BarryBandrepReview,QuantumChemistry,Bandrep1,Bandrep2,Bandrep3,JenFragile1,BarryFragile}, which represent all possible [magnetic] trivial atomic limits.  To obtain the MEBRs, we first in Appendix~\ref{sec:magWannier} introduce the minimal magnetic atomic orbitals [\emph{e.g.} $p_{x}+ ip_{y}$] that correspond to the (co)reps of the magnetic site-symmetry groups, which are isomorphic to MPGs.  In Appendix~\ref{sec:magWannier}, we additionally detail the~\href{http://www.cryst.ehu.es/cryst/corepresentationsPG}{CorepresentationsPG} tool on the BCS, which we have implemented for this work to provide access to the (co)reps of the magnetic site-symmetry groups of the MSGs.  Next, in Appendix~\ref{sec:induction}, we establish the central machinery of MTQC through which band (co)representations [band (co)reps] in momentum space are induced from site-symmetry (co)reps in position space.  We also introduce and detail in Appendix~\ref{sec:magWannier} the~\href{http://www.cryst.ehu.es/cryst/msitesym}{MSITESYM} tool, through which users may access the small (co)reps subduced from each band (co)rep of each SSG.  Finally, in Appendix~\ref{sec:mbandrep}, we complete the derivation of MTQC by enumerating the MEBRs.  In Appendix~\ref{sec:mebrStats}, we additionally detail the~\href{http://www.cryst.ehu.es/cryst/mbandrep}{MBANDREP} tool on the BCS, which we have developed for this work to compute and display both the elementary and non-elementary [\emph{i.e.} composite] band (co)reps of the MSGs.

The theory of MTQC uniquely enables us to, for the first time, enumerate all of the symmetry-based indicators of band topology (SIs)~\cite{SlagerSymmetry,AshvinIndicators,HOTIChen,ChenTCI,AshvinTCI,TMDHOTI,BarryBandrepReview,ZhidaSemimetals,AdrianSIReview,AdrianSCMagSI} [\emph{i.e.} generalized Fu-Kane-like symmetry-eigenvalue topological indices~\cite{FuKaneInversion}] for the double-valued (co)reps of the 1,651 spinful [double] magnetic and nonmagnetic space groups [SSGs].  Specifically, a (co)rep is respectively defined as single- or double-valued if the matrix representatives of time-reversal and rotation symmetries in the (co)rep square to plus or minus the identity~\cite{BigBook}.  Double groups have both single- and double-valued (co)reps, whereas single groups only have single-valued (co)reps.  Electronic [fermionic] states in solid-state materials are generically characterized by double-valued (co)reps of double symmetry groups, though in the absence of spin-dependent interactions [\emph{e.g.} spin-orbit coupling (SOC)], spin-degenerate electronic states may be labeled with single-valued (co)reps.  In Appendix~\ref{sec:topologicalBands}, we compute the SI groups and formulas for all symmetry-indicated, spinful, mean-field topological phases in the 1,651 double SSGs.  We specifically demonstrate in Appendix~\ref{sec:minimalSIProcedure} how the SI calculation can be reduced by recognizing that the SIs in all 1,651 double SSGs are dependent on the SIs in a considerably smaller subset of minimal double SSGs.  Through the minimal SI calculation, which is provided in explicit detail in Appendix~\ref{sec:34minimal}, we discover several novel, helical, magnetic higher-order topological crystalline insulators (HOTIs)~\cite{HOTIBernevig,HOTIBismuth,ChenRotation,HOTIChen,HigherOrderTIPiet,AshvinIndicators,ChenTCI,AshvinTCI,TMDHOTI,WiederAxion,DiracInsulator} whose bulk response theories do not correspond to axion electrodynamics~\cite{WilczekAxion,QHZ,VDBAxion,AndreiInversion,AshvinAxion,WiederAxion,YuanfengAXI,NicoDavidAXI1,NicoDavidAXI2,TMDHOTI,KoreanAXI,CohVDBAXI,ArisHopf,TitusRonnyKondoAXI,YoungkukMonopole,MurakamiAXI1,MurakamiAXI2,HingeStateSphereAXI,BJYangVortex,BarryBenCDW,IvoAXI1,IvoAXI2,GuidoAXI}.  The \emph{non-axionic} magnetic HOTIs discovered in this work are further detailed in Appendix~\ref{sec:newHOTIs}.  Lastly, in Appendix~\ref{sec:supplementaryTables}, we provide supplementary tables of additional data generated for this work.

\section{Magnetic Space Groups}
\label{sec:MSGs}

In this section, we list the basic group theoretic properties of the magnetic space groups (MSGs).  To begin, it was established in Refs.~\onlinecite{ZamorzaevMSG,BelovMSG} (and translated into English in Ref.~\onlinecite{ShubnikovBook}) that the Hamiltonians of 3D, periodic systems (\emph{i.e.} crystalline solids) without particle-hole symmetry are invariant under the symmetries contained in at least one of the 1,651 Shubnikov space groups (SSGs).  All of the SSGs contain the group of fundamental lattice translations:
\begin{equation}
G_{T} = T_{a} \otimes T_{b} \otimes T_{c},
\label{eq:translationGroup}
\end{equation}  
where $T_{i}$ is the group comprised of the set of lattice translations $t_{i}^{n}$, where $n\in\mathbb{Z}$ and:
\begin{equation}
t_{i} = \{E|{\bf t}_{i}\},
\label{eq:translationNotation}
\end{equation}
where $E$ is the identity operation.  Throughout this work, we will employ a notation [Eq.~(\ref{eq:translationNotation})] in which $t_{i}$ is the symmetry operation of a translation by the vector ${\bf t}_{i}$.  In Eq.~(\ref{eq:translationGroup}), the generating translations $t_{a,b,c}$ must be linearly independent, but are not necessarily orthogonal (though $t_{a,b,c}$ are indeed both linearly independent and orthogonal in many SSGs).

The 1,651 SSGs subdivide into four types, which are distinguished by their antiunitary symmetries~\cite{ZamorzaevMSG,BelovMSG,ShubnikovBook,BigBook,ChenNewPTTheory}.  Of the four types of SSGs, the 1,421 Types-I, III, and IV SSGs characterize magnetic crystals (\emph{i.e.} crystals with lattice-commensurate magnetic order); hence, in this work, we interchangeably denote Type-I, III, and IV SSGs as MSGs or SSGs.  Conversely, the 230 Type-II SSGs exclusively characterize nonmagnetic crystals; hence, in this work, we interchangeably refer to Type-II SSGs as SGs or SSGs.  Unlike in other recent works on magnetic symmetry and topology~\cite{AshvinMagnetic,MagneticBieberbach}, we will not refer to the 230 Type-II groups as MSGs, to avoid employing terminology in which the Type-II SSGs are ``nonmagnetic magnetic SGs.''  All SSGs (MSGs and SGs) are given in the notation established in Ref.~\onlinecite{BelovNotation} and reproduced on the Bilbao Crystallographic Server (BCS)~\cite{BCS1,BCS2}.  Because the set of possible $G_{T}$ in Eq.~(\ref{eq:translationGroup}) coincides with the 14 3D nonmagnetic (gray) Bravais lattices, then all 1,651 SSGs can be characterized by the 14 Bravais lattices.  However, as we will detail in Appendix~\ref{sec:type4}, the Type-IV groups -- which contain elements of the form $\mathcal{T}(t_{i}/2)$ where $\mathcal{T}$ is the operation of time-reversal -- are also frequently characterized using the 22 ``black and white'' Bravais lattices that account for the relative positions of localized spins (or classical magnetic moments) [\emph{c.f.} Chapter 7 in Ref.~\onlinecite{BigBook}].  In this work, we will refer to all 1,651 SGs by their nonmagnetic (gray) Bravais lattice (\emph{i.e.}, the Bravais lattice of their primitive, or ``magnetic'' unit cell).  This choice of Bravais lattice is naturally incorporated into the numbering and notation of Belov, Nerenova, and Smirnova~\cite{BelovNotation} (labeled as the ``BNS setting'' on the BCS), which we will employ throughout this work.  For generality and connection with other works, we also note that on the BCS, information about the SSGs can alternatively be obtained in the convention of Opechowski and Guccione~\cite{OGSetting} (labeled as the ``OG setting'' on the BCS); we will not employ, or further discuss, the OG setting in this work.

It is important to highlight the distinction between MSGs and phenomenological descriptions of magnetic order.  Specifically, while all magnetic crystals with Type-IV MSGs are antiferromagnets (see Appendix~\ref{sec:type4}), there are both ferromagnets and antiferromagnets with Type-I or Type-III MSGs~\cite{BigBook} (Appendices~\ref{sec:type1} and~\ref{sec:type3}, respectively).  For each of the three types of MSGs, we will below provide representative examples of quasi-1D chains with symmetry-allowed magnetic ordering, including phenomenologically distinct magnetic order in crystals with the same Type-I or Type-III MSG (see Appendices~\ref{sec:type1} and~\ref{sec:type3}, respectively).  Each of the quasi-1D chains shown below is invariant under a crystallographic magnetic rod group (MRG)~\cite{BigBook,MagneticBook,ITCA,subperiodicTables,HingeSM} $M_{RG}$, \emph{i.e.} a subperiodic group with 3D symmetry operations and 1D translations.  Each MRG is isomorphic to an SSG $M$ under the addition of in-plane lattice translations, where the group-subgroup relations between $M_{RG}$ and $M$ depend on the details of the additional translations.  For example, when translations in the $xy$-plane are added to Type-I MRG $(p4_{2}cm)_{RG}$, the resulting MSG is either Type-I MSG 101.179 $P4_{2}cm$ or Type-I MSG 105.211 $P4_{2}mc$, depending on whether the shortest lattice translations are respectively added in the $x$ and $y$ or $x\pm y$ directions~\cite{ITCA,subperiodicTables}.  In this work, we will refer to quasi-1D chains and rods using the terminology established in Refs.~\onlinecite{HingeSM,TMDHOTI,WiederAxion} in which a chain or rod with the translation symmetry $t_{c}=\{E|c\}$ is termed \emph{$c$-directed}.  The symbols for the MRGs referenced in this work are given in the convention employed by Litvin in Ref.~\onlinecite{MagneticBook}.

\subsection{Type-I SSGs: Ordinary (Fedorov) Groups (230 MSGs)}
\label{sec:type1}

\begin{figure}[h]
\includegraphics[width=0.7\columnwidth]{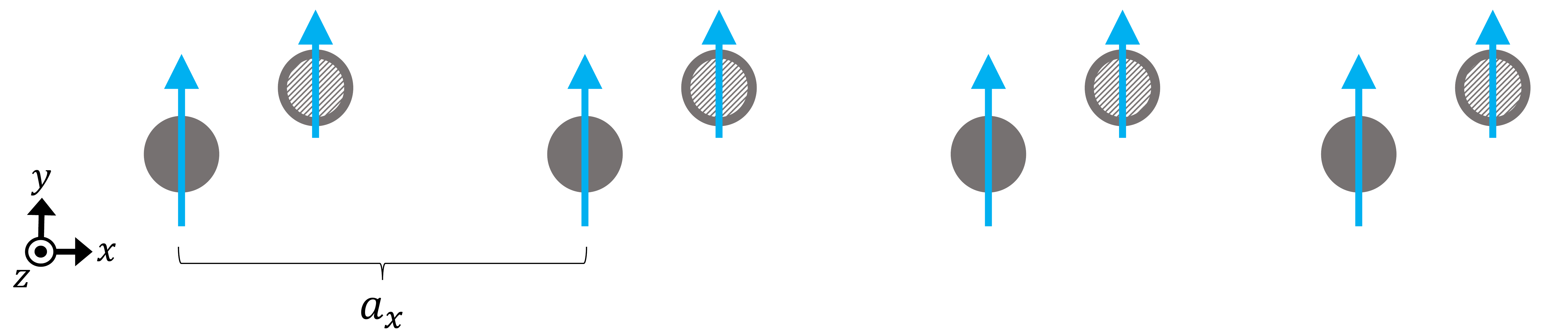}
\caption{A ferromagnetic chain with MRG $(p1)_{RG}$, which is generated by $\{E|1\}$ ($t_{x}$) where $E$ is the identity operation, and is isomorphic after the addition of perpendicular lattice translations (\emph{e.g.} $t_{y}$ and $t_{z}$) to Type-I MSG 1.1 $P1$.  There are two atoms within each unit cell, where the right-most atom in each cell (hashed circle) exhibits a weaker $y$-directed magnetic moment than the left-most atom (solid circle), lies away from $x=a_{x}/2$, and is displaced from the $xy$-plane ($z\neq 0$ for the hashed atoms).  If there was just one atom in each unit cell, if the solid and hashed atoms were moved to be coplanar, or if the magnetic moments were tuned to be the same magnitude, then the chain would respect additional symmetries, such as $\{m_{z}\times\mathcal{T}|0\}$.}
\label{fig:type1FM}
\end{figure}

\begin{figure}[h]
\includegraphics[width=0.7\columnwidth]{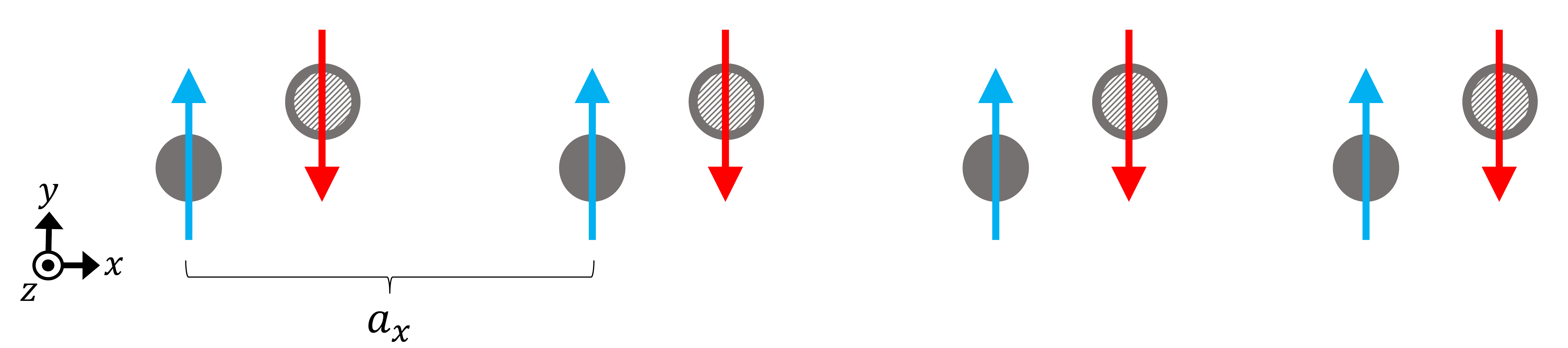}
\caption{An antiferromagnetic chain with MRG $(p1)_{RG}$, which is generated by $\{E|1\}$, and is isomorphic after the addition of perpendicular lattice translations to Type-I MSG 1.1 $P1$.  The solid and hashed circles represent magnetic atoms with distinct chemical environments [\emph{e.g.} atoms of the same species with different oxidation states or on-site (chemical) potentials] and equal and opposite magnetic moments.  The right-most atom in each cell (\emph{i.e.} the hashed atom with a red magnetic moment) lies away from $x=a_{x}/2$ and $z=0$, such that the solid and hashed atoms are neither equally spaced nor coplanar.  If the chemical environments of the solid and hashed atoms were tuned to be equivalent, if the solid and hashed atoms were moved to be coplanar, or if the atoms were shifted to be separated by a distance $a_{x}/2$ in the $x$-direction, then the chain would have additional symmetries.  For example, if the local chemical environment (\emph{i.e.} hoppings and on-site potentials) of the solid and hashed atoms were made equivalent, then the chain would respect $\{m_{z}\times\mathcal{T}|0\}$ symmetry (as well as additional symmetries), and if the atoms tuned to lie in equivalent chemical environments and shifted to be equally spaced and coplanar, then the chain would respect both $\{C_{2z}|0\}$ and $\{m_{y}\times\mathcal{T}|1/2\}$ symmetry (as well as additional symmetries).}
\label{fig:type1AFM}
\end{figure}

Each Type-I SSG $M_{I}$ is exclusively characterized by a set of unitary symmetry operations.  The simplest Type-I SSG -- MSG 1.1 $P1$ -- is isomorphic to $G_{T}$ [Eq.~(\ref{eq:translationGroup})], and is a common subgroup of all 1,651 SSGs.  The Type-I MSGs have historically been termed the \emph{ordinary groups}~\cite{BigBook}, because Type-I magnetic symmetry groups do not contain antiunitary symmetries that relate classical magnetic moments at different positions in a crystal.  Type-I MSGs can characterize a variety of magnetic configurations~\cite{BigBook}.  For example, Type-I MSG 1.1 $P1$ can characterize crystals with either ferromagnetism (Fig.~\ref{fig:type1FM}) or antiferromagnetism (Fig.~\ref{fig:type1AFM}).

\subsection{Type-II SSGs: Gray (Nonmagnetic) Groups (230 SSGs)}
\label{sec:type2}

Each Type-II SSG $M_{II}$ takes the form:
\begin{equation}
M_{II} = G \cup \{\mathcal{T}|000\}G = G\cup \mathcal{T}G,
\label{eq:type2}
\end{equation}
where $G$ is isomorphic to a Type-I SSG.  Because each Type-II SSG contains the element $\{\mathcal{T}|000\}$, then no position in the unit cell of a crystal with a Type-II SSG can host a local magnetic moment.  Therefore, crystals invariant under Type-II SSGs are necessarily $\mathcal{T}$-symmetric.  The Type-II MSGs have historically been termed the \emph{gray groups}~\cite{BigBook}, because Type-II groups do not admit the presence of localized magnetic moments, due to $\{\mathcal{T}|000\}$ symmetry at each point in each unit cell.  Unlike the symbols of the Type-I SSGs, the symbols of Type-II, III, and IV SSGs contain primes, which denote antiunitary group elements.  Because we are discussing both MSGs and (nonmagnetic) SGs in this work, we will employ the notation of Ref.~\onlinecite{MagneticBook} in which the symbols of $\mathcal{T}$-symmetric groups $M_{II}$ are followed by $1'$ to emphasize that $\{\mathcal{T}|000\}\in M_{II}$.  For example, in this work, the symbol $P4/mmm$ refers to Type-I MSG 123.339, whereas the symbol $P4/mmm1'$ refers to Type-II SSG 123.340 (which is frequently denoted in other works~\cite{QuantumChemistry,Bandrep1,Bandrep2,Bandrep3,JenFragile1,BarryFragile} using only the simplified expression ``space group 123 $P4/mmm$'').

Given a group $G$ and a subgroup $H$ of $G$, we will find it useful to define the \emph{index} of $H$ in $G$.  Here and throughout this work, we will use cosets to precisely define the group-subgroup index.  Specifically, given a group $G$ and a subgroup $H$, we can define the coset of $H$ represented by an element $g\in G$ as:
\begin{equation}
gH\equiv\{gh|h\in H\}.  
\label{eq:coset1}
\end{equation}
By construction, Eq.~(\ref{eq:coset1}) implies that every element $g\in G$ is in one (and only one) coset $gH$.  By definition, $G$ may be decomposed into cosets with respect to $H$ by the \emph{set difference} $G\setminus H$:
\begin{equation}
G = H \cup g_{1}H \cup g_{2}H \cup ...,
\end{equation}
such that:
\begin{equation}
G\setminus H = \{g|g\in G, g\notin H\} = g_{1}H \cup g_{2}H \cup ... ,
\label{eq:setDifference}
\end{equation}
where $g_{i}H$ are (unique) cosets of $H$ defined by $g_{i}H \neq g_{j}H$ for $g_{i,j}\in G,\ g_{i,j}\notin H$.  If $G$ and $H$ are groups, $E\in G$ and $E \in H$ where $E$ is the identity element, implying that $G\setminus H$ is \emph{not} a group, because $E\not\in G\setminus H$.  Similarly, there does not exist a case in which $g_{i}=E$ in Eq.~(\ref{eq:setDifference}), as this would imply that $G\setminus H = H$.  We emphasize that the choice of each $g_{i}$ in Eq.~(\ref{eq:setDifference}) is not unique; there are generically multiple, equivalent ways of expressing the decomposition of $G\setminus H$ into cosets of $H$.  Eq.~(\ref{eq:setDifference}) implies that:
\begin{equation}
G = H \cup (G\setminus H) = H \cup g_{1}H \cup g_{2}H \cup ... ,
\label{eq:uniqueCosetResum}
\end{equation}
from which we define the quotient:
\begin{equation}
G / H = \{H,g_{1}H,g_{2}H, ... \}.
\label{eq:quotientGroup}
\end{equation}
We briefly pause to note that, if $H$ is additionally a \emph{normal} subgroup of $G$, such that $gH = Hg$, then we can define a group operation on cosets:
\begin{equation}
g_{1}Hg_{2}H = g_{1}g_{2}H.
\label{eq:coset2}
\end{equation}
Finally, using Eq.~(\ref{eq:quotientGroup}), we establish the definition of the index $[G:H]$ of the subgroup $H$ of $G$ as:
\begin{equation}
[G:H] = |G / H| = |G|/|H|,
\label{eq:subsetIndex}
\end{equation}
where $|G|$, $|H|$, and $|G/H|$ are respectively the number of elements in $G$, $H$, and $G/H$ [equal to one plus the number of coset representatives $g_{i}$ in Eq.~(\ref{eq:setDifference})].  It is important to note that $|G|$ ($|H|$) in Eq.~(\ref{eq:subsetIndex}) is necessarily infinite if $G$ ($|H|$) is an infinite group.  However, if $G$ and $H$ are \emph{both} infinite, then the index $[G:H]=|G|/|H|$ may still be finite.

It is worth noting that all 1,421 MSGs are index-2 subgroups of 230 Type-II SSGs.  For the previous Type-I groups in Appendix~\ref{sec:type1}, this follows directly from Eq.~(\ref{eq:type2}), and for the Type-III and Type-IV groups, this will respectively be proved in Appendices~\ref{sec:type3} and~\ref{sec:type4}.

\subsection{Type-III SSGs: Black and White Groups without Black and White Bravais Lattices (674 MSGs)}
\label{sec:type3}

\begin{figure}[h]
\includegraphics[width=0.7\columnwidth]{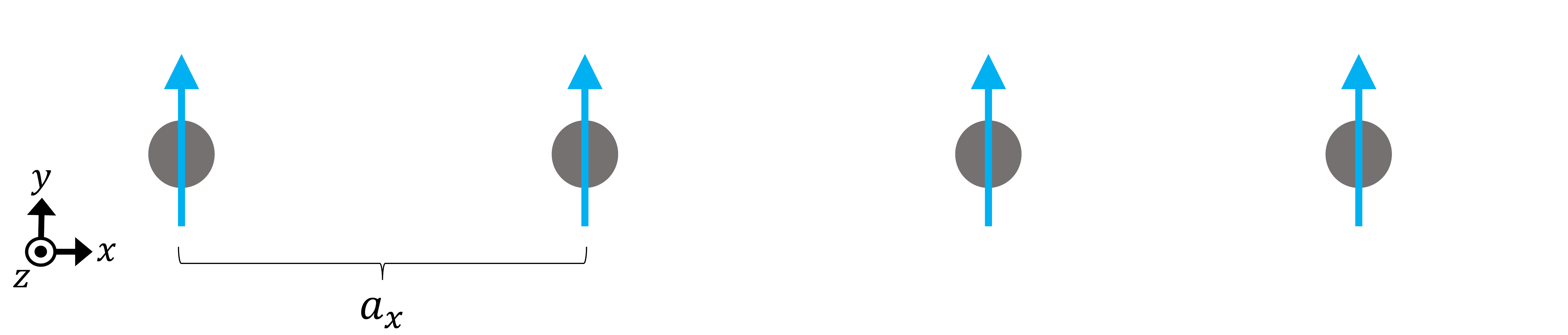}
\caption{A ferromagnetic chain with MRG $(pm'mm')_{RG}$, which is generated by $\{E|1\}$, $\{C_{2x}\times\mathcal{T}|0\}$, $\{C_{2y}|0\}$, and $\{\mathcal{I}|0\}$, and is isomorphic after the addition of perpendicular lattice translations to Type-III MSG 47.252 $Pm'm'm$.  The primes in the symbol $(pm'mm')_{RG}$ indicate that the MRG contains the symmetries $\{m_{x}\times\mathcal{T}|0\}$ and $\{m_{z}\times\mathcal{T}|0\}$.  In the decomposition in Eq.~(\ref{eq:type3}), $M_{III}=(pm'mm')_{RG}$, $G = (pmmm)_{RG}$ [isomorphic to Type-I MSG 47.249 $Pmmm$ after the addition of perpendicular lattice translations], and $H = (p12/m1)_{RG}$ [isomorphic to Type-I MSG 10.42 $P2/m$ after the addition of perpendicular lattice translations].}
\label{fig:type3FM}
\end{figure}

\begin{figure}[h]
\includegraphics[width=0.7\columnwidth]{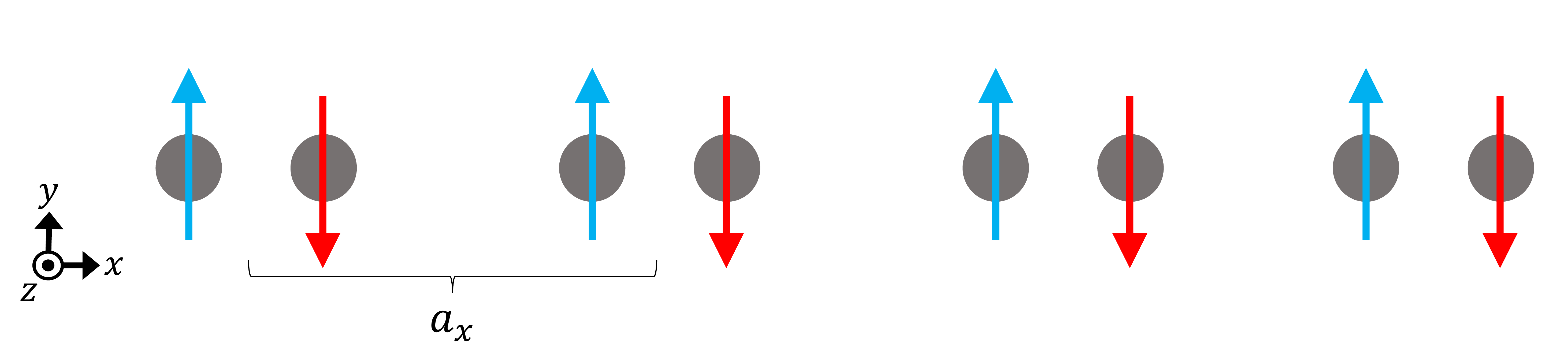}
\caption{An antiferromagnetic chain with MRG $(pmmm')_{RG}$, which is generated by $\{E|1\}$, $\{C_{2x}\times\mathcal{T}|0\}$, $\{C_{2y}\times\mathcal{T}|0\}$, and $\{\mathcal{I}\times\mathcal{T}|0\}$, and is isomorphic after the addition of perpendicular lattice translations to Type-III MSG 47.251 $Pm'mm$.  The prime in the symbol $(pmmm')_{RG}$ indicates that the MRG contains $\{m_{z}\times\mathcal{T}|0\}$ symmetry.  The red and blue magnetic moments are equal in magnitude and opposite in direction, and are related by the operation of $\{C_{2z}|0\}$ about the midpoints between adjacent atoms.  In the decomposition in Eq.~(\ref{eq:type3}), $M_{III}=(pmmm')_{RG}$, $G = (pmmm)_{RG}$ [isomorphic to Type-I MSG 47.249 $Pmmm$ after the addition of perpendicular lattice translations], and $H=(pmm2)_{RG}$ [isomorphic to Type-I MSG 25.57 $Pmm2$ after the addition of perpendicular lattice translations].}
\label{fig:type3AFM}
\end{figure}

Each Type-III SSG $M_{III}$ takes the form:
\begin{equation}
M_{III} = H \cup \mathcal{T}(G\setminus H),
\label{eq:type3}
\end{equation}
where $G$ and $H$ are isomorphic to Type-I SSGs, $H\subset G$, and $G\setminus H$ is a set that contains no elements of the form $\{E|{\bf t}\}$, where $E$ is the identity operation and ${\bf t}$ is a translation.  Hence, $G\setminus H$ in Eq.~(\ref{eq:type3}) does not include the identity element $\{E|{\bf 0}\}$, though $G\setminus H$ is free to contain elements of the form $\{f|{\bf 0}\}$ where $f$ is a unitary rotation or rotoinversion.  Because $G\setminus H$ does not contain pure translations, then it follows that $H$ is a subgroup of $G$ with the same Bravias lattice.  Following arguments recently presented in Ref.~\onlinecite{MagneticNewFermion}, we will demonstrate that $H$ is an index-2 subgroup of $G$.  To establish that $[G:H]=2$, we will first show that $H$ is an index-2 subgroup of $M_{III}$.  We begin by noting that, given an antiunitary symmetry $g_{A}\in \mathcal{T}(G\setminus H)$:
\begin{equation}
g_{A} = \mathcal{T} \times g, 
\label{eq:intermediateCosetRep}
\end{equation}
where $g$ is a unitary symmetry $g \in G$, $g\not\in H$.  Hence, $g_{A}^{2}$ is a unitary symmetry operation $g_{A}^{2}\in M_{III}$, implying that:
\begin{equation}
g_{A}^{2}\in H,\ g_{A}^{2}\not\in\mathcal{T}(G\setminus H).
\label{eq:antiUnitaryHalfwayPoint}
\end{equation}
Eqs.~(\ref{eq:type3}),~(\ref{eq:intermediateCosetRep}), and~(\ref{eq:antiUnitaryHalfwayPoint}) imply that:
\begin{equation}
M_{III} =g_{A}M_{III} = g_{A}H \cup g_{A}\mathcal{T}(G\setminus H),
\label{eq:M3andH}
\end{equation}
such that:
\begin{equation}
\mathcal{T}(G\setminus H) = g_{A}H, 
\label{eq:intermediateCosetRep2}
\end{equation}
implying that $H$ is an index-2 subgroup of $M_{III}$.  Eq.~(\ref{eq:M3andH}) also implies through Eqs.~(\ref{eq:type3}) and~(\ref{eq:intermediateCosetRep}) that:
\begin{equation}
|H|=|\mathcal{T}(G\setminus H)| = |G\setminus H|,
\label{eq:HandGType3}
\end{equation}
establishing that $H$ is also an index-2 subgroup of $G$:
\begin{equation}
G = H \cup gH,
\label{eq:missingIndex2type3}
\end{equation}
such that $gH = G\setminus H$, consistent with Eqs.~(\ref{eq:intermediateCosetRep}) and~(\ref{eq:intermediateCosetRep2}).  Finally, to see that $M_{III}$ is an index-$2$ subgroup of a Type-II SSG (specifically $M_{II} = G \cup \mathcal{T}G$), we consider the effects of restoring $\mathcal{T}$ symmetry to Eq.~(\ref{eq:type3}):
\begin{eqnarray}
M_{III} \cup \mathcal{T}M_{III} &=& H \cup \mathcal{T}(G\setminus H) \cup \mathcal{T}H \cup (G\setminus H) \nonumber \\
&=& G \cup \mathcal{T}G \nonumber \\
&=& M_{II}.
\label{eq:index2type3}
\end{eqnarray}

Like the previous Type-I MSGs in Appendix~\ref{sec:type1}, Type-III MSGs can characterize both ferromagnetic (Fig.~\ref{fig:type3FM}) and antiferromagnetic (Fig.~\ref{fig:type3AFM}) crystals.  The symbols for Type-III MSGs contain primes that denote which symmetry operations are formed from the combination of $\mathcal{T}$ and a unitary element of $G\setminus H$ [Eq.~(\ref{eq:type3})].  The Type-III MSGs have historically been termed the \emph{black and white groups without black and white Bravais lattices}, because Type-III groups contain antiunitary symmetries that relate classical magnetic moments at different positions in a crystal, but do not contain the antiferromagnetic translation symmetry $t_{0}\mathcal{T}$ common to Type-IV MSGs that generates the black and white Bravais lattices (see Appendix~\ref{sec:type4} and Chapter 7 in Ref.~\onlinecite{BigBook}.)  Representative examples demonstrating the usage of primes in Type-III magnetic group symbols are presented in Figs.~\ref{fig:type3FM} and~\ref{fig:type3AFM}.

\subsection{Type-IV SSGs: Black and White Groups with Black and White Bravais Lattices (517 MSGs)}
\label{sec:type4}

\begin{figure}[h]
\includegraphics[width=0.8\columnwidth]{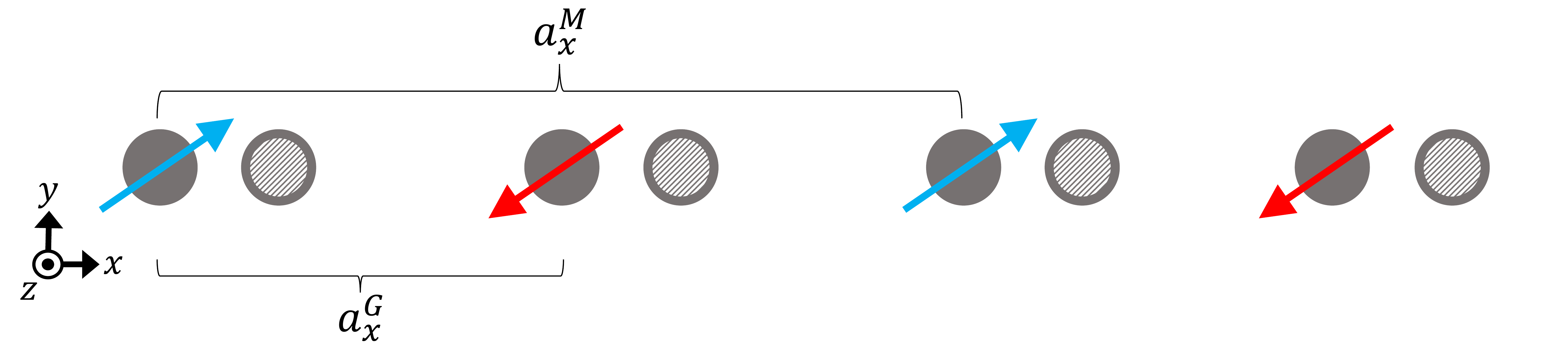}
\caption{An antiferromagnetic chain with MRG $(p_{a}1)_{RG}$, which is generated by $\{\mathcal{T}|1/2\}$ ($t_{a_{x}^{M}/2}\mathcal{T}$), and is isomorphic after the addition of perpendicular lattice translations to Type-IV MSG 1.3 $P_{S}1$.  The red and blue magnetic moments on the atoms labeled with solid circles are equal in magnitude and opposite in direction.  The two nonmagnetic atoms (hashed circles) in each magnetic unit cell are displaced out of the $xy$-plane, breaking additional symmetries such as $\{m_{z}\times\mathcal{T}|0\}$.  In terms of the black and white Bravais lattices historically employed to characterize (antiferro)magnetic structures~\cite{BigBook}, the atoms with blue magnetic moments can be taken to occupy white sites, whereas the atoms with red (time-reversed) magnetic moments can be taken to occupy black sites.  Throughout this work, we will only use the more familiar gray (nonmagnetic) Bravais lattices to characterize magnetic symmetry groups, because the black and white Bravais lattices add an additional level of complexity that does not factor into any of the group-theoretic calculations that comprise MTQC.  Further discussion and a complete enumeration of the black and white Bravais lattices is provided in Chapter 7 in Ref.~\onlinecite{BigBook}.  In the antiferromagnetic chain in this figure, the blue and red magnetic moments are related by $t_{a_{x}^{M}/2}\mathcal{T}$.  The primitive (magnetic) unit cell of the spin chain has a length $a_{x}^{M}$, whereas the nonmagnetic unit cell, which is realized by restoring $\mathcal{T}$ symmetry [Eq.~(\ref{eq:addTto4})], has a shorter length $a_{x}^{G}=a_{x}^{M}/2$.  In the decomposition in Eq.~(\ref{eq:type4}), $M_{IV} = (p_{a}1)_{RG}$, $H = (p1)_{RG}$ with the lattice constant $a = a_{x}^{M}$ [isomorphic to Type-I MSG 1.1 $P1$ after the addition of perpendicular lattice translations], and $t_{0}=t_{a_{x}^{M}/2}$.  From this, we establish the decomposition in Eqs.~(\ref{eq:alternativeType4}) and~(\ref{eq:type4G}), in which $M_{IV} = (p_{a}1)_{RG}$, $G = (p1)_{RG}$ with the lattice constant $a = a_{x}^{G}=a_{x}^{M}/2$, and $H = (p1)_{RG}$ with the lattice constant $a = a_{x}^{M}$.}
\label{fig:type4AFM}
\end{figure}

Each Type-IV SSG $M_{IV}$ takes the form:
\begin{equation}
M_{IV} = H \cup \mathcal{T}t_{0}H,
\label{eq:type4}
\end{equation}
in which $H$ is isomorphic to a Type-I SSG and $t_{0}$ is a translation whose length is half that of either $t_{a,b,c}$, $t_{a}+t_{b}$, $t_{a}+t_{c}$, $t_{b}+t_{c}$, or $t_{a}+t_{b}+t_{c}$, where $t_{a,b,c}$ are the primitive lattice translations in $H$~\cite{BigBook}.  The fractional lattice translations $t_{0}^{n}$ where $n\text{ mod }2=1$ relate positions of alternating color (classical spin orientation) in the black and white Bravais lattice of $M_{IV}$ (see Chapter 7 in Ref.~\onlinecite{BigBook}), whereas the full lattice translations $t_{a,b,c}$ relate positions with the same color in the nonmagnetic (gray) Bravais lattice of $M_{IV}$.  Hence, historically~\cite{BigBook}, the Type-IV MSGs have been termed the \emph{black and white groups with black and white Bravais lattices}.  As previously with the Type-III groups [see the text surrounding Eq.~(\ref{eq:intermediateCosetRep})], we can show that $H$ is an index-$2$ subgroup of $M_{IV}$.  To demonstrate that $[M_{IV}:H]=2$, we first rearrange Eq.~(\ref{eq:type4}) into the same form as Eq.~(\ref{eq:type3}):
\begin{equation}
M_{IV} = H \cup \mathcal{T}(G\setminus H),
\label{eq:alternativeType4}
\end{equation}
for which, by construction:
\begin{equation}
G=H\cup t_{0} H,
\label{eq:type4G}
\end{equation}
such that $G$ is isomorphic to a Type-I SSG with the gray Bravais lattice given by ignoring the colors of the black and white Bravais lattice of $M_{IV}$ (see Fig.~\ref{fig:type4AFM}).  To show that $M_{IV}$ is an index-$2$ subgroup of a Type-II group, we again restore $\mathcal{T}$ symmetry [see Eq.~(\ref{eq:index2type3})]:
\begin{eqnarray}
M_{IV} \cup \mathcal{T}M_{IV} &=& H \cup \mathcal{T}t_{0}H \cup \mathcal{T}H \cup t_{0}H \nonumber \\ 
&=& G \cup \mathcal{T}G \nonumber \\
&=& M_{II},
\label{eq:addTto4}
\end{eqnarray}
where $G$ is given in Eq.~(\ref{eq:type4G}).  As shown in the text following Eq.~(\ref{eq:subsetIndex}), Eq.~(\ref{eq:type4G}) also implies that $H$ is an index-2 subgroup of the Type-I MSG $G$.

Physically, Eq.~(\ref{eq:addTto4}) implies that the process of ``turning off'' the magnetism in a crystal with a Type-IV SSG (MSG) $M_{IV}$ generates a nonmagnetic crystal that is invariant under a Type-II group $M_{II}$ with a smaller unit cell than the magnetic unit cell of $M_{IV}$ (Fig.~\ref{fig:type4AFM}), and with the same gray Bravais lattice as $G$ in Eq.~(\ref{eq:type4G}) (as opposed to the gray Bravais lattice of $H$).  Unlike the previous Type-I and Type-III MSGs in Appendices~\ref{sec:type1} and~\ref{sec:type3}, respectively, Type-IV MSGs necessarily characterize crystals with net-zero magnetic moments, because the operation of $t_{0}\mathcal{T}\in M_{IV}$ [Eq.~(\ref{eq:type4})] relates the spin configuration of one half of the primitive (magnetic) unit cell to its time-reverse in the other half.  The symbols for Type-IV MSGs contain subscripts that denote the direction of $t_{0}$, and therefore specify the gray (nonmagnetic) Bravais lattice of $G$ in Eqs.~(\ref{eq:type4G}) and~(\ref{eq:addTto4}).

\section{Site-Symmetry Groups and Wyckoff Positions of the Magnetic Space Groups}
\label{sec:SecFullWyckoff}

Next, we will discuss the position-space action of the symmetries of the MSGs.  In Appendix~\ref{sec:siteSymmetry}, we will introduce the site-symmetry (stabilizer) groups of the MSGs, and in Appendix~\ref{sec:Wyckoff}, we will discuss how the Wyckoff positions of the MSGs are related to those of the $\mathcal{T}$-symmetric SGs.

Throughout the text below, we will provide representative 2D atomic and spin configurations highlighting properties of the site-symmetry groups and Wyckoff positions of the MSGs.  The 2D magnetic structures shown in this section each respect the symmetries of a magnetic layer group (MLG) $M_{LG}$ -- a subperiodic group with 3D symmetry operations and 2D translations~\cite{MagneticBook,ITCA,subperiodicTables,WiederLayers,DiracInsulator,SteveMagnet}.  Each MLG is isomorphic to (at least one) MSG $M$ modulo out-of-plane lattice translations.  Specifically, taking the in-plane translations to be elements of $T_{x,y}$ ($T_{a,b}$), and taking $t_{c}\in T_{c}$ ($T_{z}$) to be a lattice translation in the $z$ (out-of-plane) direction:
\begin{equation}
M = M_{LG} \cup t_{c}M_{LG}.
\label{eq:layerGroup}
\end{equation}
In this work, the symbols of MLGs are given in the convention employed by Litvin in Ref.~\onlinecite{MagneticBook}.

\subsection{Site-Symmetry Groups of the Magnetic Space Groups}
\label{sec:siteSymmetry}

In this section, we will define the \emph{site-symmetry group} $M_{\bf q}$ at a point ${\bf q}$ in crystal that is invariant under an SSG $M$.  To begin, $M$ is composed of unitary symmetry operations:
\begin{equation}
g_{U,i} = \{h_{i}|{\bf t}_{i}\},
\end{equation}
and antiunitary symmetry operations:
\begin{equation}
g_{A,j} = \{h_{j}\times\mathcal{T}|{\bf t}_{j}\},
\label{eq:antiunitaryActionforWyckoff}
\end{equation}
where each $h_{i,j}$ is a unitary symmetry operation that is either the identity, a rotation, or a rotoinversion.  Given a point ${\bf q}$ in an infinite crystal, the action of $g_{U,i}$ and $g_{A,j}$ on ${\bf q}$ is given by:
\begin{equation}
g_{U,i}{\bf q} = h_{i}{\bf q} + {\bf t}_{i},\ g_{A,j}{\bf q}=h_{j}{\bf q} + {\bf t}_{j},
\label{eq:positionSymmetryActionRedo}
\end{equation}
in which only $h_{i,j}$ and ${\bf t}_{i,j}$ act on ${\bf q}$, because $\mathcal{T}$, by definition, leaves spatial coordinates invariant~\cite{BigBook}.  As defined in Ref.~\onlinecite{QuantumChemistry}, a site-symmetry group $M_{\bf{q}}$ is spanned by the set of unitary and antiunitary symmetry operations $g\in M$ that return a site $\bf{q}$ in an infinite crystal (\emph{i.e.} a point in position space) to itself \emph{in the same unit cell}:
\begin{equation}
g{\bf q} = {\bf q},
\label{eq:gqforSiteSym}
\end{equation}
for all $g\in M_{\bf q}$.  Hence, the site-symmetry group $M_{\bf q}$ of ${\bf q}$ is finite a subgroup of the SSG $M$:
\begin{equation}
M_{\bf q}\subset M,
\label{eq:smallestSiteSymFirst}
\end{equation}
in which $M_{\bf q}$ does not contain elements of the form $\{E|{\bf t}\}$ or $\{\mathcal{T}|{\bf t}\}$, where $E$ is the identity operation and ${\bf t}$ is a translation.  Later, in Appendix~\ref{sec:Wyckoff}, we will reintroduce the Wyckoff positions of $M$ containing ${\bf q}$, as defined in Ref.~\onlinecite{Bandrep1}.

\begin{figure}[t]
\includegraphics[width=\columnwidth]{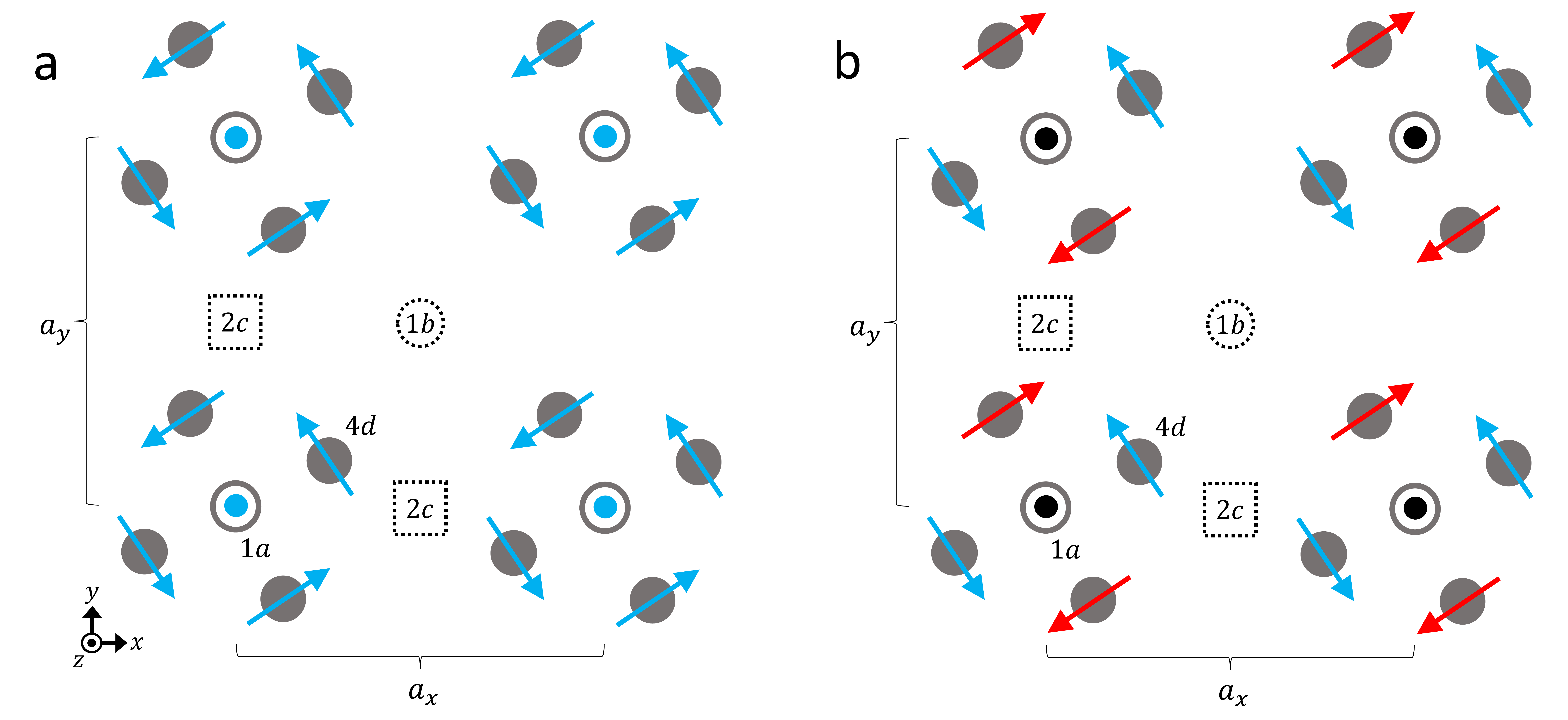}
\caption{(a) A magnetic crystal with Type-I magnetic layer group (MLG)~\cite{MagneticBook,ITCA,subperiodicTables,WiederLayers,DiracInsulator,SteveMagnet} $p4$, which is generated by $\{E|10\}$ and $\{C_{4z}|00\}$, and is isomorphic after the addition of $t_{z}$ to Type-I MSG 75.1 $P4$.  The atoms on the $1a$ sites in (a) exhibit an additional magnetic moment in the $+\hat{z}$ direction (blue dot), which we have chosen in order to break $\{m_{z}\times\mathcal{T}|00\}$ symmetry~\cite{SteveMagnet} to simplify the symmetry analysis performed in this section.  (b) A magnetic crystal with Type-III MLG $p4'$, which is generated by $\{E|10\}$ and $\{C_{4z}\times\mathcal{T}|00\}$, and is isomorphic after the addition of $t_{z}$ to Type-III MSG 75.3 $P4'$.  The red magnetic moments in (b) have the same magnitudes as the blue magnetic moments; they are only colored in red to emphasize that the red moments in (b) are related to the blue moments by the antiunitary symmetry operation ($\{C_{4z}\times\mathcal{T}|00\}$).  The atoms on the $1a$ sites in (b) do not exhibit a magnetic moment, and instead are displaced out of the $xy$-plane in the $+\hat{z}$ direction, which we have indicated with black dots.  We have chosen to displace the atoms at the $1a$ position in each unit cell in (b) out of the $xy$-plane in order to break $\{m_{z}|00\}$ symmetry, such that the MLGs in (a) and (b) share the same ``unprimed'' Type-I MLG $G=p4$ [see the text surrounding Eq.~(\ref{eq:unprimed})].  The $1a$ and $1b$ site-symmetry groups in (a) are isomorphic to Type-I MPG 9.1.29 $4$, which is generated by $C_{4z}$, whereas the $1a$ and $1b$ site-symmetry groups in (b) are isomorphic to Type-III MPG 9.3.31 $4'$, which is generated by $C_{4z}\times\mathcal{T}$.  Nevertheless, in both (a) and (b), the $2c$ site-symmetry groups are isomorphic to Type-I MPG 3.1.6 $2$, which is generated by $C_{2z}$.  In both (a) and (b), magnetic moments additionally occupy the $4d$ (general) position, where the site-symmetry groups at $4d$ in both (a) and (b) are isomorphic to Type-I MPG 1.1.1 $1$, the trivial MPG.  The MLGs in (a) and (b) [$p4$ and $p4'$, respectively] are also isomorphic to magnetic wallpaper groups~\cite{WiederLayers,DiracInsulator,SteveMagnet,ConwaySymmetries}.}
\label{fig:siteSymmetry}
\end{figure}

In Eq.~(\ref{eq:smallestSiteSymFirst}), $M_{\bf q}$ is necessarily isomorphic to one of the 122 crystallographic Shubnikov point groups (SPGs)~\cite{ShubnikovMagneticPoint,BilbaoPoint,PointGroupTables,MagneticBook,EvarestovBook,EvarestovMEBR,BCS1,BCS2,BCSMag1,BCSMag2,BCSMag3,BCSMag4}, which are listed in the~\href{http://www.cryst.ehu.es/cryst/mpoint.html}{MPOINT}~(\url{http://www.cryst.ehu.es/cryst/mpoint.html})~\cite{BCSMag1,BCSMag2,BCSMag3,BCSMag4} and~\href{http://www.cryst.ehu.es/cryst/corepresentationsPG}{CorepresentationsPG} tools on the BCS, in which the SPGs are numbered according to the convention established by Litvin in Ref.~\onlinecite{MagneticBook}.  The SPGs divide into 32 Type-I magnetic point groups (MPGs), 32 Type-II (nonmagnetic) SPGs, and 58 Type-III MPGs, where the type of an SPG is defined the same way as the type of an SSG [Appendix~\ref{sec:type1} and Eqs.~(\ref{eq:type2}) and~(\ref{eq:type3}), and~(\ref{eq:type3})].  We emphasize that, unlike in the MSGs, which subdivide into Types-I, III, and IV, there are only Type-I and Type-III MPGs.  Specifically, there are no Type-IV MPGs, because point groups, unlike space groups, cannot contain operations of the form $\{\mathcal{T}|{\bf t}\}$  [Eq.~(\ref{eq:type4})], as $\{\mathcal{T}|{\bf t}\}$ does not fix any point in position space [Eq.~(\ref{eq:positionSymmetryActionRedo})].  Following the discussions in Appendices~\ref{sec:type1} and~\ref{sec:type3}, all Type-I and Type-III MPGs are subgroups of Type-II SPGs.  For all 122 SPGs, the group-subgroup relations are provided by Ascher and Janner in Ref.~\onlinecite{AscherSubgroup}, and can be inferred by using the~\href{http://www.cryst.ehu.es/cgi-bin/cryst/programs/subgrmag1_k.pl}{KSUBGROUPSMAG} tool on the BCS~(\url{http://www.cryst.ehu.es/cgi-bin/cryst/programs/subgrmag1_k.pl})~\cite{BCSMag1,BCSMag2,BCSMag3,BCSMag4} on pairs of SSGs that are isomorphic to SPGs modulo integer lattice translations.  For example, to see that Type-III MPG 9.3.31 $4'$ and Type-I MPG 9.1.29 $4$ are both index-2 subgroups of Type-II SPG 9.2.30 $41'$, one can choose the ``List of subgroups'' option in~\href{http://www.cryst.ehu.es/cgi-bin/cryst/programs/subgrmag1_k.pl}{KSUBGROUPSMAG} for Type-II SG 75.2 $P41'$ while specifying the magnetic wavevector ${\bf k}={\bf 0}$.  Documentation and further examples of the output of~\href{http://www.cryst.ehu.es/cgi-bin/cryst/programs/subgrmag1_k.pl}{KSUBGROUPSMAG} are provided in Refs~\onlinecite{BCSMag4,BilbaoSpinWaves}.  For this work, we define Type-I MPG 1.1.1 $1$ as both the \emph{trivial MPG} and the \emph{trivial SPG}, as its only generator is the identity operation $E$, and because Type-I MPG 1.1.1 $1$ is the common subgroup of all MPGs and SPGs.

It is important to highlight that all site-symmetry groups in MSGs (\emph{i.e.} Type-I, III, and IV SSGs) are isomorphic to MPGs (\emph{i.e.} Type-I and III SPGs), and correspondingly, that all site-symmetry groups in Type-II (nonmagnetic) SSGs are isomorphic to Type-II SPGs.  To show this, we first consider the Type-II SSGs.  Because all Type-II SSGs contain the element $\{\mathcal{T}|{\bf 0}\}$, which fixes all points in space, then all site-symmetry groups in Type-II SSGs also necessarily contain $\{\mathcal{T}|{\bf 0}\}$, and are therefore isomorphic to Type-II SPGs.  Conversely, because MSGs (\emph{i.e.} Type-I, III, and IV SSGs) do not contain $\{\mathcal{T}|{\bf 0}\}$, then \emph{none} of their site-symmetry groups contain $\{\mathcal{T}|{\bf 0}\}$ (though they are free to contain antiunitary operations such as $\{C_{4z}\times\mathcal{T}|{\bf 0}\}$; hence, the site-symmetry groups in MSGs are isomorphic to either Type-I or Type-III MPGs.  In Type-I MSGs, all of the site-symmetry groups are isomorphic to Type-I MPGs, as Type-I MSGs do not contain antiunitary symmetry elements (Appendix~\ref{sec:type1}).  However, in each of the Type-III and Type-IV MSGs, site-symmetry groups can be isomorphic to either Type-I or Type-III MPGs.  For example, in Fig.~\ref{fig:siteSymmetry}(a,b), we depict atomic and spin configurations that respect the symmetries of Type-I MLG $p4$ and Type-III MLG $p4'$, respectively [see the text surrounding Eq.~(\ref{eq:layerGroup}) for the definition of an MLG].  In $p4$, the $1a$ and $1b$ site-symmetry groups are isomorphic to Type-I MPG 9.1.29 $4$, whereas in $p4'$, the $1a$ and $1b$ site-symmetry groups are isomorphic to Type-III MPG 9.3.31 $4'$.  Nevertheless, in both $p4$ and $p4'$, the $2c$ site-symmetry groups are isomorphic to Type-I MPG 3.1.6 $2$.

\subsection{Wyckoff Positions of the Magnetic Space Groups}
\label{sec:Wyckoff}

In this section, we will next reintroduce the Wyckoff positions of the MSGs.  First, we will below precisely define a Wyckoff position.  Then, in Appendix~\ref{sec:WyckoffExample}, we will apply the definitions and relations established below to illustrative 2D examples of MLGs derived from the Type-II layer group (LG) $p41'$.

To begin, as discussed in Ref.~\onlinecite{QuantumChemistry}, the Wyckoff positions of an SSG $M$ are defined using the orbits of symmetry sites.  We first select a site ${\bf q}_{\alpha}$ in a crystal that is invariant under an SSG $M$.  As defined in the text surrounding Eq.~(\ref{eq:gqforSiteSym}), the site-symmetry group $M_{{\bf q}_{\alpha}}$ contains all of the symmetries $g\in M$ that return ${\bf q}_{\alpha}$ to itself.  However, there also generically exist symmetries:
\begin{equation}
\tilde{g}_{i}\in M,\ \tilde{g}_{i}\not\in M_{{\bf q}_{\alpha}}
\label{eq:firstWyckoffEq}
\end{equation}
that act to send ${\bf q}_{\alpha}$ to other points ${\bf q}_{{\alpha}'}$ in the crystal, where ${\bf q}_{\alpha'}$ may or may not lie in the same unit cell as ${\bf q}_{\alpha}$.  We next define the set of symmetries:
\begin{equation}
\{\tilde{g}\} = M\setminus M_{{\bf q}_{\alpha}}.
\label{eq:secondWyckoff}
\end{equation}
Acting with all of the $\tilde{g}_{i}\in \{\tilde{g}\}$ on ${\bf q}_{\alpha}$ generates an infinite number of sites $\{\tilde{g}_{i}{\bf q}_{\alpha}\}$, because $M$ includes lattice translations and $M_{{\bf q}_{\alpha}}$ does not.  Additionally, it is possible for two elements $\tilde{g}_{i,j}\in \{\tilde{g}\}$ to map ${\bf q}_{\alpha}$ to the same point.  For example, if ${\bf q}_{\alpha}=(x,y,0)$, $\tilde{g}_{i} = \{C_{2z}|000\}$, and $\tilde{g}_{j} = \{\mathcal{I}|000\}$, then $\tilde{g}_{i}{\bf q}_{\alpha} = \tilde{g}_{j}{\bf q}_{\alpha} = (-x,-y,0)$.  Continuing to employ the previous definition from TQC~\cite{QuantumChemistry,Bandrep1,Bandrep2,Bandrep3,JenFragile1,BarryFragile}, we define the \emph{orbit} of ${\bf q}_{\alpha}$ to be the infinite subset of unique points $\{\tilde{g}_{i}{\bf q}_{\alpha}\}\cup{\bf q}_{\alpha}$.  We then define the \emph{Wyckoff orbit} indexed by ${\bf q}_{\alpha}$ as the finite set of points in the orbit -- including ${\bf q}_{\alpha}$ itself -- that lie in the same unit cell as ${\bf q}_{\alpha}$.  In this work, we will summarize the Wyckoff orbit containing ${\bf q}_{\alpha}$ using the notation $\{{\bf q}_{\alpha}\}$, for simplicity.  In the Wyckoff orbit of ${\bf q}_{\alpha}$, the index $\alpha$ runs from $1$ to $n$, where $n$ -- which is termed the \emph{multiplicity} of the Wyckoff orbit -- is the number of unique sites ${\bf q}_{\alpha}$ in the orbit of ${\bf q}_{\alpha}$ that lie in the same unit cell as ${\bf q}_{\alpha}$ plus one for ${\bf q}_{\alpha}$ itself.   Given a site-symmetry group $M_{{\bf q}_{\alpha}}$, all of the other site-symmetry groups in the Wyckoff orbit of ${\bf q}_{\alpha}$ are given by:
\begin{equation}
M_{{\bf q}_{\beta}}  = \tilde{g}_{\alpha\beta}M_{{\bf q}_{\alpha}}\tilde{g}_{\alpha\beta}^{-1},
\end{equation}
where $\tilde{g}_{\alpha\beta}$ is a symmetry in $M\setminus M_{{\bf q}_{\alpha}}$ [Eq.~(\ref{eq:firstWyckoffEq})] for which:
\begin{equation}
\tilde{g}_{\alpha\beta}{\bf q}_{\alpha}={\bf q}_{\beta},
\label{eq:isomorphicAndConjugate}
\end{equation}
where ${\bf q}_{\beta}$ is in same Wyckoff orbit as ${\bf q}_{\alpha}$.  Hence, all of the site-symmetry groups $M_{{\bf q}_{\beta}}$ in the same Wyckoff orbit as $M_{{\bf q}_{\alpha}}$ are isomorphic and conjugate to $M_{{\bf q}_{\alpha}}$, and to each other.  Lastly, we define the \emph{Wyckoff position} containing ${\bf q}_{\alpha}$ as the set of Wyckoff orbits with the same multiplicity in which the coordinates of the sites in the orbit $\{{\bf q}_{\alpha}\}$ can be smoothly deformed into each other without changing the Wyckoff orbit multiplicity.  For example, in Type-I MSG 2.4 $P\bar{1}$, which is generated by $\{\mathcal{I}|{\bf 0}\}$ and 3D lattice translation, the sites $[(0,0,0.1),(0,0,-0.1)]$ and $[(0,0,0.2),(0,0,-0.2)]$ define distinct Wyckoff orbits.  Nevertheless, in MSG 2.4 $P\bar{1}$, the two Wyckoff orbits $[(0,0,0.1),(0,0,-0.1)]$ and $[(0,0,0.2),(0,0,-0.2)]$ represent different parameter choices for the same Wyckoff position $[(x,y,z),(-x,-y,-z)]$ (labeled the $2i$ position on the BCS).  The coordinates, multiplicities, and site-symmetry groups of the Wyckoff positions of all 1,651 SSGs have previously been made accessible through the~\href{http://www.cryst.ehu.es/cgi-bin/cryst/programs/magget_wp.pl}{MWYCKPOS} tool on the BCS~(\url{http://www.cryst.ehu.es/cgi-bin/cryst/programs/magget_wp.pl})~\cite{BCSMag1,BCSMag2,BCSMag3,BCSMag4}.

Next, to further determine the maximal Wyckoff positions -- which we will later find to be important in calculating the magnetic elementary band (co)representations (Appendix~\ref{sec:MEBRs}) -- we follow the definition established in Ref.~\onlinecite{Bandrep1}.  First, we recognize that, for each Wyckoff position in the SSGs, there is a set of coordinates that defines the locations of atoms (magnetic atomic orbitals, see Appendix~\ref{sec:magWannier}) occupying the Wyckoff position.  In high-symmetry Wyckoff positions, some or all of the coordinates have fixed values (\emph{e.g.} $0$ or $1/2$), whereas in other, lower-symmetry positions, the coordinates have free values (\emph{e.g.} $z$) that represent distinct Wyckoff orbits in the same Wyckoff position [see the text following Eq.~(\ref{eq:isomorphicAndConjugate})].  For example, in the output of~\href{http://www.cryst.ehu.es/cgi-bin/cryst/programs/magget_wp.pl}{MWYCKPOS} on the BCS~\cite{BCSMag1,BCSMag2,BCSMag3,BCSMag4} for Type-III MSG 10.45 $P2/m'$, the $1a$ position lies at $(0,0,0)$ and has a site-symmetry group isomorphic to Type-III MPG 5.4.15 $2/m'$, whereas the $2i$ position has sites at $(0,y,0)$ and $(0,-y,0)$, each of which has a site-symmetry group isomorphic to Type-I MPG 3.1.6 $2$.  As an intermediate step towards defining a maximal Wyckoff position, we first establish a definition for connected Wyckoff positions.  We define two Wyckoff positions to be \emph{connected} if the coordinates of one of the sites in the lower-symmetry Wyckoff position [\emph{e.g.} $(0,y,0)$ in the $2i$ position in the previous example in MSG 10.45 $P2/m'$] can be adjusted to coincide with the coordinates of the higher-symmetry Wyckoff position [\emph{e.g.} the $1a$ position at $(0,0,0)$ in the previous example in MSG 10.45 $P2/m'$].  From this, we define a \emph{maximal} Wyckoff position to be a Wyckoff position that is not connected to a Wyckoff position with a higher-symmetry site-symmetry group (\emph{i.e.} the site-symmetry group of a maximal Wyckoff position must be a larger supergroup of the site-symmetry group of any Wyckoff position to which it is connected).  This definition of a maximal Wyckoff position is identical to the previous definition established in Refs.~\onlinecite{QuantumChemistry,Bandrep1} for the Type-I MSGs and Type-II SSGs; in this work, we have applied the earlier definition to the Type-III and Type-IV MSGs by incorporating the action of the antiunitary symmetries $g_{A,j}$ in Type-III and Type-IV MSGs [see Appendices~\ref{sec:type3} and~\ref{sec:type4} and the text surrounding Eq.~(\ref{eq:positionSymmetryActionRedo})].  Specifically, in both this work and in TQC, the set of site-symmetry groups in the maximal Wyckoff positions in an SSG $M$ coincide with the set of maximal site-symmetry subgroups of $M$.  In this work, the only distinction from the earlier discussion of Wyckoff positions in Refs.~\onlinecite{QuantumChemistry,Bandrep1} is the incorporation of the action of antiunitary symmetries through Eq.~(\ref{eq:positionSymmetryActionRedo}).

We will now discuss the relationship between the Wyckoff positions in the MSGs and the Wyckoff positions in the more familiar Type-II (nonmagnetic) SSGs.  First, the Wyckoff positions of the Type-I and Type-III MSGs can straightforwardly be obtained from the Wyckoff positions of the Type-II SSGs.  For the Type-I MSGs, this follows directly from the definition of a Type-II SSG (Appendix~\ref{sec:type2}).  Specifically, in a Type-II SSG $M_{II}=G \cup \mathcal{T}G$ [Eq.~(\ref{eq:type2})], all of the site-symmetry groups $M_{II,\bf{q}}$ take the form:
\begin{equation}
M_{II,{\bf q}} = G_{\bf q} \cup \mathcal{T}G_{\bf q}.
\label{eq:type2siteSubgroup}
\end{equation}
In each Wyckoff position indexed by a site ${\bf q}$ in a crystal invariant under $M_{II}$, the multiplicity of the Wyckoff position of ${\bf q}$ is only determined by the unitary symmetries $g$ of the Type-I subgroup $G$ of $M_{II}$, because $\mathcal{T}$ symmetry acts as the identity on ${\bf q}$ [see Eq.~(\ref{eq:positionSymmetryActionRedo}) and the surrounding text].  Therefore, in a Type-I MSG $G$, all of the Wyckoff positions have the same multiplicities and coordinates as the Wyckoff positions in $M_{II}=G\cup \mathcal{T}G$ [Eq.~(\ref{eq:type2})], and all of the site-symmetry groups $G_{\bf{q}}$ are isomorphic to the unitary subgroups of $M_{II,\bf{q}}$ [Eq.~(\ref{eq:type2siteSubgroup})].

Conversely, in a Type-III MSG $M_{III} = H\cup \mathcal{T}(G\setminus H)$ [Eq.~(\ref{eq:type3})], the site-symmetry groups $M_{III,{\bf q}}$ can be isomorphic to either Type-I and Type-III MPGs, as previously discussed in Appendix~\ref{sec:siteSymmetry}.  Nevertheless, we will show below that the multiplicities of the Wyckoff positions in $M_{III}$ are still inherited from the ``unprimed'' Type-I group $G$ in the definition of $M_{III}$ [Eq.~(\ref{eq:type3})].  Specifically, in this work, we define $G$ to be the ``unprimed'' group of $M_{III}$, because $G$ and $M_{III}$ share the same symbols if primes are neglected (\emph{i.e.}, under transforming group elements of the form $g'=\{h\times\mathcal{T}|{\bf t}\} \rightarrow \{h|{\bf t}\}$).  To show this, we first note that, because $\mathcal{T}$ symmetry acts as the identity on spatial coordinates, then:
\begin{equation}
\mathcal{T}{\bf q} = {\bf q},
\label{eq:unprimed}
\end{equation}
for all ${\bf q}$ in the 1,651 SSGs.  Consequently, in a Type-III MSG $M_{III}$, only the unitary parts $\{h|{\bf t}\}$ of the unitary and antiunitary symmetries in $M_{III}$ can act to send ${\bf q}$ to other positions [see Eq.~(\ref{eq:positionSymmetryActionRedo}) and the surrounding text].  As shown in Appendix~\ref{sec:type3}, the unitary parts of the unitary and antiunitary symmetries in $M_{III}$ comprise the unprimed Type-I MSG $G$ of $M_{III}$.  Additionally, as shown in Eq.~(\ref{eq:index2type3}), the unprimed group $G$ of $M_{III}$ is also the maximal unitary subgroup of the Type-II SSG $M_{III}\cup \mathcal{T}M_{III}$ (\emph{i.e.} $M_{III}\cup \mathcal{T}M_{III} = G\cup\mathcal{T}G$).  Lastly, as shown in the text surrounding Eq.~(\ref{eq:type2siteSubgroup}), the Wyckoff positions of $G$ are identical to the Wyckoff positions of $G\cup\mathcal{T}G$ [though the site-symmetry groups $G_{\bf q}$ are the unitary subgroups of the site-symmetry groups $M_{II,{\bf q}}$ in $G\cup \mathcal{T}G$].  From this, we conclude that the Type-I (unprimed) MSG $G$, the Type-II SSG $M_{III}\cup\mathcal{T}M_{III}$, and the Type-III MSG $M_{III}$ all share the same Wyckoff-position multiplicities and coordinates.  It therefore follows that each site-symmetry group $M_{III,{\bf q}}\subset M_{III}$ is an index-$2$ subgroup of $M_{II,{\bf q}} = G_{\bf q} \cup \mathcal{T}G_{\bf q}$ where:
\begin{equation}
M_{II,{\bf q}}\subset (M_{III} \cup \mathcal{T}M_{III}).  
\end{equation}
Specifically, $M_{III,\bf{q}}$ is either a Type-I site-symmetry group:
\begin{equation}
M_{III,{\bf q}} = G_{\bf q},
\end{equation}
or a Type-III site-symmetry group:
\begin{equation}
M_{III,{\bf q}} = H_{{\bf q}} \cup \mathcal{T}(G_{\bf q}\setminus H_{\bf q}),
\label{eq:type3MPGsite}
\end{equation}
where $H_{\bf q}$ is a site-symmetry group in the Type-I (maximal unitary) subgroup $H$ of $M_{III}$ [see Eq.~(\ref{eq:type3}) and the surrounding text].  We will shortly provide in Appendix~\ref{sec:WyckoffExample} an example demonstrating the relationship between $G_{\bf q}$, $M_{II,{\bf q}}$, and $M_{III,{\bf q}}$ in a Type-III magnetic symmetry group.

Unlike in Type-I and Type-III MSGs, the Wyckoff positions in Type-IV MSGs have more complicated dependencies on the Wyckoff positions in the Type-II SSGs.  This complication arises because the operation of $t_{0}\mathcal{T}$ in Eq.~(\ref{eq:type4}) enlarges the magnetic unit cell of a crystal with a Type-IV MSG (\emph{i.e.} $a_{x}^{M}$ in Fig.~\ref{fig:type4AFM}) relative to the nonmagnetic unit cell of its Type-II supergroup (\emph{i.e.} $a_{x}^{G}$ in Fig.~\ref{fig:type4AFM}).  Hence, the primitive cell of a Type-IV MSG is always larger than the primitive cell of its Type-II supergroup).  Therefore, there is no corresponding notion of an ``unprimed'' group for the Type-IV MSGs.  Instead the multiplicities, coordinates, and site-symmetry groups in Type-IV MSGs must be determined by composing the elements of the site-symmetry groups of the unitary subgroup $H$ in Eq.~(\ref{eq:type4}) with the antiunitary (antiferromagnetic) translation symmetry $t_{0}\mathcal{T}$.  An example demonstrating the composition of the unitary site-symmetry group symmetries in $H$ with $t_{0}\mathcal{T}$ in a Type-IV MSG will later be provided in Appendix~\ref{sec:exceptions}.

\subsubsection{Wyckoff Positions in Magnetic Subgroups of Type-II LG $p41'$}
\label{sec:WyckoffExample}

To demonstrate how the site-symmetry groups in Type-I and Type-III MSGs derive from those in Type-II SSGs, we will in this section analyze the examples of Type-II LG $M_{II}=p41'$ and its Type-I and Type-III magnetic subgroups Type-I MLG $G=p4$ and Type-III MLG $M_{III}=p4'$, respectively [Fig.~\ref{fig:siteSymmetry}(a,b), respectively].  $M_{II}$ is generated by $\{C_{4z}|00\}$,\ $\{\mathcal{T}|00\}$, and the lattice translation $\{E|10\}$.  Using~\href{http://www.cryst.ehu.es/cgi-bin/cryst/programs/magget_wp.pl}{MWYCKPOS} on the BCS~\cite{BCSMag1,BCSMag2,BCSMag3,BCSMag4} for Type-II SG 75.2 $P41'$, which is isomorphic to $p41'$ modulo $T_{z}$ [\emph{i.e.} after the addition of out-of-plane lattice translations, see the text surrounding Eq.~(\ref{eq:layerGroup})], we obtain the coordinates of the highest-symmetry (fourfold-symmetric) maximal Wyckoff positions of $p41'$ ($1a$ and $1b$) and the SPGs isomorphic to the fourfold-symmetric maximal site-symmetry groups:
\begin{eqnarray}
{\bf q}_{1a} &=& (0,0),\ M_{II,1a} = 41'= 4 \cup (\mathcal{T})4, \nonumber \\
{\bf q}_{1b} &=& (1/2,1/2),\ M_{II,1b} = 41' = 4 \cup (\mathcal{T})4,
\label{eq:maximalTSymmetric}
\end{eqnarray}
where the symbols $41'$ and $4$ respectively refer to Type-II SPG 9.2.30 $41'$ and Type-I MPG 9.1.29 $4$.  There is also a lower-symmetry maximal Wyckoff position in Type-II MLG $p41$ in which the site-symmetry groups do not contain fourfold rotation symmetry ($2c$).  The coordinates and site-symmetry-group-isomorphic SPGs of the $2c$ position in Type-II MLG $p41'$ are given by:
\begin{equation}
{\bf q}_{2c} = \{(1/2,0),\ (0,1/2)\},\ M_{II,2c} = 21' = 2 \cup (\mathcal{T})2,
\label{eq:lessMaximalTSymmetric}
\end{equation}
where the symbols $21'$ and $2$ respectively refer to Type-II SPG 3.2.7 $21'$ and Type-I MPG 3.1.6 $2$.

As defined in Eq.~(\ref{eq:type2}), the layer group $M_{II}=p41'$ admits a decomposition:
\begin{equation}
p41' = p4 \cup (\mathcal{T})p4,
\end{equation}
where $p4$, which is generated only by $\{C_{4z}|00\}$ and $\{E|10\}$, is the maximal unitary subgroup of $M_{II}$.  An atomic and spin configuration with MLG $p4$ is shown in Fig.~\ref{fig:siteSymmetry}(a).  Using~\href{http://www.cryst.ehu.es/cgi-bin/cryst/programs/magget_wp.pl}{MWYCKPOS} on the BCS~\cite{BCSMag1,BCSMag2,BCSMag3,BCSMag4} for MSG 75.1 $P4$, which is isomorphic to $p4$ modulo $T_{z}$, we obtain the coordinates and site-symmetry-group-isomorphic MPGs of the maximal Wyckoff positions of $p4$:
\begin{eqnarray}
{\bf q}_{1a} &=& (0,0),\ G_{1a} = 4, \nonumber \\
{\bf q}_{1b} &=& (1/2,1/2),\ G_{1b} = 4, \nonumber \\
{\bf q}_{2c} &=& \{(1/2,0),\ (0,1/2)\},\ G_{2c} = 2.
\label{eq:p4Wyckoff}
\end{eqnarray}
where the symbols $4$ and $2$ respectively refer to Type-I MPG 9.1.29 $4$ and Type-I MPG 3.1.6 $2$.  As discussed in the text following Eq.~(\ref{eq:type2siteSubgroup}), we observe that each site-symmetry group $G_{\bf q}$ in $p4$ [Eq.~(\ref{eq:p4Wyckoff})] is equivalent to the unitary subgroup of the site-symmetry group $M_{II,{\bf q}}$ of $p41'$ [Eq.~(\ref{eq:maximalTSymmetric})].

Next, we perform the analogous analysis of the Wyckoff positions and site-symmetry groups in Type-III MLG $M_{III}=p4'$, which is generated by $\{C_{4z}\times\mathcal{T}|00\}$ and $\{E|10\}$.  As discussed in the text surrounding Eq.~(\ref{eq:type3}), $M_{III}=p4'$ admits a decomposition:
\begin{equation}
p4' = p2 \cup \mathcal{T}\left[(p4)\setminus (p2)\right],
\label{eq:MLGp4prime}
\end{equation}
in which $p2$ is the Type-I MLG generated by $\{E|10\}$, $\{E|01\}$, and $\{C_{2z}|00\}=(\{C_{4z}\times\mathcal{T}|00\})^{6}$, where the exponent of $6$ is necessary to account for the possibility that $p4'$ is a double group (see Appendix~\ref{sec:appendixIntro} and Ref.~\onlinecite{BigBook}).  Because $p4$ is the unitary subgroup of $p41'$, the SSG that results from restoring $\mathcal{T}$ symmetry to $p4'$ [Eqs.~(\ref{eq:index2type3}) and~(\ref{eq:MLGp4prime})], then we refer to $p4$ as the ``unprimed'' group of $p4'$ [see Eq.~(\ref{eq:unprimed}) and the surrounding text].  An atomic and spin configuration with MLG $p4'$ is shown in Fig.~\ref{fig:siteSymmetry}(b).  Using~\href{http://www.cryst.ehu.es/cgi-bin/cryst/programs/magget_wp.pl}{MWYCKPOS} on the BCS~\cite{BCSMag1,BCSMag2,BCSMag3,BCSMag4} for MSG 75.3 $P4'$, which is isomorphic to $p4'$ modulo $T_{z}$, we obtain the coordinates and site-symmetry-group-isomorphic MPGs of the maximal Wyckoff positions of $p4'$:
\begin{eqnarray}
{\bf q}_{1a} &=& (0,0),\ M_{III,1a} = 4' = 2 \cup \mathcal{T}\left[(4)\setminus(2)\right], \nonumber \\
{\bf q}_{1b} &=& (1/2,1/2),\ M_{III,1b} = 4' = 2 \cup \mathcal{T}\left[(4)\setminus(2)\right], \nonumber \\
{\bf q}_{2c} &=& \{(1/2,0),\ (0,1/2)\},\ M_{III,2c}=G_{2c} = 2,
\label{eq:p4PrimeWyckoff}
\end{eqnarray}
where the symbols $4'$, $2$, and $4$ respectively refer to Type-III MPG 9.3.31 $4'$, Type-I MPG 3.1.6 $2$, and Type-I MPG 9.1.29 $4$.  It is important to emphasize that MLGs $p2$ and $p4'$ do not share the same Bravais lattices: the Bravis lattice of $p2$ is oblique, whereas the Bravis lattice of $p4'$ is square.  In $p2$, the sites ${\bf q}_{1b}^{p2} = (0,1/2)$ and ${\bf q}_{1c}^{p2} = (1/2,0)$ each lie in distinct, multiplicity-1, maximal Wyckoff positions.  Conversely, in $p4'$, the symmetry element $\{C_{4z}\times\mathcal{T}|00\}$ relates ${\bf q}_{1b}^{p2} = (0,1/2)$ and ${\bf q}_{1c}^{p2} = (1/2,0)$, causing the two sites to merge into a single, multiplicity-2, maximal Wyckoff position in $p4'$ [${\bf q}_{2c}$ in Eq.~(\ref{eq:p4PrimeWyckoff})].  All of the site-symmetry groups of $p4'$ are index-$2$ subgroups of the site-symmetry groups of $p41'$ [Eqs.~(\ref{eq:maximalTSymmetric}) and~(\ref{eq:lessMaximalTSymmetric})].  However, unlike previously in $p4$ [Eq.~(\ref{eq:p4Wyckoff})], some of the site-symmetry groups in Eq.~(\ref{eq:p4PrimeWyckoff}) are isomorphic to Type-III MPGs ($M_{III,1a}$ and $M_{III,1b}$), whereas others are isomorphic to Type-I MPGs ($G_{2c}$).  Crucially, for the site-symmetry groups $M_{III,{\bf q}}$ in Eq.~(\ref{eq:p4PrimeWyckoff}) that are isomorphic to Type-III MPGs, the ``unprimed'' site-symmetry groups $G_{\bf q}$ [\emph{i.e.} the site-symmetry groups that result from disregarding $\mathcal{T}$ symmetry, see Eq.~(\ref{eq:unprimed}) and the surrounding text] are still isomorphic to the unitary subgroups of the nonmagnetic site-symmetry groups $M_{II,{\bf q}}$ of $p41'$ [Eqs.~(\ref{eq:maximalTSymmetric}) and~(\ref{eq:lessMaximalTSymmetric})].  Specifically, at the $1a$ and $1b$ positions of $p4'$, the site-symmetry groups $M_{III,1a}$ and $M_{III,1b}$ are both isomorphic to Type-III MPG 9.3.31 $4'$, whose unprimed group is Type-I MPG 9.1.29 $4$.  Correspondingly, Type-I MPG 9.1.29 $4$ is also the unitary subgroup of Type-II SPG $41'$, to which the $1a$ and $1b$ site-symmetry groups of MLG $p41'$ are isomorphic [Eq.~(\ref{eq:maximalTSymmetric})].

\section{Small Coreps of the Little Groups and Full Coreps of the MSGs}
\label{sec:momentumSpace}

In this section, we will establish the analogous momentum-space description~\cite{QuantumChemistry,Bandrep1,Bandrep2,Bandrep3,JenFragile1,BarryFragile,BigBook,millerTables} of the MSGs, after having previously established a position-space description of the MSGs in Appendices~\ref{sec:MSGs} and~\ref{sec:SecFullWyckoff}.  To begin, for an infinite crystal that is invariant under an SSG $G$, the translation group $G_{T}$ [Eq.~(\ref{eq:translationGroup})] is a subgroup of $G$, where $G_{T}$ is generated by a set of three linearly-independent primitive translation operations:
\begin{equation}
t_{a} = \{E|{\bf t}_{a}\},\ t_{b}=\{E|{\bf t}_{b}\},\ t_{c}=\{E|{\bf t}_{c}\}.
\end{equation}
The shape of the unit (primitive) cell, and the (gray) Bravais lattice of $G$, are determined by the relative lengths and directions of~\cite{BigBook} ${\bf t}_{a,b,c}$.  Because the crystal is periodic and infinite, then it admits a reciprocal, Fourier-transformed description that is also periodic and infinite.  In reciprocal space, coordinates are indexed by crystal momentum ${\bf k}$, and the shapes of the reciprocal cells [\emph{i.e.} Brillouin zones (BZs)] are determined by the primitive reciprocal lattice vectors ${\bf K}_{a,b,c}$, which are defined for a $d$-dimensional crystal as a set of $d$ vectors $\{{\bf K}_{j}\}$ that satisfy:
\begin{equation}
{\bf t}_{i}\cdot {\bf K}_{j} = 2\pi\delta_{ij}.
\label{eq:defReciprocalLattice}
\end{equation}
As previously with the Bravais lattice vectors, the primitive reciprocal lattice vectors ${\bf K}_{a,b,c}$ of a 3D crystal must be linearly independent, but are not necessarily orthogonal (though ${\bf K}_{a,b,c}$ are indeed both linearly independent and orthogonal in many SSGs).  We note that, in some tools on the BCS, both ${\bf t}_{i}$ and ${\bf K}_{j}$ are expressed in reduced, dimensionless units in which factors of the Bravais lattice constants $a,b,c$ and BZ length [$2\pi$ in Eq.~(\ref{eq:defReciprocalLattice})] are suppressed (\emph{i.e.}, units in which $|{\bf t}_{a,b,c}|=|{\bf K}_{a,b,c}|=1$).  However, throughout this work, unless we are discussing the specific output of tools on the BCS, we will maintain the factor of $2\pi$ in Eq.~(\ref{eq:defReciprocalLattice}), though, like on the BCS, we will employ reduced units in which $a,b,c=1$ (\emph{i.e.}, units in which $|{\bf t}_{a,b,c}|=1$ and $|{\bf K}_{a,b,c}|=2\pi$).

Similar to the Wyckoff positions in real space, there are also sets of ${\bf k}$ points in momentum space that are related by the symmetries of the SSG $G$.  These ${\bf k}$ points subdivide into distinct sets, known as \emph{momentum stars}, which we will rigorously define in Appendix~\ref{sec:MKVEC}.  For this work, we have specifically implemented the~\href{http://www.cryst.ehu.es/cryst/mkvec}{MKVEC} tool on the BCS, through which users can access the momentum stars of the SSGs; examples of the output of~\href{http://www.cryst.ehu.es/cryst/mkvec}{MKVEC} are provided in Appendix~\ref{sec:MKVEC}.  As we will discuss in Appendix~\ref{sec:MKVEC},~\href{http://www.cryst.ehu.es/cryst/mkvec}{MKVEC} subsumes the earlier~\href{https://www.cryst.ehu.es/cryst/get_kvec.html}{KVEC} tool~(\url{https://www.cryst.ehu.es/cryst/get_kvec.html})~\cite{KVECmomentumWyckoff,BCS1,BCS2}, which was only capable of generating the momentum stars of the 230 Type-I (unitary) MSGs.  Additionally at each point ${\bf k}$ in the first BZ, energy states (Bloch wavefunctions) can be labeled by the irreducible ``small'' (co)reps of the little group~\cite{BigBook,Wigner1932,wigner1959group} $G_{\bf k}$, which are defined in Appendix~\ref{sec:coreps}.  One of the largest obstacles in constructing MTQC was the previous absence of a complete tabulation of the single-valued (spinless) and double-valued (spinful) small coreps of the little groups of all 1,651 SSGs.  Specifically, we cannot calculate the MEBRs (further detailed in Appendix~\ref{sec:MEBRs}), without a complete tabulation of the full (space group) coreps, which are induced from the small coreps at each of the ${\bf k}$ points in a momentum star~\cite{Bandrep1,StokesCampbell,BCS2}.  Previously, Miller and Love in Ref.~\onlinecite{millerTables} computed the single- and double-valued irreducible small (co)reps of the little groups of each MSG at high-symmetry points and along high-symmetry lines, but not along high-symmetry planes or in the BZ interior, which are required to complete the insulating compatibility relations for each MSG (Appendix~\ref{sec:compatibilityRelations}) and to compute the MEBRs (Appendix~\ref{sec:MEBRs}).  Additionally, the magnetic small (co)reps computed in Ref.~\onlinecite{millerTables} are not publicly available, are displayed in difficult-to-read tables outputted directly from computer code, and are hence difficult to verify.  For this work, building on a prescription outlined by Bradley and Cracknell in Ref.~\onlinecite{BigBook}, we have performed the first ever complete tabulation of the over 100,000 single- and double-valued small coreps at all ${\bf k}$ points and full coreps in all momentum stars of all 1,651 SSGs, which we have made freely accessible through the newly available~\href{http://www.cryst.ehu.es/cryst/corepresentations}{Corepresentations} tool on the BCS.  Representative examples of the output of~\href{http://www.cryst.ehu.es/cryst/corepresentations}{Corepresentations} are provided in Appendices~\ref{sec:corepExampleNoAnti} and~\ref{sec:corepExampleYesAnti}.  Combined with the small and full coreps previously calculated for the Type-I and II SSGs for TQC~\cite{QuantumChemistry,Bandrep1,Bandrep2,Bandrep3,JenFragile1,BarryFragile}, which can still be obtained through the~\href{http://www.cryst.ehu.es/cgi-bin/cryst/programs/representations.pl?tipogrupo=dbg}{REPRESENTATIONS DSG} tool on the BCS~(\url{http://www.cryst.ehu.es/cgi-bin/cryst/programs/representations.pl?tipogrupo=dbg}), the tools documented in this section represent the completion of over 70 years~\cite{ZamorzaevMSG,BelovMSG,ShubnikovBook,BigBook,millerTables} of group-theoretic efforts to exhaustively enumerate the coreps of the 1,651 SSGs.

Additionally, using the small coreps of the little groups of the SSGs, we can further derive the compatibility relations~\cite{Bandrep2,Bandrep3,ZakException1,ZakException2,ArisMagneticBlochOscillation,ZakCompatibility} that constrain the coreps at adjacent ${\bf k}$ points throughout the BZ.  For this work, we have implemented a new tool --  \href{https://www.cryst.ehu.es/cryst/mcomprel}{MCOMPREL} --  through which the compatibility relations between pairs of ${\bf k}$ points in any of the 1,651 SSGs can be obtained, including, for the first time, the Type-III and Type-IV MSGs.  In Appendix~\ref{sec:compatibilityRelations}, we detail the methodology employed to implement~\href{https://www.cryst.ehu.es/cryst/mcomprel}{MCOMPREL}, as well as outline some of the subtleties that arise when calculating compatibility relations in the MSGs.

\vspace{-0.16in}
\subsection{Little (Co)Groups, Momentum Stars, and the~\textsc{MKVEC} Tool}
\label{sec:MKVEC}

In this section, we will introduce the concepts of little groups, little co-groups, and momentum stars.  We will then demonstrate how the little (co)groups and momentum stars of all 1,651 SSGs can be obtained using the newly available~\href{http://www.cryst.ehu.es/cryst/mkvec}{MKVEC} tool on the BCS.  To begin, we define two points ${\bf k}$ and ${\bf k}'$ to be \emph{equivalent} if:
\begin{equation}
{\bf k} - {\bf k}' = {\bf K}_{\nu},
\label{eq:dependentKFull}
\end{equation}
where ${\bf K}_{\nu}$ is an integer-valued linear combination of the reciprocal lattice vectors ${\bf K}_{a,b,c}$ defined in Eq.~(\ref{eq:defReciprocalLattice}).  In this work, we will employ a condensed notation in which two equivalent points ${\bf k}$ and ${\bf k}'$ satisfy:
\begin{equation}
{\bf k}\equiv{\bf k}'.
\label{eq:dependentKs}
\end{equation}
Through Eqs.~(\ref{eq:dependentKFull}) and~(\ref{eq:dependentKs}), we establish a definition of \emph{inequivalent} ${\bf k}$ points in which two points ${\bf k}$ and ${\bf k}'$ are inequivalent if:
\begin{equation}
{\bf k} - {\bf k}' \neq {\bf K}_{\nu},
\label{eq:independentKFull}
\end{equation}
for all possible linear combinations of reciprocal lattice vectors ${\bf K}_{\nu}$.  We summarize Eq.~(\ref{eq:independentKFull}) with a condensed notation in which two inequivalent points ${\bf k}$ and ${\bf k}'$ satisfy:
\begin{equation}
{\bf k}\not\equiv{\bf k}'.
\label{eq:independentKs}
\end{equation}
Consider a symmetry:
\begin{equation}
g=\{\tilde{R}|{\bf v}\},
\end{equation}
where $g$ is an element of an SSG $G$.  In this work, $\tilde{R}$ denotes an operator, whereas $P_{\tilde{R}}$ denotes the $3\times 3$ matrix representation of the action of the unitary part of $\tilde{R}$ on coordinates in the basis of reciprocal lattice vectors.  Hence, $\tilde{R}$ is basis-independent, where as $P_{\tilde{R}}$ is basis-dependent.  We note that in earlier works~\cite{BigBook,BCS2}, symmetry actions have been formulated in terms of $P_{\tilde{R}}$, rather than $\tilde{R}$, requiring the introduction of distinct symmetry actions for unitary and antiunitary symmetries $g$.  As an example of $\tilde{R}$ and $P_{\tilde{R}}$, consider $h=\{m_{z}\times\mathcal{T}|{\bf 0}\}$, for which $\tilde{R} = m_{z}\times\mathcal{T}$ and $P_{\tilde{R}}=\text{diag}(1,1,-1)$ in the coordinate basis $(x,y,z)$.  At each of the ${\bf k}$ points in the first BZ of $G$, the symmetry operations $g$ act on ${\bf k}$ as:
\begin{equation}
g{\bf k} = \tilde{R}{\bf k},
\label{eq:kAction}
\end{equation}
where the tilde on $\tilde{R}$ is used to indicate that $\tilde{R}$ can be either a unitary symmetry of the form $\tilde{R}=R$ or an antiunitary symmetry of the form $\tilde{R}=R\times\mathcal{T}$.  In this work, we define two points ${\bf k}$ and ${\bf k}'$ to be \emph{dependent} if:
\begin{equation}
{\bf k}'\equiv g{\bf k},
\label{eq:newDefDependent}
\end{equation}
for any symmetry $g\in G$.  Given a point ${\bf k}$, we then define the subgroup $G_{\bf k} \subseteq G$, as the group of symmetries $g \in G_{\bf k}$ that act to return ${\bf k}$ to itself modulo reciprocal lattice vectors:
\begin{equation}
g{\bf k} \equiv {\bf k}.
\label{eq:littleGroupDef}
\end{equation}
Specifically, if $\tilde{R}$ is unitary ($\tilde{R}=R$), then Eq.~(\ref{eq:littleGroupDef}) is satisfied if:
\begin{equation}
\tilde{R}{\bf k} = R{\bf k} \equiv {\bf k},
\end{equation}
and if $\tilde{R}$ is antiunitary ($\tilde{R}=R\times\mathcal{T}$), then Eq.~(\ref{eq:littleGroupDef}) is satisfied if:
\begin{equation}
\tilde{R}{\bf k} = -R{\bf k}\equiv {\bf k}.
\label{eq:finalLuisConditionforLittle}
\end{equation} 
$G_{\bf k}$ is defined as the \emph{little group}~\cite{BigBook} of ${\bf k}$.  Because Bravais lattice translations 
 $\{E|{\bf t}_{a,b,c}\}$ leave ${\bf k}$ points invariant [Eqs.~(\ref{eq:kAction})], then $G_{\bf k}$ necessarily contains the group of lattice translations $G_{T}$ [Eq.~(\ref{eq:translationGroup})] at any point ${\bf k}$; hence, $G_{\bf k}$ is isomorphic to an SSG.  We may also define a \emph{little co-group} $\bar{G}_{\bf k}$, which is given by the (Shubnikov) point group of $G_{\bf k}$.  Because translation operations ${\bf v}$ leave ${\bf k}$ points invariant [Eq.~(\ref{eq:kAction})], then symmetries with and without translations [\emph{e.g.} twofold screw symmetry $\{C_{2z}|00\frac{1}{2}\}$ and twofold rotation symmetry $\{C_{2z}|000\}$, respectively] have the same action on ${\bf k}$ points.  However, as we will shortly discuss in Appendix~\ref{sec:coreps}, the momentum-space [small] (co)reps of $G_{\bf k}$, conversely, can differ depending on whether $G_{\bf k}$ contains symmetries with or without fractional lattice translations ${\bf v}$ [\emph{e.g.} in nonsymmorphic and symmorphic symmetry groups, respectively]~\cite{BigBook}.

\begin{figure}[b]
\includegraphics[width=0.8\columnwidth]{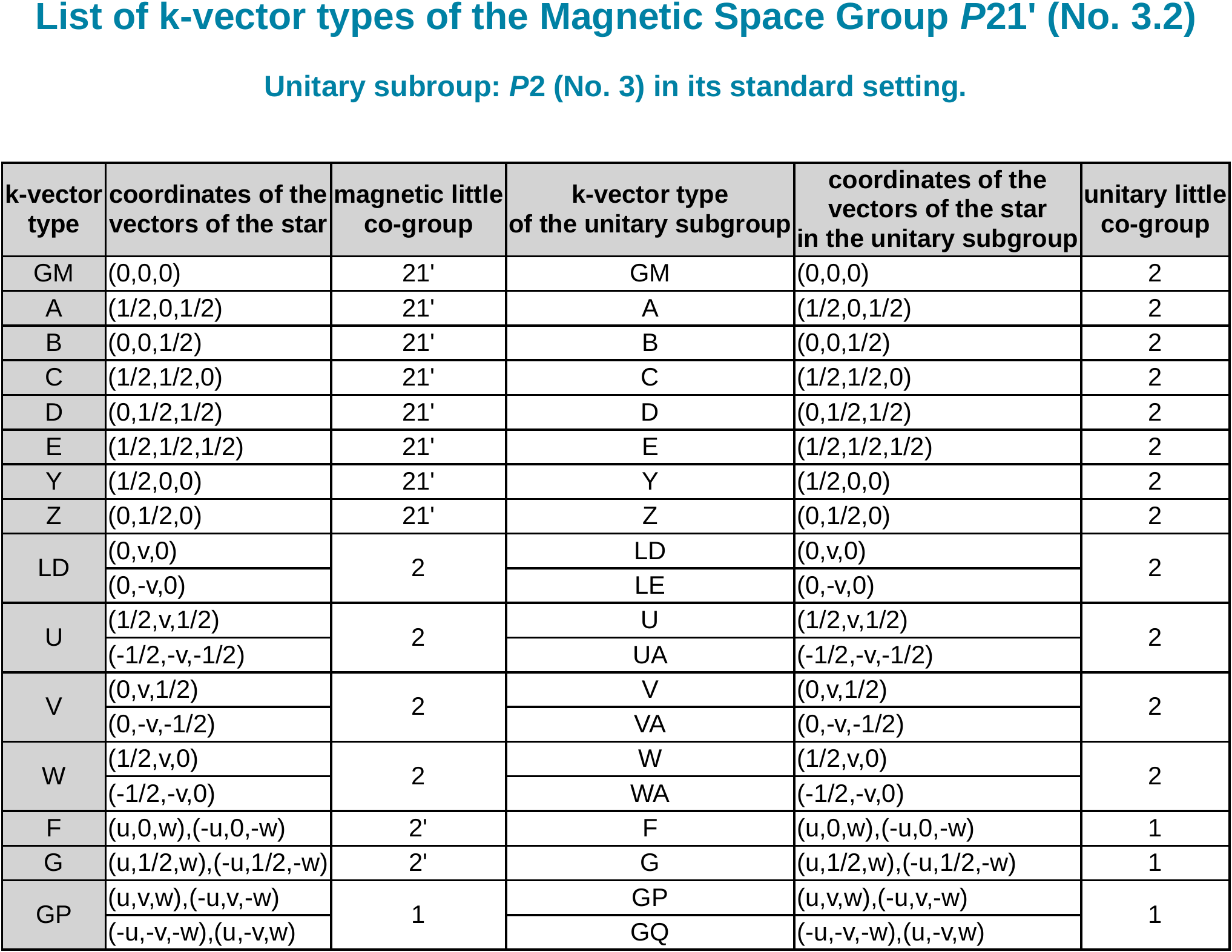}
\caption{The output of the~\href{http://www.cryst.ehu.es/cryst/mkvec}{MKVEC} tool on the BCS for Type-II SSG 3.2 $P21'$.  \href{http://www.cryst.ehu.es/cryst/mkvec}{MKVEC}, which we introduce in this work, outputs the momentum stars of all 1,651 SSGs, representing an extension of the earlier~\href{https://www.cryst.ehu.es/cryst/get_kvec.html}{KVEC} tool, which was only capable of generating the momentum stars of the 230 Type-I (unitary) MSGs.  From left to right, the columns in the output of~\href{http://www.cryst.ehu.es/cryst/mkvec}{MKVEC} list the name  (``k-vector type'') of each momentum star indexed by a point ${\bf k}$ in the first BZ of the specified SSG $G$, the coordinates of the arms of the star containing ${\bf k}$ in the standard setting (conventional cell), the little co-group $\bar{G}_{\bf k}$, the name of the vectors in the star of ${\bf k}$ in the unitary subgroup $H$ of $G$ (Appendix~\ref{sec:MSGs}), the coordinates of the arms of the star(s) in $H$ that combine to form the star of ${\bf k}$ in $G$, and the Type-I (unitary) magnetic little co-group $\bar{H}_{\bf k}$ of ${\bf k}$ in $H$.  For the labels and coordinates of the arms of each star, we have employed the convention of Stokes, Campbell, and Cordes~\cite{StokesCampbell} to be consistent with the~\href{https://stokes.byu.edu/iso/isotropy.php}{ISOTROPY} Software Suite, which was developed by Stokes, Hatch, and Campbell.  In the example of SSG 3.2 $P21'$ shown in this figure, the unitary subgroup $H$ of $G$ is isomorphic to Type-I MSG 3.1 $P2$.  In $H=P2$, there are more momentum stars (right-most three columns) than in $G=P21'$ (left-most three columns), due to the absence of $\{\mathcal{T}|{\bf 0}\}$ symmetry in $H$.  For example, in $H$ -- which is generated by $\{C_{2y}|{\bf 0}\}$ and lattice translation symmetry, LD $(0,v,0)$ and LE $(0,-v,0)$ are distinct, multiplicity-1 momentum stars; however, in $G$, LD and LE merge into a single multiplicity-2 momentum star (also named LD) $[(0,v,0),(0,-v,0)]$.}
\label{fig:simpleMKVEC}
\end{figure}

In general, given an SSG $G$ and little group $G_{\bf k} \subseteq G$, if $G\neq G_{\bf k}$, then there exists a set of symmetry elements in the subset:
\begin{equation}
\tilde{g} \in G\setminus G_{\bf k},
\label{eq:coGroupEq1}
\end{equation}
for which: 
\begin{equation}
\tilde{g}{\bf k}\not\equiv {\bf k}.
\label{eq:coGroupEq2}
\end{equation} 
Eqs.~(\ref{eq:coGroupEq1}) and~(\ref{eq:coGroupEq2}) define a set of $m$ ${\bf k}$ points $\{{\bf k}_{\gamma}\}$ in the first BZ consisting of ${\bf k}$ and all ${\bf k}'$ that are dependent on each other and on ${\bf k}$ [defined in Eq.~(\ref{eq:newDefDependent})], where the index $\gamma$ of ${\bf k}_{\gamma}$ runs from $1$ to $m$.  The set of points $\{{\bf k}_{\gamma}\}$ is known as the \emph{momentum star} of ${\bf k}$ in $G$, for which $m$ indicates the number of inequivalent ${\bf k}$ points in the star.  $m$ can alternatively be defined as the number of ${\bf k}$ points in the \emph{orbit} of ${\bf k}$, in analogy to the discussion of Wyckoff positions and symmetry sites in Appendix~\ref{sec:Wyckoff}.  In this work, to distinguish orbits in position space from symmetry-related ${\bf k}$ points in momentum space, we will refer to $m$ as the number of \emph{arms} in the star of ${\bf k}$, following the convention of Refs.~\onlinecite{BigBook,KVECmomentumWyckoff}.  From Eqs.~(\ref{eq:coGroupEq1}) and~(\ref{eq:coGroupEq2}), it follows that, for a point ${\bf k}' \equiv \tilde{g}{\bf k}$,
\begin{equation}
G_{{\bf k}'} = \tilde{g}G_{\bf k}\tilde{g}^{-1},
\label{eq:coGroupEq3}
\end{equation}
such that $G_{{\bf k}'}$ is isomorphic (and in fact conjugate) to $G_{\bf k}$.  Continuing to follow the definitions for position-space quantities established in Appendix~\ref{sec:Wyckoff}, we define two momentum stars respectively indexed by arms at ${\bf k}$ and ${\bf k}'$ to be \emph{connected} if the coordinates of any of the arms in the star of ${\bf k}$ [\emph{e.g.} the coordinate $v$ in the LD star with two arms at $(0,v,0)$ and $(0,-v,0)$ in SSG 3.2 $P21'$] can be adjusted to coincide with the coordinates of any of the arms in the star of ${\bf k}'$ [\emph{e.g.} the $\Gamma$ point $(0,0,0)$ in SSG 3.2 $P21'$, which is the only arm in its star], or vice versa.  From this, we then further define a \emph{maximal} momentum star as a momentum star indexed by an arm at ${\bf k}$ (also known as a ${\bf k}$ vector of maximal symmetry~\cite{QuantumChemistry,Bandrep1,Bandrep2,Bandrep3,JenFragile1,LuisBasicBands}) for which all connected momentum stars indexed by arms at ${\bf k}'$ have little groups $G_{{\bf k}'}$ that are proper subgroups of $G_{\bf k}$:
\begin{equation}
G_{{\bf k}'}\subset G_{\bf k},
\label{eq:maximalKvec}
\end{equation}
for all ${\bf k}'$ connected to ${\bf k}$.  We emphasize that a maximal momentum star may still have arms that lie along high-symmetry lines or planes, rather than high-symmetry ${\bf k}$ points; for example, there are maximal momentum stars with arms lying along lines and planes in SSGs that are respectively isomorphic to magnetic rod~\cite{BigBook,MagneticBook,ITCA,subperiodicTables,HingeSM} and wallpaper~\cite{WiederLayers,DiracInsulator,SteveMagnet} groups modulo translations [see the text following Eq.~(\ref{eq:translationNotation}) and the text surrounding Eq.~(\ref{eq:layerGroup})].  Because Eqs.~(\ref{eq:coGroupEq1}),~(\ref{eq:coGroupEq2}),~(\ref{eq:coGroupEq3}), and~(\ref{eq:maximalKvec}) are closely related to the definitions for real-space Wyckoff positions (Appendix~\ref{sec:Wyckoff}), then the momentum stars are sometimes also known as the ``momentum-space Wyckoff positions'' of $G$ (see Refs.~\onlinecite{KVECmomentumWyckoff,CDWWeyl} and the~\href{https://www.cryst.ehu.es/cryst/get_kvec.html}{KVEC} tool on the BCS for more information).

Prior to the completion of this work, the momentum stars and little (co)groups of the Type-I MSGs were made available on the BCS through the~\href{https://www.cryst.ehu.es/cryst/get_kvec.html}{KVEC} tool~\cite{KVECmomentumWyckoff}.  However, the earlier tool -- \href{https://www.cryst.ehu.es/cryst/get_kvec.html}{KVEC} -- only incorporated the action of unitary crystal symmetries.  In this work, we introduce a new tool -- \href{http://www.cryst.ehu.es/cryst/mkvec}{MKVEC} -- which additionally incorporates the action of the antiunitary symmetries present in Type-II, III, and IV SSGs (Appendix~\ref{sec:MSGs}).  As an example, consider the lowest-symmetry momentum stars (general momentum-space Wyckoff positions~\cite{KVECmomentumWyckoff}) in Type-I MSG 3.1 $P2$ and Type-II SSG 3.2 $P21'$ [Fig.~\ref{fig:simpleMKVEC}].  MSG 3.1 $P2$ is generated by $\{C_{2y}|{\bf 0}\}$ and 3D lattice translations, whereas SSG 3.2 $P21'$ is generated by $\{C_{2y}|{\bf 0}\}$, $\{\mathcal{T}|{\bf 0}\}$, and 3D lattice translations.  In MSG 3.1 $P2$, the lowest-symmetry star sits at generic momenta in the BZ interior, and has two arms that lie at ${\bf k}$ and $C_{2y}{\bf k}$ [Eq.~(\ref{eq:kAction})].  Conversely, in SSG 3.2 $P21'$, the lowest-symmetry star (GP in Fig.~\ref{fig:simpleMKVEC}) has \emph{four} arms, which lie at ${\bf k}$, $C_{2y}{\bf k}$, $\mathcal{T}{\bf k}$, and $C_{2y}\mathcal{T}{\bf k}$.

\begin{figure}[h]
\includegraphics[width=0.85\columnwidth]{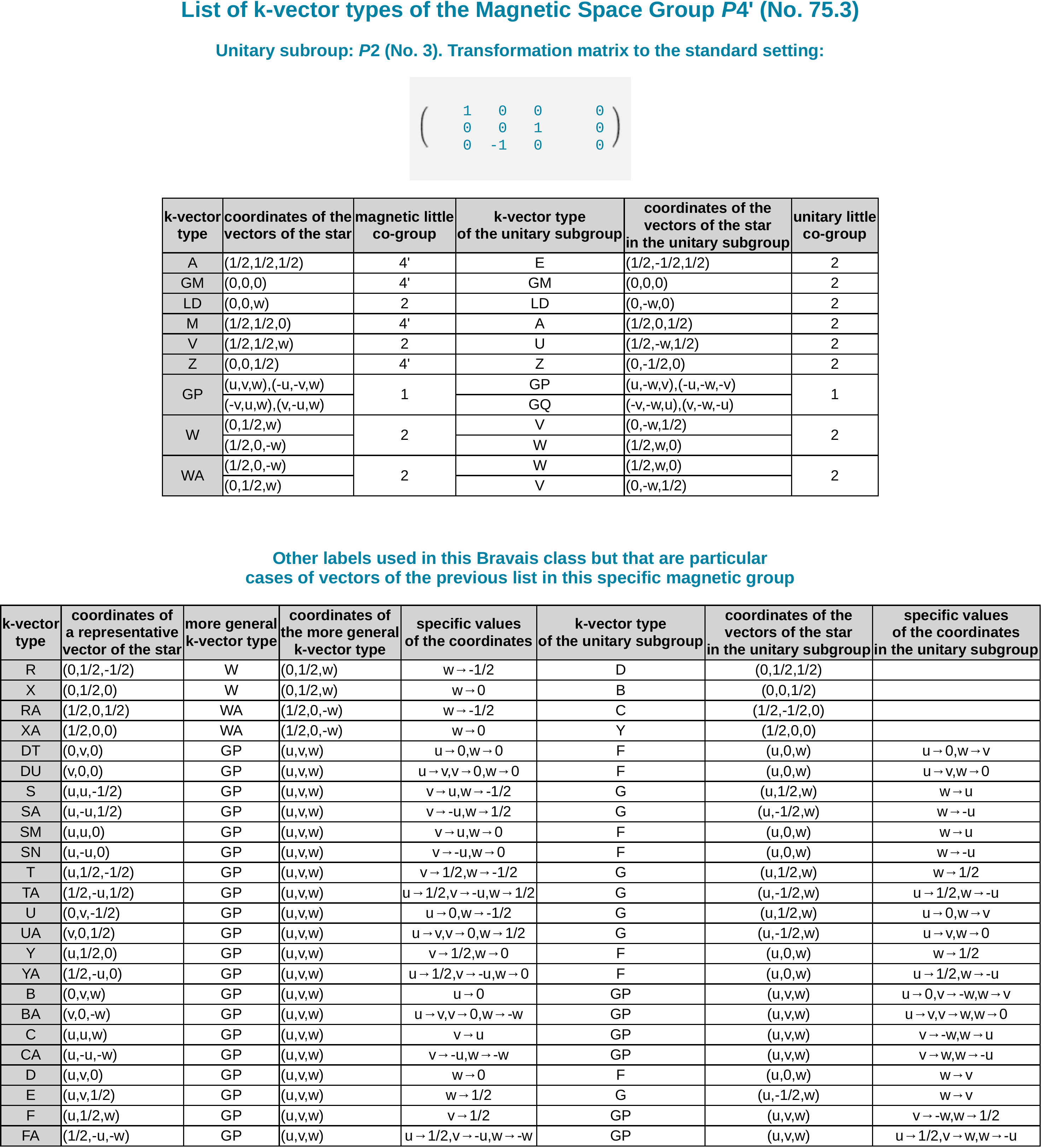}
\caption{The output of the~\href{http://www.cryst.ehu.es/cryst/mkvec}{MKVEC} tool on the BCS for Type-III MSG 75.3 $P4'$.  Unlike the previous example of SSG 3.2 $P21'$ in Fig.~\ref{fig:simpleMKVEC}, $G=P4'$ and the unitary subgroup $H$ of $G$ have different Bravais lattices.  Additionally, the unitary subgroup $H$ is generated by $\{C_{2z}|{\bf 0}\}$ and 3D lattice translation [Eq.~(\ref{eq:standardSetting2})], and is therefore isomorphic to Type-I 3.1 $P2$ in a non-standard ($z$-oriented) setting that differs from the standard ($y$-oriented) setting used throughout the BCS [see Eq.~(\ref{eq:standardSetting1}) for the definitions of standard and non-standard settings].  In~\href{http://www.cryst.ehu.es/cryst/mkvec}{MKVEC}, we account for the difference in the orientation of the twofold rotation axis between $H$ and the standard setting of MSG 3.1 $P2$ by using the $3\times 3$ $P$ matrix given by the left three columns of the gray box [Eq.~(\ref{eq:Rmatrix})].  After using the $P$ matrix to reorient the twofold rotation symmetry in MSG 3.1 $P2$ to align with the twofold axis in $H$, we then determine which of the momentum stars (\emph{e.g.} GP and GQ) in MSG 3.1 $P2$ (the three right-most columns in the top table) merge into the same momentum star (\emph{e.g.} GP) in MSG 75.3 $P4'$ (the three left-most columns in the top table).~\href{http://www.cryst.ehu.es/cryst/mkvec}{MKVEC} also refers to the \emph{Bravais classes}, which are defined in the text following Eq.~(\ref{eq:standardSetting2}).  For SSGs $G$ with fewer momentum stars than in the Bravais class of $G$,~\href{http://www.cryst.ehu.es/cryst/mkvec}{MKVEC} also outputs the bottom table, which lists additional ${\bf k}$ points that represent specific parameters for the same momentum stars in the top table chosen to coincide with distinct momentum stars in the Bravias class of $G$.  For example, in some SSGs $G$ -- such as Type-III MSG 75.3 $P4'$ in this figure -- two ${\bf k}$ points represent different parameter choices for the same star [\emph{e.g.} the X point at $(0,1/2,0)$ and the R point at $(0,1/2,-1/2)$ in the lower table represent different parameter choices for the W star in the upper table], even though the two ${\bf k}$ points lie in distinct momentum stars in the Bravais class of $G$ [which, for the example of Type-III MSG 75.3 $P4'$, is the primitive tetragonal Type-II SSG 123.340 $P4/mmm1'$].  To summarize, in Type-III MSG 75.3 $P4'$, the R and X points and W lines are all mutually \emph{connected} [defined in the text following Eq.~(\ref{eq:coGroupEq3})], and therefore appear as a single entry (W) in the top table, but the R and X points are not connected in the Bravais class of $G$ (Type-II SSG 123.340 $P4/mmm1'$), and therefore appear as distinct entries (R and X) in the bottom table.  Further details for obtaining the Bravais class of each SSG are provided in the documentation for~\href{http://www.cryst.ehu.es/cryst/mkvec}{MKVEC} on the BCS.}
\label{fig:complicatedMKVEC}
\end{figure}

In Fig.~\ref{fig:complicatedMKVEC}, we also show the output of~\href{http://www.cryst.ehu.es/cryst/mkvec}{MKVEC} for the more complicated example of Type-III MSG 75.3 $P4'$.  To explain the output of~\href{http://www.cryst.ehu.es/cryst/mkvec}{MKVEC} in Fig.~\ref{fig:complicatedMKVEC}, we must first define additional terminology.  First, in many cases, there exist multiple symmetry groups that are isomorphic to the same SSG.  For example, the MSG generated by: 
\begin{equation}
\{C_{2y}|{\bf 0}\},\ \{E|100\},\ \{E|010\},\ \{E|001\},
\label{eq:standardSetting1}
\end{equation} 
is isomorphic to the symmetry group generated by:
\begin{equation}
\{C_{2z}|{\bf 0}\},\ \{E|100\},\ \{E|010\},\ \{E|001\}.
\label{eq:standardSetting2}
\end{equation} 
Furthermore, the symmetry groups generated by the elements in Eqs.~(\ref{eq:standardSetting1}) and~(\ref{eq:standardSetting2}) are both isomorphic to Type-I MSG 3.1 $P2$.  In BCS applications, unless otherwise specified, all of the properties associated to a symmetry group are generated in a \emph{standard setting} in which the choice of rotation axes and mirror planes is fixed throughout the BCS.  For each symmetry group on the BCS, the standard setting is chosen to be the setting of the symmetry group in the \emph{International Tables for Crystallography} (Refs.~\onlinecite{ITCA,subperiodicTables}).  For example, unless otherwise specified, the properties of Type-I MSG 3.1 $P2$ are provided on the BCS in the (standard) setting in which MSG 3.1 $P2$ is generated by $\{C_{2y}|{\bf 0}\}$ and 3D lattice translations [Eq.~(\ref{eq:standardSetting1})].  In the nomenclature of this work and the BCS, the symmetry group generated by $\{C_{2z}|{\bf 0}\}$ and lattice translation [Eq.~(\ref{eq:standardSetting2})] is termed a \emph{non-standard} setting of MSG 3.1 $P2$.  Next, given an SSG $G$, we define the \emph{Bravais class} of $G$ to be the highest-symmetry, symmorphic~\cite{BigBook}, Type-II SSG with the same gray (nonmagnetic) Bravais lattice as $G$ (see Appendix~\ref{sec:MSGs}).  As discussed in Fig.~\ref{fig:complicatedMKVEC},~\href{http://www.cryst.ehu.es/cryst/mkvec}{MKVEC} compares the momentum stars of $G$ to the momentum stars of the Bravais class of $G$, and, when there is a discrepancy, outputs an additional table [the lower table in Fig.~\ref{fig:complicatedMKVEC}] indicating the specific parameters for which the momentum stars in $G$ coincide with the momentum stars in the Bravais class of $G$.

Having established definitions for standard and non-standard SSG settings and Bravais classes [Eq.~(\ref{eq:standardSetting2}) and the surrounding text], we will now analyze the output of~\href{http://www.cryst.ehu.es/cryst/mkvec}{MKVEC} for Type-III MSG 75.3 $P4'$ in Fig.~\ref{fig:complicatedMKVEC}.  $G=P4'$ is generated by:
\begin{equation}
\{C_{4z}\times\mathcal{T}|{\bf 0}\},\ \{E|100\},\ \{E|001\},
\label{eq:standardSettingC4}
\end{equation}
such that the unitary subgroup $H$ of $G$ is generated by $\{C_{2z}|{\bf 0}\}$ and 3D lattice translations, and is therefore isomorphic to Type-I MSG 3.1 in a non-standard ($z$-oriented) setting [Eq.~(\ref{eq:standardSetting2})].  Unlike in the previous example in Fig.~\ref{fig:simpleMKVEC}, there are two complications that we must consider in generating the momentum stars of the Type-III MSG $G=P4'$ from the momentum stars of a unitary (Type-I) MSG, whose momentum stars were previously computed for the earlier BCS tool~\href{https://www.cryst.ehu.es/cryst/get_kvec.html}{KVEC}~\cite{KVECmomentumWyckoff,BCS1,BCS2}.  First, in the standard setting, MSG 3.1 $P2$ is generated by $\{C_{2y}|{\bf 0}\}$ and 3D lattice translations [Eq.~(\ref{eq:standardSetting1})], as opposed to the unitary subgroup $H$ of $G=P4'$, which is isomorphic to MSG 3.1 $P2$ in a non-standard setting [see the text following Eq.~(\ref{eq:standardSettingC4})].  To begin to generate the momentum stars in $G$, we first employ a transformation matrix $P$ to convert the ${\bf k}$ points in the standard ($y$-oriented) setting of MSG 3.1 $P2$ into the non-standard ($z$-oriented) basis of $H$:
\begin{equation}
{\bf k}_{H} = P{\bf k}_{P2},
\label{eq:Rmatrix}
\end{equation}
where $P$ is the $3\times 3$ matrix in the left three columns of the gray box at the top of Fig.~\ref{fig:complicatedMKVEC}.  Next, we account for the difference in Bravais lattice between $G$ and $H$.  Specifically, $G=P4'$ has a primitive tetragonal Bravais lattice, whereas $H$ has a primitive monoclinic Bravais lattice.  Because of this, high-symmetry ${\bf k}$ points (lines) that were independent in $H$ [\emph{e.g.} $(0,1/2,w)$ and $(1/2,0,-w)$ in the upper table in Fig.~\ref{fig:complicatedMKVEC}] become merged by the symmetry $\{C_{4z}\times\mathcal{T}|{\bf 0}\}\in G$ into the same star in $G$ [\emph{e.g.} W in the left-most column of the upper table in Fig.~\ref{fig:complicatedMKVEC}].

The need for a transformation matrix $P$ [Eq.~(\ref{eq:Rmatrix}) and Fig.~\ref{fig:complicatedMKVEC}] and the difference in Bravais lattice lead to a potential ambiguity in the momentum-star labeling, namely whether we should employ the labels of an MSG (here 75.3 $P4'$) or those of the unitary subgroup [here 3.1 $P2$ in the non-standard ($z$-oriented) setting, see the text surrounding Eq.~(\ref{eq:standardSetting1})].  We note that this ambiguity does not arise in all MSGs, or at all ${\bf k}$ points -- a point ${\bf k}$ in an MSG $G$ only carries a labeling ambiguity if the ${\bf k}$ point has a different label in the Bravais lattice of $G$ than in the Bravais lattice of the unitary subgroup $H$ of $G$.  In the new tools on the BCS created for this work, we resolve a ${\bf k}$-point labeling ambiguity by continuing to label the ${\bf k}$ point using the momentum stars of $G$, while labeling the little group (small) coreps at ${\bf k}$ (which we will shortly introduce in Appendix~\ref{sec:coreps}) with \emph{both} the momentum star labels in $G$ and with the momentum star labels of the unitary (and possibly rotated) subgroup $H$ (see Fig.~\ref{fig:noAntiCoreps1} for an example of magnetic small corep labeling on the BCS).

\clearpage

\subsection{Small and Full Coreps and the~\textsc{Corepresentations} Tool}
\label{sec:coreps}

Having established the definitions of little (co)groups and momentum stars (Appendix~\ref{sec:MKVEC}), we will now in this section detail our tabulation of the small and full (co)reps of the MSGs.  At each ${\bf k}$ point in a crystal, the representations of the little group $G_{\bf k}$ can be used to characterize electronic (Bloch) wavefunctions~\cite{BigBook,Wigner1932,wigner1959group,StepanVASPIrreps}, superconducting- and magnetic-transition order parameters~\cite{CracknellMagneticTrans,samokhin2002,NormanNonsymmorphicSC,YanaseNonsymmorphicSC}, magnons~\cite{cao2015}, and Raman scattering tensors~\cite{benfatto2006}.  For the specific purposes of MTQC, we cannot derive the magnetic elementary band (co)representations without knowledge of the set of irreducible full [\emph{i.e.} space group] (co)reps in each momentum star induced from the irreducible small (co)reps in one arm of the star (see Appendix~\ref{sec:MEBRs}).  Therefore, before we can continue towards characterizing energy bands and enumerating band (co)representations across the SSGs, we must tabulate all of the small (co)reps [defined below] of each little group $G_{\bf k}$ of each ${\bf k}$ point in each of the 1,651 SSGs, which we must then use to generate the irreducible full (co)reps in each momentum star of each SSG.  Though a partial tabulation consisting of the magnetic small (co)reps at high-symmetry BZ points and along high-symmetry BZ lines was performed by Miller and Love in Ref.~\onlinecite{millerTables}, we have in this work performed the first \emph{complete} tabulation of the small (co)reps of $G_{\bf k}$ at all ${\bf k}$ points for all 1,651 single and double SSGs.

To begin, because $G_{\bf k}$ is isomorphic to an SSG [text following Eq.~(\ref{eq:littleGroupDef})], then $G_{\bf k}$ is an infinite group, and does not have a finite set of irreducible (co)reps.  Historically, several methods have been employed to extract a physically meaningful finite set of (co)reps from $G_{\bf k}$.  One option is to form a finite group from $G_{\bf k}$.  If ${\bf k}$ is an isolated high-symmetry point, then we can form the group:
\begin{equation}
^{H} G_{\bf k} = G_{\bf k} / T_{\bf k},
\label{eq:HerringsLittleGroup}
\end{equation}
where $T_{\bf k}$ is the group of translations $\{E|{\bf t}_{\mu}\}\in T_{\bf k}$ for which $\exp(-i{\bf k}\cdot {\bf t}_{\mu})=1$, and where we recall that $/$ is the set quotient [Eq.~(\ref{eq:quotientGroup})], as opposed to the set difference $\setminus$ [Eq.~(\ref{eq:setDifference})].  $^{H} G _{\bf k}$ is known as ``Herring's little group''~\cite{HerringReality,BigBook}.  At high-symmetry ${\bf k}$ points in Type-I MSGs or Type-II SSGs, it is shown in Ref.~\onlinecite{BigBook} that $^{H} G_{\bf k}$ is either isomorphic to an abstract finite point group, or to the direct product of an abstract finite point group and a 3D group of lattice translations that is a subgroup of the lattice translations of $G_{\bf k}$.  Hence, a finite number of coreps can be generated from $^{H} G_{\bf k}$ by either encountering the case in which $^{H} G_{\bf k}$ is already a finite group, or by taking  $^{H} G_{\bf k}$ modulo the remaining integer lattice translations.  The (co)reps of the abstract point subgroups of $^{H} G_{\bf k}$ for all of the ${\bf k}$ points in the single and double Type-I MSGs and Type-II SGs were exhaustively tabulated in Ref.~\onlinecite{BigBook}.  However, the abstract point subgroups of $^{H} G_{\bf k}$ for all of the ${\bf k}$ points in the single and double Type-III and Type-IV MSGs have not been calculated to date.  Additionally, when generalizing to high-symmetry BZ lines and planes, we can no longer rely on Eq.~(\ref{eq:HerringsLittleGroup}), because $G_{\bf k}/T_{\bf k}$ cannot simply be reduced to a finite group by modding out lattice translations for values of ${\bf k}$ away from high-symmetry points; a more complicated procedure involving the central extension of the little co-group $\bar{G}_{\bf k}$ may instead be employed, as detailed in Chapter 5 of Ref.~\onlinecite{BigBook}.

In this work, to avoid the complications involved with reducing $G_{\bf k}$ to a finite group, we will instead employ an alternative approach in which a finite set of (co)reps can be generated for each $G_{\bf k}$ in each SSG, regardless of whether ${\bf k}$ is a high-symmetry BZ point.  To begin, because $G_{\bf k}$ is a space group, then $G_{\bf k}$ can be expressed as a left coset decomposition with respect to the group of Bravais lattice translations $G_{T}$ [Eq.~(\ref{eq:translationGroup})]:
\begin{equation}
G_{\bf k} = \bigcup_{i}g_{i}G_{T} = G_{T} \cup \bigcup_{g_{i}\notin G_{T}}g_{i}G_{T} = G_{T} \cup \{\tilde{R}_{1}|{\bf v}_{1}\}G_{T} \cup \{\tilde{R}_{2}|{\bf v}_{2}\}G_{T} + ... , 
\label{eq:cosetOverallDecomposition}
\end{equation}
where the index $i$ in Eq.~(\ref{eq:cosetOverallDecomposition}) runs over a set of coset representatives $g_{i}=\{\tilde{R}_{i}|{\bf v}_{i}\}$ of $G_{\bf k}$ for which $g_{i}G_{T} \neq g_{j}G_{T}$ for $g_{i,j}\in G_{\bf k}$, such that each coset $g_{i}G_{T}$ is unique.  In Eq.~(\ref{eq:cosetOverallDecomposition}), we use the tilde symbol to emphasize that the symmetry operation $\tilde{R}_{i}$ can be either unitary ($\tilde{R}_{i} = R_{i}$) or antiunitary ($\tilde{R}_{i}=R_{i}\times\mathcal{T}$).  In the coset decomposition in Eq.~(\ref{eq:cosetOverallDecomposition}), $g_{i}\neq\{E|{\bf 0}\}$ in the second equality, because $\{E|{\bf 0}\}\in G_{T}$.  To motivate the coset decomposition in Eq.~(\ref{eq:cosetOverallDecomposition}), we can compare $G_{\bf k}$ to $\bar{G}_{\bf k}G_{T}$, where $\bar{G}_{\bf k}$ is the little co-group [\emph{i.e.} $\bar{G}_{\bf k}$ is the SPG obtained by setting all of the ${\bf v}_{i}\rightarrow {\bf 0}$ in Eq.~(\ref{eq:cosetOverallDecomposition}), see text following Eq.~(\ref{eq:finalLuisConditionforLittle})].  First, we define a \emph{symmorphic} SSG~\cite{BigBook} to be an SSG $G$ in which there exists a choice of origin for which each symmetry $g\in G$ takes the form $g=\{\tilde{R}|{\bf t}\}$, where $\{E|{\bf t}\}\in G_{T}$ (using the same origin for each symmetry $g$)~\cite{BigBook,ITCA}.  This implies that $G_{\bf k}=\bar{G}_{\bf k}G_{T}$ at all ${\bf k}$ points.  Hence, in symmorphic symmetry groups, we could in principle obtain a finite set of (co)reps of $G_{\bf k}$ by restricting consideration to the (co)reps of $\bar{G}_{\bf k}$.  However, in an SSG that is not symmorphic (\emph{i.e.} a \emph{nonsymmorphic} SSG), there exist ${\bf k}$ points at which $G_{\bf k}\neq\bar{G}_{\bf k}G_{T}$, providing an obstacle towards generically using $\bar{G}_{\bf k}$ to obtain finite sets of (co)reps of $G_{\bf k}$.  For example, at ${\bf k}=(0,\pi,0)$ in nonsymmorphic Type-I MSG 4.7 $P2_{1}$ -- which is generated by screw symmetry $\{C_{2y}|0\frac{1}{2}0\}$ and the lattice translations $\{E|100\}$ and $\{E|001\}$ -- $G_{\bf k}\neq\bar{G}_{\bf k}G_{T}$.  We further note that, because all Type-IV SSGs necessarily contain elements of the form $t_{0}\mathcal{T}=\{\mathcal{T}|{\bf t}_{0}\}$ for which $\{E|{\bf t}_{0}\}\notin G_{T}$ (\emph{i.e.} $t_{0}$ is a fractional lattice translation, see Appendix~\ref{sec:type4}), then all Type-IV SSGs are nonsymmorphic.

Instead, we will show below that, unlike $\bar{G}_{\bf k}$, Eq.~(\ref{eq:cosetOverallDecomposition}) will allow us to construct a prescription for obtaining a finite set of (co)reps at all ${\bf k}$ points in both symmorphic and nonsymmorphic SSGs.  First, we recognize that, even though $G_{\bf k}$ in Eq.~(\ref{eq:cosetOverallDecomposition}) is an infinite group, the number of unique cosets $g_{i}G_{T}$ of $G_{\bf k}$ is \emph{finite}.  This can be seen by recognizing that $E$ and the finite set $\{\tilde{R}_{i}\}$ in Eq.~(\ref{eq:cosetOverallDecomposition}) comprise the finite little co-group $\bar{G}_{\bf k}$.  Next, we recall that $G_{\bf k}$ is isomorphic to an SSG, implying that, in principle, there exist infinitely many (co)reps of $G_{\bf k}$.  We therefore impose an additional restriction to (co)reps $\sigma$ of $G_{\bf k}$ (not necessarily irreducible) for which lattice translations $t_{\mu}=\{E|{\bf t}_{\mu}\}$ have the matrix representatives:
\begin{equation}
\Delta_{\sigma}(t_{\mu}) =  e^{-i{\bf k}\cdot{\bf t}_{\mu}}\mathds{1}_{\chi_{\sigma}(\{E|{\bf 0}\})},
\label{eq:translationRestriction}
\end{equation}
where $\mathds{1}_{\chi_{\sigma}(\{E|{\bf 0}\})}$ is the $\chi_{\sigma}(\{E|{\bf 0}\})$-dimensional identity matrix.  Eq.~(\ref{eq:translationRestriction}) implies that, given two symmetries $g_{i}\in G_{\bf k}$ and $t_{\mu}g_{i}\in G_{\bf k}$ in the same coset $g_{i}G_{T}$, where $t_{\mu}=\{E|{\bf t}_{\mu}\}$ and $t_{\mu}\in G_{T}$, the matrix representatives $\Delta_{\sigma}(g_{i})$ and $\Delta_{\sigma}(t_{\mu}g_{i})$ in $\sigma$ in Eq.~(\ref{eq:translationRestriction}) -- which is termed a \emph{small}~\cite{BCS1,BigBook,StokesCampbell,Bandrep1} (co)rep of $G_{\bf k}$ -- are related by an overall (Bloch) phase~\cite{BigBook}.  Specifically:
\begin{equation}
\Delta_{\sigma}(t_{\mu}g_{i}) = e^{-i{\bf k}\cdot{\bf t}_{\mu}}\Delta_{\sigma}(g_{i}),
\label{eq:smallCorepDef}
\end{equation}
such that $\Delta_{\sigma}(t_{\mu}g_{i})$ and $\Delta_{\sigma}(g_{i})$ are unitarily equivalent.  Using Eqs.~(\ref{eq:cosetOverallDecomposition}),~(\ref{eq:translationRestriction}), and~(\ref{eq:smallCorepDef}), we can then extract a finite set of irreducible small (co)reps from $G_{\bf k}$ by restricting focus to the indecomposable small (co)reps whose matrix representatives are not related by an overall phase, or any other unitary transformation.  Specifically, we first define two (co)reps $\sigma$ and $\sigma'$ of a little group $G_{\bf k}$ to be \emph{equivalent} if there exists a unitary matrix $N$ that relates the matrix representatives $\Delta_{\sigma}(g)$ and $\Delta_{\sigma'}(g)$:
\begin{equation}
\Delta_{\sigma}(g) = N\Delta_{\sigma'}(g)N^{\dag},
\label{eq:Nequivalence}
\end{equation}
for all $g\in G_{\bf k}$ (in which the same matrix $N$ is used for all $g\in G_{\bf k}$).  Then, using Eq.~(\ref{eq:Nequivalence}), we define the irreducible small (co)reps of $G_{\bf k}$ as the finite set of inequivalent (co)reps of $G_{\bf k}$ that cannot be expressed as direct sums of each other and for which the matrix representatives of integer lattice translations take the form of Eq.~(\ref{eq:translationRestriction}).  We further note that, at high-symmetry ${\bf k}$ points, the small (co)reps of $G_{\bf k}$ are equivalent to the (co)reps of $^{H}G_{\bf k}$ [modulo lattice translations, see the text following Eq.~(\ref{eq:HerringsLittleGroup})], and, along high-symmetry BZ lines, the small (co)reps of $G_{\bf k}$ are equivalent to the (co)reps of the central extension of the little co-group $\bar{G}_{\bf k}$ (see Chapter 5 in Ref.~\onlinecite{BigBook} for a detailed discussion of the role of the central extension in the group theory of crystalline solids).  For Type-I and Type-II SSGs, the little group small (co)reps were previously tabulated by Bradley and Cracknell~\cite{BigBook}, and were reconstructed in the~\href{http://www.cryst.ehu.es/cgi-bin/cryst/programs/representations.pl?tipogrupo=dbg}{REPRESENTATIONS DSG} tool on the BCS for TQC~\cite{QuantumChemistry,Bandrep1,Bandrep2,Bandrep3,JenFragile1,BarryFragile}.  Conversely, there have been relatively few previous attempts to exhaustively tabulate the small coreps of the Type-III and Type-IV MSGs in an accessible form, though a partial tabulation consisting of the magnetic small (co)reps at high-symmetry BZ points and along high-symmetry BZ lines was performed by Miller and Love in Ref.~\onlinecite{millerTables} using little group decompositions of the form of Eq.~(\ref{eq:cosetOverallDecomposition}).  In this work, we have, for the first time, performed a complete tabulation of the small (co)reps of the little group $G_{\bf k}$ at each ${\bf k}$ point in each of the 1,651 SSGs, which we have made accessible through the~\href{http://www.cryst.ehu.es/cryst/corepresentations}{Corepresentations} tool on the BCS.  Across all of the momentum stars of the 1,651 single and double SSGs, the completion of~\href{http://www.cryst.ehu.es/cryst/corepresentations}{Corepresentations} required the computation of over 100,000 single- and double-valued small (co)reps.  In the text below, we will detail our methodology for tabulating the small (co)reps; in Appendices~\ref{sec:corepExampleNoAnti} and~\ref{sec:corepExampleYesAnti}, we will additionally provide representative examples of the output of~\href{http://www.cryst.ehu.es/cryst/corepresentations}{Corepresentations}.

To complete our derivation of the little group small (co)reps, we return to the coset decomposition in Eq.~(\ref{eq:cosetOverallDecomposition}).  First, we recognize that, if $G_{\bf k}$ is isomorphic to a Type-I MSG, then its small (co)reps can already be obtained from either the tables in Ref.~\onlinecite{BigBook} or through the earlier~\href{http://www.cryst.ehu.es/cgi-bin/cryst/programs/representations.pl?tipogrupo=dbg}{REPRESENTATIONS DSG} tool on the BCS~\cite{QuantumChemistry,Bandrep1,Bandrep2,Bandrep3,JenFragile1,BarryFragile}, and no further calculations are required.  Next, we consider the more complicated case in which $G_{\bf k}$ is isomorphic to a Type-II, III, or IV SSG.  In this case, $G_{\bf k}$ necessarily contains antiunitary elements, and therefore admits a decomposition of the form:
\begin{equation}
G_{\bf k} = H_{\bf k} \cup \tilde{g}_{A}H_{\bf k},
\label{eq:antiBreakdown}
\end{equation}
where $H_{\bf k}$ is the maximal unitary (index-2, see Appendices~\ref{sec:type2},~\ref{sec:type3}, and~\ref{sec:type4}) subgroup of $G_{\bf k}$, and $\tilde{g}_{A}$ is an antiunitary symmetry operation of the form:
\begin{equation}
\tilde{g}_{A} = \{R\times\mathcal{T}|{\bf v}\},
\label{eq:representativeAnti}
\end{equation}
where $\tilde{g}_{A}$ is known as the ``representative'' antiunitary symmetry operation, $R$ is a unitary point-group symmetry element (proper or improper rotation or the identity $E$), and either ${\bf v}={\bf 0}$ or ${\bf v}$ is a fractional lattice translation.  As discussed earlier in Appendices~\ref{sec:type2},~\ref{sec:type3}, and~\ref{sec:type4} and summarized in Table~\ref{tb:antiSym}, Type-II, III, and IV SSGs are distinguished by the form of $R$ and ${\bf v}$ in Eq.~(\ref{eq:representativeAnti}).

\begin{table}[h]
\begin{tabular}{|c|c|c|}
\hline
\multicolumn{3}{|c|}{SSG Definitions in Terms of Eqs.~(\ref{eq:antiBreakdown}) and~(\ref{eq:representativeAnti})} \\ 
\hline
SSG Type & Condition on $R$ & Condition on ${\bf v}$ \\
\hline
\hline
Type-II SSG & $R=E$ & ${\bf v} = {\bf t}_{\mu}$ \\
\hline 
Type-III MSG & $R\neq E$ & No constraint \\
\hline
Type-IV MSG & $R=E$ & ${\bf v}\neq {\bf t}_{\mu}$, ${\bf v}^{2} = {\bf t}_{\mu}$ \\
\hline
\end{tabular}
\caption{Definitions of the SSGs with antiunitary symmetry operations (Types-II, III, and IV, respectively defined in Appendices~\ref{sec:type2},~\ref{sec:type3}, and~\ref{sec:type4}). $E$ is the identity operation, and ${\bf t}_{\mu}$ is a Bravais lattice vector, such that $\{E|{\bf t}_{\mu}\}\in G_{T}$ [Eq.~(\ref{eq:translationGroup})].}
\label{tb:antiSym}
\end{table}

Next, for each of the cosets on the right-hand side of Eq.~(\ref{eq:cosetOverallDecomposition}) [including $G_{T}$ itself], we choose one element to place into a set $\tilde{G}_{\bf k}$.  In this work, we specifically choose the identity element $\{E|{\bf 0}\}$ from $G_{T}$, and then, from each coset $g_{i}G_{T}$, we choose one element $g_{i} = \{\tilde{R}_{i}|{\bf v}_{i}\}$ for which each component of the translation ${\bf v}_{i}$ is chosen to satisfy $|{\bf v}_{i}\cdot {\bf t}_{a,b,c}|<1$ (in reduced units where the lattice constants $a,b,c=1$), such that either ${\bf v}={\bf 0}$ or ${\bf v}_{i}$ is a specific fractional lattice translation for which $g_{i} = \{\tilde{R}_{i}|{\bf v}_{i}\}$ is an element of the little group $G_{\bf k}$.  We note that, if $G_{\bf k}$ is isomorphic to a symmorphic SSG [defined in the text following Eq.~(\ref{eq:cosetOverallDecomposition})], then $\tilde{G}_{\bf k}$ becomes a finite group [specifically, $\tilde{G}_{\bf k}=\bar{G}_{\bf k}$ if $G_{\bf k}$ is symmorphic, where $\bar{G}_{\bf k}$ is the little co-group, see the text following Eq.~(\ref{eq:finalLuisConditionforLittle})].  We note that, in this section, we will always consider the more general case in which $\tilde{G}_{\bf k}$ is a set, and not necessarily a group.  Using $\tilde{H}_{\bf k}$ -- the maximal unitary subset of $\tilde{G}_{\bf k}$ -- we can re-express Eq.~(\ref{eq:cosetOverallDecomposition}) for a Type-II, III, or IV little group $G_{\bf k}$ as:
\begin{equation}
G_{{\bf k}} = \tilde{G}_{\bf k}G_{{\bf k}} = \bigcup_{i}h_{i}G_{T} \cup \bigcup_{i}g_{A,i}G_{T} = \left(\tilde{H}_{{\bf k}} \cup \tilde{g}_{A}\tilde{H}_{\bf k}\right)G_{T},
\label{eq:CorepCoset}
\end{equation}
where $\tilde{g}_{A}$ is the representative antiunitary symmetry operation in Eq.~(\ref{eq:representativeAnti}), and where the index $i$ in Eq.~(\ref{eq:CorepCoset}) runs over all unique unitary ($h_{i}G_{T}$) and antiunitary ($g_{A,i}G_{T}$) cosets of $G_{\bf k}$.  Bradley and Cracknell outline a convention~\cite{BigBook} for choosing $\tilde{g}_{A}$ (for example, in Type-II little groups, the most natural choice is $\tilde{g}_{A} = \{\mathcal{T}|{\bf 0}\}$); however, below, we will employ a more general procedure that is independent of the form of $\tilde{g}_{A}$.  Because all $g_{A,i}\in \tilde{g}_{A}\tilde{H}_{\bf k}$ in Eq.~(\ref{eq:CorepCoset}) are antiunitary, and therefore do not have well-defined characters in any small corep of $G_{\bf k}$ (where the \emph{character} $\chi_{\sigma}(h)$ of a unitary symmetry $h$ in the corep $\sigma$ is defined~\cite{BigBook} as $\Tr[\Delta_{\sigma}(h)]$), then it is straightforward to see that the set of small coreps of $G_{\bf k}$ can only be formed from the small irreps of its unitary subgroup $H_{\bf k}$, which may become paired by the action of the elements $g_{A,i}\in \tilde{g}_{A}\tilde{H}_{\bf k}$.  We note that it is not possible for the irreducible small coreps of $G_{\bf k}$ to be composed of more than two irreps of $H_{\bf k}$, because $H_{\bf k}$ is either isomorphic to $G_{\bf k}$ (\emph{i.e.} $G_{\bf k}$ is isomorphic to a Type-I MSG, see Appendix~\ref{sec:type1}), or $H_{\bf k}$ is an index-2 subgroup of $G_{\bf k}$ (\emph{i.e.} $G_{\bf k}$ is isomorphic to a Type-II, III, or IV SSG, see Appendices~\ref{sec:type2},~\ref{sec:type3}, and~\ref{sec:type4}, respectively).

Given a small irrep $\sigma$ of $H_{\bf k}$ with a matrix representative $\Delta_{\sigma}(h)$ for each symmetry $h\in \tilde{H}_{\bf k}$, we next define a matrix:
\begin{equation}
\bar{\Delta}_{\sigma}(h) = \left[\Delta_{\sigma}(\tilde{g}_{A}^{-1}h\tilde{g}_{A})\right]^{*}. 
\label{eq:otherCorepMatrix}
\end{equation}
As shown by Bradley and Cracknell~\cite{BigBook}, the small coreps $\tilde{\sigma}$ of $G_{\bf k}$ can only take one of three forms, which we designate as ``types'' (a), (b), and (c): 
\begin{itemize}
\item{Type (a):  $\bar{\Delta}_{\sigma}(h)$ is equivalent to $\Delta_{\sigma}(h)$, such that $\Delta_{\sigma}(h) = N\bar{\Delta}_{\sigma}(h)N^{\dag}$ for all $h\in\tilde{H}_{\bf k}$ [Eq.~(\ref{eq:Nequivalence})].  Additionally, for coreps of type (a), the antiunitary matrix representative $\Delta_{\sigma}(\tilde{g}_{A}) = NK$, where $K$ is complex conjugation, carries the property that $\Delta_{\sigma}(\tilde{g}_{A}^{2})=[\Delta_{\sigma}(\tilde{g}_{A})]^{2}=NN^{*}=N^{2}$ [which is well defined, because $\tilde{g}_{A}^{2} \in \tilde{H}_{\bf k}G_{T}$ in Eq.~(\ref{eq:CorepCoset})].  For coreps of type (a), this implies that:
\begin{equation}
\tilde{\sigma}\equiv \sigma, 
\label{eq:typeA}
\end{equation}
such that the small corep $\tilde{\sigma}$ of $G_{\bf k}$ is equivalent to a small irrep $\sigma$ of $H_{\bf k}$.  However, because $G_{\bf k}$ and $H_{\bf k}$ are different symmetry groups, then the equivalence between $\tilde{\sigma}$ and $\sigma$ is defined differently than the equivalence that we previously defined between (co)reps of the same symmetry group [see the text surrounding Eq.~(\ref{eq:Nequivalence})].  Specifically, in this work, we define an irrep $\sigma$ of a Type-I (unitary) symmetry group $H_{\bf k}$ and a corep $\tilde{\sigma}$ of an index-2 Type-II, III, or IV (antiunitary) supergroup $G_{\bf k}$ of $H_{\bf k}$ to be equivalent if there exists a unitary matrix $N$ that relates the matrix representatives $\Delta_{\sigma}(h)$ and $\Delta_{\tilde{\sigma}}(h)$ by $\Delta_{\sigma}(h) = N\Delta_{\tilde{\sigma}}(h)N^{\dag}$ for all of the \emph{unitary} symmetries $h\in H_{\bf k}$, $h\in G_{\bf k}$ (where the same matrix $N$ is used for all $h\in H_{\bf k}$, $h\in G_{\bf k}$).  In nonmagnetic (Type-II) SSGs, type (a) coreps are most familiarly encountered at ${\bf k}$ points with real symmetry eigenvalues in the absence of SOC.  For example, at ${\bf k}={\bf 0}$ in Type-II SSG 2.5 $P\bar{1}1'$ in the absence of SOC, which is generated by $\{\mathcal{I}|{\bf 0}\}$, $\{\mathcal{T}|{\bf 0}\}$, and 3D lattice translations, $G_{\bf k}$ has two, one-dimensional, single-valued small coreps that each correspond to a singly degenerate, $\mathcal{T}$-invariant Bloch state (per spin)~\cite{BigBook}.  Type (a) coreps also exist in nonmagnetic SSGs $G$ in the presence of SOC at $\mathcal{T}$-invariant ${\bf k}$ points with complex-conjugate pairs of spinful symmetry eigenvalues that are already paired by unitary crystal symmetries in the unitary subgroup $H_{\bf k}$ of $G_{\bf k}$.  For example, at ${\bf k}={\bf 0}$ in Type-II SSG 25.57 $Pmm2$ in the presence of SOC, which is generated by $\{m_{x}|{\bf 0}\}$, $\{m_{y}|{\bf 0}\}$,  $\{\mathcal{T}|{\bf 0}\}$, and 3D lattice translations, $G_{\bf k}$ has one, two-dimensional small corep that is equivalent to a two-dimensional small irrep $\sigma$ of $H_{\bf k}$ with complex-conjugate pairs of $m_{x,y}$ eigenvalues due to the anticommutation relation $\{\Delta_{\sigma}(m_{x}),\Delta_{\sigma}(m_{y})\}=0$.}
\item{Type (b): $\bar{\Delta}_{\sigma}(h)$ is equivalent to $\Delta_{\sigma}(h)$ for all $h\in \tilde{H}_{\bf k}$, where equivalence continues to be defined by Eq.~(\ref{eq:Nequivalence}).  However, for coreps of type (b), $\Delta_{\sigma}(\tilde{g}_{A}^{2}) =  NN^{*} = -N^{2}$, implying through Kramers' theorem that:
\begin{equation}
\tilde{\sigma} = \sigma \oplus \sigma\equiv\sigma\sigma,
\label{eq:typeB}
\end{equation}
such that the small corep $\tilde{\sigma}$ of $G_{\bf k}$ is formed from pairing two copies of the same small irrep $\sigma$ of $H_{\bf k}$.  We further note that, because $\tilde{g}_{A}$ exchanges the two irreps $\sigma$ that comprise $\tilde{\sigma}$ in Eq.~(\ref{eq:typeB}), then the matrix representative $\Delta_{\sigma}(\tilde{g}_{A})$ is itself undefined for a single irrep $\sigma$.  Instead for coreps $\tilde{\sigma}$ of type (b), the antiunitary matrix representative $\Delta_{\tilde{\sigma}}(\tilde{g}_{A})$ is only well-defined in the larger space of the two irreps $\sigma$ in $\tilde{\sigma}$, in which the unitary part of $\Delta_{\tilde{\sigma}}(\tilde{g}_{A})$ is block-off-diagonal.  In nonmagnetic (Type-II) SSGs, type (b) coreps are most familiarly encountered at ${\bf k}$ points with real symmetry eigenvalues in the presence of SOC.  For example, at ${\bf k}={\bf 0}$ in Type-II SSG 2.5 $P\bar{1}1'$ in the presence of SOC, which is generated by $\{\mathcal{I}|{\bf 0}\}$, $\{\mathcal{T}|{\bf 0}\}$, and 3D lattice translations, $G_{\bf k}$ has two, two-dimensional, double-valued small coreps that each correspond to a doubly-degenerate (Kramers) pair of Bloch states with two parity $(\mathcal{I})$ eigenvalues of the same sign~\cite{BigBook}.}
\item{Type (c): $\bar{\Delta}_{\sigma}(h)$ is \emph{not} equivalent to $\Delta_{\sigma}(h)$ [\emph{i.e.}, there does not exist a matrix $N$ that satisfies Eq.~(\ref{eq:Nequivalence}) for all of the symmetries $h\in \tilde{H}_{\bf k}$].  Instead, $\bar{\Delta}_{\sigma}(h)$ is equivalent to $\Delta_{\sigma'}(h)$, where $\sigma'$ is a \emph{different} small irrep of $H_{\bf k}$ than $\sigma$.  This implies that:
\begin{equation}
\tilde{\sigma}=\sigma\oplus\sigma'\equiv\sigma\sigma',
\label{eq:typeC}
\end{equation}
such that the small corep $\tilde{\sigma}$ of $G_{\bf k}$ is formed from pairing two different small irreps $\sigma$ and $\sigma'$ of $H_{\bf k}$.  Unlike in coreps of type (a) or type (b), there is no constraint on the form of the matrix representative $\Delta_{\sigma}(\tilde{g}_{A}^{2})$ in coreps of type (c).  However, like previously in Eq.~(\ref{eq:typeB}), the unitary part of $\Delta_{\tilde{\sigma}}(\tilde{g}_{A})$ for a type (c) corep $\tilde{\sigma}$ [Eq.~(\ref{eq:typeC})] is off-diagonal in the block basis of $\sigma$ and $\sigma'$, and $\Delta_{\sigma}(\tilde{g}_{A})$ cannot by itself be defined for a single irrep $\sigma$ or $\sigma'$.  In nonmagnetic (Type-II) SSGs, type (c) coreps are most familiarly encountered at ${\bf k}$ points with complex symmetry characters in $H_{\bf k}$, whether or not SOC is taken into consideration.  For example, at ${\bf k}={\bf 0}$ in Type-II SSG 6.19 $Pm1'$ in the presence of SOC, which is generated by $\{m_{y}|{\bf 0}\}$, $\{\mathcal{T}|{\bf 0}\}$, and 3D lattice translations, $G_{\bf k}$ has one, two-dimensional, double-valued small corep that corresponds to a doubly-degenerate (Kramers) pair of Bloch states with a complex-conjugate ($\pm i$) pair of $m_{y}$ eigenvalues~\cite{BigBook}.}
\end{itemize}

The above definitions seem to imply that the type of small corep $\tilde{\sigma}$ induced in $G_{\bf k}$ can only be determined through a careful selection of $\tilde{g}_{A}$ in Eq.~(\ref{eq:CorepCoset}), followed by an exhaustive search for equivalence matrices $N$ that satisfy Eq.~(\ref{eq:Nequivalence}).  However, as shown by Bradley and Cracknell~\cite{BigBook}, we can also diagnose the type of the induced corep simply by calculating the modified Frobenius-Schur indicator~\cite{frobenius1906,FultonHarris,DimmockWheelerFrobenius} [\emph{c.f.} Eq.~(7.3.48) in Ref.~\onlinecite{BigBook}]:
\begin{equation}
J_{\sigma} = \sum_{i}\chi_{\sigma}(g_{A,i}^{2}),
\label{eq:Jtest}
\end{equation}
where $\chi_{\sigma}(h)$ is the character of the unitary symmetry operation $h_{i}=g_{A,i}^{2}$, $h_{i}\in H_{\bf k}$ in the small irrep $\sigma$ of $H_{\bf k}$ [which is equal to the trace of the matrix representative $\Delta_{\sigma}(h_{i})$], and where the summation in Eq.~(\ref{eq:Jtest}) runs over all of the antiunitary coset representatives in Eq.~(\ref{eq:CorepCoset}) (\emph{i.e.}, all of the distinct elements $g_{A,i}\in \tilde{g}_{A}\tilde{H}_{\bf k}$).  Because $\sigma$ in Eq.~(\ref{eq:Jtest}) is an irrep, $J_{\sigma}$ can only assume one of three values~\cite{frobenius1906,BigBook,FultonHarris}:
\begin{equation}
J_{\sigma} = \begin{cases} &\ \ |\tilde{H}_{\bf k}|,\ \sigma\text{ induces a small corep $\tilde{\sigma}$ of type (a) in }G_{\bf k} \\
&-|\tilde{H}_{\bf k}|,\ \sigma\text{ induces a small corep $\tilde{\sigma}$ of type (b) in }G_{\bf k} \\
&\ \ \ \ \ \ 0,\ \sigma\text{ induces a small corep $\tilde{\sigma}$ of type (c) in }G_{\bf k} \\
\end{cases},
\label{eq:Jtypes}
\end{equation}
where $|\tilde{H}_{\bf k}|$ is the number of elements [see the text following Eq.~(\ref{eq:quotientGroup})] in the set $\tilde{H}_{\bf k}$ [Eq.~(\ref{eq:CorepCoset})].  In a Type-II little group, $\tilde{g}_{A}H_{\bf k} = \{\mathcal{T}|{\bf 0}\}H_{\bf k}$, such that:
\begin{equation}
J^{II}_{\sigma} = \sgn\left[\chi_{\sigma}(\mathcal{T}^{2})\right]\sum_{i}\chi_{\sigma}(h_{i}^{2}),
\label{eq:Herring}
\end{equation}
where the summation in Eq.~(\ref{eq:Herring}) runs over all of the unitary coset representatives in Eq.~(\ref{eq:CorepCoset}) (\emph{i.e.}, all of the elements $h_{i}\in \tilde{H}_{\bf k}$).  We note that Eq.~(\ref{eq:Herring}) is the well-established Herring test~\cite{HerringReality,BigBook} (\emph{i.e.} the  standard Frobenius-Schur indicator~\cite{frobenius1906,BigBook,FultonHarris}) for determining the ``reality'' of $\sigma$ in a nonmagnetic (Type-II) symmetry group.  However, for little groups that are isomorphic to Type-III and Type-IV MSGs, there is no analogous simple relationship between the reality of $\sigma$ and the type of $\tilde{\sigma}$, and the more general formulas in Eqs.~(\ref{eq:Jtest}) and~(\ref{eq:Jtypes}) must be employed to determine the type of $\tilde{\sigma}$.  To confirm our complete calculation of all of the small coreps $\tilde{\sigma}$ of the SSGs, we have performed both of the independent analyses detailed in this section.  Specifically, for all of the unitary subgroup small irreps $\sigma$ and induced small coreps $\tilde{\sigma}$ of the little groups $G_{\bf k}$ at all ${\bf k}$ points in all 1,651 single and double SSGs, we have checked for all possible equivalences between $\Delta_{\sigma}(h)$ and $\bar{\Delta}_{\sigma}(h)$ [Eq.~(\ref{eq:otherCorepMatrix}) and the surrounding text], and we have confirmed that the results agree with the values of $J_{\sigma}$ [Eqs.~(\ref{eq:Jtest}) and~(\ref{eq:Jtypes})].  We will shortly provide in Appendix~\ref{sec:corepExampleYesAnti} an example of the explicit computation of $J_{\sigma}$ [Eq.~(\ref{eq:Jtypes})] in a magnetic little group.

In addition to calculating the small (co)reps of the little groups of the MSGs, we have also calculated, for the first time, the \emph{full} (co)reps of each momentum star of each MSG.  Whereas each small (co)rep is a representation of the little group $G_{\bf k}$ at a point ${\bf k}$, each full (co)rep is a representation of the entire SSG $G$ in the momentum star indexed by ${\bf k}$ (Appendix~\ref{sec:MKVEC}).  To calculate the full (co)reps, we adapt the procedure employed in Refs.~\onlinecite{Bandrep1,StokesCampbell,BCS2} to the most general case of a magnetic or nonmagnetic SSG $G$.  First, we recognize that, given a little group $G_{\bf k}\subseteq G$, there may exist a set of symmetries:
\begin{equation}
\tilde{g}\in G\setminus G_{\bf k},
\label{eq:missingSymmetriesForFull}
\end{equation}
for which:
\begin{equation}
\tilde{g}{\bf k}\equiv{\bf k}'\not\equiv {\bf k},
\end{equation}
such that ${\bf k}$ and ${\bf k}'$ lie in different arms of the same momentum star in $G$.  Because the little group $G_{{\bf k}'}$ is conjugate to $G_{\bf k}$ [Eq.~(\ref{eq:coGroupEq3})], then the (co)reps at ${\bf k}$ and ${\bf k}'$ are not independent.  Specifically, if there exists a Bloch eigenstate at ${\bf k}$ labeled by a (co)rep $\tilde{\sigma}_{\bf k}$ of $G_{\bf k}$, then there must also exist a Bloch eigenstate at ${\bf k}'$ labeled by a (co)rep $\tilde{\sigma}_{{\bf k}'}$ of $G_{{\bf k}'}$.  For $\tilde{\sigma}_{\bf k}$ and $\tilde{\sigma}_{{\bf k}'}$, the matrix representatives of each unitary symmetry $h\in \tilde{H}_{\bf k}$ and $\tilde{g}h\tilde{g}^{-1}\in\tilde{g}\tilde{H}_{\bf k}\tilde{g}^{-1}$ are related by the symmetries $\tilde{g} \in G\setminus G_{\bf k}$.  If $\tilde{g}$ is unitary, then:
\begin{equation}
\Delta_{\tilde{\sigma}_{{\bf k}'}}(\tilde{g}h\tilde{g}^{-1}) = \Delta_{\sigma_{\bf k}}(h),
\label{eq:unitaryFullRep}
\end{equation}
and if $\tilde{g}$ is antiunitary, then:
\begin{equation}
\Delta_{\tilde{\sigma}_{{\bf k}'}}(\tilde{g}h\tilde{g}^{-1}) = [\Delta_{\sigma_{\bf k}}(h)]^{*}.
\label{eq:antiunitaryFullRep}
\end{equation}

Finally, we will use Eqs.~(\ref{eq:unitaryFullRep}) and~(\ref{eq:antiunitaryFullRep}) for each of the symmetries $\tilde{g}\in G\setminus G_{\bf k}$, to compute the matrix representatives of the full (co)rep $\tilde{\Sigma}_{\bf k}$ of $G$ in the star indexed by  ${\bf k}$.  First, we define the full (co)rep of $G$ in the star of ${\bf k}$ to be:
\begin{equation}
\tilde{\Sigma}_{\bf k} = \bigoplus_{i=1}^{m}\tilde{\sigma}_{{\bf k}_{i}},
\label{eq:finalFullCorep}
\end{equation}
in which ${\bf k}_{i}$ is the $i^\text{th}$ arm of the multiplicity-$m$ momentum star of ${\bf k}$.  In Eq.~(\ref{eq:finalFullCorep}), $\tilde{\Sigma}_{\bf k}$ is an $m\times \chi_{\tilde{\sigma}_{\bf k}}(\{E|{\bf 0}\})$-dimensional full (co)rep of $G$.  The matrix representatives $\Delta_{\tilde{\Sigma}_{\bf k}}(h)$ of the unitary SSG symmetries $h\in G$ are not necessarily block-diagonal, because $\tilde{\sigma}_{\bf k}$ and $\tilde{\sigma}_{{\bf k}'}$ in Eqs.~(\ref{eq:unitaryFullRep}) and~(\ref{eq:antiunitaryFullRep}) may not be equivalent [defined in Eq.~(\ref{eq:Nequivalence}) and the surrounding text].  Instead we may choose a basis in which $\Delta_{\tilde{\Sigma}_{\bf k}}(h)$ is block-diagonal if the unitary symmetry $h\in H_{{\bf k}_{i}}$ for all of the points ${\bf k}_{i}$ in the momentum star indexed by ${\bf k}$, and is otherwise not block-diagonal.

Rather than list the over 100,000 small and full (co)reps computed for this work in paper-format tables, we have implemented the~\href{http://www.cryst.ehu.es/cryst/corepresentations}{Corepresentations} tool on the BCS, through which the irreducible small and full (co)reps at any ${\bf k}$ point and in any momentum star in any SSG can respectively be accessed.  Representative examples demonstrating the output of~\href{http://www.cryst.ehu.es/cryst/corepresentations}{Corepresentations} are provided below in Appendices~\ref{sec:corepExampleNoAnti} and~\ref{sec:corepExampleYesAnti}.

\subsubsection{Small and Full Coreps at the $X$ and $XA$ Points in Type-III MSG 75.3 $P4'$}
\label{sec:corepExampleNoAnti}

In this section, we will determine the small coreps of the little group $G_{X}$ of the $X$ point in Type-III MSG 75.3 $P4'$, as well as the full coreps induced in the momentum star of $X$ consisting of $X$ (which in some works is alternatively labeled as $X'$ or $Y$) and $XA$ (which in some works is alternatively labeled $X$).  MSG 75.3 $P4'$ is generated by:
\begin{equation}
\{C_{4z}\times\mathcal{T}|000\},\ \{E|100\},\ \{E|001\},
\label{eq:AntiunitaryStarDecomposition}
\end{equation} 
and the maximal unitary subgroup $H$ of $G=P4'$ [see Eq.~(\ref{eq:type3}) and the surrounding text] is generated by:
\begin{equation}
\{C_{2z}|000\},\ \{E|100\},\ \{E|010\},\ \{E|001\}.
\label{eq:UnitaryStarDecomposition}
\end{equation}
Hence, $H$ is isomorphic to Type-I MSG 3.1 $P2$ in a non-standard ($z$-oriented) setting [see the text surrounding Eq.~(\ref{eq:standardSetting2}) for the definitions of standard and non-standard symmetry-group settings].  Eqs.~(\ref{eq:AntiunitaryStarDecomposition}) and~(\ref{eq:UnitaryStarDecomposition}) imply the decomposition:
\begin{equation}
G=P4' = H\cup \{C_{4z}\times\mathcal{T}|000\}H,
\label{eq:noAntiBreakdown}
\end{equation}
in which $H$ is isomorphic to the $z$-oriented (non-standard) setting of Type-I MSG 3.1 $P2$.

\begin{figure}[t]
\includegraphics[width=\columnwidth]{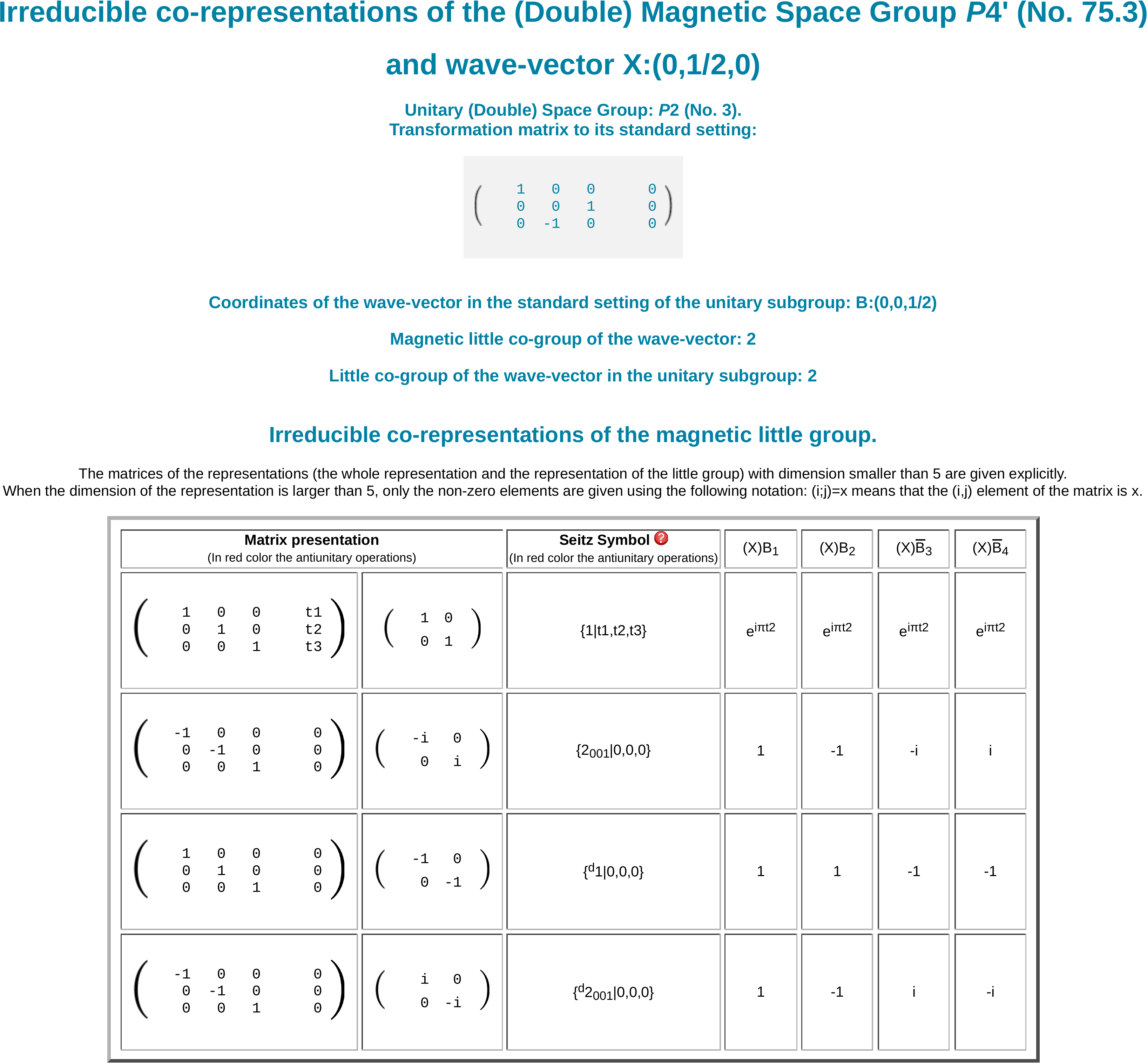}
\caption{The output of the~\href{http://www.cryst.ehu.es/cryst/corepresentations}{Corepresentations} tool for the $X$ point in Type-III MSG 75.3 $P4'$, part 1.  Even though MSG 75.3 $P4'$ contains antiunitary symmetries, the little group $G_{X}$ at ${\bf k}_{X}=(0,\pi,0)$ does not contain antiunitary symmetries, and is therefore isomorphic to $H_{X}$, its maximal unitary subgroup [see the text following Eq.~(\ref{eq:fullGroupinNoAnti})].  At the top of this figure, the $3\times 3$ matrix in the left-most three columns of the gray box is the transformation matrix $P$ that converts ${\bf k}$ points into the standard setting of the unitary subgroup.  Specifically, in $G=P4'$ [Eq.~(\ref{eq:AntiunitaryStarDecomposition})], the unitary subgroup $H$ [Eq.~(\ref{eq:UnitaryStarDecomposition})] is isomorphic to Type-I MSG 3.1 $P2$ in a non-standard ($z$-oriented) setting [see the text surrounding Eq.~(\ref{eq:standardSetting2})]; as discussed in Fig.~\ref{fig:complicatedMKVEC} and in the text surrounding Eq.~(\ref{eq:Rmatrix}), the $P$ matrix in the gray box allows quantities -- such as momentum stars and small irreps --  previously computed on the BCS for Type-I MSGs (here MSG 3.1 $P2$) to be transformed and adapted to the computation of the analogous quantities in SSGs with antiunitary symmetries (Type-II, III, and IV SSGs, see Appendices~\ref{sec:type2},~\ref{sec:type3}, and~\ref{sec:type4}, respectively).  The table in this figure shows the matrix representatives of the small coreps $\tilde{\sigma}$ of the little group $G_{X}$, for which the coreps with (without) overbars are double- (single-) valued.  Because $G_{X}$ in $P4'$ is isomorphic to $G_{B}$ in Type-I MSG 3.1 $P2$, then the coreps in this figure are labeled $(X)B_{i}$, and the table in this figure contains the same entries as the table returned by~\href{http://www.cryst.ehu.es/cryst/corepresentations}{Corepresentations} for the $B$ point in $P2$ [up to the orientation of the twofold axis, see the text following Eq.~(\ref{eq:fullGroupinNoAnti}) and Fig.~\ref{fig:complicatedMKVEC}].  We note that throughout this work, a translation ${\bf t}$ is represented at a crystal momentum ${\bf k}$ by $\exp(-i {\bf k}\cdot {\bf t})$ [\emph{i.e.}, in reduced units in which the lattice constants $a,b,c=1$], whereas on the BCS, ${\bf t}$ is represented at ${\bf k}$ by the phase $\exp(2\pi i{\bf k}\cdot {\bf t})$ [\emph{i.e.} with the opposite sign as employed in this work, and in different reduced units in which ${\bf t}$ and ${\bf k}$ are respectively expressed as multiples of the lattice and reciprocal lattice constants $a,b,c$ and $2\pi/(a,b,c)$].  We additionally note that the output of~\href{http://www.cryst.ehu.es/cryst/corepresentations}{Corepresentations} for the $X$ point in Type-III MSG 75.3 $P4'$ contains an additional table, which is shown in Fig.~\ref{fig:noAntiCoreps2}.}
\label{fig:noAntiCoreps1}
\end{figure}

The $X$ point in $G=P4'$ is one arm of a multiplicity-2 momentum star.  In the convention of the BCS, the $X$ point lies at:
\begin{equation}
{\bf k}_{X} = 2\pi(0,1/2,0),
\label{eq:XPointwithoutAnti}
\end{equation}
where the other arm of the momentum star indexed by ${\bf k}_{X}$ lies at:
\begin{equation}
{\bf k}_{XA} \equiv (C_{4z}\times\mathcal{T}){\bf k}_{X} \equiv 2\pi(1/2,0,0).
\label{eq:C4TonKx}
\end{equation}

For all of the unitary elements $h\in H$:
\begin{equation}
h{\bf k}_{X} \equiv {\bf k}_{X}.
\label{eq:littleGroupinNoAntiEx}
\end{equation}
However, for all of the antiunitary elements $\tilde{g}\in \{C_{4z}\times\mathcal{T}|000\}H$ in Eq.~(\ref{eq:noAntiBreakdown}):
\begin{equation}
\tilde{g}{\bf k}_{X} \not\equiv {\bf k}_{X}.
\label{eq:fullGroupinNoAnti}
\end{equation}

Eqs.~(\ref{eq:littleGroupinNoAntiEx}) and~(\ref{eq:fullGroupinNoAnti}) imply that the little group $G_{X}$ is isomorphic to its maximal unitary subgroup $H_{X}$.  In turn, $H_{X}$ at the point ${\bf k}_{X}=(0,\pi,0)$ in $H$ is isomorphic to $H_{B}$ at the point ${\bf k}_{B}=(0,0,\pi)$ in Type-I MSG 3.1 $P2$ in its standard ($y$-oriented setting, see Fig.~\ref{fig:complicatedMKVEC}).  Therefore, the small coreps of $G_{X}$ are simply equivalent to the small irreps of $H_{X}$, which are equivalent to the small irreps of $H_{B}$ in MSG 3.1 $P2$, where representation equivalence is defined in the text surrounding Eq.~(\ref{eq:Nequivalence}).

\begin{figure}[t]
\includegraphics[width=\columnwidth]{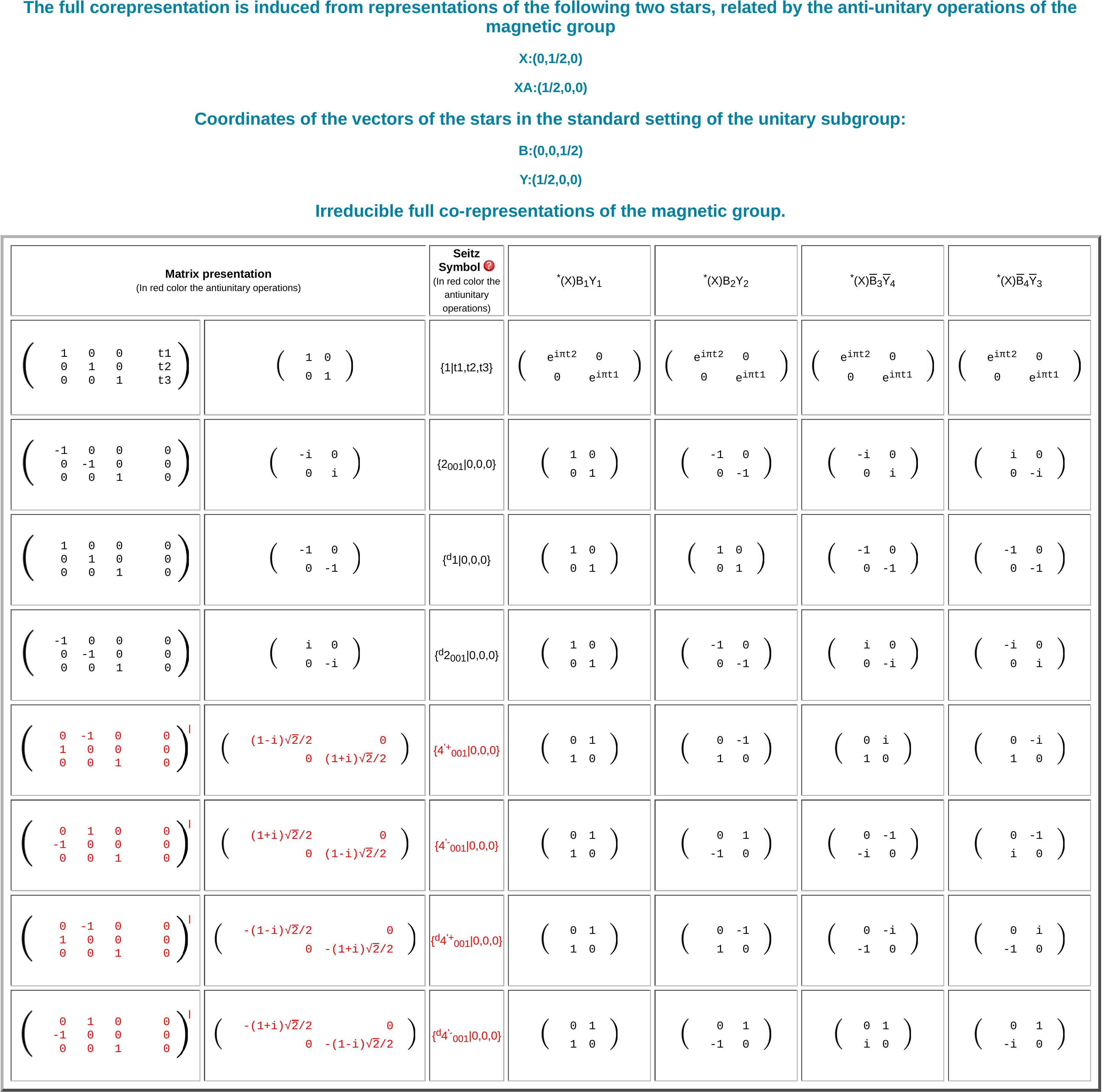}
\caption{The output of the~\href{http://www.cryst.ehu.es/cryst/corepresentations}{Corepresentations} tool for the $X$ point in Type-III MSG 75.3 $P4'$ [Eq.~(\ref{eq:XPointwithoutAnti})], part 2.  The table shown in this figure contains the \emph{full} (SSG) coreps of $G=P4'$ in the star containing ${\bf k}_{X}$ [$(0,\pi,0),(\pi,0,0)$], see Eqs.~(\ref{eq:XPointwithoutAnti}) and~(\ref{eq:C4TonKx})].  For one coset representative in each of the little group cosets in Eq.~(\ref{eq:cosetOverallDecomposition}), as well as the SSG symmetries $G\setminus G_{X}$ [Eq.~(\ref{eq:missingSymmetriesForFull}) and the surrounding text],~\href{http://www.cryst.ehu.es/cryst/corepresentations}{Corepresentations} outputs the matrix representatives in each of the irreducible full (co)reps of $G$ in the star indexed by a point ${\bf k}$.  In the table shown in this figure, the matrix representatives of antiunitary symmetries $g_{A}$ are labeled in red text, and the matrices listed for each full (co)rep $\tilde{\Sigma}_{\bf k}$ indicate the unitary part $U$ of the antiunitary matrix representative $\Delta_{\tilde{\Sigma}}(g_{A})=UK$, where $K$ is complex conjugation.  Each of the full coreps for $G=P4'$ in the star of ${\bf k}_{X}$ is labeled with both $(X)$ as well as $B_{i}Y_{j}$, to indicate that the small coreps in each arm ${\bf k}_{X}$ and ${\bf k}_{XA}$ in $G$ are respectively equivalent to the small irreps at ${\bf k}_{B}$ and ${\bf k}_{Y}$ in Type-I MSG 3.1 $P2$ [see Figs.~\ref{fig:complicatedMKVEC} and~\ref{fig:noAntiCoreps1} and the text surrounding Eqs.~(\ref{eq:Rmatrix}) and~(\ref{eq:standardSetting2})].  For each $g\in G$, each $\Delta_{\tilde{\Sigma}_{X}}(g)$ shown in this figure is a $2\times 2$ matrix, because each small corep $\tilde{\sigma}_{X}$ in Fig.~\ref{fig:noAntiCoreps1} is one-dimensional.  Additionally, each $\Delta_{\tilde{\Sigma}_{X}}(g)$ is diagonal for each symmetry $g\in G_{X}$, $g\in G_{XA}$ [\emph{e.g.} $\{C_{2z}|000\}$], but is non-diagonal for each symmetry $g \notin G_{X}$ or $g \notin G_{XA}$ [\emph{e.g.} $\{C_{4z}\times\mathcal{T}|000\}$].}
\label{fig:noAntiCoreps2}
\end{figure}

In Figs.~\ref{fig:noAntiCoreps1} and~\ref{fig:noAntiCoreps2}, we show the output of the~\href{http://www.cryst.ehu.es/cryst/corepresentations}{Corepresentations} tool for the $X$ point in Type-III MSG 75.3 $P4'$, which has been split into two figures in order to preserve the legibility of the output text.  First, in Fig.~\ref{fig:noAntiCoreps1}, we show the matrix representatives $\Delta_{\tilde{\sigma}_{X}}(h)$ of the symmetries $h\in \tilde{H}_{X}$ [Eq.~(\ref{eq:cosetOverallDecomposition})] in each of the small coreps $\tilde{\sigma}_{X}$ of $H_{X}$.  Then, in Fig.~\ref{fig:noAntiCoreps2}, we show the matrix representatives $\Delta_{\tilde{\Sigma}_{X}}(g)$ of the symmetries $g\in G$ in each of the \emph{full} coreps $\tilde{\Sigma}_{X}$ of $G$ in the star indexed by ${\bf k}_{X}$ [$\{{\bf k}_{X},{\bf k}_{XA}\}$].  Specifically, as shown in Eq.~(\ref{eq:C4TonKx}), ${\bf k}_{X}$ and ${\bf k}_{XA}$ are related by the antiunitary symmetry $\{C_{4z}\times\mathcal{T}|000\}$, for which:
\begin{equation}
\{C_{4z}\times\mathcal{T}|000\} \{C_{2z}|000\} \{(C_{4z}\times\mathcal{T})^{-1}|000\} = \{C_{2z}|000\}.
\label{eq:XPointCommutation}
\end{equation}
Eqs.~(\ref{eq:antiunitaryFullRep}) and~(\ref{eq:XPointCommutation}) imply that:
\begin{equation}
\Delta_{\tilde{\sigma}_{XA}}(h) = [\Delta_{\tilde{\sigma}_{X}}(h)]^{*},
\label{eq:fullCorepExp1}
\end{equation}
for each unitary symmetry $h\in \tilde{H}_{X}$ [see the text surrounding Eq.~(\ref{eq:CorepCoset})], which is given by:
\begin{equation}
\tilde{H}_{X} = \bigg\{\{E|000\},\ \{C_{2z}|000\},\ \{\bar{E}|000\},\ \{\bar{E}C_{2z}|000\}\bigg\}.
\label{eq:fullCorepExp2}
\end{equation}
In Eq.~(\ref{eq:fullCorepExp2}), $\bar{E}=C_{1n}$ is the symmetry operation of $360^{\circ}$ rotation about an arbitrary axis $n$, which distinguishes single-valued (spinless) and double-valued (spinful) (co)reps.  Throughout the BCS, $\bar{E}$ is also sometimes denoted with the Seitz symbol $^{d}1$, as it is in Figs.~\ref{fig:noAntiCoreps1} and~\ref{fig:noAntiCoreps2}.

Eqs.~(\ref{eq:fullCorepExp1}) and~(\ref{eq:fullCorepExp2}) imply that the full coreps $\tilde{\Sigma}_{X}$ consist of pairs of single-valued coreps at ${\bf k}_{X}$ and ${\bf k}_{XA}$ with the same real (spinless) $C_{2z}$ eigenvalues [labeled $^{*}(X)B_{1}Y_{1}$ and $^{*}(X)B_{2}Y_{2}$ in Fig.~\ref{fig:noAntiCoreps2}], and pairs of double-valued coreps with opposite imaginary (spinful) $C_{2z}$ eigenvalues [labeled $^{*}(X)\bar{B}_{3}\bar{Y}_{4}$ and $^{*}(X)\bar{B}_{4}\bar{Y}_{3}$ in Fig.~\ref{fig:noAntiCoreps2}].  Additionally, because the momentum star $\{{\bf k}_{X},{\bf k}_{XA}\}$ is left invariant under all of the symmetries $g\in P4'$, then the matrix representatives $\Delta_{\tilde{\Sigma}_{X}}(g)$ are well-defined for all of the symmetries $g\in P4'$.  This implies that $\Delta_{\tilde{\Sigma}_{X}}(g)$ is well defined for both the unitary symmetries $h \in G_{X}$, as well as the \emph{antiunitary symmetries} $\tilde{g} \in P4'\setminus H$, where $H$ is the maximal unitary subgroup of $G=P4'$ [Eq.~(\ref{eq:noAntiBreakdown})], and where $H$ is isomorphic to Type-I MSG 3.1 $P2$ in a non-standard ($z$-oriented) setting [see the text surrounding Eq.~(\ref{eq:standardSetting2})].

For each full corep $\tilde{\Sigma}_{\bf k}$ of an SSG $G$ in a momentum star indexed by an arm ${\bf k}$,~\href{http://www.cryst.ehu.es/cryst/corepresentations}{Corepresentations} outputs the matrix representative $\Delta_{\tilde{\Sigma}_{\bf k}}(g)$ for each of the unitary and antiunitary symmetries $g\in G$.  For example, unlike the table in Fig.~\ref{fig:noAntiCoreps1} for the small coreps of $G_{X}$, the table in Fig.~\ref{fig:noAntiCoreps2} for the full coreps of $G$ in the star of ${\bf k}_{X}$ contains the antiunitary matrix representatives $\Delta_{\tilde{\Sigma}_{X}}(\{C_{4z}\times\mathcal{T}|000\})$.  For each full corep $\tilde{\Sigma}_{X}$ and antiunitary symmetry $g_{A}\in G$, the full (co)rep table in~\href{http://www.cryst.ehu.es/cryst/corepresentations}{Corepresentations} displays the unitary part of the matrix representative $\Delta_{\tilde{\Sigma}_{X}}(g_{A})$, which is colored in red to indicate that $\Delta_{\tilde{\Sigma}_{X}}(g_{A})$ is antiunitary.  In general, $\Delta_{\tilde{\Sigma}_{\bf k}}(g)$ for each of the unitary and antiunitary symmetries $g\in G$ is block-diagonal if $g\in G_{{\bf k}_{i}}$ for all of the points ${\bf k}_{i}$ in the momentum star indexed by ${\bf k}$, and is otherwise non-diagonal.  For example, in Fig.~\ref{fig:noAntiCoreps2}, each $\Delta_{\tilde{\Sigma}_{X}}(g)$ is a $2\times 2$ matrix, because each small corep $\tilde{\sigma}_{X}$ in Fig.~\ref{fig:noAntiCoreps1} is one-dimensional.  Additionally, in Fig.~\ref{fig:noAntiCoreps2}, each $\Delta_{\tilde{\Sigma}_{X}}(g)$ is diagonal for each symmetry $g\in G$, $g\in G_{X}$ and $g\in G_{XA}$ [\emph{e.g.} $\{C_{2z}|000\}$], but is non-diagonal for each symmetry $g \in G$, $g \not\in G_{X}$ or $g\not\in G_{XA}$ [\emph{e.g.} $\{C_{4z}\times\mathcal{T}|000\}$].

\clearpage

\clearpage

\subsubsection{Small and Full Coreps at the $S$ Point in Type-IV MSG 25.63 $P_{C}mm2$}
\label{sec:corepExampleYesAnti}

In this section, we will determine the small coreps of the little group $G_{S}$ of the $S$ point in Type-IV MSG 25.63 $P_{C}mm2$.  We will also show that the small coreps of $G_{S}$ coincide with the full coreps induced in $G=P_{C}mm2$ in the momentum star of $S$, because the $S$ point in Type-IV MSG 25.63 $P_{C}mm2$ is the only arm of a multiplicity-1 star (see Appendix~\ref{sec:MKVEC} and the~\href{http://www.cryst.ehu.es/cryst/mkvec}{MKVEC} tool for more information).  To begin, MSG 25.63 $P_{C}mm2$ is generated by:
\begin{equation}
M_{x}=\bigg\{m_{x}\bigg|000\bigg\},\ M_{y}=\bigg\{m_{y}\bigg|000\bigg\},\ \theta=\bigg\{\mathcal{T}\bigg|\frac{1}{2}\frac{1}{2}0\bigg\},\ t_{x}=\{E|100\},\ t_{z}=\{E|001\}.
\label{eq:DiracPointGenerators}
\end{equation}
The $S$ point in MSG 25.63 $P_{C}mm2$ lies at:
\begin{equation}
{\bf k}_{S} = 2\pi(1/2,1/2,0).
\label{eq:SPointwithAnti}
\end{equation}
Unlike in the previous example in Appendix~\ref{sec:corepExampleNoAnti}, all of the symmetries in MSG 25.63 $P_{C}mm2$ return ${\bf k}_{S}$ to itself modulo reciprocal lattice vectors ($g{\bf k}_{S}\equiv {\bf k}_{S}$ for all $g\in P_{C}mm2$).  Therefore, the little group $G_{S}$ is isomorphic to MSG 25.63 $P_{C}mm2$ itself, and the set $\tilde{G}_{S}$ [defined in the text surrounding Eq.~(\ref{eq:CorepCoset})] is given by
\begin{equation}
G_{S} = \tilde{G}_{S}G_{T} = \left(\tilde{H}_{S} \cup \theta\tilde{H}_{S}\right)G_{T}.
\label{eq:SpointBreakdown}
\end{equation}
with $\tilde{g}_{A}=\theta$.  In Eq.~(\ref{eq:SpointBreakdown}), the maximal unitary subset of $\tilde{G}_{S}$ is given by $\tilde{H}_{S}$ [in the specific case of the $S$ point in Type-IV MSG 25.63 $P_{C}mm2$, $\tilde{H}_{S}$ is in fact a finite group, see the text preceding Eq.~(\ref{eq:CorepCoset}) for more information]:
\begin{equation}
\bar{H}_{S} = \bigg\{\{E|{\bf 0}\},\ \{m_{x}|{\bf 0}\},\ \{m_{y}|{\bf 0}\},\ \{C_{2z}|{\bf 0}\},\ \{\bar{E}|{\bf 0}\},\ \{\bar{E}m_{x}|{\bf 0}\},\ \{\bar{E}m_{y}|{\bf 0}\},\ \{\bar{E}C_{2z}|{\bf 0}\}\bigg\},
\label{eq:unitaryDiracSub1}
\end{equation}
where $\bar{E}$ is defined in the text following Eq.~(\ref{eq:fullCorepExp2}).  The symmetry operations in $\tilde{H}_{S}$ in Eq.~(\ref{eq:unitaryDiracSub1}) satisfy:
\begin{eqnarray}
m_{x,y}m_{y,x}m_{x,y}^{-1} &=& \bar{E}m_{y,x},\ m_{x,y}C_{2z}m_{x,y}^{-1} = \bar{E}C_{2z},\ m_{x,y}\bar{E}m_{x,y}^{-1} = C_{2z}\bar{E}C_{2z}^{-1} = \bar{E}, \nonumber \\
m_{x}m_{y} &=& C_{2z},\ \bar{E}^{2} = E,\ m_{x,y}^{2} = C_{2z}^{2} = \bar{E}.
\label{eq:unitaryDiracSub2}
\end{eqnarray}

\begin{table}[b]
\begin{tabular}{|c|c|c|c|c|c|}
\hline
\multicolumn{6}{|c|}{Matrix Representatives $\Delta_{\sigma}(h)$ of the Small Irreps of $H_{S}$} \\
\multicolumn{6}{|c|}{at the $S$ point [${\bf k}_{S} = (\pi,\pi,0)$] in Type-I MSG 25.57 $Pmm2$,} \\
\multicolumn{6}{|c|}{the Unitary Subgroup of Type-IV MSG 25.63 $P_{C}mm2$} \\ 
\hline
$h$ & $S_{1}$ & $S_{2}$ & $S_{3}$ & $S_{4}$ & $\bar{S}_{5}$ \\
\hline
\hline
$\{E|{\bf t}_{1}+{\bf t}_{2}+{\bf t}_{3}\}$ & $e^{-i\pi({\bf t}_{1}+{\bf t}_{2})}$ & $e^{-i\pi({\bf t}_{1}+{\bf t}_{2})}$ & $e^{-i\pi({\bf t}_{1}+{\bf t}_{2})}$ & $e^{-i\pi({\bf t}_{1}+{\bf t}_{2})}$ & $\left(\begin{array}{cc} e^{-i\pi({\bf t}_{1}+{\bf t}_{2})} & 0 \\ 0 & e^{-i\pi({\bf t}_{1}+{\bf t}_{2})} \end{array}\right)$ \\
\hline
$\{C_{2z}|000\}$ & $1$ & $1$ & $-1$ & $-1$ & $\left(\begin{array}{cc} 0 & -1 \\ 1 & 0 \end{array}\right)$ \\
\hline
$\{m_{y}|000\}$ & $1$ & $-1$ & $-1$ & $1$ & $\left(\begin{array}{cc} 0 & -i \\ -i & 0 \end{array}\right)$ \\
\hline
$\{m_{x}|000\}$ & $1$ & $-1$ & $1$ & $-1$ & $\left(\begin{array}{cc} -i & 0 \\ 0 & i \end{array}\right)$ \\
\hline
$\{\bar{E}|000\}$ & $1$ & $1$ & $1$ & $1$ & $\left(\begin{array}{cc} -1 & 0 \\ 0 & -1 \end{array}\right)$ \\
\hline
$\{\bar{E}C_{2z}|000\}$ & $1$ & $1$ & $-1$ & $-1$ & $\left(\begin{array}{cc} 0 & 1 \\ -1 & 0 \end{array}\right)$ \\
\hline
$\{\bar{E}m_{y}|000\}$ & $1$ & $-1$ & $-1$ & $1$ & $\left(\begin{array}{cc} 0 & i \\ i & 0 \end{array}\right)$ \\
\hline
$\{\bar{E}m_{x}|000\}$ & $1$ & $-1$ & $1$ & $-1$ & $\left(\begin{array}{cc} i & 0 \\ 0 & -i \end{array}\right)$ \\
\hline
\end{tabular}
\caption{The matrix representatives $\Delta_{\sigma}(h)$ of the small irreps $\sigma$ of the little group $H_{S}$ of the $S$ point [${\bf k}_{S} = (\pi,\pi,0)$] in Type-I MSG 25.57 $Pmm2$, the unitary subgroup of Type-IV MSG 25.63 $P_{C}mm2$.  Because MSG 25.57 $Pmm2$ is a Type-I MSG (Appendix~\ref{sec:type1}), then $\tilde{H}_{S}$ is isomorphic to its maximal unitary subset.  The values of $\Delta_{\sigma}(h)$ in this table have been reproduced from the output of the~\href{http://www.cryst.ehu.es/cryst/corepresentations}{Corepresentations} tool, and adapted to the notation employed throughout this work in which a translation ${\bf t}$ is represented at a crystal momentum ${\bf k}$ by $\exp(-i {\bf k}\cdot {\bf t})$ [\emph{i.e.}, in reduced units in which the lattice constants $a,b,c=1$, see Fig.~\ref{fig:noAntiCoreps1} for further details].  We note that in~\href{http://www.cryst.ehu.es/cryst/corepresentations}{Corepresentations} and in this table, the matrix representatives $\Delta_{\sigma}(h)$ are shown for each symmetry $h\in \tilde{H}_{\bf k}$ except for the element $\{E|000\}$ with $\exp(-i{\bf k}_{S}\cdot {\bf t}_{\mu})=1$ [see Eqs.~(\ref{eq:translationRestriction}) and~(\ref{eq:smallCorepDef})]; instead the first element $h$ in this table, and in the output of~\href{http://www.cryst.ehu.es/cryst/corepresentations}{Corepresentations}, is chosen to be $\{E|{\bf t}_{1}+{\bf t}_{2}+{\bf t}_{3}\}$, where ${\bf t}_{1,2,3}$ are respectively integer-valued multiples of the lattice vectors ${\bf t}_{x,y,z}$ (see Figs.~\ref{fig:noAntiCoreps1} and~\ref{fig:magDiracCoreps}).  We make this substitution of $\{E|{\bf t}_{1}+{\bf t}_{2}+{\bf t}_{3}\}$ for $\{E|000\}$ to provide users with information regarding the representations (phases) of translations at ${\bf k}$ (here specifically at ${\bf k}_{S}$ [Eq.~(\ref{eq:SPointwithAnti})]), which contribute towards determining the matrix representatives of all of the symmetries in $H_{\bf k}$ (as opposed to just the symmetries in $\tilde{H}_{\bf k}$), and towards determining the pairing of unitary subgroup small irreps into little group small coreps [see the text surrounding Eqs.~(\ref{eq:simplifyModifyHerring2}) and~(\ref{eq:simplestJ}), for example].  The overbar on $\sigma=\bar{S}_{5}$ is used to indicate that $\bar{S}_{5}$ is double-valued, whereas the irreps without overbars ($S_{1-4}$) are single-valued.}
\label{tb:diracIrreps}
\end{table}

Because all of the symmetries $h\in \tilde{H}_{S}$ are of the form $\{R|{\bf 0}\}$, then Eqs.~(\ref{eq:unitaryDiracSub1}) and~(\ref{eq:unitaryDiracSub2}) imply that the small irreps of $H_{S}$ are equivalent to the irreps of an abstract finite group [see Ref.~\onlinecite{BigBook} and the text following Eq.~(\ref{eq:HerringsLittleGroup})] that is isomorphic~\cite{ShubnikovMagneticPoint,BilbaoPoint,PointGroupTables,MagneticBook,EvarestovBook,EvarestovMEBR,BCS1,BCS2,BCSMag1,BCSMag2,BCSMag3,BCSMag4} to Type-I MPG 7.1.20 $mm2$, which has five irreps $\sigma$.  In Table~\ref{tb:diracIrreps}, we reproduce the matrix representatives $\Delta_{\sigma}(h)$ of the small irreps of ${H}_{S}$ from the output of the~\href{http://www.cryst.ehu.es/cryst/corepresentations}{Corepresentations} tool for the $S$ point in Type-I MSG 25.57 $Pmm2$, which is the unitary subgroup of Type-IV MSG 25.63 $P_{C}mm2$ (adjusting for the differences in convention between how translations are represented in this work and on the BCS, see Fig.~\ref{fig:noAntiCoreps1}).  The five irreps in Table~\ref{tb:diracIrreps} subdivide into four single-valued, one-dimensional irreps ($S_{1-4}$) that are distinguished by their spinless $m_{x,y}$ eigenvalues and one double-valued irrep ($\bar{S}_{5}$) that is two-dimensional because of the anticommutator $\{\Delta_{\bar{S}_{5}}(\{m_{x}|{\bf 0}\}),\Delta_{\bar{S}_{5}}(\{m_{y}|{\bf 0}\})\}=0$.

To determine the type of the small corep $\tilde{\sigma}$ induced in $G_{S}$, we calculate the indicator $J_{\sigma}=\sum_{i}\chi_{\sigma}(g_{A,i}^{2})$ [Eqs.~(\ref{eq:Jtest}) and~(\ref{eq:Jtypes})] for each irrep $\sigma$ in Table~\ref{tb:diracIrreps}.  Using Eq.~(\ref{eq:SpointBreakdown}), we determine that there are eight $g_{A,i}$ to consider:
\begin{equation}
g_{A,i}\in\theta\tilde{H}_{S},
\end{equation}
where $\theta$ is defined in Eq.~(\ref{eq:DiracPointGenerators}), and where $\tilde{H}_{S}$ is defined in Eq.~(\ref{eq:unitaryDiracSub1}) and in Table~\ref{tb:diracIrreps}.  First, we use Eqs.~(\ref{eq:unitaryDiracSub1}) and~(\ref{eq:unitaryDiracSub2}) to determine that $\theta = t_{x}t_{y}\bar{E}\theta^{-1}$, $\bar{E}^{2} = E$, and that $[\bar{E},h_{i}]=0$ for all $h_{i}\in\tilde{H}_{S}$, where $t_{x}=\{E|100\}$ and $t_{y}=\{E|010\}$.  We then determine that, in the case of $\tilde{G}_{S}$ in Type-IV MSG 25.63 $P_{C}mm2$, $\chi_{\sigma}(g_{A,i}^{2})$ can be simplified as:
\begin{equation}
\chi_{\sigma}(g_{A,i}^{2}) = \chi_{\sigma}(\theta h_{i} \theta h_{i}) = \sgn\left[\chi_{\sigma}(\bar{E})\right]\chi_{\sigma}\left([\theta h_{i}\theta^{-1}t_{x}t_{y}]h_{i}\right),
\label{eq:simplifyModifyHerring1}
\end{equation}
where $h_{i}\in\tilde{H}_{S}$ in Eq.~(\ref{eq:unitaryDiracSub1}).  Next, we use Eq.~(\ref{eq:DiracPointGenerators}) to obtain the relations:
\begin{eqnarray}
\theta C_{2z} \theta^{-1}t_{x}t_{y} &=& \theta \{C_{2z}|000\}\theta^{-1}\{E|110\} = \{C_{2z}|110\}\{E|110\} = (t_{x}t_{y}C_{2z})t_{x}t_{y} = (C_{2z}t_{x}^{-1}t_{y}^{-1})t_{x}t_{y} = C_{2z}, \nonumber \\
\theta M_{x} \theta^{-1}t_{x}t_{y} &=& \theta \{m_{x}|000\}\theta^{-1}\{E|110\} = \{m_{x}|100\}\{E|110\} = (t_{x}M_{x})t_{x}t_{y} = (M_{x}t_{x}^{-1})t_{x}t_{y} = t_{y}M_{x}, \nonumber \\
\theta M_{y} \theta^{-1}t_{x}t_{y} &=& \theta \{m_{y}|000\}\theta^{-1}\{E|110\} = \{m_{y}|010\}\{E|110\} = (t_{y}M_{y})t_{x}t_{y} = (M_{y}t_{y}^{-1})t_{x}t_{y} = t_{x}M_{y}, \nonumber \\
\theta \bar{E} \theta^{-1} &=& \theta \{\bar{E}|000\} \theta^{-1} = \{\bar{E}|000\} = \bar{E}.
\label{eq:simplifyModifyHerring2}
\end{eqnarray}    
Eqs.~(\ref{eq:simplifyModifyHerring1}) and~(\ref{eq:simplifyModifyHerring2}) imply that $J_{\sigma}$ [Eq.~(\ref{eq:Jtest})] can be further simplified before specifying a value of $\sigma$:
\begin{eqnarray}
J_{\sigma} &=& \sum_{i}\chi_{\sigma}(g_{A,i}^{2}) = \sgn\left[\chi_{\sigma}(\bar{E})\right]\sum_{i}\chi_{\sigma}\left([\theta h_{i}\theta^{-1}t_{x}t_{y}]h_{i}\right) \nonumber \\
&=& 2\sgn\left[\chi_{\sigma}(\bar{E})\right]\left(\chi_{\sigma}(E) - \chi_{\sigma}(\bar{E})\right) \nonumber \\
&=& 2\left[\chi_{\sigma}(\bar{E}) - \chi_{\sigma}(E)\right].
\label{eq:simplestJ}
\end{eqnarray} 
Remarkably, we find that Eq.~(\ref{eq:simplestJ}) only depends on whether $\sigma$ is single- or double-valued:
\begin{equation}
J_{\sigma} = \begin{cases} \ \ \ \ \ \ 0, &\text{ for }\sigma = S_{1-4}\\
-|\tilde{H}_{S}|, &\text{ for }\sigma = \bar{S}_{5}
\end{cases},
\label{eq:diracJTest}
\end{equation}
where $|\tilde{H}_{S}| = 8$ [Eq.~(\ref{eq:unitaryDiracSub1})].  Using Eq.~(\ref{eq:Jtypes}), we determine that the single-valued, one-dimensional irreps $S_{1-4}$ induce paired, two-dimensional coreps of type (c) [Eq.~(\ref{eq:typeC})], whereas the double-valued, two-dimensional irrep $\bar{S}_{5}$ induces a paired, four-dimensional corep of type (b) [Eq.~(\ref{eq:typeB})].

To complete the calculation of the small coreps of $G_{S}$ in Type-IV MSG 25.63 $P_{C}mm2$, we must determine which of the single-valued irreps $S_{1-4}$ become paired into coreps of type (c).  This can be accomplished by computing the matrix representative $\bar{\Delta}_{\sigma}(h) = \left[\Delta_{\sigma}(\tilde{g}_{A}^{-1}h\tilde{g}_{A})\right]^{*}$ [Eq.~(\ref{eq:otherCorepMatrix})].  Choosing $g_{A}=\theta$ and using Eq.~(\ref{eq:simplifyModifyHerring2}), we find that, for the single-valued irreps $\sigma=S_{1-4}$:
\begin{equation}
\bar{\Delta}_{\sigma}(C_{2z}) = \Delta_{\sigma}(C_{2z}) = \Delta_{\sigma'}(C_{2z}),\ \bar{\Delta}_{\sigma}(M_{x,y}) = -\Delta_{\sigma}(M_{x,y}) = \Delta_{\sigma'}(M_{x,y}).
\label{eq:corepDiracTypeC}
\end{equation}
Along with Eq.~(\ref{eq:diracJTest}), which implies that $\bar{S}_{5}$ induces a paired corep of type (b), Eq.~(\ref{eq:corepDiracTypeC}) implies that $G_{S}$ in Type-IV MSG 25.63 $P_{C}mm2$ has three small coreps:
\begin{equation}
\tilde{\sigma} = S_{1}S_{2},\ S_{3}S_{4},\ \bar{S}_{5}\bar{S}_{5},
\label{eq:FinalCorepsDirac}
\end{equation}
where $S_{1}S_{2}$ and $S_{3}S_{4}$ are single-valued, two-dimensional coreps and $\bar{S}_{5}\bar{S}_{5}$ is a double-valued, four-dimensional corep.  Below, we will shortly formulate a $k\cdot p$ Hamiltonian demonstrating that $\bar{S}_{5}\bar{S}_{5}$ corresponds to a 3D fourfold Dirac fermion~\cite{SteveMagnet,SteveDirac} that is enforced by spinful mirrors that anticommute with each other $\{\Delta_{\bar{S}_{5}\bar{S}_{5}}(M_{x}),\Delta_{\bar{S}_{5}\bar{S}_{5}}(M_{y})\}=0$, and with the matrix representative of $\theta$ $\{\Delta_{\bar{S}_{5}\bar{S}_{5}}(M_{x,y}),\Delta_{\bar{S}_{5}\bar{S}_{5}}(\theta)\}=0$.

\begin{figure}[h]
\includegraphics[width=\columnwidth]{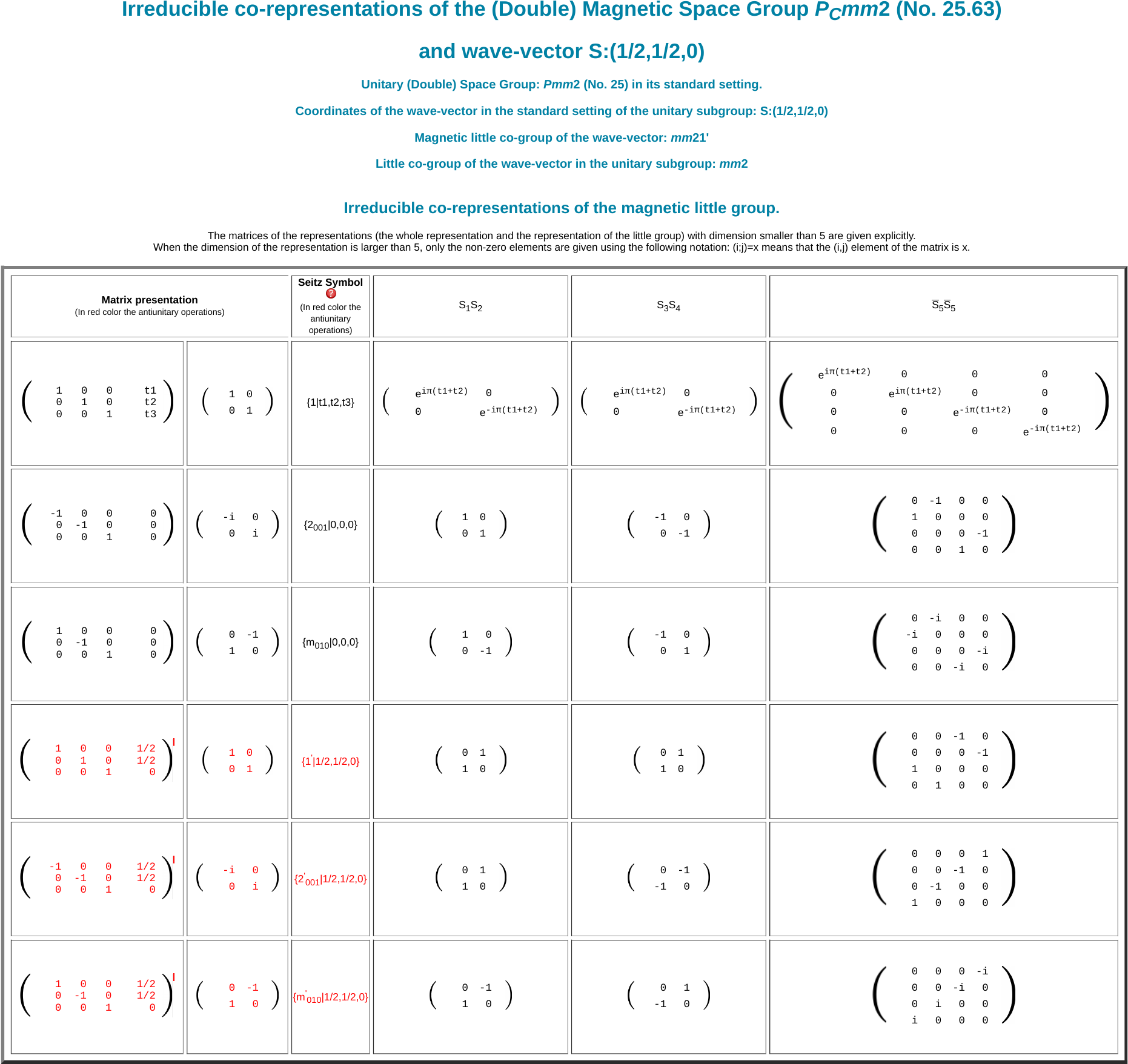}
\caption{The output of the~\href{http://www.cryst.ehu.es/cryst/corepresentations}{Corepresentations} tool for the $S$ point in Type-IV MSG 25.63 $P_{C}mm2$.  The table in this figure shows the matrix representatives of the small coreps $\tilde{\sigma}$ of the little group $G_{S}$ [${\bf k}_{S}=(\pi,\pi,0)$], for which the coreps with (without) overbars are double- (single-) valued.  As discussed in Fig.~\ref{fig:noAntiCoreps1}, throughout this work, a translation ${\bf t}$ is represented at a crystal momentum ${\bf k}$ by $\exp(-i {\bf k}\cdot {\bf t})$ [\emph{i.e.}, in reduced units in which the lattice constants $a,b,c=1$], whereas on the BCS, ${\bf t}$ is represented at ${\bf k}$ by the phase $\exp(2\pi i{\bf k}\cdot {\bf t})$ [\emph{i.e.} with the opposite sign as employed in this work, and in different reduced units in which ${\bf t}$ and ${\bf k}$ are respectively expressed as multiples of the lattice and reciprocal lattice constants $a,b,c$ and $2\pi/(a,b,c)$].  In the table shown in this figure, the matrix representatives of antiunitary symmetries $g_{A}$ are labeled in red text, and the matrices listed for each small corep $\tilde{\sigma}_{S}$ indicate the unitary part $U$ of the antiunitary matrix representative $\Delta_{\tilde{\sigma}_{S}}(g_{A})=UK$, where $K$ is complex conjugation.  We note that the output of~\href{http://www.cryst.ehu.es/cryst/corepresentations}{Corepresentations} for the $S$ point in Type-IV MSG 25.63 $P_{C}mm2$ also includes an additional table containing the matrix representatives of the full coreps in the momentum star containing ${\bf k}_{S}$ (see Fig.~\ref{fig:noAntiCoreps2}).  However, because, ${\bf k}_{S}$ in MSG 25.63 $P_{C}mm2$ is the only arm of a multiplicty-1 momentum star (Appendix~\ref{sec:MKVEC} and~\href{http://www.cryst.ehu.es/cryst/mkvec}{MKVEC}), then the second table outputted by~\href{http://www.cryst.ehu.es/cryst/corepresentations}{Corepresentations} is identical to the table shown in this figure; for concision have therefore omitted the second table outputted by~\href{http://www.cryst.ehu.es/cryst/corepresentations}{Corepresentations}.}
\label{fig:magDiracCoreps}
\end{figure}

In Fig.~\ref{fig:magDiracCoreps}, we show the output of the~\href{http://www.cryst.ehu.es/cryst/corepresentations}{Corepresentations} tool for the $S$ point in Type-IV MSG 25.63 $P_{C}mm2$, which agrees with the calculation performed in this section to obtain Eq.~(\ref{eq:FinalCorepsDirac}).  As previously discussed in Fig.~\ref{fig:noAntiCoreps1} and in the text surrounding Eq.~(\ref{eq:C4TonKx}), the table in Fig.~\ref{fig:magDiracCoreps} contains the matrix representatives of the small coreps $\tilde{\sigma}$ [Eq.~(\ref{eq:FinalCorepsDirac})] of the little group $G_{S}$ in Type-IV MSG 25.63 $P_{C}mm2$.  We note that, like in Fig.~\ref{fig:noAntiCoreps2}, the~\href{http://www.cryst.ehu.es/cryst/corepresentations}{Corepresentations} tool also outputs a second table containing the matrix representatives of the full coreps in the momentum star indexed by ${\bf k}_{S}$ [see Appendix~\ref{sec:MKVEC} and the text surrounding Eqs.~(\ref{eq:finalFullCorep}) and~(\ref{eq:SPointwithAnti})].  However, because, ${\bf k}_{S}$ in MSG 25.63 $P_{C}mm2$ is the only arm of a multiplicty-1 momentum star (Appendix~\ref{sec:MKVEC} and~\href{http://www.cryst.ehu.es/cryst/mkvec}{MKVEC}), then the second table outputted by~\href{http://www.cryst.ehu.es/cryst/corepresentations}{Corepresentations} is identical to the table shown in Fig.~\ref{fig:magDiracCoreps}.  Therefore, for concision, we have omitted the second table outputted by~\href{http://www.cryst.ehu.es/cryst/corepresentations}{Corepresentations}.

We can gain some physical intuition for the small coreps $\tilde{\sigma}$ in Eq.~(\ref{eq:FinalCorepsDirac}) by forming a $k\cdot p$ Hamiltonian characterized by one of the $\tilde{\sigma}$.  Focusing on the double-valued, four-dimensional corep $\tilde{\sigma}=\bar{S}_{5}\bar{S}_{5}$, which characterizes spinful electronic states, we can re-express the symmetry representation of the generating elements of $G_{S}$ in Table~\ref{tb:diracIrreps} and Eq.~(\ref{eq:FinalCorepsDirac}) as acting on a four-band Hamiltonian $\mathcal{H}({\bf q})=\mathcal{H}({\bf k} - {\bf k}_{S})$:
\begin{eqnarray}
M_{x}\mathcal{H}(q_{x},q_{y},q_{z})M_{x}^{-1} &=& \tau^{z}\sigma^{x}\mathcal{H}(-q_{x},q_{y},q_{z})\tau^{z}\sigma^{x}, \nonumber \\
M_{y}\mathcal{H}(q_{x},q_{y},q_{z})M_{y}^{-1} &=& \tau^{z}\sigma^{y}\mathcal{H}(q_{x},-q_{y},q_{z})\tau^{z}\sigma^{y}, \nonumber \\
\theta\mathcal{H}(q_{x},q_{y},q_{z})\theta^{-1} &=& \tau^{x}\sigma^{y}\mathcal{H}^{*}(-q_{x},-q_{y},-q_{z})\tau^{x}\sigma^{y},
\label{eq:diracSymmetryRep}
\end{eqnarray}
where $\tau^{i}$ and $\sigma^{j}$ are $2\times 2$ Pauli matrices, and where we have employed a shorthand in which $\tau^{i}\sigma^{j} = \tau^{i} \otimes \sigma^{j}$,  $\tau^{0}\otimes \sigma^{j}  = \sigma^{j}$, and $\tau^{i} \otimes \sigma^{0} = \tau^{i}$.  We note that we have not included the generating translations of $G_{S}$ in Eq.~(\ref{eq:diracSymmetryRep}), because translations are represented as phases in momentum space, and therefore do not by themselves impose constraints on $\mathcal{H}(q_{x},q_{y},q_{z})$.  The symmetry representation in Eq.~(\ref{eq:diracSymmetryRep}) admits a Hamiltonian:
\begin{equation}
\mathcal{H}({\bf q}) = [v_{1x}\sigma^{y} + v_{2x}\tau^{x}\sigma^{x}]q_{x} + [v_{1y}\sigma^{x} + v_{2y}\tau^{x}\sigma^{y}]q_{y} + [v_{1z}\tau^{z} + v_{2z}\tau^{x}\sigma^{z}]q_{z},
\label{eq:DiracHamiltonian}
\end{equation}
that characterizes a linearly dispersing, fourfold condensed matter Dirac fermion with non-degenerate bands away from ${\bf q}={\bf 0}$.  Specifically, in the $q_{z}=0$ plane, Eq.~(\ref{eq:DiracHamiltonian}) coincides with the Hamiltonian of the 2D filling-enforced~\cite{WPVZ,WiederLayers} fourfold magnetic Dirac fermion introduced in Ref.~\onlinecite{SteveMagnet}.  Most recently, the methods employed in this section -- which we have adapted from Refs.~\onlinecite{manesWeyl,SteveDirac,DDP,NewFermion,RhSi,CoSi,KramersWeyl,SteveMagnet} -- were used by the authors of Ref.~\onlinecite{MagneticNewFermion} to construct a complete list of high-symmetry-point multifold fermions in the MSGs.  Where there is overlap, the results of Ref.~\onlinecite{MagneticNewFermion} agree with the output of the~\href{http://www.cryst.ehu.es/cryst/corepresentations}{Corepresentations} tool introduced in this work.

\clearpage

\subsection{Compatibility Relations in the MSGs and the~\textsc{MCOMPREL} Tool}
\label{sec:compatibilityRelations}

In this section, building upon the definition of the small coreps of the magnetic little groups established in Appendix~\ref{sec:coreps}, we will now discuss the concept of compatibility relations (defined in detail in the text below), which relate the coreps at different ${\bf k}$ points throughout the BZ.  To begin, at a given point ${\bf k}$ in the first BZ of an SSG, the set of occupied Bloch eigenstates can be labeled by the small coreps of the little group $G_{\bf k}$.  As shown in previous works~\cite{ZakBandrep1,ZakBandrep2,EvarestovBook,EvarestovMEBR,BarryBandrepReview,QuantumChemistry,Bandrep1,Bandrep2,Bandrep3,JenFragile1,BarryFragile,ZakException1,ZakException2,ArisMagneticBlochOscillation,ZakCompatibility}, given knowledge of all of the coreps at ${\bf k}$, the possible coreps present at a point ${\bf k}'$ that is \emph{connected} to ${\bf k}$ [defined in the text following Eq.~(\ref{eq:coGroupEq3})] can be inferred from the group-subgroup relations between $G_{\bf k}$ and $G_{{\bf k}'}$.  In this section, we will review how the compatibility relations between the coreps at connected ${\bf k}$ points throughout the BZ can be reformulated using the language of graph theory.  Finally, we will conclude this section by discussing how the graph-theory interpretation of the compatibility relations can be exploited to determine if a given set of coreps at a small number of high-symmetry ${\bf k}$ vectors [specifically, the arms of the maximal momentum stars, see the text surrounding Eq.~(\ref{eq:maximalKvec})] are incompatible with the presence of an energy (band) gap at all ${\bf k}$ points in the BZ.

To begin, consider two connected points ${\bf k}$ and ${\bf k}'$ for which the little group $G_{\bf k}$ is of higher symmetry than the little group $G_{{\bf k}'}$, such that $G_{{\bf k}'}\subset G_{\bf k}$.  Next, consider a set of occupied Bloch eigenstates to be present at ${\bf k}$.  The Bloch states at ${\bf k}$ can be labeled with a small, \emph{generically-reducible} corep $\tilde{\varsigma}_{\bf k}$ of $G_{\bf k}$:
\begin{equation}
\tilde{\varsigma}_{\bf k} = \bigoplus_{i} a^{\bf k}_{i} \tilde{\sigma}_{i,{\bf k}},
\label{eq:reducibleCompatibility}
\end{equation}
where $\tilde{\sigma}_{i,{\bf k}}$ is the $i^{\text{th}}$ small (irreducible) corep of $G_{\bf k}$ (Appendix~\ref{sec:coreps}).  In Eq.~(\ref{eq:reducibleCompatibility}), $a^{\bf k}_{i}$ is a non-negative integer~\cite{SerreLinearReps}, known as the \emph{multiplicity} of $\tilde{\sigma}_{i,{\bf k}}$, that indicates the number of times that $\tilde{\sigma}_{i,{\bf k}}$ appears in the decomposition of $\tilde{\varsigma}_{\bf k}$.  The multiplicities $\{a^{\bf k}_{i}\}$ are known as the \emph{symmetry data} for each ${\bf k}$ point, and the set $\{\tilde{\varsigma}_{\bf k}\}$ over all of the arms ${\bf k}$ of the maximal momentum stars in an SSG [defined in the text surrounding Eq.~(\ref{eq:maximalKvec})] is known as the \emph{symmetry data vector}~\cite{AndreiMaterials}.  In Eq.~(\ref{eq:reducibleCompatibility}), each small corep $\tilde{\sigma}_{i,{\bf k}}$ can be further subduced onto the lower-symmetry little group $G_{{\bf k}'}$ of a point ${\bf k}'$ that is connected to ${\bf k}$:
\begin{equation}
\tilde{\sigma}_{i,{\bf k}}\downarrow G_{{\bf k}'} = \bigoplus_{j} m_{i,j}^{{\bf k}, {\bf k}'}\tilde{\sigma}_{j,{\bf k}'},
\label{eq:fullCompatibility}
\end{equation}
where $\tilde{\sigma}_{j,{\bf k}'}$ is the $j^{\text{th}}$ small (irreducible) corep of $G_{{\bf k}'}$ and $m_{i,j}^{{\bf k},{\bf k}'}$ is the multiplicity of $\tilde{\sigma}_{j,{\bf k}'}$ in $\tilde{\sigma}_{i,{\bf k}}\downarrow G_{{\bf k}'}$.  The values of $m_{i,j}^{{\bf k},{\bf k}'}$ are known as the \emph{compatibility relations}~\cite{Bandrep2,Bandrep3,ZakException1,ZakException2,ArisMagneticBlochOscillation,ZakCompatibility} between $\tilde{\sigma}_{i,{\bf k}}$ and $\tilde{\sigma}_{j,{\bf k}'}$, and are required to be non-negative integers, because they originate from group-subgroup subduction [Eq.~(\ref{eq:fullCompatibility}) and Ref.~\onlinecite{SerreLinearReps}].  For future calculations, it will be useful to re-express Eqs.~(\ref{eq:reducibleCompatibility}) and~(\ref{eq:fullCompatibility}) as:
\begin{equation}
m^{{\bf k},{\bf k}'}{\bf \tilde{{\boldsymbol \varsigma}}}_{\bf k} = {\bf \tilde{{\boldsymbol \varsigma}}}_{{\bf k}'},
\label{eq:fullCompatibilityMatrix}
\end{equation}
in which ${\bf \tilde{{\boldsymbol \varsigma}}}_{\bf k}$ (${\bf \tilde{{\boldsymbol \varsigma}}}_{{\bf k}'}$) is an $w\times 1$- ($z\times 1$-) dimensional column vector where $w$ ($z$) is the number of small coreps of $G_{\bf k}$ ($G_{{\bf k}'}$).  In the notation of Eq.~(\ref{eq:fullCompatibilityMatrix}), ${\bf \tilde{{\boldsymbol \varsigma}}}_{\bf k}$ and ${\bf \tilde{{\boldsymbol \varsigma}}}_{{\bf k}'}$ contain symmetry data [\emph{i.e.} the multiplicities $a_{i}^{\bf k}$ in Eq.~(\ref{eq:reducibleCompatibility}) and the corresponding multiplicities $a_{j}^{{\bf k}'}$ at ${\bf k}'$] indicating the number of Bloch wavefunctions that transform in the $i^{\text{th}}$ ($j^{\text{th}}$) small corep $\tilde{\sigma}_{i,{\bf k}}$ ($\tilde{\sigma}_{j,{\bf k}'}$) of $G_{\bf k}$ ($G_{{\bf k}'}$) in an energetically isolated group of Bloch states at ${\bf k}$ (${\bf k}'$).  Hence, in Eq.~(\ref{eq:fullCompatibilityMatrix}), $m^{{\bf k},{\bf k}'}$ is a $z\times w$-dimensional matrix whose entries are the compatibility relations $m_{i,j}^{{\bf k},{\bf k}'}$ in Eq.~(\ref{eq:fullCompatibility}).

If $G_{\bf k}$ and $G_{{\bf k}'}$ are Type-I little groups in a Type-I MSG (Appendix~\ref{sec:type1}), then the compatibility relations $m_{i,j}^{{\bf k},{\bf k}'}$ for any irrep pair $\tilde{\sigma}_{i,{\bf k}}$ and $\tilde{\sigma}_{j,{\bf k}'}$ at any pair of connected points ${\bf k}$ and ${\bf k}'$ can be obtained through the existing~\href{https://www.cryst.ehu.es/cgi-bin/cryst/programs/dcomprel.pl}{DCOMPREL} program on the BCS~(\url{https://www.cryst.ehu.es/cgi-bin/cryst/programs/dcomprel.pl})~\cite{QuantumChemistry,Bandrep1,Bandrep2,Bandrep3,JenFragile1,BarryFragile}.  However, if $G_{\bf k}$ or $G_{{\bf k}'}$ is isomorphic to an SSG with antiunitary symmetries (Type-II, III, or IV, Appendices~\ref{sec:type2},~\ref{sec:type3}, and~\ref{sec:type4}, respectively), then we must perform several additional steps to determine $m_{i,j}^{{\bf k}, {\bf k}'}$.  Specifically, if $G_{\bf k}$ is isomorphic to a Type-II, III, or IV SSG, then, for each small corep $\tilde{\sigma}_{i,{\bf k}}$ of $G_{\bf k}$, we first calculate the subduction:
\begin{equation}
\tilde{\sigma}_{i,{\bf k}}\downarrow H_{\bf k} = \bigoplus_{l}b_{i,l}^{\bf k}\sigma_{l,{\bf k}},
\label{eq:corepToIrrepSameK}
\end{equation}
where $H_{\bf k}$ is the maximal unitary subgroup of $G_{\bf k}$, $\sigma_{l,{\bf k}}$ is the $l^{\text{th}}$ small irrep of $H_{\bf k}$, and where each coefficient $b_{i,l}^{\bf k}=0$, $1$, or $2$, depending on whether $\tilde{\sigma}_{i,{\bf k}}$ is a type (a), (b), or (c) small corep [respectively defined in the text surrounding Eqs.~(\ref{eq:typeA}),~(\ref{eq:typeB}), and~(\ref{eq:typeC})].  Specifically, if $\tilde{\sigma}_{i,{\bf k}}$ is a type (a) [(b)] corep, then, for each value of $i$, $b_{i,l}^{\bf k}=1$ [$b_{i,l}^{\bf k}=2$] for one value of $l$, and $b_{i,l}^{\bf k}=0$ for all of the other values of $l$ at fixed $i$; conversely, if $\tilde{\sigma}_{i,{\bf k}}$ is a type (c) corep, then $b_{i,l}^{\bf k}=1$ for \emph{two} values of $l$, and $b_{i,l}^{\bf k}=0$ for all of the other values of $l$ at fixed $i$ (see Appendix~\ref{sec:coreps}).  This occurs because, if $G_{\bf k}$ is isomorphic to a Type-II, III, or IV SSG, then $H_{\bf k}$ is necessarily an index-2 subgroup of $G_{\bf k}$ (see Appendices~\ref{sec:type2},~\ref{sec:type3}, and~\ref{sec:type4}), and because $G_{\bf k} = H_{\bf k}\cup g_{A}H_{\bf k}$ where $g_{A}$ is an antiunitary symmetry $g_{A}\in G_{\bf k},\ g_{A}\not\in H_{\bf k}$.  Hence, as shown in Appendix~\ref{sec:coreps}, each of the small coreps $\tilde{\sigma}_{i,{\bf k}}$ of $G_{\bf k}$ is either equivalent to a small irrep $\sigma_{l,{\bf k}}$ of $H_{\bf k}$ such that $b_{i,l}^{\bf k}=1$ for only one value of $l$	for each $i$ [type (a) corep, see Eq.~(\ref{eq:typeA})], $\tilde{\sigma}_{i,{\bf k}}$ is equivalent to the direct sum $\sigma_{l,{\bf k}}\oplus \sigma_{l,{\bf k}}$ such that $b_{i,l}^{\bf k}=2$ for only one value of $l$ for each $i$ [type (b) corep, see Eq.~(\ref{eq:typeB})], or $\tilde{\sigma}_{i,{\bf k}}$ is equivalent to the direct sum $\sigma_{l1,{\bf k}}\oplus \sigma_{l2,{\bf k}}$ such that $b_{i,l}^{\bf k}=1$ for only two values $l=l1,l2$ for each $i$ [type (c) corep, see Eq.~(\ref{eq:typeC})].  The values of $b_{i,l}^{\bf k}$ in Eq.~(\ref{eq:corepToIrrepSameK}) can be obtained from the~\href{http://www.cryst.ehu.es/cryst/corepresentations}{Corepresentations} tool introduced in this work, which we previously detailed in Appendix~\ref{sec:coreps}.  In the notation of Eq.~(\ref{eq:fullCompatibilityMatrix}), Eq.~(\ref{eq:corepToIrrepSameK}) can be re-expressed as:
\begin{equation}
b^{\bf k}{\bf \tilde{{\boldsymbol \varsigma}}}_{\bf k} = {\boldsymbol \varsigma}_{\bf k},
\label{eq:matrixCorepToIrrepSameK}
\end{equation}
in which ${\boldsymbol \varsigma}_{\bf k}$ is an $x\times 1$-dimensional column vector whose $l^{\text{th}}$ entry is the multiplicity of $\sigma_{l,{\bf k}}$ in $\tilde{\varsigma}_{\bf k}\downarrow H_{\bf k}$, where $x$ is the number of small irreps of $H_{\bf k}$, and where $b^{\bf k}$ is a $x\times w$-dimensional matrix whose entries are $b_{i,l}^{\bf k}$ in Eq.~(\ref{eq:corepToIrrepSameK}).  Next, for each small irrep $\sigma_{l,{\bf k}}$ of $H_{\bf k}$, we further subduce onto $H_{{\bf k}'}$, the maximal unitary subgroup of $G_{{\bf k}'}$:
\begin{equation}
\sigma_{l,{\bf k}}\downarrow H_{{\bf k}'} = \bigoplus_{s} n_{l,s}^{{\bf k}, {\bf k}'}\sigma_{s,{\bf k}'},
\label{eq:irrepToIrrepOtherK}
\end{equation}
where $\sigma_{s,{\bf k}'}$ is the $s^{\text{th}}$ small irrep of $H_{{\bf k}'}$ and $n_{l,s}^{{\bf k},{\bf k}'}$ is the multiplicity of $\sigma_{s,{\bf k}'}$ in $\sigma_{l,{\bf k}}\downarrow H_{{\bf k}'}$.  As with $m_{i,j}^{{\bf k},{\bf k}'}$ in Eq.~(\ref{eq:fullCompatibility}), the values of $n_{l,s}^{{\bf k},{\bf k}'}$ in Eq.~(\ref{eq:irrepToIrrepOtherK}) are required to be non-negative integers, because they originate from group-subgroup subduction~\cite{SerreLinearReps}.  Crucially, because $H_{\bf k}$ and $H_{{\bf k}'}$ are both isomorphic to Type-I MSGs, then the compatibility relations $n_{l,s}^{{\bf k},{\bf k}'}$ for all possible connected points ${\bf k}$ and ${\bf k}'$ in all 1,651 SSGs can be determined using the earlier~\href{https://www.cryst.ehu.es/cgi-bin/cryst/programs/dcomprel.pl}{DCOMPREL} tool, which is documented in Ref.~\onlinecite{Bandrep2}.  Following Eq.~(\ref{eq:fullCompatibilityMatrix}), Eq.~(\ref{eq:irrepToIrrepOtherK}) can be re-expressed as:
\begin{equation}
n^{{\bf k},{\bf k}'}{\boldsymbol \varsigma}_{\bf k} = {\boldsymbol \varsigma}_{{\bf k}'},
\label{eq:irrepToIrrepMatrix}
\end{equation}
in which ${\boldsymbol \varsigma}_{{\bf k}'}$ is a $y\times 1$-dimensional column vector whose $s^{\text{th}}$ entry is the multiplicity of $\sigma_{s,{\bf k}'}$ in $\tilde{\varsigma}_{{\bf k}'}\downarrow H_{{\bf k}'}$, where $y$ is the number of small irreps of $H_{{\bf k}'}$, and where $n^{{\bf k},{\bf k}'}$ is an $y\times x$-dimensional matrix whose entries are the unitary subgroup compatibility relations $n_{l,s}^{{\bf k},{\bf k}'}$ in Eq.~(\ref{eq:irrepToIrrepOtherK}).  As a last step towards calculating the compatibility relations $m_{i,j}^{{\bf k},{\bf k}'}$ in Eq.~(\ref{eq:fullCompatibility}), we calculate the subduction onto $H_{{\bf k}'}$ for each small corep $\tilde{\sigma}_{j,{{\bf k}'}}$ of $G_{{\bf k}'}$:
\begin{equation}
\tilde{\sigma}_{j,{{\bf k}'}}\downarrow H_{{\bf k}'} = \bigoplus_{s} c_{j,s}^{{\bf k}'}\sigma_{s,{{\bf k}'}},
\label{eq:corepToIrrepKPrime}
\end{equation}
where $\sigma_{s,{{\bf k}'}}$ is the $s^{\text{th}}$ small irrep of $H_{{\bf k}'}$, and where, as detailed in the text following Eq.~(\ref{eq:corepToIrrepSameK}), each coefficient $c_{j,s}^{{\bf k}'}=0$, $1$, or $2$, depending on whether $\tilde{\sigma}_{j,{{\bf k}'}}$ is a type (a), (b), or (c) small corep [defined in the text surrounding Eqs.~(\ref{eq:typeA}),~(\ref{eq:typeB}), and~(\ref{eq:typeC}), respectively].  As previously with $b_{i,l}^{\bf k}$ in Eq.~(\ref{eq:corepToIrrepSameK}), the values of $c_{j,s}^{{\bf k}'}$ in Eq.~(\ref{eq:corepToIrrepKPrime}) can also be obtained from the~\href{http://www.cryst.ehu.es/cryst/corepresentations}{Corepresentations} tool introduced in this work (Appendix~\ref{sec:coreps}).  Like Eq.~(\ref{eq:corepToIrrepSameK}), Eq.~(\ref{eq:corepToIrrepKPrime}) can be re-expressed in the form of Eq.~(\ref{eq:matrixCorepToIrrepSameK}):
\begin{equation}
c^{{\bf k}'}{\bf \tilde{{\boldsymbol \varsigma}}}_{{\bf k}'} = {\boldsymbol \varsigma}_{{\bf k}'},
\label{eq:matrixCorepToIrrepKPrime}
\end{equation}
where $c^{{\bf k}'}$ is a $y\times z$-dimensional matrix whose entries are $c_{j,s}^{{\bf k}'}$ in Eq.~(\ref{eq:corepToIrrepKPrime}).  Finally, by combining Eqs.~(\ref{eq:fullCompatibilityMatrix}),~(\ref{eq:matrixCorepToIrrepSameK}),~(\ref{eq:irrepToIrrepMatrix}), and~(\ref{eq:matrixCorepToIrrepKPrime}), we determine that:
\begin{equation}
c^{{\bf k}'}m^{{\bf k},{\bf k}'}  = n^{{\bf k},{\bf k}'}b^{\bf k}.
\label{eq:nearFinalCompatibility}
\end{equation}

To solve for $m^{{\bf k},{\bf k}'}$ in Eq.~(\ref{eq:nearFinalCompatibility}), we need to obtain a left inverse for $c^{{\bf k}'}$ [\emph{i.e.} a matrix $(c^{{\bf k}'})^{-1}$ for which $(c^{{\bf k}'})^{-1}c^{{\bf k}'}=\mathds{1}_{z}$], where $(c^{{\bf k}'})^{-1}$ is guaranteed to exist (though not necessarily be unique), because of Frobenius reciprocity~\cite{SerreLinearReps,Bandrep1}.  Conversely, because $c^{{\bf k}'}$ in Eq.~(\ref{eq:matrixCorepToIrrepKPrime}) is generically non-square and left-invertible, then a right inverse for $c^{{\bf k}'}$ does not generically also exist.  Frobenius reciprocity specifically implies that we can obtain a left inverse for $c^{{\bf k}'}$ through \emph{induction}:
\begin{equation}
\sigma_{s,{\bf k}'} \uparrow G_{{\bf k}'} = \bigoplus_{j} d^{{\bf k}'}_{s,j} \tilde{\sigma}_{j,{\bf k}'},
\label{eq:compatibilityInduction}
\end{equation}
where each coefficient $d^{{\bf k}'}_{s,j}=0$ or $1$, independent of whether $\tilde{\sigma}_{j,{{\bf k}'}}$ is a type (a), (b), or (c) small corep [defined in the text surrounding Eqs.~(\ref{eq:typeA}),~(\ref{eq:typeB}), and~(\ref{eq:typeC}), respectively].  Specifically, because $G_{{\bf k}'}=H_{{\bf k}'}\cup g_{A}H_{{\bf k}'}$ where $g_{A}$ is an antiunitary symmetry, then regardless of the type of the corep $\tilde{\sigma}_{j,{{\bf k}'}}$, $d^{{\bf k}'}_{s,j}=1$ for one value of $j$, and $d^{{\bf k}'}_{s,j}=0$ for all other values of $j$ at fixed $s$ (see Appendix~\ref{sec:coreps}).  We next re-express  Eq.~(\ref{eq:compatibilityInduction}) in the form of an inverse of Eq.~(\ref{eq:matrixCorepToIrrepKPrime}):
\begin{equation}
\left(c^{{\bf k}'}\right)^{-1}{\boldsymbol \varsigma}_{{\bf k}'} = {\bf \tilde{{\boldsymbol \varsigma}}}_{{\bf k}'},
\label{eq:matrixInverseCompatibility}
\end{equation}
in which $(c^{{\bf k}'})^{-1}$ is the left inverse of $c^{{\bf k}'}$ and, crucially:
\begin{equation}
\left[\left(c^{{\bf k}'}\right)^{-1}\right]_{sj} = \frac{d^{{\bf k}'}_{s,j}}{[G_{{\bf k}'}:H_{{\bf k}'}]},
\label{eq:elementsOfInductionCompatibility}
\end{equation}
where $[G_{{\bf k}'}:H_{{\bf k}'}]$ is the index of the subgroup $H^{{\bf k}'}$ of $G^{{\bf k}'}$ [Eq.~(\ref{eq:subsetIndex})], which is present in Eq.~(\ref{eq:elementsOfInductionCompatibility}) because induction ($\uparrow$), unlike subduction ($\downarrow$), \emph{does not} preserve dimensionality (\emph{i.e.}, the character of the identity element $E$)~\cite{Bandrep1,SerreLinearReps}.  Therefore, independent of the SSG (little group) type of $G_{{\bf k}'}$, $(c^{{\bf k}'})^{-1}$ in Eq.~(\ref{eq:matrixInverseCompatibility}) is necessarily well-defined, and its entries [Eq.~(\ref{eq:elementsOfInductionCompatibility})] are non-negative, though they are not necessarily integers.  Specifically, if $G_{{\bf k}'}$ is isomorphic to a Type-II, III, or IV SSG (Appendices~\ref{sec:type2},~\ref{sec:type3}, and~\ref{sec:type4}, respectively), then $G_{{\bf k}'}$ is necessarily an index-2 supergroup of $H_{{\bf k}'}$, such that $[G_{{\bf k}'}:H_{{\bf k}'}]=2$, implying that the elements $[(c^{{\bf k}'})^{-1}]_{sj}$ in Eq.~(\ref{eq:elementsOfInductionCompatibility}) are non-negative multiples of $1/2$.  Nevertheless, we have verified that, for all connected little group pairs $G_{{\bf k}'}\subset G_{\bf k}$ in all 1,651 single and double SSGs, the elements of $m^{{\bf k},{\bf k}'}$ in the expression:
\begin{equation}
m^{{\bf k},{\bf k}'} = \left(c^{{\bf k}'}\right)^{-1}n^{{\bf k},{\bf k}'}b^{\bf k},
\label{eq:finalCompatibility}
\end{equation}
formed from Eqs.~(\ref{eq:nearFinalCompatibility}),~(\ref{eq:compatibilityInduction}),~(\ref{eq:matrixInverseCompatibility}), and~(\ref{eq:elementsOfInductionCompatibility}) are non-negative integers, as required by subduction [see the text following Eq.~(\ref{eq:fullCompatibility})].  Eq.~(\ref{eq:finalCompatibility}) implies that the multiplicities $b_{i,l}^{\bf k}$ and $c_{j,s}^{{\bf k}'}$ obtained from~\href{http://www.cryst.ehu.es/cryst/corepresentations}{Corepresentations} and the unitary subgroup compatibility relations $n_{l,s}^{{\bf k},{\bf k}'}$ obtained from~\href{https://www.cryst.ehu.es/cgi-bin/cryst/programs/dcomprel.pl}{DCOMPREL} determine the compatibility relations $m_{i,j}^{{\bf k},{\bf k}'}$ between any two small coreps $\tilde{\sigma}_{i,{\bf k}}$ and $\tilde{\sigma}_{j,{\bf k}'}$ at any two connected points ${\bf k}$ and ${\bf k}'$ in any of the 1,651 SSGs.  To simplify this procedure, we have implemented a new tool -- \href{https://www.cryst.ehu.es/cryst/mcomprel}{MCOMPREL} -- through which the values of $m_{i,j}^{{\bf k},{\bf k}'}$ can be directly obtained without using additional programs on the BCS.  Further specific details of the implementation of~\href{https://www.cryst.ehu.es/cryst/mcomprel}{MCOMPREL} are available in the documentation provided on the BCS.

We will now briefly present an example demonstrating the derivation of the multiplicities and compatibility relations at two connected ${\bf k}$ points for the double-valued small coreps of Type-III double MSG 83.45 $P4'/m$, which is generated by:
\begin{equation}
\{C_{4z}\times\mathcal{T}|000\},\ \{\mathcal{I}|000\},\ \{E|100\},\ \{E|001\}.
\end{equation}
In this example, we will specifically obtain the small corep compatibility relations [Eq.~(\ref{eq:fullCompatibility})] for $G=P4'/m$ at the connected points:
\begin{equation}
{\bf k}_{\Gamma} = (0,0,0),\ {\bf k}_{LD} = (0,0,w).
\end{equation}
First, using~\href{http://www.cryst.ehu.es/cryst/corepresentations}{Corepresentations}, we determine that the little group $G_{\Gamma}$ is isomorphic to Type-III MSG 83.45 $P4'/m$, and has two, two-dimensional, double-valued small coreps $\tilde{\sigma}_{1,\Gamma}$ and $\tilde{\sigma}_{2,\Gamma}$, which are distinguished by their $\{\mathcal{I}|{\bf 0}\}$ eigenvalues:
\begin{equation}
\chi_{\tilde{\sigma}_{1,\Gamma}}(\{\mathcal{I}|{\bf 0}\}) = 2,\ \chi_{\tilde{\sigma}_{2,\Gamma}}(\{\mathcal{I}|{\bf 0}\}) = -2.
\end{equation}
Next, continuing to employ~\href{http://www.cryst.ehu.es/cryst/corepresentations}{Corepresentations}, we focus on the maximal unitary subgroup $H_{\Gamma}$ of $G_{\Gamma}$.  $H_{\Gamma}$ is isomorphic to Type-I MSG 10.42 $P2/m$, and has \emph{four}, one-dimensional, double-valued small irreps $\sigma_{1-4,\Gamma}$, which are also distinguished by their $\{\mathcal{I}|{\bf 0}\}$ and $\{C_{2z}|{\bf 0}\}=(\{C_{4z}\times\mathcal{T}|{\bf 0}\})^{6}$ eigenvalues:
\begin{eqnarray}
\chi_{\sigma_{1,\Gamma}}(\{\mathcal{I}|{\bf 0}\}) &=& 1,\ \chi_{\sigma_{2,\Gamma}}(\{\mathcal{I}|{\bf 0}\}) = 1,\ \chi_{\sigma_{3,\Gamma}}(\{\mathcal{I}|{\bf 0}\}) = -1,\ \chi_{\sigma_{4,\Gamma}}(\{\mathcal{I}|{\bf 0}\}) = -1,\ \nonumber \\
\chi_{\sigma_{1,\Gamma}}(\{C_{2z}|{\bf 0}\}) &=& -i,\ \chi_{\sigma_{2,\Gamma}}(\{C_{2z}|{\bf 0}\}) = i,\ \chi_{\sigma_{3,\Gamma}}(\{C_{2z}|{\bf 0}\}) = -i,\ \chi_{\sigma_{4,\Gamma}}(\{C_{2z}|{\bf 0}\}) = i.\ \nonumber \\
\end{eqnarray}
We next subduce the small coreps $\tilde{\sigma}_{i,\Gamma}$ of $G_{\Gamma}$ onto $H_{\Gamma}$ [Eq.~(\ref{eq:corepToIrrepSameK})]:
\begin{equation}
\tilde{\sigma}_{1,\Gamma}\downarrow H_{\Gamma} = \sigma_{1,\Gamma}\oplus\sigma_{2,\Gamma},\ \tilde{\sigma}_{2,\Gamma}\downarrow H_{\Gamma} = \sigma_{3,\Gamma}\oplus\sigma_{4,\Gamma}, 
\end{equation}
which may be summarized by introducing the multiplicity matrix [Eq.~(\ref{eq:matrixCorepToIrrepSameK})]:
\begin{equation}
b^{\Gamma} = \left(\begin{array}{cc}
1 & 0 \\
1 & 0 \\
0 & 1 \\
0 & 1 \end{array}\right).
\end{equation}
We then focus on the little group $G_{LD}$, which is isomorphic to Type-III MSG 75.3 $P4'$, and is generated by:
\begin{equation}
\{C_{4z}\times\mathcal{T}|000\},\ \{E|100\},\ \{E|001\}.
\end{equation}
$G_{LD}$ has only one, two-dimensional double-valued small corep $\tilde{\sigma}_{1,LD}$. The maximal unitary subgroup $H_{DT}$ of $G_{DT}$ is isomorphic to Type-I MSG $3.1$ $P2$, and has two, one-dimensional, double-valued small irreps $\sigma_{1,LD}$ and $\sigma_{2,LD}$, which are distinguished by their $\{C_{2z}|{\bf 0}\}$ eigenvalues:
\begin{equation}
\chi_{\sigma_{1,LD}}(\{C_{2z}|{\bf 0}\}) = -i,\ \chi_{\sigma_{2,LD}}(\{C_{2z}|{\bf 0}\}) = i.
\end{equation}
Hence, through subduction [Eq.~(\ref{eq:corepToIrrepKPrime})], we obtain:
\begin{equation}
\tilde{\sigma}_{1,LD} = \sigma_{1,LD}\oplus\sigma_{2,LD},
\end{equation}
which may be summarized through the multiplicity matrix [Eq.~(\ref{eq:matrixCorepToIrrepKPrime})]:
\begin{equation}
c^{LD} = \left(\begin{array}{c}
1 \\
1\end{array}\right). 
\end{equation}
Next, we obtain a left inverse for $c^{LD}$ by establishing that $[G_{LD}:H_{LD}]=2$ [see Eq.~(\ref{eq:intermediateCosetRep2}) and the surrounding text], and that:
\begin{equation}
\sigma_{1,LD}\uparrow G_{LD} = \sigma_{2,LD}\uparrow G_{LD}= \tilde{\sigma}_{1,LD}.
\end{equation}
Through Eq.~(\ref{eq:elementsOfInductionCompatibility}), this implies that:
\begin{equation}
\left(c^{LD}\right)^{-1} = \frac{1}{2}\left(\begin{array}{cc}
1 & 1\end{array}\right). 
\end{equation}
As a final step towards computing the corep compatibility relations $m^{\Gamma,LD}$, we use subduction to obtain the unitary subgroup compatibility relations [Eq.~(\ref{eq:irrepToIrrepOtherK})]:
\begin{eqnarray}
\sigma_{1,\Gamma}\downarrow H_{LD} &=& \sigma_{3,\Gamma}\downarrow H_{LD} = \sigma_{1,LD}, \nonumber \\
\sigma_{2,\Gamma}\downarrow H_{LD} &=& \sigma_{4,\Gamma}\downarrow H_{LD} = \sigma_{2,LD}.
\label{eq:trivialCompatibility}
\end{eqnarray}
consistent with the output of the earlier~\href{https://www.cryst.ehu.es/cgi-bin/cryst/programs/dcomprel.pl}{DCOMPREL} tool.  Eq.~(\ref{eq:trivialCompatibility}) may be summarized by the multiplicity matrix [Eq.~(\ref{eq:irrepToIrrepMatrix})]:
\begin{equation}
n^{\Gamma,LD} = \left(\begin{array}{cccc}
1 & 0 & 1 & 0 \nonumber \\
0 & 1 & 0 & 1 \end{array}\right).
\end{equation}
Lastly, we compute the small corep compatibility relations $m^{\Gamma,LD}$ using Eq.~(\ref{eq:finalCompatibility}):
\begin{equation}
m^{\Gamma,LD} = \left(c^{LD}\right)^{-1}n^{\Gamma,LD}b^{\Gamma} = \frac{1}{2}\left(\begin{array}{cc}
1 & 1\end{array}\right)\left(\begin{array}{cccc}
1 & 0 & 1 & 0 \nonumber \\
0 & 1 & 0 & 1 \end{array}\right) \left(\begin{array}{cc}
1 & 0 \\
1 & 0 \\
0 & 1 \\
0 & 1 \end{array}\right) = \left(\begin{array}{cc}
1 & 1 \end{array}\right),
\end{equation}
in agreement with the subduction relations:
\begin{equation}
\tilde{\sigma}_{1,\Gamma}\downarrow G_{LD} = \tilde{\sigma}_{2,\Gamma}\downarrow G_{LD} = \tilde{\sigma}_{1,LD},
\end{equation}
as well as the output of the~\href{https://www.cryst.ehu.es/cryst/mcomprel}{MCOMPREL} tool introduced in this work.

\begin{figure}[t]
\includegraphics[width=\columnwidth]{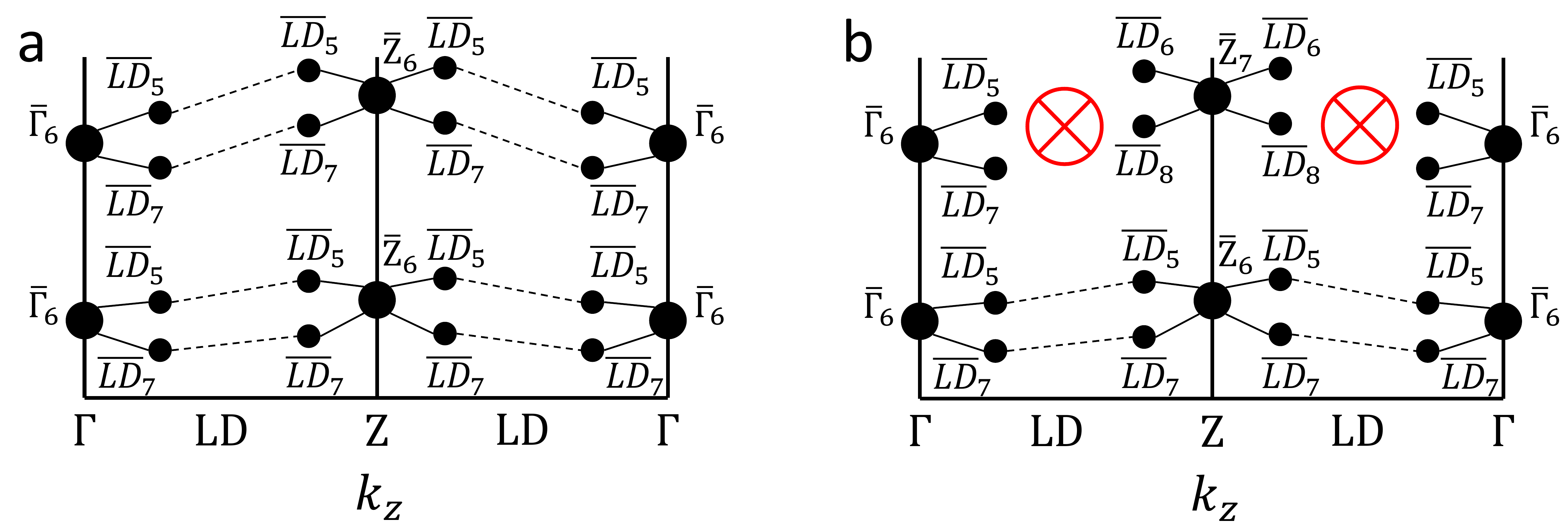}
\caption{Compatibility relations and graphs for magnetic rod group (MRG)  $(p422)_{RG}$, which is generated by $\{E|1\}$, $\{C_{4z}|0\}$, and $\{C_{2x}|0\}$, and is isomorphic after the addition of perpendicular lattice translations to Type-I MSG 89.87 $P422$ [see Refs.~\onlinecite{BigBook,MagneticBook,ITCA,subperiodicTables,HingeSM} and the text following Eq.~(\ref{eq:translationNotation})].  Using the~\href{http://www.cryst.ehu.es/cryst/mkvec}{MKVEC} tool on the BCS for the $k_{x}=k_{y}=0$ line in MSG 89.87 $P422$, we deduce that there are only three momentum stars (Appendix~\ref{sec:MKVEC}) in MRG $(p422)_{RG}$: $\Gamma$ ($k_{z}=0$), $Z$ ($k_{z} = \pi$), and LD ($k_{z} = \pm w$).  Next, using~\href{https://www.cryst.ehu.es/cryst/mcomprel}{MCOMPREL}, we obtain the compatibility relations for MRG $(p422)_{RG}$ [\emph{i.e.}, the values of $m_{i,j}^{{\bf k},{\bf k}'}$ in Eq.~(\ref{eq:fullCompatibility})], which, restricting to double-valued (spinful) coreps, are given by $\bar{\Gamma}_{6}\downarrow G_{\text{LD}} = \bar{Z}_{6}\downarrow G_{\text{LD}} = \overline{\text{LD}}_{5}\oplus \overline{\text{LD}}_{7}$ and $\bar{\Gamma}_{7}\downarrow G_{\text{LD}} = \bar{Z}_{7} \downarrow G_{\text{LD}} = \overline{\text{LD}}_{6}\oplus \overline{\text{LD}}_{8}$.  (a)  For a set of four spinful Bloch eigenstates at each $k_{z}$ point with a symmetry data vector [see Refs.~\onlinecite{AndreiMaterials,MTQCmaterials} and the text following Eq.~(\ref{eq:reducibleCompatibility})] given by $\tilde{\varsigma}_{\Gamma} = \bar{\Gamma}_{6}\oplus\bar{\Gamma}_{6}$ and $\tilde{\varsigma}_{Z} = \bar{Z}_{6}\oplus\bar{Z}_{6}$, a separated pair of connected graphs can be formed from the coreps at $\Gamma$ and $Z$ using the TQC graph-theory methodology detailed in Refs.~\onlinecite{Bandrep2,Bandrep3}.  The symmetry data in (a) is therefore compatible with an insulating (band) gap at a filling $\nu = 4$.  (b) Conversely, for a set of four spinful Bloch eigenstates at each $k_{z}$ point with a symmetry data vector given by $\tilde{\varsigma}_{\Gamma}= \bar{\Gamma}_{6}\oplus\bar{\Gamma}_{6}$ and $\tilde{\varsigma}_{Z}= \bar{Z}_{6}\oplus\bar{Z}_{7}$, there does not exist a graph for the coreps at $\Gamma$ and $Z$ that satisfies the compatibility relations.  The symmetry data in (b) is therefore \emph{incompatible} with a band gap at a filling $\nu=4$, implying that the Bloch eigenstates at $\Gamma$ and $Z$ are connected to other, unoccupied states (bands) not described by the symmetry data.  In the nomenclature of Refs.~\onlinecite{AndreiMaterials,MTQCmaterials}, the symmetry data in (b) consequently corresponds to an ``enforced semimetal'' (ES).}
\label{fig:compatibility}
\end{figure}

\begin{figure}[t]
\includegraphics[width=0.57\columnwidth]{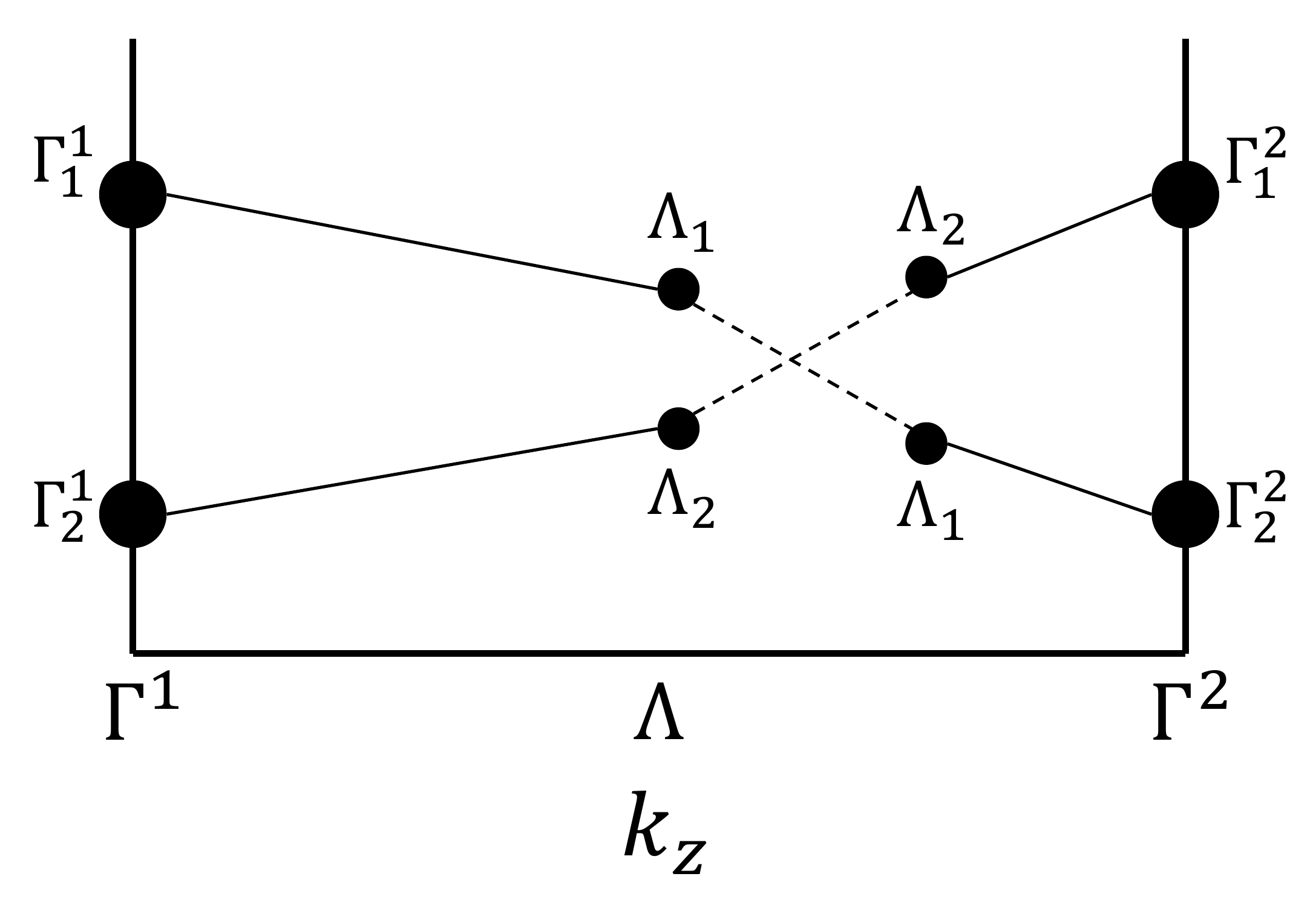}
\caption{Compatibility relations and graphs for MRG  $(p2_{1})_{RG}$, which is generated by a twofold screw operation ($s_{2_{1}}=\{C_{2z}|1/2\}$), and is isomorphic after the addition of perpendicular lattice translations to Type-I MSG 4.7 $P2_{1}$ [see Refs.~\onlinecite{BigBook,MagneticBook,ITCA,subperiodicTables,HingeSM} and the text following Eq.~(\ref{eq:translationNotation})].  Using the~\href{http://www.cryst.ehu.es/cryst/mkvec}{MKVEC} tool on the BCS for the $k_{x}=k_{z}=0$ line in MSG 4.7 $P2_{1}$, we deduce that there is only one, multiplicity-1 momentum star (Appendix~\ref{sec:MKVEC}) in MRG $(p2_{1})_{RG}$, which is labeled LD and lies at ${\bf k}_{\text{LD}} = v{\bf\hat{z}}$.  In an example of the representation monodromy discussed in this section, the matrix representatives of the small irreps of $G_{\text{LD}}$ are $4\pi$-periodic in $v$ [Eq.~(\ref{eq:LDmatReps})].  Specializing to three values of $v$ respectively given by $v=0$ ($\Gamma^{1}$), $0<v<2\pi$ ($\Lambda$), and $v=2\pi$ ($\Gamma^{2}$), we observe that the compatibility relations [Eq.~(\ref{eq:specificMonodromyMatReps})] imply that the small irreps $\Lambda_{1,2}$ connect to different irreps at the $\Gamma$ point, depending on whether $\Gamma$ lies in an odd-numbered BZ (\emph{e.g.} $\Gamma^{1}$ in the first BZ) or in an even-numbered BZ (\emph{e.g.} $\Gamma^{2}$ in the second BZ).  Crucially, the irrep labels ($\Gamma^{i}_{1,2}$) are the same at the $\Gamma$ point in each BZ, consistent with the restriction that physical observables in pristine crystals are $2\pi$-periodic (\emph{i.e.}, any physical observable in an infinite, periodic system must be the same at any two points ${\bf k}$ and ${\bf k}'$ that differ by a linear combination of reciprocal lattice vectors ${\bf K}_{\nu}$)~\cite{BigBook}.  Specifically restricting to spinless Bloch eigenstates, this implies that a pair of states (bands) with the symmetry data vector $\tilde{\varsigma}_{\Gamma^{1}}= \Gamma^{1}_{1}\oplus\Gamma^{1}_{2}$ [see Refs.~\onlinecite{AndreiMaterials,MTQCmaterials} and the text following Eq.~(\ref{eq:reducibleCompatibility})], will be connected at an odd number of ${\bf k}$ points in each BZ, where one of the crossing points in each BZ (\emph{i.e.} the intersection of the dashed lines in this figure) is movable, but unremovable~\cite{ZakException1,ZakCompatibility,WPVZ,Steve2D,WiederLayers,SteveMagnet,Bandrep2,Bandrep3,HourglassInsulator,Cohomological,DiracInsulator}.}
\label{fig:monodromy}
\end{figure}

One of the key advances of TQC~\cite{QuantumChemistry,Bandrep1,Bandrep2,Bandrep3,JenFragile1,BarryFragile} and related works~\cite{SlagerSymmetry,AshvinIndicators} was to recognize that, for each Type-I MSG and Type-II SSG, there existed a small number of maximal ${\bf k}$ vectors [Eq.~(\ref{eq:maximalKvec})] from which the connectivity of Bloch eigenstates (\emph{i.e.} energy bands) throughout the entire BZ could be inferred from the symmetry data [$\tilde{\varsigma}_{\bf k}$ in Eq.~(\ref{eq:reducibleCompatibility})].  Specifically, given a symmetry data vector $\{\tilde{\varsigma}_{\bf k}\}$, the set of small coreps at each ${\bf k}$ point can be re-expressed as the nodes of a weighted graph whose edges are required to be consistent with the compatibility relations [\emph{i.e.}, the values of $m_{i,j}^{{\bf k},{\bf k}'}$ in Eq.~(\ref{eq:fullCompatibility})].  If such a graph cannot be constructed without violating the compatibility relations, then the bands characterized by the symmetry data vector $\{\tilde{\varsigma}_{\bf k}\}$ are necessarily connected to other bands, implying that the bulk is a form of topological semimetal [an ``enforced semimetal'' (ES) in the nomenclature of Ref.~\onlinecite{AndreiMaterials} (see Fig.~\ref{fig:compatibility})].  However, if a graph can be constructed, then it may further be separated into disconnected subgraphs.  As we will discuss in Appendix~\ref{sec:MEBRs}, by collecting the symmetry data induced from (magnetic) atomic orbitals located at maximal Wyckoff positions (Appendix~\ref{sec:Wyckoff}) and using the compatibility relations in~\href{https://www.cryst.ehu.es/cryst/mcomprel}{MCOMPREL} to construct graphs (which may be additionally separable into disconnected subgraphs), we obtain the EBRs of all 1,651 SSGs (specifically the PEBRs of the Type-II SSGs, and the MEBRs of the Type-I, III, and IV MSGs)~\cite{ZakBandrep1,ZakBandrep2,EvarestovBook,EvarestovMEBR,BarryBandrepReview,QuantumChemistry,Bandrep1,Bandrep2,Bandrep3,JenFragile1,BarryFragile}.  In this work, we will not provide further specific details of the TQC graph theory implementation~\cite{Bandrep2,Bandrep3}; we will instead simply note that, for a given SSG, once the BCS tools introduced in this work have been used to obtain the momentum stars [\href{http://www.cryst.ehu.es/cryst/mkvec}{MKVEC}, see Appendix~\ref{sec:MKVEC}], small coreps [\href{http://www.cryst.ehu.es/cryst/corepresentations}{Corepresentations}, see Appendix~\ref{sec:coreps}], and compatibility relations [\href{https://www.cryst.ehu.es/cryst/mcomprel}{MCOMPREL}, see the text following Eq.~(\ref{eq:fullCompatibility}) in this section] then the previous graph theory construction from TQC can be used without further modification.  Concurrently with the preparation of this work, the MSG compatibility relations in~\href{https://www.cryst.ehu.es/cryst/mcomprel}{MCOMPREL} were employed to perform a high-throughput analysis~\cite{MTQCmaterials} of band connectivity and topology in the $\sim 500$ magnetic materials on the BCS with well-characterized MSGs~\cite{BCSMag1,BCSMag2,BCSMag3,BCSMag4}.

As a final note, there are additional subtleties that come into play in determining the compatibility relations [Eq.~(\ref{eq:fullCompatibility})] and constructing connectivity graphs (Refs.~\onlinecite{Bandrep2,Bandrep3} and Fig.~\ref{fig:compatibility}) in non-primitive SSGs [defined as SSGs whose gray Bravais lattices are not primitive~\cite{BigBook}], SSGs without orthogonal lattice vectors [\emph{e.g.} hexagonal SSGs], and nonsymmorphic SSGs [defined in the text following Eq.~(\ref{eq:cosetOverallDecomposition})].  First, in non-primitive SSGs, and in SSGs whose generating translations [Eq.~(\ref{eq:translationGroup})] are not orthogonal, the construction of a graph (or failure to construct a graph) may depend on the compatibility relations along two distinct paths between the same maximal ${\bf k}$ points.  For example, in Type-I MSG 209.48 $F432$, given symmetry data at the maximal ${\bf k}$ points $\Gamma$ [${\bf k}_{\Gamma}=(0,0,0)$, $G_{\Gamma}$ is isomorphic to Type-I MSG 209.48 $F432$] and $X$ [${\bf k}_{X}=(\pi,\pi,0)$, $G_{X}$ is isomorphic to Type-I MSG 97.151 $I422$], the possibility of constructing a graph depends on the compatibility relations along \emph{both} of the lines $DT$ [$k_{DT}=(0,v,0)$, $G_{DT}$ is isomorphic to Type-I MSG 79.25 $I4$] and $SM$ [${\bf k}_{SM}=(u,u,0)$, $G_{SM}$ is isomorphic to Type-I MSG 5.13 $C2$].  This occurs because, for generic values of $v$ and $u$, ${\bf k}_{DT}$ and ${\bf k}_{SM}$ are not related by any of the symmetries $g\in F432$ -- if ${\bf k}_{DT}$ and ${\bf k}_{SM}$ were instead related by symmetries, then ${\bf k}_{DT}$ and ${\bf k}_{SM}$ would be arms of the same momentum star, and the compatibility relations across the BZ would only depend on the compatibility relations along either $DT$ or $SM$.  We note that, throughout the BCS, ${\bf k}$ points are labeled in some applications with Greek letters (\emph{e.g.} $\Gamma$), whereas in other applications, the same ${\bf k}$ point is labeled with an English abbreviation (\emph{e.g.} $GM$).  Hence, in this work, we will in general employ a mixed notation in which Greek letters and English abbreviations are consistently used throughout each example, where specific labels are chosen to maximize consistency with previous works and with the output of the BCS tools introduced in this work.

As mentioned above, an additional subtlety occurs in nonsymmorphic SSGs.  Specifically, as discussed in Refs.~\onlinecite{ZakException1,ZakCompatibility,Steve2D,WiederLayers,HourglassInsulator,Cohomological,DiracInsulator}, because of the \emph{monodromy} of representations throughout the BZ, the compatibility relations in a nonsymmorphic SSG can even differ at two ${\bf k}$ points that are related by a reciprocal lattice vector [${\bf K}_{\nu}$ in Eq.~(\ref{eq:dependentKFull})].  For example, consider Type-I magnetic rod group (MRG)  $(p2_{1})_{RG}$ [Fig.~\ref{fig:monodromy}], which is generated by the twofold screw symmetry:
\begin{equation}
s_{2_{1}} = \{C_{2z}|1/2\},
\label{eq:defScrewSym}
\end{equation} 
and is isomorphic after the addition of perpendicular lattice translations to Type-I MSG 4.7 $P2_{1}$ [see Refs.~\onlinecite{BigBook,MagneticBook,ITCA,subperiodicTables,HingeSM} and the text following Eq.~(\ref{eq:translationNotation})].  Using the~\href{http://www.cryst.ehu.es/cryst/mkvec}{MKVEC} tool on the BCS for the $k_{x}=k_{z}=0$ line in MSG 4.7 $P2_{1}$, we deduce that there is only one, multiplicity-1 momentum star (Appendix~\ref{sec:MKVEC}) in MRG $(p2_{1})_{RG}$, which is labeled LD and lies at ${\bf k}_{\text{LD}} = v{\bf\hat{z}}$.  To see the effect of the representation monodromy on the compatibility relations, we will calculate the values of $m_{i,j}^{{\bf k},{\bf k}'}$ in Eq.~(\ref{eq:fullCompatibility}) at three specific ${\bf k}$ points along the rod axis corresponding to different values of $v$ in the same star (LD):
\begin{equation}
{\bf k}_{\Gamma^{1}} = {\bf 0},\ {\bf k}_{\Lambda} = v{\bf\hat{z}},\ {\bf k}_{\Gamma^{2}} = 2\pi{\bf\hat{z}},
\label{eq:kVecsForMonodromy}
\end{equation}
where $\Gamma^{1,2}$ are related by a reciprocal lattice vector:
\begin{equation}
{\bf k}_{\Gamma^{2}} - {\bf k}_{{\Gamma}^{1}} = 2\pi{\bf\hat{z}}.
\label{eq:kVecsReciprocalMonodromy}
\end{equation}

For simplicity, in the current demonstration of the role of representation monodromy in the compatibility relations of MRG $(p2_{1})_{RG}$, we will restrict to the case of spinless Bloch eigenstates, which transform in single-valued small coreps.  Using the~\href{http://www.cryst.ehu.es/cryst/corepresentations}{Corepresentations} tool (Appendix~\ref{sec:coreps}), we determine that, at generic points in the LD star (${\bf k}=v{\bf\hat{z}}$), there are two, one-dimensional small coreps $\text{LD}_{1,2}$, for which the matrix representatives [and characters, see the text surrounding Eq.~(\ref{eq:CorepCoset}) for more information] of the twofold screw symmetry $s_{2_{1}}$ [Eq.~(\ref{eq:defScrewSym})] are given by:
\begin{equation}
\Delta_{\text{LD}_{1}}(s_{2_{1}}) = \chi_{\text{LD}_{1}}(s_{2_{1}}) = e^{iv/2},\ \Delta_{\text{LD}_{2}}(s_{2_{1}}) = \chi_{\text{LD}_{2}}(s_{2_{1}}) = -e^{iv/2}.
\label{eq:LDmatReps}
\end{equation}
Evaluated at the ${\bf k}$ points in Eq.~(\ref{eq:kVecsForMonodromy}), the matrix representatives of twofold screw in Eq.~(\ref{eq:LDmatReps}) become:
\begin{equation}
\Delta_{\Gamma^{1}_{1}}(s_{2_{1}}) = 1,\ \Delta_{\Gamma^{1}_{2}}(s_{2_{1}}) = -1,\ \Delta_{\Lambda_{1}}(s_{2_{1}}) = e^{iv/2},\ \Delta_{\Lambda_{2}}(s_{2_{1}}) = -e^{-iv/2},\ \Delta_{\Gamma^{2}_{1}}(s_{2_{1}}) = -1,\ \Delta_{\Gamma^{2}_{2}}(s_{2_{1}}) = 1, 
\label{eq:specificMonodromyMatReps}
\end{equation}
where we have employed a notation for the small irreps at the $\Gamma^{1,2}$ points in which $\Gamma^{i}_{j}$ denotes the $j^{\text{th}}$ small irrep of the little group $G_{\Gamma^{i}}$ at the ${\bf k}$ point $\Gamma^{i}$ (\emph{i.e.}, at the $\Gamma$ point in the $i^\text{th}$ BZ).  Though $G_{\Gamma^{1}} = G_{\Lambda}=G_{\Gamma^{2}}$, we can still calculate compatibility relations of the form of Eq.~(\ref{eq:fullCompatibility}):
\begin{equation}
\Gamma^{1}_{1}\downarrow G_{\Lambda} = \Lambda_{1},\ \Gamma^{1}_{2}\downarrow G_{\Lambda} = \Lambda_{2},\ \Gamma^{2}_{1} \downarrow G_{\Lambda} = \Lambda_{2},\ \Gamma^{2}_{2} \downarrow G_{\Lambda} = \Lambda_{1}.
\label{eq:finalMonodromy}
\end{equation}
In Eq.~(\ref{eq:finalMonodromy}), we find that, because of the $4\pi$-periodicity of the matrix representatives in Eq.~(\ref{eq:LDmatReps}), the compatibility relations at ${\bf k}_{\Gamma^{1,2}}$ are \emph{different}, despite ${\bf k}_{\Gamma^{1,2}}$ differing by a reciprocal lattice vector [Eq.~(\ref{eq:kVecsReciprocalMonodromy})].  This implies that, as shown in Fig.~\ref{fig:monodromy}, a pair of spinless Bloch states at $\Gamma^{1}$ with the symmetry data $\tilde{\varsigma}_{\Gamma^{1}} = \Gamma^{1}_{1}\oplus\Gamma^{1}_{2}$ [see Refs.~\onlinecite{AndreiMaterials,MTQCmaterials} and the text following Eq.~(\ref{eq:reducibleCompatibility})] will connect with each other, specifically forming a pair of spinless bands that cross at an odd number of ${\bf k}$ points in each BZ, where one of the crossing points in each BZ is movable, but unremovable~\cite{ZakException1,ZakCompatibility,WPVZ,Steve2D,WiederLayers,SteveMagnet,Bandrep2,Bandrep3,HourglassInsulator,Cohomological,DiracInsulator}.

\begin{table}[!b]	
	\begin{tabular}{|l|ll|ll|ll|ll|}
	\hline
\multicolumn{9}{|c|}{SSGs in Which the Monodromy of Representations Provides} \\
\multicolumn{9}{|c|}{Necessary Constraints on Band Connectivity} \\ 
		\hline
Type & Symbol & Number & Symbol & Number & Symbol & Number & Symbol & Number \\
		\hline
Type-I&$P2_1$&4.7&$Pc$&7.24&$Cc$&9.37&$C222_1$&20.31\\
&$Pmc2_1$ (S)&26.66&$Pcc2$ (S)&27.78&$Pca2_1$&29.99&$Pnc2$&30.111\\
&$Pmn2_1$&31.123&$Pna2_1$&33.144&$Cmc2_1$&36.172&$Ccc2$ (S)&37.180\\
&$P4_1$&76.7&$P4_2$&77.13&$P4_3$&78.19&$I4_1$&80.29\\
&$P4_2cm$&101.179&$P4_2nm$&102.187&$P4cc$ (S)&103.195&$P4nc$ (S)&104.203\\
&$P4_2mc$&105.211&$P4_2bc$&106.219&$I4_1md$&109.239&$I4_1cd$&110.245\\
&$P3_1$&144.4&$P3_2$&145.7&$P3_112$&151.29&$P3_212$&153.37\\
&$P3c1$&158.57&$R3c$&161.69&$P6_1$&169.113&$P6_5$&170.117\\
&$P6_2$&171.121&$P6_4$&172.125&$P6_3$&173.129&$P6cc$ (S)&184.191\\
&$P6_3cm$&185.197&$P6_3mc$&186.203&&&&\\
		\hline
Type-II&$P3_11'$&144.5&$P3_21'$&145.8&$P3_1121'$&151.30&$P3_2121'$&153.38\\
\hline
Type-III&$P2_1/m'$ (S)&11.53&$P2'/c$ (S)&13.67&$P2_1'/c$&14.77&$P2_1/c'$&14.78\\
&$C2'/c$ (S)&15.87&$P2'2'2_1$&17.9&$P2_12_1'2'$&18.19&$P2_1'2_1'2_1$&19.27\\
&$C2'2'2_1$&20.33&$Pm'a2'$&28.89&$Pc'a2_1'$&29.101&$Pb'a2'$&32.137\\
&$Pn'a2_1'$&33.146&$Cm'c'2_1$&36.176&$Am'a2'$&40.205&$Ab'a2'$&41.213\\
&$Pccm'$ (S)&49.268&$Pb'an$ (S)&50.279&$Pm'ma$ (S)&51.291&$Pn'na$ (S)&52.307\\
&$Pnn'a$&52.308&$Pm'na$&53.323&$Pmna'$&53.325&$Pc'ca$&54.339\\
&$Pcca'$ (S)&54.341&$Pb'am$ (S)&55.355&$Pc'cn$&56.367&$Pccn'$ (S)&56.368\\
&$Pbc'm$ (S)&57.380&$Pbcm'$&57.381&$Pn'nm$&58.395&$Pm'mn$ (S)&59.407\\
&$Pb'cn$&60.419&$Pbc'n$&60.420&$Pbcn'$&60.421&$Pb'ca$&61.435\\
&$Pn'ma$ (S)&62.443&$Pnm'a$ (S)&62.444&$Pnma'$&62.445&$Cmcm'$ (S)&63.461\\
&$Cm'c'm'$ (S)&63.465&$Cmca'$&64.473&$Cm'c'a'$ (D)&64.477&$Cccm'$ (S)&66.494\\
&$Ccca'$ (S)&68.514&$P4_1'$&76.9&$P4_3'$&78.21&$P4_2/m'$&84.54\\
&$P4_2/n'$&86.70&$I4_1/a'$&88.84&$P4_12'2'$&91.106&$P4_1'2'2$&91.107\\
&$P4_12_1'2'$&92.114&$P4_22'2'$&93.122&$P4_22_1'2'$&94.130&$P4_32'2'$&95.138\\
&$P4_3'2'2$&95.139&$P4_32_1'2'$&96.146&$I4_12'2'$&98.160&$P\bar{4}'2'c$ (S)&112.261\\
&$P\bar{4}'2_1'c$ (S)&114.277&$P\bar{4}'c2'$ (S)&116.294&$P4/m'cc$ (S)&124.353&$P4/n'nc$ (S)&126.377\\
&$P4/m'nc$ (S)&128.401&$P4/n'cc$ (S)&130.425&$P4_2/m'mc$&131.437&$P4_2/m'cm$&132.449\\
&$P4_2/n'bc$&133.461&$P4_2/n'nm$&134.473&$P4_2/m'bc$&135.485&$P4_2/m'nm$&136.497\\
&$P4_2/n'mc$&137.509&$P4_2/n'cm$&138.521&$I4_1/a'md$&141.553&$I4_1/a'cd$&142.563\\
&$P3_112'$&151.31&$P3_12'1$&152.35&$P3_212'$&153.39&$P3_22'1$&154.43\\
&$P\bar{3}'c1$ (S)&165.93&$R\bar{3}'c$ (S)&167.105&$P6_3/m'$&176.146&$P6_12'2'$&178.159\\
&$P6_52'2'$&179.165&$P6_22'2'$&180.171&$P6_42'2'$&181.177&$P6_32'2'$&182.183\\
&$P\bar{6}'c2'$&188.218&$P\bar{6}'2'c$&190.229&$P6/m'cc$ (S)&192.245&$P6_3/m'cm$&193.255\\
&$P6_3/m'mc$&194.265&&&&&&\\
\hline
Type-IV&$P_c3_1$&144.6&$P_c3_2$&145.9&$P_c3_112$&151.32&$P_c3_212$&153.40\\
\hline
	\end{tabular}
\caption{List of SSGs for which the monodromy of representations imposes additional restrictions on small corep (band) connectivity beyond the constraints imposed by the symmetries of the SSG.  The letters (S) and (D) after the symbol of an SSG respectively indicate that the representation monodromy only provides necessary constraints on the connectivity of single- and double-valued coreps of that SSG.  In all of the other SSGs listed in this table, the representation monodromy provides necessary constraints on the connectivity of both single- and double-valued small coreps.}
\label{tb:monodromy}
\end{table}

If additional symmetries are present in an SSG, such as $\{\mathcal{T}|000\}$ in Type-II SSGs (Appendix~\ref{sec:type2}), then the effects of representation monodromy on the compatibility relations may be redundant with the constraints imposed by the additional symmetries.  Specifically, in Type-II SSGs, $\mathcal{T}$ symmetry relates half of a high-symmetry line to its time-reversal partner, providing further restrictions on corep connectivity that can be used in lieu of comparing the compatibility relations at ${\bf k}$ points that differ by a reciprocal lattice vector (\emph{e.g.} $\Gamma^{1,2}$ in Fig.~\ref{fig:monodromy})~\cite{Bandrep3}.  For example, adding $\mathcal{T}$ symmetry to an $s_{2_{1}}$-symmetric rod [see the text surrounding Eq.~(\ref{eq:defScrewSym})] both doubles the band connectivity and introduces pinned degeneracies at the high-symmetry points ${\bf k}_{\Gamma^{1}}={\bf 0}$ and ${\bf k}_{X^{1}} = \pi\hat{\bf z}$, obviating the need to consider the compatibility relations at ${\bf k}_{\Gamma^{2}}$.  In the case of a rod with $\mathcal{T}$ and $s_{2_{1}}$ screw symmetry, the pinned degeneracies at high-symmetry points specifically occur at odd electronic fillings [\emph{e.g.} $\nu=1,3$], and groups of bands connect in ``hourglass''-like patterns~\cite{Steve2D,WiederLayers,HourglassInsulator,Cohomological,DiracInsulator} with odd numbers of moveable-but-unremovable twofold degeneracies in each half of the BZ at fillings $\nu=2+4n$,\ $n\in\{\mathbb{Z}^{+},0\}$ [\emph{e.g.} $\nu=2$].  Consequently, there are only 4 Type-II single and double SSGs in which monodromy constraints must be considered in addition to those imposed by the symmetries of the SSG.  In Table~\ref{tb:monodromy}, we list the single and double SSGs in which the monodromy of representations provides necessary constraints on small corep (band) connectivity.  The Type-I and Type-II SSGs listed in Table~\ref{tb:monodromy} were previously calculated for TQC~\cite{QuantumChemistry,Bandrep1,Bandrep2,Bandrep3,JenFragile1,BarryFragile}, whereas the Type-III and Type-IV MSGs listed in Table~\ref{tb:monodromy} are a new result that we have calculated for the present work.  Surprisingly, in Table~\ref{tb:monodromy}, we find that there are only 4 Type-IV single and double MSGs in which representation monodromy must be taken into account to determine corep connectivity, despite the fact that \emph{all} Type-IV MSGs are nonsymmorphic [see the text following Eq.~(\ref{eq:cosetOverallDecomposition})].  This occurs because each Type-IV MSG [Eq.~(\ref{eq:type4})] necessarily contains a symmetry of the form:
\begin{equation}
\theta=\{\mathcal{T}|{\bf t}_{0}\}, 
\end{equation}
which acts the same as $\mathcal{T}$ symmetry ($\{\mathcal{T}|000\}$) on points in ${\bf k}$ space [Eq.~(\ref{eq:kAction})]:
\begin{equation}
\theta{\bf k} = \mathcal{T}{\bf k} = -{\bf k}.
\end{equation}
Conversely, in Type-I MSGs, which only contain unitary symmetries (Appendix~\ref{sec:type1}), and in Type-III MSGs, which only contain unitary symmetries and antiunitary symmetries of the form $\{h\times\mathcal{T}|{\bf v}\}$ in which $h$ is a unitary symmetry $h\neq E$ (Appendix~\ref{sec:type3}), we find that representation monodromy frequently provides necessary constraints on corep connectivity.  As shown in Table~\ref{tb:monodromy}, we specifically find that there are 38 Type-I single MSGs, 32 Type-I double MSGs, 92 Type-III single MSGs, and 65 Type-III double MSGs in which the connectivity of small coreps can only be fully determined by considering the effects of representation monodromy on the compatibility relations.

\section{Elementary Band Corepresentations of the MSGs (MEBRs)}
\label{sec:MEBRs}

In the sections below, we will adapt the procedure previously employed in Refs.~\onlinecite{QuantumChemistry,Bandrep3} to obtain the magnetic elementary band corepresentations (MEBRs) of the Type-III and Type-IV single and double MSGs.  Along with the Type-I MEBRs of the Type-I MSGs and the physical EBRs (PEBRs) of the Type-II SSGs previously tabulated in Refs.~\onlinecite{QuantumChemistry,Bandrep3}, the MEBRs of the Type-III and IV MSGs form the foundation of MTQC.  More generally, in this work, we will consider PEBRs and Type-III and Type-IV MEBRs to both be \emph{elementary band corepresentations} (EBRs), because they derive from Type-I MEBRs of Type-I (unitary) MSGs that are related by the action of antiunitary symmetries (Appendix~\ref{sec:coreps}).  We note that, previously in TQC~\cite{QuantumChemistry,Bandrep1,Bandrep2,Bandrep3,JenFragile1,BarryFragile}, the Type-I MEBRs of the Type-I MSGs were termed EBRs, to draw contrast with the PEBRs of the Type-II SSGs.  However, in this work, we will revise the previous terminology to accomodate the elementary band corepresentations of the Type-III and IV MSGs -- in this work, all elementary band (co)representations are in general termed \emph{EBRs}, the elementary band corepresentations of Type-II SSGs remain termed \emph{PEBRs}, and the elementary band (co)representations of Type-I, III, and IV MSGs are respectively termed \emph{Type-I, III, and IV MEBRs}.

Below, we will show that the EBRs provide a basis for all Wannierizable~\cite{MarzariDavidWannier,MarzariWannierReview,QuantumChemistry,BarryFragile}, mean-field crystalline insulators, with or without magnetism.  First in Appendix~\ref{sec:magWannier}, we will introduce the concept of (magnetic) atomic orbitals, which we will then relate to maximally (exponentially) localized, symmetric Wannier functions~\cite{MarzariDavidWannier,MarzariWannierReview}.  Importantly, in Appendix~\ref{sec:magWannier}, we will establish a rigorous correspondence between (magnetic) atomic orbitals and the (co)reps of Shubnikov point groups (SPGs)~\cite{ShubnikovMagneticPoint,BilbaoPoint,PointGroupTables,MagneticBook,EvarestovBook,EvarestovMEBR,BCS1,BCS2,BCSMag1,BCSMag2,BCSMag3,BCSMag4} (as well as site-symmetry groups, see Appendix~\ref{sec:siteSymmetry}).  Next, in Appendix~\ref{sec:induction}, we will adapt the central machinery of band induction and small corep subduction from TQC to MTQC.  Specifically, in Appendix~\ref{sec:induction}, we will use the magnetic atomic orbitals introduced in Appendix~\ref{sec:magWannier} to induce band corepresentations, which we will then Fourier transform and subduce onto little groups to obtain dependencies between small coreps in momentum space (Appendix~\ref{sec:coreps}) and site-symmetry group coreps in position space (Appendix~\ref{sec:magWannier}).  In Appendix~\ref{sec:mbandrep}, we will then enumerate the MEBRs by inducing band coreps from maximal Wyckoff positions and then excluding the \emph{exceptional} cases (Refs.~\onlinecite{Bandrep2,Bandrep3,ZakException1,ZakException2,ArisMagneticBlochOscillation,ZakCompatibility} and Appendix~\ref{sec:exceptions}) of band coreps induced from maximal Wyckoff positions that are non-elementary (\emph{i.e.} composite).  Finally, in Appendix~\ref{sec:mebrStats}, we will provide detailed statistics for the EBRs of all 1,651 SSGs, as well as introduce and detail the~\href{http://www.cryst.ehu.es/cryst/mbandrep}{MBANDREP} tool on the BCS, which we have implemented for this work to access both the EBRs and the composite band coreps induced from each Wyckoff position in each SSG.  We note that prior to this work, Evarestov Smirnov, and Egorov in Ref.~\onlinecite{EvarestovMEBR} introduced a method for obtaining the MEBRs of the MSGs and computed representative examples, but did not perform a large-scale tabulation of MEBRs or establish a connection to magnetic band topology.  As will be detailed in this section, in this work, we have employed a method equivalent to the procedure in Ref.~\onlinecite{EvarestovMEBR} to perform the first \emph{complete} tabulation of the single- and double-valued MEBRs of the 1,421 MSGs.  Furthermore, as detailed in the main text, in this work, we have used the MEBRs to construct the first complete position-space theory of mean-field magnetic band topology -- MTQC.

\subsection{Magnetic Atomic Orbitals and the \textsc{CorepresentationsPG} Tool}
\label{sec:magWannier}

One of the fundamental advances of TQC was to introduce a \emph{predictive} theory of bulk topology that derived from the \emph{position-space} chemistry of a material or model~\cite{QuantumChemistry}, instead of momentum-space quantities such as (nested) Wilson loops and Berry phases~\cite{ZakPhase,VDBpolarization,AlexeyVDBTI,AndreiXiZ2,Fidkowski2011,ArisInversion,Cohomological,HourglassInsulator,DiracInsulator,BarryFragile,multipole,WladTheory,HingeSM,WiederAxion,TMDHOTI,WiederDefect,KooIPartialNestedBerry,ArisFragile}.    Specifically, in TQC, trivial bands in momentum-space are induced from the position-space (co)reps of the site-symmetry groups of the Wyckoff positions in a pristine crystal that is invariant under a particular SSG.  As previously discussed in Appendix~\ref{sec:siteSymmetry}, site-symmetry groups in SSGs, magnetic or otherwise, are necessarily isomorphic to Shubnikov point groups (SPGs).

In the Type-II (nonmagnetic) SGs first analyzed with TQC, the authors of Ref.~\onlinecite{QuantumChemistry} exploited a correspondence between the coreps of the site-symmetry groups in solid-state materials and the eigenstates of the Schr\"{o}dinger Hamiltonian for a hydrogen atom (hydrogenic ion).  Specifically, because the Schr\"{o}dinger Hamiltonian for an ion with a single electron is spherically symmetric (isotropic) and nonmagnetic, then the Hamiltonian is invariant under the action of \emph{any} point group, crystallographic or otherwise~\cite{PointGroupTables}.  In the language of group theory, the Schr\"{o}dinger Hamiltonian for a hydrogenic ion is invariant under the action of the symmetries of the nonmagnetic (Type-II) group $\text{Pin}^{-}(3)\cup \mathcal{T}\times\text{Pin}^{-}(3)$ [see Refs.~\onlinecite{WPVZ,PinGroup} for a detailed discussion of the relationship between $\text{Pin}^{-}(3)$, $\text{SO}(3)$, and $\text{SU}(2)$ in condensed matter physics].  For the purposes of this work, it is sufficient to note that $\text{Pin}^{-}(3)\cup \mathcal{T}\times\text{Pin}^{-}(3)$ is composed of spinful rotations [\emph{e.g.} $C_{2z}$, for which $(C_{2z})^{2}=-1$], rotoinversions of the form of the product of spinful rotations and spinless inversion $\mathcal{I}$ [\emph{e.g.} $m_{z} = C_{2z}\times\mathcal{I}$, for which $(m_{z})^{2}=-1$, $(\mathcal{I})^{2}=+1$], and antiunitary elements of the form of $\mathcal{T}$ multiplied by rotation or rotoinversion [\emph{e.g.} $C_{2z}\times\mathcal{T}$, for which $(\mathcal{T})^{2}=-1$, such that $(C_{2z}\times\mathcal{T})^{2}=+1$].  Consequently, the infinite group $\text{Pin}^{-}(3)\cup \mathcal{T}\times\text{Pin}^{-}(3)$ is a supergroup of \emph{any} finite single or double 3D point group~\cite{PointGroupTables,mcQuarriePchem} [see the text following Eq.~(\ref{eq:gqforSiteSym})].  Returning to the hydrogenic ion, the eigenstates of the Schr\"{o}dinger Hamiltonian are given by ${\boldsymbol \psi}^{\sigma}(r,\theta,\phi)=R(r)Y(\theta,\phi){\bf S}^{\sigma}_{1/2}$, where ${\bf S}^{\sigma}_{1/2}$ is a two-level, fermionic spinor for which $\sigma=\uparrow,\downarrow$.  In ${\boldsymbol \psi}^{\sigma}$, the angular part $Y(\theta,\phi)$ can be expressed in either the basis of spherical or cubic harmonics~\cite{mcQuarriePchem,harmonics1,harmonics2}; therefore, in this section, we will denote $Y(\theta,\phi)$ with suppressed angular ($l$, $m_{l}$) or orbital (\emph{e.g.} $s$, $d_{xy}$) indices whenever $Y(\theta,\phi)$ appears in a basis-independent expression or statement.  Across the set of wavefunctions $\{{\boldsymbol \psi}^{\sigma}(r,\theta,\phi)\}$, the infinite set of angular and spin parts $\{Y(\theta,\phi)\}\otimes\{{\bf S}^{\sigma}_{1/2}\}$ spans both the infinite set of basis functions of $\text{Pin}^{-}(3)\cup\mathcal{T}\times\text{Pin}^{-}(3)$, as well as the infinite set of basis functions of $\text{Pin}^{-}(3)$, the maximal unitary (Type-I) magnetic subgroup of $\text{Pin}^{-}(3)\cup\mathcal{T}\times\text{Pin}^{-}(3)$.  We further note that the hydrogenic ion wavefunctions can also be expressed in a basis of coupled spinorbitals  ${\boldsymbol \psi}^{\sigma}(r,\theta,\phi)=R(r){\bf J}^{\sigma}(\theta,\phi)$.  However, the set of all spinful basis functions (spin-orbit-coupled angular parts) $\{{\bf J}^{\sigma}(\theta,\phi)\}$ can be generated using only spinless angular parts and two-level (spin-1/2) spinors, 
\begin{equation}
\bigg\{{\bf J}^{\sigma}(\theta,\phi)\bigg\}=\bigg\{Y(\theta,\phi)\bigg\}\otimes\bigg\{{\bf S}^{\sigma}_{1/2}\bigg\},
\end{equation}
in which appropriately chosen Clebsch-Gordan coefficients (\emph{c.f.} the tables in Ref.~\onlinecite{PointGroupTables}) are required to relate ${\bf J}_{l}^{j,m_{j}}(\theta,\phi)$ and $Y_{l}^{m_{l}}(\theta,\phi){\bf S}^{\sigma}_{1/2}$ for specific values of $j,\ m_{j},\ m_{l},$ and $\sigma$.  Therefore, for the purposes of this work, we are free to simplify notation by restricting consideration to hydrogenic ion wavefunctions of the form ${\boldsymbol \psi}^{\sigma}(r,\theta,\phi)=R(r)Y(\theta,\phi){\bf S}^{\sigma}_{1/2}$.  Hence, we may subduce the infinitely many irreducible coreps of $\text{Pin}^{-}(3)\cup\mathcal{T}\times\text{Pin}^{-}(3)$ onto any finite SPG $G_{\bf q}$ [which can either be a magnetic point group (MPG) or a nonmagnetic SPG, see the text following Eq.~(\ref{eq:gqforSiteSym})], yielding the established result~\cite{mcQuarriePchem,harmonics1,harmonics2,PointGroupTables} that the finite set of irreducible (co)reps of $G_{\bf q}$ are spanned by the [infinitely overcomplete] set of irreducible coreps of $\text{Pin}^{-}(3)\cup\mathcal{T}\times\text{Pin}^{-}(3)$ subduced onto $G_{\bf q}$.  Specifically, there always exists at least one [and in fact, are infinitely many] corep[s] of $\text{Pin}^{-}(3)\cup\mathcal{T}\times\text{Pin}^{-}(3)$ that subduce[s] to each irreducible (co)rep of $G_{\bf q}$.  We therefore conclude that the set $\{Y(\theta,\phi)\}\otimes\{{\bf S}^{\sigma}_{1/2}\}$ necessarily spans the basis functions of the single- and double-valued (co)reps of any $G_{\bf q}$, because the (co)reps of a particular $G_{\bf q}$ are formed from the irreps of its maximal unitary subgroup $H_{\bf q}$, which is a subgroup of $\text{Pin}^{-}(3)$.

This establishes a correspondence between appropriately chosen linear combinations of the basis functions in $\{Y(\theta,\phi)\}\otimes\{{\bf S}^{\sigma}_{1/2}\}$ and the (co)reps of $G_{\bf q}$.  For the nonmagnetic (Type-II) SPGs (site-symmetry groups) studied in TQC~\cite{QuantumChemistry}, the correspondence is intuitive.  Specifically, given a Type-II SPG $G_{\bf q}$ and a hydrogenic ion wavefunction ${\boldsymbol \psi}^{\sigma}(r,\theta,\phi)=R(r)Y(\theta,\phi){\bf S}^{\sigma}_{1/2}$ whose angular part $Y(\theta,\phi)$ is expressed in the basis of atomic orbitals in which it is real-valued (\emph{i.e.} the basis of cubic harmonics~\cite{mcQuarriePchem,harmonics1,harmonics2}), one can first determine if ${\boldsymbol \psi}^{\sigma}(r,\theta,\phi)$ is an eigenstate of the unitary symmetries (\emph{i.e.} proper rotations and rotoinversions) $h\in H_{\bf q}$, where $H_{\bf q}$ is the maximal unitary subgroup of $G_{\bf q}$, and where $h$ includes $\text{SU}(2)$ spin rotations if $H_{\bf q}$ is a single group.  First, if ${\boldsymbol \psi}^{\sigma}(r,\theta,\phi)$ is an eigenstate of each $h\in H_{\bf q}$, then ${\boldsymbol \psi}^{\sigma}(r,\theta,\phi)$ can be classified by the phase $\lambda_{h}$ that it acquires under the action of each $h\in H_{\bf q}$ [\emph{i.e.}, by the eigenvalue $\lambda_{h}$ of $h$: $h{\boldsymbol \psi}^{\sigma}(r,\theta,\phi) = \lambda_{h}{\boldsymbol \psi}^{\sigma}(r,\theta,\phi)$].  Conversely, if ${\boldsymbol \psi}^{\sigma}(r,\theta,\phi)$ is \emph{not} an eigenstate of any of the unitary operations $h\in H_{\bf q}$, then one can instead form an orthonormal set of symmetrized wavefunctions $\tilde{{\boldsymbol \psi}}^{\sigma}(r,\theta,\phi)$ from linear combinations of the wavefunctions in the set $\{h{\boldsymbol \psi}^{\sigma}(r,\theta,\phi) h^{-1}\}$ taken over all $h\in H_{\bf q}$.  Using the values of $\lambda_{h}$ for each symmetry $h\in H_{\bf q}$ acting on ${\boldsymbol \psi}^{\sigma}(r,\theta,\phi)$ [or on the orthonormal set of symmetrized $\tilde{{\boldsymbol \psi}}^{\sigma}(r,\theta,\phi)$ formed from $\{h{\boldsymbol \psi}^{\sigma}(r,\theta,\phi) h^{-1}\}$], each atomic [ionic] orbital [or symmetric set of atomic orbitals] can then be uniquely labeled by a (co)rep of $G_{\bf q}$~\cite{HydrogenIrrep1,HydrogenIrrep2}.  Specifically, for each atomic orbital or symmetric set of orbitals, there is only one (co)rep $\tilde{\rho}$ of $G_{\bf q}$ whose characters [see the text following Eq.~(\ref{eq:CorepCoset})] satisfy $\chi_{\tilde{\rho}}(h) = \sum_{i}\lambda_{h,i}$ for each $h\in H_{\bf q}$ and wavefunction $\tilde{{\boldsymbol \psi}}^{\sigma}_{i}(r,\theta,\phi)$ in the symmetrized, orthonormal basis of $\{h{\boldsymbol \psi}^{\sigma}(r,\theta,\phi)h^{-1}\}$.  Following the terminology employed in TQC~\cite{QuantumChemistry}, we refer to the correspondence between [a set of] atomic orbital[s] and a (co)rep $\tilde{\rho}$ by stating that the atomic orbital [or set of orbitals] ``transforms in'' the (co)rep $\tilde{\rho}$.

If $\mathcal{T}$ symmetry is relaxed, however, then $G_{\bf q}$ necessarily becomes isomorphic to a Type-I or Type-III magnetic point group [MPG, see the text following Eq.~(\ref{eq:gqforSiteSym})].  In the case in which $G_{\bf q}$ is isomorphic to an MPG, the correspondence between (co)reps and atomic orbitals is more opaque.  Specifically, the basis functions of the (co)reps of the MPGs are still spanned by the set $\{Y(\theta,\phi)\}\otimes\{{\bf S}^{\sigma}_{1/2}\}$, which occurs because each MPG is a subgroup of a Type-II SPG [see the text text following Eq.~(\ref{eq:gqforSiteSym})], which is itself a subgroup of $\text{Pin}^{-}(3)\cup \mathcal{T}\times\text{Pin}^{-}(3)$.  However, as we will show in this section, for some MPG (co)reps, the corresponding ${\boldsymbol \psi}^{\sigma}(r,\theta,\phi)$ is only an eigenstate of the unitary symmetries $h$ in the MPG if the angular part $Y(\theta,\phi)$ is expressed in the complex basis of spherical harmonics~\cite{mcQuarriePchem,harmonics1,harmonics2}.  Therefore, for this work, we introduce the term \emph{magnetic atomic orbital} to reference the basis functions that transform in the lowest-dimensional [\emph{i.e.} in one-dimensional] MPG (co)reps~\cite{PointGroupTables}.  As we will show in the examples below (Appendices~\ref{sec:singleFirstSiteSym},~\ref{sec:singleSecondSiteSym}, and~\ref{sec:singleThirdSiteSym}), the angular parts $Y(\theta,\phi)$ of some magnetic atomic orbitals can be expressed in the real basis of the familiar cubic harmonics (\emph{i.e.} atomic orbitals, such as $s$ and $d_{xy}$), whereas the angular parts of other magnetic orbitals necessarily take the form of $\mathcal{T}$-breaking linear combinations of cubic harmonics (\emph{i.e.} spherical harmonics, such as $p_{x}\pm i p_{y}$ magnetic atomic orbitals).

Because the 3D magnetic atomic orbitals are relatively esoteric, especially when considering the combined effects of SOC and magnetism, then we will leave the complete tabulation of the magnetic atomic orbitals that transform in each (co)rep of each SPG for future works.  However, we will still in this work detail representative examples of MPG (co)reps and their corresponding magnetic atomic orbitals.  In Appendices~\ref{sec:singleFirstSiteSym},~\ref{sec:singleSecondSiteSym}, and~\ref{sec:singleThirdSiteSym}, we will respectively determine the lowest-angular-momentum, spin-degenerate pair of magnetic atomic orbitals that transforms in each single-valued (co)rep of Type-I MPG 9.1.29 $4$, Type-III MPG 9.3.31 $4'$, and Type-II SPG 9.2.30 $41'$ [as was previously done in Appendix~\ref{sec:siteSymmetry}, we will continue to label SPGs employing the notation of the~\href{http://www.cryst.ehu.es/cryst/mpoint.html}{MPOINT} tool on the BCS~\cite{BCSMag1,BCSMag2,BCSMag3,BCSMag4} in which an SPG is labeled by its number, followed by its symbol].

Lastly, we note that double-valued MPG (co)reps in general correspond to less intuitive tensor products of fermionic spinors and real-space wavefunctions~\cite{PointGroupTables,BigBook} [\emph{i.e.}, linear combinations of the basis functions in $\{Y(\theta,\phi)\}\otimes\{{\bf S}^{\sigma}_{1/2}\}$].  For example, a $(d_{xy}+ id_{x^{2}-y^{2}})\otimes {\bf S}_{z}^{\up}$ magnetic atomic spinorbital is less familiar than a nonmagnetic spinless $d_{xy}$ orbital.  Conversely, single-valued MPG (co)reps correspond to spin-degenerate linear combinations of the basis functions in $\{Y(\theta,\phi)\}\otimes\mathds{1}_{\sigma}$, where $\mathds{1}_{\sigma}$ is the $2\times 2$ identity in the space of ${\bf S}^{\sigma}_{1/2}$.  Hence, for simplicity, in the examples in Appendices~\ref{sec:singleFirstSiteSym},~\ref{sec:singleSecondSiteSym}, and~\ref{sec:singleThirdSiteSym}, we will restrict focus to the single-valued (co)reps of single SPGs and their corresponding [spin-degenerate pairs of] magnetic atomic orbitals.

\begin{figure}[h]
\includegraphics[width=0.85\columnwidth]{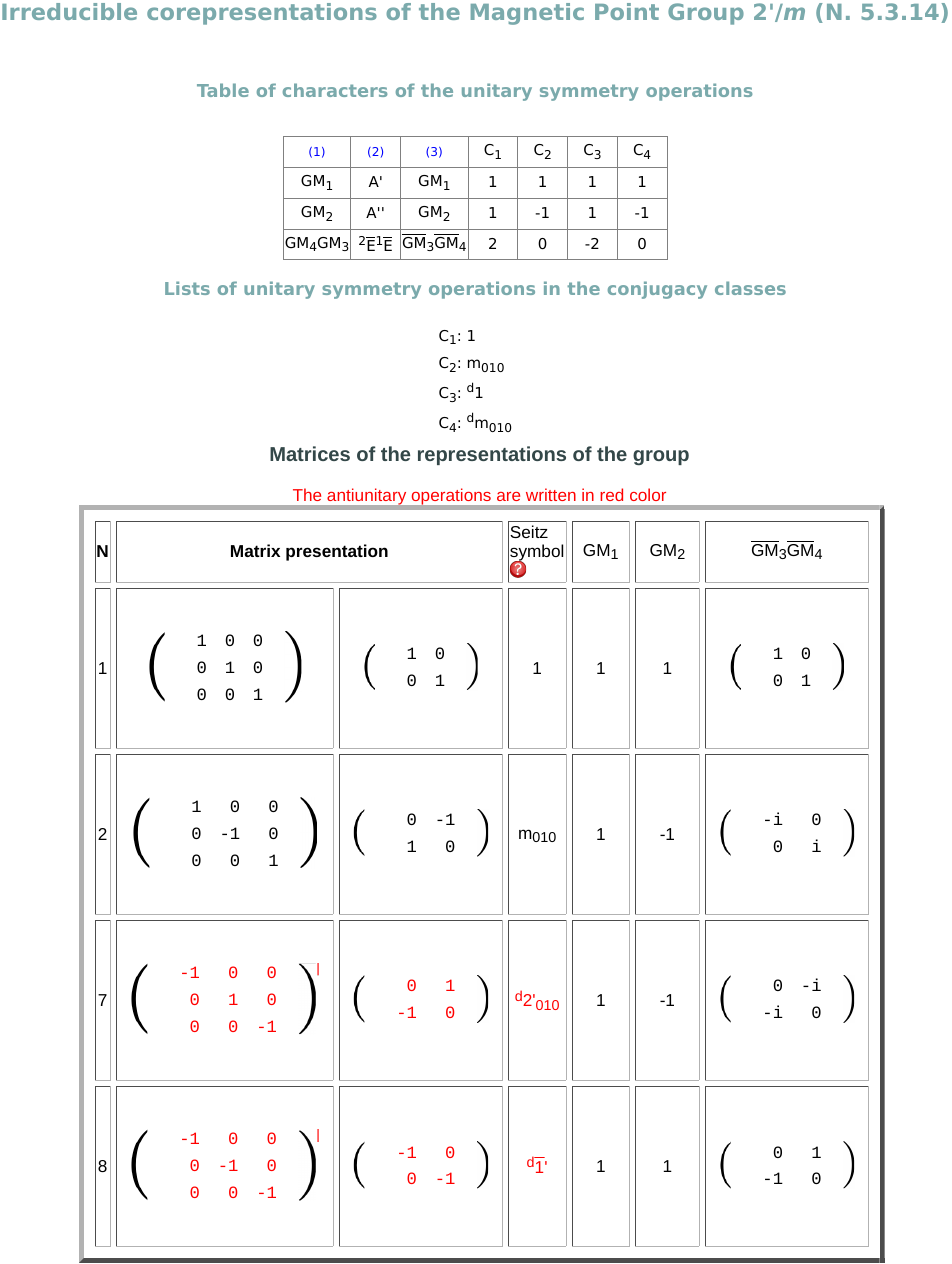}
\caption{The output of the~\href{http://www.cryst.ehu.es/cryst/corepresentationsPG}{CorepresentationsPG} tool on the BCS for Type-III MPG 5.3.14 $2'/m$.  For each of the 122 crystallographic SPGs (see Appendix~\ref{sec:siteSymmetry} and Refs.~\onlinecite{ShubnikovMagneticPoint,BilbaoPoint,PointGroupTables,MagneticBook,EvarestovBook,EvarestovMEBR,BCS1,BCS2,BCSMag1,BCSMag2,BCSMag3,BCSMag4}),~\href{http://www.cryst.ehu.es/cryst/corepresentationsPG}{CorepresentationsPG} outputs the irreducible (co)reps of the SPG, the unitary symmetry operations in the SPG, and the matrix representatives of both the unitary and antiunitary symmetry elements in the SPG.  For each antiunitary symmetry $g_{A,i}$ in the SPG, entries in the table of matrix representatives are labeled in red text, and the matrices listed for each (co)rep $\tilde{\rho}$ indicate the unitary part $U$ of the antiunitary matrix representative $\Delta_{\tilde{\rho}}(g_{A,i})=UK$, where $K$ is complex conjugation.  We note that the bottom table only contains a representative subset of the output of~\href{http://www.cryst.ehu.es/cryst/corepresentationsPG}{CorepresentationsPG} for Type-III MPG 5.3.14 $2'/m$, in order to preserve the legibility of the text in this figure.}
\label{fig:CorepsPGOutput}
\end{figure}

Throughout this section, we will obtain the (co)reps of SPGs through character tables reproduced from the~\href{http://www.cryst.ehu.es/cryst/corepresentationsPG}{CorepresentationsPG} tool on the BCS, which we have implemented for this work.  For each of the 122 crystallographic SPGs,~\href{http://www.cryst.ehu.es/cryst/corepresentationsPG}{CorepresentationsPG} outputs the single- and double-valued (co)reps, character tables, and symmetry matrix representatives.~~\href{http://www.cryst.ehu.es/cryst/corepresentationsPG}{CorepresentationsPG} subsumes the earlier~\href{https://www.cryst.ehu.es/cgi-bin/cryst/programs/representations_point.pl?tipogrupo=dbg}{REPRESENTATIONS DPG} tool~(\url{https://www.cryst.ehu.es/cgi-bin/cryst/programs/representations_point.pl?tipogrupo=dbg}), which was implemented for TQC~\cite{QuantumChemistry,Bandrep1} to output the irreps and character tables of the 32 single and double Type-I MPGs.  In Fig.~\ref{fig:CorepsPGOutput}, we show the output of~\href{http://www.cryst.ehu.es/cryst/corepresentationsPG}{CorepresentationsPG} for Type-III MPG 5.3.14 $2'/m$.

\clearpage

\subsubsection{Irreps and Magnetic Atomic Orbitals in Type-I Single MPG 9.1.29 $4$}
\label{sec:singleFirstSiteSym}

\begin{table}[h]
\begin{tabular}{|c|c|c|c|c|}
\hline
\multicolumn{5}{|c|}{Character Table for} \\ 
\multicolumn{5}{|c|}{Type-I Single MPG 9.1.29 $4$} \\
\hline
Irrep & $E$ & $C_{2z}$ & $C_{4z}$ & $C_{4z}^{-1}$ \\
\hline
\hline
$A$ & $1$ & $1$ & $1$ & $1$ \\
\hline
$B$ & $1$ & $1$ & $-1$ & $-1$ \\
\hline
$^{2}E$ & $1$ & $-1$ & $i$ & $-i$ \\
\hline
$^{1}E$ & $1$ & $-1$ & $-i$ & $i$ \\
\hline 
\end{tabular}
\caption{Single-valued irreps and characters for Type-I single MPG 9.1.29 $4$, reproduced from~\href{http://www.cryst.ehu.es/cryst/corepresentationsPG}{CorepresentationsPG} on the BCS.  For each irrep $\rho$ and unitary symmetry element $h$, elements in the table correspond to the character $\chi_{\rho}(h)=\Tr[\Delta_{\rho}(h)]$, where $\Delta_{\rho}(h)$ is the matrix representative of $h$ in the irrep $\rho$ [see the text following Eq.~(\ref{eq:cosetOverallDecomposition})].  Following the nomenclature established in Appendix~\ref{sec:coreps}, we use the symbol $E$ for the identity element.  Because $\chi_{\rho}[(C_{2z})^{2}] = \chi_{\rho}[(C_{4z})^{4}]=\chi_{\rho}(E)$ for all of the single-valued irreps $\rho$ of Type-I single MPG 9.1.29 $4$, then the irreps in this table can only correspond to 0D spinless (spin-degenerate) electronic (bosonic) states.}
\label{tb:MPG4}
\end{table}

We begin by examining the magnetic atomic orbitals that transform in irreps of Type-I single MPG 9.1.29 $4$.  In Table~\ref{tb:MPG4}, we reproduce the characters for single MPG 9.1.29 $4$, obtained from~\href{http://www.cryst.ehu.es/cryst/corepresentationsPG}{CorepresentationsPG} on the BCS.  In Table~\ref{tb:MPG4}, and for all of the SPGs discussed in this work, we have labeled (co)reps in the notation of Ref.~\onlinecite{BigBook}, which is based on the notation employed by Mulliken in Ref.~\onlinecite{MullikenSPG}.  For each irrep $\rho$ and unitary symmetry element $h$ in Table~\ref{tb:MPG4}, we list the character $\chi_{\rho}(h)=\Tr[\Delta_{\rho}(h)]$, where $\Delta_{\rho}(h)$ is the matrix representative of $h$ in the irrep $\rho$ [see the text following Eq.~(\ref{eq:cosetOverallDecomposition})].

\begin{figure}[h]
\includegraphics[width=0.81\columnwidth]{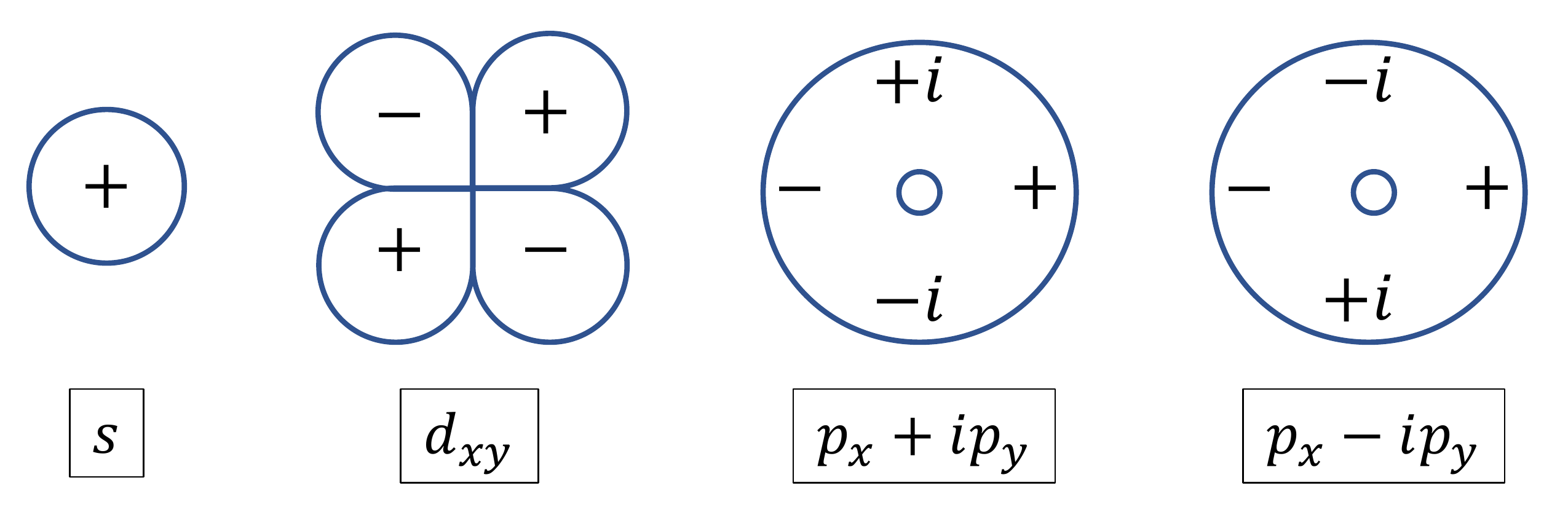}
\caption{The lowest-angular-momentum spinless (\emph{i.e.} spin-degenerate pairs of) magnetic atomic orbitals that transform in~\cite{HydrogenIrrep1,HydrogenIrrep2} single-valued irreps of Type-I single MPG 9.1.29 $4$ (Table~\ref{tb:MPG4}).  From left to right, the orbitals specifically transform in the $A$, $B$, $^{2}E$, and $^{1}E$ single-valued irreps of MPG 9.1.29 $4$.  While the spinless $s$ ($A$) and $d_{xy}$ ($B$) orbitals are the same as their familiar nonmagnetic counterparts, the spinless $p_{x}\pm ip_{y}$ ($^{2,1}E$) orbitals correspond to $\mathcal{T}$-breaking linear combinations of nonmagnetic, spinless $p_{x,y}$ orbitals.  Most precisely, the angular parts of the wavefunctions of the spinless $s$ and $d_{xy}$ orbitals are respectively given by the $s$ and $d_{xy}$ cubic harmonics, whereas the angular parts of the spinless $p_{x}\pm ip_{y}$ orbitals are given by the $l=1$, $m_{l}=\pm 1$ spherical harmonics~\cite{mcQuarriePchem,harmonics1,harmonics2}.}
\label{fig:magPoints}
\end{figure}

For each irrep of single-valued Type-I MPG 9.1.29 $4$ in Table~\ref{tb:MPG4}, we obtain the corresponding lowest-angular-momentum spinless magnetic atomic orbital through the following procedure.  First, because we are characterizing electronic states labeled by single-valued irreps, we restrict consideration to spin-degenerate pairs of orbitals $\{{\boldsymbol \psi}^{\uparrow}(r,\theta,\phi),{\boldsymbol \psi}^{\downarrow}(r,\theta,\phi)\}$, which we label by the spinless angular part of each orbital in the pair $Y(\theta,\phi)$.  Next, we search for the circular harmonics [$Y(\theta,\phi)=Y_{l}^{m_{l}}(\theta,\phi)$] or cubic harmonics [$Y(\theta,\phi)\propto Y_{l}^{m_{l}}\pm Y_{l}^{-m_{l}}$]~\cite{mcQuarriePchem,harmonics1,harmonics2} that are eigenstates of all of the unitary symmetries $h\in H_{\bf q}$ while carrying the lowest possible values of $l$ and $|m_{l}|$.  This procedure returns four (spin-degenerate pairs of) orbitals -- one for each single-valued irrep in Table~\ref{tb:MPG4} -- which we depict in Fig.~\ref{fig:magPoints}.  While the (spin-degenerate pairs) of $s$ ($A$) and $d_{xy}$ ($B$) orbitals shown in Fig.~\ref{fig:magPoints} are the same as their familiar nonmagnetic counterparts, the $p_{x}\pm ip_{y}$ ($^{2,1}E$) orbitals in Fig.~\ref{fig:magPoints} correspond to $\mathcal{T}$-breaking linear combinations of nonmagnetic $p_{x,y}$ orbitals.  Specifically, the angular parts of the wavefunctions of the $s$ and $d_{xy}$ orbitals are respectively given by the $s$ and $d_{xy}$ cubic harmonics, whereas the angular parts of the $p_{x}\pm ip_{y}$ orbitals are given by the $l=1$, $m_{l}=\pm 1$ spherical harmonics~\cite{mcQuarriePchem,harmonics1,harmonics2}.

\subsubsection{Coreps and Magnetic Atomic Orbitals in Type-III Single MPG 9.3.31 $4'$}
\label{sec:singleSecondSiteSym}

Next, in this section, we will determine the lowest-angular-momentum magnetic atomic orbitals that transform in single-valued coreps of Type-III single MPG 9.3.31 $4'$.  As discussed in the text surrounding Eq.~(\ref{eq:type3MPGsite}), a Type-III group $G_{\bf q}$ can be re-expressed as a coset decomposition with respect to its maximal index-2 unitary subgroup $H_{\bf q}$. In the case of $G_{\bf q}=4'$, the maximal unitary subgroup $H_{\bf q}$ is isomorphic to Type-I MPG 3.1.6 $2$, such that the coset decomposition is given by:
\begin{equation}
G_{\bf q} = 4' = 2 \cup \mathcal{T}(41'\setminus 2) = (E)2 \cup (C_{4z}\times\mathcal{T})2,
\label{eq:type3example4primebreakdown}
\end{equation}
where $2$ and $41'$ respectively refer to Type-I MPG 3.1.6 $2$ and Type-II SPG 9.2.30 $41'$.  Eq.~(\ref{eq:type3example4primebreakdown}) implies that, unlike the previous example of Type-I MPG 3.1.6 $2$ in Appendix~\ref{sec:singleFirstSiteSym}, Type-III MPG 9.3.31 $4'$ contains antiunitary symmetries, which comprise the coset $(C_{4z}\times\mathcal{T})2$.

\begin{table}[h]
\begin{tabular}{|c|c|c|c|c|}
\hline
\multicolumn{3}{|c|}{Character Table for} \\ 
\multicolumn{3}{|c|}{Type-I Single} \\
\multicolumn{3}{|c|}{MPG 3.1.6 $2$} \\
\hline
Irrep & $E$ & $C_{2z}$ \\
\hline
\hline
$A$ & $1$ & $1$ \\
\hline
$B$ & $1$ & $-1$ \\
\hline
\end{tabular}
\caption{Single-valued irreps and characters for Type-I single MPG 3.1.6 $2$, reproduced from~\href{http://www.cryst.ehu.es/cryst/corepresentationsPG}{CorepresentationsPG} on the BCS.  For each irrep $\rho$ and unitary symmetry in the MPG $h\in H_{\bf q}$ [Eq.~(\ref{eq:unitarySubof4Prime})], the table lists the character $\chi_{\rho}(h) = \Tr[\Delta_{\rho}(h)]$, where $\Delta_{\rho}(h)$ is the matrix representative of $h$ in $\rho$  [see the text following Eq.~(\ref{eq:cosetOverallDecomposition})].  Following the nomenclature established in Appendix~\ref{sec:coreps}, we use the symbol $E$ for the identity element.  Additionally, as previously emphasized in Table~\ref{tb:MPG4}, we again note that, because $\chi_{\rho}[(C_{2z})^{2}] = \chi_{\rho}(E)$ for all of the single-valued $\rho$ in this table, then the irreps $\rho$ can only correspond to spinless (spin-degenerate) electronic (bosonic) states.}
\label{tb:MPG4primeUnitarySubgroup}
\end{table}

To determine the single-valued coreps of Type-III single MPG 9.3.31 $4'$, we begin by examining the single-valued irreps of the maximal unitary subgroup:
\begin{equation}
H_{\bf q}=2 = \bigg\{E,\ C_{2z}\bigg\},
\label{eq:unitarySubof4Prime}
\end{equation}
where $2$ refers to Type-I single MPG 3.1.6 $2$.  In Table~\ref{tb:MPG4primeUnitarySubgroup}, we reproduce the characters for Type-I single MPG 3.1.6 $2$ from~\href{http://www.cryst.ehu.es/cryst/corepresentationsPG}{CorepresentationsPG} on the BCS.  To obtain the single-valued coreps of $G_{\bf q}$ (Type-III MPG 9.3.31 $4'$), we use the characters in Table~\ref{tb:MPG4primeUnitarySubgroup} to calculate the indicator $J_{\rho}$, adapted from the modified Frobenius-Schur indicator~\cite{frobenius1906,FultonHarris,DimmockWheelerFrobenius,BigBook} $J_{\sigma}$ for little group small coreps discussed in the text surrounding Eqs.~(\ref{eq:Jtest}) and~(\ref{eq:Jtypes}):
\begin{equation}
J_{\rho} = \sum_{i}\chi_{\rho}(g_{A,i}^{2}),
\label{eq:JtestforOrbitals}
\end{equation}
where the sum in Eq.~(\ref{eq:JtestforOrbitals}) runs over the two antiunitary elements $g_{A,i}$ in the coset $(C_{4z}\times\mathcal{T})2$ in Eq.~(\ref{eq:type3example4primebreakdown}).  For the specific case of Type-III MPG 9.3.31 $4'$, Eqs.~(\ref{eq:type3example4primebreakdown}),~(\ref{eq:unitarySubof4Prime}), and~(\ref{eq:JtestforOrbitals}) imply that:
\begin{eqnarray}
J_{\rho} &=& \chi_{\rho}(C_{4z}^{2}\times\mathcal{T}^{2}) + \chi_{\rho}(C_{4z}^{6}\times\mathcal{T}^{2}) \nonumber \\
&=& 2\chi_{\rho}(C_{2z}), 
\label{eq:JtestType3Point}
\end{eqnarray} 
where we have exploited that~\cite{BigBook} $C_{4z}C_{2z}=C_{4z}^{3}$, and that $\mathcal{T}^{2}=(C_{2z})^{2} = (C_{4z})^{4} = E$ for single groups.  Inserting $\rho=A,B$ and the characters $\chi_{A,B}(h)$ from Table~\ref{tb:MPG4primeUnitarySubgroup} into Eq.~(\ref{eq:JtestType3Point}), we determine that: 
\begin{equation}
J_{A} = |H_{\bf q}|,\ J_{B} = -|H_{\bf q}|, 
\label{eq:oneTypeAoneTypeB}
\end{equation}
where $|H_{\bf q}|=2$ is the number of elements [see the text following Eq.~(\ref{eq:quotientGroup})] in Type-I single MPG 3.1.6 $2$ [Eq.~(\ref{eq:unitarySubof4Prime})].  Following the discussion surrounding Eqs.~(\ref{eq:typeA}),~(\ref{eq:typeB}), and~(\ref{eq:Jtypes}), Eqs.~(\ref{eq:JtestType3Point}) and~(\ref{eq:oneTypeAoneTypeB}) imply that, in Type-III single MPG 9.3.31 $4'$, $\rho=A$ forms an undoubled, one-dimensional corep of type (a), whereas $\rho=B$ forms a doubled, two-dimensional corep of type (b).  The single-valued coreps of Type-III single MPG 9.3.31 $4'$ are therefore given by:
\begin{equation}
\tilde{\rho} = A,\ BB. 
\label{eq:c4TdoublesSingleIrrepsSometimes}
\end{equation}

\begin{table}[b]
\begin{tabular}{|c|c|c|c|c|}
\hline
\multicolumn{3}{|c|}{Character Table for} \\ 
\multicolumn{3}{|c|}{Type-III Single} \\
\multicolumn{3}{|c|}{MPG 9.3.31 $4'$} \\
\hline
Corep & $E$ & $C_{2z}$ \\
\hline
\hline
$A$ & $1$ & $1$ \\
\hline
$BB$ & $2$ & $-2$ \\
\hline
\end{tabular}
\caption{Single-valued coreps and characters for Type-III single MPG 9.3.31 $4'$, reproduced from~\href{http://www.cryst.ehu.es/cryst/corepresentationsPG}{CorepresentationsPG} on the BCS.  For each corep $\tilde{\rho}$ and unitary symmetry element $h\in H_{\bf q}$, where $H_{\bf q}$ is the maximal unitary subgroup of MPG 9.3.31 $4'$ [Table~\ref{tb:MPG4primeUnitarySubgroup} and Eqs.~(\ref{eq:type3example4primebreakdown}) and~(\ref{eq:unitarySubof4Prime})], the table lists the character $\chi_{\tilde{\rho}}(h)=\Tr[\Delta_{\tilde{\rho}}(h)]$, where $\Delta_{\tilde{\rho}}(h)$ is the matrix representative of $h$ in the corep $\tilde{\rho}$ [see the text following Eq.~(\ref{eq:cosetOverallDecomposition})].  Following the nomenclature established in Appendix~\ref{sec:coreps}, we use the symbol $E$ for the identity element.  Because the matrix representatives of the antiunitary symmetries in Type-III MPG 9.3.31 $4'$ [\emph{i.e.}, the antiunitary elements of the coset $(C_{4z}\times\mathcal{T})2$ in Eq.~(\ref{eq:type3example4primebreakdown})] are also antiunitary, then they do not have well-defined traces, and do not appear in the character table.  In Eqs.~(\ref{eq:JtestType3Point}),~(\ref{eq:oneTypeAoneTypeB}),~(\ref{eq:c4TdoublesSingleIrrepsSometimes}), we show that Type-III single MPG 9.3.31 $4'$ has two single-valued coreps: there is one, one-dimensional, single-valued corep $A$, which is equivalent [defined in the text surrounding Eq.~(\ref{eq:typeA})] to an irrep ($A$) of $H_{\bf q}$ (Table~\ref{tb:MPG4primeUnitarySubgroup}), and there is one, two-dimensional, single-valued corep $BB\equiv B\oplus B$, which is equivalent [defined in the text surrounding Eq.~(\ref{eq:typeB})] to two copies of the same irrep ($B$) of $H_{\bf q}$.}
\label{tb:MPG4prime}
\end{table}

In Table~\ref{tb:MPG4prime}, we reproduce the characters for Type-III single MPG 9.3.31 $4'$, obtained from~\href{http://www.cryst.ehu.es/cryst/corepresentationsPG}{CorepresentationsPG} on the BCS.  To obtain the lowest-angular-momentum (spin-degenerate pairs of) magnetic atomic orbitals that transform in each corep in Table~\ref{tb:MPG4prime}, we follow the procedure previously described at the beginning of this section (Appendix~\ref{sec:magWannier}) and in the previous section (Appendix~\ref{sec:singleFirstSiteSym}).  For the corep $A$ in Table~\ref{tb:MPG4prime}, we find that the corresponding lowest-angular-momentum atomic orbital is a spinless (\emph{i.e.} spin-degenerate pair of spinful) $s$ orbital(s) (Fig.~\ref{fig:magPoints}).  Conversely, there is no individual spinless magnetic atomic orbital that transforms in the corep $BB$ in Table~\ref{tb:MPG4prime}, because $BB$ is two-dimensional [\emph{i.e.}, because $\chi_{BB}(E)=2$].  Instead, we find that the smallest set of magnetic atomic orbitals with the lowest angular momenta that transform in $BB$ are a pair of spinless $p$ orbitals whose lobes are oriented at $C_{4z}\times\mathcal{T}$-related angles in the $xy$-plane.  An example of a pair of orbitals that transform in $BB$ is one spinless $p_{x}$ plus one spinless $p_{y}$ orbital, which span the same two-dimensional space (four-dimensional, including spin) as one spinless $p_{x}+ip_{y}$ orbital plus one spinless $p_{x}-ip_{y}$ orbital (Fig.~\ref{fig:magPoints}).  Intuitively, this can be understood by recognizing that the lowest-angular-momentum magnetic atomic orbital that transforms in the irrep $A$ ($B$) of Type-I single MPG 3.1.6 $2$ is a spinless $s$ ($p_{x}\pm i p_{y}$) orbital.  Under the action of $C_{4z}\times\mathcal{T}$ in Eq.~(\ref{eq:type3example4primebreakdown}), an $s$ orbital is transformed to itself, whereas a $p_{x}\pm ip_{y}$ orbital is transformed into a $(p_{y}\mp ip_{x})^{*}=i(p_{x} \mp ip_{y})\propto p_{x} \mp ip_{y}$ orbital.  Hence, $A\uparrow G_{\bf q}=A$ [\emph{i.e.}, the irrep $A$ of $H_{\bf q}$ induces a type (a) corep $A$ in $G_{\bf q}$, see the text surrounding Eq.~(\ref{eq:typeA})], whereas $B\uparrow G_{\bf q} = BB$ [\emph{i.e.}, the irrep $B$ of $H_{\bf q}$ induces a type (b) corep $BB$ in $G_{\bf q}$, see the text surrounding Eq.~(\ref{eq:typeB})].

\subsubsection{Coreps and Atomic Orbitals in Type-II Single SPG 9.2.30 $41'$}
\label{sec:singleThirdSiteSym}

As a final example, in this section, we will determine the lowest-angular-momentum, \emph{nonmagnetic} atomic orbitals that transform in single-valued coreps of Type-II single SPG 9.2.30 $41'$, the $\mathcal{T}$-symmetric supergroup of the MPGs previously analyzed in Appendices~\ref{sec:singleFirstSiteSym} and~\ref{sec:singleSecondSiteSym} (Type-I MPG 9.1.29 $4$ and Type-III MPG 9.3.31 $4'$, respectively).  Like a Type-II SSG [Eq.~(\ref{eq:type2})], a Type-II MPG $G_{\bf q}$ can be re-expressed as a coset decomposition with respect to its maximal index-2 unitary subgroup $H_{\bf q}$.  In the case of $G_{\bf q}=41'$, the decomposition is:
\begin{equation}
41' = 4 \cup (\mathcal{T})4,
\label{eq:type2example41primebreakdown}
\end{equation}
where $H_{\bf q}=4$ refers to Type-I single MPG 9.1.29 $4$, which we previously analyzed in Appendix~\ref{sec:singleFirstSiteSym}.  $H_{\bf q}=4$ contains four elements (Table~\ref{tb:MPG4}):
\begin{equation}
H_{\bf q} = \bigg\{E,\ C_{2z},\ C_{4z},\ C_{4z}^{-1}\bigg\}.
\label{eq:fullGroupofHq4}
\end{equation}
Using the character table for Type-I MPG 9.1.29 $4$ (Table~\ref{tb:MPG4}), we previously determined in Appendix~\ref{sec:singleFirstSiteSym} that the four single-valued irreps of $H_{\bf q}=4$ given by $\rho=A,\ B,\ ^{2}E,\ ^{1}E$ respectively correspond to spinless (\emph{i.e.} spin-degenerate pairs of) $s$, $d_{xy}$, $p_{x}+ip_{y}$, and $p_{x}-ip_{y}$ magnetic atomic orbitals.

\begin{table}[h]
\begin{tabular}{|c|c|c|c|c|}
\hline
\multicolumn{5}{|c|}{Character Table for} \\
\multicolumn{5}{|c|}{Type-II Single SPG 9.2.30 $41'$} \\
\hline
Corep & $E$ & $C_{2z}$ & $C_{4z}$ & $C_{4z}^{-1}$ \\
\hline
\hline
$A$ & $1$ & $1$ & $1$ & $1$ \\
\hline
$B$ & $1$ & $1$ & $-1$ & $-1$ \\
\hline
$^{1}E\ ^{2}E$ & $2$ & $-2$ & $0$ & $0$ \\
\hline 
\end{tabular}
\caption{Single-valued coreps and characters for Type-II single SPG 9.2.30 $41'$, reproduced from~\href{http://www.cryst.ehu.es/cryst/corepresentationsPG}{CorepresentationsPG} on the BCS.  For each corep $\tilde{\rho}$ and unitary symmetry element $h\in H_{\bf q}$, where $H_{\bf q}$ is the maximal unitary subgroup of SPG 9.2.30 ($41'$) [Table~\ref{tb:MPG4} and Eqs.~(\ref{eq:type2example41primebreakdown}) and~(\ref{eq:fullGroupofHq4})], the table lists the character $\chi_{\tilde{\rho}}(h)=\Tr[\Delta_{\tilde{\rho}}(h)]$, where $\Delta_{\tilde{\rho}}(h)$ is the matrix representative of $h$ in the corep $\tilde{\rho}$ [see the text following Eq.~(\ref{eq:cosetOverallDecomposition})].  Following the nomenclature established in Appendix~\ref{sec:coreps}, we use the symbol $E$ for the identity element.  Because the matrix representatives of the antiunitary symmetries in Type-II SPG 9.2.30 $41'$ [\emph{i.e.}, the antiunitary elements of the coset $(\mathcal{T})4$ in Eq.~(\ref{eq:type2example41primebreakdown})] are also antiunitary, then they do not have well-defined traces, and do not appear in the character table.  In Eqs.~(\ref{eq:JrhoType2Simplify}),~(\ref{eq:twoTypeAoneTypeC}), and~(\ref{eq:Tdoubledsingle41primeCoreps}), we show that Type-II single MPG 9.2.30 $41'$ has three single-valued coreps:  there are two, one-dimensional, single-valued coreps $A$ and $B$, which are equivalent [defined in the text surrounding Eq.~(\ref{eq:typeA})] to irreps ($A$ and $B$, respectively) of $H_{\bf q}$ (Table~\ref{tb:MPG4}), and there is one, two-dimensional, single-valued corep $^{1}E\ ^{2}E\equiv\ ^{1}E\ \oplus\ ^{2}E$, which is formed [defined in the text surrounding Eq.~(\ref{eq:typeC})] from pairing two different irreps ($^{1}E$ and $^{2}E$) of $H_{\bf q}$.  We note that, because $\chi_{\tilde{\rho}}[(C_{2z})^{2}] = \chi_{\tilde{\rho}}[(C_{4z})^{4}]=\chi_{\tilde{\rho}}(E)$ for all of the single-valued $\tilde{\rho}$ in this table, then the coreps $\tilde{\rho}$ can only correspond to 0D spinless (spin-degenerate) electronic (bosonic) states.}
\label{tb:SPG41prime}
\end{table}

To determine the single-valued coreps $\tilde{\rho}$ of Type-II single SPG 9.2.30 $41'$, we again calculate the indicator $J_{\rho}$ discussed in the text surrounding Eq.~(\ref{eq:JtestforOrbitals}):
\begin{equation}
J_{\rho} = \sum_{i}\chi_{\rho}(g_{A,i}^{2}),
\label{eq:JrhoType2}
\end{equation}
where the sum in Eq.~(\ref{eq:JrhoType2}) runs over the four antiunitary elements $g_{A,i}$ in the coset $(\mathcal{T})4$ in Eq.~(\ref{eq:type2example41primebreakdown}).  In the specific case of Type-II single SPG 9.2.30 $41'$, Eqs.~(\ref{eq:type2example41primebreakdown}),~(\ref{eq:fullGroupofHq4}), and~(\ref{eq:JrhoType2}) imply that:
\begin{eqnarray}
J_{\rho} &=& \chi_{\rho}(\mathcal{T}^{2}) + \chi_{\rho}(C_{2z}^{2}\times\mathcal{T}^{2}) + \chi_{\rho}(C_{4z}^{2}\times\mathcal{T}^{2}) + \chi_{\rho}(C_{4z}^{-2}\times\mathcal{T}^{2}) \nonumber \\
&=& 2\left[\chi_{\rho}(E) + \chi_{\rho}(C_{2z})\right],
\label{eq:JrhoType2Simplify}
\end{eqnarray}
where we have exploited that~\cite{BigBook} $C_{4z}^{2} = C_{2z}$, and that $C_{4z}^{-2}=C_{2z}^{-1}=C_{2z}$ and $\mathcal{T}^{2}=(C_{2z})^{2}=(C_{4z})^{4} = E$ for single groups.  Inserting $\rho=A,B, ^{2}E, ^{1}E$ and the characters from Table~\ref{tb:MPG4} into Eq.~(\ref{eq:JrhoType2Simplify}), we determine that:
\begin{equation}
J_{A} = J_{B} =  |H_{\bf q}|,\ J_{^{2}E}=J_{^{1}E}=0,
\label{eq:twoTypeAoneTypeC}
\end{equation}
where $|H_{\bf q}|=4$ is the number of elements [see the text following Eq.~(\ref{eq:quotientGroup})] in Type-I single MPG 9.1.29 $4$ [Eq.~(\ref{eq:fullGroupofHq4})].  Following the discussion surrounding Eqs.~(\ref{eq:typeA}),~(\ref{eq:typeC}), and~(\ref{eq:Jtypes}), Eqs.~(\ref{eq:JrhoType2Simplify}) and~(\ref{eq:twoTypeAoneTypeC}) imply that, in Type-II single MPG 9.2.30 $41'$, $\rho=A$ and $\rho=B$ each form undoubled, one-dimensional coreps of type (a), whereas $\rho=\ ^{2}E$ and $\rho=\ ^{1}E$ together form a paired, two-dimensional corep of type (c).  The single-valued coreps of Type-II single MPG 9.2.30 $41'$ are therefore given by:
\begin{equation}
\tilde{\rho} = A,\ B,\ ^{1}E\ ^{2}E. 
\label{eq:Tdoubledsingle41primeCoreps}
\end{equation}

In Table~\ref{tb:SPG41prime}, we reproduce the characters for Type-II SPG 9.2.30 $41'$, obtained from~\href{http://www.cryst.ehu.es/cryst/corepresentationsPG}{CorepresentationsPG} on the BCS.  Like in $H_{\bf q}=4$, the maximal unitary subgroup of SPG 9.2.30 $41'$ [see Appendix~\ref{sec:singleFirstSiteSym} and Eqs.~(\ref{eq:type2example41primebreakdown}) and~(\ref{eq:fullGroupofHq4})], the lowest-angular-momentum (spin-degenerate pairs of) atomic orbitals that transform in the single-valued coreps $A$ and $B$ of Type-II SPG 9.2.30 $41'$ are respectively spinless $s$ and spinless $d_{xy}$ orbitals (Fig.~\ref{fig:magPoints}).  Conversely, there is no individual spinless atomic orbital that transforms in the corep $^{1}E\ ^{2}E$ in Table~\ref{tb:SPG41prime}, because $^{1}E\ ^{2}E$ is two-dimensional [\emph{i.e.}, because $\chi_{^{1}E\ ^{2}E}(E)=2$].  Instead, we find that the smallest set of atomic orbitals with the lowest angular momenta that together transform in $^{1}E\ ^{2}E$ are a Kramers pair of spinless $p_{x}\pm ip_{y}$ magnetic atomic orbitals [\emph{i.e.} one spinless $p_{x}+ip_{y}$ plus one spinless $p_{x}-ip_{y}$ orbital (Fig.~\ref{fig:magPoints})], which are usually denoted more succinctly in other works~\cite{QuantumChemistry} as ``spinless $p_{x}$ and $p_{y}$ orbitals'', because they span the same two-dimensional space (four-dimensional, including spin) as one spinless $p_{x}$ orbital plus one spinless $p_{y}$ orbital.  Intuitively, this can be understood by recognizing that the lowest-angular-momentum magnetic atomic orbitals that transform in the irreps $A,\ B,\ ^{2}E,\ ^{1}E$ of Type-I MPG 9.1.29 $4$ are respectively spinless $s$, $d_{xy}$, $p_{x}+ip_{y}$, and $p_{x}-ip_{y}$ magnetic atomic orbitals (Fig.~\ref{fig:magPoints}).  Under the action of $\mathcal{T}$ symmetry in Eq.~(\ref{eq:type2example41primebreakdown}), an $s$ or $d_{xy}$ orbital is transformed to itself, whereas a $p_{x}\pm ip_{y}$ orbital is transformed into a $p_{x} \mp ip_{y}$ orbital.  Hence, $A\uparrow G_{\bf q}=A$ and $B\uparrow G_{\bf q}=B$ [\emph{i.e.}, the irreps $A$ and $B$ of $H_{\bf q}$ respectively induce the type (a) coreps $A$ and $B$ in $G_{\bf q}$, see the text surrounding Eq.~(\ref{eq:typeA})], whereas $^{1,2}E\uparrow G_{\bf q} =\ ^{1}E\ ^{2}E$ [\emph{i.e.}, the irreps $^{1,2}E$ of $H_{\bf q}$ each induce a type (c) corep $^{1}E\ ^{2}E$ in $G_{\bf q}$, see the text surrounding Eq.~(\ref{eq:typeC})].

\vspace{0.05in}

\subsection{Inducing Band Corepresentations from Magnetic Atomic Orbitals and the~\textsc{MSITESYM} Tool}
\label{sec:induction}

Building upon the earlier definitions of site-symmetry groups [Appendix~\ref{sec:siteSymmetry}], Wyckoff positions [Appendix~\ref{sec:Wyckoff}], little groups [Appendix~\ref{sec:MKVEC}], small (co)reps of the SSGs [Appendix~\ref{sec:coreps}], and magnetic atomic orbitals that transform in (co)reps of the site-symmetry groups [Appendix~\ref{sec:magWannier}], we will now in this section define the band (co)representations of the SSGs, which are induced from exponentially localized [Wannier~\cite{MarzariDavidWannier,MarzariWannierReview}] orbitals in position space.  We will also introduce and detail the~\href{http://www.cryst.ehu.es/cryst/msitesym}{MSITESYM} tool, through which users may access the small (co)reps subduced from each band (co)representation of each SSG.  This section is largely a review of previous works that discuss induced band (co)representations -- most notably Ref.~\onlinecite{Bandrep3} -- though throughout this section, we will employ a more general terminology than in Ref.~\onlinecite{Bandrep3} that encompasses both magnetic and nonmagnetic band (co)representations.  In particular, in this section, we will introduce the term \emph{band corepresentation} to refer to a band representation in an SSG with antiunitary symmetries [\emph{i.e.}, a Type-II, III, or IV SSG (Appendices~\ref{sec:type2},~\ref{sec:type3}, and~\ref{sec:type4}, respectively)].  Specific examples demonstrating usage of the theoretical machinery established in this section are provided in Appendices~\ref{sec:exceptions} and~\ref{sec:SIexP2} for cases of magnetic band (co)representations, and are provided in Refs.~\onlinecite{QuantumChemistry,Bandrep1,Bandrep2,Bandrep3} for cases of nonmagnetic band corepresentations.

To begin, consider an infinite crystal whose unit cells are furnished with initially decoupled (magnetic) atomic orbitals.  The set of atomic orbitals respects the symmetries of the SSG of the crystal $G$, and, by definition, each orbital at ${\bf q}$ occupies a site in a Wyckoff position of $G$ with a site-symmetry group $G_{\bf q}\in G$ (Appendix~\ref{sec:SecFullWyckoff}).  As discussed in Appendix~\ref{sec:Wyckoff}, $G_{\bf q}$ is a subgroup of $G$ ($G_{\bf q}\subset G$) that is isomorphic to a Shubnikov point group (SPG) (Appendix~\ref{sec:siteSymmetry}) containing a set of symmetries $g\in G_{\bf q}$, $g\in G$.  Generically, there also exist a set of symmetries:
\begin{equation}
\tilde{g} \in G\setminus G_{\bf q},
\label{eq:MEBRorbit1}
\end{equation}
for which:
\begin{equation}
\tilde{g}{\bf q} = {\bf q}',
\label{eq:MEBRorbit2}
\end{equation}
where ${\bf q}'$ is a different site than ${\bf q}$ in the same unit cell.  The set of all sites $\{{\bf q}_{\alpha}\}$ in the same unit cell as ${\bf q}$ (including ${\bf q}$ itself) form the Wyckoff orbit of ${\bf q}$, where the index $\alpha$ on ${\bf q}_{\alpha}$ runs from $1$ to $n$, where $n$ is the multiplicity of the Wyckoff orbit indexed by ${\bf q}$ (see Appendix~\ref{sec:Wyckoff}).  We emphasize that the choice of $\tilde{g}$ in Eqs.~(\ref{eq:MEBRorbit1}) and~(\ref{eq:MEBRorbit2}) is not generically unique -- for example, in Type-I MSG 10.42 $P2/m$, which is generated by $\{\mathcal{I}|{\bf 0}\}$, $\{C_{2y}|{\bf 0}\}$, and 3D lattice translations, the sites ${\bf q} = (u,0,w)$ and ${\bf q}'=(-u,0,-w)$ are related by both $\tilde{g}=\{\mathcal{I}|{\bf 0}\}$ and $\tilde{g}=\{C_{2y}|{\bf 0}\}$.  We additionally emphasize that the restriction to ${\bf q}'$ that lie in the same unit cell as ${\bf q}$ is a convention choice that was employed previously in TQC~\cite{QuantumChemistry,Bandrep1,Bandrep2,Bandrep3,JenFragile1,BarryFragile} that we will continue to employ in MTQC to obtain MEBRs consistent with the PEBRs previously calculated for TQC.  More generally, a set of EBRs can be still be computed as long as each ${\bf q}'$ is unique and is not related to ${\bf q}$ or to any other ${\bf q}'$ by an integer-valued linear combination of primitive lattice vectors.

We will find it convenient in this section to initially restrict to the case in which the crystal is furnished by a set of (magnetic) atomic orbitals at each site of a single Wyckoff position indexed by ${\bf q}$ that transforms in one and only one (\emph{i.e.} in an irreducible) (co)rep $\tilde{\rho}_{\bf q}$ of the site-symmetry group $G_{\bf q}$.  Because reducible [composite] site-symmetry [band] (co)representations may be expressed as direct sums of irreducible [elementary] site-symmetry [band] (co)representations, then, at the end of this section, we will straightforwardly relax this restriction and consider the more general case in which the unit cell contains larger sets of atomic orbitals that transform in direct sums of site-symmetry (co)reps.  In the language of Refs.~\onlinecite{MarzariDavidWannier,MarzariWannierReview}, each magnetic atomic orbital at ${\bf q}$ (including spin) corresponds to an exponentially (maximally) localized (spinful), symmetric Wannier orbital.  Specifically, while maximally localized, symmetric Wannier and magnetic atomic orbitals are not required to have the same radial parts [aside from the Wannier orbital exhibiting exponential or sharper localization], we can establish a correspondence between Wannier and atomic orbitals by restricting focus to the angular parts, which, for symmetrized [sets of] orbitals, necessarily transform in (co)reps of the 122 crystallographic SPGs [see Appendix~\ref{sec:magWannier} and Refs.~\onlinecite{ShubnikovMagneticPoint,BilbaoPoint,PointGroupTables,MagneticBook,EvarestovBook,EvarestovMEBR,BCS1,BCS2,BCSMag1,BCSMag2,BCSMag3,BCSMag4}].

Next, because a Wyckoff position generically contains more than one site [\emph{i.e.} the multiplicity of the Wyckoff position $n\geq 1$], then, given a [set of] Wannier orbital[s] that transform in a single irreducible, $D$-dimensional (co)rep of the site-symmetry group $G_{\bf q}$, in order to preserve the symmetry of the SSG $G$, there must additionally be $D$-dimensional [sets of] Wannier orbitals on each of the $n-1$ additional sites in the Wyckoff position, leading to a total of $n\times D$ Wannier orbitals in each unit cell.  For each [set of] orbital[s] at ${\bf q}$ that transforms in the (co)rep $\tilde{\rho}_{\bf q}$ of $G_{\bf q}$, there is therefore also an orbital [or set of orbitals] at each site ${\bf q}' = \tilde{g}{\bf q}\text{ mod }{\bf t}_{a,b,c}$ for each symmetry $\tilde{g}\in G\setminus G_{\bf q}$ that transform[s] in an irreducible (co)rep $\tilde{\rho}_{{\bf q}'}$ of:
\begin{equation}
G_{{\bf q}'} = \tilde{g}G_{\bf q}\tilde{g}^{-1},
\end{equation}
where $G_{{\bf q}'}$ is isomorphic and conjugate to $G_{\bf q}$.  It is important to note that even though $G_{{\bf q}'}$ is isomorphic to $G_{\bf q}$, and even though $G_{\bf q}$ and $G_{{\bf q}'}$ are both isomorphic to the same Shubnikov point group (SPG, see Appendix~\ref{sec:siteSymmetry}) $M$, the symmetries $\tilde{g} \in G \setminus G_{\bf q}$ require that the orbitals [and (co)reps] at ${\bf q}'$ are conjugate to those at ${\bf q}$.  For example, if ${\bf q}$ and ${\bf q}'$ are related by the symmetry $\{C_{4z}|{\bf 0}\}$ in an SSG $G$, then a $p_{x}$ orbital at ${\bf q}$ must be accompanied by a $p_{y}=C_{4z}p_{x}C_{4z}^{-1}$ orbital at ${\bf q}'$ in order to preserve $\{C_{4z}|{\bf 0}\}\in G$.  Employing the terminology previously established in Refs.~\onlinecite{QuantumChemistry,Bandrep1,Bandrep2,Bandrep3,JenFragile1,BarryFragile}, this can be summarized by stating that the orbital[s] that transform in $\tilde{\rho}_{\bf q}$ -- along with the orbital[s] that transform[s] in the conjugate (co)reps $\tilde{\rho}_{{\bf q}'}$ of each of the other $n-1$ sites in the Wyckoff position of ${\bf q}$ -- \emph{occupy} the Wyckoff position indexed by ${\bf q}$.  To formally define the conjugate site-symmetry (co)reps $\tilde{\rho}_{{\bf q}'}$, we first establish that, given a unitary symmetry $h\in H_{\bf q}$ -- the maximal unitary subgroup of $G_{\bf q}$ -- the matrix representative of $h$ in $\tilde{\rho}_{\bf k}$ is denoted as $\Delta_{{\tilde{\rho}_{\bf q}}}(h)$, for which the character of $h$ in $\tilde{\rho}_{\bf k}$ is given by $\Tr[\Delta_{{\tilde{\rho}_{\bf q}}}(h)]$.  In this notation, it is clear that the matrix representative $\Delta_{\tilde{\rho}_{{\bf q}'}}(ghg^{-1})$ of the conjugate symmetry $\tilde{g}h\tilde{g}^{-1}\in G_{{\bf q}'}$ does not generically equal $\Delta_{\tilde{\rho}_{{\bf q}'}}(h)$ (which itself may not be well defined, because $h$ is not required to be an element of both $G_{\bf q}$ and $G_{{\bf q}'}$).  Instead, the matrix representatives of the conjugate symmetries $\tilde{g}h\tilde{g}^{-1}\in G_{{\bf q}'}$ are conjugate to the matrix representatives of $h\in G_{\bf q}$; specifically, if $\tilde{g}$ is unitary, then:
\begin{equation}
\Delta_{\tilde{\rho}_{{\bf q}'}}(\tilde{g}h\tilde{g}^{-1})=\Delta_{{\tilde{\rho}_{\bf q}}}(h),
\label{eq:unitaryConjugateSiteCorep}
\end{equation}
and if $\tilde{g}$ is antiunitary, then:
\begin{equation}
\Delta_{\tilde{\rho}_{{\bf q}'}}(\tilde{g}h\tilde{g}^{-1})=\left[\Delta_{{\tilde{\rho}_{\bf q}}}(h)\right]^{*}.
\label{eq:antiunitaryConjugateSiteCorep}
\end{equation}

The central principle of TQC, which we will here extend to MTQC, is that, when a set of of magnetic atomic orbital[s] that transform in an irreducible site-symmetry (co)rep $\tilde{\rho}_{\bf q}$ occupy the Wyckoff position of ${\bf q}$, the orbitals induce a (co)rep of the SSG $G$: 
\begin{equation}
\tilde{\rho}_{\bf q}\uparrow G = \tilde{\rho}^{G}_{\bf q},
\label{eq:mainInductionMTQC}
\end{equation}
where $\tilde{\rho}^{G}_{\bf q}$ is a \emph{band (co)representation} [band (co)rep].  Crucially, the action of induction ($\uparrow$), unlike subduction ($\downarrow$), \emph{does not} preserve dimensionality (\emph{i.e.} the character of the identity element $E$), such that $\chi_{\tilde{\rho}^{G}_{\bf q}}(E) \neq \chi_{\tilde{\rho}_{\bf q}}(E)$.  Instead:
\begin{equation}
\chi_{\tilde{\rho}^{G}_{\bf q}}(E) = \chi_{\tilde{\rho}_{\bf q}}(E) \times [G : G_{\bf q}] = \chi_{\tilde{\rho}_{\bf q}}(E) \times n \times N,
\label{eq:finiteSitetoInfinite}
\end{equation}
where $n$ is the multiplicity of the Wyckoff position indexed by ${\bf q}$ and $N$ is the number of unit cells in the crystal.  We next take $N$ to be very large (\emph{i.e.} countably infinite), reflecting our goal of applying MTQC to theoretical models of infinite crystals to predict the topology of mesoscopic solid-state systems.  The (now infinite) factor of $N$ on the right-hand side of Eq.~(\ref{eq:finiteSitetoInfinite}) originates from the infinite subgroup index $[G:G_{\bf q}]$ of $G_{\bf q}$ in $G$ [defined in the text surrounding Eq.~(\ref{eq:subsetIndex})], which occurs because the site-symmetry group $G_{\bf q}$ is finite, whereas the SSG $G$ is infinite.  This can be seen by recognizing that $G_{T}\not\subset G_{\bf q}$, $G_{T}\subseteq G$, in which $G_{T}$ is the infinite group of 3D lattice translations [Eq.~(\ref{eq:translationGroup})].

Most importantly, as shown in Ref.~\onlinecite{Bandrep3}, Eq.~(\ref{eq:mainInductionMTQC}) can be decomposed into a sum of \emph{full} (co)reps:
\begin{equation}
\tilde{\rho}_{\bf q}^{G} = \bigoplus_{\tilde{\bf k}} \tilde{\Sigma}_{\tilde{\bf k},{\bf q}}^{G},
\label{eq:fourierFullCorepMTQC}
\end{equation}
where the sum in Eq.~(\ref{eq:fourierFullCorepMTQC}) instead runs over each of the points $\tilde{\bf k}$ in the \emph{irreducible wedge} of the first BZ~\cite{SandCKpaths,yuCardonaWedge}, which is defined as the set of points $\tilde{\bf k}$ in the first BZ containing one and only one arm of each momentum star [see Appendix~\ref{sec:MKVEC}].  In Eq.~(\ref{eq:fourierFullCorepMTQC}), $\tilde{\Sigma}_{\tilde{\bf k},{\bf q}}^{G}$ is a generically \emph{reducible} full [\emph{i.e.} space group] (co)rep of the star of the SSG $G$ indexed by $\tilde{\bf k}$ [Eq.~(\ref{eq:finalFullCorep})].  Hence:
\begin{equation}
\tilde{\Sigma}_{\tilde{\bf k},{\bf q}}^{G} = \bigoplus_{i}b_{i}^{\tilde{\bf k},{\bf q}}\tilde{\Sigma}_{i,\tilde{\bf k}},
\label{eq:fullCorepReduction}
\end{equation}
where $\tilde{\Sigma}_{i,\tilde{\bf k}}$ is the $i^{\text{th}}$ irreducible full (co)rep of the star of $G$ indexed by $\tilde{\bf k}$, and where $b_{i}^{\tilde{\bf k},{\bf q}}$ is the multiplicity of $\tilde{\Sigma}_{i,\tilde{\bf k}}$ in the decomposition of $\tilde{\Sigma}_{\tilde{\bf k},{\bf q}}^{G}$ [\emph{i.e.}, $b^{\tilde{\bf k},{\bf q}}_{i}$ is a non-negative integer that indicates the number of times that the irreducible full (co)rep $\tilde{\Sigma}_{i,\tilde{\bf k}}$ appears in $\tilde{\Sigma}_{\tilde{\bf k},{\bf q}}^{G}$, see the text surrounding Eq.~(\ref{eq:reducibleCompatibility})].  Using Eq.~(\ref{eq:finalFullCorep}), Eq.~(\ref{eq:fourierFullCorepMTQC}) can be further re-expressed in terms of the generically reducible small (co)reps $\tilde{\sigma}_{\tilde{\bf k},{\bf q}}^{G}$ at each ${\bf k}$ point:
\begin{equation}
\tilde{\rho}_{\bf q}^{G} = \bigoplus_{\tilde{\bf k}} \tilde{\Sigma}_{\tilde{\bf k},{\bf q}}^{G} = \bigoplus_{\tilde{\bf k}}\bigoplus_{{\bf k}=\tilde{\bf k}}^{{\bf k}_{m_{\tilde{\bf k}}}}\tilde{\sigma}_{{\bf k},{\bf q}}^{G} = \bigoplus_{\bf k} \tilde{\sigma}_{{\bf k},{\bf q}}^{G}
\label{eq:fourierMainMTQC}
\end{equation}
where $m_{\tilde{\bf k}}$ is the number of arms in the star of $\tilde{\bf k}$ [see the text surrounding Eq.~(\ref{eq:coGroupEq3})], such that ${\bf k}$ runs from $\tilde{\bf k}$ to ${\bf k}_{m_{\tilde{\bf k}}}$ for each star indexed by $\tilde{\bf k}$ in the sum in the second equality, and where the sum on the right-hand side of Eq.~(\ref{eq:fourierMainMTQC}) runs over each of the $N$ (infinitely many) points ${\bf k}$ in the first BZ.

Further intuition for Eqs.~(\ref{eq:finiteSitetoInfinite}),~(\ref{eq:fourierFullCorepMTQC}), and~(\ref{eq:fourierMainMTQC}), can be obtained by comparing the relative dimensionality of $\tilde{\rho}_{\bf q}$, $\tilde{\rho}^{G}_{\bf q}$, $\tilde{\Sigma}_{{\bf k},{\bf q}}^{G}$, and $\tilde{\sigma}^{G}_{{\bf k},{\bf q}}$.  First, while $\chi_{\tilde{\rho}_{\bf q}^{G}}(E)$ is infinite [Eq.~(\ref{eq:finiteSitetoInfinite})], the component $\tilde{\Sigma}_{\tilde{\bf k},{\bf q}}^{G}$ in the Fourier decomposition of the band (co)rep $\tilde{\rho}_{\bf q}^{G}$ in Eq.~(\ref{eq:fullCorepReduction}) is \emph{finite-dimensional}, and there are instead an infinite number of [generically reducible] full (co)reps $\tilde{\Sigma}_{\tilde{\bf k},{\bf q}}^{G}$ -- one at each of the infinitely many $\tilde{\bf k}$ points in the irreducible wedge of the first BZ [defined in the text following Eq.~(\ref{eq:fourierFullCorepMTQC})].  To see this, we compute the dimensionality of $\tilde{\Sigma}_{\tilde{\bf k},{\bf q}}^{G}$, which is defined as the character of the identity operation $E$:
\begin{equation}
\chi_{\tilde{\Sigma}_{\tilde{\bf k},{\bf q}}^{G}}(E) = \chi_{\tilde{\rho}_{\bf q}}(E) \times n \times m_{\tilde{\bf k}},
\label{eq:armsForSubduction}
\end{equation}
in which $n$ is the multiplicity of the Wyckoff position indexed by ${\bf q}$ (Appendix~\ref{sec:Wyckoff}), and $m_{\tilde{\bf k}}$ is the number of arms in the star of $\tilde{\bf k}$ [see the text surrounding Eq.~(\ref{eq:coGroupEq3})].  Conversely, the [generically reducible] small (co)rep $\tilde{\sigma}_{{\bf k},{\bf q}}^{G}$ in Eq.~(\ref{eq:fourierMainMTQC}) generically has a smaller (finite) dimensionality than $\tilde{\Sigma}_{\tilde{\bf k},{\bf q}}^{G}$. To see this, we first subduce $\tilde{\sigma}_{{\bf k},{\bf q}}^{G}$ onto the little group $G_{\bf k}$:
\begin{equation}
\tilde{\sigma}_{{\bf k},{\bf q}}^{G}\downarrow G_{\bf k} = \tilde{\varsigma}_{{\bf k},{\bf q}},
\label{eq:BRsymmetryData}
\end{equation}
where $\tilde{\varsigma}_{{\bf k},{\bf q}}$ is the symmetry data [see the text following Eq.~(\ref{eq:reducibleCompatibility})] induced by the (co)rep $\tilde{\rho}_{\bf q}$ of the site-symmetry group $G_{\bf q}$ into the SSG G [Eq.~(\ref{eq:mainInductionMTQC})] and then subduced onto the little group $G_{\bf k}$ of the point ${\bf k}$ in the first BZ.  We note that, because $\tilde{\sigma}_{{\bf k},{\bf q}}^{G}$ is already a [generically reducible] small (co)rep of $G_{\bf k}$ [Eq.~(\ref{eq:fourierMainMTQC})], then $\tilde{\sigma}_{{\bf k},{\bf q}}^{G}\downarrow G_{\bf k}$ in Eq.~(\ref{eq:BRsymmetryData}) is a redundant expression.  However, in this work, we will continue to employ the expression $\tilde{\sigma}_{{\bf k},{\bf q}}^{G}\downarrow G_{\bf k}$ on the left-hand side of Eq.~(\ref{eq:BRsymmetryData}) for consistency with earlier works on TQC~\cite{QuantumChemistry,Bandrep3}.  Next, we compute the dimensionality of the subduced symmetry data $\tilde{\varsigma}_{{\bf k},{\bf q}}$:
\begin{equation}
\chi_{\tilde{\varsigma}_{{\bf k},{\bf q}}}(E) = \chi_{\tilde{\rho}_{\bf q}}(E) \times n,
\label{eq:finiteSubductionDimension}
\end{equation}
where $n$ continues to be the multiplicity of the Wyckoff position indexed by ${\bf q}$ (Appendix~\ref{sec:Wyckoff}).  Physically, because the set of site-symmetry (co)reps $\{\tilde{\rho}_{{\bf q}_{\alpha}}\}$ corresponds to $\chi_{\tilde{\rho}_{\bf q}}(E) \times n$ magnetic atomic [Wannier] orbitals [Appendix~\ref{sec:magWannier}] occupying the $n$ sites ${\bf q}_{\alpha}$ in the Wyckoff position of ${\bf q}$, and therefore characterizes $\chi_{\tilde{\rho}_{\bf q}}(E) \times n$ bands in momentum space, then the subduced symmetry data $\tilde{\varsigma}_{{\bf k},{\bf q}}$ [Eq.~(\ref{eq:BRsymmetryData})] correspond to a (set of) $\chi_{\tilde{\rho}_{\bf q}}(E) \times n$ \emph{Bloch} states at ${\bf k}$.  This can be summarized by the statement that the $\chi_{\tilde{\rho}_{\bf q}}(E) \times n$ Bloch states at ${\bf k}$ \emph{transform} in $\tilde{\varsigma}_{{\bf k},{\bf q}}$, analogous to the correspondence between orbitals and position-space SPG [site-symmetry group] (co)reps established in Appendix~\ref{sec:magWannier}.

Though $\tilde{\rho}_{\bf q}$ is an irreducible (co)rep of the site-symmetry group $G_{\bf q}$, this \emph{does not} imply that $\tilde{\varsigma}_{{\bf k},{\bf q}} = \tilde{\sigma}_{{\bf k},{\bf q}}^{G}\downarrow G_{\bf k}$ in [Eq.~(\ref{eq:fourierMainMTQC})] is an \emph{irreducible} small (co)rep of $G_{\bf k}$.  In fact, generically, $\tilde{\varsigma}_{{\bf k},{\bf q}}$ is a \emph{reducible} small (co)rep of $G_{\bf k}$, such that:
\begin{equation}
\tilde{\sigma}_{{\bf k},{\bf q}}^{G}\downarrow G_{\bf k} = \tilde{\varsigma}_{{\bf k},{\bf q}} = \bigoplus_{j}a^{{\bf k},{\bf q}}_{j}\tilde{\sigma}_{j,{\bf k}},
\label{eq:corepSubduction}
\end{equation}
where $\tilde{\sigma}_{j,{\bf k}}$ is the $j^{\text{th}}$ irreducible small (co)rep of $G_{\bf k}$ and $a^{{\bf k},{\bf q}}_{j}$ is a non-negative integer corresponding to the multiplicity of $\tilde{\sigma}_{j,{\bf k}}$ in the decomposition of $\tilde{\varsigma}_{{\bf k},{\bf q}}$.  To obtain the multiplicities $a^{{\bf k},{\bf q}}_{j}$ in Eq.~(\ref{eq:corepSubduction}), we can re-express Eq.~(\ref{eq:corepSubduction}) in terms of the characters $\chi_{\tilde{\varsigma}_{{\bf k},{\bf q}}}(h_{i})$ and $\chi_{\tilde{\sigma}_{j,{\bf k}}}(h_{i})$ of each of the unitary symmetries $h_{i}\in |\tilde{H}_{\bf k}|$, the maximal unitary subset of the set of coset representatives $\tilde{G}_{\bf k}$ [text preceding Eq.~(\ref{eq:CorepCoset})] of $G_{\bf k}$ with respect to the group of lattice translations $G_{T}$ [Eq.~(\ref{eq:translationGroup})]:
\begin{equation}
\chi_{\tilde{\varsigma}_{{\bf k},{\bf q}}}(h_{i}) = \sum_{j}a^{{\bf k},{\bf q}}_{j}\chi_{\tilde{\sigma}_{j,{\bf k}}}(h_{i}).
\label{eq:characterCorepSubduction}
\end{equation}

As we will show below, it is important to emphasize that the values of $a^{{\bf k},{\bf q}}_{j}$ in Eqs.~(\ref{eq:corepSubduction}) and~(\ref{eq:characterCorepSubduction}) are determined by the choice of the (co)rep $\tilde{\rho}_{\bf q}$ of the site-symmetry group $G_{\bf q}$ [\emph{i.e.} the (magnetic) atomic orbitals occupying the Wyckoff position indexed by ${\bf q}$] in Eq.~(\ref{eq:mainInductionMTQC}).  This can be seen by first recognizing the values of $\tilde{\sigma}_{j,{\bf k}}$ in Eqs.~(\ref{eq:corepSubduction}) and~(\ref{eq:characterCorepSubduction}) are limited to the finite set of small (co)reps of $G_{\bf k}$, which can be obtained through the~\href{http://www.cryst.ehu.es/cryst/corepresentations}{Corepresentations} tool, as previously described in Appendix~\ref{sec:coreps}.  Next, we recognize that $\tilde{\varsigma}_{{\bf k},{\bf q}}$ is a component of the Fourier decomposition of the induced band (co)rep $\tilde{\rho}_{\bf q}^{G} = \tilde{\rho}_{\bf q}\uparrow G$ [Eqs.~(\ref{eq:mainInductionMTQC}) and~(\ref {eq:fourierMainMTQC})].  Specifically, Eqs.~(\ref{eq:mainInductionMTQC}) and~(\ref {eq:fourierMainMTQC}) imply that, for a given little group $G_{\bf k}$, the characters $\chi_{\tilde{\varsigma}_{{\bf k},{\bf q}}}(h_{i})$ for each unitary symmetry $h_{i}\in \tilde{H}_{\bf k}$ [the maximal unitary subset of $\tilde{G}_{\bf k}$, see the text preceding Eq.~(\ref{eq:CorepCoset})] are given by:
\begin{equation}
\chi_{\tilde{\varsigma}_{{\bf k},{\bf q}}}(h_{i}) =\sum_{\alpha=1}^{n}\chi_{\tilde{\varsigma}_{{\bf k},{\bf q}_{\alpha}}}(h_{i}),
\label{eq:finalCharacters1}
\end{equation}
where $\alpha$ runs over each of the $n$ sites ${\bf q}_{\alpha}$ in the Wyckoff position of ${\bf q}$ (including ${\bf q}$ itself, see Appendix~\ref{sec:Wyckoff}), and where, as will shortly be detailed below:  
\begin{equation}
\chi_{\tilde{\varsigma}_{{\bf k},{\bf q}_{\alpha}}}(h_{i}) =  \begin{cases} e^{-i{\bf k}\cdot\left({\bf q}_{\alpha} - h_{i}{\bf q}_{\alpha}\right)}\chi_{\tilde{\rho}_{{\bf q}_{\alpha}}}\left(\left\{E|{\bf q}_{\alpha} - h_{i}{\bf q}_{\alpha}\right\}h_{i}\right) &,\text{ if } \left\{E|{\bf q}_{\alpha} - h_{i}{\bf q}_{\alpha}\right\}h_{i} \in G_{{\bf q}_{\alpha}} \\
0 &,\text{ if } \left\{E|{\bf q}_{\alpha} - h_{i}{\bf q}_{\alpha}\right\}h_{i} \not\in G_{{\bf q}_{\alpha}} \\
\end{cases}.
\label{eq:finalCharacters2}
\end{equation}
When $\chi_{\tilde{\varsigma}_{{\bf k},{\bf q}_{\alpha}}}(h_{i})\neq 0$ in Eq.~(\ref{eq:finalCharacters2}), the vectors ${\bf q}_{\alpha} - h_{i}{\bf q}_{\alpha}$ are necessarily integer-valued linear combinations of lattice vectors [\emph{i.e.} $\{E|{\bf q}_{\alpha} - h_{i}{\bf q}_{\alpha}\}\in G_{T}$, where $G_{T}$ is defined in Eq.~(\ref{eq:translationGroup})]~\cite{Bandrep3}.  This occurs because the symmetries $h_{i}\in \tilde{H}_{\bf k}$ may shift the location of a site ${\bf q}_{\alpha}$ in the Wyckoff position of ${\bf q}$ to a site $h_{i}{\bf q}_{\alpha}$ in an adjacent unit cell that only differs from ${\bf q}_{\alpha}$ by a linear combination of lattice vectors (if ${\bf q}_{\alpha} - h_{i}{\bf q}_{\alpha}$ were not a lattice vector, then $\{E|{\bf q}_{\alpha} - h_{i}{\bf q}_{\alpha}\}h_{i}$ would instead be one of the symmetries $\{E|{\bf q}_{\alpha} - h_{i}{\bf q}_{\alpha}\}h_{i}\notin G_{{\bf q}_{\alpha}}$ that exchanges sites within the Wyckoff position of ${\bf q}$, and $\chi_{\tilde{\rho}_{{\bf q}_{\alpha}}}(\{E|{\bf q}_{\alpha} - h_{i}{\bf q}_{\alpha}\}h_{i})$ in Eq.~(\ref{eq:finalCharacters2}) would not be well defined).  The (co)reps $\tilde{\rho}_{{\bf q}_{\alpha}}$ of the sites ${\bf q}_{\alpha}$ in Eqs.~(\ref{eq:finalCharacters1}) and~(\ref{eq:finalCharacters2}) are determined from the site-symmetry (co)rep $\tilde{\rho}_{\bf q}$ by conjugation with the symmetries $\tilde{g} \in G$, $\tilde{g}\notin G_{\bf q}$, as described in the text surrounding Eqs.~(\ref{eq:unitaryConjugateSiteCorep}) and~(\ref{eq:antiunitaryConjugateSiteCorep}).

Finally, using Eqs.~(\ref{eq:finalCharacters1}) and~(\ref{eq:finalCharacters2}) for each of the unitary symmetries $h_{i}\in\tilde{H}_{\bf k}$ [Eq.~(\ref{eq:CorepCoset})], we obtain $|\tilde{H}_{\bf k}|$ equations of the form of Eq.~(\ref{eq:characterCorepSubduction}) for the multiplicities $a^{{\bf k},{\bf q}}_{j}$, which can be condensed into a matrix equation in which the summation over $j$ in Eq.~(\ref{eq:characterCorepSubduction}) is implicit:
\begin{equation}
{\boldsymbol \chi}_{\tilde{\varsigma}_{{\bf k},{\bf q}}} = \mathcal{G}_{\bf k}{\bf a}^{{\bf k},{\bf q}},
\label{eq:matrixPreMagicSchur}
\end{equation}
where ${\boldsymbol \chi}_{\tilde{\varsigma}_{{\bf k},{\bf q}}}$ is an $|\tilde{H}_{\bf k}|\times 1$-dimensional column vector whose $i^{\text{th}}$ entry is the value of $\chi_{\tilde{\varsigma}_{{\bf k},{\bf q}}}(h_{i})$ inherited from the site-symmetry group (co)rep $\tilde{\rho}_{\bf q}$ through Eqs.~(\ref{eq:finalCharacters1}) and~(\ref{eq:finalCharacters2}), and where ${\bf a}^{{\bf k},{\bf q}}$ is an $l\times 1$-dimensional column vector whose $j^{\text{th}}$ entry is the multiplicity $a^{{\bf k},{\bf q}}_{j}$ of the small (co)rep $\tilde{\sigma}_{j,{\bf k}}$ of the little group $G_{\bf k}$ in the decomposition of the subduced symmetry data $\tilde{\varsigma}_{{\bf k},{\bf q}}$, where $l$ is the number of small (co)reps of $G_{\bf k}$.  In Eq.~(\ref{eq:matrixPreMagicSchur}), $\mathcal{G}_{\bf k}$ is an $|\tilde{H}_{\bf k}|\times l$-dimensional, generically non-square matrix whose $ij^\text{th}$ element is given by the character of the unitary symmetry $h_{i}\in G_{\bf k}$ in the small (co)rep $\tilde{\sigma}_{j,{\bf k}}$ of $G_{\bf k}$:
\begin{equation}
\left[\mathcal{G}_{\bf k}\right]_{ij} = \chi_{\tilde{\sigma}_{j, {\bf k}}}(h_{i}).
\label{eq:characterTableMagic}
\end{equation}
Consequently, $\mathcal{G}_{\bf k}$ is simply the transpose of the character table for $G_{\bf k}$ (see Figs.~\ref{fig:noAntiCoreps1},~\ref{fig:noAntiCoreps2}, and~\ref{fig:magDiracCoreps} and Table~\ref{tb:diracIrreps}, for example).  Crucially, because the rows (and columns) of character tables are orthogonal~\cite{BigBook,mcQuarriePchem}, then the columns (and rows) of $\mathcal{G}_{\bf k}$ are also orthogonal.  This implies that the left inverse $\mathcal{G}^{-1}_{\bf k}$ of $\mathcal{G}_{\bf k}$ is simply given by:
\begin{equation}
\mathcal{G}^{-1}_{\bf k} = \frac{1}{|\tilde{H}_{\bf k}|}\mathcal{G}^{\dag}_{\bf k},
\label{eq:partialLeftInverseMagic}
\end{equation}
such that:
\begin{equation}
\mathcal{G}_{\bf k}^{\dag}\mathcal{G}_{\bf k} = |\tilde{H}_{\bf k}|\mathds{1},
\label{eq:leftInverseMagic}
\end{equation}
where $\mathds{1}$ in Eq.~(\ref{eq:leftInverseMagic}) is the $l \times l$ identity.  As a final step, we left-multiply Eq.~(\ref{eq:matrixPreMagicSchur}) by $\mathcal{G}^{-1}_{\bf k}$ [Eq.~(\ref{eq:partialLeftInverseMagic})] to solve for ${\bf a}^{{\bf k},{\bf q}}$:
\begin{equation}
{\bf a}^{{\bf k},{\bf q}} = \frac{1}{|\tilde{H}_{\bf k}|}\mathcal{G}^{\dag}_{\bf k}{\boldsymbol \chi}_{\tilde{\varsigma}_{{\bf k},{\bf q}}},
\label{eq:magicFormulaSchur}
\end{equation} 
thus obtaining the multiplicities $a^{{\bf k},{\bf q}}_{j}$ in Eqs.~(\ref{eq:corepSubduction}) and~(\ref{eq:characterCorepSubduction}).  We note that Eq.~(\ref{eq:magicFormulaSchur}) is in fact the matrix form of the Schur orthogonality relation (\emph{i.e.} the so-called ``magic formula'')~\cite{Bandrep1}.

\begin{figure}[b]
\includegraphics[width=\columnwidth]{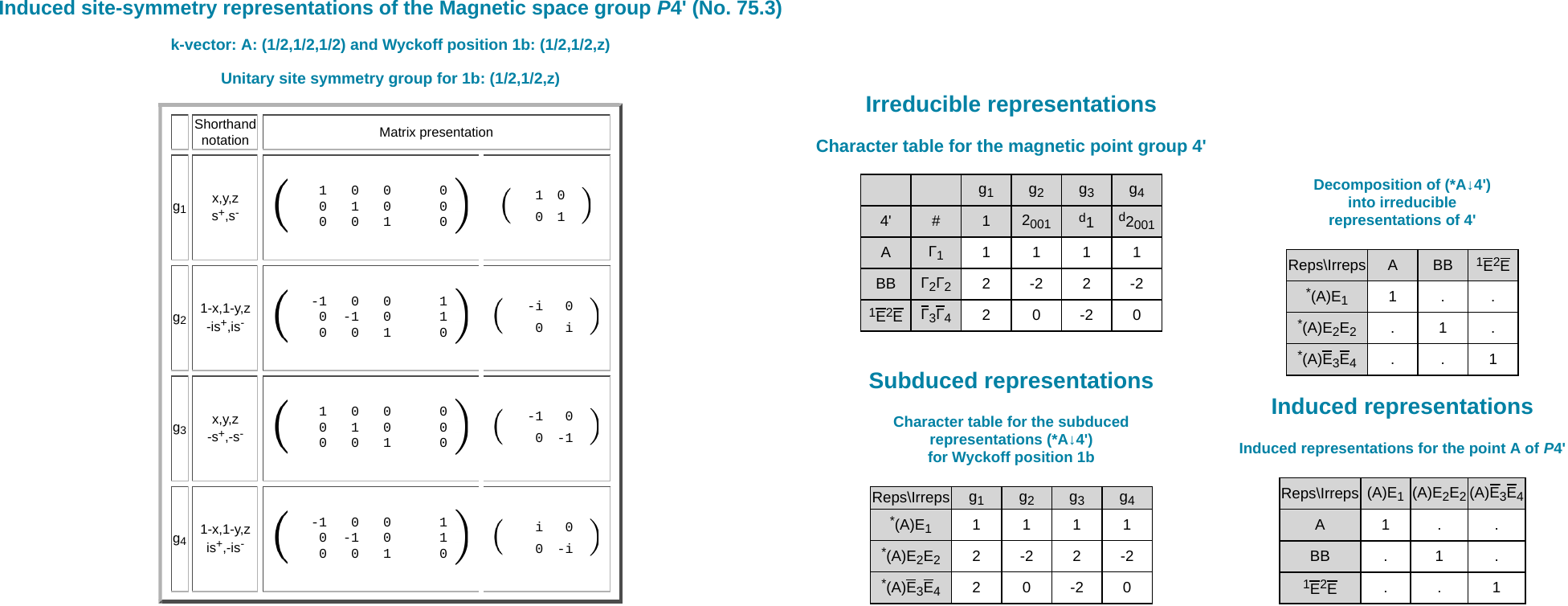}
\caption{The output of the~\href{http://www.cryst.ehu.es/cryst/msitesym}{MSITESYM} tool on the BCS for Type-III MSG 75.3 $P4'$ at the $A$ point in momentum space and the $1b$ Wyckoff position in position space.  For one ${\bf k}$ point in each momentum star (see Appendix~\ref{sec:MKVEC}) and one site ${\bf q}$ in each Wyckoff position in each SSG (see Appendix~\ref{sec:Wyckoff}),~\href{http://www.cryst.ehu.es/cryst/msitesym}{MSITESYM} outputs the irreducible (co)reps of the site-symmetry group $G_{\bf q}$ (see Appendix~\ref{sec:magWannier}), the small (co)reps of the little group $G_{\bf k}$ (see Appendix~\ref{sec:coreps}), and the multiplicities $a^{{\bf k},{\bf q}}_{j}$ in Eqs.~(\ref{eq:corepSubduction}) and~(\ref{eq:characterCorepSubduction}).  ~\href{http://www.cryst.ehu.es/cryst/msitesym}{MSITESYM} subsumes the earlier~\href{https://www.cryst.ehu.es/cgi-bin/cryst/programs/dsitesym.pl}{DSITESYM} tool~\cite{QuantumChemistry,Bandrep1,Bandrep2}, which was previously implemented for TQC to provide direct access to the single- and double-valued small irreps subduced onto a given $G_{\bf k}$ from the band rep induced from each site-symmetry irrep in each of the 230 Type-I MSGs.}
\label{fig:msitesym}
\end{figure}

For this work, we have implemented the~\href{http://www.cryst.ehu.es/cryst/msitesym}{MSITESYM} tool on the BCS to output the multiplicities [$a^{{\bf k},{\bf q}}_{j}$ in Eqs.~(\ref{eq:corepSubduction}) and~(\ref{eq:characterCorepSubduction})] of the small (co)reps $\tilde{\sigma}_{j,{\bf k}}$ subduced in the little group $G_{\bf k}$ of each ${\bf k}$ point [Eq.~(\ref{eq:BRsymmetryData})] from the band (co)rep $\tilde{\rho}_{\bf q}^{G}$ induced into each SSG $G$ [Eq.~(\ref{eq:mainInductionMTQC})] from each irreducible (co)rep $\tilde{\rho}_{\bf q}$ of one site-symmetry group $G_{\bf q}$ in each Wyckoff position of $G$.  ~\href{http://www.cryst.ehu.es/cryst/msitesym}{MSITESYM} subsumes the earlier~\href{https://www.cryst.ehu.es/cgi-bin/cryst/programs/dsitesym.pl}{DSITESYM} tool (\url{https://www.cryst.ehu.es/cgi-bin/cryst/programs/dsitesym.pl})~\cite{QuantumChemistry,Bandrep1,Bandrep2}, which was previously implemented for TQC to provide direct access to the single- and double-valued small irreps subduced onto a given $G_{\bf k}$ from the band rep $\rho_{\bf q}^{G}$ induced from each site-symmetry irrep $\rho_{\bf q}$ in each of the 230 Type-I MSGs.  In Fig.~\ref{fig:msitesym}, we show the output of~\href{http://www.cryst.ehu.es/cryst/msitesym}{MSITESYM} for Type-III MSG 75.3 $P4'$ at the $A$ point in momentum space and the $1b$ Wyckoff position in position space.

In summary, we have demonstrated in this section how decoupled Wannier orbitals that transform in site-symmetry (co)reps in position space induce band (co)reps [Eq.~(\ref{eq:mainInductionMTQC})], which in turn subduce small (co)reps at each point in momentum space that correspond to Bloch states (bands) [Eq.~(\ref{eq:BRsymmetryData})].  It is straightforward to see that, if additional Wannier orbitals are added that either transform in different (co)reps of site-symmetry groups in the same Wyckoff position, or occupy a different Wyckoff position, then additional bands will also be present in the energy spectrum, corresponding to additional small (co)reps in the symmetry data at each ${\bf k}$ point.  Therefore, we have also shown that arbitrary sets of bands induced from Wannier orbitals transform in a linear combination of band (co)reps.  In the subsequent section, Appendix~\ref{sec:mbandrep}, we will determine the minimal, or \emph{elementary}, band (co)reps [EBRs, composed of PEBRs in Type-II SSGs and MEBRs in Type-I, III, and IV MSGs]~\cite{ZakBandrep1,ZakBandrep2,EvarestovBook,EvarestovMEBR,BarryBandrepReview,QuantumChemistry,Bandrep1,Bandrep2,Bandrep3,JenFragile1,BarryFragile} that span all linear combinations of band (co)reps induced from maximally localized, symmetric Wannier orbitals.

\subsection{MEBRs, Exceptional Cases, and the~\textsc{MBANDREP} Tool}
\label{sec:mbandrep}

In this section, we will use the results of Appendix~\ref{sec:induction} to determine which of the induced band (co)reps in each SSG are \emph{elementary} -- which we will rigorously define in this section -- thus establishing the complete theory of MTQC.  We will specifically obtain the MEBRs of the Type-III and Type-IV MSGs, which, along with the MEBRs of the Type-I MSGs and the PEBRs of the Type-II SSGs previously tabulated in Refs.~\onlinecite{QuantumChemistry,Bandrep3}, form the complete set of EBRs of all of the 1,651 single and double SSGs.  We note that previously in TQC~\cite{QuantumChemistry,Bandrep1,Bandrep2,Bandrep3,JenFragile1,BarryFragile}, the Type-I MEBRs of the Type-I MSGs were termed EBRs, to draw contrast with the PEBRs of the Type-II SSGs.  However, in this work, we will revise the previous terminology to accomodate the elementary band coreps of the Type-III and IV MSGs -- in this work, all elementary band (co)reps are in general termed EBRs, the elementary band coreps of Type-II SSGs remain termed PEBRs, and the elementary band (co)reps of Type-I, III, and IV MSGs are respectively termed Type-I, III, and IV MEBRs.  Finally, we note that prior to this work, Evarestov Smirnov, and Egorov in Ref.~\onlinecite{EvarestovMEBR} introduced a method for obtaining the MEBRs of the MSGs and computed representative examples, but did not perform a large-scale tabulation of MEBRs -- the calculations performed in this section represent the first \emph{complete} tabulation of the MEBRs of the 1,421 single and double MSGs.

To begin, we previously established in Appendix~\ref{sec:induction} that, if a set of [magnetic] atomic orbitals transforming in an irreducible (co)rep $\tilde{\rho}_{{\bf q},1}$ of a site-symmetry group $G_{\bf q}$ is placed at ${\bf q}$ in each unit cell of a crystal that is invariant under an SSG $G$, then $\tilde{\rho}_{{\bf q},1}$ induces a band (co)rep $\tilde{\rho}_{{\bf q},1}^{G}=\tilde{\rho}_{{\bf q},1}\uparrow G$ [Eq.~(\ref{eq:mainInductionMTQC})].  From this, we may then consider the case in which additional orbitals are subsequently added at ${\bf q}$ that transform in the (co)rep $\tilde{\rho}_{{\bf q},2}$, such that the total set of Wannier orbitals at ${\bf q}$ transforms in the \emph{reducible} site-symmetry (co)rep $\tilde{\rho}_{{\bf q},T} = \tilde{\rho}_{{\bf q},1}\oplus\tilde{\rho}_{{\bf q},2}$.  Because representation induction is distributive~\cite{Bandrep3}, then it follows that:
\begin{equation}
\tilde{\rho}_{{\bf q},T}\uparrow G = (\tilde{\rho}_{{\bf q},1}\oplus\tilde{\rho}_{{\bf q},2})\uparrow G = \tilde{\rho}_{{\bf q},T}^{G},
\end{equation}
such that:  
\begin{equation}
\tilde{\rho}_{{\bf q},T}^{G} = (\tilde{\rho}_{{\bf q},1}\uparrow G)\oplus(\tilde{\rho}_{{\bf q},2}\uparrow G) = \tilde{\rho}_{{\bf q},1}^{G}\oplus\tilde{\rho}_{{\bf q},2}^{G}.
\label{eq:EBRdefBJW}
\end{equation}
Eq.~(\ref{eq:EBRdefBJW}) implies that $\tilde{\rho}_{{\bf q},T}^{G}$ is a \emph{composite band (co)rep}, because $\tilde{\rho}_{{\bf q},T}^{G}$ is equivalent to a sum of two other band (co)reps [$\tilde{\rho}_{{\bf q},1}^{G}$ and $\tilde{\rho}_{{\bf q},2}^{G}$].  In this work, we define two band (co)reps $\tilde{\rho}_{{\bf q},T}^{G}$ and $\tilde{\rho}_{{\bf q},1}^{G}\oplus\tilde{\rho}_{{\bf q},2}^{G}$ to be equivalent through the existence of a relation of the form of Eq.~(\ref{eq:EBRdefBJW}).  If two band (co)reps $\tilde{\rho}_{{\bf q},1}^{G}$ and $\tilde{\rho}_{{\bf q},2}^{G}$ are equivalent, then this also implies the existence of a unitary matrix-valued function $S({\bf k},t,h)$ that is smooth and non-singular in ${\bf k}$ and continuous in $t$ that interpolates for each unitary symmetry $h\in G$ between the full [space group] (co)rep matrix representatives $\Delta_{\tilde{\Sigma}_{{\bf k},{\bf q},T}^{G}}(h)$ [$t=0$] and $\Delta_{\tilde{\Sigma}_{{\bf k},{\bf q},1}^{G} \oplus \tilde{\Sigma}_{{\bf k},{\bf q},2}^{G}}(h)$ [$t=1$] in the decomposition [see the text surrounding Eqs.~(\ref{eq:finalFullCorep}) and~(\ref{eq:fourierFullCorepMTQC}) and Refs.~\onlinecite{ZakBandrep1,ZakBandrep2,EvarestovBook,EvarestovMEBR,BarryBandrepReview,QuantumChemistry,Bandrep1,Bandrep2,Bandrep3,JenFragile1,BarryFragile} for further details]:
\begin{equation}
\tilde{\rho}_{{\bf q},T}^{G} = \bigoplus_{\bf k} \tilde{\Sigma}_{{\bf k},{\bf q},T}^{G},\ \tilde{\rho}_{{\bf q},1}^{G}\oplus\tilde{\rho}_{{\bf q},2}^{G}= \bigoplus_{\bf k} \tilde{\Sigma}_{{\bf k},{\bf q},1}^{G} \oplus \tilde{\Sigma}_{{\bf k},{\bf q},2}^{G}.
\end{equation}
If a band (co)rep is not equivalent to a direct sum of other band reps, then we define the band (co)rep to be \emph{elementary} [\emph{i.e.}, an EBR]~\cite{Bandrep3,ZakException1,ZakException2,ArisMagneticBlochOscillation,QuantumChemistry}.

In order to complete the theory of MTQC, we must perform a complete enumeration of the EBRs in all of the 1,651 single and double SSGs.  Specifically, because EBRs are induced from (magnetic) Wannier orbitals (Appendix~\ref{sec:induction}), then any set of bands that transforms in a direct sum of EBRs is Wannierizable, and therefore, does not exhibit stable or fragile~\cite{AshvinFragile,AshvinFragile2,AdrianFragile,KoreanFragile,ZhidaFragile,FragileFlowMeta,ZhidaBLG,ZhidaFragile2,KoreanFragileInversion,ArisFragileNoGo,DelicateAris,SlagerMagFragile,SlagerMagFragile2} topology~\cite{QuantumChemistry,Bandrep1,Bandrep2,Bandrep3,JenFragile1,BarryFragile}.  With complete knowledge of the EBRs, we will then be able to identify the bands that do not transform in linear combinations of EBRs, which, as we will show in Appendix~\ref{sec:topologicalBands} correspond to stable topological (crystalline) insulators and topological semimetals.

To obtain an initial bound on the sites in each SSG from which EBRs may be induced, we first recognize that, if a site ${\bf q}_{0}$ indexes a Wyckoff position that is non-maximal, then $G_{{\bf q}_{0}}\subset G_{{\bf q}}$ where ${\bf q}$ is a site in a maximal Wyckoff position that is connected to the Wyckoff position containing ${\bf q}_{0}$ (see Appendix~\ref{sec:Wyckoff} for definitions of connected and maximal Wyckoff positions).  Taking $\tilde{\rho}_{{\bf q}_{0}}$ to be (co)rep of the site-symmetry group $G_{{\bf q}_{0}}$, then, through the transitive property of induction~\cite{Bandrep3}:
\begin{equation}
\tilde{\rho}_{{\bf q},0}\uparrow G = \tilde{\rho}_{{\bf q},0}^{G} = \tilde{\rho}_{{\bf q},0}\uparrow G_{\bf q}\uparrow G = \left(\bigoplus_{i=1}^{z}b^{{\bf q}_{0},{\bf q}}_{i}\tilde{\rho}_{{\bf q},i}\right)\uparrow G = \bigoplus_{i=1}^{z}b^{{\bf q}_{0},{\bf q}}_{i}\tilde{\rho}_{{\bf q},i}^{G},
\label{eq:EBCRmaximal}
\end{equation}
where $z$ is the number of unique irreducible (co)reps $\tilde{\rho}_{{\bf q},i}$ in $G_{\bf q}$, $b^{{\bf q}_{0},{\bf q}}_{i}$ is a non-negative integer, and where at least one $b^{{\bf q}_{0},{\bf q}}_{i}$ is nonzero.  Eq.~(\ref{eq:EBCRmaximal}) implies that any band (co)rep $\tilde{\rho}_{{\bf q},0}^{G}$ induced from a site ${\bf q}_{0}$ in a non-maximal Wyckoff position is equivalent to a sum of band (co)reps induced from a site ${\bf q}$ in a maximal Wyckoff position; therefore $\tilde{\rho}_{{\bf q},0}^{G}$ is either a composite band (co)rep, or is equivalent to an EBR induced from ${\bf q}$.  Consequently, the complete set of EBRs is contained within the set of band (co)reps induced from the sites of the maximal Wyckoff positions of each SSG.

Hence, in this work, we will obtain the EBRs of all single and double SSGs in two steps.  First, we will restrict consideration to the band (co)reps induced by the irreducible (co)reps of the site-symmetry groups of the maximal Wyckoff positions of each SSG.  We will then in Appendix~\ref{sec:exceptions} filter out the composite band (co)reps induced from sites in maximal Wyckoff positions, which are known as the \emph{exceptional cases}~\cite{Bandrep2,Bandrep3,ZakException1,ZakException2,ArisMagneticBlochOscillation,ZakCompatibility}; the remaining band (co)reps comprise the EBRs.  In Appendix~\ref{sec:mebrStats}, we will then provide additional statistics for the EBRs of all SSGs -- including the MEBRs of the Type-III and IV MSGs introduced in this work -- as well as detail the~\href{http://www.cryst.ehu.es/cryst/mbandrep}{MBANDREP} tool on the BCS that we have implemented for this work to access the EBRs and composite band (co)reps induced from each Wyckoff position in each of the 1,651 single and double SSGs.

\subsubsection{Exceptional Cases in the MSGs}
\label{sec:exceptions}

In most cases, when a (co)rep $\tilde{\rho}_{\bf q}$ of a site-symmetry group $G_{\bf q}$ in a maximal Wyckoff position [see Appendix~\ref{sec:Wyckoff}] is induced into an SSG $G$, the resulting band (co)rep $\tilde{\rho}_{\bf q}^{G} = \tilde{\rho}_{\bf q}\uparrow G$ [Eq.~(\ref{eq:mainInductionMTQC})] is an EBR [defined in the text following Eq.~(\ref{eq:EBRdefBJW})].  However, in some \emph{exceptional} cases, $\tilde{\rho}_{\bf q}^{G} = \tilde{\rho}_{\bf q}\uparrow G$ is instead a \emph{composite} band (co)rep.  In Ref.~\onlinecite{Bandrep3}, it was determined that exceptional cases specifically occur under the following conditions:
\begin{enumerate}
\item{Two maximal Wyckoff positions indexed by ${\bf q}$ and ${\bf q}'$ in an SSG $G$ are both connected to the same site ${\bf q}_{0}$ in a non-maximal Wyckoff position.  In Ref.~\onlinecite{Bandrep3}, $G_{{\bf q}'}$ is termed the \emph{reducing group}, and $G_{{\bf q}_{0}}=G_{\bf q}\cap G_{{\bf q}'}$ is termed the \emph{intersection group}.}
\item{There exists an irreducible (co)rep $\tilde{\rho}_{{\bf q}_0}$ of $G_{{\bf q}_{0}}$ for which $\tilde{\rho}_{{\bf q}_{0}}\uparrow G_{\bf q}$ is equivalent to an irreducible (co)rep of $G_{\bf q}$.}
\item{For the same irreducible (co)rep $\tilde{\rho}_{{\bf q}_{0}}$ of $G_{{\bf q}_{0}}$, $\tilde{\rho}_{{\bf q}_{0}}\uparrow G_{{\bf q}'}$ is equivalent to a \emph{reducible} (co)rep of $G_{{\bf q}'}$.}
\end{enumerate}
These three conditions may be summarized through the equivalence relations:
\begin{equation}
\tilde{\rho}_{{\bf q}_0}\uparrow G_{\bf q}\uparrow G = \tilde{\rho}_{\bf q} \uparrow G = \tilde{\rho}_{\bf q}^{G} = \tilde{\rho}_{{\bf q}_0}\uparrow G_{{\bf q}'}\uparrow G = \tilde{\rho}_{{\bf q}'}\uparrow G = \tilde{\rho}_{{\bf q}'}^{G},
\label{eq:mainException}
\end{equation}
in which $\tilde{\rho}_{{\bf q}'}$ is a reducible (co)rep of $G_{{\bf q}'}$, such that $\tilde{\rho}_{{\bf q}'}^{G}$ is a composite band (co)rep, implying that the equivalent band (co)rep $\tilde{\rho}_{\bf q}^{G}$ is also a composite band (co)rep, despite $\tilde{\rho}_{\bf q}$ being an irreducible (co)rep of $G_{\bf q}$.

\begin{figure}[b]
\includegraphics[width=0.85\columnwidth]{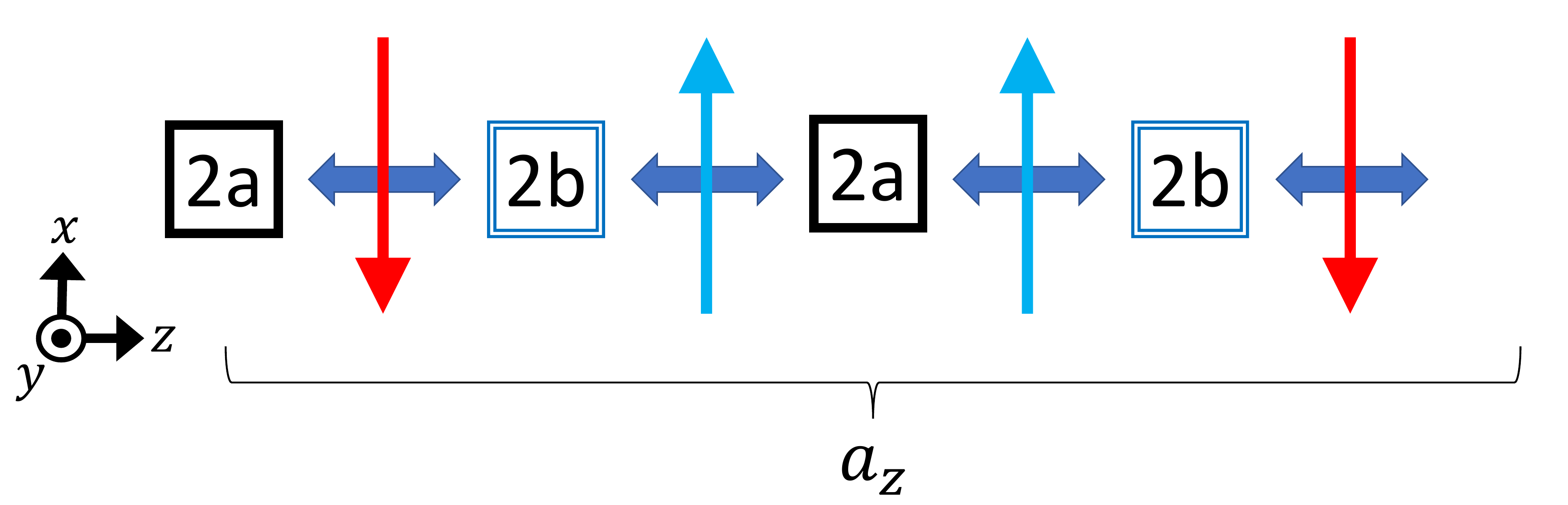}
\caption{An antiferromagnetic chain with magnetic rod group (MRG)  $(p_{c}\bar{1})_{RG}$, which is generated by $\{\mathcal{I}|0\}$ and $\{\mathcal{T}|1/2\}$ ($t_{a_{z}/2}\mathcal{T}$) and is isomorphic after the addition of perpendicular lattice translations to Type-IV MSG 2.7 $P_{S}\bar{1}$ [see Refs.~\onlinecite{BigBook,MagneticBook,ITCA,subperiodicTables,HingeSM} and the text following Eq.~(\ref{eq:translationNotation})].  There are three Wyckoff positions in MRG $(p_{c}\bar{1})_{RG}$ -- $2a$, $2b$, and $4c$ -- of which only $2a$ and $2b$ are maximal [Eq.~(\ref{eq:exceptionalRGWyckoff})].  The site-symmetry group $G_{2a}$ of sites in the maximal $2a$ position contains $\{\mathcal{I}|0\}$ [Eq.~(\ref{eq:exceptionGroup2a})], whereas the site-symmetry group $G_{2b}$ of sites in the maximal $2b$ position instead contains $\{\mathcal{I}\times\mathcal{T}|1/2\}$ [Eq.~(\ref{eq:exceptionGroup2b})]; the site-symmetry group $G_{4c}$ of sites in the general $4c$ position does not contain either $\{\mathcal{I}|0\}$ or $\{\mathcal{I}\times\mathcal{T}|1/2\}$ [Eq.~(\ref{eq:exceptionGroup4c})].  Four $\{\mathcal{I}\times\mathcal{T}|1/2\}$-related spinful $s$ orbitals occupying the $2b$ position in $G=(p_{c}\bar{1})_{RG}$ divide into two pairs that each transform in the two-dimensional irreducible double-valued corep $\left(\bar{A}\bar{A}\right)_{2b}$ of $G_{2b}$ [Eq.~(\ref{eq:exceptionCorep2b})], which is a necessary -- but crucially not sufficient -- condition for the four-dimensional band corep $\left(\bar{A}\bar{A}\right)_{2b}^{G} = \left(\bar{A}\bar{A}\right)_{2b}\uparrow G$ to be an EBR [see Eq.~(\ref{eq:EBCRmaximal}) and the surrounding text].  Indeed, in MRG $(p_{c}\bar{1})_{RG}$, we find that the four spinful $s$ orbitals at $2b$ can be moved through the $4c$ position to $2a$ without breaking a symmetry or closing a gap.   When the four $s$ orbitals are moved to $2a$, the four orbitals form two pairs of spinful bonding and antibonding orbitals that each transform in the two-dimensional \emph{reducible} corep $\left(\bar{A}_{g}\right)_{2a}\oplus\left(\bar{A}_{u}\right)_{2a}$ of $G_{2a}$ [Eq.~(\ref{eq:exceptionCorep2b})], and induce a four-dimensional \emph{composite} band corep $\left(\bar{A}_{g}\right)_{2a}^{G}\oplus\left(\bar{A}_{u}\right)_{2a}^{G}$.  Because $\left(\bar{A}_{g}\right)_{2a}^{G} \oplus \left(\bar{A}_{u}\right)_{2a}^{G} = \left(\bar{A}\right)_{4c}^{G} = \left(\bar{A}\bar{A}\right)_{2b}^{G}$ [Eqs.~(\ref{eq:exceptionBandCorepComposite1}) and~(\ref{eq:exceptionBandCorepComposite2})], then we conclude that $\left(\bar{A}\bar{A}\right)_{2b}^{G}$ is an \emph{exceptional} case of a composite band corep induced from an irreducible corep of a site-symmetry group in a maximal Wyckoff position.}
\label{fig:exceptionalAFM}
\end{figure}

In the Type-I and Type-II SSGs previously analyzed in TQC~\cite{QuantumChemistry,Bandrep1,Bandrep2,Bandrep3,JenFragile1,BarryFragile}, the exceptional cases all occurred in SSGs with point groups that were either isomorphic to Type-I MPG 8.1.24 $mmm$ or to MPGs with higher-fold rotation, rotoinversion, or $\mathcal{T}$ symmetries [\emph{c.f.} Tables~S10,~S11, and~S12 in Ref.~\onlinecite{QuantumChemistry}].  Conversely, in this work, we find there are exceptional composite band coreps in some of the lowest-symmetry Type-III and Type-IV MSGs.  Previously in TQC~\cite{QuantumChemistry,Bandrep1,Bandrep2,Bandrep3,JenFragile1,BarryFragile}, it was specifically recognized that if two maximal Wyckoff positions in the same symmetry group have the same multiplicity, but the band (co)reps induced from the Wyckoff positions have different dimensionality, then it is possible that at least one of the induced band (co)reps is composite.  In this section we will consider the example of double magnetic rod group [MRG] $(p_{c}\bar{1})_{RG}$ [Fig.~\ref{fig:exceptionalAFM}], which we have selected because the $2a$ and $2b$ Wyckoff positions both have a multiplicity of 2, but the band coreps induced from $2a$ are two-dimensional, whereas the band corep induced from $2b$ is four-dimensional [and indeed exceptional-case composite].

MRG  $(p_{c}\bar{1})_{RG}$ is generated by:
\begin{equation}
\{\mathcal{I}|0\},\ \{\mathcal{T}|1/2\},
\end{equation}
and is isomorphic after the addition of perpendicular lattice translations to Type-IV double MSG 2.7 $P_{S}\bar{1}$ [see Refs.~\onlinecite{BigBook,MagneticBook,ITCA,subperiodicTables,HingeSM} and the text following Eq.~(\ref{eq:translationNotation})].  Using~\href{http://www.cryst.ehu.es/cgi-bin/cryst/programs/magget_wp.pl}{MWYCKPOS} on the BCS~\cite{BCSMag1,BCSMag2,BCSMag3,BCSMag4} for Type-IV MSG 2.7 $P_{S}\bar{1}$ and restricting to Wyckoff positions with $x=y=0$ in the reduced notation of~\href{http://www.cryst.ehu.es/cgi-bin/cryst/programs/magget_wp.pl}{MWYCKPOS}, we obtain the coordinates and site-symmetry-group-isomorphic MPGs of the Wyckoff positions of MRG $(p_{c}\bar{1})_{RG}$:
\begin{eqnarray}
{\bf q}_{2a} &=& 0,\ 1/2,\ G_{2a} = \bar{1}, \nonumber \\
{\bf q}_{2b} &=& 1/4,\ 3/4,\ G_{2b} = \bar{1}', \nonumber \\
{\bf q}_{4c} &=& z,\ (1/2)-z,\ (1/2)+z,\ 1-z,\ G_{4c} = 1,
\label{eq:exceptionalRGWyckoff}
\end{eqnarray}
where we have employed units in which $a_{z}=1$ in Fig.~\ref{fig:exceptionalAFM}.  In Eq.~(\ref{eq:exceptionalRGWyckoff}), the symbols $\bar{1}$, $\bar{1}'$, and $1$ respectively refer to Type-I MPG 2.1.3 $\bar{1}$, Type-III MPG 2.3.5 $\bar{1}'$, and Type-I MPG 1.1.1 $1$ [the trivial MPG, see the text following Eq.~(\ref{eq:gqforSiteSym})].  In Eq.~(\ref{eq:exceptionalRGWyckoff}), the $2a$ and $2b$ positions are maximal, whereas $4c$ is the (non-maximal) general position.  First, we will examine the site-symmetry groups of the ${\bf q}_{2a}$ maximal Wyckoff position, which are isomorphic to Type-I double MPG 2.1.3 $\bar{1}$.  $G_{2a}$ contains only four symmetry operations and is equal to its maximal unitary subgroup $H_{2a}$:
\begin{equation}
G_{2a} = H_{2a} = \bigg\{\{E|0\},\ \{\mathcal{I}|0\},\ \{\bar{E}|0\},\ \{\bar{E}\mathcal{I}|0\}\bigg\},
\label{eq:exceptionGroup2a}
\end{equation}
where $E$ is the identity operation, and $\bar{E}=C_{1n}$ is the symmetry operation of $360^{\circ}$ rotation about an arbitrary axis $n$, which distinguishes single-valued (spinless) and double-valued (spinful) coreps.  Using the~\href{http://www.cryst.ehu.es/cryst/corepresentationsPG}{CorepresentationsPG} tool on the BCS for MPG 2.1.3 $\bar{1}$, we determine that there are only two double-valued irreducible coreps of $G_{2a}$:
\begin{equation}
\tilde{\rho}_{2a} = \left(\bar{A}_{g}\right)_{2a},\ \left(\bar{A}_{u}\right)_{2a},
\label{eq:exceptionCorep2a}
\end{equation}
for which:
\begin{eqnarray}
\chi_{\left(\bar{A}_{g}\right)_{2a}}\big(\{E|0\}\big) &=& \chi_{\left(\bar{A}_{u}\right)_{2a}}\big(\{E|0\}\big) = -\chi_{\left(\bar{A}_{g}\right)_{2a}}\big(\{\bar{E}|0\}\big) = -\chi_{\left(\bar{A}_{u}\right)_{2a}}\big(\{\bar{E}|0\}\big) = 1, \nonumber \\
\chi_{\left(\bar{A}_{g}\right)_{2a}}\big(\{\mathcal{I}|0\}\big) &=& -\chi_{\left(\bar{A}_{g}\right)_{2a}}\big(\{\bar{E}\mathcal{I}|0\}\big)=1,\ \chi_{\left(\bar{A}_{u}\right)_{2a}}\big(\{\mathcal{I}|0\}\big)=-\chi_{\left(\bar{A}_{u}\right)_{2a}}\big(\{\bar{E}\mathcal{I}|0\}\big)=-1,
\end{eqnarray}
implying that the lowest-angular-momentum spinful magnetic atomic orbitals (see Appendix~\ref{sec:magWannier}) that transform in $\left(\bar{A}_{g}\right)_{2a}$ and $\left(\bar{A}_{u}\right)_{2a}$ are spin-split (singly-degenerate) $s$ and $p$ orbitals, respectively.  We next examine the site-symmetry groups of the ${\bf q}_{2b}$ maximal Wyckoff position in Eq.~(\ref{eq:exceptionalRGWyckoff}) and Fig.~\ref{fig:exceptionalAFM}, which are isomorphic to Type-III double MPG 2.3.5 $\bar{1}'$.  $G_{2b}$ also contains four symmetry operations:
\begin{equation}
G_{2b} = \bigg\{\{E|0\},\ \{\mathcal{I}\times\mathcal{T}|1/2\},\ \{\bar{E}|0\},\ \{\bar{E}\mathcal{I}\times\mathcal{T}|1/2\}\bigg\},
\label{eq:exceptionGroup2b}
\end{equation}
in which only $\{E|0\}$ and $\{\bar{E}|0\}$ are unitary.  Hence the maximal unitary subgroup $H_{2b}$ of $G_{2b}$ is given by:
\begin{equation}
H_{2b} = \bigg\{\{E|0\},\ \{\bar{E}|0\}\bigg\},
\label{eq:exceptionGroup2bUnitary}
\end{equation}
such that $H_{2b}$ is isomorphic to the trivial MPG [Type-I MPG 1.1.1 $1$, see the text following Eq.~(\ref{eq:gqforSiteSym})].  As discussed in Ref.~\onlinecite{BigBook}, $(\mathcal{I}\times\mathcal{T})^{2}=\bar{E}$ in double SPGs, and $\chi_{\tilde{\rho}}(\{\bar{E}|0\})=-\chi_{\tilde{\rho}}(\{E|0\})$ for double-valued coreps $\tilde{\rho}$.  From this, in agreement with the output of the~\href{http://www.cryst.ehu.es/cryst/corepresentationsPG}{CorepresentationsPG} tool on the BCS for Type-III double MPG 2.3.5 $\bar{1}'$, we determine that $G_{2b}$ has only one, two-dimensional, double-valued irreducible corep [see Eq.~(\ref{eq:JtestforOrbitals}) and the surrounding text]:
\begin{equation}
\tilde{\rho}_{2b} = \left(\bar{A}\bar{A}\right)_{2b},
\label{eq:exceptionCorep2b}
\end{equation}
for which:
\begin{equation}
\chi_{\left(\bar{A}\bar{A}\right)_{2b}}\big(\{E|0\}\big) = -\chi_{\left(\bar{A}\bar{A}\right)_{2b}}\big(\{\bar{E}|0\}\big) = 2,
\end{equation}
implying that the lowest-angular-momentum spinful magnetic atomic orbitals that transform in $\left(\bar{A}\bar{A}\right)_{2b}$ are an $\{\mathcal{I}\times\mathcal{T}|1/2\}$-related pair of spinful $s$ orbitals, which are twofold-degenerate because $\chi_{\left(\bar{A}\bar{A}\right)_{2b}}([\{\mathcal{I}\times\mathcal{T}|1/2\}]^{2})=-\chi_{\left(\bar{A}\bar{A}\right)_{2b}}(\{E|0\})=-2$.  Lastly, the site-symmetry groups in the ${\bf q}_{4c}$ position in Eq.~(\ref{eq:exceptionalRGWyckoff}) and Fig.~\ref{fig:exceptionalAFM} are isomorphic to the trivial MPG [Type-I MPG 1.1.1 $1$, see the text following Eq.~(\ref{eq:gqforSiteSym})], and are thus equal to their maximal unitary subgroups $H_{4c}$:
\begin{equation}
G_{4c} = H_{4c} = \bigg\{\{E|0\},\ \{\bar{E}|0\}\bigg\}.
\label{eq:exceptionGroup4c}
\end{equation}
There is only one, one-dimensional, double-valued irreducible corep of $G_{4c}$:
\begin{equation}
\tilde{\rho}_{4c} = \left(\bar{A}\right)_{4c},
\label{eq:exceptionCorep4c}
\end{equation}
for which:
\begin{equation}
\chi_{\left(\bar{A}\right)_{4c}}\big(\{E|0\}\big)=-\chi_{\left(\bar{A}\right)_{4c}}\big(\{\bar{E}|0\}\big)=1.
\label{eq:extraStepExceptionCorep4c}
\end{equation}
Eq.~(\ref{eq:extraStepExceptionCorep4c}) implies that the lowest-angular-momentum spinful magnetic atomic orbital that transforms in $\left(\bar{A}\right)_{4c}$ is a spin-split (singly-degenerate) $s$ orbital.

Next, to determine if any of the band coreps induced from the maximal $2a$ and $2b$ Wyckoff positions in Eq.~(\ref{eq:exceptionalRGWyckoff}) and Fig.~\ref{fig:exceptionalAFM} are exceptional cases (\emph{i.e.} composite), we induce band coreps from the intermediate $4c$ position that is connected to $2a$ and $2b$ [Eq.~(\ref{eq:mainException}) and the surrounding text].  First, we focus on band coreps induced from $4c$ through $2b$.  Because $G_{4c}$ is an index-2 subgroup of $G_{2b}$ ($[G_{2b}:G_{4c}]=2$, see Eqs.~(\ref{eq:subsetIndex}),~(\ref{eq:exceptionGroup2b}), and~(\ref{eq:exceptionGroup4c})), and because $G_{4c}$ and $G_{2b}$ have isomorphic unitary subgroups $H_{4c}=H_{2b}$ [Eq.~(\ref{eq:exceptionGroup2bUnitary}) and~(\ref{eq:exceptionGroup4c})], then:
\begin{equation}
\left(\bar{A}\right)_{4c}\uparrow G_{2b} = \left(\bar{A}\bar{A}\right)_{2b}.
\label{eq:exceptionInduction1}
\end{equation}
where $\left(\bar{A}\bar{A}\right)_{2b}$ is the irreducible corep of $G_{2b}$ [Eq.~(\ref{eq:exceptionCorep2b})].  Eq.~(\ref{eq:exceptionInduction1}) implies that, for:
\begin{equation}
G= (p_{c}\bar{1})_{RG},
\end{equation}
it is possible for $\left(\bar{A}\right)_{4c}^{G}=\left(\bar{A}\right)_{4c}\uparrow G$ to be an EBR, because:
\begin{equation}
\left(\bar{A}\right)_{4c}^{G} = \left(\bar{A}\right)_{4c}\uparrow G_{2b}\uparrow G = \left(\bar{A}\bar{A}\right)_{2b} \uparrow G = \left(\bar{A}\bar{A}\right)_{2b}^{G},
\label{eq:exceptionBandCorepComposite1}
\end{equation}
such that $\left(\bar{A}\bar{A}\right)_{2b}^{G}$ is a band corep induced from an irreducible corep of a site-symmetry group in a maximal Wyckoff position [see Eq.~(\ref{eq:EBCRmaximal}) and the surrounding text].

However, to determine if $\left(\bar{A}\bar{A}\right)_{2b}^{G}$ is indeed an EBR, we must also calculate the band coreps induced from $4c$ through $2a$, which are equivalent to $\left(\bar{A}\bar{A}\right)_{2b}^{G}$ [Eq.~(\ref{eq:mainException})].  Because $G_{4c}$ is an index-2 subgroup of $G_{2a}$ ($[G_{2a}:G_{4c}]=2$, see Eqs.~(\ref{eq:subsetIndex}),~(\ref{eq:exceptionGroup2a}), and~(\ref{eq:exceptionGroup4c})), because $\{E|0\}\in G_{2a}$, $\{E|0\}\in G_{4c}$, and because $\{\mathcal{I}|0\}\in G_{2a}$, $\{\mathcal{I}|0\}\not\in G_{4c}$, then:
\begin{equation}
\left(\bar{A}\right)_{4c}\uparrow G_{2a} = \left(\bar{A}_{g}\right)_{2a} \oplus \left(\bar{A}_{u}\right)_{2a},
\label{eq:exceptionInduction2}
\end{equation}
where $\left(\bar{A}_{g}\right)_{2a}$ and $\left(\bar{A}_{u}\right)_{2a}$ are the irreducible coreps of $G_{2a}$ [Eq.~(\ref{eq:exceptionCorep2a})], implying that $\left(\bar{A}_{g}\right)_{2a} \oplus \left(\bar{A}_{u}\right)_{2a}$ is a \emph{reducible} corep of $G_{2a}$.  Eq.~(\ref{eq:exceptionInduction2}) indicates that $\left(\bar{A}\right)_{4c}^{G}=\left(\bar{A}\right)_{4c}\uparrow G$ is not an EBR, but is instead a composite band corep, because 
\begin{equation}
\left(\bar{A}\right)_{4c}^{G} = \left(\bar{A}\right)_{4c}\uparrow G_{2a}\uparrow G = \bigg[\left(\bar{A}_{g}\right)_{2a} \oplus \left(\bar{A}_{u}\right)_{2a}\bigg] \uparrow G =\left(\bar{A}_{g}\right)_{2a}^{G} \oplus \left(\bar{A}_{u}\right)_{2a}^{G}.
\label{eq:exceptionBandCorepComposite2}
\end{equation}
Because $\left(\bar{A}_{g}\right)_{2a}^{G} \oplus \left(\bar{A}_{u}\right)_{2a}^{G} = \left(\bar{A}\right)_{4c}^{G} = \left(\bar{A}\bar{A}\right)_{2b}^{G}$ [Eqs.~(\ref{eq:exceptionBandCorepComposite1}) and~(\ref{eq:exceptionBandCorepComposite2})], then we conclude that $\left(\bar{A}\bar{A}\right)_{2b}^{G}$ is an exceptional case of a composite band corep induced from an irreducible corep of a site-symmetry group in a maximal Wyckoff position.

We can gain physical intuition for why $\left(\bar{A}\bar{A}\right)_{2b}^{G}$ is an exceptional-case composite band corep from the orbitals and spins depicted in Fig.~\ref{fig:exceptionalAFM}.  We begin with two $\{\mathcal{I}\times\mathcal{T}|1/2\}$-related pairs of spin-up and spin-down $s$ orbitals that occupy $2b$ (\emph{i.e.} four total spinful $s$ orbitals separated into $\{\mathcal{I}\times\mathcal{T}|1/2\}$-reversed pairs at each of the two sites in the $2b$ position), where each pair transforms in the two-dimensional irreducible site-symmetry corep $\left(\bar{A}\bar{A}\right)_{2b}$.  We are then free to move the four orbitals to $2a$ without breaking a symmetry of $(p_{c}\bar{1})_{RG}$ or closing a gap to introduce additional Wannier orbitals (which, conversely, is required in the closely-related obstructed-atomic-limit Wannier-sliding transitions discussed in Refs.~\onlinecite{QuantumChemistry,HingeSM,JenOAL}).  When the four spinful $s$ orbitals reach $2a$, the four orbitals form two bonding and antibonding pairs that each transform in the two-dimensional reducible site-symmetry corep $\left(\bar{A}_{g}\right)_{2a}\oplus\left(\bar{A}_{u}\right)_{2a}$ of $G_{2a}$ [Eq.~(\ref{eq:exceptionCorep2a})], which induces a four-dimensional composite band corep $\left(\bar{A}_{g}\right)_{2a}^{G}\oplus\left(\bar{A}_{u}\right)_{2a}^{G}$ of $(p_{c}\bar{1})_{RG}$, indicating that $\left(\bar{A}\bar{A}\right)_{2a}^{G}$ is an exceptional composite band corep.

In Appendix~\ref{sec:exceptionalTables}, we provide a complete enumeration of all of the exceptional cases in the 1,651 single and double SSGs.  For the Type-I MSGs and Type-II SGs previously analyzed in TQC~\cite{QuantumChemistry,Bandrep1,Bandrep2,Bandrep3,JenFragile1,BarryFragile}, the exceptional cases listed in Appendix~\ref{sec:exceptionalTables} agree with the previous tabulations performed in Refs.~\onlinecite{QuantumChemistry,Bandrep3}.  As shown in the text following Eq.~(\ref{eq:EBCRmaximal}), any band (co)rep induced from an irreducible (co)rep of a site in a maximal Wyckoff position that is not listed in the tables in Appendix~\ref{sec:exceptionalTables} is an EBR.  Hence, by calculating all of the band (co)reps induced from the irreducible (co)reps of the site-symmetry groups of the maximal Wyckoff positions of the 1,651 single and double SSGs, and then subsequently excluding the exceptional cases listed in Appendix~\ref{sec:exceptionalTables}, we obtain the complete list of single- and double-valued EBRs of the SSGs, completing the theory of MTQC.

\subsubsection{Statistics for the MEBRs and the \textsc{MBANDREP} Tool}
\label{sec:mebrStats}

In this section, we provide general statistics for the EBRs previously obtained in Appendix~\ref{sec:exceptions} [which include the MEBRs of the Type-I MSGs and PEBRs of the Type-II SSGs previously tabulated for TQC~\cite{QuantumChemistry,Bandrep1,Bandrep2,Bandrep3,JenFragile1,BarryFragile}, as well as the MEBRs of the Type-III and Type-IV MSGs calculated for the present work].  We additionally detail in this section the~\href{http://www.cryst.ehu.es/cryst/mbandrep}{MBANDREP} tool on the BCS, which we have implemented for this work to access both the elementary and non-elementary band (co)reps of all 1,651 single and double SSGs.

To begin, in Tables~\ref{tb:MEBRsSingle} and~\ref{tb:MEBRsDouble}, we provide the number of elementary and composite band (co)reps of the 1,651 single and double SSGs, respectively.  Tables~\ref{tb:MEBRsSingle} and~\ref{tb:MEBRsDouble} include the number of \emph{exceptional} cases [Appendices~\ref{sec:exceptions} and~\ref{sec:exceptionalTables}] in which an irreducible (co)rep of a site-symmetry group of a site in a maximal Wyckoff position does not induce an EBR.  For the Type-I MSGs and Type-II SGs analyzed in TQC~\cite{QuantumChemistry,Bandrep1,Bandrep2,Bandrep3,JenFragile1,BarryFragile}, the band (co)rep statistics in Tables~\ref{tb:MEBRsSingle} and~\ref{tb:MEBRsDouble} agree with the calculations previously performed in Refs.~\onlinecite{QuantumChemistry,Bandrep3}.  In Tables~\ref{tb:MEBRsSingle} and~\ref{tb:MEBRsDouble}, we also list the number of EBRs that can be decomposed into disconnected branches [\emph{i.e.} \emph{decomposable} or ``split'' EBRs with disconnected subgraphs, see Appendix~\ref{sec:compatibilityRelations} and Refs.~\onlinecite{QuantumChemistry,Bandrep2,JenFragile1,BarryFragile,AndreiMaterials,LuisBasicBands,AshvinFragile}].  As shown in Refs.~\onlinecite{QuantumChemistry,JenFragile1,BarryFragile,AndreiMaterials,LuisBasicBands}, at least one disconnected piece of each decomposable EBR is topologically nontrivial, either in a stable or fragile sense~\cite{AshvinFragile,AshvinFragile2,AdrianFragile,KoreanFragile,ZhidaFragile,FragileFlowMeta,ZhidaBLG,ZhidaFragile2,KoreanFragileInversion,ArisFragileNoGo,DelicateAris,SlagerMagFragile,SlagerMagFragile2}.  In Appendix~\ref{sec:topologicalBands}, we will provide a complete enumeration of the symmetry-based indicators of stable band topology~\cite{SlagerSymmetry,AshvinIndicators,HOTIChen,ChenTCI,AshvinTCI,BarryBandrepReview,ZhidaSemimetals,AdrianSIReview,AdrianSCMagSI} in the 1,651 double SSGs, which can be used to diagnose the stable topological indices of the disconnected branches of the decomposable double-valued EBRs in Table~\ref{tb:MEBRsDouble}.  Lastly, to provide complete statistics for all of the band (co)reps that can be induced by \emph{any} set of magnetic atomic orbitals in any Wyckoff position in a magnetic crystal, we additionally list in Tables~\ref{tb:MEBRsSingle} and~\ref{tb:MEBRsDouble} the number of composite band (co)reps that can be induced from the unique irreducible (co)reps of the site-symmetry groups of the \emph{non-maximal} Wyckoff positions in SSGs of the same type.  Specifically, we obtain the numbers listed in the ``Unique Non-Maximal Band (Co)reps'' columns in Tables~\ref{tb:MEBRsSingle} and~\ref{tb:MEBRsDouble} by summing over the composite band (co)reps induced from each unique irreducible (co)rep of one site-symmetry group in each non-maximal Wyckoff position in each SSG of the same type.

\begin{table}[!h]	
\begin{tabular}{|c|c|c|c|c|}
\hline
Single SSG Type & Number of SSGs & Number of EBRs & Exceptional & Unique Non-Maximal \\
 & & [Decomposable EBRs] & Cases & Band (Co)reps \\
\hline
Type-I & 230 & 3,383 & 40 & 1,931\\
	& & [219] & & \\
\hline
Type-II & 230 & 3,141 & 39 & 1,852 \\
	& & [156] & & \\
\hline
Type-III & 674 & 7,492 & 151 & 5,279\\
	& & [833] & & \\
\hline
Type-IV & 517 & 6,190 & 130 & 4,501 \\
	& & [699] & & \\
\hline
Total & 1,651 & 20,206 & 360 & 13,563 \\
	& & [1,907] & & \\
\hline
\end{tabular}
\caption{Single-valued band (co)reps of the 1,651 single SSGs.  In order, the columns in this table list the type of the single SSG (Appendix~\ref{sec:MSGs}), the number of single SSGs of each type, the total number of single-valued elementary band (co)reps [EBRs] of the SSGs of the same type [see the text surrounding Eq.~(\ref{eq:EBRdefBJW})], the total number of exceptional composite single-valued band (co)reps of the SSGs of the same type (Appendices~\ref{sec:exceptions} and~\ref{sec:exceptionalTables}), and the total number of composite single-valued band (co)reps induced from unique irreducible (co)reps of the site-symmetry groups of the non-maximal Wyckoff positions in SSGs of the same type.}
\label{tb:MEBRsSingle}
\end{table}

\begin{table}[!h]	
\begin{tabular}{|c|c|c|c|c|}
\hline
Double SSG Type & Number of SSGs & Number of EBRs & Exceptional & Unique Non-Maximal \\
 & & [Decomposable EBRs] & Cases & Band (Co)reps \\
\hline
Type-I & 230 & 2,258 & 107 & 1,589 \\
	& & [355] & & \\
\hline
Type-II & 230 & 1,616 & 0 & 1,001 \\
	& & [426] & & \\
\hline
Type-III & 674 & 5,047 & 591 & 4,882 \\
	& & [662] & & \\
\hline
Type-IV & 517 & 3,882 & 556 & 3,984 \\
	& & [639] & & \\
\hline
Total & 1,651 & 12,803 & 1,254 & 11,456 \\
	& & [2,082] & & \\
\hline
\end{tabular}
\caption{Double-valued band (co)reps of the 1,651 double SSGs.  In order, the columns in this table list the type of the double SSG (Appendix~\ref{sec:MSGs}), the number of double SSGs of each type, the total number of double-valued EBRs of the SSGs of the same type [see the text surrounding Eq.~(\ref{eq:EBRdefBJW})], the total number of exceptional composite double-valued band (co)reps of the SSGs of the same type (Appendices~\ref{sec:exceptions} and~\ref{sec:exceptionalTables}), and the total number of composite double-valued band (co)reps induced from unique irreducible (co)reps of the site-symmetry groups of the non-maximal Wyckoff positions in SSGs of the same type.}
\label{tb:MEBRsDouble}
\end{table}

\begin{figure}[h]
\includegraphics[width=\columnwidth]{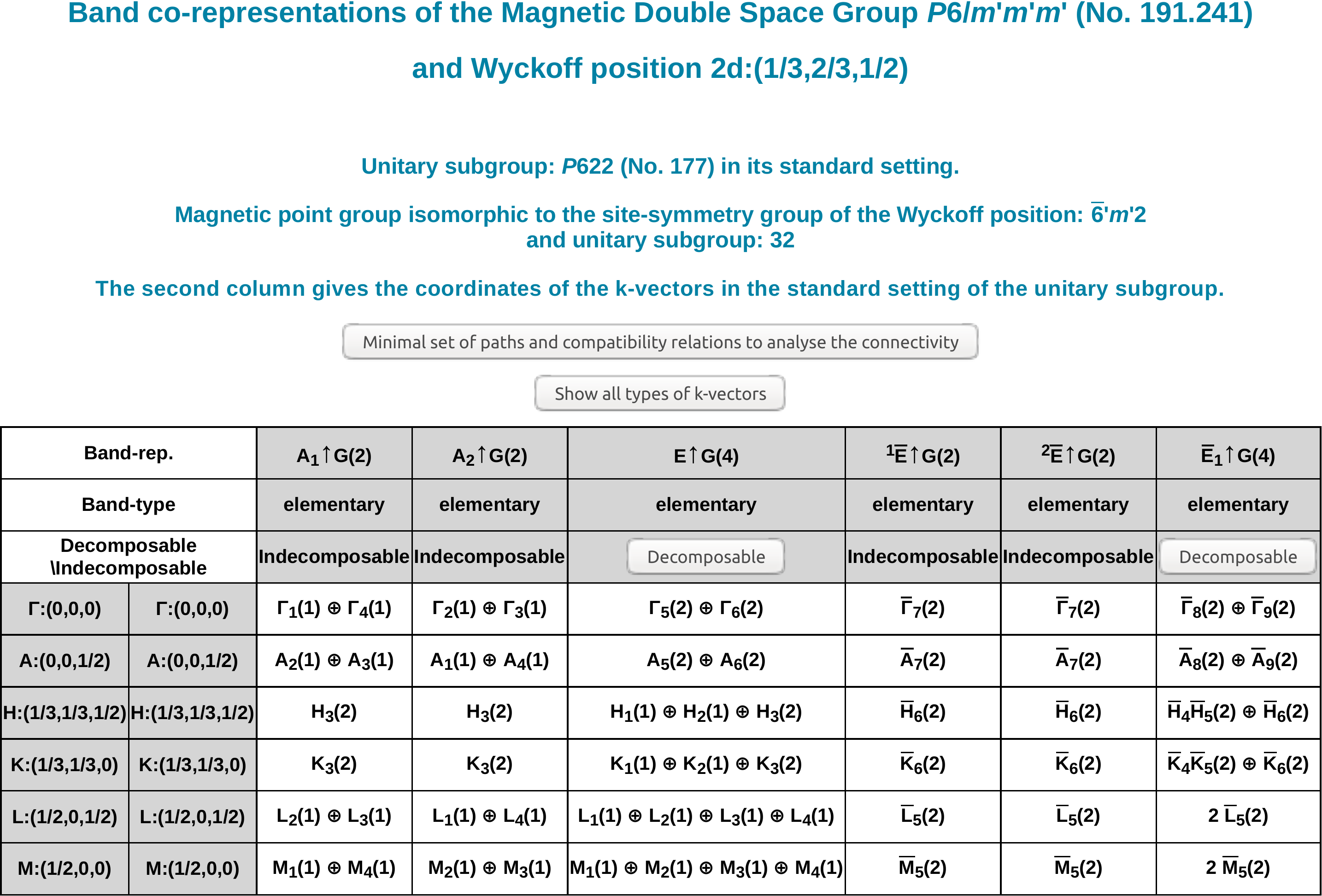}
\caption{The output of the~\href{http://www.cryst.ehu.es/cryst/mbandrep}{MBANDREP} tool for the $2d$ Wyckoff position in Type-III MSG 191.241 $P6/m'm'm'$.  Similar to the earlier~\href{https://www.cryst.ehu.es/cgi-bin/cryst/programs/bandrep.pl}{BANDREP} tool implemented for TQC~\cite{QuantumChemistry,Bandrep1,Bandrep2,Bandrep3,JenFragile1,BarryFragile},~\href{http://www.cryst.ehu.es/cryst/mbandrep}{MBANDREP} allows users to choose between the EBRs of each SSG and the band (co)reps induced from each Wyckoff position in the SSG.  When the Wyckoff position option is selected in~\href{http://www.cryst.ehu.es/cryst/mbandrep}{MBANDREP}, users can additionally select non-maximal Wyckoff positions to access the unique composite band (co)reps discussed in Tables~\ref{tb:MEBRsSingle} and~\ref{tb:MEBRsDouble} and the surrounding text [though we have only shown the output of~\href{http://www.cryst.ehu.es/cryst/mbandrep}{MBANDREP} for a maximal Wyckoff position in this figure].  Specifically, to generate this figure, we have selected the Wyckoff position option in~\href{http://www.cryst.ehu.es/cryst/mbandrep}{MBANDREP} for the $2d$ position in Type-III MSG 191.241 $P6/m'm'm'$.  For each irreducible (co)rep $\tilde{\rho}_{\bf q}$ of one site-symmetry group $G_{\bf q}$ in each Wyckoff position in each SSG,~\href{http://www.cryst.ehu.es/cryst/mbandrep}{MBANDREP} outputs whether the induced band (co)rep $\tilde{\rho}_{\bf q}^{G}=\tilde{\rho}_{\bf q}\uparrow G$ is elementary, indicates whether $\tilde{\rho}_{\bf q}^{G}$ is decomposable~\cite{QuantumChemistry,Bandrep2,JenFragile1,BarryFragile,AndreiMaterials,LuisBasicBands,AshvinFragile}, and lists the subduced small (co)reps in $\tilde{\sigma}^{G}_{{\bf k},{\bf q}}\downarrow G_{\bf k}$ [Eq.~(\ref{eq:corepSubduction})] for each maximal ${\bf k}$ vector [Eq.~(\ref{eq:maximalKvec}) and the surrounding text] in the notation of the~\href{http://www.cryst.ehu.es/cryst/corepresentations}{Corepresentations} tool introduced in this work [see Appendix~\ref{sec:coreps}].  If an EBR is decomposable, users may click on the ``Decomposable'' button in~\href{http://www.cryst.ehu.es/cryst/mbandrep}{MBANDREP} to access a list of the allowed decompositions [branches] of the band (co)rep.}
\label{fig:mbandrep}
\end{figure}

Next, in Appendix~\ref{sec:EBCRdimension}, we provide tables of the minimum and maximum EBR dimension in each single and double SSG.  In particular, the minimum EBR dimensions in the double SSGs in Appendix~\ref{sec:EBCRdimension} provide an upper bound on the \emph{minimal insulating filling} of each double SSG~\cite{WPVZ,WiederLayers,DDP,SteveMagnet,WPVZprl,FillingEnforcedQBI,AshvinMagnetic,MagneticBieberbach}, which is defined as the set of electronic fillings at which a short-range-entangled insulating phase is permitted for arbitrarily strong, SSG-symmetry-preserving interactions, analogous to the Lieb-Schultz-Mattis filling constraints for a 1D spin chain~\cite{LSMTheorem}.  In the cases in which the minimum-dimension EBRs in an SSG are decomposable, a tighter bound on the minimal insulating filling can be further obtained by determining the minimum disconnected branch dimension of each decomposable EBR~\cite{FillingEnforcedQBI,LuisBasicBands}.  Hence, the minimum double-valued EBR dimensions of the Type-III and Type-IV MSGs listed in Appendix~\ref{sec:EBCRdimension} provide upper bounds on the minimal electronic fillings at which short-range-entangled magnetic insulating phases are permitted in each Type-III and Type-IV MSG -- at fillings that violate these bounds, any gapped, MSG-symmetric insulator must therefore exhibit long-range-entangled, magnetic topological order.  Due to complications arising from the antiunitary symmetries of Type-III and Type-IV MSGs (see Appendices~\ref{sec:type3} and~\ref{sec:type4}, respectively), the search for $\mathcal{T}$-breaking, long-range-entangled MSG-symmetric, insulating topological phases has thus far only been addressed from the perspective of minimal insulating filling in a handful of recent works~\cite{MagneticBieberbach,AshvinMagnetic,ThorngrenLSM}.  For each single and double SSG, we have specifically confirmed that the minimum EBR dimension listed in Appendix~\ref{sec:EBCRdimension} is consistent with the minimum atomic insulator dimension previously calculated in Ref.~\onlinecite{AshvinMagnetic}.  In summary, the Type-III and Type-IV MEBRs computed in this work provide new information -- including small (co)rep characters [Appendix~\ref{sec:coreps}] and compatibility relations [Appendix~\ref{sec:compatibilityRelations}] -- applicable to the search for novel long-range-entangled topological phases with magnetic crystal symmetries.

Finally, for this work, we have implemented the~\href{http://www.cryst.ehu.es/cryst/mbandrep}{MBANDREP} tool on the BCS to access both the elementary and non-elementary band (co)reps of all the 1,651 single and double SSGs.  \href{http://www.cryst.ehu.es/cryst/mbandrep}{MBANDREP} thus subsumes the earlier~\href{https://www.cryst.ehu.es/cgi-bin/cryst/programs/bandrep.pl}{BANDREP} tool~(\url{https://www.cryst.ehu.es/cgi-bin/cryst/programs/bandrep.pl})~\cite{QuantumChemistry,Bandrep1}, which was previously implemented for TQC~\cite{QuantumChemistry,Bandrep1,Bandrep2,Bandrep3,JenFragile1,BarryFragile} to access the band (co)reps of the Type-I and Type-II SSGs.  Unlike the earlier~\href{https://www.cryst.ehu.es/cgi-bin/cryst/programs/bandrep.pl}{BANDREP} tool,~\href{http://www.cryst.ehu.es/cryst/mbandrep}{MBANDREP} does not provide separate options for accessing band  (co)reps with and without $\mathcal{T}$ symmetry, which are instead separately listed in~\href{http://www.cryst.ehu.es/cryst/mbandrep}{MBANDREP} under Type-II and Type-I SSGs, respectively [see Appendices~\ref{sec:type2} and~\ref{sec:type1}, respectively].  In Fig.~\ref{fig:mbandrep}, we reproduce the output of~\href{http://www.cryst.ehu.es/cryst/mbandrep}{MBANDREP} for the $2d$ Wyckoff position in Type-III MSG 191.241 $P6/m'm'm'$.

\section{Symmetry-Indicated Magnetic Topological Bands}
\label{sec:topologicalBands}

In the previous sections of this supplement, we established the theory of MTQC.  The building blocks of MTQC are topologically trivial bands that transform in direct sums of EBRs [defined in the text surrounding Eq.~(\ref{eq:EBRdefBJW})], and, consequently, can be inverse-Fourier-transformed into (magnetic) Wannier orbitals in position space [see Appendix~\ref{sec:magWannier}].  Generically, however, energetically isolated bands [specifically, bands that satisfy the insulating compatibility relations along all high-symmetry BZ lines and planes, see Appendix~\ref{sec:compatibilityRelations}] are not required to be equivalent [defined in the text following Eq.~(\ref{eq:EBRdefBJW})] to integer-valued linear combinations of EBRs.  As we will show in this section, if a band $B$ is not equivalent to an integer linear combination of EBRs, then $B$ either corresponds to a topological semimetal whose nodal points lie away from the high-symmetry BZ lines and planes (along which bands satisfy the insulating compatibility relations)~\cite{ZhidaSemimetals,YoungkukLineNode}, or is the Fourier-transformed description of a stable topological insulator or topological crystalline insulator (TI or TCI, respectively)~\cite{ThoulessWannier,QWZ,QHZ,AlexeyVDBWannier,AlexeyVDBTI,TRPolarization,MooreBalentsTI,FuKaneInversion,WiederAxion,HOTIChen,TMDHOTI}.

When unitary crystal symmetries -- such as spatial inversion ($\mathcal{I}$) or fourfold rotoinversion ($C_{4}\times\mathcal{I}$) -- are present in the SSG of the 3D bulk, then the stable topology of a set of energetically-isolated bands (along all high-symmetry BZ lines and planes) may be diagnosed using symmetry eigenvalues through a \emph{symmetry-based indicator (SI) formula}.  By exhaustion, it has been demonstrated~\cite{FuKaneMele,FuKaneInversion,QHZ,HsiehTCI,TeoFuKaneTCI,HourglassInsulator,DiracInsulator,AshvinIndicators,LiangTCIOriginal,ChenBernevigTCI,ChenRotation,HOTIChen,ChenTCI,AshvinTCI,HOTIBismuth,TMDHOTI,WladTheory,HOTIBernevig} that $\mathcal{T}$-symmetric, symmetry-indicated, stable 3D TIs and TCIs necessarily exhibit anomalous 2D surface and 1D hinge states crossing the bulk gap, where the surface and hinge states are respectively protected by the symmetries of Type-II surface wallpaper groups and hinge frieze or line groups~\cite{WiederLayers,ConwaySymmetries,BalkanLineGroups2,BalkanLineGroups3}.  The quintessential SI formula in 3D is the Fu-Kane parity ($\mathcal{I}$) criterion for diagnosing 3D $\mathcal{T}$-symmetric TIs~\cite{FuKaneInversion}.  More recently, it was shown in Refs.~\onlinecite{SlagerSymmetry,AshvinIndicators,HOTIChen,ChenTCI,AshvinTCI,TMDHOTI,BarryBandrepReview,ZhidaSemimetals,AdrianSIReview,AdrianSCMagSI} that the compatibility relations and EBRs in an SSG can be used to generate a set of linearly independent SI formulas for stable topological bands that respect the symmetries of the SSG.  The procedure introduced in Refs.~\onlinecite{SlagerSymmetry,AshvinIndicators,HOTIChen} returns the SI group (\emph{e.g.} $\mathbb{Z}_{4}\times\mathbb{Z}_{2}^{3}$) as well as the SI formula for the SSG in an arbitrary basis.  Previously, in Ref.~\onlinecite{AshvinMagnetic}, the authors derived the SI groups of all 1,651 single and double SSGs, but not the SI formulas or the physical interpretation (\emph{i.e.} the bulk topology and anomalous boundary states) of the magnetic bands with nontrivial SIs.

In the sections below, restricting consideration to the double-valued (co)reps of the 1,651 double SSGs, which characterize spinful electronic states in solid-state materials~\cite{BigBook}, we will go beyond the analysis in Ref.~\onlinecite{AshvinMagnetic} and generate the SI formulas in a consistent and physically-motivated basis.  In the \emph{physical} SI formula basis introduced in this work, all previously identified nonmagnetic double SI formulas correspond to established nonmagnetic semimetallic, TI, and TCI phases.  Additionally, in the physical SI formula basis, the SIs of symmetry-indicated TIs and TCIs with the same bulk topology (\emph{e.g.} 3D TIs and AXIs with the common nontrivial axion angle $\theta=\pi$) are related by intuitive SI subduction relations.  We will also introduce layer constructions~\cite{ChenTCI,HermelePointGroup,ZhidaHermeleCrystal} in the minimal double SSGs (defined in Appendix~\ref{sec:34minimal}) for each TCI phase that admits a decomposition into layered 2D Chern insulators, TIs and mirror TCIs, which we will then use to determine symmetry-respecting bulk and anomalous surface and hinge states for all topological bands in the minimal double SSGs.  First, in Appendix~\ref{sec:reviewTopologicalBands}, we will review the method employed in Refs.~\onlinecite{SlagerSymmetry,ChenTCI,AshvinTCI,HOTIBismuth,TMDHOTI} in which the multiplicities of small (co)reps are used to determine the symmetry-indicated topology of energetically isolated bands.  Next, in Appendix~\ref{sec:smithForm}, we will introduce the Smith normal form~\cite{SmithForm} decomposition of the EBR matrix of an SSG $G$, through which one can infer the SIs in $G$.  Then, in Appendix~\ref{sec:minimalSIProcedure}, we will detail a procedure for obtaining a set of minimal SIs on which the SIs in all 1,651 double SSGs are dependent.  In the following section -- Appendix~\ref{sec:34minimal} -- we will then compute the minimal SI formulas for spinful topological phases in the 34 minimal double SSGs containing the minimal SIs in the self-consistent, physical basis described above.  In Appendix~\ref{sec:34minimal}, we will also formulate layer constructions -- where possible -- for the symmetry-indicated TI and TCI phases in the minimal double SSGs.  We have confirmed that the spinful SI groups obtained in this work agree with the previous tabulation of magnetic and nonmagnetic SI groups in the 1,651 double SSGs performed in Ref.~\onlinecite{AshvinMagnetic}.  The results of the calculations that we will perform in Appendix~\ref{sec:34minimal} will be summarized in Appendix~\ref{sec:summaryDoubleSIs}.  Lastly, in Appendix~\ref{sec:newHOTIs}, we will further detail the helical (\emph{i.e.} non-axionic) magnetic higher-order TCI (HOTI) phases~\cite{HOTIBernevig,HOTIBismuth,ChenRotation,HOTIChen,HigherOrderTIPiet,AshvinIndicators,ChenTCI,AshvinTCI,TMDHOTI,WiederAxion,DiracInsulator} discovered in this work through the SI calculations performed in Appendix~\ref{sec:34minimal}.  For the spinful helical magnetic HOTI phases discovered in this work, we will specifically detail symmetry-enhanced fermion doubling theorems~\cite{DiracInsulator,ChenRotation,SteveMagnet,Steve2D} in Appendix~\ref{sec:HOTIfermionDoubling}, and will provide tight-binding models in Appendix~\ref{sec:HOTItbModel}.

\subsection{Diagnosing Band Topology from Symmetry Eigenvalues}
\label{sec:reviewTopologicalBands}

In this section, we will review the procedure by which a symmetry data vector $B$ [see Refs.~\onlinecite{AndreiMaterials,MTQCmaterials} and the text following Eq.~(\ref{eq:reducibleCompatibility})] derived from the band structure of a material or model can be evaluated for nontrivial topology.  The discussion in this section is largely a review of previous works on stable~\cite{ChenTCI,AshvinTCI,AshvinIndicators,AndreiMaterials,AshvinMaterials,ChenMaterials,MTQCmaterials,BarryBandrepReview,ZhidaSemimetals,AdrianSIReview,AdrianSCMagSI} and fragile~\cite{JenFragile1,BarryFragile,AshvinFragile,AshvinFragile2,AdrianFragile,KoreanFragile,KoreanFragileInversion,ArisFragileNoGo,DelicateAris,SlagerMagFragile,SlagerMagFragile2,ZhidaFragile,FragileFlowMeta,ZhidaFragile2} topology.  To begin, in a given SSG $G$, if a set of bands is energetically isolated from all of the other bands in the spectrum at all high-symmetry ${\bf k}$ points and along all high-symmetry BZ lines and planes, then we may extract the symmetry data $B_{\bf k}$ at each point ${\bf k}$.  As discussed in Appendix~\ref{sec:compatibilityRelations}, the symmetry data $B_{\bf k}$ is composed of the multiplicities of the irreducible small (co)reps of the little group $G_{\bf k}$ that correspond to the set of energetically isolated Bloch eigenstates at ${\bf k}$ [see the text following Eq.~(\ref{eq:finiteSubductionDimension})].  Given symmetry data $B_{\bf k}$ at a point ${\bf k}$, the symmetry data $B_{{\bf k}'}$ at a point ${\bf k}'$ that is connected to ${\bf k}$ [defined in the text following Eq.~(\ref{eq:coGroupEq3})] is fully determined by $B_{\bf k}$ through the compatibility relations $m^{{\bf k},{\bf k}'}$ [Eq.~(\ref{eq:fullCompatibilityMatrix})] if the bands that transform in the symmetry data vector $B$ are energetically isolated at all high-symmetry ${\bf k}$ points and along all high-symmetry BZ lines and planes.  Hence, we may summarize the complete set of $B_{\bf k}$ in $B$ with the symmetry data at a smaller number of ${\bf k}$ vectors consisting of one point ${\bf k}$ within each of the maximal momentum stars in $G$ [defined in the text surrounding Eq.~(\ref{eq:maximalKvec})]:
\begin{equation}
B = (m(\tilde{\sigma}_{1,\kk_1}), 
     m(\tilde{\sigma}_{2,\kk_1}),\cdots,
     m(\tilde{\sigma}_{1,\kk_2}),
     m(\tilde{\sigma}_{2,\kk_2}),\cdots)^T,
\end{equation}
where $m(\tilde{\sigma}_{l,\kk_n})$ denotes the multiplicity of the $l^\text{th}$ small (co)rep of $G_{{\bf k}_{n}}$, and where $B$ contains $N_{B}$ entries.  The multiplicities $m(\tilde{\sigma}_{l,{\bf k}_{n}})$ in $B$ must obey a set of linear constraints imposed by the compatibility relations $\CR$, such that:
\begin{equation}
\CR \cdot B = 0,
\label{eq:compRelForB}
\end{equation}
in which each row of $\CR$ provides a linear constraint on $B$, and where the entries in $\CR$ are given by $m^{{\bf k},{\bf k}'}(m^{{\bf k}'',{\bf k}'})^{-1}$ taken over all pairs ${\bf k}$ and ${\bf k}''$ of maximal ${\bf k}$ vectors in $G$ and all symmetry-unrelated ${\bf k}$ vectors ${\bf k}'$ that are mutually connected [defined in the text following Eq.~(\ref{eq:coGroupEq3})] to ${\bf k}$ and ${\bf k}''$.  We emphasize that $(m^{{\bf k}'',{\bf k}'})^{-1}$, like $(c^{{\bf k}'})^{-1}$ in Eq.~(\ref{eq:matrixInverseCompatibility}), is guaranteed to exist (though not necessarily be unique) through Frobenius reciprocity~\cite{SerreLinearReps,Bandrep1}, because the elements of $m^{{\bf k}'',{\bf k}'}$ are defined through subduction in Eq.~(\ref{eq:fullCompatibility}) [see the text surrounding Eqs.~(\ref{eq:fullCompatibilityMatrix}) and~(\ref{eq:nearFinalCompatibility})].

In particular, the symmetry data of an EBR contain the multiplicities of small coreps that are induced from site-symmetry coreps in position space [see the text surrounding Eq.~(\ref{eq:mainInductionMTQC})].  For each SSG $G$, we may define an \emph{EBR matrix}:
\begin{equation}
\EBR = ( B^{\tilde{\rho}_{1,\qq_1}}, 
         B^{\tilde{\rho}_{2,\qq_1}}, \cdots
         B^{\tilde{\rho}_{1,\qq_2}},
         B^{\tilde{\rho}_{2,\qq_2}}, \cdots ), 
\label{eq:EBR-def}
\end{equation}
in which each column $B^{\tilde{\rho}_{j,\qq_i}}$ contains the symmetry data vector of the EBR of $G$ induced from the $j^{\text{th}}$ (co)rep $\tilde{\rho}_{j,{\bf q}_i}$ of the site-symmetry group $G_{{\bf q}_{i}}$ in the maximal Wyckoff position indexed by ${\bf q}_{i}$ (see Appendix~\ref{sec:Wyckoff}).  In the SSG $G$, we define the number of EBRs as $N_{\EBR}$, such that $\EBR$ in Eq.~(\ref{eq:EBR-def}) is an $N_{B}\times N_{\EBR}$-dimensional matrix.  By definition, an EBR must correspond to a set of Bloch states that are energetically isolated at all high-symmetry ${\bf k}$ points and along all high-symmetry BZ lines and planes, such that each $B^{\tilde{\rho}_{j,\qq_i}}$ in Eq.~(\ref{eq:EBR-def}) satisfies the compatibility relations:
\begin{equation}
\CR\cdot\EBR = 0.
\label{eq:CREBRKernel}
\end{equation}
We find that, in each of the 1,651 single and double SSGs, the rank of $\EBR$ is always equal to the dimension of the kernel of $\CR$ over the rational numbers, implying that the columns of $\EBR$ are at least a complete -- and are in general an overcomplete -- basis set of the kernel of $\CR$.

Given a set of bands that is energetically isolated at all high-symmetry ${\bf k}$ points and along all high-symmetry BZ lines and planes, the symmetry data vector $B$ of the bands can be expressed in terms of $\EBR$:
\begin{equation}
[B]_a = \sum_b [\EBR]_{ab} p_b(B) = [\EBR\cdot p(B)]_a, 
\label{eq:B-EBR}
\end{equation}
in which $p(B)$ is a vector of EBR multiplicities:
\begin{equation}
p(B) =  ( p({\tilde{\rho}_{1,\qq_1}}), 
    p({\tilde{\rho}_{2,\qq_1}}), \cdots
    p({\tilde{\rho}_{1,\qq_2}}),
    p({\tilde{\rho}_{2,\qq_2}}), \cdots )^{T}, 
\label{eq:pVectorForTopology}
\end{equation}
where $p(\tilde{\rho}_{j,\qq_i})$ indicates the multiplicity of the EBR symmetry data vector $B^{\tilde{\rho}_{j,\qq_i}}$ in $B$ [see the text folowing Eq.~(\ref{eq:EBR-def})], and where each $p(\tilde{\rho}_{j,\qq_i})$ is rational, but not necessarily integer-valued.  For all possible symmetry data vectors $B$ that satisfy the compatibility relations, a decomposition of the form of Eqs.~(\ref{eq:B-EBR}) and~(\ref{eq:pVectorForTopology}) is always permitted, because the symmetry data of the EBRs spans the set of symmetry data vectors that satisfy the compatibility relations in each SSG [\emph{i.e.} because $\EBR$ spans the kernel of $\CR$, see Eq.~(\ref{eq:CREBRKernel}) and the surrounding text]~\cite{ChenTCI,AshvinIndicators,AndreiMaterials,AshvinMaterials,ChenMaterials,MTQCmaterials,Bandrep2,Bandrep3}.  When $\mrm{rank}(\EBR)= N_{\EBR}$, the multiplicities $p(\tilde{\rho}_{j,\qq_i})$ in Eq.~(\ref{eq:pVectorForTopology}) are unique; however, when $\mrm{rank}(\EBR)< N_{\EBR}$, then $p(B)$ is not unique.

As discussed in several previous works~\cite{ChenTCI,AshvinIndicators,AndreiMaterials,AshvinMaterials,ChenMaterials,MTQCmaterials,Bandrep2,Bandrep3,SlagerSymmetry,HOTIChen,JenFragile1,BarryFragile,AshvinFragile,AshvinFragile2,KoreanFragile,KoreanFragileInversion,ArisFragileNoGo,DelicateAris,SlagerMagFragile,SlagerMagFragile2,ZhidaFragile,FragileFlowMeta,ZhidaFragile2,BarryBandrepReview,ZhidaSemimetals,AdrianSIReview,AdrianSCMagSI}, the values of $p(\tilde{\rho}_{j,\qq_i})$ can be used to infer the topology of the bands that transform in $B$.  Specifically, given a symmetry data vector $B$ that satisfies the compatibility relations, there are three possibilities for the components of $p(B)$ in Eq.~(\ref{eq:pVectorForTopology}):
\begin{enumerate}
\item{In each of the possible $p(B)$-vector solutions to Eq.~(\ref{eq:B-EBR}), at least one of the multiplicities $p(\tilde{\rho}_{j,\qq_i})$ is not an integer (but is still rational)~\cite{AshvinIndicators}.}
\item{There exists at least one solution to Eq.~(\ref{eq:B-EBR}) in which all of the multiplicities $p(\tilde{\rho}_{j,\qq_i})\in\mathbb{Z}$, though there do not exist solutions in which all of the multiplicities $p(\tilde{\rho}_{j,\qq_i})\in\{\mathbb{Z}^{+},0\}$; therefore, at least one $p(\tilde{\rho}_{j,\qq_i})$ is negative in the solution in which $p(\tilde{\rho}_{j,\qq_i})\in\mathbb{Z}$ for all $i$ and $j$.}
\item{There exists at least one solution to Eq.~(\ref{eq:B-EBR}) in which all of the multiplicities $p(\tilde{\rho}_{j,\qq_i})\in\{\mathbb{Z}^{+},0\}$.}
\end{enumerate}
In case 3, $B$ contains the same small (co)reps as a direct sum of EBRs, such that the bands that transform in $B$ exhibit the same symmetry eigenvalues as a trivial insulator.  We note that this does not exclude the possibility that the bands that transform in $B$ exhibit non-symmetry-indicated topology~\cite{BarryFragile,HingeSM,WiederAxion,TMDHOTI,JenOAL,WiederDefect}.  In case 2, it is possible to add EBRs to the bands that transform in $B$ until the direct sum of the bands that transform in $B$ and the added EBRs realizes a set of bands with a symmetry data vector $B'$ classified by case 3.  Therefore, as shown in Refs.~\onlinecite{JenFragile1,BarryFragile,AshvinFragile,AshvinFragile2,AdrianFragile,KoreanFragile,KoreanFragileInversion,ArisFragileNoGo,DelicateAris,SlagerMagFragile,SlagerMagFragile2,ZhidaFragile,FragileFlowMeta,ZhidaFragile2},  in case 2, the bands that transform in $B$ exhibit symmetry-indicated \emph{fragile} topology.  In the nomenclature of Refs.~\onlinecite{AndreiMaterials,MTQCmaterials,AndreiMaterials2}, the symmetry data vectors in cases 2 and 3 correspond to ``linear combinations of EBRs'' [LCEBR].  Finally, in case 1, there does not exist an integer-valued linear combination of EBRs that can be added to the bands that transform in $B$ to produce a set of bands with integer-valued $p(\tilde{\rho}_{j,\qq_i})$.  Consequently, as shown in Refs.~\onlinecite{ChenTCI,AshvinIndicators,AshvinTCI,AndreiMaterials,AshvinMaterials,ChenMaterials,MTQCmaterials,AndreiMaterials2}, the bands that transform in $B$ in case 1 are not Wannierizable, and either correspond to a topological semimetal that satisfies the compatibility relations~\cite{ZhidaSemimetals}, or to a symmetry-indicated \emph{stable} TI or TCI with anomalous surface or hinge states.

\subsection{Symmetry-Based Indicator (SI) Groups and Formulas from the Smith Normal Form}
\label{sec:smithForm}

In this section, we will introduce the method employed in this work to calculate the SI groups and formulas for spinful stable topological phases in all 1,651 double SSGs.  In Appendix~\ref{sec:SIexP2}, we will then as an example provide an explicit calculation of the SI groups and formulas for double-valued irreps in Type-I double MSG 3.1 $P2$.  Variants of the method described in this section were previously introduced in Refs.~\onlinecite{SlagerSymmetry,AshvinIndicators,HOTIChen,ChenTCI,AshvinTCI,TMDHOTI,BarryBandrepReview,ZhidaSemimetals,AdrianSIReview,AdrianSCMagSI}.  We will leave the enumeration of the symmetry-indicated fragile bands in the 1,651 single and double SSGs for future works.  To begin, if the entries of a matrix are integer-valued, then the matrix carries a unique Smith normal form~\cite{SmithForm}.  Consequently, given an SSG $G$, the EBR matrix $\EBR$ [defined in Eq.~(\ref{eq:EBR-def})] -- whose entries are the integer-valued multiplicities of induced small (co)reps  -- can be decomposed into the Smith normal form:
\begin{equation}
\EBR = L_\EBR \Lambda_\EBR R_\EBR, 
\label{eq:EBR-snf}
\end{equation}
where $L_{\EBR}$ is an $N_{B}\times N_{B}$-dimensional unimodular matrix with integer-valued entries, $R_{\EBR}$ is an $N_\EBR \times N_\EBR$-dimensional unimodular matrix, and $\Lambda_\EBR$ is an $N_{B}\times N_\EBR$-dimensional (\emph{i.e.} generically non-square) matrix with integer-valued entries $[\Lambda_\EBR]_{ij}$ for which:
\begin{equation}
[\Lambda_\EBR]_{ij}=0\text{ if }i\neq j.
\end{equation}
Consequently, $\Lambda_\EBR$ -- which is \emph{the Smith normal form} of $\EBR$ -- generically appears as:
\begin{equation}
\Lambda_\EBR = \begin{pmatrix}
\lambda_1 & \cdots & 0     & 0 & \cdots & 0\\
\vdots & \ddots  & \vdots & \vdots & \ddots & \vdots\\
0   & \cdots & \lambda_{r} & 0 & \cdots & 0\\
0 & \cdots & 0 & 0 & \cdots & 0\\
\vdots & \ddots & \vdots & \vdots & \ddots & \vdots\\
0 & \cdots & 0 & 0 & \cdots & 0
\end{pmatrix},
\label{eq:lambdaForSI}
\end{equation}
in which $1\le \lambda_1 \le \lambda_2 \le \cdots \lambda_r$ are positive integers and $r=\mrm{rank}(\EBR)$.  We note that, in contrast to $\Lambda_\EBR$, $L_\EBR$ and $R_\EBR$ in Eq.~(\ref{eq:EBR-snf}) are not unique.  For example, given $\EBR$ and $\Lambda_\EBR$, for any choice of $L_\EBR$ and $R_\EBR$, $L'_{\EBR} = -L_{\EBR}$ and $R'_{\EBR} = -R_{\EBR}$ always also satisfy the decomposition in Eq.~(\ref{eq:EBR-snf}).

Next, we consider all possible bands that transform in the most general symmetry data vector $B$ in $G$ that satisfies the compatibility relations [Eq.~(\ref{eq:compRelForB})]:
\begin{equation}
B = (m(\tilde{\sigma}_{1,\kk_1}), 
     m(\tilde{\sigma}_{2,\kk_1}),\cdots,
     m(\tilde{\sigma}_{1,\kk_2}),
     m(\tilde{\sigma}_{2,\kk_2}),\cdots)^T,
\label{eq:mostGeneralBforAllOfG}
\end{equation}
where $m(\tilde{\sigma}_{l,\kk_n})$ denotes the multiplicity of the $l^\text{th}$ small (co)rep of the little group $G_{{\bf k}_{n}}$.  Previously, in the text following Eq.~(\ref{eq:pVectorForTopology}), we described a procedure for diagnosing whether bands that satisfy the compatibility relations exhibit symmetry-indicated stable topology.  In the following text, we will now additionally describe a method for \emph{classifying} stable band topology, which we will accomplish by parameterizing the space of solutions to Eq.~(\ref{eq:B-EBR}). First, we act on both sides of Eq.~(\ref{eq:B-EBR}) with the left inverse $L_{\EBR}^{-1}$, which is guaranteed to exist, because $L_{\EBR}$ is an integer, unimodular matrix [see the text following Eq.~(\ref{eq:EBR-snf})]:
\begin{equation}
L_\EBR^{-1} B = \Lambda_\EBR R_\EBR\cdot p(B).
\label{eq:B-p}
\end{equation}
Because only the first $r$ rows of $\Lambda_\EBR$ are nonzero, then, in order for a solution $p(B)$ to exist in Eq.~(\ref{eq:B-p}), the $(r+1)^\text{th}$ to the $N_\EBR^\text{th}$ rows of $L_\EBR^{-1} B$ must be zero.  However, the $(r+1)^\text{th}$ to the $N_\EBR^\text{th}$ rows of $L_\EBR^{-1} B$ are guaranteed to be zero, because $\EBR$ spans the kernel of $\CR$ [defined in Eq.~(\ref{eq:compRelForB})], and because $B$ satisfies the compatibility relations.  Hence, we obtain a solution for $p(B)$ in Eq.~(\ref{eq:B-p}).

For each nonzero $\lambda_{i}$ in Eq.~(\ref{eq:lambdaForSI}), we next construct an $r$-dimensional vector $y(B)$ by multiplying $B$ by $L_\EBR^{-1}$ and the pseudoinverse of $\Lambda_\EBR$ [Eq.~(\ref{eq:lambdaForSI})]:
\begin{equation}
[y]_i(B) = \frac{1}{\lambda_i}[L_\EBR^{-1} \cdot B]_i = [R_\EBR \cdot p(B)]_i,\ i=1\cdots r, 
\label{eq:y-def}
\end{equation}
in which the entries $[y]_{i}(B)$ are rational numbers.  We then re-express $B$ in terms of $y(B)$ using Eq.~(\ref{eq:y-def}):
\begin{equation}
[B]_j = \sum_{i=1}^{r} [L_\EBR]_{ji} [y]_i(B) \lambda_i.
\label{eq:B-y}
\end{equation}
Because $L_{\EBR}$ is unimodular, then the correspondance between the components of $B$ and $y(B)$ is one-to-one.  Conversely, the correspondence between $y(B)$ and $p(B)$ is generically one-to-many.  Specifically, given $y(B)$, the most general solution for $p(B)$ takes the form:
\begin{equation}
p(B) = R_\EBR^{-1} \cdot ( y_1(B), y_2(B), \cdots, y_r(B),
k_1,k_2,\cdots, k_{N_\EBR -r } )^T, 
\label{eq:p-y-k}
\end{equation}
in which $k_{i}$ are rational-valued free parameters.

To diagnose the stable topology of bands whose symmetry data satisfy the compatibility relations in $G$, we therefore restrict focus to the first $r$ components of $p(B)$.  Because $R_\EBR$ is a unimodular matrix, then the components of $p(B)$ are integer-valued if and only if $y_{i}(B)$ and $k_{i}$ are integer-valued for all $i$, which reduces to the requirement that the values of $y_{i}(B)$ are integer-valued, because the values of $k_{i}$ are free parameters in Eq.~(\ref{eq:p-y-k}).  Finally, using the values of $y_{i}(B)$, we define:
\begin{equation}
z_i(B) = (y_i (B)\lambda_i)\text{ mod }\lambda_i =[L_{\EBR}^{-1} \cdot B]_i\text{ mod }\lambda_i,\  i=i_0\cdots r,
\label{eq:stable-index}
\end{equation} 
in which we have defined $i_{0}$ to be the smallest value of $i$ for which $\lambda_{i_0}>1$, and where each $y_{i}(B)$ is integer-valued if and only if $z_{i}(B)=0$.  When $B$ is expressed in terms of the most general small (co)rep multiplicities that satisfy the compatibility relations [\emph{i.e.} in the form of Eq.~(\ref{eq:mostGeneralBforAllOfG})], then the $z_{i}(B)$ -- which are implicitly functions of the small (co)rep multiplicities $m(\tilde{\sigma}_{l,\kk_n})$ -- are known as the \emph{SI formulas} of $G$~\cite{SlagerSymmetry,AshvinIndicators}.  Correspondingly, the \emph{representative $B_{i}$ vector} for each $i$ is defined as the $i^\text{th}$ column of $L_\EBR$ for which $z_{i}(B_i)=(L^{-1}_\EBR\cdot L_\EBR)_{ii}\text{ mod }\lambda_i=1$.

Next, given a specific symmetry data vector $B'$ with fixed values of $m(\tilde{\sigma}_{l,\kk_n})$ that satisfy the compatibility relations, we may calculate the values $z_{i}(B')$, which necessarily satisfy $\{z_{i}(B')\in\mathbb{Z}|0\leq z_{i}(B')\leq \lambda_{i}-1\}$.  Hence, given $B'$, the appearance of nonzero $z_{i}(B')$ in Eq.~(\ref{eq:stable-index}) implies that the components of $y(B')$ and $p(B')$ are not integer-valued, and that the bands that transform in $B'$ exhibit stable topology.  From this, we define the \emph{SI vector of $B'$} as:  
\begin{equation}
\mbf{z}^{G}(B')= (z_{i_0}(B'), z_{i_0+1}(B'),\cdots, z_r(B'))^T,
\label{eq:SIvector}
\end{equation}
where $z_{i}(B')\in\mathbb{Z}_{\lambda_{i}}$.  Notably, the SI vectors of the representative $B_{i}$ vectors satisfy $\mathbf{z}^G_j(B_i) = (L_\EBR^{-1}L_\EBR)_{ji}\text{ mod }\lambda_j = \delta_{ji}\text{ mod }\lambda_j $.  
Lastly, using the values of $\lambda_{i}$ obtained from Eqs.~(\ref{eq:lambdaForSI}),~(\ref{eq:p-y-k}), and~(\ref{eq:stable-index}), we define the \emph{SI group} of $G$:
\begin{equation}
Z^{G} = \bigotimes_{i=i_0}^r \mathbb{Z}_{\lambda_i}.
\label{eq:SIgroup}
\end{equation}
Consequently, in $G$, the bands that transform in the representative $B_{i}$ vectors may be summed with each other and with the EBRs of $G$ to generate $|Z^{G}|-1$ classes of stable topological bands that are not related by linear combinations of EBRs, as well as one class of (generically trivial) bands whose symmetry data vectors $\tilde{B}$ map to the trivial (identity) element of the SI group [${\bf z}^{G}(\tilde{B})={\bf 0}$ in Eqs.~(\ref{eq:stable-index}) and~(\ref{eq:SIvector})].  Specifically, the SI group is spanned by summing the representative topological bands (\emph{e.g.} $2B_{i} = B_{i}\oplus B_{i}$), such that $\mathbf{z}^G_i(nB_i) = n \text{ mod }\lambda_i$ where $n\in\mathbb{Z}^{+}$.  One stable topological band from each of the $|Z^{G}|-1$ classes of stable topological bands and one integer-valued linear combination of EBRs that transforms in one $\tilde{B}$ vector together form a nonunique set of $|Z^{G}|$ bands that we designate in this work as the \emph{SI topological bands}.

Using the method described in this section, we have obtained the SI formulas and groups for the double-valued (co)reps of all 1,651 double SSGs, which we term the \emph{double SIs}.  We have confirmed that the SI groups obtained in our calculations agree with the previous tabulation performed in Ref.~\onlinecite{AshvinMagnetic}.  However, in general, both the SI formulas and the representative $B_{i}$ vectors are computed in an arbitrary basis that is generically not the natural (physical) basis for classifying topological phases.  Specifically, additional bulk- and boundary-state~\cite{AshvinTCI,HOTIChen} or layer-construction~\cite{ChenTCI} calculations must be performed to determine the semimetallic, TI, or TCI phases that correspond to each possible value of $z_{i}(B)$.  Later, in Appendices~\ref{sec:minimalSIProcedure},~\ref{sec:minimalSSGTables}, and~\ref{sec:34minimal}, we will determine a self-consistent, physically motivated basis and the corresponding bulk topology for the double SIs in all 1,651 double SSGs.

\subsubsection{Double SI Group and Formulas in Type-I Double MSG 3.1 $P2$}
\label{sec:SIexP2}

As an example of the Smith normal form calculation described in Refs.~\onlinecite{SlagerSymmetry,AshvinIndicators,HOTIChen,ChenTCI,AshvinTCI,TMDHOTI,BarryBandrepReview,ZhidaSemimetals,AdrianSIReview,AdrianSCMagSI} and in the text following Eq.~(\ref{eq:EBR-snf}), we will in this section calculate the double SI group and formulas of Type-I double MSG 3.1 $P2$.

\begin{figure}[h]
\centering
\includegraphics[width=0.6\linewidth]{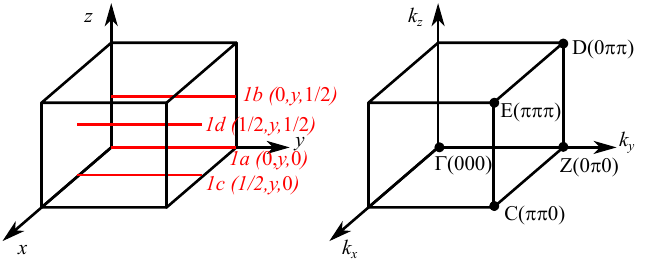}
\caption{The unit cell and BZ of Type-I MSG 3.1 $P2$.  (Left panel) The unit cell of MSG 3.1 $P2$ with the maximal Wyckoff positions [Eq.~(\ref{eq:maximalWyckoffSmithForm})] labeled with red lines.  (Right panel) The BZ of MSG 3.1 $P2$ with the maximal ${\bf k}$ vectors in Eq.~(\ref{eq:independentMomentaEx}), as well as the $\Gamma$ point [${\bf k}_{\Gamma}=(0,0,0)$], labeled with black circles.}
\label{fig:P2BZ}
\end{figure}

First, using~\href{https://www.cryst.ehu.es/cgi-bin/cryst/programs/magget_gen.pl}{MGENPOS} on the BCS~\cite{BCSMag1,BCSMag2,BCSMag3,BCSMag4}, we determine that MSG 3.1 $P2$ is generated by:
\begin{equation}
\{C_{2y}|000\},\ \{E|100\},\ \{E|010\},\ \{E|001\}.
\end{equation}
Next, using the~\href{http://www.cryst.ehu.es/cryst/mkvec}{MKVEC} tool (see Appendix~\ref{sec:MKVEC}), we determine that there are four maximal momentum stars in $M=P2$ [defined in the text surrounding Eq.~(\ref{eq:maximalKvec})].  Using~\href{https://www.cryst.ehu.es/cryst/mcomprel}{MCOMPREL} (see Appendix~\ref{sec:compatibilityRelations}), we then find that, due to the compatibility relations, the small irrep multiplicities throughout the BZ of $M$ are entirely determined by the irrep multiplicities at only one of the high-symmetry points in each of the four maximal momentum stars (Fig.~\ref{fig:P2BZ}):
\begin{equation}
{\bf k}_{Z} = 2\pi(0,1/2,0),\ {\bf k}_{E} = 2\pi(1/2,1/2,1/2),\ {\bf k}_{D}=2\pi(0,1/2,1/2),\ {\bf k}_{C} = 2\pi(1/2,1/2,0).
\label{eq:independentMomentaEx}
\end{equation} 
At each of the four ${\bf k}$ points in Eq.~(\ref{eq:independentMomentaEx}), there are only two double-valued small irreps $\bar{\sigma}_{\bf k}^{\pm \frac{1}{2}}$ for which:
\begin{equation}
\chi_{\bar{\sigma}^{\mp \frac{1}{2}}_{\bf k}}(\{C_{2y}|\mathbf{0}\}) = \pm i.
\end{equation}
In the notation of the~\href{http://www.cryst.ehu.es/cryst/corepresentations}{Corepresentations} tool on the BCS (Appendix~\ref{sec:coreps}):
\begin{eqnarray}
\bar{\sigma}_{\bf k}^{\frac{1}{2}} &=& \ovl{Z}_{3},\ \ovl{E}_{3},\ \ovl{D}_{3},\text{ and }\ovl{C}_{3}\text{ for }{\bf k} = {\bf k}_{Z},\ {\bf k}_{E},\ {\bf k}_{D},\text{ and }{\bf k}_{C},\text{ respectively}, \nonumber \\ 
\bar{\sigma}_{\bf k}^{-\frac{1}{2}} &=& \ovl{Z}_{4},\ \ovl{E}_{4},\ \ovl{D}_{4},\text{ and }\ovl{C}_{4}\text{ for }{\bf k} = {\bf k}_{Z},\ {\bf k}_{E},\ {\bf k}_{D},\text{ and }{\bf k}_{C},\text{ respectively},
\label{eq:BCStoZhidaNotation}
\end{eqnarray}
such that the most general symmetry data vector $B$ that satisfies the compatibility relations of $M$ is given by:
\begin{equation}
B = ( m(\ovl{Z}_3), m(\ovl{Z}_4),
    m(\ovl{E}_3), m(\ovl{E}_4),
    m(\ovl{D}_3), m(\ovl{D}_4),
    m(\ovl{C}_3), m(\ovl{C}_4)
     )^{T}.
\label{eq:B-P2}
\end{equation}

To calculate the Smith normal form of $M$ described in the text surrounding Eq.~(\ref{eq:EBR-snf}), we next determine the symmetry data vectors of the EBRs of $M$.  Using~\href{http://www.cryst.ehu.es/cgi-bin/cryst/programs/magget_wp.pl}{MWYCKPOS} on the BCS~\cite{BCSMag1,BCSMag2,BCSMag3,BCSMag4}, we find that $M$ has four, multiplicity-1 maximal Wyckoff positions (defined in Appendix~\ref{sec:Wyckoff}), which are indexed by the sites (Fig.~\ref{fig:P2BZ}):
\begin{equation}
{\bf q}_{1a} = (0,y,0),\ {\bf q}_{1b} = (0,y,1/2),\ {\bf q}_{1c} = (1/2,y,0),\ {\bf q}_{1d} = (1/2,y,1/2),
\label{eq:maximalWyckoffSmithForm}
\end{equation}
where $y\in[-1/2,1/2)$, such that each of the sites in Eq.~(\ref{eq:maximalWyckoffSmithForm}) lies along a line of $\{C_{2y}|{\bf 0}\}$ symmetry (modulo integer lattice translations).  At each of the four sites in Eq.~(\ref{eq:maximalWyckoffSmithForm}), the site-symmetry group $G_{\bf q}$ is isomorphic to Type-I double MPG 3.1.6 $2$, which is generated by $C_{2y}$.  Using the~\href{http://www.cryst.ehu.es/cryst/corepresentationsPG}{CorepresentationsPG} tool (Appendix~\ref{sec:magWannier}), we determine that each site-symmetry group $G_{\bf q}$ in $M$ has two double valued irreps $({}^1\ovl{E})_{\bf q}$ and $({}^2\ovl{E})_{\bf q}$, where:
\begin{equation}
\chi_{({}^1\ovl{E})_{\bf q}}(C_{2y}) = i,\ \chi_{({}^2\ovl{E})_{\bf q}}(C_{2y}) = -i.
\end{equation}

To obtain the symmetry data vectors of the EBRs of $M$, we use the~\href{http://www.cryst.ehu.es/cryst/mbandrep}{MBANDREP} tool introduced in this work [Appendix~\ref{sec:mbandrep}, see also Eqs.~(\ref{eq:mainInductionMTQC}) and~(\ref{eq:finalCharacters2})], the output of which is reproduced below in the condensed notation of Refs.~\onlinecite{ZhidaBLG,WiederAxion}:
\begin{equation}
({}^2\ovl{E})_{1a} \up M= 
\ovl{Z}_3 \oplus \ovl{E}_3 \oplus \ovl{D}_3 \oplus \ovl{C}_3,\qquad
({}^1\ovl{E})_{1a} \up M=
\ovl{Z}_4 \oplus \ovl{E}_4 \oplus \ovl{D}_4\oplus \ovl{C}_4, \nonumber
\end{equation}
\begin{equation}
({}^2\ovl{E})_{1b} \up M= 
\ovl{Z}_3\oplus \ovl{E}_4\oplus \ovl{D}_4\oplus \ovl{C}_3,\qquad
({}^1\ovl{E})_{1b} \up M=
\ovl{Z}_4\oplus \ovl{E}_3\oplus \ovl{D}_3\oplus \ovl{C}_4, \nonumber 
\end{equation}
\begin{equation}
({}^2\ovl{E})_{1c} \up M=
\ovl{Z}_3\oplus \ovl{E}_4\oplus \ovl{D}_3\oplus \ovl{C}_4,\qquad
({}^1\ovl{E})_{1c} \up M=
\ovl{Z}_4\oplus \ovl{E}_3\oplus \ovl{D}_4\oplus \ovl{C}_3, \nonumber
\end{equation}
\begin{equation}
({}^2\ovl{E})_{1d} \up M=
\ovl{Z}_3\oplus \ovl{E}_3\oplus \ovl{D}_4\oplus \ovl{C}_4,\qquad
({}^1\ovl{E})_{1d} \up M=
\ovl{Z}_4\oplus \ovl{E}_4\oplus \ovl{D}_3\oplus \ovl{C}_3.
\label{eq:exampleOfEBRSymmetryData}
\end{equation}

Using Eq.~(\ref{eq:exampleOfEBRSymmetryData}), we next construct the $\EBR$ matrix [Eq.~(\ref{eq:EBR-def})]:
\begin{eqnarray}
\EBR &=&(B^{({}^2\ovl{E})_{1a}},B^{({}^1\ovl{E})_{1a}},B^{({}^2\ovl{E})_{1b}},B^{({}^1\ovl{E})_{1b}},B^{({}^2\ovl{E})_{1c}},B^{({}^1\ovl{E})_{1c}},B^{({}^2\ovl{E})_{1d}},B^{({}^1\ovl{E})_{1d}}), \nonumber \\ \nonumber \\
&\ & \ \ \ \ \ \ \ \ \ \ \ \ \ \ \ \ \EBR  = \begin{pmatrix}
1 & 0 & 1 & 0 & 1 & 0 & 1 & 0\\
0 & 1 & 0 & 1 & 0 & 1 & 0 & 1\\
1 & 0 & 0 & 1 & 0 & 1 & 1 & 0\\
0 & 1 & 1 & 0 & 1 & 0 & 0 & 1\\
1 & 0 & 0 & 1 & 1 & 0 & 0 & 1\\
0 & 1 & 1 & 0 & 0 & 1 & 1 & 0\\
1 & 0 & 1 & 0 & 0 & 1 & 0 & 1\\
0 & 1 & 0 & 1 & 1 & 0 & 1 & 0\\
\end{pmatrix},
\label{eq:EBRmatrixforExampleSmith}
\end{eqnarray}
in which the eight columns respectively correspond to the eight EBR symmetry data vectors of $M$ given in the order of Eq.~(\ref{eq:exampleOfEBRSymmetryData}), and the eight rows respectively correspond to small irrep multiplicities given in the order of Eq.~(\ref{eq:B-P2}).

$\EBR$ in Eq.~(\ref{eq:EBRmatrixforExampleSmith}) admits a Smith normal decomposition [Eq.~(\ref{eq:EBR-snf})]:
\begin{equation}
L_\EBR = \left(\begin{array}{rrrrrrrr}
1 & -1 & 0 & -1 & -1 & 0 & 0 & 0 \\
0 & 1 & 1 & 1 & 1 & 0 & 0 & 0 \\
1 & 0 & 1 & -1 & 0 & 0 & 0 & 0 \\
0 & 0 & 0 & 1 & 0 & 0 & 0 & 1 \\
1 & -1 & 1 & 0 & 0 & 0 & 0 & 0 \\
0 & 1 & 0 & 0 & 0 & 0 & 1 & 0 \\
1 & 0 & 0 & 0 & 0 & 1 & 0 & 0 \\
0 & 0 & 1 & 0 & 0 & 0 & 0 & 0
\end{array}\right),
\Lambda_\EBR = \left(\begin{array}{rrrrrrrr}
1 & 0 & 0 & 0 & 0 & 0 & 0 & 0 \\
0 & 1 & 0 & 0 & 0 & 0 & 0 & 0 \\
0 & 0 & 1 & 0 & 0 & 0 & 0 & 0 \\
0 & 0 & 0 & 1 & 0 & 0 & 0 & 0 \\
0 & 0 & 0 & 0 & 2 & 0 & 0 & 0 \\
0 & 0 & 0 & 0 & 0 & 0 & 0 & 0 \\
0 & 0 & 0 & 0 & 0 & 0 & 0 & 0 \\
0 & 0 & 0 & 0 & 0 & 0 & 0 & 0
\end{array}\right),
R_\EBR = \left(\begin{array}{rrrrrrrr}
1 & 0 & 1 & 0 & 0 & 1 & 0 & 1 \\
0 & 1 & 1 & 0 & 0 & 1 & 1 & 0 \\
0 & 1 & 0 & 1 & 1 & 0 & 1 & 0 \\
0 & 1 & 1 & 0 & 1 & 0 & 0 & 1 \\
0 & -1 & -1 & 0 & -1 & 0 & -1 & 0 \\
0 & 0 & 0 & 0 & 0 & 0 & 0 & 1 \\
0 & 0 & 0 & 0 & 0 & 1 & 0 & 0 \\
0 & 0 & 0 & 1 & 0 & 0 & 0 & 0
\end{array}\right),
\label{eq:exampleOfTheDecompositionP2}
\end{equation}
in which the left inverse of $L_{\EBR}$ is given by:
\begin{equation}
L_\EBR^{-1} = \left(\begin{array}{rrrrrrrr}
    1 & 1 & 0 & 0 & 0 & 0 & 0 & -1 \\
    1 & 1 & 0 & 0 & -1 & 0 & 0 & 0 \\
    0 & 0 & 0 & 0 & 0 & 0 & 0 & 1 \\
    1 & 1 & -1 & 0 & 0 & 0 & 0 & 0 \\
    -2 & -1 & 1 & 0 & 1 & 0 & 0 & -1 \\
    -1 & -1 & 0 & 0 & 0 & 0 & 1 & 1 \\
    -1 & -1 & 0 & 0 & 1 & 1 & 0 & 0 \\
    -1 & -1 & 1 & 1 & 0 & 0 & 0 & 0
    \end{array}\right).    
\end{equation}
As described in the text surrounding Eq.~(\ref{eq:stable-index}), we first examine the nonzero values in $\Lambda_{\EBR}$ to isolate the rows of $L_{\EBR}^{-1}$ that contain SI formulas for $M$.  There is only a single entry $\lambda_{i} > 1$ in $\Lambda_{\EBR}$ in Eq.~(\ref{eq:exampleOfTheDecompositionP2}): $\lambda_{5}=2$.  This implies that the double SI group of $M$ [Eq.~(\ref{eq:SIgroup})] is:
\begin{equation}
Z^{M} = \mathbb{Z}_{2},
\end{equation}
and that the fifth row of $L_{\EBR}^{-1}$ contains the formula for a $\mathbb{Z}_{2}$-valued double SI:
\begin{equation}
z_{2R}(B) = -2m(\ovl{Z}_3) - m(\ovl{Z}_4) + m(\ovl{E}_3) + m(\ovl{D}_3) - m(\ovl{C}_4)\text{ mod }2,
\end{equation}
which can be re-expressed using the modulo $2$ operation as:
\begin{equation}
z_{2R}(B) = m(\ovl{Z}_4) + m(\ovl{E}_3) + m(\ovl{D}_3) + m(\ovl{C}_4)\text{ mod }2.
\label{eq:tempZ2Rex}
\end{equation}
Recognizing that $z_{2R}(B')=0$ for any EBR symmetry data vector $B'$, we next add the symmetry data vectors $({}^1\ovl{E})_{1a} \up M$ and $({}^2\ovl{E})_{1b} \up M$ from Eq.~(\ref{eq:exampleOfEBRSymmetryData}) to Eq.~(\ref{eq:tempZ2Rex}) to rotate $z_{2R}(B)$ into a more recognizable form:
\begin{equation}
z_{2R}(B) = m(\ovl{Z}_3) + m(\ovl{E}_3) + m(\ovl{D}_3) + m(\ovl{C}_3)\text{ mod }2.
\end{equation}
Specifically, using the small irrep label substitution in Eq.~(\ref{eq:BCStoZhidaNotation}), we recognize $z_{2R}(B)$ as the formula from Refs.~\onlinecite{QWZ,ChenBernevigTCI} for the Chern number modulo 2 in the $k_{y}=\pi$ plane:
\begin{equation}
z_{2R} = \sum_{K=Z,E,D,C} n^{\frac12}_K\text{ mod }2,
\end{equation}
in which we have substituted ${\bf k}\rightarrow K$ for notational consistency with previous works~\cite{ZhidaSemimetals,ChenTCI}.  Because $\{C_{2y}|{\bf 0}\}$ is a symmetry of every BZ plane of constant $k_{y}$ (see Fig.~\ref{fig:P2BZ}), then in an insulating phase, the compatibility relations require that the $\{C_{2y}|{\bf 0}\}$ eigenvalues of the occupied bands along each of the $\{C_{2y}|{\bf 0}\}$-invariant lines $k_{x,y}=0,\pi$ are the same at each $k_{y}$.  Hence $z_{2R}=1$ implies that the $k_{y}=\pi$ and $k_{y}=0$ planes \emph{both} exhibit odd Chern numbers, such that the occupied bands either correspond to a 3D quantum anomalous Hall (QAH) insulator with an odd number of chiral modes per $k_{y}$ on surfaces whose normal vectors point in the $xz$-plane, or to a Weyl semimetal with an even number of Weyl points between $k_{y}=0,\pi$.

\subsection{Minimal Double SIs in the 1,651 Double SSGs}
\label{sec:minimalSIProcedure}

Because there are 805 double SSGs $G$ for which the double SI group $|Z^{G}|> 1$ (see Table~\ref{tb:doubleSIStats}), then individually calculating the bulk and anomalous surface and hinge states and physical basis for each stable topological symmetry data vector in each SSG is a practically intractable task.  However, in this section, we will detail a procedure for identifying a considerably smaller set of \emph{minimal SSGs} with \emph{minimal double SIs}, on which the double SIs in all 1,651 double SSGs are dependent.  Specifically, by recognizing that the symmetry-indicated spinful topological semimetals, TIs, and TCIs in non-minimal double SSGs are indicated by the same bulk symmetries as spinful topological semimetals, TIs, and TCIs in the minimal double SSGs, we will reduce the calculation of the physical double-SI-formula bases and symmetry-respecting bulk and boundary states to a smaller, tractable problem.  

\begin{table}[h]	
\begin{tabular}{|l|c|c|}
\hline
\multicolumn{3}{|c|}{Statistics of the Double SIs} \\ 
\hline
Type & SSGs with $|Z^{G}|>1$ & Minimal SSGs \\
\hline
\hline
Type-I & 126 & 18 \\
\hline
Type-II & 117 & 5 \\
\hline
Type-III & 286 & 11 \\
\hline
Type-IV & 276 & 0 \\
\hline
\hline
Total & 805 & 34 \\
\hline
\end{tabular}
\caption{Statistics for the double SIs of the 1,651 double SSGs.  In order, each row of this table contains the type of the double SSG [see Appendix~\ref{sec:MSGs}], the number of double SSGs with nontrivial double SI groups [$|Z^{G}|>1$, see Eq.~(\ref{eq:SIgroup}) and the surrounding text], and the number of minimal double SSGs with minimal double SIs.}
\label{tb:doubleSIStats}
\end{table}

To begin, consider a double SSG $G$ and a subgroup $M$ of $G$ that is isomorphic to an SSG.  Using the procedure detailed in Appendix~\ref{sec:smithForm}, we then calculate the double SI groups $Z^{G,M}$, double SI formulas (in their original, arbitrary bases), and the symmetry data vectors $B^{G}_{i}$ and $B^{M}_{j}$ of the SI topological bands in $G$ and $M$, respectively.  We next restrict consideration to the case in which the double SI groups $Z^{G,M}$ are both nontrivial (\emph{i.e.} $|Z^{G,M}|\neq 1|$).   Lastly, we determine whether the SI topological bands in $G$ subduce to inequivalent SI topological bands in $M$, in which case, we consider the double SIs in $G$ to be \emph{dependent} on the double SIs in $M$.  Specifically, for an SSG $G$ and a subgroup $M$ of $G$ that is isomorphic to an SSG (but not necessarily an SSG with the same Bravais lattice as $G$), the double SIs in $G$ are dependent on the double SIs in $M$ if and only if:
\begin{enumerate}
\item{$|Z^{G}|\leq |Z^{M}|$.}  
\item{For each SI topological band in $G$ with a symmetry data vector $B_{i}^{G}$ [defined in the text following Eq.~(\ref{eq:SIgroup})], the \emph{subduced SI vector} ${\bf z}^{M}(B_{i}^{G}\downarrow M)$ [Eq.~(\ref{eq:SIvector})] exhibits a distinct value for each choice of $i$.  Specifically, given any two SI topological bands $B^G_{i_1}$ and $B^G_{i_2}$ in $G$ for which $\mathbf{z}^G(B^G_{i_1})\neq \mathbf{z}^G(B^G_{i_2})$, the SIs in $G$ can only be dependent on the SIs in $M$ if $\mathbf{z}^M(B^G_{i_1}\downarrow M)\neq \mathbf{z}^M(B^G_{i_2}\downarrow M)$ for all choices of $B^G_{i_1}$ and $B^G_{i_2}$.}
\end{enumerate}

The above requirements indicate the conditions under which the double SIs in $G$ are dependent on the double SIs in $M$.  However, there may also exist subgroups $M'$ of $M$ where the double SIs in \emph{both} $G$ and $M$ are dependent on the double SIs in $M'$.  Hence, given an SSG $M$ for which $|Z^{M}|>1$, if there does not an exist a subgroup $M'$ of $M$ for which the double SIs in $M$ are dependent on the double SIs in $M'$, then we define $M$ as a \emph{minimal double SSG}.  Correspondingly, we define the \emph{minimal double SIs} of the 1,651 double SSGs as the double SIs of the minimal double SSGs.

We note that in this work, we have employed a more narrow definition than in other previous works~\cite{ZhidaSemimetals,ChenTCI} for minimal SIs.  Specifically, in Refs.~\onlinecite{ZhidaSemimetals,ChenTCI} the authors considered cases in which the SIs in $G$ in are neither dependent on the SIs in the subgroups $M\subset G$ and $M'\subset G$ (where $M$ is not isomorphic to $M'$), but where the SIs in $G$ are still spanned by the \emph{combined} SIs in $M$ and $M'$.  As we will show below, using our narrower definition of minimal SIs, we still obtain a manageable number of minimal double SSGs.

Next, given a minimal SSG $M$ and an SSG $G$ in which the SIs are dependent on the SIs in $M$, it follows that all of the symmetry-indicated stable topological semimetals, TIs, and TCIs in $G$ are indicated by the same bulk symmetries that indicate the bulk topology in $M$.  Specifically, this dependency occurs because the set of SI topological bands in $G$ subduced onto $M$ is spanned by the SI topological bands in $M$ modulo EBRs of $M$, and because the EBRs of $M$ do not exhibit topological bulk, surface, or hinge states~\cite{AshvinTCI,ChenTCI,WiederAxion}, as they are Wannierizable~\cite{QuantumChemistry,Bandrep1,Bandrep2,Bandrep3,JenFragile1,BarryFragile}.

Conversely, if the bulk bands of a symmetry-indicated TI or TCI in $G$ are subduced onto an SSG $M$ where the SIs in $G$ are dependent on the SIs in $M$, the subduced topological insulating phase in $M$ may exhibit different anomalous boundary states.  For example, when symmetry-indicated 3D TIs -- such as an insulator with $z_{2}=1$ in Type-II double SG 81.34 $P\bar{4}1'$ (see Ref.~\onlinecite{ChenTCI}) -- are subduced to magnetic axion insulator (AXI)~\cite{WilczekAxion,QHZ,VDBAxion,AndreiInversion,AshvinAxion,WiederAxion,YuanfengAXI,NicoDavidAXI1,NicoDavidAXI2,TMDHOTI,KoreanAXI,CohVDBAXI,ArisHopf,TitusRonnyKondoAXI,YoungkukMonopole,MurakamiAXI1,MurakamiAXI2,HingeStateSphereAXI,BJYangVortex,BarryBenCDW,IvoAXI1,IvoAXI2,GuidoAXI} phases in minimal MSGs (in this case, Type-I double MSG 81.33 $P\bar{4}$, see Appendix~\ref{sec:minimalSSGTables}), the twofold surface Dirac cones of the parent 3D TI become gapped on surfaces in which the Dirac cones are only protected by $\mathcal{T}$ symmetry (see Refs.~\onlinecite{ChenRotation,WiederAxion,FangFuMobius,WladTheory,HOTIBernevig}), revealing a symmetric-sample-spanning network of chiral hinge modes.  More generally, given a TI or TCI that respects the symmetries in the bulk SSG $G$, the anomalous 2D surface states on a surface with a Miller index vector $\hat{\bf n}$ are necessarily protected by the symmetries of a wallpaper subgroup of $G$~\cite{WiederLayers,ConwaySymmetries} that leaves $\hat{\bf n}$ invariant.  However, when the occupied topological bands are subduced onto a subgroup $M\subset G$ where $M$ is isomorphic to an SSG, it is not generically guaranteed that the 2D surface states on the $\hat{\bf n}$-normal surface are still gapless, because the $\hat{\bf n}$-normal surface only respects the symmetries of a wallpaper subgroup of $M$.  Nevertheless, we find that a finite (0D) geometry can in many cases be chosen for a symmetry-indicated TI or TCI that respects the symmetries of a bulk SSG $G$ such that, upon subducing the bulk bands onto a subgroup $M\subset G$, the boundary states do not become gapped.  Importantly, 3D TIs that subduce to magnetic AXIs~\cite{FuKaneMele,FuKaneInversion,QHZ,HsiehDiracInsulator} represent a notable exception, because \emph{all} 2D surfaces of 3D TIs exhibit odd numbers of twofold Dirac cones, whereas there do not exist magnetic AXIs in which all 2D surfaces are gapless~\cite{WilczekAxion,QHZ,VDBAxion,AndreiInversion,AshvinAxion,WiederAxion,YuanfengAXI,NicoDavidAXI1,NicoDavidAXI2,TMDHOTI,KoreanAXI,CohVDBAXI,ArisHopf,TitusRonnyKondoAXI,YoungkukMonopole,MurakamiAXI1,MurakamiAXI2,HingeStateSphereAXI,BJYangVortex,BarryBenCDW,IvoAXI1,IvoAXI2,GuidoAXI}.

Furthermore, we note that it is also possible for an SI topological band $B_{i}^{G}$ in $G$ to correspond to a gapless (semimetallic) phase even if a subduced SI topological band $B_{i}^{G}\downarrow M$ corresponds to a gapped (TI or TCI) phase.  An example occurs in Type-IV double MSG 75.5 $P_{C}4$ and its minimal double subgroup Type-I MSG 75.1 $P4$.  As we will show below in Appendix~\ref{subsec:P4}, all of the double SIs in $M=P4$ are compatible with rotation-symmetry-indicated QAH states.  However, because:
\begin{equation}
G = P_{C}4 = P4\cup \{\mathcal{T}|{\bf t}_{c}/2\}P4,
\end{equation}
then $G$ contains the antiunitary symmetry $\{C_{2z}\times\mathcal{T}|{\bf t}_{c}/2\}$, which enforces the presence of gapless (Weyl) points in the $k_{z}=0,\pi$ planes for all nontrivial values of the SIs in $G$~\cite{ChenRotation,ChenTCI,ZhijunS4WeylHighThroughput}.  This can be seen by recognizing that $\{C_{2z}\times\mathcal{T}|{\bf t}_{c}/2\}$ symmetry can protect gapless points in 2D systems (\emph{e.g.} high-symmetry BZ-planes), and that the Chern numbers of the occupied bands in $\{C_{2z}\times\mathcal{T}|{\bf t}_{c}/2\}$-invariant planes (\emph{e.g.} $k_{z}=0,\pi$) are required by symmetry to vanish.  After the submission of this work, the authors of Ref.~\onlinecite{ChenMagneticCrystalSIs} performed a complete enumeration of the cases in which an SI topological band $B_{i}^{G}$ in $G$ corresponds to a gapless phase while the SI topological band $B_{i}^{G}\downarrow M$ in the subgroup $M\subset G$ is compatible with a gapped topological phase.

In this work, we have exhaustively calculated the double SI groups and formulas of all 1,651 double SSGs, and have determined that, remarkably, there are only 34 minimal double SSGs (see Table~\ref{tb:doubleSIStats}):
\begin{enumerate}
\item{Minimal Type-I Double MSGs (18 MSGs): 
     2.4 $P\bar1$, 
     3.1 $P2$, 
     10.42 $P2/m$, 
     47.249 $Pmmm$, 
     75.1 $P4$, 
     77.13 $P4_2$, 
     81.33 $P\bar4$, 
     83.43 $P4/m$, 
     84.51 $P4_2/m$, 
     88.81 $I4_1/a$, 
     123.339 $P4/mmm$, 
     143.1 $P3$, 
     147.13 $P\bar3$, 
     168.109 $P6$, 
     174.133 $P\bar6$, 
     175.137 $P6/m$, 
     176.143 $P6_3/m$, 
     191.233 $P6/mmm$.}
\item{Minimal Type-II Double SGs (5 SGs):  
     2.5 $P\bar11'$,  
     83.44 $P4/m1'$,  
     87.76 $I4/m1'$
     175.138 $P6/m1'$
     176.144 $P6_{3}/m1'$.}
\item{Minimal Type-III Double MSGs (11 MSGs):  
     27.81 $Pc'c'2$,
     41.215 $Ab'a'2$,
     54.342 $Pc'c'a$,
     56.369 $Pc'c'n$,
     60.424 $Pb'cn'$,
     83.45 $P4'/m$,
     103.199 $P4c'c'$, 
     110.249 $I4_1c'd'$, 
     130.429 $P4/nc'c'$, 
     135.487 $P4_2'/mbc'$, 
     184.195 $P6c'c'$.}
\end{enumerate}

Interestingly, we observe that there are no minimal Type-IV double MSGs (see Table~\ref{tb:doubleSIStats}).  As discussed in the main text, this implies that symmetry-indicated spinful topological phases in Type-IV MSGs are actually enforced by the symmetries of lower-symmetry Type-I or Type-III double MSGs.  For example, we find that the inversion- ($\mathcal{I}$-) symmetric antiferromagnetic (AFM) TCIs introduced in Ref.~\onlinecite{AFMMooreTI}, which respect the symmetries of Type-IV MSGs containing $\{\mathcal{I}|\mathbf{0}\}$, in fact subduce to $\mathcal{I}$-symmetric AXIs~\cite{WilczekAxion,QHZ,VDBAxion,AndreiInversion,AshvinAxion,WiederAxion,YuanfengAXI,NicoDavidAXI1,NicoDavidAXI2,TMDHOTI,KoreanAXI,CohVDBAXI,ArisHopf,TitusRonnyKondoAXI,YoungkukMonopole,MurakamiAXI1,MurakamiAXI2,HingeStateSphereAXI,BJYangVortex,BarryBenCDW,IvoAXI1,IvoAXI2,GuidoAXI} in Type-I double MSG 2.4 $P\bar{1}$ (see Appendix~\ref{sec:P-1} for the double SI group and formulas of double MSG 2.4 $P\bar{1}$).  Previously, in Ref.~\onlinecite{ChenTCI}, the authors determined that the double SIs in all Type-II double SGs are dependent on the double SIs in one of six Type-II double SGs: 2.5 $P\bar11'$, 81.34 $P\bar41'$, 83.44 $P4/m1'$, 174.134 $P\bar61'$, 175.138 $P6/m1'$, and 176.144 $P6_3/m1'$.  However, in this work, we find that Type-II SGs 81.34 $P\bar41'$ and 174.134 $P\bar61'$ are no longer minimal double SSGs after including magnetic subgroups of Type-II SGs, because their double SIs are respectively dependent on the double SIs in Type-I double MSGs 83.33 $P\bar4$ and 174.133 $P\bar6$.  Additionally, for the purposes of this work, we have included Type-II SG 87.76 I4/m1' in our list of minimal double SSGs, because its double SI formulas can only be spanned by subducing SI topological bands onto two different minimal double SSGs (Type-II SGs 2.5 $P\bar11'$ and 83.44 $P4/m1'$), rather than one.  In the Supplementary Table in Appendix~\ref{sec:minimalSSGTables}, we provide a complete enumeration of the minimal double SSGs with the minimal double SIs on which the double SIs in each double SSGs are dependent.

\vspace{0.1in}

\subsection{Double SI Formulas for Spinful Bands with Stable Topology in the 34 Minimal Double SSGs}
\label{sec:34minimal}

Previously, in Appendix~\ref{sec:minimalSIProcedure}, we determined that the SIs in each of the 1,651 double SSGs are fully dependent on the minimal double SIs in one of 34 minimal double SSGs (the minimal double SSG associated to each double SSG is listed in the Supplementary Table in Appendix~\ref{sec:minimalSSGTables}).  In this section, we will present the minimal double SIs in all 34 minimal double SSGs, and hence, the minimal double SIs of spinful band topology in all 1,651 double SSGs.  We will additionally transform the double SI formulas into a unified basis -- which we term a \emph{physical} basis -- in which the double SIs for previously established spinful topological semimetals (SMs), TIs, and TCIs~\cite{AshvinWeyl,AndreiWeyl,SYWeyl,WeylReview,BurkovBalentsWeyl,FuKaneMele,FuKaneInversion,QHZ,LiangTCIOriginal,HsiehTCI,TeoFuKaneTCI,HourglassInsulator,DiracInsulator,ChenRotation,WladTheory} take the same form as the double SIs introduced in previous works~\cite{SlagerSymmetry,AshvinIndicators,HOTIChen,ChenTCI,AshvinTCI,TMDHOTI,BarryBandrepReview,ZhidaSemimetals,AdrianSIReview,AdrianSCMagSI,ChenBernevigTCI,HOTIBismuth,HOTIBernevig}.  In a physical basis, the SIs for topological phases with the same response theories [\emph{e.g.} a $z_{8}=1$ 3D TI in Type-II SG 123.340 $P4/mmm1'$ and an $\eta_{4I}=2$ magnetic AXI in MSG 2.4 $P\bar{1}$, see Ref.~\onlinecite{ChenTCI} and Appendix~\ref{sec:P-1}] are related through simple relations obtained from group-subgroup subduction [\emph{e.g.} the relation $\eta_{4I}=2(z_{8}\text{ mod }2)$ introduced in this work].

Below, for each minimal double SSG $G$, we will list the SI group $Z^{G}$ [Eq.~(\ref{eq:SIgroup})] and the SI formula(s).  We will additionally formulate layer constructions for the gapped (TI and TCI) phases, where admitted (see Refs.~\onlinecite{ChenTCI,ZhidaHermeleCrystal} for further discussions of cases in which TI and TCI phases do not admit layer constructions).  
For the symmetry-indicated 3D QAH phases that we identify in the 34 minimal double SSGs, the anomalous boundary states are chiral modes along surfaces perpendicular to the Chern-layer stacking direction~\cite{AshvinWeyl,AshvinMagnetic,BurkovBalentsWeyl}.  We will show that the remaining 3D symmetry-indicated, spinful, gapped topological phases in the 34 minimal double SSGs are 3D TI and TCI phases -- which we will show to consist of AXIs~\cite{WilczekAxion,QHZ,VDBAxion,AndreiInversion,AshvinAxion,WiederAxion,YuanfengAXI,NicoDavidAXI1,NicoDavidAXI2,TMDHOTI,KoreanAXI,CohVDBAXI,ArisHopf,TitusRonnyKondoAXI,YoungkukMonopole,MurakamiAXI1,MurakamiAXI2,HingeStateSphereAXI,BJYangVortex,BarryBenCDW,IvoAXI1,IvoAXI2,GuidoAXI} with chiral hinge states, 3D TIs with twofold-degenerate, $\mathcal{T}$-symmetry-protected surface Dirac cones~\cite{FuKaneMele,FuKaneInversion,QHZ,HsiehDiracInsulator}, helical mirror TCIs with mirror-protected surface states~\cite{HsiehTCI,TeoFuKaneTCI}, and higher-order TCIs (HOTIs) with mirror- or $\mathcal{T}$-protected helical hinge states~\cite{HOTIBernevig,HOTIBismuth,ChenRotation,HOTIChen,HigherOrderTIPiet,AshvinIndicators,ChenTCI,AshvinTCI,TMDHOTI,WiederAxion,DiracInsulator}.  We emphasize that, employing the convention of Refs.~\onlinecite{DiracInsulator,HOTIBernevig}, a 2D crystal surface can only respect the symmetries of a wallpaper group, whereas a 1D hinge may either respect the symmetries of a frieze group or a line group [defined in Refs.~\onlinecite{WiederLayers,ConwaySymmetries,BalkanLineGroups2,BalkanLineGroups3}], depending on how the finite sample is cut from an infinite crystal.  In this work, we define a helical (\emph{i.e.} non-axionic) TCI phase to be higher-order topological if the TCI phase, when cut into a nanorod geometry, exhibits anomalous helical states that run along nanorod edges that are parallel to bulk rotation axes, where each edge is left invariant under a frieze or line group that contains either $\mathcal{T}$ symmetry or a mirror line parallel to the nanorod edge.

For each of the 34 minimal double SSGs, we will additionally identify the minimal layer constructions necessary to span the subset of SI topological bands [defined in the text following Eq.~(\ref{eq:SIgroup})] corresponding to gapped (TI and TCI) phases; however, as we will detail below, we find that some of the symmetry-indicated spinful TI and TCI phases in the 34 minimal double SSGs are not layer-constructable.  Specifically, as demonstrated in Ref.~\onlinecite{ChenTCI}, a large subset of the previously identified TI and TCI phases in each Type-II SG $G$ can be modeled by placing decoupled, flat layers of Chern insulators, 2D TIs, and 2D TCIs in each unit cell of a crystal that respects the symmetries of $G$.  In this work, we find that a subset of the AXI phases in the minimal double SSGs cannot be constructed from layers of 2D TIs and TCIs.  We conjecture that the AXI phases without layer constructions can still be constructed using the ``topological crystal'' framework discussed in Ref.~\onlinecite{ZhidaHermeleCrystal}, which incorporates cell complexes of 2D TIs and TCIs.

Throughout this section, we will obtain the properties of each minimal double SSG using tools on the BCS.  Specifically, we will obtain the generators for each minimal double SSG using the~\href{https://www.cryst.ehu.es/cgi-bin/cryst/programs/magget_gen.pl}{MGENPOS} tool~\cite{BCSMag1,BCSMag2,BCSMag3,BCSMag4}, the maximal Wyckoff positions using the~\href{https://www.cryst.ehu.es/cgi-bin/cryst/programs/magget_wp.pl}{MWYCKPOS} tool~\cite{BCSMag1,BCSMag2,BCSMag3,BCSMag4}, the maximal momentum stars using the~\href{http://www.cryst.ehu.es/cryst/mkvec}{MKVEC} tool (see Appendix~\ref{sec:MKVEC}), the small (co)reps using the~\href{http://www.cryst.ehu.es/cryst/corepresentations}{Corepresentations} tool (see Appendix~\ref{sec:coreps}), and the EBRs using the~\href{http://www.cryst.ehu.es/cryst/mbandrep}{MBANDREP} tool (see Appendix~\ref{sec:mbandrep}).

In this work, we will provide each double SI formula in the notation of Appendix~\ref{sec:SIexP2} and Refs.~\onlinecite{ChenTCI,ZhidaSemimetals}.  For centrosymmetric SSGs (\emph{i.e.} SSGs that contain $\{\mathcal{I}|{\bf 0}\}$ in at least one definition of the unit cell origin), we will use the symbols $n^{\pm}_K$ to respectively indicate the number of Bloch eigenstates at the inversion-invariant point ${\bf k} =  K$ with the parity ($\{\mathcal{I}|{\bf 0}\}$) eigenvalues $\pm 1$ in a given energy range (which is typically runs over the occupied bands).  For SSGs that contain rotation symmetries of the form $\{C_{n}|{\bf 0}\}$ or screw symmetries of the form $\{C_{nl}|{\bf t}_{l}/b\}$ in at least one definition of the unit cell origin, we will use the symbol $n^j_K$ to indicate the number of Bloch eigenstates at the $C_{n}$-rotation(or screw)-invariant point ${\bf k}=K$ with the rotation eigenvalue $e^{-i\frac{2\pi}{n}j}$ in a given energy range.  Because we are restricting focus in this work to the double SIs of spinful band topology in the 1,651 SSGs, the factor of $j$ in each rotation eigenvalue $e^{-i\frac{2\pi}{n}j}$ is half-integer-valued; in this work, we term $j$ the \emph{angular momentum} (taken modulo $n$) of the rotation- or screw-invariant Bloch eigenstates at $K$.  Next, for SSGs that contain fourfold rotoinversion symmetries of the form $\{S_{4}|{\bf 0}\} = \{C_{4}\times\mathcal{I}|{\bf 0}\}$ (but not fourfold rotation symmetries of the form $\{C_{4}|{\bf 0}\}$) in at least one definition of the unit cell origin, we will use the symbol $n^j_K$ to indicate the number of Bloch eigenstates at the $S_{4}$-invariant point ${\bf k}=K$ with the $\{S_{4}|{\bf 0}\}$ rotoinversion eigenvalues $e^{-i\frac{2\pi}{4}j}$ in a given energy range.  Generically, $n_K^j$ ($j=\pm\frac12,\pm\frac32$) is defined using $\{S_4|\bf 0\}$ eigenvalues \emph{only if} the point $K$ is $\{S_4|\bf 0\}$-invariant, but not $\{C_4|\bf 0\}$-invariant.  Conversely, if $K$ is $\{C_4|\bf 0\}$-invariant, then $n_K^j$ ($j=\pm\frac12,\pm\frac32$) is always defined using the eigenvalues of $\{C_4|\bf 0\}$.  Lastly, for SSGs that contain both mirror symmetries of the form $\{m_{l}|{\bf 0}\}$ and rotation symmetries of the form $\{C_{nl}|{\bf 0}\}$ or screw symmetries of the form $\{C_{nl}|{\bf t}_{l}/b\}$ in at least one definition of the unit cell origin, we will use the symbols  $n^{j,\pm i}_K$ to respectively indicate the number of Bloch eigenstates at the rotation- or screw-invariant point ${\bf k}=K$ with the rotation or screw eigenvalue $e^{-i\frac{2\pi}{n}j}$ and the mirror eigenvalue $\pm i$ in a given energy range.

Before we will derive the double SIs in the 34 minimal double SSGs, we will first summarize our labeling convention for double SIs.  First, for double SIs that have the same SI formulas as the nonmagnetic double SIs introduced in Refs.~\onlinecite{ChenTCI}, we have followed the labeling convention established in Ref.~\onlinecite{ChenTCI}:
\begin{enumerate}
\item{$z_{2w,i}$ ($i=1,2,3$) are the weak TI SIs in the $k_{i}=\pi$ planes, or the weak mirror Chern numbers modulo $2$ in the $k_{i}=\pi$ planes in the absence of $\{\mathcal{T}|{\bf 0}\}$ symmetry.}
\item{$z_{nm,k}$ ($n=4,3,6$, $k=0,\pi$) are the mirror Chern numbers (modulo $n$) in the $k_z=k$ plane indicated by rotation eigenvalues in SSGs 83.44 $P4/m1'$, 174.134 $P\bar61'$, 175.138 $P6/m1'$ for $n=4,3,6$, respectively.  In this work, we will use the symbol $z_{nm,k}^\pm$ to represent the Chern numbers of sets of bands with mirror eigenvalues $\pm i$, respectively.}
\item{$z_{4},\ z_2, \ z_{8},\ z_{12}$, and $z_{12}'$ indicate strong 3D TIs and helical TCIs and HOTIs in SSGs 2.5 $P\bar11'$, 81.34 $P\bar41'$, 83.44 $P4/m1'$, 175.138 $P6/m1'$, 176.144 $P6_3/m1'$, respectively.  Odd values of $z_{4},\ z_2, \ z_{8},\ z_{12}$, and $z_{12}'$ correspond to strong TIs. $z_4=2$, $z_8=4$, $z_{12}=6$, $z_{12}'=6$ correspond to non-axionic HOTI phases with helical hinge states or mirror TCIs with even mirror Chern numbers (see Appendices~\ref{sec:P-11p} through~\ref{sec:P63/m1p} and Ref.~\onlinecite{ChenTCI}).}
\end{enumerate}
If the double SIs in a Type-II SSG $G$ continue to indicate stable topological phases in a magnetic subgroup $M$ of $G$, then we will use the same double SI labels and formulas in $G$ and $M$.

We additionally find that there are Type-I and Type-III double MSGs with new double SIs that are not subduced from Type-II SSGs (see Table~\ref{tb:doubleSIStats}).  For these minimal \emph{magnetic} double SIs, we have adopted a convention in which:
\begin{enumerate}
\item{$z_{nR}$ ($n=2,3,4,6$) represent Chern numbers (modulo $n$) indicated by rotation eigenvalues.}
\item{$z_{nR}'$ and $z_{nR}''$ ($n=2,3,4,6$) represent doubled Chern numbers indicated by rotation eigenvalues [\emph{i.e.} $z_{nR}'=(C/2)\text{ mod }n$] in nonsymmorphic MSGs.} 
\item{$\eta_{4I}$ is defined in MSG 2.4 $P\bar1$.  Odd values of $\eta_{4I}$ correspond to Weyl semimetals, and $\eta_{4I}=2$ corresponds to an AXI provided that the net Chern numbers are zero and there are no Weyl points in the BZ interior.  We use the symbol ``$\eta$'' rather than ``$z$'' to distinguish $\eta_{4I}$ from the double SI $z_{4}$ in the minimal double SSGs 2.5 $P\bar11'$, 47.249 $Pmmm$, and 83.45 $P4'/m$ and from the double SI $z_{4}'$ in double MSG 135.487 $P4_2'/mbc'$.}
\item{$z_{2I,i}$ ($i=1,2,3$) are defined in double MSG 2.4 $P\bar1$, and respectively represent the Chern numbers modulo $2$ in the $k_{i}=\pi$ planes indicated by $\mathcal{I}$ (parity) eigenvalues.  We have used the subscript ``$I$'' to distinguish $z_{2I,i}$ from $z_{2R}$ (the Chern number modulo $2$ indicated by $C_2$ rotation eigenvalues) and $z_{2w,i}$ (the weak TI and TCI parity indices discussed above).}
\item{$\eta_{2I}'=\frac12\eta_{4I}$ represents a doubled variant of $\eta_{4I}$ that is present in SSGs in which symmetry requires $\eta_{4I}$ to be even.}
\item{$\delta_{nm}$ ($n=2,3,4,6$) represent the differences between the mirror Chern numbers in the $k_z=0,\pi$ planes (modulo $n$).}
\item{$z_{4S}$ and $\delta_{2S}$ are defined in MSG 81.33 $P\bar4$.  Respectively, $z_{4S}$ and $\delta_{2S}$ represent the total Chern number (modulo 4) in the $k_z=\pi$ plane and twice the difference of the total Chern numbers in the $k_z=0,\pi$ planes (\emph{i.e.} $\delta_{2S}=[(C_{k_z=\pi}-C_{k_z=0})/2]\text{ mod }2$).}
\item{$z_{4}'$ in MSG 135.487 $P4_2'/mbc'$ represents a different $\mathbb{Z}_{4}$-valued doubled variant of the double SI $\eta_{4I}$ than the double SI $z_{4}$ discussed above [\emph{e.g.} $\eta_{4I} = (2z_{4}')\text{ mod }4$].}
\item{For all of the symbols of the double SIs, the first number ($n$) in the subscript indicates that the corresponding double SI takes integer values in the range $[0,n-1]$.}
\end{enumerate}

\vspace{-0.3in}
\subsubsection{Double SIs in Type-I Double MSG 2.4 $P\bar{1}$} 
\label{sec:P-1}

The double MSG 2.4 $P\bar1$ is generated by $\{E|100\}$, $\{E|010\}$, $\{E|001\}$, and $\{\INV|\mathbf{0}\}$.  The SIs of MSG 2.4 $P\bar{1}$ were previously analyzed in Refs.~\onlinecite{AshvinMagnetic,EslamPRBHOTI}; the previous analyses performed in Refs.~\onlinecite{AshvinMagnetic,EslamPRBHOTI} agree with the analysis performed in this section.

\textit{Double SIs} -- The double MSG 2.4 $P\bar1$ has the SI group $\ZZ_4\times \ZZ_2^3$.  We define the four SIs of double MSG 2.4 $P\bar{1}$ to be $(\eta_{4I},z_{2I,1},z_{2I,2},z_{2I,3})$, and we define the four SI formulas to be:
\begin{equation}
\eta_{4I} = \sum_K n_K^-\text{ mod }4 = \sum_{K} \frac12(n_K^- - n_K^+)\text{ mod }4, 
\label{eq:eta4I}
\end{equation}
and:
\begin{equation}
z_{2I,i=1,2,3} =C_{k_i=\pi}\text{ mod }2= \sum_{K, K_i=\pi} n_K^-\text{ mod }2,
\label{eq:z2I}
\end{equation}
where $K$ runs over the eight $\mathcal{I}$-invariant momenta in the first BZ, and $n^{\pm}_K$ are the number of Bloch states with $\pm1$ parity ($\mathcal{I}$) eigenvalues at $K$ in the group of bands under consideration (typically the occupied bands).  We find that $z_{2I,i}$ indicates the parity of the momentum-space Chern number in the $k_i=\pi$ plane, in agreement with the Chern number SI formulas previously introduced in Refs.~\onlinecite{QWZ,ChenBernevigTCI}.  Correspondingly, we find that $\eta_{4I}\text{ mod } 2$ is the parity of the difference between the Chern numbers in the $k_z=0$ and $k_z=\pi$ planes.  Because a 3D $|C|=1$ Weyl point is equivalent to the quantum critical point~\cite{AshvinWeyl} between 2D Chern insulating phases with $|\Delta C|=1$, then this implies that $\eta_{4I}=1,3$ correspond to Weyl SM (WSM) phases that satisfy the insulating compatibility relations (see Appendix~\ref{sec:compatibilityRelations}), similar to the WSM and nodal-line SM phases previously analyzed in Refs.~\onlinecite{AshvinWeyl,ZhidaSemimetals,YoungkukMonopole,YoungkukLineNode,TMDHOTI,ChenWithWithout,ChenNodalLineMagnon}.  The boundary states of the $\eta_{4I}=1,3$ WSM phases differ from each other by a chiral hinge state or gapless surface states, because, as we will show below, the SI difference $\Delta\eta_{4I}=3-1=2$ either corresponds to an AXI or a 3D QAH state.  In this work, we refer to symmetry-indicated SM phases that satisfy the insulating compatibility relations as \emph{Smith-index SMs} (SISMs).

\textit{Layer constructions} -- We will now formulate layer constructions of the symmetry-indicated spinful TI and TCI phases in double MSG 2.4 $P\bar{1}$.  In each unit cell, we will use the relative 3D coordinates $(x,y,z)$ to index layer positions, where the unit cell is defined as lying within $0\le x,y,z<1$.  In position space, an $\mathcal{I}$ center at $(0,0,0)$ transforms the coordinates $(x,y,z)$ to $(-x,-y,-z)$.  For a position ${\bf r}$ to be considered $\mathcal{I}$-invariant, we require that:
\begin{equation}
\mathcal{I}{\bf r} = {\bf r}\text{ mod }(1,0,0)\text{ mod }(0,1,0)\text{ mod }(0,0,1).
\end{equation}
Consequently, the eight maximal Wyckoff positions (\emph{i.e.} the $\mathcal{I}$ centers) in MSG 2.4 $P\bar{1}$ lie at $x,y,z = 0,1/2$.

We next study the layer constructions of the insulating subset of the SI topological bands (\emph{i.e.} the symmetry-indicated topological phases that do not correspond to Weyl SISMs with odd $\eta_{4I}$ indices).  We first introduce the layer construction generators, each of which is equivalent to a 3D QAH insulator~\cite{AndreiInversion,AshvinAxion,AshvinMagnetic,BurkovBalentsWeyl}, where the double SIs for each layer construction are given in the order $(\eta_{4I},z_{2I,1},z_{2I,2},z_{2I,3})$:
\begin{enumerate}
\item An $\hat{\bf x}$-normal Chern layer with $C_{x}=\pm1$ in the $x=0$ plane has the SIs $(2100)$.
\item An $\hat{\bf x}$-normal Chern layer with $C_{x}=\pm1$ in the $x=\frac12$ plane has the SIs $(0100)$.
\item A $\hat{\bf y}$-normal Chern layer with $C_{y}=\pm1$ in the $y=0$ plane has the SIs $(2010)$.
\item A $\hat{\bf y}$-normal Chern layer with $C_{y}=\pm1$ in the $y=\frac12$ plane has the SIs $(0010)$.
\item A $\hat{\bf z}$-normal Chern layer with $C_{z}=\pm1$ in the $z=0$ plane has the SIs $(2001)$.
\item A $\hat{\bf z}$-normal Chern layer with $C_{z}=\pm1$ in the $z=\frac12$ plane has the SIs $(0001)$.
\end{enumerate}
For the double SIs of the above layer constructions, we have adopted the convention used in Refs.~\onlinecite{ChenTCI,ZhidaSemimetals} in which commas are suppressed for specific values of the SIs [\emph{e.g.} ($\eta_{4I},z_{2I,1},z_{2I,2},z_{2I,3})=(2100)$].  Below, we will detail the explicit calculations that we have performed to calculate the SIs of each layer construction, focusing on the cases of $\hat{\bf z}$-normal Chern layers with $C_{z}=1$ respectively placed at $z=0$ and $z=\frac{1}{2}$.  In this work, we will only consider layer constructions of stable topological phases (as opposed to fragile phases, see Appendix~\ref{sec:reviewTopologicalBands} and Refs.~\onlinecite{JenFragile1,BarryFragile,AshvinFragile,AshvinFragile2,AdrianFragile,KoreanFragile,KoreanFragileInversion,ArisFragileNoGo,DelicateAris,SlagerMagFragile,SlagerMagFragile2,ZhidaFragile,FragileFlowMeta,ZhidaFragile2}), which do not depend of the positions of layers with trivial 2D stable topological invariants [\emph{i.e.} layers of 2D fragile phases or (obstructed) atomic limits]~\cite{ChenTCI}.  Hence, for stable topological phases that admit layer constructions, the stable SIs are fully determined by the positions, orientations, and 2D stable topology of the layers.

First, we consider a crystal in double MSG 2.4 $P\bar{1}$ that is constructed of layered, $\hat{\bf z}$-normal Chern insulators with $C_{z}=1$ that lie at $z=0$ in each unit cell.  We assume, without loss of generality, that each Chern insulator originates from placing one valence (occupied) spinful $s$ orbital at $(x,y)=(0,0)$, placing one conduction (unoccupied) spinful $p$ orbital at $(x,y)=(0,0)$, and then inverting bands at $(k_{x},k_{y})=(0,0)$, resulting in the occupied parity eigenvalues:
\begin{equation}
\lambda'(0,0) = -1,\ \lambda'(\pi,0) = 1,\ \lambda'(0,\pi) = 1,\ \lambda'(\pi,\pi) = 1.
\label{eq:temppbar1}
\end{equation}
As shown in Refs.~\onlinecite{QWZ,ChenBernevigTCI} the Chern number $C_{z}$ of each layer satisfies:
\beq
(-1)^{C_{z}} = \prod_{K} \prod_{n\in\rm occ} \lambda_{n}^\pr(K), 
\eeq
where $K$ runs over the four $\mathcal{I}$-invariant momenta in Eq.~(\ref{eq:temppbar1}), and $\lambda^\pr_{K,n}$ is the parity eigenvalue of the $n^\text{th}$ energetically isolated band at $K$ [though for the specific case that we are discussing, there is only one isolated (valence) band $n=1$].  The parity eigenvalues shown in Eq.~(\ref{eq:temppbar1}) indicate that each layer carries a nontrivial Chern number $C_{z}\text{ mod }2=1$.

Next, we express the occupied band of each Chern layer in a basis of hybrid Wannier functions~\cite{MarzariDavidWannier,MarzariWannierReview} in which states within the layers are exponentially localized in $z$ and depend on the crystal momenta $k_{x,y}$.  We then return to momentum space by Fourier-transforming the $z$ component of the hybrid Bloch-Wannier wavefunction of the occupied band:
\beq
\ket{ \psi_{\kk} } = \frac1{\sqrt{N_z}}\sum_{z=0,\pm1\cdots} e^{-i z k_z} \ket{ \psi_{k_x,k_y,z} },
\label{eq:Fourier-z0}
\eeq
in which $N_{z}$ is the number of unit cells in the crystal in the $z$-direction.  In the hybrid basis of $(k_{x},k_{y},z)$:
\begin{equation}
\mathcal{I}(k_{x},k_{y},z) = (-k_{x},-k_{y},-z),
\end{equation}
and hybrid coordinates ${\bf h}$ are considered to be $\mathcal{I}$-invariant if:
\begin{equation}
\mathcal{I}{\bf h} = {\bf h}\text{ mod }(2\pi,0,0)\text{ mod }(0,2\pi,0)\text{ mod }(0,0,1).
\end{equation}

However, it is important to emphasize that the hybrid wavefunction $\ket{ \psi_{k_x,k_y,z} }$ of each layer, unlike the Bloch wavefunction $\ket{ \psi_{\kk} }$, is generically not an eigenstate of $\mathcal{I}$:
\begin{equation}
\mathcal{I}\ket{ \psi_{k_x,k_y,z}}=\lambda'(k_{x},k_{y})\ket{ \psi_{k_x,k_y,-z}},
\end{equation} 
in which $\lambda'(k_{x},k_{y})$ is the parity eigenvalue of the occupied band in the 2D BZ of a single Chern insulator.  For a crystal in double MSG 2.4 $P\bar{1}$ furnished by $\hat{\bf z}$-normal, $C_{z}=1$ Chern layers in the $z=0$ plane of each cell, this implies that the parity eigenvalues at the $\mathcal{I}$-invariant ${\bf k}$ points are given by:
\begin{eqnarray}
\INV \ket{ \psi_{\kk} } &=& \frac1{\sqrt{N_z}}\sum_{z=0,\pm1\cdots} e^{-i z k_z}\mathcal{I}\ket{ \psi_{k_x,k_y,z} }  \nonumber \\
&=& \frac1{\sqrt{N_z}}\sum_{z=0,\pm1\cdots} e^{-i z k_z}\lambda^\pr(k_x,k_y) \ket{ \psi_{k_x,k_y,-z} }  \nonumber \\
&=& \lambda^\pr(k_x,k_y)\left[\frac1{\sqrt{N_z}}\sum_{z=0,\pm1\cdots} e^{i z k_z}\ket{ \psi_{k_x,k_y,z} }\right]  \nonumber \\
&=& \lambda^\pr(k_x,k_y)\left[\frac1{\sqrt{N_z}}\sum_{z=0,\pm1\cdots} e^{-i z k_z}\left[e^{ik_{z}}\right]^{2z}\ket{ \psi_{k_x,k_y,z} }\right]  \nonumber \\
&=& \lambda^\pr(k_x,k_y)\left[\frac1{\sqrt{N_z}}\sum_{z=0,\pm1\cdots} e^{-i z k_z}\ket{ \psi_{k_x,k_y,z} }\right]  \nonumber \\
&=&  \lambda^\pr(k_x,k_y) \ket{ \psi_{\kk} },
\label{eq:parityFormulaForAllLayers}
\end{eqnarray}
where in the fifth line, we have used the relation $[e^{ik_{z}}]^{2z}=1$ for $\mathcal{I}$-invariant momenta $k_{z}=0,\pi$ and $z\in\mathbb{Z}$.  Through Eq.~(\ref{eq:parityFormulaForAllLayers}), we determine that the 3D parity eigenvalues $\lambda(k_x,k_y,k_z)$ satisfy $\lambda(k_{x},k_{y},k_{z}) = \lambda^\pr(k_x,k_y)$.  From the parity eigenvalues of each layer listed in Eq.~(\ref{eq:temppbar1}), this implies that:
\begin{eqnarray}
\lambda(0,0,0) &=& -1,\ \lambda(\pi,0,0) = 1,\ \lambda(0,\pi,0) = 1,\ \lambda(\pi,\pi,0) = 1, \nonumber \\
\lambda(0,0,\pi) &=& -1,\ \lambda(\pi,0,\pi) = 1,\ \lambda(0,\pi,\pi) = 1,\ \lambda(\pi,\pi,\pi) = 1.
\label{eq:temppbar2}
\end{eqnarray}
Substituting the parity eigenvalues from Eq.~(\ref{eq:temppbar2}) into Eqs.~(\ref{eq:eta4I}) and~(\ref{eq:z2I}), we obtain the SIs (2001) for a 3D crystal in double MSG 2.4 $P\bar{1}$ with $\hat{\bf z}$-normal $C_{z}=1$ Chern insulators placed at $z=0$ in each cell.

We next consider the case in which the 3D crystal is furnished with layers of $\hat{\bf z}$-normal, $C_{z}=1$ Chern insulators that lie at $z\text{ mod }1=\frac{1}{2}$.  The Bloch wavefunction of the occupied band of the 3D crystal takes the form:
\beq
\ket{ \psi_{\kk} } = \frac1{\sqrt{N_z}}\sum_{z=\pm\frac12,\pm\frac32\cdots} e^{-i z k_z} \ket{ \psi_{k_x,k_y,z} }.
\eeq
Unlike previously in Eq.~(\ref{eq:parityFormulaForAllLayers}), for a crystal in double MSG 2.4 $P\bar{1}$ furnished by $\hat{\bf z}$-normal, $C_{z}=1$ Chern insulator in the $z=\frac{1}{2}$ plane of each unit cell, the parity eigenvalues at the $\mathcal{I}$-invariant ${\bf k}$ points are given by:
\begin{eqnarray}
\INV \ket{ \psi_{\kk} } &=& \frac1{\sqrt{N_z}}\sum_{z=\pm\frac12,\pm\frac32\cdots} e^{-i z k_z}\mathcal{I}\ket{ \psi_{k_x,k_y,z} } \nonumber \\
&=& \frac1{\sqrt{N_z}}\sum_{z=\pm\frac12,\pm\frac32\cdots} e^{-i z k_z}\lambda^\pr(k_x,k_y) \ket{ \psi_{k_x,k_y,-z} } \nonumber \\
&=& \lambda^\pr(k_x,k_y) \left[\frac1{\sqrt{N_z}}\sum_{z=\pm\frac12,\pm\frac32\cdots} e^{i z k_z}\ket{ \psi_{k_x,k_y,z} }\right] \nonumber \\
&=& \lambda^\pr(k_x,k_y) \left[\frac1{\sqrt{N_z}}\sum_{z=\pm\frac12,\pm\frac32\cdots} e^{-i z k_z}\left[e^{ik_{z}}\right]^{2z}\ket{ \psi_{k_x,k_y,z} }\right] \nonumber \\
&=& \lambda^\pr(k_x,k_y)e^{ik_{z}}\left[\frac1{\sqrt{N_z}}\sum_{z=\pm\frac12,\pm\frac32\cdots} e^{-i z k_z}\ket{ \psi_{k_x,k_y,z} }\right] \nonumber \\
&=&  \lambda^\pr(k_x,k_y) e^{ik_z} \ket{ \psi_{\kk} },
\label{eq:parityFormulaForAllLayers2}
\end{eqnarray}
where in the fifth line, we have exploited that the summation is taken over half-integer values of $z$.  Eq.~(\ref{eq:parityFormulaForAllLayers2}) implies that the 3D parity eigenvalues $\lambda(k_x,k_y,k_z)$ satisfy $\lambda(k_{x},k_{y},k_{z}) = e^{ik_{z}}\lambda^\pr(k_x,k_y)$.  From the parity eigenvalues of each layer listed in Eq.~(\ref{eq:temppbar1}), this indicates that:
\begin{eqnarray}
\lambda(0,0,0) &=& -1,\ \lambda(\pi,0,0) = 1,\ \lambda(0,\pi,0) = 1,\ \lambda(\pi,\pi,0) = 1, \nonumber \\
\lambda(0,0,\pi) &=& 1,\ \lambda(\pi,0,\pi) = -1,\ \lambda(0,\pi,\pi) = -1,\ \lambda(\pi,\pi,\pi) = -1.
\label{eq:temppbar3}
\end{eqnarray}
Substituting the parity eigenvalues from Eq.~(\ref{eq:temppbar3}) into Eqs.~(\ref{eq:eta4I}) and~(\ref{eq:z2I}), we obtain the SIs (0001) for a 3D crystal of $\hat{\bf z}$-normal $C_{z}=1$ Chern insulators placed at $z=\frac{1}{2}$ in double MSG 2.4 $P\bar{1}$.

For the remainder of this work, we will not explicitly calculate the 3D symmetry eigenvalues that are implied by each layer construction.  However, because the unitary symmetries of magnetic crystals are drawn from the same set as the unitary symmetries of nonmagnetic crystals [\emph{i.e.} because the unitary subgroups of both Type-II SGs and Type-III and IV MSGs are isomorphic to Type-I MSGs, see Appendix~\ref{sec:MSGs}], then the symmetry eigenvalues of the magnetic layer constructions introduced in this work can be extrapolated from the analogous analyses of nonmagnetic layer constructions in Ref.~\onlinecite{ChenTCI}.

\textit{The inversion $\ZZ_2$ invariant and AXIs} -- We find that $\eta_{4I}=2$ if and only if the $\mathcal{I}$-center at the origin $(000)$ is occupied by a layer with an odd Chern number.  For 3D QAH states (\emph{i.e.} 3D insulators with nonzero Chern numbers), the $\eta_{4I}=0,2$ phases have the same bulk response.  For example, layer constructions 1 and 2 for double MSG 2.4 $P\bar{1}$ -- which exhibit $\eta_{4I}=2,0$, respectively -- are related by a shift of origin from (000) to $(00\frac12)$.  Nevertheless, the boundary states of insulators with $\eta_{4I}=0,2$ are distinct.  For a finite-size sample with an $\mathcal{I}$ center at $(000)$, the state with $\eta_{4I}=2$ has a single Chern layer passing through the $\mathcal{I}$ center, and pairs of Chern layers at positions $(00,\pm z)$.  In the finite sample, the total Chern number is therefore odd, and there is an $\mathcal{I}$-symmetric chiral hinge (or surface) mode surrounding the sample guaranteed by the net-odd Chern number.  However, in the state with $\eta_{4I}=0,$ all of the Chern layers appear in pairs at the positions $(00,\pm z)$, such that the total Chern number is even.  This implies the possibility of a completely gapped finite sample (\emph{i.e.} a total sample Chern number of zero).

For 3D insulators with vanishing Chern numbers, $\eta_{4I}=0,2$ correspond to trivial insulators and AXIs, respectively.  For example, the TCI constructed by one $C_{z}=1$ layer in the $z=0$ plane and one $C_{z}=-1$ layer in the $z=\frac12$ plane is an AXI with the double SIs $(2000)$~\cite{AshvinMagnetic,IvoAXI1,NicoDavidAXI2,BarryBenCDW,EslamPRBHOTI,MurakamiAXI1,MurakamiAXI2}.  The chiral hinge states of the AXI can be understood by observing that the chiral modes on the boundary of the layered crystal alternate in direction, and can hence pairwise annihilate -- when the finite-sized crystal is $\mathcal{I}$-symmetric, there is an unpaired chiral mode that is equivalent to a boundary-encircling chiral hinge state.  Specializing to the even sector of $\eta_{4I}$, this implies that an inversion $\mathbb{Z}_{2}$ invariant may be defined as:
\begin{equation}
\eta_{2I}^\pr = \frac12 \eta_{4I} \text{ mod }2.
\label{eq:eta2Ip}
\end{equation}
We emphasize that it is $\eta_{4I}$ -- as opposed to $\eta_{2I}'$ -- that is returned by the Smith normal form calculation (see Appendix~\ref{sec:smithForm}) for double MSG 2.4 $P\bar{1}$.  The non-minimal index $\eta_{2I}'$ is integer-valued only for 3D insulators or WSMs with even numbers of Weyl points in each half of the bulk BZ.  Nevertheless, as we will show below, in many higher-symmetry double SSGs in which the SIs depend on the SIs in double MSG 2.4 $P\bar{1}$ (see Appendix~\ref{sec:minimalSSGTables}), the Smith normal form calculation \emph{does} return $\eta_{2I}'$.  From previous works~\cite{WilczekAxion,QHZ,VDBAxion,AndreiInversion,AshvinAxion,WiederAxion,YuanfengAXI,NicoDavidAXI1,NicoDavidAXI2,TMDHOTI,KoreanAXI,CohVDBAXI,ArisHopf,TitusRonnyKondoAXI,YoungkukMonopole,MurakamiAXI1,MurakamiAXI2,HingeStateSphereAXI,BJYangVortex,BarryBenCDW,IvoAXI1,IvoAXI2,GuidoAXI}, we recognize that $\eta_{2I}'=1$ is related to the axion angle $\theta=\pi$:
\begin{equation}
\theta\text{ mod }2\pi = \pi \eta_{2I}'.
\label{eq:thetaForAXIInv}
\end{equation}

However, it is crucial to note that the axion angle $\theta=\pi$ does not always indicate an axionic band-insulating phase (\emph{i.e.} an AXI or 3D TI, see Refs.~\onlinecite{WilczekAxion,QHZ,VDBAxion,AndreiInversion,AshvinAxion,WiederAxion,YuanfengAXI,NicoDavidAXI1,NicoDavidAXI2,TMDHOTI,KoreanAXI,CohVDBAXI,ArisHopf,TitusRonnyKondoAXI,YoungkukMonopole,MurakamiAXI1,MurakamiAXI2,HingeStateSphereAXI,BJYangVortex,BarryBenCDW,IvoAXI1,IvoAXI2,GuidoAXI,FuKaneMele,FuKaneInversion}).  For example, consider the case of a crystal in MSG 2.4 $P\bar{1}$ furnished by one $C_{z}=1$ layer in the $z=0$ plane and one $C_{z}=1$ layer in the $z=\frac{1}{2}$ plane -- the bulk topological phase is \emph{not} an AXI, but is instead a 3D QAH insulator with $C_{z}=2$ per unit cell.  For the $C_{z}=2$ QAH insulator, the SIs [$(2000$)] are the same as those of the AXI discussed in the text surrounding Eqs.~(\ref{eq:eta2Ip}) and~(\ref{eq:thetaForAXIInv}), indicating that $\theta = \pi \eta_{2I}' = \pi$, despite the fact that the bulk is not an AXI.  This can be understood by recognizing that the axion $\theta$ angle is origin-dependent when the $\mathbb{Z}$-valued, \emph{non-symmetry-indicated} Chern numbers of a 3D crystal do not vanish~\cite{IvoAXI1,NicoDavidAXI2,BarryBenCDW,zilberberg2018photonic,zilberbergHOTI}, and hence $\theta$ can still be nonzero in a 3D QAH phase depending on the choice of origin.  Therefore, in order for Eq.~(\ref{eq:thetaForAXIInv}) to indicate the \emph{origin-independent} $\theta$ angle of an AXI, it is additionally required that the total Chern numbers $C_{x,y,z}$ vanish in each unit cell.  Lastly, we note that Eq.~(\ref{eq:thetaForAXIInv}) differs by $\pi$ from the definition of $\theta$ as a ``Chern number polarization'' employed in Refs.~\onlinecite{NicoDavidAXI2,zilberberg2018photonic,BarryBenCDW}.  Hence, in 3D insulators with non-vanishing position-space Chern numbers (\emph{i.e.} nonzero total Chern numbers in any direction summed across the layers in each position-space unit cell) and origin- (\emph{i.e.} convention-) dependent $\theta$ angles, the SIs introduced in this work [\emph{e.g.} Eq.~(\ref{eq:thetaForAXIInv})] return values of $\theta$ that are shifted from the values in Refs.~\onlinecite{NicoDavidAXI2,zilberberg2018photonic,BarryBenCDW} by $\pi$.  Importantly, however, both Eq.~(\ref{eq:thetaForAXIInv}) and the Chern number polarization in Refs.~\onlinecite{NicoDavidAXI2,zilberberg2018photonic,BarryBenCDW} correctly diagnose the convention-independent bulk $\theta$ angle of AXIs to be $\theta=\pi$.

\textit{Relationship with the SIs in other double SSGs} -- As shown in Refs.~\onlinecite{AshvinIndicators,AshvinTCI,ChenTCI}, the double SIs in double SG 2.5 $P\bar11^\pr$ take the same form as Eqs.~(\ref{eq:z2I}) and~(\ref{eq:eta4I}) under the replacement of $n^-_K$ with $n^-_K/2$ [\emph{i.e.} the number of energetically isolated \emph{Kramers pairs} of Bloch states at $K$].  The SI topological bands in double SG 2.5 $P\bar{1}1'$ subduced onto double MSG 2.4 $P\bar{1}$ imply the double SI dependencies:
\begin{equation}
(z_4,z_{2w,1}, z_{2w,2}, z_{2w,3})_{P\bar11^\pr} \to (\eta_{4I},z_{2I,1},z_{2I,2},z_{2I,3})_{P\bar1}= (2(z_4\text{ mod }2),000)_{P\bar1}.
\end{equation}

\subsubsection{Double SIs in Type-I Double MSG 3.1 $P2$}
\label{subsec:P2}

The double MSG 3.1 $P2$ is generated by $\{E|100\}$, $\{E|010\}$, $\{E|001\}$, and $\{C_{2y}|\mathbf{0}\}$, and has the double SI group $\ZZ_2$.  We first recall the formula established in Refs.~\onlinecite{QWZ,AndreiInversion,ChenBernevigTCI} for the parity of the Chern number in a $\hat{\bf y}$-normal 2D insulator with $\{C_{2y}|00\}$ symmetry:
\begin{equation}
(-1)^{C_{y}} = \prod_{n\in occ} \prod_{K} \zeta_n(K),
\label{eq:tempC2ChernParity}
\end{equation}
where $C_{y}$ is the Chern number in the $y$-direction, $\zeta_n(K)$ is the $\{C_{2y}|00\}$ eigenvalue of the $n^\text{th}$ energetically isolated state at $K$, and $K$ runs over the four $\{C_{2y}|00\}$-invariant momenta in 2D.  Using Eq.~(\ref{eq:tempC2ChernParity}), we define the double SI $z_{2R}$ of MSG 3.1 $P2$ to be the parity of the Chern number $C_{y}$ in the $k_y=\pi$ plane:
\begin{equation}
z_{2R} = C_{k_{y}=\pi} \text{ mod } 2= \sum_{K=Z,D,C,E} n^{\frac12}_{K} \text{ mod } 2, 
\label{eq:z2R}
\end{equation}
where $n^{\frac12}_K$ is the number of energetically isolated states with the $\{C_{2y}|\mathbf{0}\}$ eigenvalue $-i$ [corresponding to an angular momentum (modulo 2) of $j=\frac12$] at $K$.  For 3D insulating phases, the Chern numbers in all of the BZ planes of constant $k_y$ for $-\pi \le k_y\ < \pi$ must be the same (otherwise, there would be bulk Weyl points, and the bulk would not be an insulator).  Hence, a 3D insulator with $z_{2R}=1$ is a 3D QAH state with $C_y\text{ mod }2=1$.

If the symmetry operation $\{\mathcal{T}|\mathbf{0}\}$ were added, a crystal in double MSG 3.1 $P2$ would become invariant under Type-II double SG 3.2 $P21'$.  In $P21'$, states at the four TRIM points $K$ in Eq.~(\ref{eq:z2R}) form Kramers pairs with opposite $\{C_{2y}|\mathbf{0}\}$ eigenvalues, causing $n^{\frac12}_{K}$ to be even, and $z_{2R}$ to be zero.  This agrees with the absence of double SIs in Type-II double SG 3.2 $P21'$ (see Appendix~\ref{sec:minimalSSGTables}), and the requirement that the position-space Chern numbers $C_{x,y,z}$ vanish in a nonmagnetic ($\mathcal{T}$-symmetric) crystal~\cite{HaldaneModel,TitusFractionalHaldane}.

\subsubsection{Double SIs in Type-I Double MSG 10.42 $P2/m$}

The double MSG 10.42 $P2/m$ is generated by $\{E|100\}$, $\{E|010\}$, $\{E|001\}$, $\{C_{2y}|\mathbf{0}\}$, and $\{m_y|\mathbf{0}\}$.

\textit{SIs} -- The double MSG 10.42 $P2/m$ has the SI group $\ZZ_2^3$.  In the physical basis, the three double SIs of double MSG 10.42 $P2/m$ ($\delta_{2m}, z_{2m,\pi}^+, z_{2m,\pi}^-$) have the respective SI formulas:
\begin{equation}
\delta_{2m} = C_\pi^+-C_0^-\text{ mod }2=\sum_{K=Z,D,C,E} n^{\frac12,+i}_K - \sum_{K=\Gamma,A,B,Y} n^{\frac12,-i}_K \text{ mod }2,
\label{eq:d2m}
\end{equation}
\begin{equation}
z_{2m,\pi}^+ = C_\pi^+\text{ mod }2 = \sum_{K=Z,D,C,E} n^{\frac12,+i}_K \text{ mod }2,
\label{eq:z2mp+}
\end{equation}
\begin{equation}
z_{2m,\pi}^- = C_\pi^-\text{ mod }2 =\sum_{K=Z,D,C,E} n^{\frac12,-i}_K \text{ mod }2,
\label{eq:z2mp-}
\end{equation}
where $n^{j,\pm i}_K$ is the number of occupied states with angular momentum $j$, the $\{C_{2y}|0\}$ eigenvalue $e^{-i\pi j}$, and the $\{m_y|0\}$ eigenvalue $\pm i$.  Because the matrix representative of $\{m_y|0\}$ commutes with the matrix representative of $\{C_{2y}|0\}$ in all double-valued small irreps at each of the $\mathcal{I}$-invariant ${\bf k}$ points in double MSG 10.42 $P2/m$, then $z_{2m,\pi}^{\pm}$, respectively indicate the Chern number parities in the mirror sector of the $k_{y}=\pi$ plane with $\{m_y|0\}$ eigenvalue $\pm i$.  As discussed in Appendix~\ref{sec:SIexP2}, if the bulk is a 3D insulator, then the occupied states in the $k_y=0,\pi$ planes have the same total Chern numbers (\emph{i.e} the sum of the Chern numbers over the two mirror sectors in each of the $k_{y}=0,\pi$ planes is the same), because the insulating compatibility relations require that the occupied bands in the $k_{y}=0,\pi$ planes have the same $\{C_{2y}|0\}$ eigenvalues.

\textit{Layer constructions} -- To diagnose the topology associated to each nontrivial value of the double SIs, we employ the layer construction method.  We denote the Chern numbers of the occupied bands in each mirror sector -- which we term the \emph{mirror sector Chern numbers} -- as $(C^+_{k_y=0},C^{-}_{k_y=0}, C^+_{k_y=\pi},C^-_{k_y=\pi})$.  The insulating compatibility relations require that $C^+_{k_y=0}+C^-_{k_y=0}\text{ mod }2 = C^+_{k_y=\pi}+C^-_{k_y=\pi} \text{ mod } 2$.  We emphasize that the double SIs in Eqs.~(\ref{eq:d2m}),~(\ref{eq:z2mp+}), and~(\ref{eq:z2mp-}) are fully determined by the above mirror sector Chern numbers $(C^+_{k_y=0},C^{-}_{k_y=0}, C^+_{k_y=\pi},C^-_{k_y=\pi})$.  We next calculate the minimal double SIs of double MSG 10.42 $P2/m$ in the order ($\delta_{2m}, z_{2m,\pi}^+, z_{2m,\pi}^-$), as well as the subduced double SIs $(\eta_{4I},z_{2I,1},z_{2I,2},z_{2I,3})_{P\bar1}$ in the subgroup double MSG 2.4 $P\bar1$ for a physical comparison and to identify symmetry-indicated AXI phases in MSG 10.42 $P2/m$.
\begin{enumerate}
\item A $\hat{\bf y}$-normal layer with $C^+_y=1$, $C^-_y=0$ in the $y=0$ plane has the mirror sector Chern numbers =$(1010)$ and the SIs $(110)$. The subgroup SIs are $(\eta_{4I},z_{2I,1},z_{2I,2},z_{2I,3})_{P\bar1}=(2010)_{P\bar1}$.
\item A $\hat{\bf y}$-normal layer with $C^+_y=0$, $C^-_y=1$ in the $y=0$ plane has the mirror sector Chern numbers $(0101)$ and the SIs $(101)$. The subgroup SIs are $(2010)_{P\bar1}$.
\item A $\hat{\bf y}$-normal layer with $C^+_y=1$, $C^-_y=0$ in the $y=\frac12$ plane has the mirror sector Chern numbers $(1001)$ and the SIs $(001)$. The subgroup SIs are $(0010)_{P\bar1}$.
\item A $\hat{\bf y}$-normal layer with $C^+_y=0$, $C^-_y=1$ in the $y=\frac12$ plane has the mirror sector Chern numbers $(0110)$ and the SIs $(010)$. The subgroup SIs are $(0010)_{P\bar1}$.
\end{enumerate}

\textit{Relationship with the SIs in other double SSGs} -- To identify the AXI phases in double MSG 10.42 $P2/m$, we subduce the SIs onto the SIs of double MSG 2.4 $P\bar1$:
\begin{equation}
\pare{\delta_{2m}, z_{2m,\pi}^+, z_{2m,\pi}^-}_{P2/m} \to (\eta_{4I},z_{2I,1},z_{2I,2},z_{2I,3})_{P\bar1}= \pare{2\delta_{2m}, 0,\; z_{2m,\pi}^+ + z_{2m,\pi}^-,\; 0}_{P\bar1}.
\end{equation}
We find that both the $(100)$ and $(111)$ states in double MSG 10.42 $P2/m$ are consistent with AXI phases [but may also, for example, be 3D QAH phases, see Appendix~\ref{sec:P-1}].  We label the four layer constructions as $L_{1,2,3,4}$.  The $(100)$ and $(111)$ states in double MSG 10.42 $P2/m$ can be constructed as $L_1 - L_4$ and $L_1 - L_3$, respectively.  Lastly, $-L_{3}$ ($-L_4$) has the same construction as $L_{3}$ ($L_4$), except for a difference in the position-space mirror sector Chern number $C^{+}_y=-1$ ($C^-_y=-1$).

Lastly, Type-II double SSG 10.43 $P2/m1^\pr$ -- the double SSG that results from adding $\{\mathcal{T}|{\bf 0}\}$ symmetry to Type-I double MSG 10.42 $P2/m$ -- has the SI group $\ZZ_4\times \ZZ_2^3$.  The subduction relations between the double SIs in double SSG 10.43 $P2/m1^\pr$ and double MSG 10.42 $P2/m$ are given by:
\begin{equation}
(z_4,z_{2w,1},z_{2w,2},z_{2w,3})_{P2/m1^\pr} \to (\delta_{2m},z_{2m,\pi}^+,z_{2m,\pi}^-)_{P2/m}= (z_4\text{ mod }2,z_{2w,2},z_{2w,2})_{P2/m}.
\end{equation}

\subsubsection{Double SIs in Type-I Double MSG 47.249 $Pmmm$}
\label{subsec:Pmmm}

The double MSG 47.249 $Pmmm$ is generated by $\{E|100\}$, $\{E|010\}$, $\{E|001\}$, $\{m_x|\mathbf{0}\}$, $\{m_y|\mathbf{0}\}$, and $\{\INV|\mathbf{0}\}$.

\textit{SIs} -- The double MSG 47.249 $Pmmm$ has the SI group $\ZZ_4\times \ZZ_2^3$.  In double-valued small irreps of the little groups at the $\mathcal{I}$-invariant ${\bf k}$ points, the matrix representatives of perpendicular mirror symmetries  (\emph{e.g.} $\{m_x|\mathbf{0}\}$ and $\{m_y|\mathbf{0}\}$) anticommute.  Hence, Bloch states at the eight $\mathcal{I}$-invariant momenta must be at least twofold degenerate (and in fact are exactly twofold degenerate in double MSG 47.249 $Pmmm$).  The double SIs can be chosen to be the same as the double SIs of SSG 47.250 $Pmmm1^\pr$, because the addition of $\TRS$ symmetry to double MSG 47.249 $Pmmm$ does not change the dimensions and characters of the small irreps at the high-symmetry BZ points or the compatibility relations between the high-symmetry-point small irreps.  In the physical basis, the $\ZZ_4$ double SI is:
\begin{equation}
z_4 = \sum_{K} \frac14 (n_K^- - n_K^+) \text{ mod } 4,
\label{eq:z4}
\end{equation}
where $K$ indexes all $\mathcal{I}$-invariant momenta and $n_K^\pm$ is the number of occupied states with $\pm 1$ parity ($\mathcal{I}$) eigenvalues at $K$.  $z_4$ has the same form as $\eta_{4I}$ [Eq.~(\ref{eq:eta4I})], but carries an additional prefactor of $1/2$.  The extra factor of $1/2$ in Eq.~(\ref{eq:z4}) can be understood from the double degeneracy of the Bloch states at the $\mathcal{I}$-invariant TRIM points, where the two states in each doublet have the same parity eigenvalues and complex-conjugate pairs of spinful mirror eigenvalues $\pm i$, due to the anticommutation relations discussed above.  Hence, the SI formula for $z_{4}$ [Eq.~(\ref{eq:z4})] is simply one half of the SI formula for $\eta_{4I}$ [Eq.~(\ref{eq:eta4I})] (before applying the modulo $4$ operation).  The three $\mbb{Z}_2$ SIs are the mirror Chern number parities in the $k_{1,2,3}=\pi$ planes:
\begin{equation}
z_{2w,i=1,2,3} = \sum_{K, K_i=\pi} \frac12 n_{K}^{-}\text{ mod }2.  
\label{eq:z2w}
\end{equation}
Specifically, because an in-plane mirror operation reverses the sign of a 2D Chern number, and because all of the mirror planes in the bulk BZ have additional in-plane mirror symmetries (\emph{e.g.} the Hamiltonian in each BZ mirror plane must respect the symmetries of magnetic layer group~\cite{MagneticBook,ITCA,subperiodicTables,WiederLayers,DiracInsulator,SteveMagnet} $pmmm$), then the net Chern number in each BZ mirror plane in double MSG 47.249 $Pmmm$ must vanish.  For a group of bands in a mirror-invariant BZ (position-space) plane for which $C^{+}_{k_{i}}=-C^{-}_{k_{i}}$ ($C^{+}=-C^{-}$), we then define the \emph{mirror Chern number}~\cite{TeoFuKaneTCI,HsiehTCI} to be $|C^{+}_{k_{i}}|$ ($|C^{+}|$).

\textit{Layer constructions} -- To diagnose the topology associated to each nontrivial value of the double SIs, we employ the layer construction method.  In the layer constructions below, $C^+=-C^-$ due to the net-zero Chern numbers enforced by the mirror symmetries.  Hence, we will omit $C^-$ in further discussions of the topology in double MSG 47.249 $Pmmm$.  The layer constructions for the double SIs $(z_{4},z_{2w,1},z_{2w,2},z_{2w,3})$ of MSG 47.249 $Pmmm$ are given by:
\begin{enumerate}
\item An $\hat{\bf x}$-normal mirror Chern layer with $C^+_x=1$ in the $x=0$ plane has the mirror sector Chern numbers $(C^+_{k_x=0},C^+_{k_x=\pi})=(11)$ and the SIs $(2100)$.
\item An $\hat{\bf x}$-normal mirror Chern layer with $C^+_x=1$ in the $x=\frac12$ plane has the mirror sector Chern numbers $(C^+_{k_x=0},C^+_{k_x=\pi})=(1,-1)$ and the SIs $(0100)$.
\item A $\hat{\bf y}$-normal mirror Chern layer with $C^+_y=1$ in the $y=0$ plane has the mirror sector Chern numbers $(C^+_{k_y=0},C^+_{k_y=\pi})=(11)$ and the SIs $(2010)$.
\item A $\hat{\bf y}$-normal mirror Chern layer with $C^+_y=1$ in the $y=\frac12$ plane has the mirror sector Chern numbers $(C^+_{k_y=0},C^+_{k_y=\pi})=(1,-1)$ and the SIs $(0010)$.
\item A $\hat{\bf z}$-normal mirror Chern layer with $C^+_z=1$ in the $z=0$ plane has the mirror sector Chern numbers $(C^+_{k_z=0},C^+_{k_z=\pi})=(11)$ and the SIs $(2001)$.
\item A $\hat{\bf z}$-normal mirror Chern layer with $C^+_z=1$ in the $z=\frac12$ plane has the mirror sector Chern numbers $(C^+_{k_z=0},C^+_{k_z=\pi})=(1,-1)$ and the SIs $(0001)$.
\end{enumerate}
The layer-construction calculations in this section parallel with the previous calculations in Appendix~\ref{sec:P-1} of the layer constructions of the insulating phases in double MSG 2.4 $P\bar{1}$.  Hence, we will only consider layer construction $5$ as an example of the generalization from the layer constructions and bulk topology in double MSG 2.4 $P\bar{1}$ to that in double MSG 47.249 $Pmmm$.

In layer construction $5$, we take each layer to consist of a $\hat{\bf z}$-normal 2D mirror Chern insulator ($C^+_z=-C^-_z=1$) with the occupied parity ($\mathcal{I}$) eigenvalues $\lambda_{1,2}'(k_x,k_y)=--,++,++,++$ at $(k_x,k_y)=(00),(0\pi),(\pi0),(\pi\pi)$, respectively.  The subscripts $1,2$ on $\lambda_{1,2}'(k_x,k_y)$ represent the $\{m_z|\mathbf{0}\}$ eigenvalue sectors $i$ and $-i$, respectively.  Applying the Fourier transformation in Eq.~(\ref{eq:Fourier-z0}), we find that the parity eigenvalues of the 3D system are given by $\lambda_{1,2}(k_x,k_y,k_z)=\lambda'_{1,2}(k_x,k_y)$ [Eq.~(\ref{eq:parityFormulaForAllLayers})].  This implies that $\lambda_{1,2}(k_x,k_y,k_z)=--,++,++,++,--,++,++,++$ for $(k_x,k_y,k_z)=(000), (0\pi0),(\pi00), (\pi\pi0), (00\pi), (0\pi\pi), (\pi0\pi), (\pi\pi\pi)$, respectively.  Substituting the parity eigenvalues of layer construction 5 into Eqs.~(\ref{eq:z4}) and~(\ref{eq:z2w}), we obtain the SIs (2001).

\textit{Axion insulators} -- We find that states with odd $z_4$ SIs cannot be constructed from layers of 2D stable topological phases.  However, we may still use subduction relations to determine the bulk topology of insulators with odd values of $z_{4}$.  First, as we will show below, $(1000)$ and $(3000)$ subduce to $(2000)_{P\bar{1}}$ in MSG 2.4 $P\bar1$.  Hence, if the $(1000)$ and $(3000)$ phases in double MSG 47.249 $Pmmm$ are insulating, then the bulk insulator must either be an AXI or a 3D QAH state.  Because the net Chern numbers $C_{x,y,z}=0$ must vanish if the bulk is gapped, due to the mirror symmetries of double MSG 47.249 $Pmmm$, then the $(1000)$ and $(3000)$ states must be AXIs.  This result can also be understood by subducing from a $\mathcal{T}$-symmetric SSG.  Specifically, because $(1000)$ and $(3000)$ in MSG 47.249 $Pmmm$ can respectively be subduced from $(1000)_{Pmmm1'}$ and $(3000)_{Pmmm1'}$ in Type-II SG 47.250 $Pmmm1^\pr$, which correspond to $\mathcal{T}$-symmetric 3D TIs with $\theta=\pi$~\cite{AshvinIndicators,ChenTCI,AshvinTCI}, then $(1000)$ and $(3000)$ are compatible with bulk-gapped states.  Hence, we conclude that 3D insulators with $(1000)$ and $(3000)$ in double MSG 47.249 $Pmmm$ are AXIs, \emph{without ambiguity}.  We conjecture that the $(1000)$ and $(3000)$ AXIs in MSG 47.249 $Pmmm$ can be constructed using the topological crystal method~\cite{ZhidaHermeleCrystal}, which additionally incorporates cell complexes of 2D Chern insulators, TIs, and TCIs.

\begin{figure}[h]
\centering
\includegraphics[width=\linewidth]{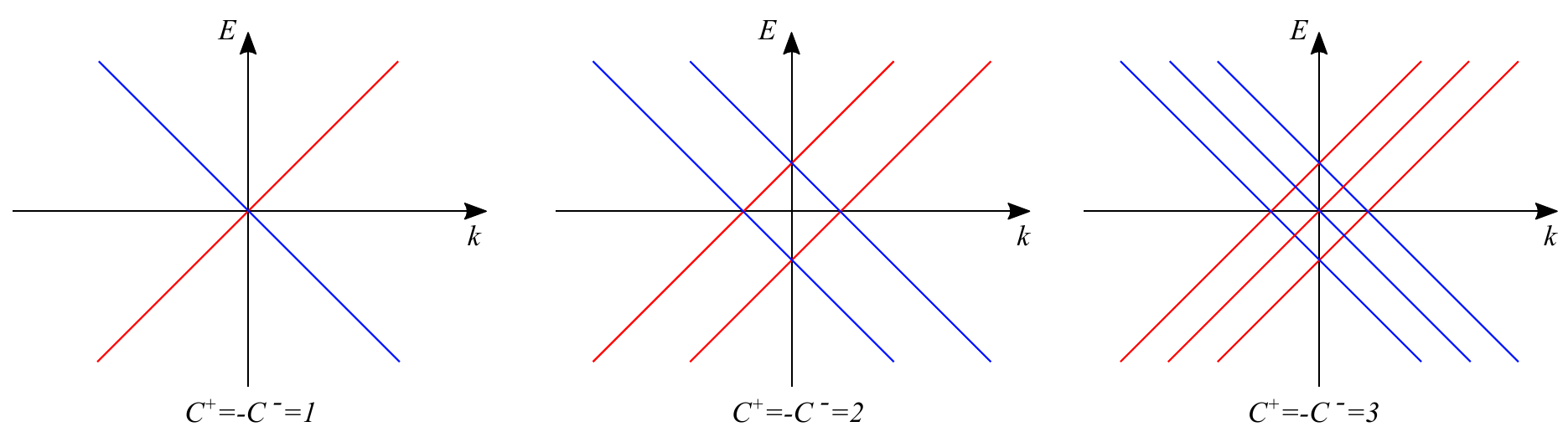}
\caption{Surface Dirac points protected by mirror Chern numbers.  In this figure, we respectively depict the surface states of insulators with the bulk mirror Chern numbers $C^+=-C^-=1,2,3$, where $C^\pm$ respectively refer to the Chern number in the mirror sector with eigenvalue $\pm i$.  In each panel, we depict a topological surface band structure along a mirror-invariant surface BZ line, where the red and blue lines respectively indicate bands with the mirror eigenvalues $i$ and $-i$.  At half-filling, the number of twofold surface Dirac points is given by the mirror Chern number $|C^+|$, where $|C^+|=|C^-|$.}
\label{fig:MirrorChern}
\end{figure}

\begin{figure}[h]
\centering
\includegraphics[width=0.6\linewidth]{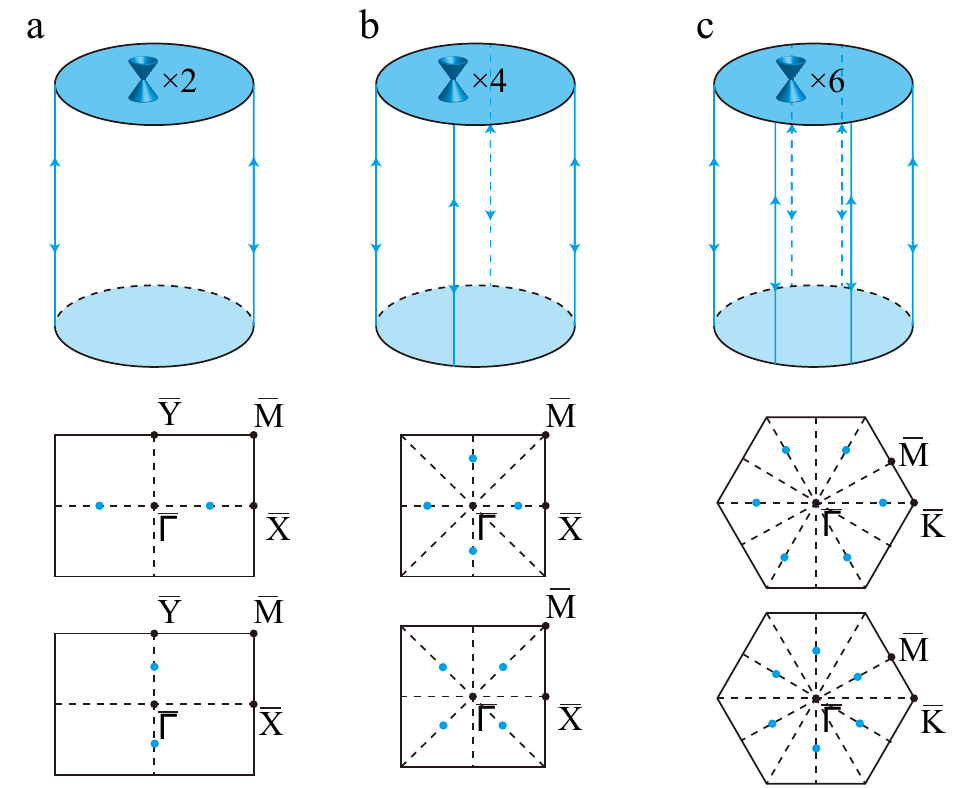}    
\caption{The boundary states of the non-axionic magnetic HOTI phases in double MSGs (a) 47.249 $Pmmm$, (b) 123.339 $P4/mmm$, (c) 191.233 $P6/mmm$.  In the top panel, we show the mirror-protected surface twofold Dirac cones and side-surface helical hinge modes of each non-axionic magnetic HOTI (see Appendix~\ref{sec:newHOTIs} for further details).  The symmetry groups of the top ($\hat{\bf z}$-normal) surfaces are Type-I magnetic wallpaper groups (a) $pmm$, (b) $p4m$, and (c) $p6m$ (see Refs.~\onlinecite{DiracInsulator,WiederLayers,ChenRotation,SteveMagnet,ConwaySymmetries} and Appendix~\ref{sec:newHOTIs}).  We note that, in this work, we have labeled wallpaper groups -- which are also sometimes termed \emph{plane groups} -- using the short notation previously employed in Refs.~\onlinecite{DiracInsulator,WiederLayers,HingeSM}; in the long notation of the~\href{https://www.cryst.ehu.es/plane/get_plane_gen.html}{Get Plane Gen} tool on the BCS~\cite{BCS1,BCS2}, the magnetic wallpaper groups in (a-c) $pmm$, $p4m$, and $p6m$ are respectively labeled by the symbols $p2mm$, $p4mm$, and $p6mm$.  In (a-c), the helical hinge states are pinned to the hinge projections of the bulk mirror planes, and therefore originate from nontrivial bulk mirror Chern numbers.  In the middle and bottom panels, we depict two possible configurations of anomalous twofold Dirac points in the top-surface BZ, where the dashed lines represent the top-surface projections of bulk mirror planes.  In Appendix~\ref{sec:HOTIfermionDoubling}, we will introduce magnetic Dirac fermion doubling theorems for 2D insulators with the magnetic wallpaper groups of the top surfaces in (a-c).  The fermion doubling theorems for the top-surface wallpaper groups in (a-c) are respectively circumvented by the non-axionic magnetic HOTI phases discovered in this work.}
\label{fig:SS}
\end{figure}

\textit{Helical HOTI phases protected by mirror} -- First, the double SIs $(2000)_{Pmmm1^\pr}$ of Type-II double SSG 47.250 $Pmmm1^\pr$ correspond to a helical (non-axionic, \emph{i.e.} $\theta\text{ mod }2\pi=0$) HOTI protected by $\mathcal{I}$ and $\TRS$ symmetries.  In the $\mathcal{I}$- and $\mathcal{T}$-symmetric HOTI phase, an odd number of helical modes encircle a finite sample with $\mathcal{I}$-symmetry.  Because double SSG 47.250 $Pmmm1^\pr$ contains $\{m_{x,y,z}|{\bf 0}\}$ symmetries, then a single helical hinge mode on a boundary must also be pinned to the hinge projection of a bulk mirror plane, and must indicate a bulk mirror Chern number $C_{m}=2$ (because a mirror-invariant hinge is a 1D domain wall between two 2D surfaces with two massive twofold Dirac cones with oppositely-signed masses related by mirror symmetry, see Refs.~\onlinecite{ChenTCI,HOTIBernevig,DiracInsulator}).

Returning to the magnetic subgroup Type-I MSG 47.249 $Pmmm$ of Type-II SG 47.250 $Pmmm1'$, we denote the six layer constructions introduced in this section as $L_a$ ($a=1\cdots 6$), respectively.  Without loss of generality, we consider $(2n+1)L_1 \oplus (2m+1) L_2$.  We next consider a $90^{\circ}$ hinge of a $z$-directed, $mmm$-invariant rod that lies between $x+y>0$, $x-y<0$, where the rod is centered at the origin.  On the 1D hinge, $(2n+1)L_1$, which has the mirror sector Chern numbers $C^+_x=-C^-_y=2n+1$ and has 2D TCI layers at $x=0,\pm1\cdots$, will contribute $2n+1$ helical modes to the hinge at $x=y=0$.

We next note that the bulk mirror Chern numbers $(C^+_{k_x=0},C^+_{k_x=\pi})$ are $(2n+2m+2, 2n-2m)$.  If $n-m\text{ mod }2=0$, then $C^+_{k_x=0}\text{ mod }4=2$ and $C^+_{k_x=\pi}\text{ mod }4=0$, and if $n-m\text{ mod }2=1$, then $C^+_{k_x=0}\text{ mod }4=0$ and $C^+_{k_x=\pi}\text{ mod }4=2$.  In general, the SIs (2000) can be constructed as $(2n+1)L_1 \oplus (2m+1) L_2$, or $ (2n+1) L_3 \oplus (2m+1) L_4$, or $(2n+1) L_5 \oplus (2m+1) L_6$ ($m,n\in\mathbb{Z}$), or through any superposition of an odd number of the aforementioned layer constructions.  Hence, there exists a direction $i\in\{x,y,z\}$ such that the mirror Chern numbers in the $i$ direction are either $(C^+_{k_i=0}\text{ mod }4, C^+_{k_i=\pi}\text{ mod }4)=(2,0)$ or $(C^+_{k_i=0}\text{ mod }4, C^+_{k_i=\pi}\text{ mod }4)=(0,2)$.  Therefore, the bulk of the $(2000)$ is a mirror TCI. Nevertheless, in this work, we refer to the $(2000)$ phase in double MSG 47.249 $Pmmm$ as a helical HOTI, because the $(2000)$ phase of double MSG 47.249 can be connected to a $(z_4,z_{2w,1},z_{2w,2},z_{2w,3})_{Pmmm1'}=(2000)_{Pmmm1'}$ mirror TCI phase in the $\mathcal{T}$-symmetric supergroup Type-II double SG 47.250 $Pmmm1'$ without closing a bulk or surface gap.  In turn, the $(2000)_{Pmmm1'}$ TCI phase subduces to an $\mathcal{I}$- and $\mathcal{T}$-protected $(z_4,z_{2w,1},z_{2w,2},z_{2w,3})_{P\bar{1}1'}=(2000)_{P\bar{1}1'}$ helical HOTI in Type-II double SG 2.5 $P\bar{1}1'$ [see Appendix~\ref{sec:P-11p} and Refs.~\onlinecite{ChenTCI,AshvinTCI,AshvinIndicators,TMDHOTI}].  To summarize, there exists at least one mirror-symmetric surface in the $(2000)$ HOTI state that has $2+4n$ ($n\in\{\mathbb{Z}^{+},0\}$) twofold Dirac points, in agreement with nontrivial even bulk mirror Chern number.  We depict the anomalous surface and hinge states of the $(2000)$ HOTI phase in Fig.~\ref{fig:SS}(a).

For completeness, we next consider the boundary states of the layer construction $(2n+1)L_1 \oplus (2m+1) L_2$.  The Chern numbers in the $\{m_x|\mathbf{0}\}$ mirror sectors are $C_{k_x=0}^+=-C_{k_x=0}^-=2n+2m+2$, $C_{k_x=\pi}^+=-C_{k_x=\pi}^-=2n-2m$.  We consider either a $\hat{\bf y}$- or a $\hat{\bf z}$-normal surface, either of which preserves $\{m_x|\mathbf{0}\}$ mirror symmetry.  In the 2D surface BZ, the bulk Chern number $C^+_{k_x=0}$ mandates the presence of $|2n+2m+2|$ twofold Dirac points on the $k_x=0$ line (see Fig.~\ref{fig:MirrorChern}), and $C^+_{k_x=\pi}$ mandates the presence of $|2n-2m|$ Dirac points on the $k_x=\pi$ line.  Hence the total number of twofold surface Dirac points is $|2n+2m+2|+|2n-2m|\text{ mod }4=2$.  Similarly, $ (2n+1) L_3 \oplus (2m+1) L_4$ and $(2n+1) L_5 \oplus (2m+1) L_6$ will exhibit $2+4n$ ($n\in\{\mathbb{Z}^{+},0\}$) twofold Dirac points on $\{m_y|\mathbf{0}\}$- and $\{m_z|\mathbf{0}\}$-preserving surfaces, respectively.  In Appendix~\ref{sec:HOTIfermionDoubling}, we will prove that, on surfaces of the $(2000)$ state that respect the symmetries of Type-I double magnetic wallpaper group~\cite{DiracInsulator,WiederLayers,ChenRotation,SteveMagnet,ConwaySymmetries} $pmm$, the presence of $2+4n$ ($n\in\{\mathbb{Z}^{+},0\}$) twofold surface Dirac points circumvents the fermion multiplication theorem for 2D lattices with double magnetic wallpaper group $pmm$.

\textit{Relationship with the SIs in other double SSGs} -- To identity the AXI phases, we subduce the SIs onto double MSG 2.4 $P\bar1$.  As explained in the text following Eq.~(\ref{eq:z4}), because each doublet of Bloch states at an $\mathcal{I}$-invariant ${\bf k}$ point in MSG 47.249 $Pmmm$ has the same parity eigenvalues (and complex-conjugate mirror eigenvalues), then $z_4$ is simply a doubling of $\eta_{4I}$.  Hence $\eta_{4I}=2z_4\text{ mod }4$.  Similarly, $z_{2I,i}=2z_{2w,i}\text{ mod }2=0$.  In summary:
\begin{equation}
(z_4, z_{2w,1}, z_{2w,2}, z_{2w,3})_{Pmmm} \to (\eta_{4I},z_{2I,1},z_{2I,2},z_{2I,3})_{P\bar1}= (2 z_4\text{ mod }4,000)_{P\bar1}.
\end{equation}
Hence, the (1000) and (3000) states in double MSG 47.249 $Pmmm$, if gapped, correspond to AXIs.

Lastly, the correspondence between the double SIs of Type-I double MSG 47.249 $Pmmm$ and the double SIs of Type-II double SSG 47.250 $Pmmm1^\pr$ is one-to-one.

\subsubsection{Double SIs in Type-I Double MSG 75.1 $P4$}
\label{subsec:P4}

The double MSG 75.1 $P4$ is generated by $\{E|100\}$, $\{E|010\}$, $\{E|001\}$, and $\{C_{4z}|\mathbf{0}\}$.

The double SI group of double MSG 75.1 $P4$ is $\ZZ_4$.  To determine the physical basis for the double SIs, we first recall the formula for the Chern number in the presence of fourfold rotation symmetry~\cite{ChenBernevigTCI}:
\begin{equation}
i^C = (-1)^{N_\mrm{occ}} \prod_{n\in occ} \xi_n(00) \xi_n(\pi\pi) \zeta_n(0\pi),
\end{equation}
where $\xi_n(K)$ is the $\{C_{4z}|\mathbf{0}\}$ eigenvalue of the $n^{\rm th}$ occupied state at $K$, and $\zeta_n(K)$ is the $\{C_{2z}|\mathbf{0}\}$ eigenvalue of the $n^{\rm th}$ occupied state at $K$.
We can define the SI as the Chern number in the $k_z=\pi$ plane modulo $4$:
\begin{align}
z_{4R} &= C_{k_z=\pi} \text{ mod } 4 = 2N_\mrm{occ} + 
\sum_{K=Z,A} \pare{
-\frac12 n^{\frac12}_K + \frac12 n^{-\frac12}_K
-\frac32 n^{\frac32}_K + \frac32 n^{-\frac32}_K}
- n^{\frac12}_R + n^{-\frac12}_R \text{ mod }4\nono\\
&= \sum_{K=Z,A} \pare{
-\frac12 n^{\frac12}_K + \frac12 n^{-\frac12}_K
-\frac32 n^{\frac32}_K + \frac32 n^{-\frac32}_K}
+ n^{\frac12}_R - n^{-\frac12}_R \text{ mod } 4, 
\label{eq:z4R}
\end{align}
where $n^{\frac12,-\frac12,\frac32,-\frac32}_{Z,A}$ are the number of occupied states with $\{C_{4z}|\mathbf{0}\}$ eigenvalues $e^{-i\frac{\pi}4}$, $e^{i\frac{\pi}4}$, $e^{-i\frac{3\pi}4}$, $e^{i\frac{3\pi}4}$, respectively, and $n^{\frac12,-\frac12}_R$ are the number of occupied states with $\{C_{2z}|\mathbf{0}\}$ eigenvalues $e^{-i\frac{\pi}2}$, $e^{i\frac{\pi}2}$, respectively.  In deriving Eq.~(\ref{eq:z4R}), we have used the relation $N_{\rm occ} = n^{\frac12}_R + n^{-\frac12}_R$.

Due to the compatibility relations and the fact that a chiral fermion in 3D occurs when there is a change in a momentum-space Chern number, a 3D insulator must satisfy $C_{k_z=\pi}=C_{k_z}$ for all $k_{z}$.  Hence, we may have equivalently defined the SI $z_{4R}$ using the occupied $\{C_{4z}|{\bf 0}\}$ and $\{C_{2z}|{\bf 0}\}$ eigenvalues in $k_z=0$ plane, or in any other BZ plane of constant $k_{z}$.  In general, in this work, in order to match the convention employed in Ref.~\onlinecite{ChenTCI}, we will use the rotation eigenvalues in the $k_i=\pi$ plane to define double SIs in the physical basis.  To summarize, if a 3D system is insulating and exhibits $z_{4R}\neq0$, then the system is in a 3D QAH state with $C_{k_z=0}=C_{k_z=\pi}$ and $z_{4R} = C_{k_z=0}\text{ mod }4$.

Because the physical meaning of the double SIs is straightforward (\emph{i.e.} the nontrivial phases are 3D QAH states composed of stacks of Chern insulators), then will not provide explicit layer constructions for double MSG 75.1 $P4$.

If we impose $\mathcal{T}$-symmetry, then the position-space Chern numbers must vanish, which enforces $z_{4R}$ to be zero.  Furthermore, if we add $\mathcal{T}$ symmetry to a system that respects double MSG 75.1 $P4$, we specifically find that the SI group becomes trivial.

\vspace{0.2in}
\subsubsection{Double SIs in Type-I Double MSG 77.13 $P4_{2}$}

The double MSG 77.13 $P4_2$ is generated by $\{E|100\}$, $\{E|010\}$, $\{E|001\}$, and $\{C_{4z}|00\frac12\}$.

The SI group of double MSG 77.13 $P4_{2}$ is $\ZZ_2$.  We can define the SI as half of the Chern number $C_{0}$ in the $k_z=0$ plane modulo $2$ (where we will show below that $C_{0}$ is always even due to the screw symmetry $\{C_{4z}|00\frac12\}$):
\begin{align}
z_{2R}^\pr =& \frac{C_0}{2} \text{ mod }2 \nono\\
=& N_\mrm{occ} - \frac14 n^{\frac12}_\Gamma + \frac14 n^{-\frac12}_\Gamma
-\frac34 n^{\frac32}_\Gamma + \frac34 n^{-\frac32}_\Gamma
-\frac14 n^{\frac12}_M + \frac14 n^{-\frac12}_M
-\frac34 n^{\frac32}_M + \frac34 n^{-\frac32}_M
- \frac12 n^{\frac12}_X + \frac12 n^{-\frac12}_X \text{ mod }2, 
\label{eq:z2Rp}
\end{align}
where $n_{\Gamma,M}^{\frac12,-\frac12,\frac32,-\frac32}$ are the number of occupied states with  $\{C_{4z}|00\frac12\}$ eigenvalues $e^{-i\frac{\pi}4}$, $e^{i\frac{\pi}4}$, $e^{-i\frac{3\pi}4}$, $e^{i\frac{3\pi}4}$, respectively, and $n^{\frac12,-\frac12}_X$ are the number of occupied states with $\{C_{2z}|\mathbf{0}\}$ eigenvalues $e^{-i\frac{\pi}2}$, $e^{i\frac{\pi}2}$, respectively.  For Chern number SIs determined by screw symmetry eigenvalues, we note that we may, in general, either define the SI using the screw eigenvalues in the $k_i=0$ plane or the eigenvalues in the $k_i=\pi$ plane.  However, as we will shortly see in the case of double MSG 84.51 $P4_{2}/m$ in Appendix~\ref{sec:P4_2/m}, if a mirror symmetry is also present whose matrix representatives do not commute with those of screw symmetry at all ${\bf k}$ points where both symmetries are in the little group $G_{\bf k}$, then additional constraints are imposed on the small (co)rep characters of screw.  Hence, in this work, we will only use screw eigenvalues in the $k_i=0$ plane to define double SIs.

Due to the monodromy of small (co)reps in nonsymmorphic SSGs [see Appendix~\ref{sec:compatibilityRelations}], the overall sign of each eigenvalue of $\{C_{4z}|00\frac12\}$ changes when $k_z$ is advanced through a period of the reciprocal lattice.  This implies the compatibility relations: 
$n_{\Gamma}^{\frac12}=n_{\Gamma}^{-\frac32}$, 
$n_{\Gamma}^{-\frac12}=n_{\Gamma}^{\frac32}$,
$n_{M}^{\frac12}=n_{M}^{-\frac32}$, 
$n_{M}^{-\frac12}=n_{M}^{\frac32}$.
Imposing the compatibility relation constraints on Eq.~(\ref{eq:z4R}) [and substituting $\Gamma,M,X$ for $Z,A,R$, respectively], we find that the Chern number $C_{0}$ is always even.

We therefore define $(C_0/2)\text{ mod }2$ [as opposed to $C_0\text{ mod }4$] to be the SI $z_{2R}'$ of double MSG 77.13 $P4_{2}$.  Using $N_{\rm occ} = n^{\frac12}_X + n^{-\frac12}_X$, and the compatibility relations, we then simplify Eq.~(\ref{eq:z2Rp}) to be:
\begin{equation}
z_{2R}^\pr =
-\frac12 n^{\frac32}_\Gamma + \frac12 n^{-\frac32}_\Gamma
-\frac12 n^{\frac32}_M + \frac12 n^{-\frac32}_M
+ \frac12 n^{\frac12}_X - \frac12 n^{-\frac12}_X \text{ mod }2.
\end{equation}
If an insulating state has $z_{2R}'=1$, then the state is a 3D QAH phase with $C_0\text{ mod }4=2$ in the $z$-direction.

We can also understand the even Chern number from the perspective of layer constructions.  Specifically, if a Chern layer is placed in the $z=z_0$ plane, then, because of the $\{C_{4z}|00\frac12\}$ symmetry, there must be another Chern layer with the same Chern number in the $z=z_0+\frac12$ plane.

If we impose $\mathcal{T}$ symmetry then the position-space Chern numbers must vanish, which enforces $z_{2R}'$ to be zero.

\vspace{-0.1in}
\subsubsection{Double SIs in Type-I Double MSG 81.33 $P\bar{4}$}
\label{sec:P-4}

The double MSG 81.33 $P\bar4$ is generated by $\{E|100\}$, $\{E|010\}$, $\{E|001\}$, and $\{S_{4z}|\mathbf{0}\}$.

\textit{SIs} -- The double MSG 81.33 $P\bar4$ has the SI group $\ZZ_4\times\ZZ_2^2$.  We choose the $\ZZ_4$ SI to be the Chern number in the $k_z=\pi$ plane modulo 4:
\begin{equation}
z_{4S} = C_{k_z=\pi} \text{ mod } 4 = - \frac12 n^{\frac12}_Z + \frac12 n^{-\frac12}_Z
-\frac32 n^{\frac32}_Z + \frac32 n^{-\frac32}_Z
-\frac12 n^{\frac12}_A + \frac12 n^{-\frac12}_A
-\frac32 n^{\frac32}_A + \frac32 n^{-\frac32}_A
+ n^{\frac12}_R - n^{-\frac12}_R \text{ mod }4,
\label{eq:z4S}
\end{equation}
where $n^{\frac12,-\frac12,\frac32,-\frac32}_{Z,A}$ are the number of occupied states with $\{S_{4z}|\mathbf{0}\}$ eigenvalues $e^{-i\frac{\pi}4}$, $e^{i\frac{\pi}4}$, $e^{-i\frac{3\pi}4}$, $e^{i\frac{3\pi}4}$, respectively, and $n^{\frac12,-\frac12}_R$ are the number of occupied states with $\{C_{2z}|\mathbf{0}\}$ eigenvalues $e^{-i\frac{\pi}2}$, $e^{i\frac{\pi}2}$, respectively.  Due to the compatibility relations, the occupied bands in the $k_z=0,\pi$ planes must have the same $\{C_{2z}|\mathbf{0}\}$ rotation eigenvalues; hence, the Chern numbers $C_{k_z=0}$ and $C_{k_z=\pi}$ have the same parity ($C_{k_z=0}\text{ mod }2=C_{k_z=\pi}\text{ mod }2$).  We define the first $\ZZ_2$ SI to be half of the difference between the Chern numbers in the $k_z=0,\pi$ planes, taken modulo 2:
\begin{align}
\delta_{2S} =& \frac{C_{k_z=\pi}-C_{k_z=0}}{2} \text{ mod }2 \nono\\
=&- \frac14 n^{\frac12}_Z + \frac14 n^{-\frac12}_Z
-\frac34 n^{\frac32}_Z + \frac34 n^{-\frac32}_Z
-\frac14 n^{\frac12}_A + \frac14 n^{-\frac12}_A
-\frac34 n^{\frac32}_A + \frac34 n^{-\frac32}_A
+\frac12 n^{\frac12}_R - \frac12 n^{-\frac12}_R \nono\\
&+\frac14 n^{\frac12}_\Gamma - \frac14 n^{-\frac12}_\Gamma
+ \frac34 n^{\frac32}_\Gamma - \frac34 n^{-\frac32}_\Gamma
+ \frac14 n^{\frac12}_M - \frac14 n^{-\frac12}_M
+ \frac34 n^{\frac32}_M - \frac34 n^{-\frac32}_M
- \frac12 n^{\frac12}_X + \frac12 n^{-\frac12}_X \text{ mod } 2.
\end{align}
Next using the relations:
\begin{equation}
n_{\Gamma,M}^{\frac12} +n_{\Gamma,M}^{-\frac32} =n_{Z,A}^{\frac12} +n_{Z,A}^{-\frac32},\ n_{\Gamma,M}^{-\frac12} +n_{\Gamma.M}^{\frac32} =n_{Z,A}^{-\frac12} +n_{Z,A}^{\frac32},\ n_X^{\frac12} = n_R^{\frac12},\ n_X^{-\frac12} = n_R^{-\frac12},
\end{equation}
we simplify $\delta_{2S}$:
\begin{equation}
\delta_{2S}=
- n^{\frac32}_Z +  n^{-\frac32}_Z
- n^{\frac32}_A +  n^{-\frac32}_A 
+ n^{\frac32}_\Gamma - n^{-\frac32}_\Gamma
+ n^{\frac32}_M - n^{-\frac32}_M \text{ mod } 2. 
\label{eq:d2S}
\end{equation}
Because of the difference of 2 (modulo 4) in the Chern numbers in the $k_{z}=0,\pi$ planes indicated by $\delta_{2S}=1$, we deduce that $\delta_{2S}=1$ indicates a WSM with $2+4n$ ($n\in\{\mathbb{Z}^{+},0\}$) Weyl points between $k_z=0$ and $k_z=\pi$.  Additionally, the Chern number in the $k_z=0$ plane  (modulo 4) is completely determined by the compatibility relations and the SIs -- specifically, $C_{k_z=0}\text{ mod }4 = z_{4S} - 2\delta_{2S}\text{ mod }4$.  We note that during the preparation of this work, an SI equivalent to $\delta_{2S}$ was introduced in Ref.~\onlinecite{ZhijunS4WeylHighThroughput} as an intermediate quantity relevant to the high-throughput numerical identification of nonmagnetic solid-state WSMs.

The second $\ZZ_2$ SI in double MSG 81.33 $P\bar4$ is given by:
\begin{equation}
z_2 = \sum_{K=\Gamma, M, Z, A}\frac{n^{\frac12}_K-n^{-\frac32}_K}2  \text{ mod }2.
\label{eq:z2}
\end{equation}
Below, we will show that $z_{2}$ is in one-to-one correspondence with the WSM and 3D TI invariant $(z_{2})_{P\bar{4}1'}$ in the Type-II double SSG 81.34 $P\bar{4}1'$ generated by adding $\{\mathcal{T}|{\bf 0}\}$ symmetry to double MSG 81.33 $P\bar{4}$, where $(z_{2})_{P\bar{4}1'}$ was previously introduced in Ref.~\onlinecite{ChenTCI}.  Hence, we will show below that a gapped state with $z_{2}=1$ is compatible with a fourfold-rotoinversion- ($S_{4}$) -protected AXI phase if $z_{4S}=\delta_{2S}=0$.

\textit{Layer constructions} -- We next employ the layer construction method to diagnose the topology of the symmetry-indicated topological insulating phases in double MSG 81.33 $P\bar4$, where the double SIs of each layer construction are given in the order $(z_{4S},\delta_{2S},z_2)$:
\begin{enumerate}
\item A $\hat{\bf z}$-normal Chern layer with $C_z=1$ in the $z=0$ plane can be realized by a 3D insulator whose occupied bands transform in the small irreps $\frac12,\frac32,-\frac12,\frac12,\frac32$, and $-\frac12$ at $\Gamma,M,X,Z,A$, and $R$ ($N_\mrm{occ}=1$), respectively.  The SIs of this layer construction are (101).
\item A $\hat{\bf z}$-normal Chern layer with $C_z=1$ in the $z=\frac12$ plane can be realized by a 3D insulator whose occupied bands transform in the small irreps $\frac12,\frac32,-\frac12,-\frac32,-\frac12$, and $-\frac12$ at $\Gamma,M,X,Z,A$, and $R$ ($N_\mrm{occ}=1$), respectively.  The SIs of this layer construction are (100). 
\end{enumerate}
Both of the layer constructions are 3D QAH states with $C=1$ in the $z$-direction.  Because $S_{4z}=\INV C_{4z}^{-1}$, and $\INV$ leads to an additional minus sign in the occupied $S_{4z}=\INV C_{4z}^{-1}$ eigenvalue in the $k_z=\pi$ plane contributed by the layer $z=\frac12$ (\emph{i.e.} $e^{-i\frac{2\pi}{4}j}\rightarrow -e^{-i\frac{2\pi}{4}j}$, see Appendix~\ref{sec:P-1}), then the $\{S_{4z}|\mathbf{0}\}$ eigenvalues at $Z$ and $A$ in the $z=\frac12$ layer construction have opposite signs compared to the occupied $\{S_{4z}|\mathbf{0}\}$ eigenvalues at $\Gamma$ and $M$, respectively.  We additionally note that the occupied $C_{2z}$ eigenvalues are required to be the same at $R$ and $X$ due to the compatibility relations.

\textit{The $S_4$ $\mbb{Z}_2$ invariant and axion insulators} -- When the total Chern number is zero and the bulk is insulating, the axion angle $\theta$ is given by $\theta\text{ mod }2\pi = \pi z_2$, where $z_2$ is termed the $S_4$ $\mbb{Z}_2$ invariant.  We may construct an AXI phase by placing a Chern layer with $C_z=1$ in the $z=0$ plane and a Chern layer with $C_z=-1$ in the $z=\frac12$ plane.  The AXI phase has the SIs (001).  However, we emphasize that the total Chern number cannot be completely determined by the SIs.  For example, the 3D QAH state consisting of a Chern layer with $C_z=3$ in the $z=0$ plane and a Chern layer with $C_z=1$ in the $z=\frac12$ plane also has the SIs (001).

\textit{Relationship with the SIs in other double SSGs} -- Double SSG 83.34 $P\bar41^\pr$, which is the double SSG that results from adding $\{\mathcal{T}|{\bf 0}\}$ symmetry to Type-I double MSG 81.33 $P\bar{4}$ -- has the SI group $\ZZ_2$.  The $\ZZ_2$ double SI in double SSG 83.34 $P\bar41^\pr$ either corresponds to a $\mathcal{T}$-invariant WSM, or to a $\mathcal{T}$-symmetric 3D TI~\cite{ChenTCI}.  Consequently, a 3D TI phase in double SSG 83.34 $P\bar41^\pr$ must subduce to an AXI in double MSG 83.33 $P\bar{4}$ if $\{S_{4z}|{\bf 0}\}$ and primitive lattice translation symmetries are preserved while breaking $\mathcal{T}$, because both insulators share the common nontrivial axion angle $\theta=\pi$.  Hence, the double SI subduction relations are given by:
\begin{equation}
(z_2)_{P\bar4 1^\pr} \to (z_{4S},\delta_{2S},z_2)_{P\bar4}=(00,z_2)_{P\bar4}.
\end{equation}

\subsubsection{Double SIs in Type-I Double MSG 83.43 $P4/m$}
\label{subsec:P4/m}

The double MSG 83.43 $P4/m$ is generated by $\{E|100\}$, $\{E|010\}$, $\{E|001\}$, $\{C_{4z}|\mathbf{0}\}$, and $\{m_z|\mathbf{0}\}$.

\textit{SIs} -- The double MSG 83.43 $P4/m$ has the SI group $\ZZ_4^3$.  We choose the three $\mathbb{Z}_{4}$-valued SIs to be:
\begin{align}
\delta_{4m} =& - C_{k_z=\pi}^+ + C_{k_z=0}^-\text{ mod }4 \nono\\
=&-\sum_{K=Z,A} \pare{-\frac12 n^{\frac12,+i}_K + \frac12 n^{-\frac12,+i}_K
-\frac32 n^{\frac32,+i}_K + \frac32 n^{-\frac32,+i}_K}
+ n^{\frac12,+i}_R - n^{-\frac12,+i}_R \nono\\
&+\sum_{K=\Gamma,M} \pare{-\frac12 n^{\frac12,-i}_K + \frac12 n^{-\frac12,-i}_K
-\frac32 n^{\frac32,-i}_K + \frac32 n^{-\frac32,-i}_K}
- n^{\frac12,-i}_X + n^{-\frac12,-i}_X \text{ mod } 4,
\label{eq:d4m}
\end{align}
\begin{equation}
z_{4m,\pi}^+ = C_{k_z=\pi}^+\text{ mod }4
= \sum_{K=Z,A} \pare{-\frac12 n^{\frac12,+i}_K + \frac12 n^{-\frac12,+i}_K
-\frac32 n^{\frac32,+i}_K + \frac32 n^{-\frac32,+i}_K}
+ n^{\frac12,+i}_R - n^{-\frac12,+i}_R \text{ mod } 4,
\label{eq:z4mp+}
\end{equation}
\begin{equation}
z_{4m,\pi}^- = C_{k_z=\pi}^-\text{ mod }4=
\sum_{K=Z,A} \pare{-\frac12 n^{\frac12,-i}_K + \frac12 n^{-\frac12,-i}_K
-\frac32 n^{\frac32,-i}_K + \frac32 n^{-\frac32,-i}_K}
+ n^{\frac12,-i}_R - n^{-\frac12,-i}_R \text{ mod } 4,
\label{eq:z4mp-}
\end{equation}
where the $\pm i$ superscripts indicate the signs of the mirror eigenvalues.  In Eqs.~(\ref{eq:d4m}),~(\ref{eq:z4mp+}), and~(\ref{eq:z4mp-}), we have defined $\delta_{4m}$ to be $-C_{k_z=\pi}^+ + C_{k_z=0}^-$, rather than $C_{k_z=\pi}^+ - C_{k_z=0}^-$, such that the double SI $z_8$ in double MSG 123.339 $P4/mmm$, which we will shortly define in Appendix~\ref{subsec:P4/mmm}, is related to $\delta_{4m}$ through the subduction relation $\delta_{4m} = z_8\text{ mod } 4$.

\textit{Layer constructions} -- To diagnose the topology associated to each nontrivial value of the double SIs $(\delta_{4m}, z_{4m,\pi}^+, z_{4m,\pi}^-)$, we employ the layer construction method.  We denote the Chern number in each mirror sector in the $k_z=0,\pi$ planes as $(C^+_{k_z=0},C^{-}_{k_z=0}, C^+_{k_z=\pi},C^-_{k_z=\pi})$, respectively.  We will also calculate the subduced SIs in the subgroups double MSG 2.4 $P\bar1$ and double MSG 81.33 $P\bar4$, which we will shortly use to determine the double SI subduction relations.  The layer constructions for Type-I double MSG 83.43 $P4/m$ are given by:
\begin{enumerate}
\item A $\hat{\bf z}$-normal layer with $C^+_z=1$, $C^-_z=0$ in the $z=0$ plane has the mirror sector Chern numbers $(1010)$ and the SIs $(310)$. The subduced subgroup SIs are $(\eta_{4I},z_{2I,1},z_{2I,2},z_{2I,3})_{P\bar1}=(2001)_{P\bar1}$, $(z_{4S},\delta_{2S},z_2)_{P\bar4}=(101)_{P\bar4}$.
\item A $\hat{\bf z}$-normal layer with $C^+_z=0$, $C^-_z=1$ in the $z=0$ plane has the mirror sector Chern numbers $(0101)$ and the SIs $(101)$. The subduced subgroup SIs are $(2001)_{P\bar1}$, $(101)_{P\bar4}$.
\item A $\hat{\bf z}$-normal layer with $C^+_z=1$, $C^-_z=0$ in the $z=\frac12$ plane has the mirror sector Chern numbers $(1001)$ and the SIs $(001)$. The subduced subgroup SIs are $(0001)_{P\bar1}$, $(100)_{P\bar4}$.
\item A $\hat{\bf z}$-normal layer with $C^+_z=0$, $C^-_z=1$ in the $z=\frac12$ plane has the mirror sector Chern numbers $(0110)$ and the SIs $(010)$. The subduced subgroup SIs are $(0001)_{P\bar1}$, $(100)_{P\bar4}$.
\end{enumerate}
We emphasize that Chern insulators whose normal vectors lie in the $xy$-plane are disallowed by $\{m_z|\mathbf{0}\}$ symmetry.

\textit{Relationship with the SIs in other double SSGs} -- In order to identify the AXI phases, we will subduce the SIs in double MSG 83.43 $P4/m$ onto the SIs in double MSG 2.4 $P\bar1$ and double MSG 81.33 $P\bar4$.  The subduction relations are given by:
\begin{equation}
\pare{\delta_{4m}, z_{4m,\pi}^+, z_{4m,\pi}^-}_{P4/m} \to (\eta_{4I},z_{2I,1},z_{2I,2},z_{2I,3})_{P\bar1} = \pare{2(\delta_{4m} \text{ mod }2), 0,0,\; z_{4m,\pi}^+ + z_{4m,\pi}^-\text{ mod }2}_{P\bar1},
\end{equation}
\begin{equation}
\pare{\delta_{4m}, z_{4m,\pi}^+, z_{4m,\pi}^-}_{P4/m} \to (z_{4S}, \delta_{2S}, z_{2})_{P\bar4} = \pare{z_{4m,\pi}^+ + z_{4m,\pi}^-\text{ mod }4,\; 0,\; \delta_{4m} \text{ mod }2}_{P\bar4},
\end{equation}
which imply that $\eta_{2I}^\pr=\frac12\eta_{4I}=z_2=\delta_{4m}\text{ mod }2$ [see Eqs.~(\ref{eq:eta2Ip}) and~(\ref{eq:z2})].  In MSG 2.4 $P\bar1$ and MSG 81.33 $P\bar4$, we previously found in Appendices~\ref{sec:P-1} and~\ref{sec:P-4} that the $\eta_{2I}'=1$ and $z_2=1$ states are AXIs protected by $\{\INV|\mathbf{0}\}$ and $\{S_{4z}|\mathbf{0}\}$, respectively (provided that the non-symmetry-indicated net Chern numbers are zero).  Hence, the AXI phases in MSG 83.43 $P4/m$ are simultaneously protected by $\{\INV|\mathbf{0}\}$ and $\{S_{4z}|\mathbf{0}\}$.

Lastly, we will study the effects of imposing $\TRS$ symmetry.  Adding $\{\mathcal{T}|{\bf 0}\}$ symmetry to Type-I double MSG 83.43 $P4/m$ generates the Type-II double SSG 83.44 $P4/m1^\pr$, which has the SI group $\ZZ_8\times \ZZ_4 \times \ZZ_2$.  The SIs in double SSG 83.44 $P4/m1^\pr$ are related to the SIs in double MSG 83.43 ${P4/m}$ through the subduction relations:
\begin{equation}
\pare{z_8, z_{4m,\pi},z_{2w,1}}_{P4/m1^\pr} \to  \pare{\delta_{4m}, z_{4m,\pi}^+, z_{4m,\pi}^-}_{P4/m}= \pare{z_8\text{ mod }4,-z_{4m,\pi},z_{4m,\pi} }_{P4/m}.
\end{equation}

The subduction relations imply that strong 3D TIs in double SSG 83.44 $P4/m1^\pr$ indicated by odd $z_8$ and mirror TCIs indicated by $z_8$ mod 4 and $z_{4m,\pi}$ will continue to exhibit symmetry-indicated nontrivial topology if $\{\TRS|{\bf 0}\}$ is broken while preserving the symmetries of double MSG 83.43 $P4/m$.  Conversely, the weak TI phases indicated by $z_{2w,1}$ and the rotation-anomaly HOTI indicated by $z_8=4$ in double SSG 83.44 $P4/m1^\pr$ no longer exhibit symmetry-indicated stable topology when subduced onto double MSG 83.43 ${P4/m}$.  Specifically, the SIs $(400)_{P4/m1'}$ correspond to either a mirror TCI phase with $C^+_{k_z=0}\text{ mod }4 = 8$ or $C^+_{k_z=\pi}\text{ mod }8=4$ or a HOTI with vanishing mirror Chern numbers~\cite{ChenTCI}. The HOTI phase has a gapless top ($\hat{\bf z}$-normal) surface~\cite{ChenRotation} with $4+8n$ ($n\in\{\mathbb{Z}^{+},0\}$) twofold Dirac cones that are locally protected by $\{C_{2z}\times\mathcal{T}|{\bf 0}\}$ symmetry and are anomalous due to surface and bulk $\{C_{4z}|{\bf 0}\}$ symmetry (see Appendix~\ref{sec:HOTIfermionDoubling} and Ref.~\onlinecite{ChenRotation}).  The HOTI phase, when cut into a $4/m1'$-symmetric rod geometry, exhibits $4+8n$ helical hinge states that are locally protected by $\TRS$ symmetry and globally protected by $\{C_{4z}|\mathbf{0}\}$ symmetry.  If $\mathcal{T}$ symmetry is relaxed, then the HOTI hinge states must become gapped, because there are no side-surface mirror lines to protect helical spectral flow in the absence of $\mathcal{T}$ symmetry in MSG 83.33 $P4/m$ (see Appendix~\ref{sec:newHOTIs}).  We leave the finer question of whether any non-symmetry-indicated crystalline topology in MSG 83.33 $P4/m$ is subduced from the $(400)_{P4/m1'}$ HOTI phase in double SSG 83.44 $P4/m1^\pr$ for future works.

\clearpage

\subsubsection{Double SIs in Type-I Double MSG 84.51 $P4_{2}/m$} 
\label{sec:P4_2/m}

The double MSG 84.51 $P4_2/m$ is generated by $\{E|100\}$, $\{E|010\}$, $\{E|001\}$, $\{C_{4z}|00\frac12\}$, and $\{m_z|\mathbf{0}\}$.

\textit{SIs} -- The double MSG 84.51 $P4_2/m$ has the SI group $\ZZ_4\times \ZZ_2$.  We define the two SIs to be:
\begin{align}
z_{4m,0}^+ =& C_{k_z=0}^+\text{ mod } 4 \nono\\
=&  \sum_{K=\Gamma,M} \pare{
-\frac12 n^{\frac12,+i}_K + \frac12 n^{-\frac12,+i}_K
-\frac32 n^{\frac32,+i}_K + \frac32 n^{-\frac32,+i}_K}
+ n^{\frac12,+i}_X - n^{-\frac12,+i}_X \text{ mod } 4, 
\label{eq:z4m0+}
\end{align}
\begin{equation}
\delta_{2m} = C_{k_z=\pi}^+-C_{k_z=0}^-\text{ mod }2,
\end{equation}
where an explicit formula for $\delta_{2m}$ was previously provided in Eq.~(\ref{eq:d2m}).  Because the matrix representatives of $\{C_{4z}|00\frac12\}$ and $\{m_z|\mathbf{0}\}$ do not commute in all of the small irreps at the ${\bf k}$ points in the $k_{z}=\pi$ plane at which $\{C_{4z}|00\frac12\}$ and $\{m_z|\mathbf{0}\}$ are both elements of the little group, then we cannot determine the mirror sector Chern numbers (modulo 4) in the $k_z=\pi$ plane using $\{C_{4z}|00\frac12\}$ eigenvalues.  Conversely, because the matrix representatives of $\{C_{2z}|\mathbf{0}\}$ and $\{m_z|\mathbf{0}\}$ \emph{commute} in all of the small irreps at the ${\bf k}$ points in the $k_{z}=\pi$ plane at which $\{C_{2z}|\mathbf{0}\}$ and $\{m_z|\mathbf{0}\}$ are both elements of the little group, then we \emph{can} determine the mirror sector Chern numbers (modulo 2) in the $k_z=\pi$ plane using the occupied $\{C_{2z}|\mathbf{0}\}$ eigenvalues.  We thus specifically determine that $\delta_{2m} = C_{k_z=\pi}^+-C_{k_z=0}^-\text{ mod }2$.

\textit{Layer constructions} -- We find that all of the double SIs in double MSG 84.51 $P4_2/m$ can be realized by layer constructions.  Before introducing the layer constructions, we first note that the mirror planes in double MSG 84.51 $P4_2/m$ lie at $z=0,\frac12$.  However, the $\mathcal{I}$ centers lie in the $z=0,\frac12$ planes, whereas, conversely, the $S_4$ centers lie in the $z=\frac14,\frac34$ planes.  For each layer construction, we also compute the subduced SIs in the subgroup MSG 2.4 $P\bar1$, which we will shortly use to determine the SI subduction relations.  The layer constructions of the double SIs $(z_{4m,0}^+,\delta_{2m})$ in double MSG 84.51 $P4_2/m$ are given by:
\begin{enumerate}
\item A $\hat{\bf z}$-normal layer with $C^+_z=1$, $C^-_z=0$ in the $z=0$ plane. Due to the $\{C_{4z}|00\frac12\}$ symmetry, there is another $C^+_z=1$, $C^-_z=0$ layer in the $z=\frac12$ plane. The mirror sector Chern numbers in momentum space are $(C_{k_z=0}^+,C_{k_z=0}^-,C_{k_z=\pi}^+,C_{k_z=\pi}^-)=(2011)$, where the subscripts 0 and $\pi$ indicate values of $k_z$. The SIs are (21). The subduced subgroup SIs are $(\eta_{4I},z_{2I,1},z_{2I,2},z_{2I,3})_{P\bar1}=(2000)_{P\bar1}$, $(z_{4S},\delta_{2S},z_2)_{P\bar4}=(200)_{P\bar4}$.
\item A $\hat{\bf z}$-normal layer with $C^+_z=0$, $C^-_z=1$ in the $z=0$ plane. Due to the $\{C_{4z}|00\frac12\}$ symmetry, there is another $C^+_z=0$, $C^-_z=1$ layer in the $z=\frac12$ plane. The mirror sector Chern numbers in momentum space are $(C_{k_z=0}^+,C_{k_z=0}^-,C_{k_z=\pi}^+,C_{k_z=\pi}^-)=(0211)$. The SIs are (01). The subduced subgroup SIs are $(2000)_{P\bar1}$, $(200)_{P\bar4}$.
\item A $\hat{\bf z}$-normal layer with $C_z=1$ in the $z=\frac14$ plane. Due to the $\{C_{4z}|00\frac12\}$ symmetry, there is another $C_z=1$ layer in the $z=\frac34$ plane. The mirror sector Chern numbers in momentum space are $(C_{k_z=0}^+,C_{k_z=0}^-,C_{k_z=\pi}^+,C_{k_z=\pi}^-)=(1111)$. The SIs are (10). The subduced subgroup SIs are $(0000)_{P\bar1}$, $(201)_{P\bar4}$.
\end{enumerate}

\textit{Relationship with the SIs in other double SSGs} -- In order to later identify the AXI phases, we subduce the SIs onto double MSG 2.4 $P\bar1$ and double MSG 81.33 $P\bar4$:
\begin{equation}
(z_{4m,0}^+,\delta_{2m})_{P4_2/m} \to (\eta_{4I},z_{2I,1},z_{2I,2},z_{2I,3})_{P\bar1}= (2\delta_{2m},000)_{P\bar1},
\end{equation}
\begin{equation}
(z_{4m,0}^+,\delta_{2m})_{P4_2/m} \to (z_{4S},\delta_{2S},z_2)_{P\bar4}=(2z_{4m,0}-2\delta_{2m}\text{ mod }4,\; 0,\; z_{4m,0}^+\text{ mod }2)_{P\bar4}.
\end{equation}

We next study the effects of imposing $\TRS$ symmetry.  The double SSG 84.52 $P4_2/m1^\pr$ -- the SSG generated by adding $\{\mathcal{T}|{\bf 0}\}$ symmetry to MSG 84.51 $P4_{2}/m$ -- has the SI group $\ZZ_4\times \ZZ_2$. The $\ZZ_4$ SI is the parity index $z_4$ (\emph{i.e.} the $\mathcal{I}$ $z_{4}$ index), and the $\ZZ_2$ SI is the $\mathcal{I}$ $z_{2w,1}$ index.  Hence, the subduction relations are given by:
\begin{equation}
(z_4,z_{2w,1})_{P4_2/m1^\prime} \to  (z_{4m,0}^+,\delta_{2m})_{P4_2/m} = ( z_4, z_4\text{ mod }2 )_{P4_2/m},
\end{equation} 
implying that adding $\{\TRS|{\bf 0}\}$ to an insulating phase in double MSG 84.51 $P4_2/m$ results in an insulator with the SIs $\delta_{2m}=z_4\text{ mod }2$.  Furthermore, in an insulator, it is required that $C_{k_z=0}^+ + C_{k_z=0}^- = C_{k_z=\pi}^+ + C_{k_z=\pi}^- = 2 C_{k_z=\pi}^+$.  $\{\TRS|{\bf 0}\}$ further enforces $C_{k_z=\pi}^+=0$, $C_{k_z=0}^+ = -C_{k_z=0}^-$, such that $\delta_{2m}=C_{k_z=\pi}^+ - C_{k_z=0}^-\text{ mod }2 = C_{k_z=0}^+\text{ mod }2 = z_4\text{ mod }2$.

\textit{Axion insulators} -- Because the $S_4$ centers in position space do not coincide with the $\mathcal{I}$ centers in MSG 84.51 $P4_{2}/m$, then the $S_4$ invariant $z_2 = z_{4m,0}^+\text{ mod }2$ is free to differ from the $\mathcal{I}$ invariant $\eta_{2I}^\pr=\delta_{2m}$.  An AXI phase must have vanishing position-space Chern numbers, as well as $z_2=\eta_{2I}^\pr=1$, due to the definitions $\theta\text{ mod }2\pi=\pi \eta_{2I}^\pr$ and $\theta\text{ mod }2\pi=\pi z_{2}$ (see Appendices~\ref{sec:P-1} and~\ref{sec:P-4}, respectively).  Thus, in order to guarantee that the net Chern numbers vanish, we may, for example, only construct an AXI phase with $C=1$ layers at $z=0,\frac12$ if $C=-1$ layers are additionally placed at $z=\frac14,\frac34$.  In this example of an AXI, the $C=1$ ($C=-1$) Chern layers occupy the $\mathcal{I}$ ($S_{4}$ centers).

\subsubsection{Double SIs in Type-I Double MSG 88.81 $I4_{1}/a$}
\label{subsec:88.81}

The double MSG 88.81 $I4_1/a$ is generated by 
    $\{E|-\frac12,\frac12,\frac12\}$, 
    $\{E|\frac12,-\frac12,\frac12\}$,
    $\{E|\frac12,\frac12,-\frac12\}$,
    $\{C_{4z}|\frac34 \frac14 \frac14\}$, 
and $\{\INV|\mathbf{0}\}$.

\textit{SIs} -- The double MSG 88.81 $I4_1/a$ has the SI group $\ZZ_2^2$.  As we will explicitly derive later in this section, the first $\ZZ_2$ SI $\eta^\pr_{2I}$ is related by subduction to the $\mathcal{I}$ invariant $\eta_{2I}^\prime$ in double MSG 2.4 $P\bar1$ [Eq.~(\ref{eq:eta2Ip})]:
\begin{equation}
\eta_{2I}^\pr = \frac{\eta_{4I}}2\text{ mod }2=
\frac12 n_\Gamma^- + \frac12 n_M^- + \frac12 n_X^- + \frac12 n_X^+ +\frac32n_N^- + \frac12 n_N^+ \text{ mod }2.
\end{equation}
The second $\ZZ_2$ SI $z_2$ is related by subduction to the $S_{4}$ invariant $z_2$ in double MSG 81.33 $P\bar4$ [Eq.~(\ref{eq:z2})]:
\begin{equation}
z_2 = \frac{n_{\Gamma}^\frac12-n_{\Gamma}^{-\frac32}}2 
+ \frac{n_P^\frac12-n_P^{-\frac32} + n_P^{\frac32}-n_P^{-\frac12} }2.
\end{equation}

\textit{Layer constructions} -- The double MSG 88.81 has a body-centered lattice generated by:
\begin{equation}
\mbf{a}_1=(-\frac12,\frac12,\frac12),\qquad
\mbf{a}_2=(\frac12,-\frac12,\frac12),\qquad
\mbf{a}_3=(\frac12,\frac12,-\frac12).
\end{equation}
There are two types of maximal Wyckoff positions: $\mathcal{I}$ centers: 
\begin{equation}
8c:\qquad 
(0,0,0),\qquad 	(\frac12,0,\frac12),\qquad 	
(\frac34,\frac14,\frac14),\qquad (\frac34,\frac34,\frac34),
\end{equation}
\begin{equation}
8d:\qquad
(0,0,\frac12),\qquad 	(\frac12,0,0),\qquad 	
(\frac34,\frac14,\frac34),\qquad 	(\frac34,\frac34,\frac14),
\end{equation}
and $S_4$ ($\{S_{4z}|\frac14\frac34\frac34\}=\{C_{4z}|\frac14\frac34\frac34\}\{\INV|\mathbf{0}\}$) centers: 
\begin{equation}
4a:\qquad
(0,\frac14,\frac18),\qquad 	(\frac12,\frac14,\frac38),
\end{equation}
\begin{equation}
4b:\qquad
(0,\frac14,\frac58),\qquad 	(\frac12,\frac14,\frac78),
\end{equation}
using the notation of the~\href{https://www.cryst.ehu.es/cgi-bin/cryst/programs/magget_wp.pl}{MWYCKPOS} tool on the BCS~\cite{BCSMag1,BCSMag2,BCSMag3,BCSMag4}, and where all coordinates are given in the conventional cell.  We consider the following layer constructions:
\begin{enumerate}
\item A $\hat{\bf z}$-normal Chern layer with $C_z=1$ in the $z=0$ plane. The screw symmetry operation additionally generates Chern layers with $C_z=1$ in the $z=\frac14,\frac12,\frac34\cdots$ planes.  All of the $\mathcal{I}$ centers are occupied, all of the $S_4$ centers are unoccupied, and the total Chern number in each unit cell is $C_{z}=2$, such that $\eta_{2I}^\pr=1$, $z_2=0$.
\item A $\hat{\bf z}$-normal Chern layer with $C_z=1$ in the $z=\frac18$ plane. The screw symmetry operation additionally generates Chern layers with $C_z=1$ in the $z=\frac38,\frac58,\frac78\cdots$ planes.  All of the $\mathcal{I}$ centers are unoccupied, all of the $S_4$ centers are occupied, and the total Chern number in each unit cell is $C_{z}=2$, such that $\eta_{2I}^\pr=0$, $z_2=1$. 
\end{enumerate}

\textit{Axion insulators} -- Because the $S_4$ centers do not coincide with the $\mathcal{I}$ centers in position space in double MSG 88.81 $I4_1/a$, then the $S_4$ invariant $z_2$ is free to differ from the $\mathcal{I}$ invariant $\eta_{2I}^\pr$.  An AXI phase must have vanishing position-space Chern numbers, as well as $z_2=\eta_{2I}^\pr=1$, due to the definitions $\theta\text{ mod }2\pi=\pi \eta_{2I}^\pr$ and $\theta\text{ mod }2\pi=\pi z_{2}$ (see Appendices~\ref{sec:P-1} and~\ref{sec:P-4}, respectively).  Hence, to generate an AXI with vanishing position-space Chern numbers by placing $C=1$ layers at $z=0,\frac14,\frac12,\frac34$, we must also place $C=-1$ layers at $z=\frac18,\frac38,\frac58,\frac78$, such that the Chern layers with $C=1$ ($C=-1$) occupy the $\mathcal{I}$ ($S_{4}$) centers.

\textit{Relationship with the SIs in other double SSGs} -- We will now study the effects of imposing $\TRS$ symmetry.  The double SSG 88.82 $I4_1/a1^\pr$ -- which is generated by adding $\{\mathcal{T}|{\bf 0}\}$ to Type-I double MSG 88.81 $I4_{1}/a$ -- has the SI group $\ZZ_4$.  The subduction relations for the double SIs are given by: \begin{equation}
(z_4)_{I4_1/a1^\pr} \to (\eta_{2I}',z_{2})_{I4_1/a}= (z_4\text{ mod }2,z_4\text{ mod }2)_{I4_1/a}.
\end{equation} 
Hence, a symmetry-indicated 3D TI in $I4_1/a1^\pr$ will necessarily become an $\mathcal{I}$- or $S_4$-protected AXI if $\TRS$ symmetry is relaxed while preserving the symmetries of MSG 88.81 $I4_{1}/a$, because infinitesimal $\mathcal{T}$-breaking in a 3D insulator cannot change the momentum-space Chern numbers of the occupied bands in any 2D BZ plane.

\textit{Subduction of $\eta_{2I}^\pr$ onto double MSG 2.4 $P\bar{1}$} -- In MSG 88.81 $I4_{1}/a$, the reciprocal lattice is generated by:
\begin{equation}
\mbf{b}_1 = (0,2\pi,2\pi),\qquad
\mbf{b}_2 = (2\pi,0,2\pi),\qquad
\mbf{b}_3 = (2\pi,2\pi,0).
\end{equation}
There are four inequivalent, $\mathcal{I}$-invariant momenta:
\begin{equation}
\Gamma(0,0,0),\qquad M(2\pi,0,0),\qquad X(\pi,\pi,0),\qquad N(\pi,0,\pi),
\label{eq:88conventional}
\end{equation}
where the equivalence between ${\bf k}$ points is defined in Eq.~(\ref{eq:independentKFull}) and the surrounding text, and where the coordinates of $\Gamma$, $M$, $X$, and $N$ in Eq.~(\ref{eq:88conventional}) are given in the conventional cell.

The star of $X$ has two arms -- $X_1(\pi,\pi,0)$ and $X_2(\pi,-\pi,0)$, which are related by the screw operation $\{C_{4z}|\frac34 \frac14 \frac14\}$.
If $\ket{\psi_{X_1}}$ is a Bloch state at $X_1$, then $\ket{\psi_{X_2}}=\{C_{4z}|\frac34 \frac14 \frac14\} \ket{\psi_{X_1}}$ is a state at $X_2$.  Taking $\ket{\psi_{X_1}}$ to have the parity ($\mathcal{I}$) eigenvalue $\xi$, we will now determine the parity eigenvalue of $\ket{\psi_{X_2}}$.  Because:
\begin{equation}
\{\INV|\mathbf{0}\}\{C_{4z}|\frac34 \frac14 \frac14\} \{\INV|\mathbf{0}\}^{-1}=
\{E|-\frac32,-\frac12,-\frac12\}
\{C_{4z}|\frac34 \frac14 \frac14\},
\end{equation}
then:
\begin{equation}
\{\INV|\mathbf{0}\} \ket{\psi_{X_2}} = \{E|-\frac32,-\frac12,-\frac12\} \xi \ket{\psi_{X_2}}= - \xi \ket{\psi_{X_2}}.
\end{equation}
Hence, taking the parity eigenvalue of $\ket{\psi_{X_1}}$ to be $\xi$, the parity eigenvalue of $\ket{\psi_{X_2}}$ is $-\xi$.

Next, the star of $N$ has four arms: $N_1(\pi,0,\pi)$, $N_2(0,\pi,\pi)$, $N_3(-\pi,0,\pi)$, and $N_4(0,-\pi,\pi)$, which are related to $N_{1}$ by the operations, $\{E|\mathbf{0}\}$, $\{C_{4z}|\frac34 \frac14 \frac14\}$, $\{C_{2z}|\frac12 0 \frac12\}$, and $\{C_{4z}^{-1}|\frac34\frac34\frac34\}$, respectively.  Because:
\begin{equation}
\{\INV|\mathbf{0}\}\{C_{4z}|\frac34 \frac14 \frac14\}\{\INV|\mathbf{0}\}^{-1} 
=\{E|-\frac32,-\frac12,-\frac12\} \{C_{4z}|\frac34 \frac14 \frac14\},
\end{equation}
\begin{equation}
\{\INV|\mathbf{0}\} \{C_{2z}|\frac12 0 \frac12\} \{\INV|\mathbf{0}\}^{-1} 
=\{E|-1,0,-1\} \{C_{2z}|\frac12 0 \frac12\},
\end{equation}
\begin{equation}
\{\INV|\mathbf{0}\} \{C_{4z}^{-1}|\frac34\frac34\frac34\} \{\INV|\mathbf{0}\}^{-1} 
=\{E|-\frac32, -\frac32, -\frac32\} \{C_{4z}^{-1}|\frac34\frac34\frac34\},
\end{equation}
then the extra phase factor in the SI for the occupied parity eigenvalue at $\kk_{N_\alpha}$ is given by $e^{-i\mathbf{t}_\alpha\cdot \kk_{N_\alpha}}$ ($\alpha=2,3,4$), where $\mathbf{t}_\alpha$ is the extra translation determined above, and where $\kk_{N_\alpha}$ is the momentum $N_\alpha$.  The parity SI phases at $N_2$, $N_3$, and $N_4$ are thus $-1$, $1$, and $1$, respectively.

To determine the $\mathcal{I}$ double SI $\eta_{4I}$, we apply Eq.~(\ref{eq:eta4I}) to the parity eigenvalue multiplicities at the eight $\mathcal{I}$-invariant momenta $\Gamma$, $M$, $X_{1,2}$, and $N_{1,2,3,4}$, respectively:
\begin{equation}
\eta_{4I} = n_\Gamma^- + n_M^- + n_X^- + n_X^+ + 3n_N^- + n_N^+ \text{ mod } 4.
\end{equation}
We find that the parity eigenvalues enforce that $\eta_{4I}\text{ mod }2=0$.  Hence, the $\mathcal{I}$ double SI in MSG 88.81 $I4_{1}/a$ is $\eta_{2I}'=\frac{1}{2}\eta_{4I}$ [Eq.~(\ref{eq:eta2Ip})].

\textit{Subduction of $z_2$ onto double MSG 81.33 $P\bar4$} -- There are three inequivalent $S_4$-invariant momenta:
\begin{equation}
\Gamma(0,0,0),\qquad
M(2\pi,0,0),\qquad
P(\pi,\pi,\pi).
\end{equation}

First, the star of $P$ has two arms -- $P_1(\pi,\pi,\pi)$ and $P_2(-\pi,-\pi,-\pi)$, which are related by $\mathcal{I}$.  Because,
\begin{equation}
\{S_{4z}|\frac14\frac34\frac34\} \{\INV|\mathbf{0}\} \{ S_{4z}|\frac14\frac34\frac34\}^{-1} 
=\{E|\frac12, \frac32, \frac32\} \{\INV|\mathbf{0}\},
\end{equation}
and:
\begin{equation}
\exp\pare{i\pare{\frac12, \frac32, \frac32}\cdot (-\pi,-\pi,-\pi)} = \exp\pare{i\frac{\pi}2},
\end{equation}
then, if $P_1$ has a Bloch state with the $\{S_{4z}|\frac14, \frac34, \frac34\}$ eigenvalue $e^{-i\frac{\pi}2 j}$, $P_2$ is required to have a Bloch state with the $\{S_{4z}|\frac14, \frac34, \frac34\}$ eigenvalue $e^{-i\frac{\pi}2(j-1)}$.  We thus conclude that:
\begin{equation}
n_{P_1}^j = n_{P_2}^{j-1}.
\end{equation}

To determine the $S_{4}$ double SI $z_{2}$, we apply Eq.~(\ref{eq:z2}) to the $\{S_4|\frac14\frac34\frac34\}$ eigenvalue multiplicities at the four $\{S_4|\frac14\frac34\frac34\}$-invariant momenta $\Gamma$, $M$, and $P_{1,2}$:
\begin{equation}
z_2 = \frac{n_{\Gamma}^\frac12-n_{\Gamma}^{-\frac32}}2 
+ \frac{n_M^\frac12-n_M^{-\frac32}}2 
+ \frac{n_P^\frac12-n_P^{-\frac32} + n_P^{\frac32}-n_P^{-\frac12} }2 \text{ mod } 2.
\label{eq:tempBodyCenterZhida}
\end{equation}
Using the~\href{http://www.cryst.ehu.es/cryst/corepresentations}{Corepresentations},~\href{https://www.cryst.ehu.es/cryst/mcomprel}{MCOMPREL}, and~\href{http://www.cryst.ehu.es/cryst/mbandrep}{MBANDREP} tools on the BCS introduced in this work (see Appendices~\ref{sec:coreps},~\ref{sec:compatibilityRelations}, and~\ref{sec:mbandrep}, respectively), we find that $n_M^\frac12=n_M^{-\frac32}$ is required in any insulating state in double MSG 88.81 $I4_{1}/a$.  Hence, the factor of $\frac{n_M^\frac12-n_M^{-\frac32}}2$ can be omitted in Eq.~(\ref{eq:tempBodyCenterZhida}), leading to a final expression:
\begin{equation}
z_2 = \frac{n_{\Gamma}^\frac12-n_{\Gamma}^{-\frac32}}2 
+ \frac{n_P^\frac12-n_P^{-\frac32} + n_P^{\frac32}-n_P^{-\frac12} }2 \text{ mod } 2.
\end{equation}

\subsubsection{Double SIs in Type-I Double MSG 123.339 $P4/mmm$}
\label{subsec:P4/mmm}

The double MSG 123.339 $P4/mmm$ is generated by 
    $\{E|100\}$, 
    $\{E|010\}$,
    $\{E|001\}$,
    $\{C_{4z}|\mathbf{0}\}$, 
    $\{m_x|\mathbf{0}\}$,
and $\{\INV|\mathbf{0}\}$.

\textit{SIs} -- The double MSG 123.339 $P4/mmm$ has the SI group $\ZZ_8\times\ZZ_4\times\ZZ_2$.  In double-valued small irreps of the little groups at the $\mathcal{I}$-invariant ${\bf k}$ points, the matrix representatives of perpendicular mirror symmetries  (\emph{e.g.} $\{m_x|\mathbf{0}\}$ and $\{m_y|\mathbf{0}\}$) anticommute.  Hence, Bloch states at the eight $\mathcal{I}$-invariant momenta must be at least twofold degenerate (and in fact are exactly twofold degenerate in double MSG 123.339 $P4/mmm$).  The double SIs can be chosen to be the same as the double SIs of SSG 123.340 $P4/mmm1^\pr$ (previously introduced in Refs.~\onlinecite{ChenTCI,AshvinTCI,HOTIChen}), because the addition of $\TRS$ symmetry to double MSG 123.339 $P4/mmm$ does not change the dimensions and characters of the small irreps at the high-symmetry BZ points or the compatibility relations between the high-symmetry-point small irreps.  In the physical basis, the $\ZZ_8$ double SI is:
\begin{equation}
z_8 = \frac32 n^{\frac32,+} - \frac32 n^{\frac32,-} - \frac12 n^{\frac12,+} + \frac12 n^{\frac12,-} \text{ mod } 8, 
\label{eq:z8}
\end{equation}
\begin{equation}
n^{j,\pm} = \sum_{K=\Gamma,M,Z,A} n^{j,\pm}_K + \sum_{K=X,R} n^{\frac12,\pm}_K, 
\label{eq:nj-z8}
\end{equation}
where $n_K^{j,\pm}$ is the number of states with parity ($\mathcal{I}$) eigenvalue $\pm 1$ and angular momentum $j$ (modulo 4) at the momentum $K$, which corresponds to the $\{C_{4z}|{\bf 0}\}$ eigenvalue $e^{-i\frac{2\pi}4 j}$ at $K=\Gamma,M,Z,A$, and the $\{C_{2z}|{\bf 0}\}$ eigenvalue $e^{-i\frac{\pi}2 j}$ at $K=X,R$.  The $\ZZ_4$ SI is $z_{4m,\pi}^- $, which indicates the Chern number in the negative mirror sector in the $k_{z}=\pi$ plane $z_{4m,\pi}^-$, and is related by subduction to the same SI ($z_{4m,\pi}^-$) in double MSG 83.43 $P4/m$ [Eq.~(\ref{eq:z4mp-})].  The $\ZZ_2$ SI corresponds to the weak TCI $\mathcal{I}$ invariant $z_{2w,1}$ for the mirror Chern number (modulo 2) in the $k_{x,y}=\pi$ planes, and is related by subduction to the same SI ($z_{2w,1}$) in double MSG 47.249 $Pmmm$ [Eq.~(\ref{eq:z2w})].

\textit{Layer constructions} -- To diagnose the topology associated to each nontrivial value of the double SIs, we employ the layer construction method.  In the layer constructions below, $C^+=-C^-$ due to the net-zero Chern numbers enforced by the mirror symmetries.  Hence, we will omit $C^-$ in further discussions of the topology in double MSG 123.339 $P4/mmm$.  The layer constructions for the double SIs $(z_8,z_{4m,\pi}^+,z_{2w,1})$ in MSG 123.339 $P4/mmm$ are given by:
\begin{enumerate}
\item A $\hat{\bf z}$-normal layer with $C^+_z=1$ in the $z=0$ plane has the SIs (230).
\item A $\hat{\bf z}$-normal layer with $C^+_z=1$ in the $z=\frac12$ plane has the SIs (010).
\item An $\hat{\bf x}$-normal layer with $C^+_x=1$ in the $x=0$ plane has the SIs (401).  We emphasize that, in this layer construction, there is also a superposed $\hat{\bf y}$-normal layer with $C^+_y=1$ in the $y=0$ plane implied by the $\{C_{4z}|\mathbf{0}\}$ rotation symmetry.
\item An $\hat{\bf x}$-normal layer with $C^+_x=1$ in the $x=\frac12$ plane has the SIs (001).  We emphasize that, in this layer construction, there is also a superposed $\hat{\bf y}$-normal layer with $C^+_y=1$ in the $y=\frac12$ plane implied by the $\{C_{4z}|\mathbf{0}\}$ rotation symmetry.
\item An $(\hat{\bf x}+\hat{\bf y})$-normal layer with $C^+_{x+y}=1$ in the $x+y=0$ plane has the SIs (400).  We emphasize that, in this layer construction, there is also a superposed $(\hat{\bf x} - \hat{\bf y})$-normal layer with $C^+_{x-y}=1$ in the $x-y=0$ plane implied by the $\{C_{4z}|\mathbf{0}\}$ rotation symmetry.
\end{enumerate}

\textit{Axion insulators} -- We find that states with odd $z_8$ SIs cannot be constructed from layers of 2D stable topological phases.  However, we may still use subduction relations to determine the bulk topology of insulators with odd values of $z_{8}$.  First, as we will show below, $(100)$, $(300)$, $(500)$, and $(700)$ subduce to $(2000)_{P\bar{1}}$ in MSG 2.4 $P\bar1$.  Hence, if the $(100)$, $(300)$, $(500)$, and $(700)$ phases in double MSG 123.339 $P4/mmm$ are insulating, then the bulk insulator must either be an AXI or a 3D QAH state.  Because the net Chern numbers $C_{x,y,z}=0$ must vanish if the bulk is gapped, due to the mirror symmetries of MSG 123.339 $P4/mmm$, then the $(100)$, $(300)$, $(500)$, and $(700)$ states must be AXIs.  This result can also be understood by subducing from a $\mathcal{T}$-symmetric SSG.  Specifically, because $(100)$, $(300)$, $(500)$, and $(700)$ in double MSG 123.339 $P4/mmm$ can respectively be subduced from $(100)_{P4/mmm1'}$, $(300)_{P4/mmm1'}$, $(500)_{P4/mmm1'}$, and $(700)_{P4/mmm1'}$ in Type-II SG 123.340 $P4/mmm1^\pr$, which correspond to $\mathcal{T}$-symmetric 3D TIs with $\theta=\pi$~\cite{AshvinIndicators,ChenTCI,AshvinTCI}, then $(100)$, $(300)$, $(500)$, and $(700)$ are compatible with bulk-gapped states.  Hence, we conclude that 3D insulators with $(100)$, $(300)$, $(500)$, and $(700)$ in double MSG 123.349 $P4/mmm$ are AXIs, \emph{without ambiguity}.  We conjecture that $(100)$, $(300)$, $(500)$, and $(700)$ AXIs in double MSG 123.349 $P4/mmm$ can be constructed using the topological crystal method~\cite{ZhidaHermeleCrystal}, which additionally incorporates cell complexes of 2D Chern insulators, TIs, and TCIs.

\textit{Helical HOTI phases protected by mirror and $C_4$ rotation symmetry} -- First, the double SIs $(400)_{P4/mmm1^\pr}$ of Type-II double SSG 123.340 $P4/mmm1^\pr$ either correspond to a rotation-anomaly (non-axionic, \emph{i.e.} $\theta\text{ mod }2\pi=0$) HOTI protected by $C_{4}$ and $\TRS$ symmetries, or a mirror TCI with $C_{m_{z}}\text{ mod }8=4$ (\emph{c.f.} Table 7 in the Supplementary Material of Ref.~\onlinecite{ChenTCI}).  In the $C_{4}$- and $\mathcal{T}$-symmetric HOTI phase, there are $4 +8n$ ($n\in\{\mathbb{Z}^{+},0\}$) helical hinge modes on a $z$-directed, $C_{4}$- and $\mathcal{T}$-symmetric rod, and $4+8n$ twofold Dirac points on the top ($\hat{\bf z}$-normal) rod surface that are locally protected by mirror symmetry (see Appendix~\ref{sec:newHOTIs}).  Because double SSG 123.340 $P4/mmm1^\pr$ contains $\{m_{x,y}|{\bf 0}\}$ symmetries, then four of the helical hinge modes on the boundary of a $4/mmm1'$-symmetric sample must also be pinned to the hinge projections of bulk mirror planes whose normal vectors lie in the $xy$-plane, and must be indicated by bulk mirror Chern numbers.  Hence, when $\mathcal{T}$ symmetry is relaxed in a fourfold rotation-anomaly $(400)_{P4/mmm1^\pr}$ HOTI phase in Type-II double SSG 123.340 $P4/mmm1^\pr$ while preserving the symmetries of MSG 123.339 $P4/mmm$, the surface and hinge states will remain gapless and anomalous [see Fig.~\ref{fig:SS}(b) and Appendix~\ref{sec:HOTItbModel}].

We will next prove that there are $4+8n$ twofold Dirac points on the top surface of a $4/mmm$-symmetric nanorod of the $(400)$ fourfold rotation-anomaly magnetic HOTI phase in double MSG 123.339 $P4/mmm$ introduced in this work.  We denote the five layer constructions as $L_a$ ($a=1\cdots 5$), respectively.  The fourfold rotation-anomaly HOTI phase can be constructed as $(2n+1)L_3 \oplus (2m+1) L_4$, or $ (2n+1) L_5$, or through any superposition of odd number of the aforementioned layer constructions.  Adding $4n' L_1$ or $4m' L_2$, which have the SIs (000), to the layer-constructed HOTI phase will not change the top surface spectrum, because $L_1$ and $L_2$ consist of horizontal (\emph{i.e.} $\hat{\bf z}$-normal) layers, and hence only contribute surface and hinge states on boundaries whose normal vectors lie in the $xy$-plane.

We will thus focus on the top surface spectra of the $(2n+1)L_3 \oplus (2m+1) L_4$ and $(2n+1)L_5$  layer constructions.  We first consider $(2n+1)L_3 \oplus (2m+1) L_4$.  The Chern numbers in the $m_x$ mirror sectors are $C_{k_x=0}^+=-C_{k_x=0}^-=2n+2m+2$, $C_{k_x=\pi}^+=-C_{k_x=\pi}^-=2n-2m$. 
Due to the $C_4$ symmetry, the Chern numbers in the $m_y$ mirror sectors are $C_{k_y=0}^+=-C_{k_x=0}^-=2n+2m+2$, $C_{k_y=\pi}^+=-C_{k_x=\pi}^-=2n-2m$.  In the 2D top surface BZ, $C^+_{k_x=0}$ ($C^+_{k_y=0}$) mandates the presence of $|2n+2m+2|$ twofold Dirac points on the $k_x=0$ ($k_y=0$) line, and $C^+_{k_x=\pi}$ ($C^+_{k_y=\pi}$) mandates the presence of $|2n-2m|$ twofold Dirac points on the $k_x=\pi$ ($k_y=\pi$) line.  Hence, the total number of Dirac points is $2|2n+2m+2|+2|2n-2m|\text{ mod }8= 4$.

Lastly, we consider the layer construction $(2n+1)L_5$.  As shown in Supplementary Note 5 in Ref.~\onlinecite{ChenTCI} and in Table 6 of the Supplementary Material of Ref.~\onlinecite{ChenTCI}, the mirror sector Chern numbers are given by $C^+_{k_x+k_y=0}=- C^-_{k_x+k_y=0}=4n+2$, $C^+_{k_x-k_y=0}=- C^-_{k_x-k_y=0}=4n+2$, and $C^+_{k_x\pm k_y=\pi}=- C^-_{k_x\pm k_y=\pi}=0$.  To understand this result, one can enlarge the unit cell to a supercell with the lattice vectors $(1,1,0)$ and $(1,-1,0)$.  We emphasize that mirror (sector) Chern numbers do not change upon enlarging the unit cell if the number of layers per cell does not change; hence we can compute the mirror sector Chern numbers in the supercell.  We define $x'=\frac12 x+ \frac12y$, $y'=\frac12x-\frac12y$, and correspondingly define $k_x'=k_x+k_y$, $k_y'=k_x-k_y$.  As shown in Appendix~\ref{sec:P-1}, the Chern numbers of the layers at $x'=0$ ($y'=0$) and $x'=\frac12$ ($y'=\frac12$) contribute with the same signs towards $C^+_{k_x'=0}=-C^-_{k_x'=0}$ ($C^+_{k_y'=0}=-C^-_{k_y'=0}$) and with opposite signs towards $C^+_{k_x'=\pi}=-C^-_{k_x'=\pi}$ ($C^+_{k_y'=\pi}=-C^-_{k_y'=\pi}$).  Hence, $C^+_{k_x'=0}=-C^-_{k_x'=0}=4n+2$, $C^+_{k_x'=\pi}=-C^-_{k_x'=\pi}=0$, $C^+_{k_y'=0}=-C^-_{k_y'=0}=4n+2$, and $C^+_{k_y'=\pi}=-C^-_{k_y'=\pi}=0$.  In the 2D top surface BZ, $C^+_{k_x+k_y=0}$ ($C^+_{k_x-k_y=0}$) mandates the presence of $|4n+2|$ Dirac points on the $k_x+k_y=0$ ($k_x-k_y=0$) line.  We additionally note that the mirror sector Chern numbers $C^+=-C^-$ mandate the presence of $|C^+|$ twofold Dirac points on the surface, as shown in Fig.~\ref{fig:MirrorChern}.  In summary, the total number of top-surface twofold Dirac points in the first surface BZ is $2|4n+2|\text{ mod }8= 4$.

In Appendix~\ref{sec:HOTIfermionDoubling}, we will prove that, on the top surface of the $(400)$ HOTI state -- which respects the symmetries of Type-I double magnetic wallpaper group~\cite{DiracInsulator,WiederLayers,ChenRotation,SteveMagnet,ConwaySymmetries} $p4m$ -- the presence of $4+8n$ ($n\in\{\mathbb{Z}^{+},0\}$) twofold surface Dirac points circumvents the fermion multiplication theorem for 2D lattices with double magnetic wallpaper group $p4m$.

\textit{Relationship with the SIs in other double SSGs} -- To identify the AXI phases, we subduce the SIs onto double MSG 2.4 $P\bar1$:
\begin{equation}
(z_8,z_{4m,\pi}^+,z_{2w,1})_{P4/mmm} \to  (\eta_{4I},z_{2I,1},z_{2I,2},z_{2I,3})_{P\bar1}= (2(z_8\text{ mod }2), 000)_{P\bar1}.
\end{equation}
Because the AXI $\mathcal{I}$ SI $\eta_{2I'}=\frac12\eta_{4I}=z_8\text{ mod }2$ [Eq.~(\ref{eq:eta2Ip})], then we conclude that insulators with odd $z_8$ SIs in double MSG 123.339 $P4/mmm$ are AXIs.

\vspace{-0.2in}
\subsubsection{Double SIs in Type-I Double MSG 143.1 $P3$}

The double MSG 143.1 $P3$ is generated by 
    $\{E|100\}$, 
    $\{E|010\}$,
    $\{E|001\}$, and
    $\{C_{3z}|\mathbf{0}\}$, where the angle between the $\{E|100\}$ and $\{E|010\}$ translations is chosen to be $2\pi/3$ for consistency with the $\{C_{3z}|\mathbf{0}\}$ rotation symmetry.

The double MSG 143.1 $P3$ has the SI group $\ZZ_3$.  To determine the physical basis for the double SIs, we first recall the formula for the 2D Chern number in the presence of threefold rotation symmetry~\cite{ChenBernevigTCI}:
\begin{equation}
e^{i\frac{2\pi}3 C} = (-1)^{N_\mrm{occ}} \prod_{n\in\mrm{occ}} \theta_n(\Gamma) \theta_n(\mathcal{K}) \theta_n(\mathcal{KA}),
\end{equation}
where $\theta_n(\Gamma,\mathcal{K},\mathcal{KA})$ is the $\{C_{3z}|\mathbf{0}\}$ eigenvalue of the $n^{\rm th}$ occupied state at the corresponding momentum (where it is important to distinguish the $\{C_{3z}|\mathbf{0}\}$ eigenvalues $\theta_n$ from the axion angle $\theta$).  We can define the SI as the Chern number in the $k_{z}=\pi$ plane modulo 3:
\begin{equation}
z_{3R} = C_{k_z=\pi} \text{ mod } 3
= \frac32 N_\mrm{occ} + \sum_{K=A,H,HA} \pare{-\frac12 n^{\frac12}_K + 
\frac12 n^{-\frac12}_K + \frac32 n^{\frac32}_K } \text{ mod } 3,
\label{eq:Z3rTemp}
\end{equation}
where the superscripts $j=-\frac12,\frac12,\frac32$ represent the $\{C_{3z}|\mathbf{0}\}$ eigenvalues $e^{-i\frac{2\pi}3 j}$= $e^{-i\frac{\pi}3}, e^{i\frac{\pi}3},-1$, respectively, and $N_{\rm occ}$ is the number of occupied bands.  Because $\frac32 N_{\rm occ} = \sum_{K=A,H,HA} \frac12 n_K^{-\frac12} + \frac12 n_K^{\frac12} + \frac12 n_K^{\frac32}$, then Eq.~(\ref{eq:Z3rTemp}) can be simplified:
\begin{equation}
z_{3R} = \sum_{K=A,H,HA} \pare{ 
 n^{-\frac12}_K - n^{\frac32}_K } \text{ mod } 3.
 \label{eq:z3R}
\end{equation}
Due to the compatibility relations and the fact that a chiral fermion in 3D occurs when there is a change in a momentum-space Chern number, a 3D insulator must satisfy $C_{k_z=\pi}=C_{k_z}$ for all $k_{z}$.  Hence, we may have equivalently defined the SI $z_{3R}$ using the occupied $\{C_{3z}|{\bf 0}\}$ eigenvalues in $k_z=0$ plane, or in any other BZ plane of constant $k_{z}$.  To summarize, if a 3D system is insulating and exhibits $z_{3R}\neq0$, then the system is in a 3D QAH state with $C_{k_z=0}=C_{k_z=\pi}$ and $z_{3R} = C_{k_z=0}\text{ mod }3$.

Because the physical meaning of the double SIs is straightforward (\emph{i.e.} the nontrivial phases are 3D QAH states composed of stacks of Chern insulators), then will not provide explicit layer constructions for double MSG 143.1 $P3$.

If we impose $\mathcal{T}$-symmetry, then the position-space Chern numbers must vanish, which enforces $z_{3R}$ to be zero.  Furthermore, if we add $\mathcal{T}$ symmetry to a system that respects double MSG 143.1 $P3$, we specifically find that the SI group becomes trivial.

\subsubsection{Double SIs in Type-I Double MSG 147.13 $P\bar{3}$}
\label{sec:P-3}

The double MSG 147.13 $P\bar{3}$ is generated by 
    $\{E|100\}$, 
    $\{E|010\}$,
    $\{E|001\}$, 
    $\{C_{3z}|\mathbf{0}\}$, and
    $\{\INV|\mathbf{0}\}$, where the angle between the $\{E|100\}$ and $\{E|010\}$ translations is chosen to be $2\pi/3$ for consistency with the $\{C_{3z}|\mathbf{0}\}$ rotation symmetry.

\textit{SIs} -- The double MSG 147.13 $P\bar3$ has the SI group $\ZZ_{12}\times \ZZ_{2} \sim \ZZ_{4}\times \ZZ_{3}\times \ZZ_2$.  We find that the $\ZZ_{4}$, $\ZZ_{3}$, and $\ZZ_2$ double SIs in double MSG 147.13 $P\bar{3}$ all subduce to previously introduced double SIs.  First, the $\ZZ_4$ SI subduces to $\eta_{4I}$ in double MSG 2.4 $P\bar1$ [Eq.~(\ref{eq:eta4I})], where, as shown in Appendix~\ref{sec:P-1}, $\eta_{4I}=1,3$ indicate WSM phases, $\eta_{4I}=2$ indicates that an insulator is either an AXI or in a 3D QAH state, and $\eta_{4I}=0$ indicates that an insulator is either topologically trivial or in a 3D QAH state.  The $\ZZ_3$ SI subduces to $z_{3R}$ in double MSG 143.1 $P3$ [Eq.~(\ref{eq:z3R})].  In insulating states, $z_{3R}$ indicates the Chern number modulo 3 in BZ planes of constant $k_{z}$.  Lastly, the $\ZZ_2$ SI subduces to $z_{2I,3}$ in double MSG 2.4 $P\bar1$ [Eq.~(\ref{eq:z2I})], where $z_{2I,3}$ indicates the Chern number modulo 2 in the $k_z=\pi$ plane.  In summary, the double SIs in double MSG 147.13 $P\bar{3}$ in the physical basis are given by the previously-defined double SIs $(\eta_{4I},z_{3R},z_{2I,3})$.

\textit{Layer constructions} -- In Cartesian coordinates $(x,y,z)$, the primitive lattice translation vectors in double MSG 147.13 $P\bar{3}$ -- $\{E|100\}$, $\{E|010\}$, and $\{E|001\}$ -- respectively correspond to $\tt_1=(0,-1,0)$, $\tt_2=(\frac{\sqrt3}2,\frac12,0)$, and $\tt_3=(0,0,1)$.  We consider the following four layer constructions, where the double SIs of each layer construction are given in the order $(\eta_{4I},z_{3R},z_{2I,3})$:
\begin{enumerate}
\item A $\hat{\bf z}$-normal Chern layer with $C_z=1$ in the $z=0$ plane has the SIs (211).
\item A $\hat{\bf z}$-normal Chern layer with $C_z=-2$ in the $z=0$ plane has the SIs (010).
\item A $\hat{\bf z}$-normal Chern layer with $C_z=1$ in the $z=\frac12$ plane has the SIs (011).
\item An $\hat{\bf x}$-normal Chern layer with $C_x=1$ in the $x=0$ plane has the SIs (200).  We emphasize that, in this layer construction, there are also $|C|=1$ Chern layers in the $C_{3z}\hat{\bf x}$ and $C_{3z}^2 \hat{\bf x}$ directions implied by the $\{C_{3z}|\mathbf{0}\}$ rotation symmetry.
\end{enumerate}
We label the four layer constructions as $L_{1,2,3,4}$, respectively.  We note that $-2L_1$ and $-2L_3$ exhibit the same symmetry-indicated topology as $L_2$, where $-L_{1}$ ($-L_3$) has the same construction as $L_{1}$ ($L_3$), except for a sign change in the Chern number $C_{z}\rightarrow - C_{z}$.

\textit{Total Chern number modulo 6} -- The Chern number at $k_z=\pi$ can be determined modulo 6:
\begin{equation}
C_{k_z=\pi}\text{ mod }6 = -2z_{3R}+3 z_{2I,3} \text{ mod } 6.
\label{eq:sixfoldTemp}
\end{equation}
Eq.~(\ref{eq:sixfoldTemp}) takes the same form as the SI introduced in Ref.~\onlinecite{ChenBernevigTCI} for the Chern number in a 2D insulator with sixfold rotation symmetry, which occurs because the point group of double MSG 147.13 $P\bar3$ (isomorphic to Type-I MPG 17.1.62 $P\bar{3}$) exhibits sixfold symmetry generated by $C_{3z}$ and $\mathcal{I}$~\cite{BigBook,PointGroupTables}.  In general, we find that, if $\eta_{4I}=0,2$, then the Chern number of the occupied bands in the $k_z=0$ plane is the same as the Chern number of the occupied bands in the $k_z=\pi$ plane (modulo 6).  Lastly, if $\eta_{4I}=1,3$, then the Chern number of the occupied bands in the $k_z=0$ plane differs from the Chern number of the occupied bands in the $k_z=\pi$ plane by 3 (modulo 6), implying the presence of an odd number of Weyl points in the BZ between $k_{z}=0,\pi$, in agreement with the odd value of $\eta_{4I}$ (see Appendix~\ref{sec:P-1}).

\textit{Relationship with the SIs in other double SSGs} -- The double SSG 147.14 $P\bar31^\pr$ -- which is generated by adding $\{\mathcal{T}|{\bf 0}\}$ symmetry to double MSG 147.13 $P\bar{3}$ -- has the double SI group $\mbb{Z}_4\times\mbb{Z}_2$.  The SIs in double SSG 147.14 $P\bar31^\pr$ are related to the SIs in double MSG 147.13 $P\bar3$ through the subduction relations:
\begin{equation}
(z_4,z_{2w,3})_{P\bar31^\pr} \to (\eta_{4I},z_{3R},z_{2I,3})_{P\bar3}= (2(z_4\text{ mod }2),00)_{P\bar3}.
\end{equation}

\subsubsection{Double SIs in Type-I Double MSG 168.109 $P6$}
\label{subsec:P6}

The double MSG 168.109 $P6$ is generated by 
    $\{E|100\}$, 
    $\{E|010\}$,
    $\{E|001\}$, and $\{C_{6z}|\mathbf{0}\}$, where the angle between the $\{E|100\}$ and $\{E|010\}$ translations is chosen to be $2\pi/3$ for consistency with the $\{C_{3z}|\mathbf{0}\}=(\{C_{6z}|\mathbf{0}\})^{2}$ rotation symmetry.

The double MSG 168.109 $P6$ has the SI group $\ZZ_6$.  To determine the physical basis for the double SIs, we first recall the formula for the 2D Chern number in the presence of sixfold symmetries~\cite{ChenBernevigTCI}:
\begin{equation}
e^{i\frac{2\pi}6 C} = (-1)^{N_\mrm{occ}} \prod_{n\in\mrm{occ}} \eta_n(\Gamma) \theta_n(\mathcal{K}) \zeta_n(M),
\end{equation}
where $\eta_n(\Gamma)$, $\theta_n(\mathcal{K})$, $\zeta_n(M)$ are the $\{C_6|{\bf 0}\}$, $\{C_3|{\bf 0}\}$, and $\{C_2|{\bf 0}\}$ eigenvalues of the $n^\text{th}$ occupied state at $\Gamma$, $\mathcal{K}$, and $M$, respectively.  We define the SI as the Chern number in the $k_{z}=\pi$ plane modulo 6:
{\small\begin{align}
z_{6R} =& C_{k_z=\pi}\text{ mod } 6 =  3N_\mrm{occ} 
- \frac12 n^{\frac12}_A + \frac12 n^{-\frac12}_A
- \frac32 n^{\frac32}_A + \frac32 n^{-\frac32}_A
- \frac52 n^{\frac52}_A + \frac52 n^{-\frac52}_A
-  n^{\frac12}_H + n^{-\frac12}_H + 3n^{\frac32}_H
- \frac{3}{2} n^{\frac12}_L + \frac32 n^{-\frac12}_L \text{ mod } 6 \nono\\
=& - \frac12 n^{\frac12}_A + \frac12 n^{-\frac12}_A
- \frac32 n^{\frac32}_A + \frac32 n^{-\frac32}_A
- \frac52 n^{\frac52}_A + \frac52 n^{-\frac52}_A
-  n^{\frac12}_H + n^{-\frac12}_H + 3n^{\frac32}_H
+ \frac{3}{2} n^{\frac12}_L - \frac32 n^{-\frac12}_L \text{ mod } 6,
\label{eq:z6R}
\end{align}}
where the superscripts $n^{j}_{A}$ represent the $\{C_{6z}|{\bf 0}\}$ eigenvalues $e^{-i\frac{2\pi}{6}j}$ at $A$, $n^{j}_{H}$ is the number of occupied states with $\{C_{3z}|{\bf 0}\}$ eigenvalue $e^{-i\frac{2\pi}{3}j}$ at $H$, and where $n^{j}_{L}$ is the number of states with $\{C_{2z}|{\bf 0}\}$ eigenvalue $e^{-i\frac{\pi}{2}j}$ at $L$.  Due to the compatibility relations and the fact that a chiral fermion in 3D occurs when there is a change in a momentum-space Chern number, a 3D insulator must satisfy $C_{k_z=\pi}=C_{k_z}$ for all $k_{z}$.  Hence, we may have equivalently defined the SI $z_{6R}$ using the occupied rotation symmetry eigenvalues in the $k_z=0$ plane, or in any other BZ plane of constant $k_{z}$.  To summarize, if a 3D system is insulating and exhibits $z_{6R}\neq0$, then the system is in a 3D QAH state with $C_{k_z=0}=C_{k_z=\pi}$ and $z_{6R} = C_{k_z=0}\text{ mod }6$.

Because the physical meaning of the double SIs is straightforward (\emph{i.e.} the nontrivial phases are 3D QAH states composed of stacks of Chern insulators), then will not provide explicit layer constructions for double MSG 168.109 $P6$.

If we impose $\mathcal{T}$-symmetry, then the position-space Chern numbers must vanish, which enforces $z_{6R}$ to be zero.  Furthermore, if we add $\mathcal{T}$ symmetry to a system that respects double MSG 168.109 $P6$, we specifically find that the SI group becomes trivial.

\subsubsection{Double SIs in Type-I Double MSG 174.133 $P\bar{6}$}

The double MSG 174.133 $P\bar6$ is generated by 
    $\{E|100\}$, 
    $\{E|010\}$,
    $\{E|001\}$, 
    $\{C_{3z}|\mathbf{0}\}$, and $\{m_z|\mathbf{0}\}$, where the angle between the $\{E|100\}$ and $\{E|010\}$ translations is chosen to be $2\pi/3$ for consistency with the $\{C_{3z}|\mathbf{0}\}=(\{C_{6z}|\mathbf{0}\})^{2}$ rotation symmetry.

\textit{SIs} -- The double MSG 174.133 $P\bar 6$ has the SI group $\ZZ_3^3$.  In the physical basis, the three $\mathbb{Z}_{3}$-valued SIs are: ($\delta_{3m},z_{3m,\pi}^+,z_{3m,\pi}^-$), for which the SI formulas are:
\begin{align}
\delta_{3m} =  C_{k_z=\pi}^+ - C_{k_z=0}^- \text{ mod } 3
=& \sum_{K=A,H,HA} \pare{ 
n^{-\frac12,+i}_K - n^{\frac32,+i}_K } - \sum_{K=\Gamma,\mathcal{K},\mathcal{KA}} \pare{ 
n^{-\frac12,-i}_K - n^{\frac32,-i}_K }\text{ mod }3,
\label{eq:d3m}
\end{align}
\begin{equation}
z_{3m,\pi}^+ =  C_{k_z=\pi}^+\text{ mod } 3
=  
\sum_{K=A,H,HA} \pare{ 
n^{-\frac12,+i}_K - n^{\frac32,+i}_K }\text{ mod }3,
\label{eq:z3mp+}
\end{equation}
\begin{equation}
z_{3m,\pi}^- = C_{k_z=\pi}^-\text{ mod } 3
= \sum_{K=A,H,HA} \pare{ 
n^{-\frac12,-i}_K - n^{\frac32,-i}_K }\text{ mod }3,
\label{eq:z3mp-}
\end{equation}
such that a 3D insulator with $z_{3m,\pi}^{+}\neq -z_{3m,\pi}^{-}\text{ mod }3$ is in a 3D QAH state.  The compatibility relations require the $k_z=0,\pi$ planes to have the same occupied $C_{3z}$ eigenvalues, and hence the same Chern numbers (modulo 3).  In an insulating state (\emph{i.e.} in the absence of bulk Weyl points), it is further required that $C_{k_z=0}^+ + C_{k_z=0}^- = C_{k_z=\pi}^+ + C_{k_z=\pi}^-$ and $C_{k_z=0}^+ - C_{k_z=\pi}^-\text{ mod }3=   C_{k_z=\pi}^+ - C_{k_z=0}^-\text{ mod }3=\delta_{3m}$.  As we will show below, in insulators with net-zero position-space Chern numbers, AXI phases may be stabilized in double MSG 174.133 $P\bar6$ by $S_{6}$ rotoinversion symmetry, but will not be symmetry indicated, because the strong index $\delta_{3m}$ is $\mathbb{Z}_{3}$-valued, whereas the axion angle $\theta$ is $\mathbb{Z}_{2}$-valued (if quantized).

\textit{Layer constructions} -- To diagnose the topology associated to each nontrivial value of the double SIs $(\delta_{3m},z_{3m,\pi}^+,z_{3m,\pi}^-)$, we employ the layer construction method.  In each layer construction, we denote the mirror sector Chern numbers of the occupied bands at $k_z=0,\pi$ as $(C_{k_z=0}^+,C_{k_z=0}^-,C_{k_z=\pi}^+,C_{k_z=\pi}^-)$.  The layer constructions for Type-I double MSG 174.133 $P\bar6$ are given by:
\begin{enumerate}
\item A $\hat{\bf z}$-normal Chern layer with $C^+_z=1$, $C^-_z=0$ in the $z=0$ plane has the mirror sector Chern numbers (1010) and the SIs (110).
\item A $\hat{\bf z}$-normal Chern layer with $C^+_z=0$, $C^-_z=1$ in the $z=0$ plane has the mirror sector Chern numbers (0101) and the SIs (201).
\item A $\hat{\bf z}$-normal Chern layer with $C^+_z=1$, $C^-_z=0$ in the $z=\frac12$ plane has the mirror sector Chern numbers (1001) and the SIs (001).
\item A $\hat{\bf z}$-normal Chern layer with $C^+_z=0$, $C^-_z=1$ in the $z=\frac12$ plane has the mirror sector Chern numbers (0110) and the SIs (010).
\end{enumerate}

\textit{Relationship with the SIs in other double SSGs} -- We next study the effects of imposing $\TRS$ symmetry.  The double SSG 174.134 $P\bar61^\pr$ -- the SSG generated by adding $\{\mathcal{T}|{\bf 0}\}$ symmetry to MSG 174.133 $P\bar{6}$ -- has the SI group $\ZZ_3\times\ZZ_3$.  The SIs in double SSG 174.134 $P\bar{6}1'$ are related to the SIs in double MSG 174.133 $P\bar{6}$ through the subduction relations:
\begin{equation}
(z_{3m,0},z_{3m,\pi})_{P\bar61^\pr} \to (\delta_{3m},z_{3m,\pi}^+,z_{3m,\pi}^-)_{P\bar6}= (z_{3m,\pi}+z_{3m,0}\text{ mod }3,\; z_{3m,\pi},\; -z_{3m,\pi}\text{ mod }3)_{P\bar6}.
\label{eq:pbar6Subductions}
\end{equation}
In Type-II double SSG 174.134 $P\bar61^\pr$ insulators with net-odd odd mirror Chern numbers~\cite{ChenTCI} are 3D TIs.  However, because $z_{3m,0}$ and $z_{3m,\pi}$ only indicate the mirror Chern numbers in the $k_{z}=0,\pi$ planes modulo 3, then there is no relationship between $z_{3m,0}$ and $z_{3m,\pi}$ and the axion (3D TI) angle $\theta$.  Specifically, consider a 3D TI in double SSG 174.134 $P\bar61^\pr$ with $(z_{3m,0},z_{3m,\pi})_{P\bar61'}=(10)_{\bar{6}1'}$, where the bulk axion angle $\theta=\pi$.  Taking three superposed copies of the 3D TI results in an insulator with the SIs $(00)_{\bar{6}1'}$ and the axion angle $\theta=\pi$.  Hence, $z_{3m,0}$ and $z_{3m,\pi}$ are individually (and as a set) independent of $\theta$, because $z_{3m,0}$ and $z_{3m,\pi}$ are $\mathbb{Z}_{3}$-valued, whereas $\theta$ is $\mathbb{Z}_{2}$-valued (if quantized).  We thus conclude that, while axionic mirror TCI phases can be stabilized by $\{m_{z}|{\bf 0}\}$ mirror and $\{S_{6}|{\bf 0}\}$ rotoinversion symmetries in the magnetic subgroup double MSG 174.133 $P\bar{6}$ of double SSG 174.134 $P\bar61^\pr$, $\theta$ is not symmetry-indicated in double MSG 174.133 $P\bar6$.

\subsubsection{Double SIs in Type-I Double MSG 175.137 $P6/m$}
\label{subsec:P6/m}

The double MSG 175.137 $P6/m$ is generated by
    $\{E|100\}$, 
    $\{E|010\}$,
    $\{E|001\}$, 
    $\{C_{6z}|\mathbf{0}\}$, and
    $\{\INV|\mathbf{0}\}$, where the angle between the $\{E|100\}$ and $\{E|010\}$ translations is chosen to be $2\pi/3$ for consistency with the $\{C_{3z}|\mathbf{0}\}=(\{C_{6z}|\mathbf{0}\})^{2}$ rotation symmetry.  We note that double MSG 175.137 $P6/m$ additionally contains a mirror symmetry operation: $\{m_z|\mathbf{0}\} = \{C_{6z}|\mathbf{0}\}^3 \{\INV|\mathbf{0}\}$.

\textit{SIs} -- The double MSG 175.137 $P6/m$ has the SI group $\ZZ_6^3$.  In the physical basis, the three $\mathbb{Z}_{6}$-valued SIs are: ($\delta_{6m},z_{6m,\pi}^+,z_{6m,\pi}^-$), for which the SI formulas are:
\begin{align}
\delta_{6m} = C_{k_z=\pi}^+-C_{k_z=0}^- \text{ mod } 6
=& 
- \frac12 n^{\frac12,+i}_A + \frac12 n^{-\frac12,+i}_A
- \frac32 n^{\frac32,+i}_A + \frac32 n^{-\frac32,+i}_A
- \frac52 n^{\frac52,+i}_A + \frac52 n^{-\frac52,+i}_A \nono\\
&-n^{\frac12,+i}_H + n^{-\frac12,+i}_H + 3n^{\frac32,+i}_H
+ \frac{3}{2} n^{\frac12,+i}_L - \frac32 n^{-\frac12,+i}_L \nono\\
&+\frac12 n^{\frac12,-i}_\Gamma - \frac12 n^{-\frac12,-i}_\Gamma
+ \frac32 n^{\frac32,-i}_\Gamma - \frac32 n^{-\frac32,-i}_\Gamma
+ \frac52 n^{\frac52,-i}_\Gamma - \frac52 n^{-\frac52,-i}_\Gamma \nono\\
&+ n^{\frac12,-i}_{\mathcal{K}} - n^{-\frac12,-i}_{\mathcal{K}} - 3n^{\frac32,-i}_{\mathcal{K}}
- \frac{3}{2} n^{\frac12,-i}_M + \frac32 n^{-\frac12,-i}_M\text{ mod }6,
\label{eq:d6m}
\end{align}
\begin{align}
z_{6m,\pi}^+ =C_{k_z=\pi}^+ \text{ mod } 6 
=&
- \frac12 n^{\frac12,+i}_A + \frac12 n^{-\frac12,+i}_A
- \frac32 n^{\frac32,+i}_A + \frac32 n^{-\frac32,+i}_A
- \frac52 n^{\frac52,+i}_A + \frac52 n^{-\frac52,+i}_A \nono\\
&-  n^{\frac12,+i}_H + n^{-\frac12,+i}_H + 3n^{\frac32,+i}_H
+ \frac{3}{2} n^{\frac12,+i}_L - \frac32 n^{-\frac12,+i}_L \text{ mod } 6,
\label{eq:z6mp+}
\end{align}
\begin{align}
z_{6m,\pi}^- =C_{k_z=\pi}^- \text{ mod } 6 
=&
- \frac12 n^{\frac12,-i}_A + \frac12 n^{-\frac12,-i}_A
- \frac32 n^{\frac32,-i}_A + \frac32 n^{-\frac32,-i}_A
- \frac52 n^{\frac52,-i}_A + \frac52 n^{-\frac52,-i}_A \nono\\
&-  n^{\frac12,-i}_H + n^{-\frac12,-i}_H + 3n^{\frac32,-i}_H
+ \frac{3}{2} n^{\frac12,-i}_L - \frac32 n^{-\frac12,-i}_L \text{ mod } 6,
\label{eq:z6mp-}
\end{align}
such that a 3D insulator with $z_{6m,\pi}^{+}\neq -z_{6m,\pi}^{-}\text{ mod }6$ is in a 3D QAH state.  As we will show below, in insulators with net-zero position-space Chern numbers, odd values of $\delta_{6m}$ indicate mirror TCI phases with $\theta=\pi$.  The compatibility relations require that the occupied bands in the $k_z=0,\pi$ planes have the same rotation symmetry eigenvalues, and hence the same Chern numbers (modulo 6).  In an insulating state (\emph{i.e.} in the absence of bulk Weyl points), it is further required that $C_{k_z=0}^+ + C_{k_z=0}^- = C_{k_z=\pi}^+ + C_{k_z=\pi}^-$ and $C_{k_z=0}^+ - C_{k_z=\pi}^-\text{ mod }6=   C_{k_z=\pi}^+ - C_{k_z=0}^-\text{ mod }6=\delta_{6m}$.

\textit{Layer constructions} -- To diagnose the topology associated to each nontrivial value of the double SIs $(\delta_{6m},z_{6m,\pi}^+,z_{6m,\pi}^-)$, we employ the layer construction method.  In each layer construction, we denote the mirror sector Chern numbers of the occupied bands at $k_z=0,\pi$ as $(C_{k_z=0}^+,C_{k_z=0}^-,C_{k_z=\pi}^+,C_{k_z=\pi}^-)$, and additionally compute the subduced SIs $(\eta_{4I},z_{2I,1},z_{2I,2},z_{2I,3})_{P\bar1}$ in the subgroup double MSG 2.4 $P\bar{1}$ (see Appendix~\ref{sec:P-1}).  The layer constructions for Type-I double MSG 175.137 $P6/m$ are given by:
\begin{enumerate}
\item A $\hat{\bf z}$-normal Chern layer with $C^+_z=1$, $C^-_z=0$ at $z=0$ has the mirror sector Chern numbers (1010) and the SIs (110).  The subduced subgroup SIs are $(\eta_{4I},z_{2I,1},z_{2I,2},z_{2I,3})_{P\bar1}=(2001)_{P\bar1}$.
\item A $\hat{\bf z}$-normal Chern layer with $C^+_z=0$, $C^-_z=1$ at $z=0$ has the mirror sector Chern numbers (0101) and the SIs (501).  The subduced subgroup SIs are $(2001)_{P\bar1}$.
\item A $\hat{\bf z}$-normal Chern layer with $C^+_z=1$, $C^-_z=0$ at $z=\frac12$ has the mirror sector Chern numbers (1001) and the SIs (001).  The subduced subgroup SIs are $(0001)_{P\bar1}$.
\item A $\hat{\bf z}$-normal Chern layer with $C^+_z=0$, $C^-_z=1$ at $z=\frac12$ has the mirror sector Chern numbers (0110) and the SIs (010).  The subduced subgroup SIs are $(0001)_{P\bar1}$.
\end{enumerate}

\textit{Relationship with the SIs in other double SSGs} -- To identify the AXI phases in double MSG 175.137 $P6/m$, we subduce the SIs onto double MSG 2.4 $P\bar1$:
\begin{equation}
\pare{\delta_{6m}, z_{6m,\pi}^+, z_{6m,\pi}^-}_{P6/m} \to (\eta_{4I},z_{2I,1},z_{2I,2},z_{2I,3})_{P\bar1}= \pare{ 2(\delta_{6m}\text{ mod }2),0,0, z_{6m,\pi}^+ + z_{6m,\pi}^-\text{ mod }2}_{P\bar1}.
\label{eq:tempSubductionp6overM}
\end{equation}
Eq.~(\ref{eq:tempSubductionp6overM}) implies that the $\mathcal{I}$ AXI index $\eta_{2I'}$ [Eq.~(\ref{eq:eta2Ip})] is related to $\delta_{6m}$ by $\eta_{2I'}=\frac12\eta_{4I}=\delta_{6m}\text{ mod }2$, such that gapped states with $\delta_{6m}\text{ mod }2=1$ and $z_{6m,\pi}^+ + z_{6m,\pi}^-\text{ mod }2=0$ in MSG 175.137 $P6/m$ are AXIs if the non-symmetry-indicated Chern numbers vanish.

Lastly, we study the effects of imposing $\mathcal{T}$ symmetry.  The double SSG 175.138 $P6/m1^\pr$ -- the SSG generated by adding $\{\mathcal{T}|{\bf 0}\}$ to double MSG 175.137 $P6/m$ -- has the SI group $\ZZ_{12}\times \ZZ_6$.  The SIs in double SSG 175.138 $P6/m1^\pr$ are related to the SIs in double MSG 175.137 $P6/m$ through the subduction relations:
\begin{equation}
(z_{12},z_{6m,\pi})_{P6/m1^\pr} \to (\delta_{6m},z_{6m,\pi}^+,z_{6m,\pi}^-)_{P6/m}= (z_{12}\text{ mod }6,z_{6m,\pi},-z_{6m,\pi}\text{ mod 6})_{P6/m}.
\end{equation}

\subsubsection{Double SIs in Type-I Double MSG 176.143 $P6_{3}/m$}
\label{sec:P6_3/m}

The double MSG 176.143 $P6_3/m$ is generated by 
    $\{E|100\}$, 
    $\{E|010\}$,
    $\{E|001\}$, 
    $\{C_{6z}|00\frac12\}$, and
    $\{\INV|\mathbf{0}\}$, where the angle between the $\{E|100\}$ and $\{E|010\}$ translations is chosen to be $2\pi/3$ for consistency with the $\{C_{3z}|\mathbf{0}\}= \{E|00\bar1\}\{C_{6z}|00\frac12\}^2$ rotation symmetry.  We note that double MSG 176.143 $P6_3/m$ additionally contains a mirror symmetry operation: $\{m_z|00\frac12\} = \{E|00\bar1\}\{C_{6z}|00\frac12\}^3 \{\INV|\mathbf{0}\}$.

\textit{SIs} -- The double MSG 176.143 $P6_3/m$ has the SI group $\ZZ_6\times \ZZ_3$.  In the physical basis, the SIs are $(z_{6m,0}^+,\delta_{3m})$, where $\delta_{3m}=C_{k_z=\pi}^+ - C_{k_z=0}^-\text{ mod }3$ subduces to the same SI ($\delta_{3m}$) in double MSG 174.133 $P\bar{6}$ [Eq.~(\ref{eq:d3m})].  The SI formula for $z_{6m,0}^+$ is given by:
\begin{align}
z_{6m,0}^+ =C_{k_z=0}^+ \text{ mod } 6 
=& 
- \frac12 n^{\frac12,+i}_\Gamma + \frac12 n^{-\frac12,+i}_\Gamma
- \frac32 n^{\frac32,+i}_\Gamma + \frac32 n^{-\frac32,+i}_\Gamma
- \frac52 n^{\frac52,+i}_\Gamma + \frac52 n^{-\frac52,+i}_\Gamma \nono\\
&-  n^{\frac12,+i}_{\mathcal{K}} + n^{-\frac12,+i}_{\mathcal{K}} + 3n^{\frac32,+i}_{\mathcal{K}}
+ \frac{3}{2} n^{\frac12,+i}_M - \frac32 n^{-\frac12,+i}_M \text{ mod } 6,
\label{eq:z6m0+}
\end{align}
where $n^{j,+i}_K$ is the number of occupied states with mirror $\{m_z|00\frac12\}$ eigenvalue $i$ and rotation eigenvalue $ e^{-i\frac{2\pi}{n} j}$ ($n=6,3,2$ for $K=\Gamma$, $\mathcal{K}$, and $M$, respectively).  As we will show below, insulators with $z_{6m,0}^+\text{ mod }2=1$ and net-zero position-space Chern numbers in double MSG 176.143 $P6_3/m$ are AXIs -- all of the other insulators in double MSG 176.143 $P6_3/m$ with nontrivial SIs are 3D QAH states.

\textit{Layer constructions} -- To diagnose the topology associated to each nontrivial value of the double SIs $(z_{6m,0}^+,\delta_{3m})$, we employ the layer construction method.  In each layer construction, we denote the mirror sector Chern numbers of the occupied bands at $k_z=0,\pi$ as $(C_{k_z=0}^+,C_{k_z=0}^-,C_{k_z=\pi}^+,C_{k_z=\pi}^-)$, and additionally compute the subduced SIs $(\eta_{4I},z_{2I,1},z_{2I,2},z_{2I,3})_{P\bar1}$ in the subgroup double MSG 2.4 $P\bar{1}$ (see Appendix~\ref{sec:P-1}).  We note that, while the $\mathcal{I}$ centers in MSG 176.143 $P6_3/m$ lie in the $z=0,\frac12$ planes, the mirror planes lie at $z=\frac14,\frac34$ in each cell. The layer constructions for Type-I double MSG 176.143 $P6_3/m$ are given by:
\begin{enumerate}
\item A $\hat{\bf z}$-normal layer with $C_z=1$ in the $z=0,\frac12$ planes has the mirror sector Chern numbers (1111) and the SIs (10). The subduced subgroup SIs are $(\eta_{4I},z_{2I,1},z_{2I,2},z_{2I,3})_{P\bar1}=(2000)_{P\bar1}$.
\item A $\hat{\bf z}$-normal layer with $C^+_z=1$, $C^-_z=0$ in the $z=\frac14,\frac34$ planes has the mirror sector Chern numbers (2011) and the SIs (21). The subduced subgroup SIs are $(0000)_{P\bar1}$.
\item A $\hat{\bf z}$-normal layer with $C^+_z=0$, $C^-_z=1$ layer in the $z=\frac14,\frac34$ planes has the mirror sector Chern numbers (0211) and the SIs (02). The subduced subgroup SIs are $(0000)_{P\bar1}$.
\end{enumerate}

\textit{Relationship with the SIs in other double SSGs} -- To identify the AXI phases, we subduce the SIs in double MSG 176.143 $P6_3/m$ onto double MSG 2.4 $P\bar1$:
\begin{equation}
(z_{6m,0}^+,\delta_{3m}) \to  (\eta_{4I},z_{2I,1},z_{2I,2},z_{2I,3})_{P\bar1}=(2(z_{6m,0}^+\text{ mod }2),000)_{P\bar1},
\end{equation}
which implies that the $\mathcal{I}$ AXI index $\eta_{2I'}$ [Eq.~(\ref{eq:eta2Ip})] is related to $z_{6m,0}^+$ by $\eta_{2I'}=\frac12\eta_{4I}=z_{6m,0}^+\text{ mod }2$.  Hence, we conclude that insulators in double MSG 176.143 $P6_3/m$ with $z_{6m,0}^+\text{ mod }2=1$ and net-zero position-space Chern numbers are AXIs.

Lastly, we study the effects of imposing $\mathcal{T}$ symmetry.  The double SSG 176.144 $P6_3/m1^\pr$ -- the SSG generated by adding $\{\mathcal{T}|{\bf 0}\}$ to double MSG 176.143 $P6_3/m$ -- has the SI group $\ZZ_{12}$.  The SIs in double SSG 176.144 $P6_3/m1^\pr$ are related to the SIs in double MSG 176.143 $P6_3/m$ through the subduction relations:
\begin{equation}
(z_{12}^\pr)_{P6_3/m1^\pr} \to (z_{6m,0}^+,\delta_{3m})_{P6_3/m}= (z_{12}^\pr\text{ mod }6,z_{12}^\pr\text{ mod }3)_{P6_3/m}.
\end{equation}

\subsubsection{Double SIs in Type-I Double MSG 191.233 $P6/mmm$}
\label{subsec:P6/mmm}

The double MSG 191.233 $P6/mmm$ is generated by 
    $\{E|100\}$, 
    $\{E|010\}$,
    $\{E|001\}$, 
    $\{C_{6z}|\mathbf{0}\}$,
    $\{\INV|\mathbf{0}\}$, and 
    $\{m_x|\mathbf{0}\}$, where the angle between the $\{E|100\}$ and $\{E|010\}$ translations is chosen to be $2\pi/3$ for consistency with the $\{C_{3z}|\mathbf{0}\}=(\{C_{6z}|\mathbf{0}\})^{2}$ rotation symmetry.  We note that double MSG 191.233 $P6/mmm$ additionally contains a mirror symmetry operation: $\{m_z|\mathbf{0}\} = \{C_{6z}|\mathbf{0}\}^3 \{\INV|\mathbf{0}\}$.  In Cartesian coordinates $(x,y,z)$, the primitive lattice translation vectors in double MSG 191.233 $P6/mmm$ -- $\{E|100\}$, $\{E|010\}$, and $\{E|001\}$ -- respectively correspond to $\tt_1=(0,-1,0)$, $\tt_2=(\frac{\sqrt3}2,\frac12,0)$, and $\tt_3=(0,0,1)$.

\textit{SIs} -- The double MSG 191.233 $P6/mmm$ has the SI group $\ZZ_{12}\times\ZZ_6$.  In double-valued small irreps of the little groups at the $\mathcal{I}$-invariant ${\bf k}$ points, the matrix representatives of perpendicular mirror symmetries  (\emph{e.g.} $\{m_x|\mathbf{0}\}$ and $\{m_y|\mathbf{0}\}$) anticommute.  Hence, Bloch states at the eight $\mathcal{I}$-invariant momenta must be at least twofold degenerate (and in fact are exactly twofold degenerate in double MSG 191.233 $P6/mmm$).  The double SIs can be chosen to be the same as the double SIs of SSG 191.234 $P6/mmm1^\pr$ (previously introduced in Ref.~\onlinecite{ChenTCI}), because the addition of $\TRS$ symmetry to double MSG 191.233 $P6/mmm$ does not change the dimensions and characters of the small irreps at the high-symmetry BZ points or the compatibility relations between the high-symmetry-point small irreps.  In the physical basis, the $\ZZ_{12}$ double SI is:
\begin{equation}
z_{12} = \delta_{6m} + 3[( \delta_{6m} - z_4) \text{ mod }4] \text{ mod } 12,
\label{eq:z12}
\end{equation}
where $\delta_{6m}$ is computed by subduction onto double MSG 175.137 $P6/m$ [Eq.~(\ref{eq:d6m})], and $z_4$ is computed by subduction onto double MSG 2.4 $P\bar{1}$ [Eq.~(\ref{eq:z4})].  As we will show below, odd values of the strong index $z_{12}$ indicate mirror TCI phases with $\theta=\pi$ (\emph{i.e.} AXIs), and nontrivial even values indicate non-axionic (helical) magnetic TCI and HOTI phases.  Lastly, in the physical basis, the $\ZZ_6$-valued double SI is the weak TCI invariant $z_{6m,\pi}^+$ for the mirror Chern number (modulo 6) in the $k_{z}=\pi$ plane, and can also be computed by subduction onto double MSG 175.137 $P6/m$ [Eq.~(\ref{eq:z6mp+})].

\textit{Layer constructions} -- To diagnose the topology associated to each nontrivial value of the double SIs, we employ the layer construction method.  In the layer constructions below, $C^+=-C^-$ due to the net-zero Chern numbers enforced by the mirror symmetries.  Hence, we will omit $C^-$ in further discussions of the topology in double MSG 191.233 $P6/mmm$.  The layer constructions for the double SIs $(z_{12},z_{6m,\pi}^+)$ in MSG 191.233 $P6/mmm$ are given by:
\begin{enumerate}
\item A $\hat{\bf z}$-normal layer with $C^+_z=1$ in the $z=0$ plane has the SIs (21).
\item A $\hat{\bf z}$-normal layer with $C^+_z=1$ in the $z=\frac12$ plane has the SIs (05).
\item An $\hat{\bf x}$-normal layer with $C^+_x=1$ in the $x=0$ plane has the SIs (60).  We emphasize that, in this layer construction, there are also $|C^{+}|=1$ mirror Chern layers in the $C_{6z}\hat{\bf x}$, $C_{6z}^2 \hat{\bf x}$, $C_{6z}^3 \hat{\bf x}$, $C_{6z}^4 \hat{\bf x}$, and $C_{6z}^5 \hat{\bf x}$ directions implied by the $\{C_{6z}|\mathbf{0}\}$ rotation symmetry.
\item A $\hat{\bf y}$-normal layer with $C^+_y=1$ in the $y=0$ plane has the SIs (60).  We emphasize that, in this layer construction, there are also $|C^{+}|=1$ mirror Chern layers in the $C_{6z}\hat{\bf y}$, $C_{6z}^2 \hat{\bf y}$, $C_{6z}^3 \hat{\bf y}$, $C_{6z}^4 \hat{\bf y}$, and $C_{6z}^5 \hat{\bf y}$ directions implied by the $\{C_{6z}|\mathbf{0}\}$ rotation symmetry.
\end{enumerate}

\textit{Axion insulators} -- We find that states with odd $z_{12}$ SIs cannot be constructed from layers of 2D stable topological phases.  However, we may still use subduction relations to determine the bulk topology of insulators with odd values of $z_{12}$.  First, as we will show below, insulators in double MSG 191.233 $P6/mmm$ with $z_{12}\text{ mod }2 =1$ subduce to $(2000)_{P\bar{1}}$ in MSG 2.4 $P\bar1$.  Hence, if the $z_{12}\text{ mod }2 =1$ phases in double MSG 191.233 $P6/mmm$ are insulating, then the bulk insulator must either be an AXI or a 3D QAH state.  Because the net Chern numbers $C_{x,y,z}=0$ must vanish if the bulk is gapped, due to the mirror symmetries of double MSG 191.233 $P6/mmm$, then insulators with $z_{12}\text{ mod }2 =1$ in double MSG 191.233 $P6/mmm$ must be AXIs.  This result can also be understood by subducing from a $\mathcal{T}$-symmetric SSG.  Specifically, because insulators with the double SIs $z_{12}\text{ mod }2 =1$ in double MSG 191.233 $P6/mmm$ can be subduced from insulators with $(z_{12})_{P6/mmm1'}\text{ mod }2=1$ in Type-II SG 191.234 $P6/mmm1^\pr$, which correspond to $\mathcal{T}$-symmetric 3D TIs with $\theta=\pi$~\cite{AshvinIndicators,ChenTCI,AshvinTCI}, then the double SIs $z_{12}\text{ mod }2=1$ in double MSG 191.233 $P6/mmm$ are compatible with bulk-gapped states.  Hence, we conclude that 3D insulators with $z_{12}\text{ mod }2=1$ in double MSG 191.233 $P6/mmm$ are AXIs, \emph{without ambiguity}.  We conjecture that $z_{12}\text{ mod }2=1$ AXIs in double MSG 191.233 $P6/mmm$ can be constructed using the topological crystal method~\cite{ZhidaHermeleCrystal}, which additionally incorporates cell complexes of 2D Chern insulators, TIs, and TCIs.

\textit{Helical HOTI phases protected by mirror and $C_6$ rotation symmetry} -- First, the double SIs $(60)_{P6/mmm1^\pr}$ of Type-II double SSG SG 191.234 $P6/mmm1^\pr$ either correspond to a rotation-anomaly (non-axionic, \emph{i.e.} $\theta\text{ mod }2\pi=0$) HOTI protected by $C_{6}$ and $\TRS$ symmetries, or a mirror TCI with $C_{m_{z}}\text{ mod }12=6$ (\emph{c.f.} Table 7 in the Supplementary Material of Ref.~\onlinecite{ChenTCI}).  In the $C_{6}$- and $\mathcal{T}$-symmetric HOTI phase, there are $6 +12n$ ($n\in\{\mathbb{Z}^{+},0\}$) helical hinge modes on a $z$-directed, $C_{6}$- and $\mathcal{T}$-symmetric rod, and $6+12n$ twofold Dirac points on the top ($\hat{\bf z}$-normal) rod surface that are locally protected by mirror symmetry (see Appendix~\ref{sec:newHOTIs}).  Because double SSG SG 191.234 $P6/mmm1^\pr$ contains $\{m_{x,y}|{\bf 0}\}$ symmetries (as well as their conjugates under $C_{6z}$ symmetry), then six of the helical hinge modes on the boundary of a $6/mmm1'$-symmetric sample must also be pinned to the hinge projections of bulk mirror planes whose normal vectors lie in the $xy$-plane, and must be indicated by bulk mirror Chern numbers.  Hence, when $\mathcal{T}$ symmetry is relaxed in a sixfold rotation-anomaly $(60)_{P6/mmm1^\pr}$ HOTI phase in Type-II double SSG 191.234 $P6/mmm1^\pr$ while preserving the symmetries of MSG 191.233 $P6/mmm$, the surface and hinge states will remain gapless and anomalous [see Fig.~\ref{fig:SS}(c) and Appendix~\ref{sec:HOTItbModel}].

We will next prove that there are $6+12n$ twofold Dirac points on the top surface of a $6/mmm$-symmetric nanorod of the $(60)$ sixfold rotation-anomaly magnetic HOTI phase in double MSG 191.233 $P6/mmm$ introduced in this work.  We denote the four layer constructions as $L_a$ ($a=1\cdots 4$), respectively.  First, we note that the $(60)$ mirror TCI phase with $C_{m_{z}}\text{ mod }12=6$ can be constructed as $(6m+3)L_1\oplus (6m'+3) L_2$.  Next, the sixfold rotation-anomaly HOTI phase can be constructed as $(2n+1)L_3$, or $ (2n+1) L_4$, or through any superposition of an odd number of the aforementioned layer constructions.  Adding $6L_1$ or $6 L_2$, which have SIs (00), to the layer-constructed HOTI phase will not change the top surface spectrum, because $L_1$ and $L_2$ consist of horizontal (\emph{i.e.} $\hat{\bf z}$-normal) layers, and hence only contribute surface and hinge states on boundaries whose normal vectors lie in the $xy$-plane.

We will thus focus on the top surface spectra of the $(2n+1)L_3$ and $ (2n+1) L_4$ layer constructions.  We first consider $(2n+1)L_3$.  As shown in Supplementary Note 5 in Ref.~\onlinecite{ChenTCI} and in Table 6 of the Supplementary Material of Ref.~\onlinecite{ChenTCI}, the Chern numbers in the $m_{x}$ mirror sectors are given by $C_{k_x=0}^+=-C_{k_x=0}^-=4n+2$, $C_{k_x=\pi}^+=-C_{k_x=\pi}^-=0$.  In the 2D top surface BZ, $C^+_{k_x=0}$ mandates the presence of $|4n+2|$ twofold Dirac points on the $k_x=0$ line.  Due to the $C_{6z}$ symmetry, there must also be $2|4n+2|$ twofold Dirac points on the $C_{6z}$ and $C_{6z}^{-1}$ conjugates of the $k_{x}=0$ line.  Hence, the total number of top-surface Dirac points is $3|4n+2|\text{ mod }12= 6$.  Lastly, we note that performing the analogous analysis on the $(2n+1)L_4$ layer construction also returns the same number of mirror-protected twofold Dirac points on the top surface ($6+12n$).

In Appendix~\ref{sec:HOTIfermionDoubling}, we will prove that, on the top surface of the $(60)$ HOTI state -- which respects the symmetries of Type-I double magnetic wallpaper group~\cite{DiracInsulator,WiederLayers,ChenRotation,SteveMagnet,ConwaySymmetries} $p6m$ -- the presence of $6+12n$ ($n\in\{\mathbb{Z}^{+},0\}$) twofold surface Dirac points circumvents the fermion multiplication theorem for 2D lattices with double magnetic wallpaper group $p6m$.

\textit{Relationship with the SIs in other double SSGs} -- To identify the AXI phases, we subduce the SIs onto double MSG 2.4 $P\bar1$:
\begin{equation}
(z_{12},z_{6m,\pi}^+)_{P6/mmm} \to (\eta_{4I},z_{2I,1},z_{2I,2},z_{2I,3})_{P\bar1}=(2(z_{12}\text{ mod }2),000)_{P\bar1}.
\end{equation}
Because the AXI $\mathcal{I}$ SI $\eta_{2I'}=\frac12\eta_{4I}=z_{12}\text{ mod }2$ [Eq.~(\ref{eq:eta2Ip})], then we conclude that insulators with odd $z_{12}$ SIs in double MSG 191.233 $P6/mmm$ are AXIs.

\subsubsection{Double SIs in Type-II Double SG 2.5 $P\bar{1}1'$}
\label{sec:P-11p}

Using the definition of a minimal double SSG established in Appendix~\ref{sec:minimalSIProcedure}, we find that there are only five minimal Type-II double SSGs: 
    2.5 $P\bar11'$,  
    83.44 $P4/m1'$,  
    87.76 $I4/m1'$,
    175.138 $P6/m1'$, and
    176.144 $P6_{3}/m1'$.  The SIs, SI formulas, and physical interpretation of the SIs in the Type-II double SSGs were previously determined in Refs.~\onlinecite{AshvinIndicators,ChenTCI,AshvinTCI}.  In the physical basis employed in this work, the SI formulas, physical interpretations, and layer constructions of the double SIs in the above minimal Type-II double SSGs are provided in Ref.~\onlinecite{ChenTCI}.  Here and below [Appendices~\ref{sec:P4/m1p},~\ref{sec:I4/m1p},~\ref{sec:P6/m1p}, and~\ref{sec:P63/m1p}, respectively], we will briefly review the established SI formulas and physical interpretations of the double SIs in the five minimal Type-II double SSGs.

To begin, the double SSG 2.5 $P\bar11'$ is generated by
    $\{E|100\}$, 
    $\{E|010\}$,
    $\{E|001\}$, 
    $\{\INV|\mathbf{0}\}$, and $\{\TRS|{\bf 0}\}$.

The SI group is $\ZZ_4\times\ZZ_2^3$.  In the physical basis, the four double SIs $(z_4,z_{2w,1},z_{2w,2},z_{2w,3})$ of double SSG 2.5 $P\bar{1}1'$ have the respective SI formulas:
\begin{equation}
z_4 = \sum_{K} \frac12 n_K^- = \sum_{K} \frac{n_K^- - n_K^+}{4} \text{ mod } 4, 
\end{equation}
\begin{equation}
z_{2w,i} = \sum_{K,K_i=\pi} \frac12 n_K^- = \sum_{K,K_i=\pi} \frac{n_K^- - n_K^+}{4} \text{ mod } 2\qquad (i=1,2,3),
\end{equation}
where $K$ runs over the eight $\mathcal{I}$-invariant momenta in the first BZ, and $n^{\pm}_K$ are the number of Bloch states with $\pm1$ parity ($\mathcal{I}$) eigenvalues at $K$ in the group of bands under consideration.  The double SIs $(z_4,z_{2w,1},z_{2w,2},z_{2w,3})_{P\bar{1}1'}$ in double SSG 2.5 $P\bar{1}1'$ have the same SI formulas as the double SIs in $(z_4,z_{2w,1},z_{2w,2},z_{2w,3})_{Pmmm}$ in double MSG 47.249 $Pmmm$ [Eqs.~(\ref{eq:z4}) and~(\ref{eq:z2w})], which we previously analyzed in Appendix~\ref{subsec:Pmmm}.

The physical interpretations of the double SIs in Type-II double SSG 2.5 $P\bar{1}1'$ are given below:~\cite{ChenTCI}:
\begin{enumerate}
\item $z_{4}=1,3$ indicate strong 3D TIs protected by $\TRS$ symmetry.
\item For $z_{4}=0,2$, $z_{2w,i}=1$ indicates a weak TI phase that can be deformed into a stack of 2D TIs whose normal vectors point in the $i$-direction [\emph{e.g.}, the double SIs $(z_4,z_{2w,1},z_{2w,2},z_{2w,3})_{P\bar{1}1'}=(2110)_{P\bar{1}1'}$ indicate a weak TI that is equivalent to a stack of 2D TIs oriented in the $x+y$-direction].
\item For $z_{2w,1}=z_{2w,2}=z_{2w,3}=0$, $z_{4}=2$ indicates a non-axionic helical HOTI protected by $\mathcal{I}$ and $\mathcal{T}$ symmetries with a sample-encircling helical hinge mode (see Supplementary Note 5 in Ref.~\onlinecite{ChenTCI}).
\end{enumerate}

\vspace{0.2in}
\subsubsection{Double SIs in Type-II Double SG 83.44 $P4/m1'$}
\label{sec:P4/m1p}

The double SSG 83.44 $P4/m1'$ is generated by
    $\{E|100\}$, 
    $\{E|010\}$,
    $\{E|001\}$, 
    $\{\INV|\mathbf{0}\}$,
    $\{C_{4z}|\mathbf{0}\}$, and $\{\TRS|{\bf 0}\}$.

The double SSG 83.44 $P4/m1'$ has the SI group $\ZZ_8\times\ZZ_4\times\ZZ_2$.  In the physical basis, the $\ZZ_8$ double SI has the SI formula:
\begin{equation}
z_8 = \frac32 n^{\frac32,+} - \frac32 n^{\frac32,-} - \frac12 n^{\frac12,+} + \frac12 n^{\frac12,-} \text{ mod } 8,
\end{equation}
\begin{equation}
n^{j,\pm} = \sum_{K=\Gamma,M,Z,A} n^{j,\pm}_K + \sum_{K=X,R} n^{\frac12,\pm}_K,
\end{equation}
where $n_K^{j,\pm}$ ($K=\Gamma,M,Z,A$) are the number of states at the momentum $K$ with parity ($\mathcal{I}$) eigenvalue $\pm 1$ and $\{C_{4z}|\mathbf{0}\}$ eigenvalue angular momentum $j$ (modulo 4), and $n_K^{j,\pm}$ ($K=X,R$) are the number of states at the momentum $K$ with parity eigenvalue $\pm 1$ and angular momentum $j$ (modulo 2).  The $\ZZ_4$ SI subduces to the weak TCI invariant $z_{4m,\pi}^-$ in double MSG 83.43 $P4/m$ [Eq.~(\ref{eq:z4mp-})], and the $\ZZ_2$ SI subduces to the weak TI invariant $z_{2w,1}$ in double SSG 2.5 $P\bar11'$ [Eq.~(\ref{eq:z2w})].  We note that in Ref.~\onlinecite{ChenTCI}, $z_{4m,\pi}^-$ is instead labeled $z_{4m,\pi}$.  As a set, the three double SIs $(z_{8},z_{4m,\pi}^-,z_{2w,1})_{P4/m1'}$ in double SSG 83.44 $P4/m1'$ have the same SI formulas as the double SIs $(z_{8},z_{4m,\pi}^-,z_{2w,1})_{P4/mmm}$ in double MSG 123.339 $P4/mmm$, which we previously analyzed in Appendix~\ref{subsec:P4/mmm}.

The physical interpretations of the double SIs in Type-II double SSG 83.44 $P4/m1'$ are given below~\cite{ChenTCI}:
\begin{enumerate}
\item $z_{2w,1}=1$ indicates the presence of nontrivial weak TI indices in the $k_{x,y}=\pi$ planes.
\item Nonzero values of $z_{4m,\pi}^-$ indicate nontrivial mirror sector Chern numbers in the $k_z=\pi$ plane: $z_{4m,\pi}^- = C_{k_z=\pi}^-\text{ mod }4 = - C_{k_z=\pi}^+\text{ mod }4$ [see Eq.~(\ref{eq:z4mp-}) and the surrounding text].  
\item $z_8\neq 0,4$ indicate nontrivial mirror sector Chern numbers in the $k_z=0,\pi$ planes: $C_{k_z=0}^- - C_{k_z=\pi}^+\text{ mod }4= z_8$ [see Appendix~\ref{subsec:P4/m} for the subduction relations between $(z_8)_{P4/m1'}$ in double SSG 83.44 $P4/m1'$ and the double SIs in double MSG 83.43 $P4/m$].  $z_8\text{ mod }2=1$ specifically indicates strong 3D TI phases.
\item For $z_{2w,1}=z_{4m,\pi}^-=0$, $z_8 = 4$ either indicates a mirror TCI phase with $C_{m_{z}}\text{ mod }8=4$, or a non-axionic fourfold rotation-anomaly HOTI phase with $C_{4z}$- and $\mathcal{T}$-symmetry-protected bulk topology and $4 + 8n$ ($n\in\{\mathbb{Z}^{+},0\}$) $\mathcal{T}$-protected helical hinge modes (see Supplementary Note 5 in Ref.~\onlinecite{ChenTCI}).
\end{enumerate}

\subsubsection{Double SIs in Type-II Double SG 87.76 $I4/m1'$}
\label{sec:I4/m1p}

The double SSG 87.76 $I4/m1'$ is generated by
    $\{E|\bar{\frac12}\frac12\frac12\}$, 
    $\{E|\frac12\bar{\frac12}\frac12\}$,
    $\{E|\frac12\frac12\bar{\frac12}\}$, 
    $\{\INV|\mathbf{0}\}$,
    $\{C_{4z}|\mathbf{0}\}$, and $\{\TRS|{\bf 0}\}$.

The double SSG 87.76 $I4/m1'$ has the SI group $\ZZ_8\times\ZZ_2$.  In the physical basis, the $\mathbb{Z}_{8}$ double SI subduces to $(z_{8})_{P4/m1'}$ in double SSG 83.44 $P4/m1'$ (see Appendix~\ref{sec:P4/m1p}):
\begin{equation}
z_8 = \frac32 n^{\frac32,+} - \frac32 n^{\frac32,-} - \frac12 n^{\frac12,+} + \frac12 n^{\frac12,-} \text{ mod } 8,
\end{equation}
in which $n^{j,\pm}$ are given by:
\begin{equation}
n^{j,\pm} = \sum_{K=\Gamma,M} n^{j,\pm}_K + \sum_{K=X,N} n^{\frac12,\pm}_K + \sum_{K=P} n^{\pm j}_K,
\end{equation}
where $n_K^{j,\pm}$ ($K=\Gamma,M$) are the number of states at the momentum $K$ with parity eigenvalue $\pm 1$ and $\{C_{4z}|\mathbf{0}\}$ eigenvalue angular momentum $j$ (modulo 4), $n_K^{j,\pm}$ ($K=X,N$) are the number of states at the momentum $K$ with parity eigenvalue $\pm 1$ and angular momentum $j$ (modulo 2), and $n_P^{\pm j}$ are the number of states at the momentum $P$ with $\{S_{4z}|0\}$ eigenvalue $e^{\mp i \frac{2\pi}4 j}$.  The $\ZZ_2$ SI subduces to the weak TI invariant $z_{2w,1}$ in double SSG 2.5 $P\bar11'$ [Eq.~(\ref{eq:z2w})].

The physical interpretations of the double SIs in Type-II double SSG 87.76 $I4/m1'$ are closely related to the physical interpretations of the double SIs in double SSG 83.44 $P4/m1'$ previously determined in Appendix~\ref{sec:P4/m1p} and Ref.~\onlinecite{ChenTCI}:
\begin{enumerate}
\item $z_{2w,1}=1$ indicates the presence of nontrivial weak TI indices in the $k_{x,y}=\pi$ planes in the primitive-cell BZ.
\item $z_8\neq 0,4$ indicate nontrivial mirror sector Chern numbers in the $k_z=0$ plane: $C_{k_z=0}^-\text{ mod }4 = -C_{k_z=0}^+\text{ mod }4= z_8$ (noting that the $k_{z}=0,\pi$ planes are related by reciprocal lattice vectors, because the Bravais lattice of SSG 87.76 $I4/m1'$ is body-centered tetragonal~\cite{BigBook}).  $z_8\text{ mod }2=1$ specifically indicates strong 3D TI phases.
\item For $z_{2w,1}=0$, $z_8 = 4$ either indicates a mirror TCI phase with $C_{m_{z}}\text{ mod }8=4$, or a non-axionic fourfold rotation-anomaly HOTI phase with $C_{4z}$- and $\mathcal{T}$-symmetry-protected bulk topology and $4 + 8n$ ($n\in\{\mathbb{Z}^{+},0\}$) $\mathcal{T}$-protected helical hinge modes (see Supplementary Note 5 in Ref.~\onlinecite{ChenTCI}).
\end{enumerate}

\subsubsection{Double SIs in Type-II Double SG 175.138 $P6/m1'$}
\label{sec:P6/m1p}

The double SSG 175.138 $P6/m1'$ is generated by
    $\{E|100\}$, 
    $\{E|010\}$,
    $\{E|001\}$, 
    $\{\INV|\mathbf{0}\}$,
    $\{C_{6z}|\mathbf{0}\}$, and $\{\TRS|{\bf 0}\}$.

The double SSG 175.138 $P6/m1'$ has the SI group $\ZZ_{12}\times\ZZ_6$.  In the physical basis, the SI formula of the $\ZZ_{12}$ SI can be expressed in terms of previously established double SIs~\cite{ChenTCI}:
\begin{equation}
z_{12} = \delta_{6m} + 3[( \delta_{6m} - z_4) \text{ mod }4] \text{ mod } 12,
\end{equation}
where $\delta_{6m}$ is computed by subduction onto double MSG 175.137 $P6/m$ [Eq.~(\ref{eq:d6m})], and $z_4$ is computed by subduction onto double SSG 2.5 $P\bar{1}1'$ [see Appendix~\ref{sec:P-11p}].  Additionally, in the physical basis, the $\ZZ_6$ SI is the weak TCI invariant $z_{6m,\pi}^+$ for the mirror Chern number (modulo 6) in the $k_{z}=\pi$ plane, and can also be computed by subduction onto double MSG 175.137 $P6/m$ [Eq.~(\ref{eq:z6mp+})].  We note that in Ref.~\onlinecite{ChenTCI}, $z_{6m,\pi}^+$ is instead labeled $z_{6m,\pi}$.  As a set, the two double SIs $(z_{12},z_{6m,\pi}^{+})_{P6/m1'}$ in double SSG 175.138 $P6/m1'$ have the same SI formulas as the double SIs $(z_{12},z_{6m,\pi}^{+})_{P6/mmm}$ in double MSG 191.233 $P6/mmm$, which we previously analyzed in Appendix~\ref{subsec:P6/mmm}.

The physical interpretations of the double SIs in double SSG 175.138 $P6/m1'$ are given below~\cite{ChenTCI}:
\begin{enumerate}
\item Nonzero values of $z_{6m,\pi}^+$ indicate nontrivial mirror sector Chern numbers in the $k_z=\pi$ plane: $z_{6m,\pi}^+= C_{k_z=\pi}^+\text{ mod }6 = - C_{k_z=\pi}^-\text{ mod }6$ [see Eq.~(\ref{eq:z6mp+}) and the surrounding text].
\item $z_{12}\neq 0,6$ indicate nontrivial mirror sector Chern numbers in the $k_z=0,\pi$ planes: $C_{k_z=\pi}^+ - C_{k_z=0}^-\text{ mod }6 = z_{12}$ [see Appendix~\ref{subsec:P6/m} for the subduction relations between $(z_{12})_{P6/m1'}$ in double SSG 175.138 $P6/m1'$ and the double SIs in double MSG 175.137 $P6/m$].  $z_{12}\text{ mod }2=1$ specifically indicates strong 3D TI phases.
\item For $z_{6m,\pi}^+=0$, $z_{12}= 6$ either indicates a mirror TCI phase with $C_{m_{z}}\text{ mod }12=6$, or a non-axionic sixfold rotation-anomaly HOTI phase with $C_{6z}$- and $\mathcal{T}$-symmetry-protected bulk topology and $6 + 12n$ ($n\in\{\mathbb{Z}^{+},0\}$) $\mathcal{T}$-protected helical hinge modes (see Supplementary Note 5 in Ref.~\onlinecite{ChenTCI}).
\end{enumerate}

\subsubsection{Double SIs in Type-II Double SG 176.144 $P6_{3}/m1'$}
\label{sec:P63/m1p}

The double SSG 176.144 $P6_3/m1'$ is generated by
    $\{E|100\}$, 
    $\{E|010\}$,
    $\{E|001\}$, 
    $\{\INV|\mathbf{0}\}$,
    $\{C_{6z}|00\frac12\}$, and $\{\TRS|{\bf 0}\}$.

The double SSG 176.144 $P6_3/m1'$ has the SI group $\ZZ_{12}$.  In the physical basis, the SI formula of the $\mathbb{Z}_{12}$ SI can be expressed in terms of previously established double SIs~\cite{ChenTCI}:
\begin{equation}
z_{12}' = z_{6m,0}^+ + 3[( z_{6m,0}^+ - z_4) \text{ mod }4] \text{ mod } 12, 
\label{eq:z12p}
\end{equation}
where $z_{6m,0}^+$ is computed by subduction onto double MSG 176.143 $P6_3/m$ [Eq.~(\ref{eq:z6m0+})], and $z_4$ is computed by subduction onto double SSG 2.5 $P\bar{1}1'$ [see Appendix~\ref{sec:P-11p}].  We note that, unlike previously in double SSG 175.138 $P6/m1'$ (Appendix~\ref{sec:P6/m1p}), the mirror sector Chern numbers in the $k_{z}=\pi$ plane individually vanish $C^{\pm}_{k_{z}=\pi}=0$ for any group of bands in double SSG 176.144 $P6_3/m1'$.  This can be seen by first recognizing that the matrix representatives of $\{\TRS C_{6z}|00\frac12\}$ and $\{m_z|00\frac12\}$ anticommute in any small corep of any little group in the $k_{z}=\pi$ plane that contains both $\{\TRS C_{6z}|00\frac12\}$ and $\{m_z|00\frac12\}$.  Hence, if $|\psi\rangle$ is a Bloch eigenstate of $\{m_z|00\frac12\}$ at a ${\bf k}$ point in the $k_{z}=\pi$ plane with the $\{m_z|00\frac12\}$ eigenvalue $i$, then $\{\TRS C_{6z}|00\frac12\}|\psi\rangle$ is also an eigenstate of $\{m_z|00\frac12\}$ with the same eigenvalue ($i$).  Consequently, there is an effective time-reversal symmetry ($\{\TRS C_{6z}|00\frac12\}$) within each mirror sector, which enforces that the mirror sector Chern numbers in the $k_{z}=\pi$ plane individually vanish.

The physical interpretations of the double SIs in double SSG 176.144 $P6_3/m1'$ are given below~\cite{ChenTCI}:
\begin{enumerate}
\item $z_{12}'\neq 0,6$ indicate nontrivial mirror sector Chern numbers in the $k_{z}=0$ plane: $C_{k_z=0}^+\text{ mod }6 = -C_{k_z=0}^-\text{ mod }6=z_{12}'$ [see Appendix~\ref{sec:P6_3/m} for the subduction relations between $(z_{12}')_{P6_{3}/m1'}$ in double SSG 176.144 $P6_3/m1'$ and the double SIs in double MSG 176.143 $P6_3/m$].  $z_{12}'\text{ mod }2=1$ specifically indicates strong 3D TI phases.
\item $z_{12}'= 6$ either indicates a mirror TCI phase with $C_{m_{z}}\text{ mod }12=6$, or a non-axionic sixfold rotation-anomaly HOTI phase with $6_{3}$-screw- and $\mathcal{T}$-symmetry-protected bulk topology and $6 + 12n$ ($n\in\{\mathbb{Z}^{+},0\}$) $\mathcal{T}$-protected helical hinge modes (see Supplementary Note 5 in Ref.~\onlinecite{ChenTCI}).
\end{enumerate}

\subsubsection{Double SIs in Type-III Double MSG 27.81 $Pc'c'2$}

Finally, beginning here with double MSG 27.81 $Pc'c'2$ and continuing below, we will introduce the physical-basis SI formulas and the physical interpretations of the double SIs in the 11 minimal Type-III double MSGs (see Appendix~\ref{sec:minimalSIProcedure}).  To begin, the double MSG 27.81 $Pc'c'2$ is generated by 
    $\{E|100\}$, 
    $\{E|010\}$,
    $\{E|001\}$, 
    $\{C_{2z}|\mathbf{0}\}$, and $\{\TRS m_x |00\frac12\}$.

\textit{SI} -- The double MSG 27.81 $Pc^\pr c^\pr 2$ has the SI group $\ZZ_2$.  As we will shortly demonstrate, in the physical basis, the double SI $z_{2R}^\pr$ indicates the \emph{even-valued} Chern number in the $k_{z}=\pi$ plane (modulo 4): $C_{k_{z}=\pi}\text{ mod }4 = 2z_{2R}^\pr$.  Hence, insulators with $z_{2R}^\pr=1$ are 3D QAH states with $C_{z}\text{ mod }4=2$.

We will first demonstrate that Bloch states at the $\{C_{2z}|{\bf 0}\}$-invariant momenta in the $k_{z}=\pi$ plane in double MSG 27.81 $Pc'c'2$ form doubly-degenerate pairs with the same $\{C_{2z}|{\bf 0}\}$ eigenvalues.  To begin, in the $k_{z}=\pi$ plane, the matrix representative of $\{\TRS m_x|00\frac12\}$ squares to minus the identity in all double-valued small coreps.  Hence, all of the irreducible small coreps in the $k_{z}=\pi$ plane along the $\{\TRS m_x|00\frac12\}$-invariant lines $k_{y}=0,\pi$ must be at least twofold degenerate (and in fact are exactly twofold degenerate in double MSG 27.81 $Pc'c'2$).  Next, the matrix representatives of $\{\TRS m_x|00\frac12\}$ and $\{C_{2z}|\mathbf{0}\}$ anticommute in all small coreps at the $\{C_{2z}|\mathbf{0}\}$-invariant points $k_{x,y}=0,\pi$ in the $k_{z}=\pi$ plane.  This implies that, if $\ket{\psi}$ is a Bloch state at $k_{x,y}=0,\pi$ in the $k_{z}=\pi$ plane for which $\{C_{2z}|\mathbf{0}\} \ket{\psi}=i\ket{\psi}$, then:
\begin{equation}
\{C_{2z}|\mathbf{0}\} \{\TRS m_x|00\frac12 \} \ket{\psi}=- \{\TRS m_x|00\frac12 \} \{C_{2z}|\mathbf{0}\} \ket{\psi} = i \{\TRS m_x|00\frac12 \} \ket{\psi}. 
\label{eq:27.81-relation}
\end{equation}
Eq.~(\ref{eq:27.81-relation}) implies that both Bloch states in each $\{\TRS m_x|00\frac12\}$ doublet at $k_{x,y}=0,\pi$, $k_{z}=\pi$ must have the \emph{same} $\{C_{2z}|{\bf 0}\}$ eigenvalues.  We therefore define the $\ZZ_2$ SI as the parity of the number of doublets with $\{C_{2z}|{\bf 0}\}$ eigenvalue $-i$ in the $k_z=\pi$ plane:
\begin{equation}
z_{2R}^\pr = \sum_{K=Z,T,U,R} \frac12 n^{\frac12}_K \text{ mod } 2=\frac{C_{k_z=\pi}}{2} \text{ mod } 2, 
\label{eq:z2Rp-27.81}
\end{equation}
where $n^{\frac12}_{K}$ is the number of states with $\{C_{2z}|\mathbf{0}\}$ eigenvalue $-i$, such that $\frac12 n^{\frac12}_{K}$ is the number of \emph{doublets} in which both Bloch states have the $\{C_{2z}|{\bf 0}\}$ eigenvalues $-i$.

\textit{Layer constructions} -- To diagnose the topology associated to $z_{2R}^\pr=1$, we employ the layer construction method.  We begin by placing a $\hat{\bf z}$-normal Chern layer with $C_z=1$ in the $z=0$ plane.  Due to the $\{T m_x|00\frac12\}$ symmetry in double MSG 27.81 $Pc'c'2$, there must be another Chern layer with $C_z=1$ in the $z=\frac12$ plane, such that the total Chern number per cell is $C_{z}=2$, and the Chern number of the occupied bands in the $k_{z}=\pi$ plane is $C_{k_{z}=\pi}=2$.  Hence, in this layer construction of a 3D QAH state with $C_{z}=2$, the $\mathbb{Z}_{2}$ SI is nontrivial $z_{2R}^\pr=1$.

\textit{Relationship with the SIs in other double SSGs} -- We next compute the subduction relations between the SIs in double MSG 27.81 $Pc'c'2$ and the SIs in the maximal unitary subgroup double MSG 3.1 $P2$ (see Appendix~\ref{subsec:P2}):
\begin{equation}
(z_{2R}^\pr)_{Pc^\pr c^\pr 2} \to 
(z_{2R})_{P2} = (0)_{P2}.
\label{eq:cPrcPr2Subduct}
\end{equation}
Eqs.~(\ref{eq:z2Rp-27.81}) and~(\ref{eq:cPrcPr2Subduct}) imply that symmetry-indicated 3D QAH states with $z_{2R}^\pr=1$ in double MSG 27.81 $Pc'c'2$ necessarily subduce to \emph{non-symmetry-indicated} 3D QAH states with $(z_{2R})_{P2}=0$ in double MSG 3.1 $P2$, in agreement with the physical-basis double SI relations $C_{k_{z}=\pi}\text{ mod }4 = 2z_{2R}^\pr$ and $C_{k_{z}=\pi}\text{ mod }2 = z_{2R}$ [taking the twofold axis in double MSG 3.1 $P2$ to be oriented in the $z$-direction, see Eq.~(\ref{eq:z2R}) and the surrounding text].

Lastly, if we impose $\mathcal{T}$ symmetry, then the position-space Chern numbers must vanish, which enforces $z_{2R}^\pr$ to be zero.  Correspondingly, in double SSG 27.79 $Pcc21'$ -- the SSG generated by adding $\{\mathcal{T}|{\bf 0}\}$ to double MSG 27.81 $Pc'c'2$ -- the double SI group is trivial.

\subsubsection{Double SIs in Type-III Double MSG 41.215 $Ab'a'2$}

The double MSG 41.215 $Ab^\pr a^\pr 2$ is generated by 
    $\{E|100\}$, 
    $\{E|0\frac12\frac12\}$,
    $\{E|0\frac12\bar{\frac12}\}$, 
    $\{C_{2z}|\mathbf{0}\}$, and $\{\TRS m_x |\frac12\frac12 0\}$.  The primitive lattice vectors are:
\begin{equation}
\mbf{a}_1=(1,0,0),\qquad \mbf{a}_2=(0,\frac12,\frac12),\qquad \mbf{a}_3=(0,\frac12,-\frac12),
\end{equation}
and the reciprocal lattice vectors of the primitive cell are:
\begin{equation}
\mbf{b}_1 =2\pi(1,0,0),\qquad \mbf{b}_2=2\pi(0,1,1),\qquad \mbf{b}_3 = 2\pi (0,1,-1).
\end{equation}
In the conventional (super)cell, the lattice vectors are:
\begin{equation}
\mbf{a}_1'=(1,0,0),\qquad \mbf{a}_2'=\mbf{a}_2 + \mbf{a}_3=(0,1,0),\qquad \mbf{a}_3'=\mbf{a}_2 - \mbf{a}_3=(0,0,1).
\end{equation}
and the reciprocal lattice vectors of the conventional cell are:
\begin{equation}
\mbf{b}_1' =2\pi(1,0,0),\qquad \mbf{b}_2'=2\pi(0,1,0),\qquad \mbf{b}_3' = 2\pi (0,0,1).
\end{equation}
In the analysis below of the double SIs in double MSG 41.215 $Ab^\pr a^\pr 2$, we will refer to coordinates in the basis of the conventional cell for consistency with the convention employed in the BCS applications implemented for MTQC.

\textit{SI} -- The double MSG 41.215 $Ab^\pr a^\pr 2$ has the SI group $\ZZ_2$.  In the physical basis, the $\mathbb{Z}_{2}$ SI has the SI formula:
\begin{equation}
z_{2R} = n_\Gamma^{\frac12} \text{ mod }2,
\label{eq:z2R-41.215}
\end{equation}
where $n_\Gamma^{\frac12}$ is the number of occupied states with $\{C_{2z}|\mathbf{0}\}$ eigenvalues $-i$ at $\Gamma$.  Below, we will demonstrate that $z_{2R}=C_{z}\text{ mod }2$ where $C_{z}$ is the total position-space Chern number in the primitive cell (or equivalently $z_{2R}$ indicates the even-valued Chern number $C_{z}$ in the conventional cell modulo $4$), such that insulators in double MSG 41.215 $Ab^\pr a^\pr 2$ with $z_{2R}=1$ are 3D QAH states.

\textit{Layer constructions} -- To diagnose the topology associated to $z_{2R}=1$, we employ the layer construction method.  We begin by placing a $\hat{\bf z}$-normal Chern insulator with $C_z=1$ in the $z=0$ plane.  In the conventional cell, the system has $\{\TRS m_x|\frac12\frac120\}$ and $\{C_{2z}|\mathbf{0}\}$ symmetries, as well as the conventional-cell translation symmetries $\{E|100\}$ and $\{E|010\}$.  Because a minimal Chern insulator has one occupied band~\cite{QWZ}, then, in the conventional supercell -- which is twice as large as the primitive cell -- the system has two occupied bands.  Below, we will demonstrate that a set of occupied bands compatible with this layer construction exhibits $z_{2R}=1$.

We next determine the constraints imposed by symmetry on the occupied $\{C_{2z}|\mathbf{0}\}$ eigenvalues at the momenta $\Gamma(0,0,0)$, $Z(\pi,0,0)$, $(0,\pi,0)$, and $(\pi,\pi,0)$ [where we note that $(0,\pi,0)$ and $(\pi,\pi,0)$ are not high-symmetry points in MSG 41.215 $Ab^\pr a^\pr 2$, see~\href{http://www.cryst.ehu.es/cryst/mkvec}{MKVEC} (Appendix~\ref{sec:MKVEC})].  Because $\{\TRS m_x|\frac12\frac120\}^2=\{E|010\}$, then the matrix representative of $\{\TRS m_x|\frac12\frac120\}$ squares to minus the identity in all small coreps in the $k_y=\pi$ plane, and states in the $k_{y}=\pi$ plane must be at least [and are in fact exactly] twofold degenerate, whereas states in the $k_{y}=0$ plane are not required by $\{\TRS m_x|\frac12\frac120\}$ to be doubly degenerate [and are in fact singly degenerate at $\Gamma(0,0,0)$].  We then consider a Bloch eigenstate $\ket{\psi(k_x,\pi,0)}$ ($k_x\in\{0,\pi\}$) with $\{C_{2z}|\mathbf{0}\}$ eigenvalue $\xi\in \{i,-i\}$, and compute the $\{C_{2z}|{\bf 0}\}$ eigenvalue of the state $\{T m_x | \frac12\frac12 0\} \ket{\psi(k_x,\pi,0)}$:
\begin{equation}
\{C_{2z}|\mathbf{0}\}\{T m_x | \frac12\frac12 0\} \ket{\psi(k_x,\pi,0)} = -\{E|\bar1\bar1 0\} \{T m_x | \frac12\frac12 0\}\xi \ket{\psi(k_x,\pi,0)} = e^{-i k_x} \xi^* \{T m_x | \frac12\frac12 0\} \ket{\psi(k_x,\pi,0)}.
\label{eq:tempdoubletAaPrBPr2}
\end{equation}
Eq.~(\ref{eq:tempdoubletAaPrBPr2}) implies that doublets at $(\pi,\pi,0)$ consist of Bloch states with the same $\{C_{2z}|\mathbf{0}\}$ eigenvalues, but that the two states in each doublet at $(0,\pi,0)$ have oppositely-signed $\{C_{2z}|\mathbf{0}\}$ eigenvalues.  Next, we consider there to be a state $\ket{\psi(k_x,0,0)}$ ($k_x\in\{0,\pi\}$) with $\{C_{2z}|\mathbf{0}\}$ eigenvalue $\xi\in \{i,-i\}$, and compute the $\{C_{2z}|\mathbf{0}\}$ eigenvalue of $\{T m_x | \frac12\frac12 0\} \ket{\psi(k_x,0,0)}$:
\begin{equation}
\{C_{2z}|\mathbf{0}\}\{T m_x | \frac12\frac12 0\} \ket{\psi(k_x,0,0)} = -\{E|\bar1\bar1 0\} \{T m_x | \frac12\frac12 0\}\xi \ket{\psi(k_x,0,0)} = -e^{-i k_x} \xi^* \{T m_x | \frac12\frac12 0\} \ket{\psi(k_x,0,0)}.
\label{eq:tempdoubletAaPrBPr2More}
\end{equation}
Eq.~(\ref{eq:tempdoubletAaPrBPr2More}) implies that Bloch states at $Z(\pi,0,0)$ are doubly degenerate with complex-conjugate pairs of $\{C_{2z}|\mathbf{0}\}$ eigenvalues $\{i,-i\}$.

We have thus determined that Bloch states at $\Gamma(0,0,0)$ are singly degenerate, Bloch states at $Z(\pi,0,0)$ and $(0,\pi,0)$ are doubly degenerate and have opposite $\{C_{2z}|{\bf 0}\}$ eigenvalues, and that Bloch states at $(\pi,\pi,0)$ are doubly degenerate and have the same $\{C_{2z}|{\bf 0}\}$ eigenvalues.  Thus, one possible set of occupied $\{C_{2z}|{\bf 0}\}$ eigenvalues that satisfy the above constraints and the compatibility relations are $(-i,+i)$, $(-i,+i)$, $(-i,+i)$, $(+i,+i)$ at $\Gamma(0,0,0)$, $Z(\pi,0,0)$, $(0,\pi,0)$, $(\pi,\pi,0)$, respectively.  Next, we consider the $\{C_{2z}|{\bf 0}\}$ eigenvalues at the remaining two high-symmetry ${\bf k}$ points: $Y(0,2\pi,0)$ and $T(\pi,2\pi,0)$.  Because $\kk_Y - \bb_3 = (0,0,2\pi)$, then the occupied states at $Y$ must have the same $\{C_{2z}|\mathbf{0}\}$ rotation eigenvalues as the occupied states at $\Gamma(0,0,0)$ for bands that satisfy the compatibility relations.  Next, because $\kk_T - \bb_3 = (\pi,0,2\pi)$, then the occupied states at $T$ must have the same $\{C_{2z}|\mathbf{0}\}$ rotation eigenvalues as the occupied states at $Z(\pi,0,0)$ for bands that satisfy the compatibility relations.  In summary, the $\{C_{2z}|{\bf 0}\}$ eigenvalues of the occupied bands at the high-symmetry ${\bf k}$ points are given by:
\begin{equation}
\begin{tabular}{c|c|c|c|c}
    & $\Gamma(000)$ & $Z(\pi 00)$ & $Y(0,2\pi,0)$ & $T(\pi,2\pi,0)$\\
\hline
$\{C_{2z}|{\bf 0}\}$ & $-i,+i$ & $-i,+i$ & $-i,+i$ & $-i,+i$ \\
\end{tabular}.
\end{equation} 
Using Eq.~(\ref{eq:z2R-41.215}), we determine that the occupied bands have $z_{2R} =1$.  Next, using the established formula for the parity of the Chern numbers in the $k_{z}=0,\pi$ planes in terms of $\{C_{2z}|{\bf 0}\}$ rotation eigenvalues~\cite{QWZ,ChenBernevigTCI} [which is equivalent to Eq.~(\ref{eq:z2R})], we conclude that $C_{k_{z}=0,\pi}\text{ mod }2=1$, which is compatible with the layer construction for $z_{2R}=1$ introduced in the text preceding Eq.~(\ref{eq:tempdoubletAaPrBPr2}).  Thus, we conclude that insulators with $z_{2R}=1$ are 3D QAH states with $C_{z}\text{ mod }2=1$ in the primitive cell.

Lastly, if we impose $\mathcal{T}$ symmetry, then the position-space Chern numbers must vanish, which enforces $z_{2R}^\pr$ to be zero.  Correspondingly, in double SSG 41.212 $Aba21'$ -- the SSG generated by adding $\{\mathcal{T}|{\bf 0}\}$ to double MSG 41.215 $Ab^\pr a^\pr 2$ -- the double SI group is trivial.

\subsubsection{Double SIs in Type-III Double MSG 54.342 $Pc'c'a$}
\label{sec:PcPrcPra}

The double MSG 54.342 $Pc^\pr c^\pr a$ is generated by 
    $\{E|100\}$, 
    $\{E|010\}$,
    $\{E|001\}$, 
    $\{\INV|\mathbf{0}\}$,
    $\{C_{2z}|\frac1200\}$, and
    $\{\TRS m_y |00\frac12\}$.

\textit{SIs} -- The double MSG 54.342 $Pc^\pr c^\pr a$ has the SI group $\mathbb{Z}_2\times \mathbb{Z}_2$.  In the physical basis, the double SIs of double MSG 54.342 $Pc^\pr c^\pr a$ $(\eta_{2I}^\pr,z_{2R}^\pr)$ individually subduce to previously introduced double SIs.  First, the $\mathcal{I}$ AXI index $\eta_{2I}'$ subduces to the non-minimal index $(\eta_{2I}')_{P\bar{1}}$ in double MSG 2.4 $P\bar{1}$ (see Appendix~\ref{sec:P-1}).  Next, the even Chern number SI $2z_{2R}^\pr=C_{k_{z}=\pi}\text{ mod }4$ subduces to the same SI $(z_{2R}^\pr)_{Pc'c'2}$ in double MSG 27.81 $Pc'c'2$ [see Eq.~(\ref{eq:z2Rp-27.81}) and the surrounding text].  Hence, as we will show below, an insulator with $(\eta_{2I}^\pr,z_{2R}^\pr)=(10)$ in double MSG 54.342 $Pc^\pr c^\pr a$ is an $\mathcal{I}$-protected AXI if the non-symmetry-indicated Chern numbers vanish, and an insulator with $z_{2R}^\pr=1$ is a 3D QAH state with $C_{z}\text{ mod }4=2$.

\textit{Layer constructions} -- We find that all of the double SIs in double MSG 54.342 $Pc^\pr c^\pr a$ can be realized by layer constructions.  The layer constructions for the double SIs $(\eta_{2I}^\pr,z_{2R}^\pr)$ in double MSG 54.342 $Pc^\pr c^\pr a$ are given by:
\begin{enumerate}
\item A $\hat{\bf z}$-normal Chern layer with $C_z=1$ at the $z=0$ has the SIs (11).  We emphasize that, in this layer construction, there is also a $C_{z}=1$ Chern layer in the $z=\frac{1}{2}$ plane implied by the $\{\TRS m_y|00\frac12\}$ symmetry operation.  This layer construction is a 3D QAH state with $C_{k_{z}}=2$ in all BZ planes of constant $k_{z}$.
\item An $\hat{\bf x}$-normal Chern layer with $C_x=1$ at the $x=0$ plane has the SIs (10).  We emphasize that, in this layer construction, there is also a $C_{x}=-1$ Chern layer in the $x=\frac12$ plane implied by the $\{C_{2z}|\frac1200\}$ symmetry.  Because this layer construction consists of layers with alternating odd Chern numbers occupying $\mathcal{I}$ centers, then this layer construction is an $\mathcal{I}$-protected AXI (see Appendix~\ref{sec:P-1} and Refs.~\onlinecite{WilczekAxion,QHZ,VDBAxion,AndreiInversion,AshvinAxion,WiederAxion,YuanfengAXI,NicoDavidAXI1,NicoDavidAXI2,TMDHOTI,KoreanAXI,CohVDBAXI,ArisHopf,TitusRonnyKondoAXI,YoungkukMonopole,MurakamiAXI1,MurakamiAXI2,HingeStateSphereAXI,BJYangVortex,BarryBenCDW,IvoAXI1,IvoAXI2,GuidoAXI}).
\end{enumerate}

\textit{Relationship with the SIs in other double SSGs} -- The SIs in double MSG 54.342 $Pc^\pr c^\pr a$ are related to the SIs in double MSG 2.4 $P\bar1$ through the subduction relations:
\begin{equation}
(\eta_{2I}^\pr,z_{2R}^\pr)_{Pc^\pr c^\pr a} \to (\eta_{4I},z_{2I,1},z_{2I,2},z_{2I,3})_{P\bar1}=(2\eta_{2I}^\pr,000)_{P\bar1}.
\end{equation}

Lastly, we study the effects of imposing $\mathcal{T}$ symmetry.  The double SSG 54.338 $Pcca1'$ -- the SSG generated by adding $\{\mathcal{T}|{\bf 0}\}$ symmetry to double MSG 54.342 $Pc^\pr c^\pr a$ -- has the SI group $\mathbb{Z}_{4}\times\mathbb{Z}_{2}$~\cite{ChenTCI}.  The SIs in double SSG 54.338 $Pcca1'$ are related to the SIs in double MSG 54.342 $Pc^\pr c^\pr a$ through the subduction relations:
\begin{equation}
(z_{4},z_{2w,2})_{Pcca1'} \to (\eta_{2I}^\pr,z_{2R}^\pr)_{Pc^\pr c^\pr a} = (z_{4}\text{ mod }2,0)_{Pc^\pr c^\pr a}.
\end{equation}

\subsubsection{Double SIs in Type-III Double MSG 56.369 $Pc'c'n$}
\label{sec:PcPrcPrn}

The double MSG 56.369 $Pc^\pr c^\pr n$ is generated by 
    $\{E|100\}$, 
    $\{E|010\}$,
    $\{E|001\}$, 
    $\{\INV|\mathbf{0}\}$,
    $\{C_{2z}|\frac12\frac120\}$, and
    $\{\TRS m_y|0\frac12\frac12\}$.

\textit{SIs} -- The double MSG 56.369 $Pc^\pr c^\pr n$ has the SI group $\mathbb{Z}_2\times \mathbb{Z}_2$.  In the physical basis, the double SIs of double MSG 56.369 $Pc^\pr c^\pr n$ $(\eta_{2I}^\pr,z_{2R}^\pr)$ individually subduce to previously introduced double SIs.  First, the $\mathcal{I}$ AXI index $\eta_{2I}'$ subduces to the non-minimal index $(\eta_{2I}')_{P\bar{1}}$ in double MSG 2.4 $P\bar{1}$ (see Appendix~\ref{sec:P-1}).  Next, the even Chern number SI $2z_{2R}^\pr=C_{k_{z}=\pi}\text{ mod }4$ subduces to the same SI $(z_{2R}^\pr)_{Pc'c'2}$ in double MSG 27.81 $Pc'c'2$ [see Eq.~(\ref{eq:z2Rp-27.81}) and the surrounding text].  Hence, as we will show below, an insulator with $(\eta_{2I}^\pr,z_{2R}^\pr)=(10)$ in double MSG 56.369 $Pc^\pr c^\pr n$ is an $\mathcal{I}$-protected AXI if the non-symmetry-indicated Chern numbers vanish, and an insulator with $z_{2R}^\pr=1$ is a 3D QAH state with $C_{z}\text{ mod }4=2$.

\textit{Layer constructions} -- We find that all of the double SIs in double MSG 56.369 $Pc^\pr c^\pr n$ can be realized by layer constructions.  The layer constructions for the double SIs $(\eta_{2I}^\pr,z_{2R}^\pr)$ in double MSG 56.369 $Pc^\pr c^\pr n$ are given by:
\begin{enumerate}
\item A $\hat{\bf z}$-normal Chern layer with $C_z=1$ at the $z=0$ has the SIs (11).  We emphasize that, in this layer construction, there is also a $C_{z}=1$ Chern layer in the $z=\frac{1}{2}$ plane implied by the $\{\TRS m_y|0\frac12\frac12\}$ symmetry operation.  This layer construction is a 3D QAH state with $C_{k_{z}}=2$ in all BZ planes of constant $k_{z}$.
\item An $\hat{\bf x}$-normal Chern layer with $C_x=1$ at the $x=0$ plane has the SIs (10).  We emphasize that, in this layer construction, there is also a $C_{x}=-1$ Chern layer in the $x=\frac12$ plane implied by the $\{C_{2z}|\frac12\frac120\}$ symmetry.  Because this layer construction consists of layers with alternating odd Chern numbers occupying $\mathcal{I}$ centers, then this layer construction is an $\mathcal{I}$-protected AXI (see Appendix~\ref{sec:P-1} and Refs.~\onlinecite{WilczekAxion,QHZ,VDBAxion,AndreiInversion,AshvinAxion,WiederAxion,YuanfengAXI,NicoDavidAXI1,NicoDavidAXI2,TMDHOTI,KoreanAXI,CohVDBAXI,ArisHopf,TitusRonnyKondoAXI,YoungkukMonopole,MurakamiAXI1,MurakamiAXI2,HingeStateSphereAXI,BJYangVortex,BarryBenCDW,IvoAXI1,IvoAXI2,GuidoAXI}).
\end{enumerate}

\textit{Relationship with the SIs in other double SSGs} -- The SIs in double MSG 56.369 $Pc^\pr c^\pr n$ are related to the SIs in double MSG 2.4 $P\bar1$ through the subduction relations:
\begin{equation}
(\eta_{2I}^\pr,z_{2R}^\pr)_{Pc^\pr c^\pr n} \to (\eta_{4I},z_{2I,1},z_{2I,2},z_{2I,3})_{P\bar1}=(2\eta_{2I}^\pr,000)_{P\bar1}.
\end{equation}

Lastly, we study the effects of imposing $\mathcal{T}$ symmetry.  The double SSG 56.366 $Pccn1'$ -- the SSG generated by adding $\{\mathcal{T}|{\bf 0}\}$ symmetry to double MSG 56.369 $Pc^\pr c^\pr n$ -- has the SI group $\mathbb{Z}_{4}$~\cite{ChenTCI}.  The SIs in double SSG 56.366 $Pccn1'$ are related to the SIs in double MSG 56.369 $Pc^\pr c^\pr n$ through the subduction relations:
\begin{equation}
(z_{4})_{Pccn1'} \to (\eta_{2I}^\pr,z_{2R}^\pr)_{Pc^\pr c^\pr n} = (z_{4}\text{ mod }2,0)_{Pc^\pr c^\pr n}.
\end{equation}

\subsubsection{Double SIs in Type-III Double MSG 60.424 $Pb'cn'$}

The double MSG 60.424 $Pb^\pr c n'$ is generated by 
    $\{E|100\}$, 
    $\{E|010\}$,
    $\{E|001\}$, 
    $\{\INV|\mathbf{0}\}$,
    $\{C_{2y}|00\frac12\}$, and
    $\{\TRS m_x|\frac12\frac120\}$.

\textit{SIs} -- The double MSG 60.424 $Pb^\pr c n^\pr$ has the SI group $\mathbb{Z}_2\times \mathbb{Z}_2$.  In the physical basis, the double SIs of double MSG 60.424 $Pb^\pr c n^\pr$ are $(\eta_{2I}^\pr,z_{2R}^\pr)$.  As previously in double MSGs 54.342 $Pc'c'a$ and 56.369 $Pc'c'n$ (Appendices~\ref{sec:PcPrcPra} and~\ref{sec:PcPrcPrn}, respectively), $\eta_{2I}^\pr$ is the $\mathcal{I}$ AXI index, and subduces to the non-minimal index $(\eta_{2I}^\pr)_{P\bar{1}}$ in double MSG 2.4 $P\bar{1}$.

However, unlike previously in double MSGs 54.342 $Pc'c'a$ and 56.369 $Pc'c'n$, $z_{2R}^\pr$ does not subduce to a previously introduced minimal double SI.  Nevertheless, we will shortly use layer constructions to demonstrate that like in double MSGs 54.342 $Pc'c'a$ and 56.369 $Pc'c'n$, $z_{2R}^\pr$ indicates the value of a Chern number -- here the position-space Chern number $C_{y}$ -- modulo 4.  Hence, an insulator with $(\eta_{2I}^\pr,z_{2R}^\pr)=(10)$ in double MSG 60.424 $Pb^\pr c n^\pr$ is an $\mathcal{I}$-protected AXI if the non-symmetry-indicated Chern numbers vanish, and an insulator with $z_{2R}^\pr=1$ is a 3D QAH state with $C_{y}\text{ mod }4=2$.  Using the Smith normal form of the EBR matrix [see Appendix~\ref{sec:smithForm}] and the definition of the $\mathbb{Z}_{2}$ AXI parity index $\eta_{2I}^{\pr}$ obtained by subduction onto double MSG 2.4 $P\bar{1}$ [see the text surrounding Eq.~(\ref{eq:eta2Ip})], we define the second $\mathbb{Z}_{2}$ SI in double MSG 60.424 $Pb^\pr c n^\pr$ to be:
\begin{equation}
z_{2R}^\pr = \eta_{2I}^{\pr} + m(\ovl{\Gamma}_{3})\text{ mod }2,
\label{eq:z2Rp-60.424}
\end{equation}
where $m(\ovl{\bf k}_{i})$ is the multiplicity of the small corep $\ovl{\bf k}_{i}$ of the little group $G_{\bf k}$ in the symmetry data vector of the occupied bands [where the symmetry data vector of a group of bands is defined in the text following Eq.~(\ref{eq:reducibleCompatibility})].

\textit{Layer constructions} -- We find that all of the double SIs in double MSG 60.424 $Pb^\pr c n^\pr$ can be realized by layer constructions.  The layer constructions for the double SIs $(\eta_{2I}^\pr,z_{2R}^\pr)$ in double MSG 60.424 $Pb^\pr c n^\pr$ are given by:
\begin{enumerate}
\item A $\hat{\bf y}$-normal Chern layer with $C_y=1$ at the $y=0$ has the SIs (11).  We emphasize that, in this layer construction, there is also a $C_{y}=1$ Chern layer in the $y=\frac{1}{2}$ plane implied by the $\{\TRS m_x|\frac12\frac120\}$ symmetry operation.  This layer construction is a 3D QAH state with $C_{k_{y}}=2$ in all BZ planes of constant $k_{y}$.
\item An $\hat{\bf z}$-normal Chern layer with $C_z=1$ at the $z=0$ plane has the SIs (10).  We emphasize that, in this layer construction, there is also a $C_{z}=-1$ Chern layer in the $z=\frac12$ plane implied by the $\{C_{2y}|00\frac12\}$ symmetry.  Because this layer construction consists of layers with alternating odd Chern numbers occupying $\mathcal{I}$ centers, then this layer construction is an $\mathcal{I}$-protected AXI (see Appendix~\ref{sec:P-1} and Refs.~\onlinecite{WilczekAxion,QHZ,VDBAxion,AndreiInversion,AshvinAxion,WiederAxion,YuanfengAXI,NicoDavidAXI1,NicoDavidAXI2,TMDHOTI,KoreanAXI,CohVDBAXI,ArisHopf,TitusRonnyKondoAXI,YoungkukMonopole,MurakamiAXI1,MurakamiAXI2,HingeStateSphereAXI,BJYangVortex,BarryBenCDW,IvoAXI1,IvoAXI2,GuidoAXI}).
\end{enumerate}

\textit{Relationship with the SIs in other double SSGs} -- The SIs in double MSG 60.424 $Pb^\pr c n^\pr$ are related to the SIs in double MSG 2.4 $P\bar1$ through the subduction relations:
\begin{equation}
(\eta_{2I}^\pr,z_{2R}^\pr)_{Pb^\pr c n^\pr} \to (\eta_{4I},z_{2I,1},z_{2I,2},z_{2I,3})_{P\bar1}=(2\eta_{2I}^\pr,000)_{P\bar1}.
\end{equation}

Lastly, we study the effects of imposing $\mathcal{T}$ symmetry.  The double SSG 60.418 $Pbcn1'$ -- the SSG generated by adding $\{\mathcal{T}|{\bf 0}\}$ symmetry to double MSG 60.424 $Pb^\pr c n^\pr$ -- has the SI group $\mathbb{Z}_{4}$~\cite{ChenTCI}.  The SIs in double SSG 60.418 $Pbcn1'$ are related to the SIs in double MSG 60.424 $Pb^\pr c n^\pr$ through the subduction relations:
\begin{equation}
(z_{4})_{Pbcn1'} \to (\eta_{2I}^\pr,z_{2R}^\pr)_{Pb^\pr c n^\pr} = (z_{4}\text{ mod }2,0)_{Pb^\pr c n^\pr}.
\end{equation}

\subsubsection{Double SIs in Type-III Double MSG 83.45 $P4'/m$}
\label{sec:p4P/m}

The double MSG 83.45 $P4'/m$ is generated by 
    $\{E|100\}$, 
    $\{E|010\}$,
    $\{E|001\}$, 
    $\{\INV|\mathbf{0}\}$, and $\{\TRS C_{4z}|\mathbf{0}\}$.  We note that double MSG 83.45 $P4'/m$ additionally contains a mirror symmetry operation: $\{m_z|\mathbf{0}\}=\{\INV|\mathbf{0}\}\{\TRS C_{4z}|\mathbf{0}\}^2$.

\textit{SIs} -- The double MSG 83.45 $P4^\pr/m$ has the SI group $\ZZ_4\times\ZZ_2$.  In double-valued small coreps of the little groups at the $\mathcal{I}$- and $\mathcal{T}C_{4z}$-invariant ${\bf k}$ points $\Gamma$ [${\bf k}_{\Gamma} = (000)$], $M$ [${\bf k}_{M}=(\pi\pi 0)$], $Z$ [${\bf k}_{Z}=(00\pi)$], and $A$ [${\bf k}_{A}=(\pi\pi\pi)$], the matrix representatives of $\{\INV|\mathbf{0}\}$, $\{\TRS C_{4z}|\mathbf{0}\}$, and $\{m_z|\mathbf{0}\}$ commute.  Hence, Bloch states $\ket{\psi}$ at the $\Gamma$, $M$, $Z$, and $A$ points may be simultaneously labeled with both parity ($\{\INV|\mathbf{0}\}$) and mirror eigenvalues.  Taking $\ket{\psi}$ to be a state at an $\mathcal{I}$- and $\mathcal{T}C_{4z}$-invariant ${\bf k}$ point with $\{m_z|\mathbf{0}\}$ eigenvalue $i$, and parity eigenvalue $\xi\in\{-1,1\}$, we next compute the $\{m_z|\mathbf{0}\}$ eigenvalues of the state $\{\TRS C_{4z}|\mathbf{0}\}\ket{\psi}$:
\begin{equation}
\{m_z|\mathbf{0}\}\{\TRS C_{4z}|\mathbf{0}\}\ket{\psi} = \{\TRS C_{4z}|\mathbf{0}\}\{m_z|\mathbf{0}\}\ket{\psi} = -i\{\TRS C_{4z}|\mathbf{0}\}\ket{\psi},
\label{eq:type3AxionMirrorOpposite}
\end{equation}
and the $\{\INV|\mathbf{0}\}$ eigenvalues of $\{\TRS C_{4z}|\mathbf{0}\}\ket{\psi}$:
\begin{equation}
\{\INV|\mathbf{0}\}\{\TRS C_{4z}|\mathbf{0}\}\ket{\psi}=\{\TRS C_{4z}|\mathbf{0}\}\{\INV|\mathbf{0}\}\ket{\psi} = \xi\{\TRS C_{4z}|\mathbf{0}\}\ket{\psi}.
\label{eq:type3AxionParitySame}
\end{equation}
Eqs.~(\ref{eq:type3AxionMirrorOpposite}) and~(\ref{eq:type3AxionParitySame}) imply that the Bloch states at $\Gamma$, $M$, $Z$, and $A$ form doublets with complex-conjugate $\{m_z|\mathbf{0}\}$ eigenvalues and the same parity eigenvalues.  At the $\mathcal{I}$-invariant ${\bf k}$ points $X$ [${\bf k}_{X} = (0\pi 0)$], $XA$ [${\bf k}_{XA}=(\pi 0 0)$], $R$ [${\bf k}_{R}=(0\pi\pi)$], and $RA$ [${\bf k}_{RA}=(\pi 0 \pi)$] at which $\{\TRS C_{4z}|\mathbf{0}\}$ is not an element of the little group $G_{\bf k}$, Bloch states are instead singly degenerate (see the output of the~\href{http://www.cryst.ehu.es/cryst/corepresentations}{Corepresentations} tool introduced in this work for the double-valued small coreps of double MSG 83.45 $P4^\pr/m$, where~\href{http://www.cryst.ehu.es/cryst/corepresentations}{Corepresentations} is detailed in Appendix~\ref{sec:coreps}).  However, the insulating compatibility relations require that there is always an even number of singly-degenerate occupied Bloch states at $X$, $XA$, $R$, and $RA$, which are required to subdivide into pairs of states (at different energies) with complex-conjugate $\{m_z|\mathbf{0}\}$ eigenvalues and the same parity eigenvalues (see the output of the~\href{https://www.cryst.ehu.es/cryst/mcomprel}{MCOMPREL} tool introduced in this work for the double-valued small coreps of double MSG 83.45 $P4^\pr/m$, where~\href{https://www.cryst.ehu.es/cryst/mcomprel}{MCOMPREL} is detailed in Appendix~\ref{sec:compatibilityRelations}).

Therefore, like in other centrosymmetric SSGs in which insulators with nontrivial SIs have even numbers of occupied bands that subdivide at each $\mathcal{I}$-invariant ${\bf k}$ point into doublets with the same parity eigenvalues [\emph{e.g.} double MSG 47.249 $Pmmm$ and double SSG 2.5 $P\bar{1}1'$, see Appendices~\ref{subsec:Pmmm} and~\ref{sec:P-11p}, respectively], the double SIs of MSG 83.45 $P4^\pr/m$ in the physical basis ($z_{4},z_{2w,3}$) have the respective SI formulas:
\begin{equation}
z_4 = \sum_{K} \frac12 n_K^- = \sum_{K} \frac{n_K^- - n_K^+}{4} \text{ mod } 4, 
\end{equation}
\begin{equation}
z_{2w,3} = \sum_{K,K_3=\pi} \frac12 n_K^- = \sum_{K,K_3=\pi} \frac{n_K^- - n_K^+}{4} \text{ mod } 2,
\end{equation}
where $K$ runs over the eight $\mathcal{I}$-invariant momenta in the first BZ, and $n^{\pm}_K$ are the number of Bloch states with $\pm1$ parity eigenvalues at $K$ in the group of bands under consideration.  Like in double MSG 47.249 $Pmmm$ (see Appendix~\ref{subsec:Pmmm}), insulators with $z_{4}\text{ mod }2=1$ are TCIs with $\theta=\pi$ (specifically AXIs with the same $C_{4z}\times\mathcal{T}$-symmetric configuration of chiral hinge states as the magnetic HOTIs introduced in Ref.~\onlinecite{HOTIBernevig}), $z_{2w,3}$ indicates the mirror Chern number in the $k_{z}=\pi$ plane modulo 2 ($z_{2w,3}=C^{+}_{k_{z}=\pi}\text{ mod }2 = C^{-}_{k_{z}=\pi}\text{mod }2)$, and the double SIs $(z_{4},z_{2w,3})=(20)$ indicate a helical (non-axionic) magnetic mirror TCI with $C_{m_{z}}\text{ mod }4=2$.

\textit{Layer constructions} --  To diagnose the topology associated to each nontrivial value of the double SIs, we employ the layer construction method.  In the layer constructions below, $C^+=-C^-$ due to the net-zero Chern numbers enforced by the symmetries $\{\TRS C_{4z}|\mathbf{0}\}$ and $\{m_z|\mathbf{0}\}$.  Hence, we will omit $C^-$ in further discussions of the topology in double MSG 83.45 $P4^\pr/m$.  The layer constructions for the double SIs $(z_{4},z_{2w,3})$ of MSG 83.45 $P4^\pr/m$ are given by:
\begin{enumerate}
\item A $\hat{\bf z}$-normal mirror Chern layer with $C^+_z=1$ in the $z=0$ plane has the mirror sector Chern numbers $(C^+_{k_z=0},C^+_{k_z=\pi})=(11)$ and the SIs $(21)$.
\item A $\hat{\bf z}$-normal mirror Chern layer with $C^+_z=1$ in the $z=\frac12$ plane has the mirror sector Chern numbers $(C^+_{k_z=0},C^+_{k_z=\pi})=(1,-1)$ and the SIs $(01)$.
\end{enumerate}

\textit{Axion insulators} -- We find that states with odd $z_4$ SIs cannot be constructed from layers of 2D stable topological phases.  However, we may still use subduction relations to determine the bulk topology of insulators with odd values of $z_{4}$.  First, as we will show below, $(10)$ and $(30)$ subduce to $(2000)_{P\bar{1}}$ in MSG 2.4 $P\bar1$.  Hence, if the $(10)$ and $(30)$ phases in double MSG 83.45 $P4^\pr/m$ are insulating, then the bulk insulator must either be an AXI or a 3D QAH state.  Because the net Chern numbers $C_{x,y,z}=0$ must vanish if the bulk is gapped, due to the symmetries $\{\TRS C_{4z}|\mathbf{0}\}$ and $\{m_z|\mathbf{0}\}$ of double MSG 83.45 $P4^\pr/m$, then the $(10)$ and $(30)$ states must be AXIs.  As we will show below, this result can also be understood by subducing from a $\mathcal{T}$-symmetric SSG.  Specifically, because $z_{4}\text{ mod }2=1$, $z_{2w,3}=0$ states in MSG 83.45 $P4^\pr/m$ can respectively be subduced from insulators with $z_{8}\text{ mod }2=1$, $z_{4m,\pi}^{-}=z_{2w,1}=0$ in Type-II double SG 83.44 $P4/m1^\pr$, which correspond to $\mathcal{T}$-symmetric 3D TIs with $\theta=\pi$ (see Appendix~\ref{sec:P4/m1p} and Refs.~\onlinecite{AshvinIndicators,HOTIChen,ChenTCI,AshvinTCI}), then $(10)$ and $(30)$ are compatible with bulk-gapped states in double MSG 83.45 $P4^\pr/m$.  Hence, we conclude that 3D insulators with $(10)$ and $(30)$ in double MSG 83.45 $P4^\pr/m$ are AXIs, \emph{without ambiguity}.  We conjecture that the $(10)$ and $(30)$ AXIs in double MSG 83.45 $P4^\pr/m$ can be constructed using the topological crystal method~\cite{ZhidaHermeleCrystal}, which additionally incorporates cell complexes of 2D Chern insulators, TIs, and TCIs.

\textit{Relationship with the SIs in other double SSGs} -- The SIs in double MSG 83.45 $P4^\pr/m$ are related to the SIs in double MSG 2.4 $P\bar1$ through the subduction relations:
\begin{equation}
(z_{4},z_{2w,3})_{P4^\pr/m} \to (\eta_{4I},z_{2I,1},z_{2I,2},z_{2I,3})_{P\bar1}=(2(z_{4}\text{ mod }2),000)_{P\bar1}.
\end{equation}

Lastly, we study the effects of imposing $\mathcal{T}$ symmetry.  The double SSG 83.44 $P4/1m1'$ -- the SSG generated by adding $\{\mathcal{T}|{\bf 0}\}$ symmetry to double MSG 83.45 $P4^\pr/m$ -- has the SI group $\mathbb{Z}_{8}\times\mathbb{Z}_{4}\times\mathbb{Z}_{2}$ (see Appendix~\ref{sec:P4/m1p} and Refs.~\onlinecite{AshvinIndicators,HOTIChen,ChenTCI,AshvinTCI}).  The SIs in double SSG 83.44 $P4/1m1'$ are related to the SIs in double MSG 83.45 $P4^\pr/m$ through the subduction relations:
\begin{equation}
(z_{8},z_{4m,\pi}^-,z_{2w,1})_{P4/m1'} \to (z_{4},z_{2w,3})_{P4^\pr/m} = (z_{8}\text{ mod }4,z_{4m,\pi}^-\text{ mod }2)_{P4^\pr/m}.
\end{equation}

\subsubsection{Double SIs in Type-III Double MSG 103.199 $P4c'c'$}

The double MSG 103.199 $P4c'c'$ is generated by 
    $\{E|100\}$, 
    $\{E|010\}$,
    $\{E|001\}$, 
    $\{C_{4z}|\mathbf{0}\}$, and $\{\TRS m_y|00\frac12\}$.

\textit{SI} -- The double MSG 103.199 $P4c^\pr c^\pr$ has the SI group $\ZZ_4$.  As we will shortly demonstrate, in the physical basis, the double SI $z_{4R}^\pr$ indicates the \emph{even-valued} Chern number in the $k_{z}=\pi$ plane (modulo 8): $C_{k_{z}=\pi}\text{ mod }8 = 2z_{4R}^\pr$.  Hence, insulators with nontrivial values of $z_{4R}^\pr$ are 3D QAH states.

We first emphasize that Bloch states at the $\{C_{4z}|{\bf 0}\}$-invariant momenta in the $k_{z}=\pi$ plane in double MSG 103.199 $P4c'c'$ form doubly-degenerate pairs with the same $\{C_{4z}|{\bf 0}\}$ eigenvalues.  Specifically, in the $k_{z}=\pi$ plane, the matrix representative of $\{\TRS m_y|00\frac12\}$ squares to minus the identity in all double-valued small coreps.  Furthermore, using the~\href{http://www.cryst.ehu.es/cryst/corepresentations}{Corepresentations} tool introduced in this work (detailed in Appendix~\ref{sec:coreps}), we determine that, in all of $\{\TRS m_y|00\frac12\}$-paired doublets at the $\{C_{4z}|{\bf 0}\}$-invariant ${\bf k}$ points $k_{x}=k_{y}=0,\pi$ in the $k_{z}=\pi$ plane, both states have the same $\{C_{4z}|{\bf 0}\}$ (and $\{C_{2z}|{\bf 0}\}$) eigenvalues.  Additionally, using the output of~\href{http://www.cryst.ehu.es/cryst/corepresentations}{Corepresentations} for the double-valued small coreps of double MSG 103.199 $P4c'c'$, we find that, at the $\{C_{4z}|{\bf 0}\}$-invariant ${\bf k}$ points $(k_{x},k_{y})=(0\pi)$ and $(\pi 0)$, both of the Bloch states within each doublet have the same $\{C_{2z}|{\bf 0}\}$ eigenvalues.

We therefore define the $\ZZ_4$ SI to be half of the even-valued Chern number (modulo 4) of the occupied bands in the $k_{z}=\pi$ plane:
\begin{equation}
z_{4R}^\pr =  
\sum_{K=Z,A} \pare{
-\frac14 n^{\frac12}_K + \frac14 n^{-\frac12}_K
-\frac34 n^{\frac32}_K + \frac34 n^{-\frac32}_K}
+ \frac12 n^{\frac12}_R - \frac12 n^{-\frac12}_R \text{ mod }4 = \frac{C_{k_z=\pi} }{2} \text{ mod } 4.
\label{eq:z4Rp}
\end{equation}
where $n^{\frac12,-\frac12,\frac32,-\frac32}_{Z,A}$ are the number of occupied states with $\{C_{4z}|\mathbf{0}\}$ eigenvalues $e^{-i\frac{\pi}4}$, $e^{i\frac{\pi}4}$, $e^{-i\frac{3\pi}4}$, $e^{i\frac{3\pi}4}$, respectively, and $n^{\frac12,-\frac12}_R$ are the number of occupied states with $\{C_{2z}|\mathbf{0}\}$ eigenvalues $e^{-i\frac{\pi}2}$, $e^{i\frac{\pi}2}$, respectively.

\textit{Layer constructions} --  To diagnose the topology associated to nontrivial values of $z_{4R}^\pr$, we employ the layer construction method.  We begin by placing a $\hat{\bf z}$-normal Chern layer with $C_z=1$ in the $z=0$ plane.  Due to the $\{T m_y|00\frac12\}$ symmetry in double MSG 103.199 $P4c'c'$, there must be another Chern layer with $C_z=1$ in the $z=\frac12$ plane, such that the total Chern number per cell is $C_{z}=2$, and the Chern number of the occupied bands in the $k_{z}=\pi$ plane is $C_{k_{z}=\pi}=2$.  Hence, in this layer construction of a 3D QAH state, $C_{z}=2$ and $z_{4R}^\pr=1$.

\textit{Relationship with the SIs in other double SSGs} -- We next compute the subduction relations between the SIs in double MSG 103.199 $P4c'c'$ and the SIs in the maximal unitary subgroup double MSG 75.1 $P4$ (see Appendix~\ref{subsec:P4}):
\begin{equation}
(z_{4R}^\pr)_{P4c^\pr c^\pr} \to 
(z_{4R})_{P4} = (2(z_{4R}^\pr\text{ mod }2))_{P4}.
\label{eq:cPrcPr4Subduct}
\end{equation}
Eq.~(\ref{eq:cPrcPr4Subduct}) implies that symmetry-indicated 3D QAH states with $z_{4R}^\pr\text{ mod }2=1$ in double MSG 103.199 $P4c'c'$ subduce to symmetry-indicated 3D QAH states with $(z_{4R})_{P4}=2$ in double MSG 75.1 $P4$, whereas symmetry-indicated 3D QAH states with $z_{4R}^\pr=2$ in double MSG 103.199 $P4c'c'$ necessarily subduce to \emph{non-symmetry-indicated} 3D QAH states with $(z_{4R})_{P4}=0$ in double MSG 75.1 $P4$, in agreement with the physical-basis double SI relations $C_{k_{z}=\pi}\text{ mod }8 = 2z_{4R}^\pr$ and $C_{k_{z}=\pi}\text{ mod }4 = z_{4R}$ [see Eq.~(\ref{eq:z4R}) and the surrounding text].

Lastly, if we impose $\mathcal{T}$ symmetry, then the position-space Chern numbers must vanish, which enforces $z_{4R}^\pr$ to be zero.  Correspondingly, in double SSG 103.196 $P4cc1'$ -- the SSG generated by adding $\{\mathcal{T}|{\bf 0}\}$ to double MSG 103.199 $P4c'c'$ -- the double SI group is trivial.

\subsubsection{Double SIs in Type-III Double MSG 110.249 $I4_{1}c'd'$}

The double MSG 110.249 $I4_1c'd'$ is generated by 
    $\{E|\bar{\frac12}\frac12\frac12\}$, 
    $\{E|\frac12\bar{\frac12}\frac12\}$,
    $\{E|\frac12\frac12\bar{\frac12}\}$, 
    $\{C_{4z}|0\frac12\frac14\}$, and $\{\TRS m_x|\frac12\frac120\}$.  The Bravais lattice of double MSG 110.249 $I4_1c'd'$ is body-centered.  Correspondingly, in the primitive cell, the lattice vectors are:
\begin{equation}
\mbf{a}_1 = (-{\frac12},\frac12,\frac12),\quad
\mbf{a}_2 = (\frac12,-{\frac12},\frac12),\quad
\mbf{a}_3 = (\frac12,\frac12,-{\frac12}),
\end{equation} 
and the reciprocal lattice vectors are:
\begin{equation}
\mbf{b}_1 = 2\pi(0,1,1),\quad
\mbf{b}_2 = 2\pi(1,0,1),\quad
\mbf{b}_3 = 2\pi(1,1,0).
\end{equation}
We additionally note that $\{C_{4z}|0\frac12\frac14\}^2=\{C_{2z}|\frac12\frac12\frac12\} = \{E|\frac12\frac12\frac12\}\{C_{2z}|{\bf 0}\}$, where $\{E|\frac12\frac12\frac12\}$ is a primitive translation symmetry.  Hence, the $4_{1}$ screw symmetry operation $\{C_{4z}|0\frac12\frac14\}$ only contains a \emph{half} lattice translation in the $z$ direction in the primitive cell, such that double MSG 110.249 $I4_1c'd'$ also contains the rotation symmetry $\{C_{2z}|{\bf 0}\}$.

\textit{SI} -- The double MSG 110.249 $I4_1c^\pr d^\pr$ has the SI group $\ZZ_2$.  In the physical basis, the $\mathbb{Z}_{2}$ SI has the SI formula:
\begin{equation}
z_{2R}^\prpr = m(\ovl{\Gamma}_6) \text{ mod } 2=\frac{C_{k_z=0}}{2}\text{ mod }2,
\label{eq:z2Rp-110.249}
\end{equation}
where $m(\ovl{\Gamma}_6)$ is the multiplicity of the double-valued small corep $\ovl{\Gamma}_6$ in the symmetry data $\tilde{\varsigma}_{\Gamma}$ corresponding to the occupied states at $\Gamma$ [${\bf k}_{\Gamma}=(0,0,0)$], where the symmetry data at a ${\bf k}$ point is defined in the text following Eq.~(\ref{eq:reducibleCompatibility}), and where the $\{C_{4z}|0\frac12\frac14\}$ and $\{C_{2z}|\mathbf{0}\}$ characters of the irreducible small coreps $\tilde{\sigma}$ at $\Gamma$ are given by:
\begin{equation}
\begin{tabular}{c|c|c|c|c}
& $\ovl{\Gamma}_5$ & $\ovl{\Gamma}_6$ & $\ovl{\Gamma}_7$ & $\ovl{\Gamma}_8$\\
\hline
$\chi_{\tilde{\sigma}}(\{C_{4z}|0\frac12\frac14\})$ & $e^{i\frac{3\pi}4}$ & $e^{-i\frac{\pi}4}$ & $e^{-i\frac{3\pi}4}$ & $e^{i\frac{\pi}4}$\\
\hline
$\chi_{\tilde{\sigma}}(\{C_{2z}|\mathbf{0}\})$ & $-i$ & $-i$ & $i$ & $i$
\end{tabular}. 
\end{equation}
Hence, as we will show below, insulators in double MSG 110.249 $I4_1c^\pr d^\pr$ with $z_{2R}^\prpr=1$ are 3D QAH states with $C_{z}\text{ mod }4=2$ per primitive cell.

\textit{Layer constructions} -- To diagnose the topology associated to $z_{2R}^\prpr=1$, we employ the layer construction method.  We begin by placing a $\hat{\bf z}$-normal Chern insulator with $C_z=1$ in the $z=0$ plane.  Due to the $\{C_{4z}|0\frac12\frac14\}$ screw symmetry in double MSG 110.249 $I4_1c^\pr d^\pr$, there must be a second $\hat{\bf z}$-normal Chern insulator with $C_z=1$ in the $z=\frac14$ plane.  Using the established formula for the parity of the Chern number in terms of $\{C_{2z}|{\bf 0}\}$ rotation eigenvalues~\cite{QWZ,ChenBernevigTCI} and the constraints imposed by the compatibility relations on the eigenvalues of the $4_{1}$ screw symmetry $\{C_{4z}|0\frac12\frac14\}$ in an insulating state (see the output of the~\href{https://www.cryst.ehu.es/cryst/mcomprel}{MCOMPREL} tool introduced in this work for the double-valued small coreps of double MSG 110.249 $I4_1c^\pr d^\pr$, where~\href{https://www.cryst.ehu.es/cryst/mcomprel}{MCOMPREL} is detailed in Appendix~\ref{sec:compatibilityRelations}), we find that a set of symmetry data compatible with this layer construction is given by $\tilde{\varsigma}_{\Gamma} = \ovl{\Gamma}_5+\ovl{\Gamma}_6+\ovl{\Gamma}_7+\ovl{\Gamma}_8$.  Hence, $z_{2R}^\prpr=1$ in this layer construction of a 3D QAH state, in agreement with the net position-space Chern number $C_{z}=2$ per primitive cell.

Lastly, if we impose $\mathcal{T}$ symmetry, then the position-space Chern numbers must vanish, which enforces $z_{2R}^\pr$ to be zero.  Correspondingly, in double SSG 110.246 $I4_{1}cd1'$ -- the SSG generated by adding $\{\mathcal{T}|{\bf 0}\}$ to double MSG 110.249 $I4_1c^\pr d^\pr$ -- the double SI group is trivial.

\subsubsection{Double SIs in Type-III Double MSG 130.429 $P4/nc'c'$}

The double MSG 130.429 $P4/nc^\pr c^\pr$ is generated by 
    $\{E|100\}$, 
    $\{E|010\}$,
    $\{E|001\}$, 
    $\{C_{4z}|\frac1200\}$, $\{\INV|\mathbf{0}\}$, and $\{\TRS m_{1\bar10}|00\frac12\}$.

\textit{SIs} -- The double MSG 130.429 $P4/nc^\pr c^\pr$ has the SI group $\ZZ_4\times \ZZ_2$.  In the physical basis, the double SIs of double MSG 130.429 $P4/nc^\pr c^\pr$ $(z_{4R}^\pr,\eta_{2I}^\pr)$ individually subduce to previously introduced double SIs.  First, the even Chern number SI $2z_{4R}^\pr=C_{k_{z}=\pi}\text{ mod }8$ subduces to the same SI $(z_{4R}^\pr)_{P4c'c'}$ in double MSG 103.199 $P4c'c'$ [see Eq.~(\ref{eq:z4Rp}) and the surrounding text].  Next, the $\mathcal{I}$ AXI index $\eta_{2I}'$ subduces to the non-minimal index $(\eta_{2I}')_{P\bar{1}}$ in double MSG 2.4 $P\bar{1}$ (see Appendix~\ref{sec:P-1}).  Hence, as we will show below, an insulator with $(z_{4R}^\pr,\eta_{2I}^\pr)=(01)$ in double MSG 130.429 $P4/nc^\pr c^\pr$ is an $\mathcal{I}$-protected AXI if the non-symmetry-indicated Chern numbers vanish, and insulators with $z_{4R}^\pr\neq 0$ are 3D QAH states.

\textit{Layer constructions} -- We find that in double MSG 130.429 $P4/nc^\pr c^\pr$, the 3D QAH states -- but not the AXI states -- can be realized by layer constructions.  The double SIs $(z_{4R}^\pr,\eta_{2I}^\pr)$ of the symmetry-indicated 3D QAH states in double MSG 130.429 $P4/nc^\pr c^\pr$ are spanned by superpositions of the following layer construction:
\begin{enumerate}
\item A $\hat{\bf z}$-normal Chern layer with $C_z=1$ at the $z=0$ has the SIs (11).  We emphasize that, in this layer construction, there is also a $C_{z}=1$ Chern layer in the $z=\frac{1}{2}$ plane implied by the $\{\TRS m_{1\bar10}|00\frac12\}$ symmetry operation.  This layer construction is a 3D QAH state with $C_{k_{z}}=2$ in all BZ planes of constant $k_{z}$.
\end{enumerate}

\textit{Axion insulators and 3D QAH states} -- We find that states with the double SIs $(z_{4R}^\pr,\eta_{2I}^\pr)=(01)$ in double MSG 130.429 $P4/nc^\pr c^\pr$ cannot be constructed from layers of 2D stable topological phases.  However, we may still use subduction relations to determine the bulk topology of insulators with the double SIs (01).  First, as we will show below, $(01)$ subduces to $(2000)_{P\bar{1}}$ in MSG 2.4 $P\bar1$.  Hence, if a $(01)$ state in double MSG 130.429 $P4/nc^\pr c^\pr$ is insulating, then the bulk insulator must either be an AXI or a 3D QAH state, and will specifically be an AXI if the non-symmetry-indicated Chern numbers vanish.  As we will show below, this result can also be understood by subducing from a $\mathcal{T}$-symmetric SSG.  Specifically, because $(01)$ states in double MSG 130.429 $P4/nc^\pr c^\pr$ can be subduced from insulators with $(z_{4})_{P4/ncc1'}\text{ mod }2=1$ in Type-II double SG 130.424 $P4/ncc1'$, which correspond to $\mathcal{T}$-symmetric 3D TIs with $\theta=\pi$~\cite{ChenTCI}, then the double SIs $(01)$ are compatible with a bulk-gapped state in double MSG 130.429 $P4/nc^\pr c^\pr$.  This provides further evidence that 3D insulators with $(01)$ and net-zero position-space Chern numbers in double MSG 130.429 $P4/nc^\pr c^\pr$ are AXIs.  We conjecture that $(01)$ AXIs in double MSG 130.429 $P4/nc^\pr c^\pr$ can be constructed using the topological crystal method~\cite{ZhidaHermeleCrystal}, which additionally incorporates cell complexes of 2D Chern insulators, TIs, and TCIs.

\textit{Relationship with the SIs in other double SSGs} -- The SIs in double MSG 130.429 $P4/nc^\pr c^\pr$ are related to the SIs in double MSG 2.4 $P\bar1$ through the subduction relations:
\begin{equation}
(z_{4R}^\pr,\eta_{2I}^\pr)_{P4/nc^\pr c^\pr} \to (\eta_{4I},z_{2I,1},z_{2I,2},z_{2I,3})_{P\bar1}=(2\eta_{2I}^\pr,000)_{P\bar1}.
\end{equation}

Lastly, we study the effects of imposing $\mathcal{T}$ symmetry.  The double SSG 130.424 $P4/ncc1'$ -- the SSG generated by adding $\{\mathcal{T}|{\bf 0}\}$ symmetry to double MSG 130.429 $P4/nc^\pr c^\pr$ -- has the SI group $\mathbb{Z}_{4}$~\cite{ChenTCI}.  The SIs in double SSG 130.424 $P4/ncc1'$ are related to the SIs in double MSG 130.429 $P4/nc^\pr c^\pr$ through the subduction relations:
\begin{equation}
(z_{4})_{P4/ncc1'} \to (z_{4R}^\pr,\eta_{2I}^\pr)_{P4/nc^\pr c^\pr} = (0,z_{4}\text{ mod }2)_{P4/nc^\pr c^\pr}.
\end{equation}

\subsubsection{Double SIs in Type-III Double MSG 135.487 $P4_{2}'/mbc'$}

The double MSG 135.487 $P4_2^\pr/mbc^\pr$ is generated by 
    $\{E|100\}$, 
    $\{E|010\}$,
    $\{E|001\}$, 
    $\{\INV|\mathbf{0}\}$, $\{m_z|\mathbf{0}\}$, $\{C_{2x}|\frac12\frac120\}$, and $\{\TRS C_{4z}|00\frac12\}$.

\textit{SI} -- The double MSG 135.487 $P4_2^\pr/mbc^\pr$ has the SI group $\ZZ_4$.  At the $\mathcal{I}$-invariant momenta, the double-valued irreducible small coreps are either two- or four-dimensional.  An expression for the SI formula of the $\mathbb{Z}_{4}$ double SI computed from the Smith normal form of the EBR matrix (see Appendix~\ref{sec:smithForm}) is given by:
\begin{equation}
z_{4}^{\pr} = 2m(\ovl{\Gamma}_5) - m(\ovl{\Gamma}_6) - m(\ovl{M}_5) + 2 m(\ovl{X}_3),
\label{eq:z4p}
\end{equation}
where $m(\ovl{\bf k}_{i})$ is the multiplicity of the small corep $\ovl{\bf k}_{i}$ of the little group $G_{\bf k}$ in the symmetry data vector of the occupied bands [where the symmetry data vector of a group of bands is defined in the text following Eq.~(\ref{eq:reducibleCompatibility})].  As we will shortly show below through layer constructions, like in double MSGs 47.249 $Pmmm$ and 83.45 $P4'/m$, insulators with $z_{4}'\text{ mod }2=1$ are TCIs with $\theta=\pi$ (\emph{i.e.} AXIs), and $z_{4}'=2$ indicates a helical (non-axionic) magnetic mirror TCI with $C_{m_{z}}\text{ mod }4=2$.

\textit{Layer constructions} -- We find that in double MSG 135.487 $P4_2^\pr/mbc^\pr$, the non-axionic magnetic TCI phases -- but not the AXI phases -- can be realized by layer constructions.  The double SI $z_{4}'=2$ of a symmetry-indicated non-axionic TCI phase in $P4_2^\pr/mbc^\pr$ with $C_{m_{z}}=2$ is realized by the following layer construction:
\begin{enumerate}
\item{A $\hat{\bf z}$-normal mirror Chern layer with $C^+_z=-C^-_z=1$ in the $z=0$ plane has the double SI $z_{4}'=2$.  We emphasize that, in this layer construction, there is also a $\hat{\bf z}$-normal mirror Chern layer with $C^+_z=-C^-_z=1$ in the $z=\frac12$ plane implied by the $\{\TRS C_{4z}|00\frac12\}$ symmetry operation.}
\end{enumerate}

\textit{Axion insulators} -- We find that states with odd $z_{4}'$ SIs cannot be constructed from layers of 2D stable topological phases.  However, we may still use subduction relations to determine the bulk topology of insulators with odd values of $z_{4}'$.  First, as we will show below, the double SIs $z_{4}'=1,3$ in double MSG 135.487 $P4_2^\pr/mbc^\pr$ subduce to $(2000)_{P\bar{1}}$ in MSG 2.4 $P\bar1$.  Hence, if the $z_{4}'\text{ mod }2=1$ phases in double MSG 135.487 $P4_2^\pr/mbc^\pr$ are insulating, then the bulk insulator must either be an AXI or a 3D QAH state.  Because the net Chern numbers $C_{x,y,z}=0$ must vanish if the bulk is gapped, due to the symmetries $\{m_z|\mathbf{0}\}$ and $\{C_{2x}|\frac12\frac120\}$ of double MSG 135.487 $P4_2^\pr/mbc^\pr$, then the $z_{4}'=1,3$ states must be AXIs.  As we will show below, this result can also be understood by subducing from a $\mathcal{T}$-symmetric SSG.  Specifically, because $z_{4}'\text{ mod }2=1$ states in MSG 135.487 $P4_2^\pr/mbc^\pr$ can respectively be subduced from insulators with $z_{4}=1,3$ in Type-II double SG 135.484 $P4_{2}/mbc1'$, which correspond to $\mathcal{T}$-symmetric 3D TIs with $\theta=\pi$~\cite{ChenTCI}, then the double SIs $z_{4}'=1,3$ are compatible with bulk-gapped states in double MSG 135.487 $P4_2^\pr/mbc^\pr$.  Hence, we conclude that 3D insulators with $z_{4}'\text{ mod }2=1$ in double MSG 135.487 $P4_2^\pr/mbc^\pr$ are AXIs, \emph{without ambiguity}.  We conjecture that the $z_{4}'=1,3$ AXIs in double MSG 135.487 $P4_2^\pr/mbc^\pr$ can be constructed using the topological crystal method~\cite{ZhidaHermeleCrystal}, which additionally incorporates cell complexes of 2D Chern insulators, TIs, and TCIs.

\textit{Relationship with the SIs in other double SSGs} -- The SIs in double MSG 135.487 $P4_2^\pr/mbc^\pr$ are related to the SIs in double MSG 2.4 $P\bar1$ through the subduction relations:
\begin{equation}
(z_{4}')_{P4_2^\pr/mbc^\pr} \to (\eta_{4I},z_{2I,1},z_{2I,2},z_{2I,3})_{P\bar1}=(2(z_{4}'\text{ mod }2),000)_{P\bar1}.
\end{equation}

Lastly, we study the effects of imposing $\mathcal{T}$ symmetry.  The double SSG 135.484 $P4_{2}/mbc1'$ -- the SSG generated by adding $\{\mathcal{T}|{\bf 0}\}$ symmetry to double MSG 135.487 $P4_2^\pr/mbc^\pr$ -- has the SI group $\mathbb{Z}_{4}$~\cite{ChenTCI}.  The SIs in double SSG 135.484 $P4_{2}/mbc1'$ are in one-to-one correspondence with the SIs in double MSG 135.487 $P4_2^\pr/mbc^\pr$:
\begin{equation}
(z_{4})_{P4_{2}/mbc1'} \to (z_{4}')_{P4_2^\pr/mbc^\pr} = (z_{4})_{P4_2^\pr/mbc^\pr}.
\end{equation}
Nevertheless, because the EBRs in Type-II double SSG 135.484 $P4_{2}/mbc1'$ and the MEBRs in Type-III double MSG 135.487 $P4_2^\pr/mbc^\pr$ are not in one-to-one correspondence, then we will continue throughout this work to employ separate labels ($z_{4}$ and $z_{4}'$ respectively) for the double SIs in double SSGs 135.484 $P4_{2}/mbc1'$ and 135.487 $P4_2^\pr/mbc^\pr$.

\vspace{0.1in}
\subsubsection{Double SIs in Type-III Double MSG 184.195 $P6c'c'$}
\label{sec:P6cpcp}

The double MSG 184.195 $P6c^\pr c^\pr$ is generated by 
    $\{E|100\}$, 
    $\{E|010\}$,
    $\{E|001\}$, 
    $\{C_{6z}|\mathbf{0}\}$ and $\{\TRS m_x|00\frac12\}$, where the angle between the $\{E|100\}$ and $\{E|010\}$ translations is chosen to be $2\pi/3$ for consistency with the $\{C_{3z}|\mathbf{0}\}=(\{C_{6z}|\mathbf{0}\})^{2}$ rotation symmetry.

\textit{SI} -- The double MSG 184.195 $P6c^\pr c^\pr$ has the SI group $\ZZ_6$.  As we will shortly demonstrate, in the physical basis, the double SI $z_{6R}^\pr$ indicates the \emph{even-valued} Chern number in the $k_{z}=\pi$ plane (modulo 12): $C_{k_{z}=\pi}\text{ mod }12 = 2z_{6R}^\pr$.  Hence, insulators with nontrivial values of $z_{6R}^\pr$ are 3D QAH states.

First, using the~\href{http://www.cryst.ehu.es/cryst/corepresentations}{Corepresentations} tool introduced in this work (detailed in Appendix~\ref{sec:coreps}), we determine that Bloch states at the $\{C_{nz}|{\bf 0}\}$-invariant ($n=2,3,6$) ${\bf k}$ points in the $k_{z}=\pi$ plane in double MSG 184.195 $P6c^\pr c^\pr$ form doubly-degenerate pairs with the same $\{C_{nz}|{\bf 0}\}$ rotation symmetry eigenvalues.  We therefore define the $\mathbb{Z}_{6}$ SI to be half of the even-valued Chern number (modulo 6) of the occupied bands in the $k_{z}=\pi$ plane:
\begin{eqnarray}
z_{6R}^\pr &=&  \frac12 \pare{ 
- \frac12 n^{\frac12}_A + \frac12 n^{-\frac12}_A
- \frac32 n^{\frac32}_A + \frac32 n^{-\frac32}_A
- \frac52 n^{\frac52}_A + \frac52 n^{-\frac52}_A
-  n^{\frac12}_H + n^{-\frac12}_H + 3n^{\frac32}_H
+ \frac{3}{2} n^{\frac12}_L - \frac32 n^{-\frac12}_L} \text{ mod } 6 \nonumber \\
&=& \frac{C_{k_z=\pi} }{2} \text{ mod } 6,
\label{eq:z6Rp}
\end{eqnarray}
where the superscripts $n^{j}_{A}$ represent the $\{C_{6z}|{\bf 0}\}$ eigenvalues $e^{-i\frac{2\pi}{6}j}$ at $A$, $n^{j}_{H}$ is the number of occupied states with $\{C_{3z}|{\bf 0}\}$ eigenvalue $e^{-i\frac{2\pi}{3}j}$ at $H$, and where $n^{j}_{L}$ is the number of states with $\{C_{2z}|{\bf 0}\}$ eigenvalue $e^{-i\frac{\pi}{2}j}$ at $L$.

\textit{Layer constructions} --  To diagnose the topology associated to nontrivial values of $z_{6R}^\pr$, we employ the layer construction method.  We begin by placing a $\hat{\bf z}$-normal Chern layer with $C_z=1$ in the $z=0$ plane.  Due to the $\{\TRS m_x|00\frac12\}$ symmetry in double MSG 184.195 $P6c^\pr c^\pr$, there must be another Chern layer with $C_z=1$ in the $z=\frac12$ plane, such that the total Chern number per cell is $C_{z}=2$, and the Chern number of the occupied bands in the $k_{z}=\pi$ plane is $C_{k_{z}=\pi}=2$.  Hence, in this layer construction of a 3D QAH state, $C_{z}=2$ and $z_{6R}^\pr=1$.

\textit{Relationship with the SIs in other double SSGs} -- We next compute the subduction relations between the SIs in double MSG 184.195 $P6c^\pr c^\pr$ and the SIs in the maximal unitary subgroup double MSG 168.109 $P6$ (see Appendix~\ref{subsec:P6}):
\begin{equation}
(z_{6R}^\pr)_{P6c^\pr c^\pr} \to 
(z_{6R})_{P6} = (2(z_{6R}^\pr\text{ mod }3))_{P6}.
\label{eq:cPrcPr6Subduct}
\end{equation}
Eq.~(\ref{eq:cPrcPr6Subduct}) implies that symmetry-indicated 3D QAH states with $z_{6R}^\pr\text{ mod }3\neq 0$ in double MSG 184.195 $P6c^\pr c^\pr$ subduce to symmetry-indicated 3D QAH states with even values of $(z_{6R})_{P6}$ in double MSG 168.109 $P6$, whereas symmetry-indicated 3D QAH states with $z_{6R}^\pr\text{ mod }3=0$ in double MSG 184.195 $P6c^\pr c^\pr$ necessarily subduce to \emph{non-symmetry-indicated} 3D QAH states with $(z_{6R})_{P6}=0$ in double MSG 168.109 $P6$, in agreement with the physical-basis double SI relations $C_{k_{z}=\pi}\text{ mod }12 = 2z_{6R}^\pr$ and $C_{k_{z}=\pi}\text{ mod }6 = z_{6R}$ [see Eq.~(\ref{eq:z6R}) and the surrounding text].

Lastly, if we impose $\mathcal{T}$ symmetry, then the position-space Chern numbers must vanish, which enforces $z_{6R}^\pr$ to be zero.  Correspondingly, in double SSG 184.192 $P6cc1'$ -- the SSG generated by adding $\{\mathcal{T}|{\bf 0}\}$ to double MSG 184.195 $P6c^\pr c^\pr$ -- the double SI group is trivial.

\subsection{Summary of the Double SIs in the Minimal Double SSGs}
\label{sec:summaryDoubleSIs}

In this section, we will summarize and review the results of the minimal double SI calculations performed in Appendix~\ref{sec:34minimal}.  In Table~\ref{tb:completeSIs}, we present a summary of the complete, independent, minimal double SIs of spinful band topology in the 1,651 SSGs.  All symmetry-indicated spinful SISM [specifically symmetry-indicated WSM, see the text following Eq.~(\ref{eq:z2I})], TI, and TCI phases in crystalline solids necessarily exhibit nontrivial values of at least one of the double SIs listed in Table~\ref{tb:completeSIs}.

We note that, in Table~\ref{tb:completeSIs}, some minimal double SSGs $G$ are associated to a smaller set of SIs than the SI group $Z^{G}$.  This occurs because, in some cases, some -- but not all -- of the double SIs in $G$ have already been established in subgroups $M$ of $G$ (\emph{i.e.}, the double SIs in $G$ are not dependent on the double SIs in $M$, even though some of the double SIs in $G$ are the same as the double SIs in $M$, see Appendix~\ref{sec:minimalSIProcedure} for the definition of dependent SIs).  For example, the indicator group of double MSG 147.13 $P\bar{3}$ is $\mathbb{Z}_{12}\times\mathbb{Z}_{2}$, whereas double MSG 147.13 $P\bar{3}$ is only associated in Table~\ref{tb:completeSIs} to the $\mathbb{Z}_{3}$-valued index $z_{3R}$.  In the minimal double MSG 147.13 $P\bar{3}$ [Appendix~\ref{sec:P-3}], the double SIs $(\eta_{4I}$, $z_{3R}$, $z_{2I,3})$  are not dependent on the double SIs in any individual lower-symmetry double MSG.  However, the double SIs $\eta_{4I}$ and $z_{2I,3}$ also appear in the minimal double MSG 2.4 $P\bar{1}$, where the definitions of $\eta_{4I}$ and $z_{2I,3}$ [the product of the parity eigenvalues of a set of bands at all of the $\mathcal{I}$-invariant ${\bf k}$ points and in the $k_{3}=\pi$ plane, respectively, see Eq.~(\ref{eq:eta4I})] is the same in both double MSG 147.13 $P\bar{3}$ and double MSG 2.4 $P\bar{1}$.  Correspondingly, when the spinful SI topological bands of double MSG 147.13 $P\bar{3}$ are subduced onto the subgroup double MSG 2.4 $P\bar{1}$, the values of $(\eta_{4I})_{P\bar{3}}$ and $(z_{2I,3})_{P\bar{3}}$ for the spinful SI topological bands of double MSG 147.13 $P\bar{3}$ are the same as the values of $(\eta_{4I})_{P\bar{1}}$ and $(z_{2I,3})_{P\bar{1}}$ for the SI topological bands subduced onto double MSG 2.4 $P\bar{1}$.  Hence, double MSG 147.13 $P\bar{3}$ is not associated to $\eta_{4I}$ or $z_{2I,3}$ in Table~\ref{tb:completeSIs}, even though the double SIs $(\eta_{4I}$, $z_{3R}$, $z_{2I,3})$ of double MSG 147.13 $P\bar{3}$ include $\eta_{4I}$ and $z_{2I,3}$.

Additionally, in Table~\ref{tb:completeSIs}, some double SIs are associated to more than one minimal double SSG.  This occurs when minimal double SIs that indicate the same bulk topology arise in two minimal double SSGs $G$ and $M$ for which neither $G\not\subset M$ nor $M\not\subset G$.  For example, $z_{4}$ in Table~\ref{tb:completeSIs} is associated to both double SG 2.5 $P\bar{1}1'$ and double MSG 47.249 $Pmmm$.  In both double SG 2.5 $P\bar{1}1'$ and double MSG 47.249 $Pmmm$, $z_{4}=2$ indicates a non-axionic HOTI phase with helical hinge states through the $\mathbb{Z}_{4}$-valued parity eigenvalue formula introduced in Refs.~\onlinecite{AshvinIndicators,ChenTCI,AshvinTCI} [reproduced in Eq.~(\ref{eq:z4})].  We will further analyze the $z_{4}=2$ non-axionic magnetic HOTI phase protected by the symmetries of double MSG 47.249 $Pmmm$ in Appendix~\ref{sec:newHOTIs}.

\clearpage

\begin{table}[t]
\centering
\begin{scriptsize}
\begin{tabular}{|c|c|c|}
\hline
\multicolumn{3}{|c|}{Independent Minimal Double SIs of Spinful Band Topology in the 1,651 Magnetic and Nonmagnetic Double SSGs} \\
\hline
SI & Bulk Topology & Minimal Double SSG(s) [Double SI Formula(s)] \\
\hline
\hline
$\eta_{4I}$ &  WSM/QAH/AXI &  2.4 $P\bar{1}$ [Eq.~(\ref{eq:eta4I})] \\
\hline
$z_{2I,i}$ & QAH: $C_{k_i=\pi}\text{ mod }2$& 2.4 $P\bar{1}$ [Eq.~(\ref{eq:z2I})] \\
\hline
$\eta_{2I}^\pr$  & AXI & 2.4 $P\bar{1}$ [Eq.~(\ref{eq:eta2Ip})] \\
\hline
$z_{2R}$ & QAH: $C_{y} \text{ mod }2$ & 3.1 $P2$ [Eq.~(\ref{eq:z2R})], 41.215 $Ab'a'2$ [Eq.~(\ref{eq:z2R-41.215})]\\
\hline
$\delta_{2m}$  & QAH/AXI/TCI: $C_{k_{y}=\pi}^+-C_{k_{y}=0}^-\text{ mod }2$ & 10.42 $P2/m$ [Eq.~(\ref{eq:d2m})] \\
\hline
$z_{2m,\pi}^+$  & QAH/weak TI/weak TCI: $C_{k_{y}=\pi}^+\text{ mod }2$ & 10.42 $P2/m$ [Eq.~(\ref{eq:z2mp+})] \\
\hline
$z_{2m,\pi}^-$  & QAH/weak TI/weak TCI: $C_{k_{y}=\pi}^-\text{ mod }2$ & 10.42 $P2/m$ [Eq.~(\ref{eq:z2mp-})] \\
\hline
$z_{4}$ & AXI/TCI/HOTI & 2.5 $P\bar{1}1'$, 47.249 $Pmmm$, 83.45 $P4'/m$ [Eq.~(\ref{eq:z4})] \\
\hline
$z_{2w,i}$ & weak TI/weak TCI: $C_{k_i=\pi}^+\text{ mod }2$ & 2.5 $P\bar{1}1'$, 47.249 $Pmmm$, 83.45 $P4'/m$ [Eq.~(\ref{eq:z2w})] ($\dag$) \\
\hline
$z_{4R}$ & QAH: $C_{z}\text{ mod }4$ & 75.1 $P4$ [Eq.~(\ref{eq:z4R})] \\
\hline
$z_{2R}^\pr$, $z_{2R}^\prpr$ & QAH: $C_{y,z}/2\text{ mod }2$ & 77.13 $P4_{2}$ [Eq.~(\ref{eq:z2Rp})], 27.81 $Pc'c'2$ [Eq.~(\ref{eq:z2Rp-27.81})], 54.342 $Pc'c'a$ [Eq.~(\ref{eq:z2Rp-27.81})], \\
 & & 56.369 $Pc'c'n$ [Eq.~(\ref{eq:z2Rp-27.81})], 60.424 $Pb'cn'$ [Eq.~(\ref{eq:z2Rp-60.424})], 110.249 $I4_{1}c'd'$ [Eq.~(\ref{eq:z2Rp-110.249})] ($\ddag$) \\
\hline
$z_{4S}$ & QAH: $C_{z}\text{ mod }4$ & 81.33 $P\bar{4}$ [Eq.~(\ref{eq:z4S})] \\
\hline
$\delta_{2S}$ & WSM & 81.33 $P\bar{4}$ [Eq.~(\ref{eq:d2S})] \\
\hline
$z_2$ & AXI & 81.33 $P\bar{4}$ [Eq.~(\ref{eq:z2})] \\
\hline
$\delta_{4m}$ & QAH/AXI: $C_{k_{z}=\pi}^+ - C_{k_{z}=0}^-\text{ mod }4$ & 83.43 $P4/m$ [Eq.~(\ref{eq:d4m})] \\
\hline
$z_{4m,\pi}^+$ & weak TI/weak TCI: $C_{k_{z}=\pi}^+\text{ mod }4$ & 83.43 $P4/m$ [Eq.~(\ref{eq:z4mp+})] \\
\hline
$z_{4m,\pi}^-$ & weak TI/weak TCI: $C_{k_{z}=\pi}^-\text{ mod }4$ & 83.43 $P4/m$ [Eq.~(\ref{eq:z4mp-})] \\
\hline
$z_{4m,0}^+$ & QAH/weak TI/weak TCI: $C_{k_{z}=0}^+\text{ mod }4$ & 84.51 $P4_{2}/m$ [Eq.~(\ref{eq:z4m0+})] \\ 
\hline
$z_8$ & AXI/TCI/HOTI & 83.44 $P4/m1'$, 123.339 $P4/mmm$ [Eq.~(\ref{eq:z8})] \\
\hline
$z_{3R}$ & QAH: $C_{z}\text{ mod }3$ & 147.13 $P\bar{3}$ [Eq.~(\ref{eq:z3R})] \\
\hline
$z_{6R}$ & QAH: $C_{z}\text{ mod } 6$ & 168.109 $P6$ [Eq.~(\ref{eq:z6R})] \\
\hline
$\delta_{3m}$ & QAH/AXI/TCI: $C_{k_{z}=\pi}^+-C_{k_{z}=\pi}^-\text{ mod }3$ & 174.133 $P\bar{6}$ [Eq.~(\ref{eq:d3m})] \\
\hline
$z_{3m,\pi}^+$ & weak TI/weak TCI: $C_{k_{z}=\pi}^+\text{ mod }3$ & 174.133 $P\bar{6}$ [Eq.~(\ref{eq:z3mp+})] \\
\hline
$z_{3m,\pi}^-$ & weak TI/weak TCI: $C_{k_{z}=\pi}^-\text{ mod }3$ & 174.133 $P\bar{6}$ [Eq.~(\ref{eq:z3mp-})] \\
\hline
$\delta_{6m}$ & QAH/AXI/TCI: $C_{k_{z}=\pi}^+-C_{k_{z}=\pi}^-\text{ mod }6$ & 175.137 $P6/m$ [Eq.~(\ref{eq:d6m})] \\
\hline
$z_{6m,\pi}^+$ & weak TI/weak TCI: $C_{k_{z}=\pi}^+\text{ mod }6$ & 175.137 $P6/m$ [Eq.~(\ref{eq:z6mp+})] \\
\hline
$z_{6m,\pi}^-$ & weak TI/weak TCI: $C_{k_{z}=\pi}^-\text{ mod }6$ & 175.137 $P6/m$ [Eq.~(\ref{eq:z6mp-})] \\
\hline
$z_{6m,0}^+$ & QAH/weak TI/weak TCI: $C_{k_{z}=0}^+\text{ mod }6$ & 176.143 $P6_{3}/m$ [Eq.~(\ref{eq:z6m0+})] \\
\hline
$z_{12}$ & AXI/TCI/HOTI & 175.138 $P6/m1'$, 191.233 $P6/mmm$ [Eq.~(\ref{eq:z12})] \\
\hline
$z_{12}'$ & AXI/TCI/HOTI & 176.144 $P6_{3}/m1'$ [Eq.~(\ref{eq:z12p})] \\
\hline
$z_{4R}^\pr$ & QAH: $C_{z}/2\text{ mod }4$ & 103.199 $P4c'c'$ [Eq.~(\ref{eq:z4Rp})] \\
\hline
$z_{4}^\pr$ & AXI/TCI & 135.487 $P4_{2}'/mbc'$ [Eq.~(\ref{eq:z4p})] \\
\hline
$z_{6R}^\pr$ & QAH: $C_{z}/2\text{ mod }6$ & 184.195 $P6c'c'$ [Eq.~(\ref{eq:z6Rp})] \\
\hline
\end{tabular}
\end{scriptsize}
\caption{The independent minimal double SIs of spinful band topology in all 1,651 double SSGs.  In order, this table contains the symbol of each double SI, the bulk topological phase(s) associated to nontrivial values of the double SI including -- where applicable -- the momentum- or position-space Chern numbers indicated by the double SI, and the minimal double SSG(s) associated to the double SI [\emph{i.e.} the lowest-symmetry SSG(s) in which the double SI predicts nontrivial band topology, see Appendices~\ref{sec:minimalSIProcedure} and~\ref{sec:minimalSSGTables}], as well as the equation in Appendix~\ref{sec:34minimal} containing the explicit double SI formula in terms of crystal symmetry eigenvalues.  All symmetry-indicated spinful SISM [specifically symmetry-indicated WSM, see the text following Eq.~(\ref{eq:z2I})], TI, and TCI phases in crystalline solids necessarily exhibit nontrivial values of at least one of the double SIs listed in this table.  We note that, in this table, the symbol ``AXI'' refers to both magnetic AXIs and $\mathcal{T}$-symmetric 3D TIs, because AXI and 3D TI phases are both defined by the nontrivial bulk axion angle $\theta=\pi$~\cite{WilczekAxion,QHZ,VDBAxion,AndreiInversion,AshvinAxion,WiederAxion,YuanfengAXI,NicoDavidAXI1,NicoDavidAXI2,TMDHOTI,KoreanAXI,CohVDBAXI,ArisHopf,TitusRonnyKondoAXI,YoungkukMonopole,MurakamiAXI1,MurakamiAXI2,HingeStateSphereAXI,BJYangVortex,BarryBenCDW,IvoAXI1,IvoAXI2,GuidoAXI,FuKaneMele,FuKaneInversion}.  Additionally, the symbols ``TCI'' and ``HOTI'' respectively indicate helical (\emph{i.e.} non-axionic) mirror Chern insulators and HOTIs~\cite{HsiehTCI,TeoFuKaneTCI,HourglassInsulator,DiracInsulator,AshvinIndicators,ChenBernevigTCI,ChenRotation,HOTIChen,ChenTCI,AshvinTCI,HOTIBismuth,TMDHOTI,HOTIBernevig}, which include the magnetic HOTIs  introduced in this work (see Appendix~\ref{sec:newHOTIs}).  We have placed a $\dag$ symbol after MSG 83.45 $P4'/m$ in the row for $z_{2w,i}$ to emphasize that, of the three $z_{2w,i}$, only $z_{2w,3}$ is a minimal double SI in MSG 83.45 $P4'/m$ (where minimal double SIs are defined in Appendix~\ref{sec:minimalSIProcedure}).  We have placed a $\ddag$ symbol after MSG 110.249 $I4_{1}c'd'$ in the row for the indices $z_{2R}'$ and $z_{2R}''$ to emphasize that the position-space Chern number $C_{z}$ (modulo $2$) is indicated by $z_{2R}''$ only in the primitive cell of a crystal in MSG 110.249 $I4_{1}c'd'$ -- in the conventional cell, the position-space Chern number is given by $C_{z}\text{ mod }8 = 4 z_{2R}^\prpr$ [see the text surrounding Eq.~(\ref{eq:z2Rp-110.249})].}
\label{tb:completeSIs} 
\end{table}

\clearpage

\subsection{Non-Axionic Spinful Magnetic HOTIs}
\label{sec:newHOTIs}

In the sections below, we will further analyze the spinful helical magnetic HOTI phases discovered in this work.  As discussed in the main text and in Appendices~\ref{sec:34minimal} and~\ref{sec:summaryDoubleSIs}, we have discovered helical magnetic (\emph{i.e.} $\{\mathcal{T}|{\bf 0}\}$-broken) HOTI phases indicated by $z_{4}=2$ in double MSG 47.249 $Pmmm$, $z_{8}=4$ in double MSG 123.339 $P4/mmm$, and $z_{12}=6$ in double MSG 191.233 $P6/mmm$, as well as trivial values for all other independent minimal double SIs in Table~\ref{tb:completeSIs}.  In this work, we refer to the helical magnetic HOTIs that will be analyzed in this section as \emph{non-axionic}, because the helical HOTIs exhibit trivial axion angles $\theta\text{ mod }2\pi=0$ [see Refs.~\onlinecite{WilczekAxion,QHZ,VDBAxion,AndreiInversion,AshvinAxion,WiederAxion,YuanfengAXI,NicoDavidAXI1,NicoDavidAXI2,TMDHOTI,KoreanAXI,CohVDBAXI,ArisHopf,TitusRonnyKondoAXI,YoungkukMonopole,MurakamiAXI1,MurakamiAXI2,HingeStateSphereAXI,BJYangVortex,BarryBenCDW,IvoAXI1,IvoAXI2,GuidoAXI,FuKaneMele,FuKaneInversion} for further discussions of chiral HOTIs (\emph{i.e.} AXIs), which conversely exhibit nontrivial axion angles $\theta=\pi$].  When terminated in nanorod geometries, the helical magnetic HOTIs generically exhibit even numbers of massive or massless twofold surface Dirac cones, and domain walls between surfaces with oppositely-signed masses bind mirror-protected helical hinge states.  As we will show in Appendix~\ref{sec:HOTItbModel}, the helical magnetic HOTIs discovered in this work can be connected to nonmagnetic ``rotation-anomaly'' TCIs~\cite{ChenRotation,HOTIChen} without closing a bulk or surface gap or gapping the anomalous surface or hinge states.  First, in Appendix~\ref{sec:HOTIfermionDoubling}, we will introduce the symmetry-enhanced fermion doubling theorems~\cite{DiracInsulator,ChenRotation,Steve2D,SteveMagnet} for twofold Dirac fermions in the surface wallpaper groups of the helical magnetic HOTIs, which we will then use to diagnose the 2D surface states as anomalous.  Unlike in Ref.~\onlinecite{ChenRotation}, the twofold Dirac fermion doubling theorems introduced in Appendix~\ref{sec:HOTIfermionDoubling} do not require $\{\mathcal{T}|{\bf 0}\}$ to be enforced, and are instead only enforced by the spinful unitary magnetic symmetries of Type-I magnetic double wallpaper groups.  Lastly, in Appendix~\ref{sec:HOTItbModel}, we will introduce tight-binding models for the helical magnetic HOTI phases, which we will use to explicitly demonstrate the presence of anomalous, mirror-protected 2D surface and 1D hinge states.

\vspace{0.15in}
\subsubsection{Symmetry-Enhanced Fermion Doubling Theorems for Non-Axionic Magnetic HOTIs}
\label{sec:HOTIfermionDoubling}

In this section, we will derive 2D symmetry-enhanced fermion doubling theorems~\cite{DiracInsulator,ChenRotation,Steve2D,SteveMagnet} for the surface wallpaper groups~\cite{WiederLayers,ConwaySymmetries} of spinful, helical magnetic HOTIs.  Through the doubling theorems established in this section, we will demonstrate that the 2D, twofold Dirac surface states of helical magnetic HOTIs are anomalous (see Appendix~\ref{sec:HOTItbModel} for tight-binding models and surface- and hinge-state calculations for helical magnetic HOTIs).

To begin, in each BZ of a 2D crystal, the parity anomaly excludes the presence of a single (\emph{i.e.} unpaired) twofold-degenerate, linearly dispersing, ($\{\mathcal{T}|{\bf 0}\}\}$- or magnetic-) symmetry-stabilized Dirac fermion~\cite{HaldaneModel,DiracInsulator,WittenParity,RedlichParity,JackiwParity}.  However, on the 2D surfaces of 3D TIs~\cite{FuKaneMele,FuKaneInversion,MulliganAnomaly,LapaParity} (and some AXIs, see Refs.~\onlinecite{WilczekAxion,QHZ,VDBAxion,AndreiInversion,AshvinAxion,WiederAxion,YuanfengAXI,NicoDavidAXI1,NicoDavidAXI2,TMDHOTI,KoreanAXI,CohVDBAXI,ArisHopf,TitusRonnyKondoAXI,YoungkukMonopole,MurakamiAXI1,MurakamiAXI2,HingeStateSphereAXI,BJYangVortex,BarryBenCDW,IvoAXI1,IvoAXI2,GuidoAXI}),  unpaired twofold Dirac fermions are anomalously stabilized by the combination of surface wallpaper group symmetries and spectral (Wannier) flow.  As shown in Refs.~\onlinecite{ChenRotation,DiracInsulator}, for 3D crystals whose surface wallpaper groups contain additional rotation and reflection symmetries, there also exist \emph{symmetry-enhanced} fermion doubling theorems that may similarly be evaded through a combination of wallpaper group symmetry and spectral flow.

\begin{figure}[h]
\centering
\includegraphics[width=0.70\textwidth]{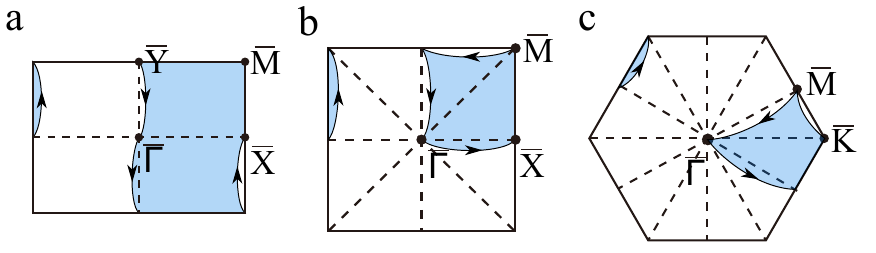}
\caption{The 2D BZs of wallpaper groups with even-fold rotation symmetries.  (a) The 2D BZ of Type-I magnetic wallpaper group $pmm$ [isomorphic to Type-I MSG 25.57 $Pmm2$ modulo out-of-plane lattice translations] or Type-II nonmagnetic wallpaper group $pmm1'$ [isomorphic to Type-II SG 25.58 $Pmm21'$ modulo out-of-plane lattice translations].  (b) The 2D BZ of Type-I magnetic wallpaper group $p4m$ [isomorphic to Type-I MSG 99.163 $P4mm$ modulo out-of-plane lattice translations] or Type-II nonmagnetic wallpaper group $p4m1'$ [isomorphic to Type-II SG 99.164 $P4mm1'$ modulo out-of-plane lattice translations].  (c) The 2D BZ of Type-I magnetic wallpaper group $p6m$ [isomorphic to Type-I MSG 183.185 $P6mm$ modulo out-of-plane lattice translations] or Type-II nonmagnetic wallpaper group $p6m1'$ [isomorphic to Type-II SG 183.186 $P6mm1'$ modulo out-of-plane lattice translations].  The dashed lines in (a-c) indicate mirror lines.  The blue patches in (a-c) respectively indicate patches of the 2D BZ whose area is one half, one quarter, and one sixth of the first 2D BZ; the boundaries of the blue patches are explicitly chosen to avoid coinciding with the mirror lines.  Eqs.~(\ref{eq:Berry-C2}),~(\ref{eq:Berry-C4}), and~(\ref{eq:Berry-C6}) respectively indicate the combinations of rotation symmetry eigenvalues that correspond to the quantized Berry phases $\Theta_{2,4,6}=0,\pi$ in the blue patches in (a-c).}
\label{fig:fermionMultiplication}
\end{figure}

In Ref.~\onlinecite{ChenRotation}, the authors specifically defined the \emph{fermion multiplication theorem} for twofold Dirac fermions in nonmagnetic [Type-II, see Appendix~\ref{sec:type2}] double wallpaper groups [which we will in this section take to be $\hat{\bf z}$-normal] that contain the symmetries $\{\mathcal{T}|{\bf 0}\}$ and $\{C_{2z}|{\bf 0}\}$, as well as, optionally, $\{C_{4z}|{\bf 0}\}$ or $\{C_{6z}|{\bf 0}\}$.  To derive the fermion multiplication for nonmagnetic double wallpaper groups, we begin by exploiting the formulas derived in Ref.~\onlinecite{ChenBernevigTCI} for Berry phase in 2D crystals with rotation symmetries.  Specifically, in Ref.~\onlinecite{ChenBernevigTCI}, it was shown that the Berry phase $\Theta_{2}$ in one half of the BZ of a 2D crystal with $\{C_{2z}|{\bf 0}\}$ symmetry [Fig.~\ref{fig:fermionMultiplication}(a)] is given by:
\begin{equation}
e^{i\Theta_2} = (-1)^{N_{occ}} \prod_{m\in occ} \zeta_m(\bar{\Gamma}) \zeta_n(\bar{X}) \zeta_m(\bar{Y}) \zeta_m(\bar{M}), 
\label{eq:Berry-C2}
\end{equation}
that the Berry phase $\Theta_{4}$ in one quarter of the BZ of a 2D crystal with $\{C_{4z}|{\bf 0}\}$ symmetry [Fig.~\ref{fig:fermionMultiplication}(b)] is given by:
\begin{equation}
e^{i\Theta_4} = (-1)^{N_{occ}} \prod_{m\in occ} \xi_m(\bar{\Gamma}) \xi_m(\bar{M}) \zeta_m(\bar{X}),
\label{eq:Berry-C4}
\end{equation}
and that the Berry phase $\Theta_{6}$ in one sixth of the BZ of a 2D crystal with $\{C_{6z}|{\bf 0}\}$ symmetry [Fig.~\ref{fig:fermionMultiplication}(c)] is given by:
\begin{equation}
e^{i\Theta_6} = (-1)^{N_{occ}} \prod_{m\in occ} \eta_m(\bar{\Gamma}) \theta_m(\bar{K}) \zeta_m(\bar{M}),
\label{eq:Berry-C6}
\end{equation}
where $\zeta_m(K)$, $\xi_m(K)$, $\eta_m(K)$, $\theta_m(K)$ respectively refer to the $C_{2z}$, $C_{4z}$, $C_{6z}$, and $C_{3z}$ eigenvalues of the $m^\text{th}$ Bloch state at $K$, and where $N_{occ}$ is the number of Bloch states at each high-symmetry ${\bf k}$ point in a given energy range.

In Ref.~\onlinecite{ChenBernevigTCI}, it was shown that $\Theta_{2,4,6}=\pi$ in Eqs.~(\ref{eq:Berry-C2}),~(\ref{eq:Berry-C4}), and~(\ref{eq:Berry-C6}) respectively indicates a nontrivial bulk Chern number.  However, in the presence of $\{\mathcal{T}|{\bf 0}\}$ or in-plane mirror symmetries, the Chern number is required to vanish~\cite{FuKaneMele,FuKaneInversion,QHZ,YoungkukLineNode,FangFuMobius,WiederAxion,ChenRotation,VDBAxion,NicoDavidAXI2}.  In wallpaper groups with $C_{2z}$, $C_{4z}$, or $C_{6z}$ rotation symmetry and either $\{\mathcal{T}|{\bf 0}\}$ or in-plane mirror lines, the disagreement between $\Theta_{2,4,6}=\pi$ and the symmetry restriction that the Chern number vanish can be resolved by recognizing that a twofold Dirac fermion respectively placed in each half, quarter, and sixth of the 2D BZ also provides a source of $\pi$ Berry phase indicated by $\Theta_{2,4,6}=\pi$~\cite{HaldaneModel,DiracInsulator,WittenParity,RedlichParity,JackiwParity,FuKaneMele,FuKaneInversion,MulliganAnomaly,LapaParity}.  In nonmagnetic (Type-II) wallpaper groups with $C_{2z}$, $C_{4z}$, or $C_{6z}$ rotation symmetry, or in Type-I magnetic wallpaper groups with mirror and $C_{2z}$, $C_{4z}$, or $C_{6z}$ rotation symmetry, $\Theta_{2,4,6}$ therefore respectively indicate the number of twofold Dirac cones in each BZ modulo $4$, $8$, and $12$.  Specifically, if $\Theta_{2,4,6}=\pi$ ($\Theta_{2,4,6}=0$), there must be an odd (even) number of twofold Dirac cones in the blue BZ patches in Fig.~\ref{fig:fermionMultiplication}(a-c), respectively implying the presence of $2+4a$ ($4a$), $4+8a$ ($8a$), or $6 + 12a$ ($12a$) twofold Dirac fermions in each BZ [where $a\in\{\mathbb{Z}^{+},0\}$].  In nonmagnetic wallpaper groups with $C_{2z}$, $C_{4z}$, or $C_{6z}$ rotation symmetry, the Dirac fermions are stabilized by $\{C_{2z}\times\mathcal{T}|{\bf 0}\}$ symmetry~\cite{FangFuMobius,YoungkukLineNode,WiederAxion,ChenRotation}, and in the magnetic wallpaper groups in Fig.~\ref{fig:fermionMultiplication}(a-c), the Dirac fermions are stabilized by mirror symmetry.

However, in Ref.~\onlinecite{ChenBernevigTCI}, it was shown that $\Theta_{2,4,6}=0$ for all 2D spinful lattice models with $\{\mathcal{T}|{\bf 0}\}$ symmetry, due to the constraints imposed by $\mathcal{T}$ symmetry on the eigenvalues of spinful rotation symmetries.  Below, we will show that $\Theta_{2,4,6}=0$ is also required in all 2D spinful lattice models that respect the symmetries of Type-I magnetic double wallpaper groups containing $\{m_{x}|{\bf 0}\}$ and $\{m_{y}|{\bf 0}\}$ and $C_{2z}$, $C_{4z}$, or $C_{6z}$ rotation symmetries (\emph{i.e.} Type-I magnetic double wallpaper groups $pmm$, $p4m$, and $p6m$, respectively~\cite{DiracInsulator,WiederLayers,ChenRotation,SteveMagnet,ConwaySymmetries}).  We note that throughout this work, the symbols of wallpaper groups -- which are also sometimes termed \emph{plane groups} -- are given in the short notation previously employed in Refs.~\onlinecite{DiracInsulator,WiederLayers,HingeSM}; in the long notation of the~\href{https://www.cryst.ehu.es/plane/get_plane_gen.html}{Get Plane Gen} tool on the BCS~\cite{BCS1,BCS2}, magnetic wallpaper groups $pmm$, $p4m$, and $p6m$ are respectively labeled by the symbols $p2mm$, $p4mm$, and $p6mm$.  For each $C_{2z}$-symmetric wallpaper group in Fig.~\ref{fig:fermionMultiplication}(a-c), we will choose a patch of the 2D BZ whose boundary intersects the rotation-invariant ${\bf k}$ points in Eqs.~(\ref{eq:Berry-C2}),~(\ref{eq:Berry-C4}), and~(\ref{eq:Berry-C6}), respectively, while avoiding the mirror lines, which may host mirror-symmetry-stabilized Dirac fermions.

First, in double magnetic wallpaper group $pmm$ [isomorphic to Type-I MSG 25.57 $Pmm2$ modulo out-of-plane lattice translations], the matrix representatives of $\{C_{2z}|{\bf 0}\}$ and $\{m_{x}|{\bf 0}\}$ anticommute at each of the four $C_{2z}$-invariant ${\bf k}$ points in Fig.~\ref{fig:fermionMultiplication}(a) and Eq.~(\ref{eq:Berry-C2}) [this result can be obtained by applying the~\href{http://www.cryst.ehu.es/cryst/corepresentations}{Corepresentations} tool detailed in Appendix~\ref{sec:coreps} to MSG 25.57 $Pmm2$].  Consequently, in 2D lattice models constructed from MEBRs, all of the small irreps $\sigma$ at the four $C_{2z}$-invariant ${\bf k}$ points are two-dimensional, and exhibit net-zero $C_{2z}$ eigenvalues: $\chi_{\sigma}(\{C_{2z}|{\bf 0}\})=0$.  For any set of energetically isolated multiplets of Bloch states at the four $C_{2z}$-invariant points in Fig.~\ref{fig:fermionMultiplication}(a) and Eq.~(\ref{eq:Berry-C2}), this implies that $\Theta_{2}\text{ mod }2\pi=0$.  Consequently, spinful lattice models in double magnetic wallpaper group $pmm$ must exhibit even numbers of twofold Dirac fermions in each half of the 2D BZ in Fig.~\ref{fig:fermionMultiplication}(a).

Similarly, for double magnetic wallpaper groups $p4m$ [isomorphic to Type-I MSG 99.163 $P4mm$ modulo out-of-plane lattice translations] and $p6m$ [isomorphic to Type-I MSG 183.185 $P6mm$ modulo out-of-plane lattice translations], it can be shown through the~\href{http://www.cryst.ehu.es/cryst/corepresentations}{Corepresentations} tool (see Appendix~\ref{sec:coreps}) that the spinful rotation eigenvalues of energetically isolated multiplets of Bloch states must also appear in complex-conjugate pairs.  This respectively implies that, for spinful lattice models in double magnetic wallpaper groups $p4m$ and $p6m$, $\Theta_{4,6}\text{ mod }2\pi=0$.  Consequently, spinful lattice models in double magnetic wallpaper group $p4m$ [$p6m$] must exhibit even numbers of twofold Dirac fermions in each quarter [sixth] of the 2D BZ in Fig.~\ref{fig:fermionMultiplication}(b) [Fig.~\ref{fig:fermionMultiplication}(c)].

As we will shortly see in Appendix~\ref{sec:HOTItbModel}, the surfaces of the helical magnetic HOTIs discovered in this work exhibit odd numbers of twofold Dirac cones in the blue regions of the 2D BZs shown in Fig.~\ref{fig:fermionMultiplication}(a-c), representing anomalous exceptions to the magnetic fermion multiplication theorem derived in this section.

\subsubsection{Tight-Binding Models and Boundary States for Non-Axionic Magnetic HOTIs}
\label{sec:HOTItbModel}

Through the double SIs computed in Appendix~\ref{sec:34minimal}, we have discovered three novel variants of spinful, helical magnetic HOTIs with trivial axion angles $\theta\text{ mod }2\pi =0$.  In this section, we will provide tight-binding models and surface- and hinge-state calculations for the three non-axionic magnetic HOTI phases discovered in this work.  For each phase, we will also demonstrate how the top ($\hat{\bf z}$-normal) surface states circumvent a magnetic fermion multiplication theorem (see Appendix~\ref{sec:HOTIfermionDoubling}).  We will leave the development of bulk (nested) Wilson loop invariants~\cite{WladTheory,WiederAxion,TMDHOTI,DiracInsulator,HourglassInsulator} for the helical magnetic HOTI phases for future works.  However, we note that, like the fourfold-rotation-anomaly HOTI phase in SnTe~\cite{HsiehTCI,TeoFuKaneTCI,HOTIBernevig,HOTIChen}, in the helical magnetic HOTIs modeled in this section, the occupied bands in half of the bulk mirror planes that project to the $\hat{\bf z}$-normal surface (\emph{e.g.} $\{m_{x\pm y}|{\bf 0}\}$) exhibit mirror Chern numbers $C_{m}\text{ mod }4=2$, whereas the other half (\emph{e.g.} $\{m_{x,y}|{\bf 0}\}$) exhibit $C_{m}\text{ mod }4=0$ (see Fig.~\ref{fig:SS}).

\textit{$D_{2h}$ HOTI in double MSG 47.249 $Pmmm$} -- We will here analyze the helical magnetic TCI phase protected by the symmetries of double MPG $mmm$ 8.1.24 [$D_{2h}$] (see Appendices~\ref{sec:siteSymmetry} and~\ref{sec:magWannier} and Refs.~\onlinecite{ShubnikovMagneticPoint,BilbaoPoint,PointGroupTables,MagneticBook,EvarestovBook,EvarestovMEBR,BCS1,BCS2,BCSMag1,BCSMag2,BCSMag3,BCSMag4}), which we term the \emph{$D_{2h}$ HOTI} (as previously in Appendices~\ref{sec:siteSymmetry} and~\ref{sec:magWannier}, we will continue to label MPGs in this section employing the notation of the~\href{http://www.cryst.ehu.es/cryst/mpoint.html}{MPOINT} tool on the BCS~\cite{BCSMag1,BCSMag2,BCSMag3,BCSMag4} in which an MPG is labeled by its number, followed by its symbol).  As discussed in Appendix~\ref{subsec:Pmmm}, the double SIs $(z_4,z_{2w,1},z_{2w,2},z_{2w,3})=(2000)$ in double MSG 47.249 $Pmmm$ indicate a mirror TCI for which the mirror Chern numbers $C_{m_{x}}\text{ mod }2=C_{m_{y}}\text{ mod }2=C_{m_{z}}\text{ mod }2=0,\ C_{m_{x}}+C_{m_{y}} + C_{m_{z}}\text{ mod }4=2$.  Nevertheless, in this work, we refer to the $(2000)$ phase in double MSG 47.249 $Pmmm$ as a helical HOTI for two reasons.  First, as discussed in Appendix~\ref{subsec:Pmmm}, the $(2000)$ phase of double MSG 47.249 can be connected to a $(z_4,z_{2w,1},z_{2w,2},z_{2w,3})_{Pmmm1'}=(2000)_{Pmmm1'}$ mirror TCI phase in the $\mathcal{T}$-symmetric supergroup Type-II double SG 47.250 $Pmmm1'$ without closing a bulk or surface gap.  In turn, the $(2000)_{Pmmm1'}$ TCI phase subduces to an $\mathcal{I}$- and $\mathcal{T}$-protected $(z_4,z_{2w,1},z_{2w,2},z_{2w,3})_{P\bar{1}1'}=(2000)_{P\bar{1}1'}$ helical HOTI in Type-II double SG 2.5 $P\bar{1}1'$ [see Appendix~\ref{sec:P-11p} and Refs.~\onlinecite{ChenTCI,AshvinTCI,AshvinIndicators,TMDHOTI}].  Second, in each of the momentum-space mirror planes in double MSG 47.249 $Pmmm$, a nontrivial mirror Chern number cannot be identified by the 2D symmetry-based mirror Chern indices implied by the Chern number SI formulas in Ref.~\onlinecite{ChenBernevigTCI}.  Specifically, each momentum-space mirror plane in double MSG 47.249 $Pmmm$ has only mirror, twofold rotation, and inversion symmetries, which can only indicate the mirror Chern number modulo $2$.  For example, in the $k_{z}=\pi$ plane of the bulk BZ in double MSG 47.249 $Pmmm$, the only SI of stable 2D topology is $z_{2w,3}$, which only indicates $C_{m_{z}}(k_{z}=\pi)\text{ mod }2$ (see Appendix~\ref{subsec:Pmmm}).  Hence, the nontrivial \emph{even} mirror Chern numbers of the $(2000)$ phase in double MSG 47.249 $Pmmm$ can only be inferred from the 3D double SIs $(z_4,z_{2w,1},z_{2w,2},z_{2w,3})$, and cannot be inferred from symmetry-indicated momentum-space mirror Chern numbers evaluated in BZ planes.  As we will show in this section, and as discussed in previous works~\cite{HOTIBernevig,AshvinTCI,ChenTCI}, TCI surface states may in general be interpreted as HOTI hinge modes if a finite sample is cut into a geometry in which the bulk mirror planes project to 1D hinges, as opposed to flat 2D surfaces.

\begin{figure}[t]
\centering
\includegraphics[width=0.9\linewidth]{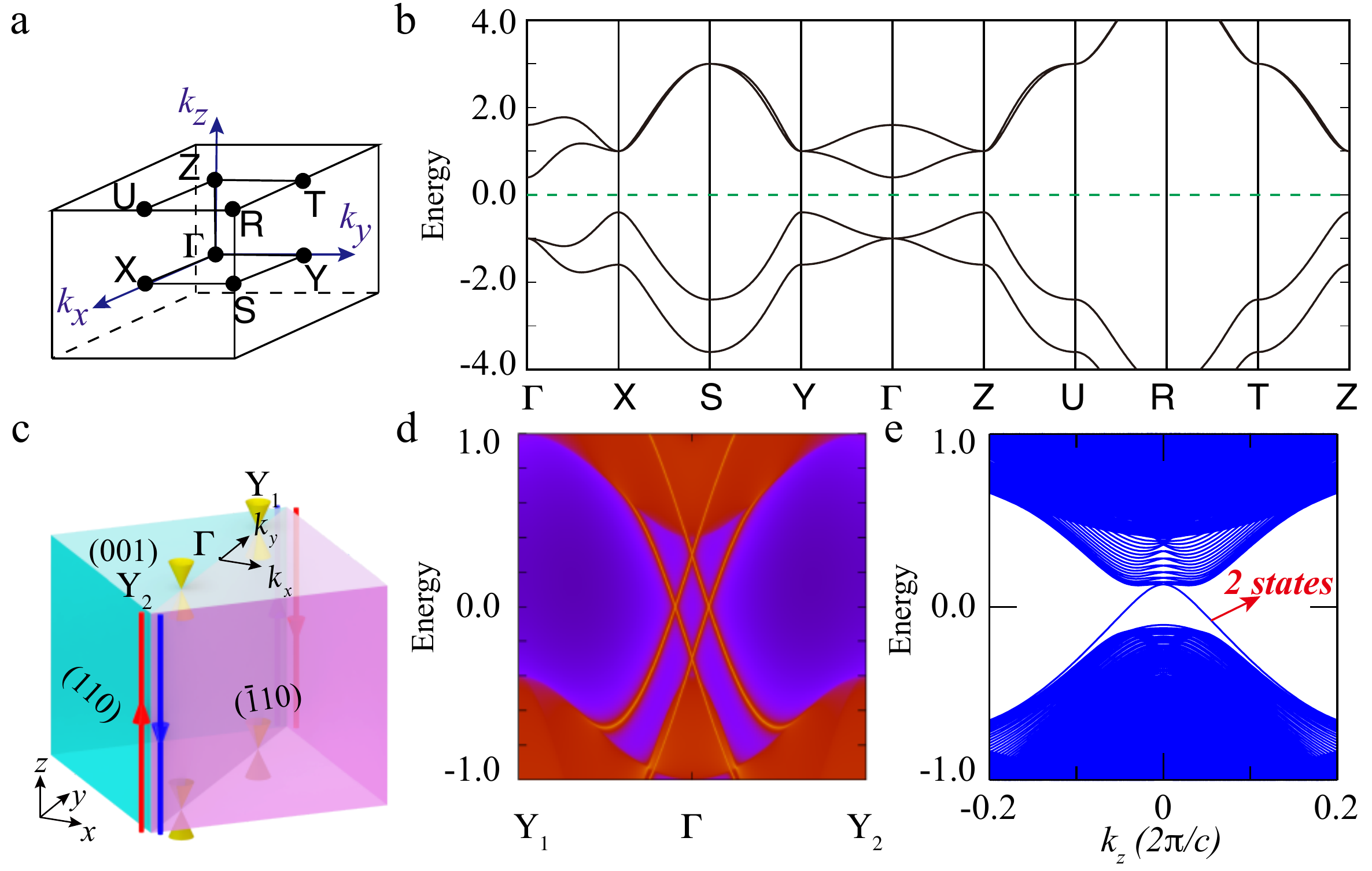}
\caption{Surface and hinge states of the helical magnetic $D_{2h}$ HOTI phase in double MSG 47.249 $Pmmm$. (a) The bulk BZ. (b) The bulk band structure obtained from Eq.~(\ref{eq:D2hHOTI}) with $\Delta_{0}=1$ and $\Delta_1=0.3$.  We note that Eq.~(\ref{eq:D2hHOTI}) contains additional, extraneous symmetries beyond those of double MSG 47.249 $Pmmm$, such that the band structure in (b) exhibits additional degeneracies away from the Fermi level -- such as the occupied fourfold degeneracy at $\Gamma$ -- that are not robust to symmetry-preserving perturbations. (c) Schematic of the top ($\hat{\bf z}$-normal) surface states and nanorod hinge states.  The top surface of the rectangular nanorod in (c) respects the symmetries of Type-I double magnetic wallpaper group $pmm$, and the hinges respect the symmetries of frieze groups that contain either $\{m_{x}|0\}$ or $\{m_{y}|0\}$ (see Appendices~\ref{sec:34minimal} and~\ref{sec:HOTIfermionDoubling} and Refs.~\onlinecite{DiracInsulator,WiederLayers,HOTIBernevig,ConwaySymmetries,SteveMagnet}).  (d) The top surface spectrum plotted along $k_{y}$, obtained from surface Green's functions calculated for the model in (b) terminated in a $z$-directed slab geometry.  In (d), the surface bands exhibit mirror Chern $C_{m_{x}}=2$ spectral flow.  We have verified through surface-state calculations that the slab surface spectrum along $k_{x}$ does not exhibit spectral flow, and that $C_{m_{z}}=0$.  Together, this implies that the top surface exhibits two twofold Dirac cones, circumventing the fermion multiplication theorem for double magnetic wallpaper group $pmm$ derived in Appendix~\ref{sec:HOTIfermionDoubling}, and implies that the bulk is a $D_{2h}$ HOTI.  (e)  The spectrum of an infinite, $z$-directed, $m_{x,y}$-symmetric nanorod of the model in (b) features two pairs of hinge-localized helical modes (four total hinge states), demonstrating that the model in (b) exhibits higher-order spectral flow.}
\label{fig:HOTI-Pmmm}
\end{figure}

To model the $D_{2h}$ HOTI phase in double MSG 47.249 $Pmmm$, we begin by introducing the Bernevig-Hughes-Zhang Hamiltonian for a 3D TI~\cite{AndreiTI,FuKaneMele,FuKaneInversion,QHZ}:
\begin{equation}
H_{\rm TI}({\bf k}) = -\tau^{z}\bigg( 2 - \sum_{i=x,y,z} \cos k_i \bigg) + \sum_{i=x,y,z} \tau^x \sigma^i \sin k_i ,
\label{eq:TRS-TI1}
\end{equation}
where $\tau^{i}$ and $\sigma^{j}$ are each $2\times 2$ Pauli matrices, and where we have employed a notation in which $\tau^{i}\sigma^{j} = \tau^{i}\otimes\sigma^{j}$ and factors of the $2\times 2$ identity matrices $\tau^{0}$ and $\sigma^{0}$ are suppressed.  Eq.~(\ref{eq:TRS-TI1}) respects $\mathcal{I}$ and spinful $\mathcal{T}$ symmetries, which are represented through the symmetry action:
\begin{equation}
\mathcal{I}H_{\rm TI}({\bf k})\mathcal{I}^{-1} = \tau^{z}H_{\rm TI}(-{\bf k})\tau^z,\ \mathcal{T}H_{\rm TI}({\bf k})\mathcal{T}^{-1} = \sigma^{y}H^{*}_{\rm TI}(-{\bf k})\sigma^{y}.
\label{eq:ITsymmetryD2h}
\end{equation}
We next construct the helical $D_{2h}$ HOTI phase by first superposing two copies of the 3D TI phase of Eq.~(\ref{eq:TRS-TI1}), and then introducing perturbative couplings to break $\mathcal{T}$ symmetry:
\begin{equation}
H^{Pmmm}_{\rm HOTI}({\bf k}) = \mu^{0}H_{\rm TI}({\bf k}) + \Delta_0 \mu^y \tau^y \sin k_x 
+ \Delta_1 [\mu^z(\tau^z+\tau^0)
    +  (\tau^z+\tau^0)\sigma^z\sin k_x \sin k_y],
\label{eq:D2hHOTI}
\end{equation}
in which $\mu^{i}$ is a $2\times 2$ Pauli matrix that indexes the two coupled 3D TI models, and where we have employed a notation in which $\mu^{i}\tau^{j}\sigma^{k} = \mu^{i}\otimes\tau^{j}\otimes\sigma^{k}$ and factors of the $2\times 2$ identity matrices $\mu^{0}$, $\tau^{0}$, $\sigma^{0}$ are suppressed in terms other than $\mu^{0}H_{\rm TI}({\bf k})$ when the identity matrices are not summed with other Pauli matrices.  $H^{Pmmm}_{\rm HOTI}({\bf k})$ in Eq.~(\ref{eq:D2hHOTI}) respects the symmetries of double MPG $mmm$ 8.1.24 [$D_{2h}$], whose generating elements are represented through the action:
\begin{eqnarray}
\mathcal{I}H^{Pmmm}_{\rm HOTI}({\bf k})\mathcal{I}^{-1} &=& \tau^{z}H^{Pmmm}_{\rm HOTI}(-{\bf k})\tau^{z},\nonumber \\
C_{2x}H^{Pmmm}_{\rm HOTI}({\bf k})C_{2x}^{-1} &=& \sigma^{x}H^{Pmmm}_{\rm HOTI}(C_{2x}{\bf k})\sigma^{x}, \nonumber \\
C_{2y}H^{Pmmm}_{\rm HOTI}({\bf k})C_{2y}^{-1} &=& \mu^{z}\sigma^{y}H^{Pmmm}_{\rm HOTI}(C_{2y}{\bf k})\mu^{z}\sigma^{y}.
\label{eq:PmmmNoT}
\end{eqnarray}
Because $H^{Pmmm}_{\rm HOTI}({\bf k})$ in Eq.~(\ref{eq:D2hHOTI}) also respects the group of 3D orthogonal lattice translations, then Eq.~(\ref{eq:PmmmNoT}) implies that $H^{Pmmm}_{\rm HOTI}({\bf k})$ respects the symmetries of double MSG 47.249 $Pmmm$.  In Eq.~(\ref{eq:D2hHOTI}), the $\Delta_0$ and $\Delta_1$ terms break $\mathcal{T}$ symmetry.  The $\Delta_{0}$ term vanishes at the eight $\mathcal{I}$-invariant ${\bf k}$ points $k_{x,y,z}=0,\pi$ [Fig.~\ref{fig:HOTI-Pmmm}(a)], whereas the $\Delta_{1}$ term is generically nonzero at all values of ${\bf k}$.

To realize the helical $D_{2h}$ HOTI phase of $H^{Pmmm}_{\rm HOTI}({\bf k})$, we choose $\Delta_{0}=1,\ \Delta_1=0.3$ in Eq.~(\ref{eq:D2hHOTI}).  We have chosen a relatively small value of $\Delta_{1}$ to ensure that the band ordering remains the same as in the $\mathcal{T}$-symmetric limit in which $\Delta_{0,1}$ vanish.  Specifically, as discussed in Appendix~\ref{subsec:Pmmm} and earlier in this section, in the $\mathcal{T}$-symmetric limit, $H^{Pmmm}_{\rm HOTI}({\bf k})$ realizes a twofold-rotation-anomaly, helical, nonmagnetic HOTI phase with a nontrivial bulk mirror Chern number~\cite{ChenRotation,TMDHOTI,AshvinIndicators,ChenTCI,AshvinTCI,HOTIBismuth} indicated by the double SIs $(z_4,z_{2w,1},z_{2w,2},z_{2w,3})=(2000)$ in the Type-II double SG 47.250 $Pmmm1'$.  In Fig.~\ref{fig:HOTI-Pmmm}(b), we plot the bulk band structure of Eq.~(\ref{eq:D2hHOTI}); we emphasize that Eq.~(\ref{eq:D2hHOTI}) contains additional, extraneous (artificial) symmetries beyond those of double MSG 47.249 $Pmmm$.  Hence, the band structure in Fig.~\ref{fig:HOTI-Pmmm}(b) exhibits additional degeneracies away from the Fermi level -- such as the occupied fourfold degeneracy at $\Gamma$ -- that are not robust to symmetry-preserving perturbations.

\begin{table}[t]
\begin{tabular}{|c|c |c|c|c|c |c|c|c|c| }
\hline
\multicolumn{2}{|c|}{Bands}  & $\Gamma(000)$ & $X(\pi00)$ & $Y(0\pi0)$ & $Z(00\pi)$ & $S(\pi\pi0)$ & $T(0\pi\pi)$ & $U(\pi0\pi)$ & $R(\pi\pi\pi)$\\
\hline
\multirow{3}{*}{1-2} & Energy & -1 & -1.6 & -1.6 & -1.6 & -3.6 & -3.6 & -3.6 & -5.6\\
& $\sigma$ & $\ovl{\Gamma}_6$ & $\ovl{X}_5$ & $\ovl{Y}_5$ & $\ovl{Z}_5$ & $\ovl{S}_5$ & $\ovl{T}_5$ & $\ovl{U}_5$ & $\ovl{R}_5$ \\
& $\Delta_{\sigma}(\mathcal{I})$ & $-\xi^0$ & $\xi^0$ & $\xi^0$ & $\xi^0$ & $\xi^0$ & $\xi^0$ & $\xi^0$ & $\xi^0$\\
\hline
\multirow{3}{*}{3-4} & Energy & -1 & -0.4 & -0.4 & -0.4 & -2.4 & -2.4 & -2.4 & -4.4\\
& $\sigma$ & $\ovl{\Gamma}_6$ & $\ovl{X}_5$ & $\ovl{Y}_5$ & $\ovl{Z}_5$ & $\ovl{S}_5$ & $\ovl{T}_5$ & $\ovl{U}_5$ & $\ovl{R}_5$ \\
& $\Delta_{\sigma}(\mathcal{I})$ & $-\xi^0$ & $\xi^0$ & $\xi^0$ & $\xi^0$ & $\xi^0$ & $\xi^0$ & $\xi^0$ & $\xi^0$\\
\hline
\end{tabular}
\caption{The double-valued small irreps corresponding to the four occupied bulk bands of the helical $D_{2h}$ magnetic HOTI phase of Eq.~(\ref{eq:D2hHOTI}) [Fig.~\ref{fig:HOTI-Pmmm}(b)].  At each of the eight $\mathcal{I}$-invariant ${\bf k}$ points in MSG 47.249 $Pmmm$ [given in the notation ${\bf k}(k_{x}k_{y}k_{z})$ and obtained through~\href{http://www.cryst.ehu.es/cryst/mkvec}{MKVEC}, see Appendix~\ref{sec:MKVEC} and Fig.~\ref{fig:HOTI-Pmmm}(a)], we list the occupied band index and energy, the label of the double-valued small irrep $\sigma$ that corresponds to each pair of occupied Bloch states at ${\bf k}$ in the notation of the~\href{http://www.cryst.ehu.es/cryst/corepresentations}{Corepresentations} tool [see Appendix~\ref{sec:coreps}], and the matrix representative $\Delta_{\sigma}(\mathcal{I})$ in the basis of the $2\times 2$ Pauli matrices $\xi^{i}$.}
\label{tb:rep-HOTI-Pmmm}
\end{table}

To diagnose the topology of Eq.~(\ref{eq:D2hHOTI}), we will perform two sets of calculations.  First, we will calculate the double SIs of the four occupied bands.  Then, we will demonstrate the presence of anomalous surface and hinge states when Eq.~(\ref{eq:D2hHOTI}) is terminated in a finite, $D_{2h}$-symmetric nanorod geometry [Fig.~\ref{fig:HOTI-Pmmm}(c)].  To begin, in Table~\ref{tb:rep-HOTI-Pmmm}, we list the double-valued small irreps that correspond to the four occupied spinful Bloch eigenstates at each of the eight $\mathcal{I}$-invariant ${\bf k}$ points in MSG 47.249 $Pmmm$ [Fig.~\ref{fig:HOTI-Pmmm}(a)].  The matrix representative $\Delta_{\sigma}(\mathcal{I})$ of each two-dimensional small irrep $\sigma$ in Table~\ref{tb:rep-HOTI-Pmmm} is diagonal, indicating that each pair of Bloch states at each $\mathcal{I}$-invariant ${\bf k}$ point has two parity eigenvalues with the same sign.  From Table~\ref{tb:rep-HOTI-Pmmm}, we obtain the occupied parity eigenvalue multiplicities:
\begin{equation}
n_\Gamma^-=4,\ n_\Gamma^+=0,\ n_K^-=0, n_K^+=4 \text{ for }K=X,Y,Z,S,T,U,R.
\label{eq:D2hEigenvalues}
\end{equation}
Substituting Eq.~(\ref{eq:D2hEigenvalues}) into the double SI formulas in Type-I double MSG 47.249 $Pmmm$ [Eqs.~(\ref{eq:z4}) and~(\ref{eq:z2w})], we find that:
\begin{equation}
z_4 =\sum_{K} \frac12 n_K^-\text{ mod }2 =  \sum_{K} \frac{n_K^- - n_K^+} 4\text{ mod }4 = \frac{4-28}4\text{ mod }4 = 2,
\end{equation}
and: 
\begin{eqnarray}
z_{2w,1} &=& \sum_{K=X,S,U,R} \frac12 n_K^-\text{ mod }2 = 0, \nonumber \\
z_{2w,2} &=& \sum_{K=Y,S,T,R} \frac12 n_K^-\text{ mod }2 =0, \nonumber \\
z_{2w,3} &=& \sum_{K=Z,T,U,R} \frac12 n_K^-\text{ mod }2= 0,
\end{eqnarray}
such that the occupied bands of Eq.~(\ref{eq:D2hHOTI}) shown in Fig.~\ref{fig:HOTI-Pmmm}(b) exhibit the double SIs $(z_4,z_{2w,1},z_{2w,2},z_{2w,3})=(2000)$.

Previously, in Appendix~\ref{subsec:Pmmm}, we showed that the double SIs $(z_4,z_{2w,1},z_{2w,2},z_{2w,3})=(2000)$ in double MSG 47.249 $Pmmm$ indicate a mirror TCI phase that we designate in this work to be a helical $D_{2h}$ HOTI.  To demonstrate that Eq.~(\ref{eq:D2hHOTI}), with the parameters used to obtain Fig.~\ref{fig:HOTI-Pmmm}(b), exhibits the anomalous surface and hinge states of a $D_{2h}$ HOTI, we have performed two boundary state calculations.  First, as shown in Fig.~\ref{fig:HOTI-Pmmm}(d), we have calculated the top ($\hat{\bf z}$-normal) surface spectrum of $H^{Pmmm}_{\rm HOTI}({\bf k})$ terminated in a $z$-directed slab geometry.  The top surface of a crystal in MSG 47.249 $Pmmm$ respects the symmetries of Type-I magnetic wallpaper group $pmm$ (see Appendices~\ref{sec:34minimal} and~\ref{sec:HOTIfermionDoubling} and Refs.~\onlinecite{DiracInsulator,WiederLayers,HOTIBernevig,ConwaySymmetries,SteveMagnet}).  The slab surface spectrum in Fig.~\ref{fig:HOTI-Pmmm}(d) exhibits mirror Chern $C_{m_{x}}=2$ spectral flow, and we have additionally verified through surface-state calculations that $C_{m_{y}}=0$.  Together, this implies that the top surface exhibits two twofold Dirac cones, circumventing the fermion multiplication theorem for double magnetic wallpaper group $pmm$ derived in Appendix~\ref{sec:HOTIfermionDoubling}.  We next calculate the spectrum of an infinite, $z$-directed, $m_{x,y}$-symmetric nanorod of $H^{Pmmm}_{\rm HOTI}({\bf k})$ [Fig.~\ref{fig:HOTI-Pmmm}(e)].  We observe two pairs of hinge-localized helical modes in the nanorod spectrum in Fig.~\ref{fig:HOTI-Pmmm}(e), confirming that $H^{Pmmm}_{\rm HOTI}({\bf k})$ exhibits the higher-order spectral flow of a $D_{2h}$ HOTI.

\textit{$D_{4h}$ HOTI in double MSG 123.339 $P4/mmm$} -- We will next analyze the helical magnetic HOTI phase protected by the symmetries of double MPG 15.1.53 $4/mmm$ [$D_{4h}$] (see Appendices~\ref{sec:siteSymmetry} and~\ref{sec:magWannier} and Refs.~\onlinecite{ShubnikovMagneticPoint,BilbaoPoint,PointGroupTables,MagneticBook,EvarestovBook,EvarestovMEBR,BCS1,BCS2,BCSMag1,BCSMag2,BCSMag3,BCSMag4}), which we term the \emph{$D_{4h}$ HOTI}.  As discussed in Appendix~\ref{subsec:P4/mmm}, the double SIs $(z_8,z_{4m,\pi}^-,z_{2w,1})=(400)$ in double MSG 123.339 $P4/mmm$ either indicate a mirror TCI with mirror Chern number $C_{m_{z}}\text{ mod }8=4$, or indicate a helical $D_{4h}$ HOTI phase in which half of the $z$-projecting mirror planes (\emph{e.g.} the $\{m_{x\pm y}|{\bf 0}\}$-invariant planes) exhibit $C_{m}\text{ mod }4=2$, the other half (\emph{e.g.} the $\{m_{x,y}|{\bf 0}\}$-invariant planes) exhibit $C_{m}\text{ mod }4=0$, and $C_{m_{z}}=0$ [see Fig.~\ref{fig:SS}(b)].  To construct the helical $D_{4h}$ HOTI phase, we first superpose two copies of the 3D TI phase of Eq.~(\ref{eq:TRS-TI1}), but crucially, in a manner in which the two 3D TIs are formed from different orbital hybridization [\emph{e.g.} $s-p_{z}$ and $s-f_{xyz}$].  As we will see, this implies that the two superposed 3D TIs exhibit different valence $C_{4z}$ eigenvalues (see Ref.~\onlinecite{HingeSM} for closely related discussions of orbital hybridization and anomalous corner modes in 2D TIs and 3D Dirac semimetals).  We next add perturbative couplings to break $\mathcal{T}$ symmetry, resulting in the 3D Hamiltonian:
\begin{equation}
H^{P4/mmm}_{\rm HOTI}({\bf k}) = \mu^{0}H_{\rm TI}({\bf k}) + 
\Delta_0  (\mu^x+\mu^y) (2\tau^y 
    -\sigma^z\sin k_z   )(\cos k_x - \cos k_y)
    + \Delta_1 \mu^z,
\label{eq:D4hHOTI}
\end{equation}
employing the notation detailed in the text following Eq.~(\ref{eq:D2hHOTI}).  $H^{P4/mmm}_{\rm HOTI}({\bf k})$ respects the symmetries of double MPG 15.1.53 $4/mmm$ [$D_{4h}$], whose generating elements are represented through the action:
\begin{eqnarray}
\mathcal{I}H^{P4/mmm}_{\rm HOTI}({\bf k})\mathcal{I}^{-1} &=& \mu^{z}\tau^{z}H^{P4/mmm}_{\rm HOTI}(-{\bf k})\mu^{z}\tau^{z},\ \nonumber \\
C_{4z}H^{P4/mmm}_{\rm HOTI}({\bf k})C_{4z}^{-1} &=& \mu^{z} e^{-i\frac{\pi}4 \sigma^z}H^{P4/mmm}_{\rm HOTI}(C_{4z}{\bf k})\mu^{z} e^{i\frac{\pi}4 \sigma^z}, \nonumber \\
C_{2x}H^{P4/mmm}_{\rm HOTI}({\bf k})C_{2x}^{-1} &=& \sigma^{x}H^{P4/mmm}_{\rm HOTI}(C_{2x}{\bf k})\sigma^{x}.
\label{eq:P4mmmNoT}
\end{eqnarray}
Because $H^{P4/mmm}_{\rm HOTI}({\bf k})$ in Eq.~(\ref{eq:D4hHOTI}) also respects the group of 3D tetragonal lattice translations, then Eq.~(\ref{eq:P4mmmNoT}) implies that $H^{P4/mmm}_{\rm HOTI}({\bf k})$ respects the symmetries of double MSG 123.339 $P4/mmm$.  In Eq.~(\ref{eq:D4hHOTI}), the $\Delta_{0}$ term breaks $\mathcal{T}$ symmetry, and the $\Delta_{1}$ term breaks the extraneous exchange symmetry represented by $\mu^{x} + \mu^{y}$ between the two superposed 3D TIs at all ${\bf k}$ points.

\begin{figure}[t]
\centering
\includegraphics[width=0.9\linewidth]{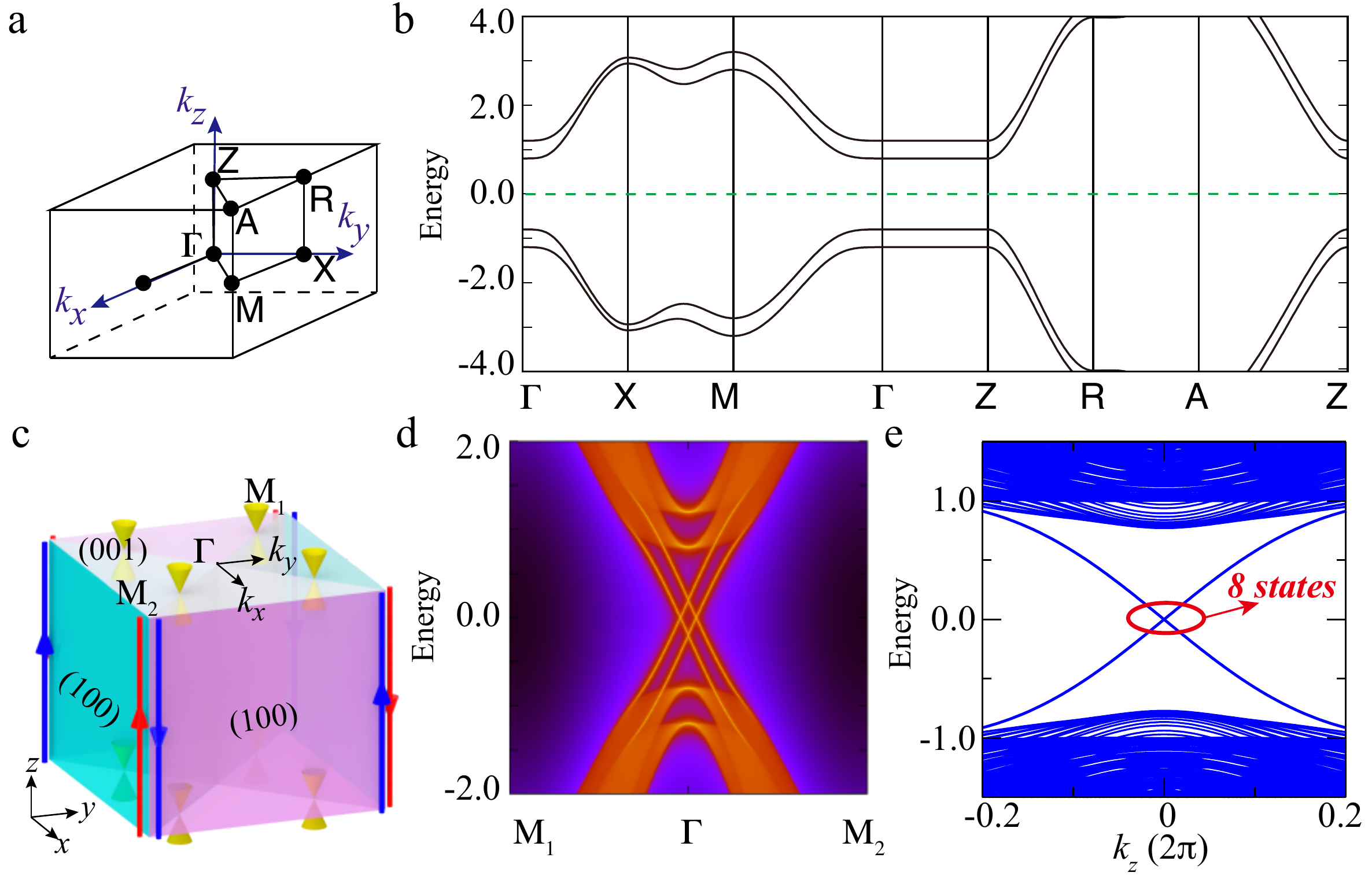}
\caption{Surface and hinge states of the helical magnetic $D_{4h}$ HOTI phase in double MSG 123.339 $P4/mmm$.  (a) The bulk BZ.  (b) The bulk band structure obtained from Eq.~(\ref{eq:D4hHOTI}) with $\Delta_{0}=0.5$ and $\Delta_{1}=0.2$.  (c) Schematic of the top ($\hat{\bf z}$-normal) surface states and nanorod hinge states.  The top surface of the square nanorod in (c) respects the symmetries of Type-I double magnetic wallpaper group $p4m$, and the hinges respect the symmetries of frieze groups that contain either $\{m_{x\pm y}|0\}$ (see Appendices~\ref{sec:34minimal} and~\ref{sec:HOTIfermionDoubling} and Refs.~\onlinecite{DiracInsulator,WiederLayers,HOTIBernevig,ConwaySymmetries,SteveMagnet}).  (d) The top surface spectrum plotted along $k_{x-y}$, obtained from surface Green's functions calculated for the model in (b) terminated in a $z$-directed slab geometry.  In (d), the surface bands exhibit mirror Chern $C_{m_{x+y}}=2$ spectral flow.  We have verified through surface-state calculations that the $C_{4z}$-related slab surface spectrum along $k_{x+y}$ also exhibits $C_{m_{x-y}}=2$ spectral flow, that the surface spectrum along $k_{x,y}$ exhibits trivial $C_{m_{y,x,}}=0$ spectral flow, and that $C_{m_{z}}=0$.  Together, this implies that the top surface exhibits four twofold Dirac cones, circumventing the fermion multiplication theorem for double magnetic wallpaper group $p4m$ derived in Appendix~\ref{sec:HOTIfermionDoubling}, and implies that the bulk is a $D_{4h}$ HOTI.  (e) The spectrum of an infinite, $C_{4z}$- and $m_{x\pm y}$-symmetric nanorod of the model in (b) features four pairs of hinge-localized helical modes (eight total hinge states), demonstrating that the model in (b) exhibits higher-order spectral flow.}
\label{fig:HOTI-P4mmm}
\end{figure}

To realize the helical $D_{4h}$ HOTI phase of $H^{P4/mmm}_{\rm HOTI}({\bf k})$, we choose $\Delta_{0}=0.5$ and $\Delta_{1}=0.2$ in Eq.~(\ref{eq:D4hHOTI}).  We have chosen a relatively small value of $\Delta_{1}$ to ensure that the band ordering remains the same as in the $\mathcal{T}$-symmetric limit in which $\Delta_{0}$ vanishes.  Specifically, as discussed in Appendix~\ref{subsec:P4/mmm}, in the $\mathcal{T}$-symmetric limit, $H^{P4/mmm}_{\rm HOTI}({\bf k})$ realizes the same fourfold-rotation-anomaly, helical, nonmagnetic HOTI phase indicated by the double SIs in $(z_8,z_{4m,\pi}^-,z_{2w,1})=(400)$ in the Type-II double SG 123.340 $P4/mmm1'$ as a tetragonal supercell of the well-studied TCI SnTe~\cite{ChenRotation,AshvinTCI,ChenTCI,HOTIBernevig,HOTIChen,HsiehTCI,TeoFuKaneTCI}.  In Fig.~\ref{fig:HOTI-P4mmm}(b), we plot the bulk band structure of Eq.~(\ref{eq:D4hHOTI}).

\begin{table}[t]
\begin{tabular}{|c|c |c|c|c|c| }
\hline
\multicolumn{2}{|c|}{Bands}  & $\Gamma (000)$ & $Z (00\pi)$ & $M(\pi\pi0)$ & $A(\pi\pi\pi)$ \\
\hline
\multirow{5}{*}{1-2} 
& Energy & -1.2                         & -1.2                          & -3.2                           & -5.2 \\
& $\sigma$  & $\ovl{\Gamma}_6$             & $\ovl{Z}_8$                   & $\ovl{M}_8$                    & $\ovl{A}_8$ \\
& $\Delta_{\sigma}(\mathcal{I})$ & $\xi^0$                   & $-\xi^0$                   & $-\xi^0$                    & $-\xi^0$ \\
& $\Delta_{\sigma}(C_{4z})$  & $e^{-i\frac{3\pi}4\xi^z}$ & $e^{-i\frac{3\pi}4\xi^z}$  & $e^{-i\frac{3\pi}4\xi^z}$   & $e^{-i\frac{3\pi}4\xi^z}$\\
& $\Delta_{\sigma}(m_z)$  & $i\xi^z$                  & $-i\xi^z$                  & $-i\xi^z$                   & $-i\xi^z$
\\
\hline
\multirow{5}{*}{3-4} 
& Energy & -0.8                         & -0.8                        & -2.8                           & -4.8 \\
& $\sigma$  & $\ovl{\Gamma}_9$             & $\ovl{Z}_7$                 & $\ovl{M}_7$                    & $\ovl{A}_7$ \\
& $\Delta_{\sigma}(\mathcal{I})$ & $-\xi^0$                  & $\xi^0$                  & $\xi^0$                     & $\xi^0$ \\
& $\Delta_{\sigma}(C_{4z})$  & $e^{-i\frac{\pi}4\xi^z}$  & $e^{-i\frac{\pi}4\xi^z}$ & $e^{-i\frac{\pi}4\xi^z}$    & $e^{-i\frac{\pi}4\xi^z}$ \\
& $\Delta_{\sigma}(m_z)$  & $i\xi^z$                  & $-i\xi^z$                & $-i\xi^z$                   & $-i\xi^z$\\
\hline
\end{tabular}
\begin{tabular}{|c|c |c|c| }
\hline
\multicolumn{2}{|c|}{Bands} & $X(0\pi0)$ & $R(0\pi\pi)$ \\
\hline
\multirow{5}{*}{1-2} 
& Energy & -3.07        & -4.27 \\
& $\sigma$  & $\ovl{X}_6$  & $\ovl{R}_6$ \\
& $\Delta_{\sigma}(\mathcal{I})$  & $-\xi^0$   & $-\xi^0$ \\
& $\Delta_{\sigma}(C_{2z})$  & $-i\xi^z$ & $-i\xi^z$  \\
& $\Delta_{\sigma}(m_z)$  & $i\xi^z$  & $i\xi^z$ \\
\hline
\multirow{5}{*}{3-4} 
& Energy & -2.94    & -3.98 \\
& $\sigma$  & $\ovl{X}_5$ & $\ovl{R}_5$ \\
& $\Delta_{\sigma}(\mathcal{I})$ & $\xi^0$    & $\xi^0$ \\
& $\Delta_{\sigma}(C_{2z})$  & $-i\xi^z$ & $-i\xi^z$\\
& $\Delta_{\sigma}(m_z)$  & $-i\xi^z$ & $-i\xi^z$\\
\hline
\end{tabular}
\caption{The double-valued small irreps corresponding to the four occupied bulk bands of the helical $D_{4h}$ magnetic HOTI phase of Eq.~(\ref{eq:D4hHOTI}) [Fig.~\ref{fig:HOTI-P4mmm}(b)].  At one ${\bf k}$ point in each of the six maximal momentum stars in MSG 123.339 $P4/mmm$ [given in the notation ${\bf k}(k_{x}k_{y}k_{z})$ and obtained through~\href{http://www.cryst.ehu.es/cryst/mkvec}{MKVEC}, see Appendix~\ref{sec:MKVEC} and Fig.~\ref{fig:HOTI-P4mmm}(a)], we list the occupied band index and energy, the label of the double-valued small irrep $\sigma$ that corresponds to each pair of occupied Bloch states at ${\bf k}$ in the notation of the~\href{http://www.cryst.ehu.es/cryst/corepresentations}{Corepresentations} tool [see Appendix~\ref{sec:coreps}], and the matrix representatives $\Delta_{\sigma}(h)$ of the representative unitary symmetries $h$ of the little group $G_{\bf k}$ [see Eq.~(\ref{eq:CorepCoset}) and the surrounding text] in the basis of the $2\times 2$ Pauli matrices $\xi^{i}$.}
\label{tb:rep-HOTI-P4/mmm}
\end{table}

To diagnose the topology of Eq.~(\ref{eq:D4hHOTI}), we will perform two sets of calculations.  First, we will calculate the double SIs of the four occupied bands.  Then, we will demonstrate the presence of anomalous surface and hinge states when Eq.~(\ref{eq:D4hHOTI}) is terminated in a finite, $D_{4h}$-symmetric nanorod geometry [Fig.~\ref{fig:HOTI-P4mmm}(c)].  To begin, in Table~\ref{tb:rep-HOTI-P4/mmm}, we list the double-valued small irreps that correspond to the four occupied spinful Bloch eigenstates at the six high-symmetry ${\bf k}$ points shown in Fig.~\ref{fig:HOTI-P4mmm}(a).  From the matrix representative $\Delta_{\sigma}(h)$ of each two-dimensional small irrep for each of the representative unitary symmetries $h$ of the little group $G_{\bf k}$ [\emph{e.g.} $C_{4z}$ and $\mathcal{I}$, see Eq.~(\ref{eq:CorepCoset}) and the surrounding text], we may infer the symmetry eigenvalues of the four occupied bands.  Using the matrix representatives in Table~\ref{tb:rep-HOTI-P4/mmm}, we then compute the auxiliary variables [see Eq.~(\ref{eq:nj-z8}) and the surrounding text]:
\begin{eqnarray}
n^{\frac32,+} &=& \sum_{K=\Gamma,M,Z,A} n^{\frac32,+}_K + \sum_{K=X,R} n^{\frac12,+} = 1 + 4 = 5, \nonumber \\
n^{\frac32,-} &=& \sum_{K=\Gamma,M,Z,A} n^{\frac32,-}_K + \sum_{K=X,R} n^{\frac12,-} = 3 + 4 = 7, \nonumber \\
n^{\frac12,+} &=& \sum_{K=\Gamma,M,Z,A} n^{\frac12,+}_K + \sum_{K=X,R} n^{\frac12,+} = 1 + 4 = 5, \nonumber \\
n^{\frac12,-} &=& \sum_{K=\Gamma,M,Z,A} n^{\frac12,-}_K + \sum_{K=X,R} n^{\frac12,-} = 3 + 4 = 7,
\label{eq:tempP4mmm}
\end{eqnarray}
where $n_{K}^{j,\pm}$ is the number of occupied states with the $C_{4z}$ eigenvalues $e^{-i\frac{\pi}2 j}$ and the parity ($\mathcal{I}$) eigenvalues $\pm 1$ at $K$ [which is only well-defined at the four $C_{4z}$-invariant momenta $K=\Gamma,M,Z,A$, see Fig.~\ref{fig:HOTI-P4mmm}(a)].  Additionally, in Eq.~(\ref{eq:tempP4mmm}), $n_{K}^{\frac12,\pm}$ is the number of occupied states with the $C_{2z}$ eigenvalues $-i$ and the parity eigenvalues $\pm 1$ at the points $K=X,R$.  Substituting Eq.~(\ref{eq:tempP4mmm}) into the double SI formula for $z_{8}$ in Type-I double MSG 123.339 $P4/mmm$ [Eq.~(\ref{eq:z8})], we obtain:
\begin{equation}
z_8 = \frac{3 (n^{\frac32,+} - n^{\frac32,-}) - (n^{\frac12,+} - n^{\frac12,-})}2\text{ mod }8
=\frac{3\times(5-7) - (5-7)}2\text{ mod }8 = 4.
\end{equation}

To complete the double SI calculation, we must also determine the values of $z_{4m,\pi}^-$ and $z_{2w,1}$ (see Appendix~\ref{subsec:P4/mmm}).  To compute $z_{4m,\pi}^-$ and $z_{2w,1}$, we first use Table~\ref{tb:rep-HOTI-P4/mmm} to calculate the number of $C_{4z}$ eigenvalues at $Z$ and $A$ in each mirror sector:
\begin{eqnarray}
n_{Z}^{\frac12,-i} &=& 1,\;
n_{Z}^{-\frac12,-i} = 0,\;
n_{Z}^{\frac32,-i} = 1,\;
n_{Z}^{-\frac32,-i} = 0, \nonumber \\
n_{A}^{\frac12,-i} &=& 1,\;
n_{A}^{-\frac12,-i} = 0,\;
n_{A}^{\frac32,-i} = 1,\;
n_{A}^{-\frac32,-i} = 0,
\label{eq:tempP4mmm2}
\end{eqnarray}
as well as the number of $C_{2z}$ eigenvalues at $R$ in each mirror sector:
\begin{equation}
n_{R}^{\frac12,-i} = 
n_{R}^{-\frac12,-i} = 1.
\label{eq:tempP4mmm3}
\end{equation}
From Eqs.~(\ref{eq:tempP4mmm2}) and~(\ref{eq:tempP4mmm3}), we then compute $z_{4m,\pi}^-$ [Eq.~(\ref{eq:z4mp-})]:
\begin{align}
z_{4m,\pi}^- =& \sum_{K=Z,A} \pare{-\frac12 n^{\frac12,-i}_K + \frac12 n^{-\frac12,-i}_K
-\frac32 n^{\frac32,-i}_K + \frac32 n^{-\frac32,-i}_K}
+ n^{\frac12,-i}_R - n^{-\frac12,-i}_R\text{ mod }4 \nono\\
=& -\frac12(1+1) + \frac12(0+0) - \frac32 (1+1) + \frac32(0+0) + 1 -1\text{ mod }4= 0.
\end{align}
Lastly, using Eqs.~(\ref{eq:tempP4mmm}),~(\ref{eq:tempP4mmm2}), and~(\ref{eq:tempP4mmm3}), we compute $z_{2w,1}$ [Eq.~(\ref{eq:z2})]:
\begin{equation}
z_{2w,1} = \sum_{K=X',R',M,A} \frac12 n_K^- \text{ mod }2 = \frac12(2 + 2 + 2 + 2)\text{ mod }2 =0,
\label{eq:c4HOTIWeak}
\end{equation}
where $X' = C_{4z}^{-1}X$ and $R' = C_{4z}^{-1}R$.  Eq.~(\ref{eq:c4HOTIWeak}) implies that the occupied bands of Eq.~(\ref{eq:D4hHOTI}) shown in Fig.~\ref{fig:HOTI-P4mmm}(b) exhibit the double SIs $(z_8,z_{4m,\pi}^-,z_{2w,1})=(400)$.

Previously, in Appendix~\ref{subsec:P4/mmm}, we showed that the double SIs $(z_8,z_{4m,\pi}^-,z_{2w,1})=(400)$ in double MSG 123.339 $P4/mmm$ either indicate a mirror TCI with mirror Chern number $C_{m_{z}}\text{ mod }8=4$, or indicate a helical $D_{4h}$ HOTI phase in which half of the $z$-projecting mirror planes (\emph{e.g.} the $\{m_{x\pm y}|{\bf 0}\}$-invariant planes) exhibit $C_{m}\text{ mod }4=2$, the other half (\emph{e.g.} the $\{m_{x,y}|{\bf 0}\}$-invariant planes) exhibit $C_{m}\text{ mod }4=0$, and $C_{m_{z}}=0$ [see Fig.~\ref{fig:SS}(b)].  To demonstrate that Eq.~(\ref{eq:D4hHOTI}), with the parameters used to obtain Fig.~\ref{fig:HOTI-P4mmm}(b), is a $D_{4h}$ HOTI, we have performed two boundary state calculations.  First, as shown in Fig.~\ref{fig:HOTI-P4mmm}(d), we have calculated the top ($\hat{\bf z}$-normal) surface spectrum of $H^{P4/mmm}_{\rm HOTI}({\bf k})$ terminated in a $z$-directed slab geometry.  The top surface of a crystal in double MSG 123.339 $P4/mmm$ respects the symmetries of Type-I magnetic wallpaper group $p4m$ (see Appendices~\ref{sec:34minimal} and~\ref{sec:HOTIfermionDoubling} and Refs.~\onlinecite{DiracInsulator,WiederLayers,HOTIBernevig,ConwaySymmetries,SteveMagnet}).  The slab surface spectrum in Fig.~\ref{fig:HOTI-P4mmm}(d) exhibits four twofold Dirac cones, circumventing the fermion multiplication theorem for double magnetic wallpaper group $p4m$ derived in Appendix~\ref{sec:HOTIfermionDoubling}.  We next calculate the spectrum of an infinite, $z$-directed, $C_{4z}$- and $m_{x\pm y}$-symmetric nanorod of $H^{P4/mmm}_{\rm HOTI}({\bf k})$ [Fig.~\ref{fig:HOTI-P4mmm}(e)].  We observe four pairs of hinge-localized helical modes in the nanorod spectrum in Fig.~\ref{fig:HOTI-P4mmm}(e), confirming that $H^{P4/mmm}_{\rm HOTI}({\bf k})$ exhibits the higher-order spectral flow of a $D_{4h}$ HOTI.

\textit{$D_{6h}$ HOTI in double MSG 191.233 $P6/mmm$} --  Finally, we will now analyze the helical magnetic HOTI phase protected by the symmetries of double MPG 27.100 $6/mmm$ [$D_{6h}$] (see Appendices~\ref{sec:siteSymmetry} and~\ref{sec:magWannier} and Refs.~\onlinecite{ShubnikovMagneticPoint,BilbaoPoint,PointGroupTables,MagneticBook,EvarestovBook,EvarestovMEBR,BCS1,BCS2,BCSMag1,BCSMag2,BCSMag3,BCSMag4}), which we term the \emph{$D_{6h}$ HOTI}.  As discussed in Appendix~\ref{subsec:P6/mmm}, the double SIs $(z_{12},z_{6m,\pi}^+)=(60)$ in double MSG 191.233 $P6/mmm$ either indicate a mirror TCI with mirror Chern number $C_{m_{z}}\text{ mod }12=6$, or indicate a helical $D_{6h}$ HOTI in which half of the $z$-projecting mirror planes (\emph{e.g.} the $\{m_{x}|{\bf 0}\}$-, $\{C_{6z}m_{x}C_{6z}^{-1}|{\bf 0}\}$-, and $\{C_{6z}^{-1}m_{x}C_{6z}|{\bf 0}\}$-invariant planes) exhibit $C_{m}\text{ mod }4=2$, the other half (\emph{e.g.} the $\{m_{y}|{\bf 0}\}$-, $\{C_{6z}m_{y}C_{6z}^{-1}|{\bf 0}\}$-, and $\{C_{6z}^{-1}m_{y}C_{6z}|{\bf 0}\}$-invariant planes) exhibit $C_{m}\text{ mod }4=0$, and $C_{m_{z}}=0$ [see Fig.~\ref{fig:SS}(c)].

\begin{figure}[t]
\centering
\includegraphics[width=0.9\linewidth]{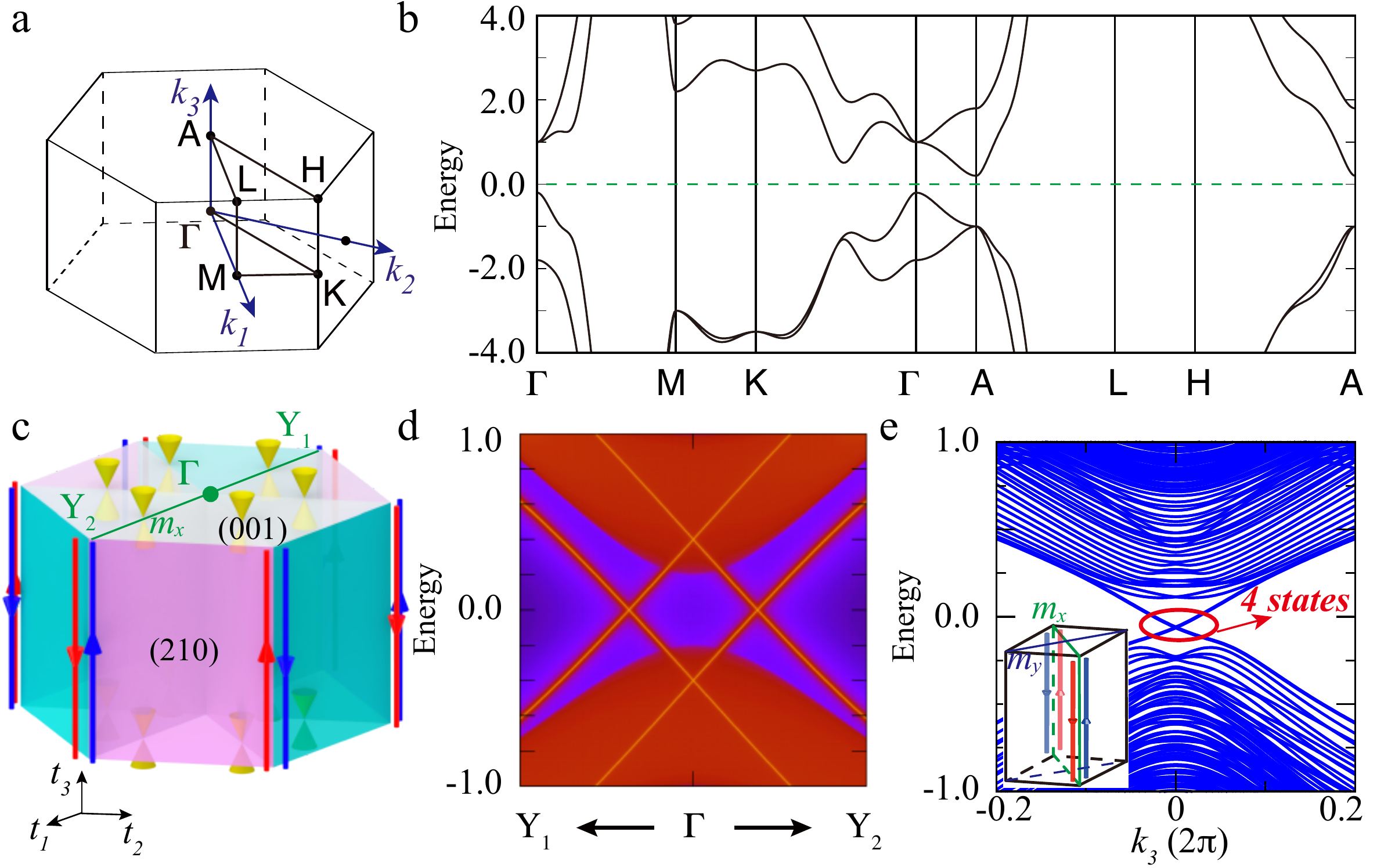}
\caption{Surface and hinge states of the helical magnetic $D_{6h}$ HOTI phase in double MSG 191.233 $P6/mmm$.  (a) The bulk BZ.  (b) The bulk band structure obtained from Eqs.~(\ref{eq:D6hHOTI}) and~(\ref{eq:D6hHOTIdelta0}) with $\Delta_0=2$ and $\Delta_1=0.4$.  We note that Eq.~(\ref{eq:D6hHOTI}) contains additional symmetries beyond those of double MSG 191.233 $P6/mmm$, such that the band structure in (b) exhibits additional degeneracies away from the Fermi level -- such as the unoccupied fourfold degeneracy at $\Gamma$ -- that are not robust to symmetry-preserving perturbations.  (c) Schematic of the top ($\hat{\bf z}$-normal) surface states and nanorod hinge states.  The top surface of the hexagonal nanorod in (c) respects the symmetries of Type-I double magnetic wallpaper group $p6m$, and the hinges respect the symmetries of frieze groups with mirror lines parallel to the hinge translation direction (see Appendices~\ref{sec:34minimal} and~\ref{sec:HOTIfermionDoubling} and Refs.~\onlinecite{DiracInsulator,WiederLayers,HOTIBernevig,ConwaySymmetries,SteveMagnet}).  (d) The top surface spectrum plotted along $k_{y}=0$ [see Eqs.~(\ref{eq:HexagonalLatVecForHOTIs}) and~(\ref{eq:HexagonalKVecForHOTIs})] obtained from surface Green's functions calculated for the model in (b) terminated in a $\hat{\bf z}$- (${\bf t}_{3}$-) normal slab geometry.  In (d), the surface bands exhibit mirror Chern $C_{m_{x}}=2$ spectral flow.  We have verified through surface-state calculations that the $C_{6z}$-related slab surface spectrum along the $C_{6}m_{x}C_{6}^{-1}$- and $C_{6}^{-1}m_{x}C_{6}$-invariant surface mirror lines in $p6m$ [see Fig.~\ref{fig:fermionMultiplication}(c)] also exhibits mirror Chern $C_{m}=2$ spectral flow, that the surface spectrum along the other three surface mirror lines [\emph{i.e.} the $m_{y}$-, $C_{6}m_{y}C_{6}^{-1}$-, and $C_{6}^{-1}m_{y}C_{6}$-invariant lines] exhibits trivial $C_{m}=0$ spectral flow, and that $C_{m_{z}}=0$.  Together, this implies that the top surface exhibits six twofold Dirac cones, circumventing the fermion multiplication theorem for double magnetic wallpaper group $p6m$ derived in Appendix~\ref{sec:HOTIfermionDoubling}, and implies that the bulk is a $D_{6h}$ HOTI.  (e) Unlike in Figs.~\ref{fig:HOTI-Pmmm}(e) and~\ref{fig:HOTI-P4mmm}(e), it is numerically simpler to implement a hinge-state calculation in which the bulk insulator in (b) is cut into a $z$-directed nanorod that preserves $\mathcal{I}$ and $m_{x,y}$ symmetries, but does not preserve $C_{3z}$ and $C_{6z}$ rotation symmetries [see the inset panel in (e)].  In (e), we show the spectrum of a $z$-directed, $m_{x,y}$-symmetric nanorod of the model in (b); the nanorod in (e) features two pairs of helical hinge states along the $m_{x}$-invariant hinges (four total hinge states), and does not exhibit any other states crossing the gap.  The nanorod spectrum in (e) implies that a $D_{6h}$-symmetric nanorod of the model in (b) [\emph{i.e.} a nanorod with $m_{x}$, $C_{6z}m_{x}C_{6z}^{-1}$, and $C_{6z}^{-1}m_{x}C_{6z}$ symmetries], would feature six pairs of hinge-localized helical modes [twelve total hinge states], demonstrating that the model in (b) exhibits the higher-order spectral flow of a $D_{6h}$ helical magnetic HOTI.}
\label{fig:HOTI-P6mmm}
\end{figure}

To construct the helical $D_{6h}$ HOTI phase, we begin by introducing the hexagonal lattice vectors:
\begin{equation}
\tt_1=(0,-1,0),\ \tt_2=(\sqrt{3}/2,1/2,0),\ \tt_3=(0,0,1),
\label{eq:HexagonalLatVecForHOTIs}
\end{equation}
and reciprocal lattice vectors:
\begin{equation}
\mbf{b}_1=({\sqrt3}/3,-1,0),\ \mbf{b}_2=({2\sqrt3}/3,0,0),\ \mbf{b}_3=(0,0,1).
\label{eq:HexagonalKVecForHOTIs}
\end{equation}
We define the first BZ to consist of the points $\kk=\sum_{i=1,2,3} k_i \mbf{b}_i $, $k_i \in [-\pi, \pi)$ [see Fig.~\ref{fig:HOTI-P6mmm}(a)].

Next, we introduce a model for a 3D TI with hexagonal lattice vectors:
\begin{equation}
H_{\rm TI}^{P6/mmm1'}({\bf k})= \tau^{z}M({\bf k}) + \tau^{x}\bar{\sigma}^{1}\sin(2k_1+k_2) + \tau^{x}\bar{\sigma}^{2}\sin(k_2-k_1) + \tau^{x}\bar{\sigma}^{3}\sin(k_1+2k_2) + \tau^{x}\sigma^{z}\sin(k_{3}),
\label{eq:hexagonalTIforP6mmm}
\end{equation}
where we have employed the notation detailed in the text following Eq.~(\ref{eq:TRS-TI1}), and where:
\begin{equation}
M({\bf k})=3-\sum_{i=1,2,3} \cos(k_i)-\cos(k_1+k_2).
\end{equation}  
In Eq.~(\ref{eq:hexagonalTIforP6mmm}), we have employed a canonical Pauli matrix transformation given by:
\begin{equation}
\bar{\sigma}^1=\frac{\sqrt{3}}{2}\sigma^x-{\frac{1}{2}}\sigma^y,\ \bar{\sigma}^2=\sigma^y,\ \bar{\sigma}^3=\frac{\sqrt{3}}{2}\sigma^x+{\frac{1}{2}}\sigma^y.
\end{equation}
Eq.~(\ref{eq:hexagonalTIforP6mmm}) respects $\mathcal{I}$ and spinful $\mathcal{T}$ symmetries, which are represented through the symmetry action:
\begin{equation}
\mathcal{I}H_{\rm TI}^{P6/mmm1'}({\bf k})\mathcal{I}^{-1} = \tau^{z}H_{\rm TI}^{P6/mmm1'}(-{\bf k})\tau^z,\ \mathcal{T}H_{\rm TI}^{P6/mmm1'}({\bf k})\mathcal{T}^{-1} = \sigma^{y}[H^{P6/mmm1'}_{\rm TI}(-{\bf k})]^{*}\sigma^{y}.
\label{eq:ITsymmetryD6h}
\end{equation}
As was done for the $D_{4h}$ HOTI earlier in this section, we next superpose two copies of the 3D TI phase of Eq.~(\ref{eq:hexagonalTIforP6mmm}), but again in a manner in which the two 3D TIs are formed from different orbital hybridization, such that the occupied bands of the two 3D TIs exhibit different $C_{6z}$ and $C_{3z}$ eigenvalues [see Ref.~\onlinecite{HingeSM} and the text preceding Eq.~(\ref{eq:D4hHOTI})].  We then add perturbative couplings to break $\mathcal{T}$ symmetry, resulting in the 3D Hamiltonian:
\begin{equation}
H_{\rm HOTI}^{P6/mmm}({\bf k}) = \mu^{0}H_{\rm TI}^{P6/mmm1'}({\bf k}) + (\mu^x+\mu^y)[(\tau^x+\tau^y) + \frac{1}{2}\sigma^z\sin(k_3)]f({\bf k})  + \Delta_1 \mu^z (\tau^z+\tau^0), 
\label{eq:D6hHOTI}
\end{equation}
where we have employed the notation detailed in the text following Eq.~(\ref{eq:D2hHOTI}), and where:
\begin{equation}
f({\bf k})=\Delta_0 [\sin(k_2-k_1)+\sin(2k_1+k_2)-\sin(k_1+2k_2)].
\label{eq:D6hHOTIdelta0}
\end{equation}
$H_{\rm HOTI}^{P6/mmm}({\bf k})$ respects the symmetries of double MPG 27.1.100 $6/mmm$ [$D_{6h}$], whose generating elements are represented through the action:
\begin{eqnarray}
\mathcal{I}H_{\rm HOTI}^{P6/mmm}({\bf k})\mathcal{I}^{-1} &=& \tau^{z}H_{\rm HOTI}^{P6/mmm}(-{\bf k})\tau^{z}, \nonumber \\
C_{6z}H_{\rm HOTI}^{P6/mmm}({\bf k})C_{6z}^{-1} &=& \mu^{z}e^{-i\frac{\pi}6 \sigma^z}H_{\rm HOTI}^{P6/mmm}(C_{6z}{\bf k})\mu^{z}e^{i\frac{\pi}6 \sigma^z}, \nonumber \\
C_{2y}H_{\rm HOTI}^{P6/mmm}({\bf k})C_{2y}^{-1} &=& \mu^{z}\sigma^{x}H_{\rm HOTI}^{P6/mmm}(C_{2y}{\bf k})\mu^{z}\sigma^{x}.
\label{eq:P6mmmNoT}
\end{eqnarray}
Because $H_{\rm HOTI}^{P6/mmm}({\bf k})$ in Eq.~(\ref{eq:D6hHOTIdelta0}) also respects the group of 3D hexagonal lattice translations, then Eq.~(\ref{eq:P6mmmNoT}) implies that $H_{\rm HOTI}^{P6/mmm}({\bf k})$ respects the symmetries of double MSG 191.233 $P6/mmm$.  In Eqs.~(\ref{eq:D6hHOTI}) and~(\ref{eq:D6hHOTIdelta0}), the $\Delta_{0}$ term breaks $\mathcal{T}$ symmetry, and the $\Delta_{1}$ term breaks the extraneous exchange symmetry represented by $\mu^{x} + \mu^{y}$ between the two superposed hexagonal 3D TIs in the $\tau^{+}=\frac{1}{2}[\tau^{z}+\tau^{0}]$ subspace at all ${\bf k}$ points.

To realize the helical $D_{6h}$ HOTI phase of $H_{\rm HOTI}^{P6/mmm}({\bf k})$, we choose $\Delta_0=2$ and $\Delta_1=0.4$ in Eqs.~(\ref{eq:D6hHOTI}) and~(\ref{eq:D6hHOTIdelta0}).  We have chosen a relatively small value of $\Delta_{1}$ to ensure that the band ordering remains the same as in the $\mathcal{T}$-symmetric limit in which $\Delta_{0}$ vanishes.  Specifically, as discussed in Appendix~\ref{subsec:P6/mmm}, in the $\mathcal{T}$-symmetric limit, $H_{\rm HOTI}^{P6/mmm}({\bf k})$ realizes a sixfold-rotation-anomaly, helical, nonmagnetic HOTI phase~\cite{ChenRotation,ChenTCI,AshvinTCI} indicated by the double SIs $(z_{12},z_{6m,\pi}^+)=(60)$ in the Type-II double SG 191.234 $P6/mmm1'$.  In Fig.~\ref{fig:HOTI-P6mmm}(b), we plot the bulk band structure of Eqs.~(\ref{eq:D6hHOTI}) and~(\ref{eq:D6hHOTIdelta0}); we emphasize that Eqs.~(\ref{eq:D6hHOTI}) and~(\ref{eq:D6hHOTIdelta0}) contain additional, extraneous (artificial) symmetries beyond those of double MSG 191.233 $P6/mmm$.  Hence, the band structure in Fig.~\ref{fig:HOTI-P6mmm}(b) exhibits additional degeneracies away from the Fermi level -- such as the unoccupied fourfold degeneracy at $\Gamma$ -- that are not robust to symmetry-preserving perturbations.

\begin{table}[t]
\begin{tabular}{|c|c |c|c| }
\hline
\multicolumn{2}{|c|}{Bands}  & $\Gamma (000)$ & $A (00\pi)$ \\
\hline
\multirow{5}{*}{1-2} 
& Energy & -1.8                         & -1                  \\
& $\sigma$  & $\ovl{\Gamma}_8$             & $\ovl{A}_{11}$      \\
& $\Delta_{\sigma}(\mathcal{I})$ & $\xi^0$                   & $-\xi^0$         \\
& $\Delta_{\sigma}(C_{6z})$  & $e^{-i\frac{5\pi}6\xi^z}$ & $e^{-i\frac{5\pi}6\xi^z}$ \\
& $\Delta_{\sigma}(m_z)$  & $-i\xi^z$                 & $i\xi^z$        \\
\hline
\multirow{5}{*}{3-4} 
& Energy & -0.2                         & -1           \\
& $\sigma$  & $\ovl{\Gamma}_9$             & $\ovl{A}_{12}$  \\
& $\Delta_{\sigma}(\mathcal{I})$ & $\xi^0$                   & $-\xi^0$   \\
& $\Delta_{\sigma}(C_{6z})$  & $e^{-i\frac{\pi}6\xi^z}$  & $e^{-i\frac{\pi}6\xi^z}$ \\
& $\Delta_{\sigma}(m_z)$  & $-i\xi^z$                 & $i\xi^z$                \\
\hline
\end{tabular}
\begin{tabular}{|c|c |c|c| }
\hline
\multicolumn{2}{|c|}{Bands}  & $K (\frac{2\pi}3\frac{2\pi}30)$ & $H (\frac{2\pi}3\frac{2\pi}3 \pi)$ \\
\hline
\multirow{5}{*}{1-2} 
& Energy & -3.5                        & -5.5                  \\
& $\sigma$  & $\ovl{K}_8$                 & $\ovl{H}_{8}$      \\
& $\Delta_{\sigma}(C_{3z})$  & $e^{-i\frac{\pi}3\xi^z}$ & $e^{-i\frac{\pi}3\xi^z}$ \\
& $\Delta_{\sigma}(m_z)$  & $-i\xi^z$                & $-i\xi^z$        \\
\hline
\multirow{5}{*}{3-4} 
& Energy & -3.5                         & -5.5           \\
& $\sigma$  & $\ovl{K}_9$                  & $\ovl{H}_{9}$  \\
& $\Delta_{\sigma}(C_{3z})$  & $e^{i\frac{\pi}3\xi^z}$   & $e^{i\frac{\pi}3\xi^z}$ \\
& $\Delta_{\sigma}(m_z)$  & $-i\xi^z$                 & $-i\xi^z$                \\
\hline
\end{tabular}
\begin{tabular}{|c|c |c|c| }
\hline
\multicolumn{2}{|c|}{Bands}  & $M (\pi 00)$ & $L (\pi 0 \pi)$ \\
\hline
\multirow{5}{*}{1-2} 
& Energy & -3                          & -5                  \\
& $\sigma$  & $\ovl{M}_6$                 & $\ovl{L}_{6}$      \\
& $\Delta_{\sigma}(\mathcal{I})$ & $-\xi^0$                 & $-\xi^0$ \\
& $\Delta_{\sigma}(m_z)$  & $i\xi^z$                 & $i\xi^z$        \\
\hline
\multirow{5}{*}{3-4} 
& Energy & -3                           & -5           \\
& $\sigma$  & $\ovl{M}_6$                  & $\ovl{L}_{6}$  \\
& $\Delta_{\sigma}(\mathcal{I})$ & $-\xi^0$                  & $-\xi^0$ \\
& $\Delta_{\sigma}(m_z)$  & $i\xi^z$                  & $i\xi^z$                \\
\hline
\end{tabular}
\caption{The double-valued small irreps corresponding to the four occupied bulk bands of the helical $D_{6h}$ magnetic HOTI phase of Eq.~(\ref{eq:D6hHOTI}) [Fig.~\ref{fig:HOTI-P6mmm}(b)].  At one ${\bf k}$ point in each of the six maximal momentum stars in MSG 191.233 $P6/mmm$ [given in the notation ${\bf k}(k_{1}k_{2}k_{3})$ and obtained through~\href{http://www.cryst.ehu.es/cryst/mkvec}{MKVEC}, see Appendix~\ref{sec:MKVEC} and Fig.~\ref{fig:HOTI-P6mmm}(a)], we list the occupied band index and energy, the label of the double-valued small irrep $\sigma$ that corresponds to each pair of occupied Bloch states at ${\bf k}$ in the notation of the~\href{http://www.cryst.ehu.es/cryst/corepresentations}{Corepresentations} tool [see Appendix~\ref{sec:coreps}], and the matrix representatives $\Delta_{\sigma}(h)$ of the representative unitary symmetries $h$ of the little group $G_{\bf k}$ [see Eq.~(\ref{eq:CorepCoset}) and the surrounding text] in the basis of the $2\times 2$ Pauli matrices $\xi^{i}$.}
\label{tb:rep-HOTI-P6/mmm}
\end{table}

To diagnose the topology of Eqs.~(\ref{eq:D6hHOTI}) and~(\ref{eq:D6hHOTIdelta0}), we will perform two sets of calculations.  First, we will calculate the double SIs of the four occupied bands.  Then, we will demonstrate the presence of anomalous surface and hinge states when Eqs.~(\ref{eq:D6hHOTI}) and~(\ref{eq:D6hHOTIdelta0}) are terminated in a finite, $D_{6h}$-symmetric nanorod geometry [Fig.~\ref{fig:HOTI-P6mmm}(c)].  To begin, in Table~\ref{tb:rep-HOTI-P6/mmm}, we list the double-valued small irreps that correspond to the four occupied spinful Bloch eigenstates at the six high-symmetry ${\bf k}$ points shown in Fig.~\ref{fig:HOTI-P6mmm}(a).  From the matrix representative $\Delta_{\sigma}(h)$ of each two-dimensional small irrep for each of the representative unitary symmetries $h$ of the little group $G_{\bf k}$ [\emph{e.g.} $C_{6z}$, $C_{3z}$, and $\mathcal{I}$, see Eq.~(\ref{eq:CorepCoset}) and the surrounding text], we may infer the symmetry eigenvalues of the four occupied bands.

In Appendix~\ref{subsec:P6/mmm}, we previously expressed the double SI $z_{12}$ in terms of other double SIs in double MSGs with lower symmetry than double MSG 191.233 $P6/mmm$ [Eq.~(\ref{eq:z12})]:
\begin{equation}
z_{12} = \delta_{6m} + 3[(\delta_{6m} - z_{4})\text{ mod }4]\text{ mod }12,
\label{eq:z12forD6hHOTI}
\end{equation}
where $z_{4}$ and $\delta_{6m}$ are respectively defined in Eqs.~(\ref{eq:z4}) and~(\ref{eq:d6m}).  Using the matrix representatives in Table~\ref{tb:rep-HOTI-P6/mmm}, we first determine the parity eigenvalue multiplicities:
\begin{equation}
n_\Gamma^-=0,\ n_\Gamma^+=4,\ n_A^-=4,\ n_A^+=0,\ n_M^-=4,\ n_M^+=0,\ n_L^-=4,\ n_L^+=0.
\label{eq:tempD6hParity}
\end{equation}
In MSG 191.233 $P6/mmm$, the $M$ and $L$ points lie within multiplicity-3 momentum stars (see Appendix~\ref{sec:MKVEC} and~\href{http://www.cryst.ehu.es/cryst/mkvec}{MKVEC}); therefore, the eight $\mathcal{I}$-invariant momenta in MSG 191.233 $P6/mmm$ are given by:
\begin{equation}
k_{\mathcal{I}} = \bigg\{\Gamma,A,M,(C_{6z})M,(C_{6z})^{2}M,L,(C_{6z})L,(C_{6z})^{2}L\bigg\}.
\label{eq:tempImomenta}
\end{equation}
Eqs.~(\ref{eq:tempD6hParity}) and~(\ref{eq:tempImomenta}) imply that:
\begin{eqnarray}
z_4 &=& \sum_{K\in K_{\mathcal{I}}} \frac{n_K^--n_K^+}4\text{ mod }4  \nonumber \\
&=& \frac{n^-_\Gamma - n^+_\Gamma }{4} + \frac{n^-_A - n^+_A }{4} + 
3\frac{n^-_M - n^+_M }{4} + 3\frac{n^-_L - n^+_L }{4}\text{ mod }4 \nonumber \\
&=& -1 + 1 + 3 + 3\text{ mod }4 = 2.
\label{eq:tempz4forD6hHOTI}
\end{eqnarray}

Next, to compute $\delta_{6m}$, we use Table~\ref{tb:rep-HOTI-P6/mmm} to obtain the rotation eigenvalues in each mirror sector:
\begin{eqnarray}
n_{A}^{\frac12,i} &=& 1,\;\;
n_{A}^{-\frac12,i} = 0,\;\;
n_{A}^{\frac32,i} = 0,\;\;
n_{A}^{-\frac32,i} = 0,\;\;
n_{A}^{\frac52,i} = 1,\;\;
n_{A}^{-\frac52,i} = 0, \nonumber \\
n_{H}^{\frac12,i} &=& 1,\;\;
n_{H}^{-\frac12,i} = 1,\;\;
n_{H}^{\frac32,i} = 0,\;\;
n_{L}^{\frac12,i} = 2,\;\;
n_{L}^{-\frac12,i} = 0, \nonumber \\
n_{\Gamma}^{\frac12,-i} &=& 1,\;\;
n_{\Gamma}^{-\frac12,-i} = 0,\;\;
n_{\Gamma}^{\frac32,-i} = 0,\;\;
n_{\Gamma}^{-\frac32,-i} = 0,\;\;
n_{\Gamma}^{\frac52,-i} = 1,\;\;
n_{\Gamma}^{-\frac52,-i} = 0, \nonumber \\
n_{K}^{\frac12,-i} &=& 1,\;\;
n_{K}^{-\frac12,-i} = 1,\;\;
n_{K}^{\frac32,-i} = 0,\;\;
n_{M}^{\frac12,-i} = 0,\;\;
n_{M}^{-\frac12,-i} = 2.
\label{eq:tempAllRotationD6h}
\end{eqnarray}
Using Eqs.~(\ref{eq:d6m}) and~(\ref{eq:tempAllRotationD6h}), we then compute $\delta_{6m}$:
\begin{eqnarray}
\delta_{6m} &=& 
- \frac12 n^{\frac12,+i}_A + \frac12 n^{-\frac12,+i}_A
- \frac32 n^{\frac32,+i}_A + \frac32 n^{-\frac32,+i}_A
- \frac52 n^{\frac52,+i}_A + \frac52 n^{-\frac52,+i}_A \nonumber \\
&-& n^{\frac12,+i}_H + n^{-\frac12,+i}_H + 3n^{\frac32,+i}_H
+ \frac{3}{2} n^{\frac12,+i}_L - \frac32 n^{-\frac12,+i}_L \nonumber \\
&+& \frac12 n^{\frac12,-i}_\Gamma - \frac12 n^{-\frac12,-i}_\Gamma
+ \frac32 n^{\frac32,-i}_\Gamma - \frac32 n^{-\frac32,-i}_\Gamma
+ \frac52 n^{\frac52,-i}_\Gamma - \frac52 n^{-\frac52,-i}_\Gamma \nonumber \\
&+& n^{\frac12,-i}_K - n^{-\frac12,-i}_K - 3n^{\frac32,-i}_K
- \frac{3}{2} n^{\frac12,-i}_M + \frac32 n^{-\frac12,-i}_M\text{ mod }6 \nonumber \\
&=& (-\frac12 -\frac52 -1+1 + 3) + (\frac12 + \frac52 +1 - 1 + 3)\text{ mod }6 =0.
\label{eq:tempd6mD6h}
\end{eqnarray}
From Eqs.~(\ref{eq:z12forD6hHOTI}),~(\ref{eq:tempz4forD6hHOTI}), and~(\ref{eq:tempd6mD6h}), we next compute $z_{12}$:
\begin{equation}
z_{12} = \delta_{6m} + 3[( \delta_{6m} -z_4)\text{ mod }4]\text{ mod }12 = 0 + 3\times 2\text{ mod }12 = 6.
\label{eq:finalz12ForD6h}
\end{equation}
Lastly, to complete the calculation of the double SIs in MSG 191.233 $P6/mmm$, we compute $z_{6m,\pi}^+$ [Eq.~(\ref{eq:z6mp+})]:
\begin{align}
z_{6m,\pi}^+ =& - \frac12 n^{\frac12,+i}_A + \frac12 n^{-\frac12,+i}_A
- \frac32 n^{\frac32,+i}_A + \frac32 n^{-\frac32,+i}_A
- \frac52 n^{\frac52,+i}_A + \frac52 n^{-\frac52,+i}_A \nono\\
& -n^{\frac12,+i}_H + n^{-\frac12,+i}_H + 3n^{\frac32,+i}_H
+ \frac{3}{2} n^{\frac12,+i}_L - \frac32 n^{-\frac12,+i}_L\text{ mod }6\nono\\
=& -\frac12 -\frac52 -1+1 + 3\text{ mod }6 =0.
\label{eq:tempWeakMirrorTCID6h}
\end{align}
From Eqs.~(\ref{eq:finalz12ForD6h}) and~(\ref{eq:tempWeakMirrorTCID6h}), we determine that the occupied bands of Eqs.~(\ref{eq:D6hHOTI}) and~(\ref{eq:D6hHOTIdelta0}) shown in Fig.~\ref{fig:HOTI-P6mmm}(b) exhibit the double SIs $(z_{12},z_{6m,\pi}^+)=(60)$.

Previously, in Appendix~\ref{subsec:P6/mmm}, we showed that the double SIs $(z_{12},z_{6m,\pi}^+)=(60)$ in double MSG 191.233 $P6/mmm$ either indicate a mirror TCI with mirror Chern number $C_{m_{z}}\text{ mod }12=6$ or indicate a helical $D_{6h}$ HOTI phase in which half of the $z$-projecting mirror planes [\emph{e.g.} the $\{m_{x}|{\bf 0}\}$-, $\{C_{6z}m_{x}C_{6z}^{-1}|{\bf 0}\}$-, and $\{C_{6z}^{-1}m_{x}C_{6z}|{\bf 0}\}$-invariant planes] exhibit $C_{m}\text{ mod }4=2$, the other half  [\emph{e.g.} the $\{m_{y}|{\bf 0}\}$-, $\{C_{6z}m_{y}C_{6z}^{-1}|{\bf 0}\}$-, and $\{C_{6z}^{-1}m_{y}C_{6z}|{\bf 0}\}$-invariant planes] exhibit $C_{m}\text{ mod }4=0$, and $C_{m_{z}}=0$ [see Fig.~\ref{fig:SS}(c)].  To demonstrate that Eq.~(\ref{eq:D6hHOTI}), with the parameters used to obtain Fig.~\ref{fig:HOTI-P6mmm}(b), is a $D_{6h}$ HOTI, we have performed two boundary state calculations.  First, as shown in Fig.~\ref{fig:HOTI-P6mmm}(d), we have calculated the top surface spectrum of $H^{P6/mmm}_{\rm HOTI}({\bf k})$ terminated in a $\hat{\bf z}$- (${\bf t}_{3}$-) normal slab geometry.  The top surface of a crystal in double MSG 191.233 $P6/mmm$ respects the symmetries of Type-I magnetic wallpaper group $p6m$ (see Appendices~\ref{sec:34minimal} and~\ref{sec:HOTIfermionDoubling} and Refs.~\onlinecite{DiracInsulator,WiederLayers,HOTIBernevig,ConwaySymmetries,SteveMagnet}).  The slab surface spectrum in Fig.~\ref{fig:HOTI-P6mmm}(d) exhibits six twofold Dirac cones, circumventing the fermion multiplication theorem for double magnetic wallpaper group $p6m$ derived in Appendix~\ref{sec:HOTIfermionDoubling}.  We then calculate the spectrum of an infinite, $z$-directed, $m_{x,y}$-symmetric nanorod of $H^{P6/mmm}_{\rm HOTI}({\bf k})$, which we find to exhibit the higher-order spectral flow of a $D_{6h}$ HOTI [Fig.~\ref{fig:HOTI-P6mmm}(e)].

\clearpage

\section{Supplementary Tables}
\label{sec:supplementaryTables}

In the sections below, we will provide further supplementary tables containing additional data generated for this work.  First, in Appendix~\ref{sec:exceptionalTables}, we will provide a complete tabulation of the exceptional composite band (co)reps of the 1,651 single and double SSGs [see Appendix~\ref{sec:exceptions}].  Then, in Appendix~\ref{sec:EBCRdimension}, we will tabulate the minimum and maximum EBR dimension in each single and double SSG.  Finally, in Appendix~\ref{sec:minimalSSGTables}, we will list the minimal double SSG with the minimal double SIs on which the double SIs in each double SSGs are dependent (see Appendix~\ref{sec:minimalSIProcedure}).

\subsection{Exceptional Composite Band Coreps Induced from Maximal Site-Symmetry Groups}
\label{sec:exceptionalTables}

In this section, we provide a complete tabulation of the exceptional cases [defined in detail in Appendix~\ref{sec:exceptions}] in the 1,651 single and double SSGs in which an irreducible (co)rep of a site-symmetry group of a site in a maximal Wyckoff position does not induce an elementary band (co)rep [EBR].  For the Type-I MSGs and Type-II SGs analyzed in TQC~\cite{QuantumChemistry,Bandrep1,Bandrep2,Bandrep3,JenFragile1,BarryFragile}, the exceptional cases listed in the tables below agree with the previous tabulations performed in Refs.~\onlinecite{QuantumChemistry,Bandrep3}.  Among the tables provided in this section, there is no table of exceptional cases in the Type-II double SGs, because, as previously shown in Refs.~\onlinecite{QuantumChemistry,Bandrep3} and in Table~\ref{tb:MEBRsDouble}, there are no exceptional composite band coreps in the Type-II double SGs.

\subsubsection{Exceptional Composite Band Reps in the Type-I Single MSGs}

$\\$

\begin{longtable}{|cc|c|cc|cc|cc|c|}
\caption{
Exceptional composite band reps induced from site-symmetry irreps in the Type-I single MSGs (Appendix~\ref{sec:type1}).  In order, the columns in this table list the number of the MSG in the BNS setting and the symbol of the MSG, the letter of the maximal Wyckoff position containing ${\bf q}$ and the single-valued irrep of the site-symmetry group $G_{\bf q}$, the symbol of the MPG isomorphic to the site-symmetry group $G_{\bf q}$ in the Hermann-Mauguin notation of the~\href{http://www.cryst.ehu.es/cryst/mpoint.html}{MPOINT} tool on the BCS~\cite{BCSMag1,BCSMag2,BCSMag3,BCSMag4} and the number of the MPG isomorphic to $G_{\bf q}$ in the convention established by Litvin in Ref.~\onlinecite{MagneticBook}, the symbol and number of the MPG isomorphic to the reducing group $G_{{\bf q}'}$, the symbol and number of the MPG isomorphic to the intersection group $G_{{\bf q}_{0}}=G_{\bf q}\cap G_{{\bf q}'}$, and the dimension $d$ of the exceptional composite band rep.  See Appendix~\ref{sec:exceptions} for further information regarding exceptional composite band reps.}
\\
\hline
\multicolumn{2}{|c|}{MSG}&Irrep&\multicolumn{2}{|c}{$G_{\mathbf{q}}$}&\multicolumn{2}{|c}{$G_{\mathbf{q}'}$}&\multicolumn{2}{|c|}{$G_{\mathbf{q}_0}$}&d\\\hline
124.351&$P4/mcc$&a,$E$&$422$&12.1.40&$4/m$&11.1.35&4&9.1.29&4\\
124.351&$P4/mcc$&c,$E$&$422$&12.1.40&$4/m$&11.1.35&4&9.1.29&4\\
131.435&$P4_2/mmc$&e,$E$&$\bar{4}2m$&14.1.48&$mmm$&8.1.24&mm2&7.1.20&4\\
131.435&$P4_2/mmc$&f,$E$&$\bar{4}2m$&14.1.48&$mmm$&8.1.24&mm2&7.1.20&4\\
132.447&$P4_2/mcm$&b,$E$&$\bar{4}2m$&14.1.48&$mmm$&8.1.24&mm2&7.1.20&4\\
132.447&$P4_2/mcm$&d,$E$&$\bar{4}2m$&14.1.48&$mmm$&8.1.24&mm2&7.1.20&4\\
\hline\end{longtable}

$\\$

\subsubsection{Exceptional Composite Band Reps in the Type-I Double MSGs}

$\\$

\begin{longtable}{|cc|c|cc|cc|cc|c|}
\caption{
Exceptional composite band reps induced from site-symmetry irreps in the Type-I double MSGs (Appendix~\ref{sec:type1}).  In order, the columns in this table list the number of the MSG in the BNS setting and the symbol of the MSG, the letter of the maximal Wyckoff position containing ${\bf q}$ and the double-valued irrep of the site-symmetry group $G_{\bf q}$, the symbol of the MPG isomorphic to the site-symmetry group $G_{\bf q}$ in the Hermann-Mauguin notation of the~\href{http://www.cryst.ehu.es/cryst/mpoint.html}{MPOINT} tool on the BCS~\cite{BCSMag1,BCSMag2,BCSMag3,BCSMag4} and the number of the MPG isomorphic to $G_{\bf q}$ in the convention established by Litvin in Ref.~\onlinecite{MagneticBook}, the symbol and number of the MPG isomorphic to the reducing group $G_{{\bf q}'}$, the symbol and number of the MPG isomorphic to the intersection group $G_{{\bf q}_{0}}=G_{\bf q}\cap G_{{\bf q}'}$, and the dimension $d$ of the exceptional composite band rep.  See Appendix~\ref{sec:exceptions} for further information regarding exceptional composite band reps.}
\\
\hline
\multicolumn{2}{|c|}{SSG}&Irrep&\multicolumn{2}{|c}{$G_{\mathbf{q}}$}&\multicolumn{2}{|c}{$G_{\mathbf{q}'}$}&\multicolumn{2}{|c|}{$G_{\mathbf{q}_0}$}&d\\\hline
49.265&$Pccm$&e,$\overline{E}$&$222$&6.1.17&$2/m$&5.1.12&2&3.1.6&4\\
49.265&$Pccm$&f,$\overline{E}$&$222$&6.1.17&$2/m$&5.1.12&2&3.1.6&4\\
49.265&$Pccm$&g,$\overline{E}$&$222$&6.1.17&$2/m$&5.1.12&2&3.1.6&4\\
49.265&$Pccm$&h,$\overline{E}$&$222$&6.1.17&$2/m$&5.1.12&2&3.1.6&4\\
51.289&$Pmma$&e,$\overline{E}$&$mm2$&7.1.20&$2/m$&5.1.12&m&4.1.9&4\\
51.289&$Pmma$&f,$\overline{E}$&$mm2$&7.1.20&$2/m$&5.1.12&m&4.1.9&4\\
\hline\end{longtable}

$\\$

\subsubsection{Exceptional Composite Band Coreps in the Type-II Single SGs}

$\\$

\begin{longtable}{|cc|c|cc|cc|cc|c|}
\caption{
Exceptional composite band coreps induced from site-symmetry coreps in the Type-II single SGs (Appendix~\ref{sec:type2}).  In order, the columns in this table list the number of the SG in the BNS setting and the symbol of the SG, the letter of the maximal Wyckoff position containing ${\bf q}$ and the single-valued corep of the site-symmetry group $G_{\bf q}$, the symbol of the SPG isomorphic to the site-symmetry group $G_{\bf q}$ in the Hermann-Mauguin notation of the~\href{http://www.cryst.ehu.es/cryst/mpoint.html}{MPOINT} tool on the BCS~\cite{BCSMag1,BCSMag2,BCSMag3,BCSMag4} and the number of the SPG isomorphic to $G_{\bf q}$ in the convention established by Litvin in Ref.~\onlinecite{MagneticBook}, the symbol and number of the SPG isomorphic to the reducing group $G_{{\bf q}'}$, the symbol and number of the SPG isomorphic to the intersection group $G_{{\bf q}_{0}}=G_{\bf q}\cap G_{{\bf q}'}$, and the dimension $d$ of the exceptional composite band corep.  See Appendix~\ref{sec:exceptions} for further information regarding exceptional composite band coreps.}
\\
\hline
\multicolumn{2}{|c|}{SG}&Corep&\multicolumn{2}{|c}{$G_{\mathbf{q}}$}&\multicolumn{2}{|c}{$G_{\mathbf{q}'}$}&\multicolumn{2}{|c|}{$G_{\mathbf{q}_0}$}&d\\\hline
84.52&$P4_2/m1'$&e,$\,^1E\,^2E$&$\bar{4}1'$&10.2.33&$2/m1'$&5.2.13&21'&3.2.7&4\\
84.52&$P4_2/m1'$&f,$\,^1E\,^2E$&$\bar{4}1'$&10.2.33&$2/m1'$&5.2.13&21'&3.2.7&4\\
87.76&$I4/m1'$&d,$\,^1E\,^2E$&$\bar{4}1'$&10.2.33&$2/m1'$&5.2.13&21'&3.2.7&4\\
112.260&$P\bar{4}2c1'$&e,$\,^1E\,^2E$&$\bar{4}1'$&10.2.33&$2221'$&6.2.18&21'&3.2.7&4\\
112.260&$P\bar{4}2c1'$&f,$\,^1E\,^2E$&$\bar{4}1'$&10.2.33&$2221'$&6.2.18&21'&3.2.7&4\\
116.292&$P\bar{4}c21'$&c,$\,^1E\,^2E$&$\bar{4}1'$&10.2.33&$2221'$&6.2.18&21'&3.2.7&4\\
\hline\end{longtable}

$\\$

\subsubsection{Exceptional Composite Band Coreps in the Type-III Single MSGs}

$\\$

\begin{longtable}{|cc|c|cc|cc|cc|c|}
\caption{
Exceptional composite band coreps induced from site-symmetry coreps in the Type-III single MSGs (Appendix~\ref{sec:type3}).  In order, the columns in this table list the number of the MSG in the BNS setting and the symbol of the MSG, the letter of the maximal Wyckoff position containing ${\bf q}$ and the single-valued corep of the site-symmetry group $G_{\bf q}$, the symbol of the MSG isomorphic to the site-symmetry group $G_{\bf q}$ in the Hermann-Mauguin notation of the~\href{http://www.cryst.ehu.es/cryst/mpoint.html}{MPOINT} tool on the BCS~\cite{BCSMag1,BCSMag2,BCSMag3,BCSMag4} and the number of the MSG isomorphic to $G_{\bf q}$ in the convention established by Litvin in Ref.~\onlinecite{MagneticBook}, the symbol and number of the MPG isomorphic to the reducing group $G_{{\bf q}'}$, the symbol and number of the MPG isomorphic to the intersection group $G_{{\bf q}_{0}}=G_{\bf q}\cap G_{{\bf q}'}$, and the dimension $d$ of the exceptional composite band corep.  See Appendix~\ref{sec:exceptions} for further information regarding exceptional composite band coreps.}
\\
\hline
\multicolumn{2}{|c|}{MSG}&Corep&\multicolumn{2}{|c}{$G_{\mathbf{q}}$}&\multicolumn{2}{|c}{$G_{\mathbf{q}'}$}&\multicolumn{2}{|c|}{$G_{\mathbf{q}_0}$}&d\\\hline
84.53&$P4_2'/m$&e,$BB$&$\bar{4}'$&10.3.34&$2/m$&5.1.12&2&3.1.6&4\\
84.53&$P4_2'/m$&f,$BB$&$\bar{4}'$&10.3.34&$2/m$&5.1.12&2&3.1.6&4\\
84.54&$P4_2/m'$&e,$BB$&$\bar{4}'$&10.3.34&$2/m'$&5.4.15&2&3.1.6&4\\
84.54&$P4_2/m'$&f,$BB$&$\bar{4}'$&10.3.34&$2/m'$&5.4.15&2&3.1.6&4\\
87.77&$I4'/m$&d,$BB$&$\bar{4}'$&10.3.34&$2/m$&5.1.12&2&3.1.6&4\\
87.78&$I4/m'$&d,$BB$&$\bar{4}'$&10.3.34&$2/m'$&5.4.15&2&3.1.6&4\\
\hline\end{longtable}

$\\$

\clearpage

\subsubsection{Exceptional Composite Band Coreps in the Type-III Double MSGs}

$\\$
 
\begin{longtable}{|cc|c|cc|cc|cc|c|}
\caption{
Exceptional composite band coreps induced from site-symmetry coreps in the Type-III double MSGs (Appendix~\ref{sec:type3}).  In order, the columns in this table list the number of the MSG in the BNS setting and the symbol of the MSG, the letter of the maximal Wyckoff position containing ${\bf q}$ and the double-valued corep of the site-symmetry group $G_{\bf q}$, the symbol of the MSG isomorphic to the site-symmetry group $G_{\bf q}$ in the Hermann-Mauguin notation of the~\href{http://www.cryst.ehu.es/cryst/mpoint.html}{MPOINT} tool on the BCS~\cite{BCSMag1,BCSMag2,BCSMag3,BCSMag4} and the number of the MSG isomorphic to $G_{\bf q}$ in the convention established by Litvin in Ref.~\onlinecite{MagneticBook}, the symbol and number of the MPG isomorphic to the reducing group $G_{{\bf q}'}$, the symbol and number of the MPG isomorphic to the intersection group $G_{{\bf q}_{0}}=G_{\bf q}\cap G_{{\bf q}'}$, and the dimension $d$ of the exceptional composite band corep.  See Appendix~\ref{sec:exceptions} for further information regarding exceptional composite band coreps.}
\\
\hline
\multicolumn{2}{|c|}{MSG}&Corep&\multicolumn{2}{|c}{$G_{\mathbf{q}}$}&\multicolumn{2}{|c}{$G_{\mathbf{q}'}$}&\multicolumn{2}{|c|}{$G_{\mathbf{q}_0}$}&d\\\hline
11.52&$P2_1'/m$&a,$\overline{A}\overline{A}$&$\bar{1}'$&2.3.5&$m$&4.1.9&1&1.1.1&4\\
11.52&$P2_1'/m$&b,$\overline{A}\overline{A}$&$\bar{1}'$&2.3.5&$m$&4.1.9&1&1.1.1&4\\
11.52&$P2_1'/m$&c,$\overline{A}\overline{A}$&$\bar{1}'$&2.3.5&$m$&4.1.9&1&1.1.1&4\\
11.52&$P2_1'/m$&d,$\overline{A}\overline{A}$&$\bar{1}'$&2.3.5&$m$&4.1.9&1&1.1.1&4\\
11.53&$P2_1/m'$&a,$\overline{A}\overline{A}$&$\bar{1}'$&2.3.5&$m'$&4.3.11&1&1.1.1&4\\
11.53&$P2_1/m'$&b,$\overline{A}\overline{A}$&$\bar{1}'$&2.3.5&$m'$&4.3.11&1&1.1.1&4\\
\hline\end{longtable}

$\\$

\subsubsection{Exceptional Composite Band Coreps in the Type-IV Single MSGs}

$\\$

\begin{longtable}{|cc|c|cc|cc|cc|c|}
\caption{
Exceptional composite band coreps induced from site-symmetry coreps in the Type-IV single MSGs (Appendix~\ref{sec:type4}).  In order, the columns in this table list the number of the MSG in the BNS setting and the symbol of the MSG, the letter of the maximal Wyckoff position containing ${\bf q}$ and the single-valued corep of the site-symmetry group $G_{\bf q}$, the symbol of the MSG isomorphic to the site-symmetry group $G_{\bf q}$ in the Hermann-Mauguin notation of the~\href{http://www.cryst.ehu.es/cryst/mpoint.html}{MPOINT} tool on the BCS~\cite{BCSMag1,BCSMag2,BCSMag3,BCSMag4} and the number of the MSG isomorphic to $G_{\bf q}$ in the convention established by Litvin in Ref.~\onlinecite{MagneticBook}, the symbol and number of the MPG isomorphic to the reducing group $G_{{\bf q}'}$, the symbol and number of the MPG isomorphic to the intersection group $G_{{\bf q}_{0}}=G_{\bf q}\cap G_{{\bf q}'}$, and the dimension $d$ of the exceptional composite band corep.  See Appendix~\ref{sec:exceptions} for further information regarding exceptional composite band coreps.}
\\
\hline
\multicolumn{2}{|c|}{MSG}&Corep&\multicolumn{2}{|c}{$G_{\mathbf{q}}$}&\multicolumn{2}{|c}{$G_{\mathbf{q}'}$}&\multicolumn{2}{|c|}{$G_{\mathbf{q}_0}$}&d\\\hline
81.36&$P_c\bar{4}$&b,$BB$&$\bar{4}'$&10.3.34&$\bar{4}$&10.1.32&2&3.1.6&4\\
81.36&$P_c\bar{4}$&d,$BB$&$\bar{4}'$&10.3.34&$\bar{4}$&10.1.32&2&3.1.6&4\\
82.42&$I_c\bar{4}$&b,$BB$&$\bar{4}'$&10.3.34&$\bar{4}$&10.1.32&2&3.1.6&4\\
82.42&$I_c\bar{4}$&c,$BB$&$\bar{4}'$&10.3.34&$\bar{4}$&10.1.32&2&3.1.6&4\\
83.48&$P_c4/m$&b,$\,^1E\,^2E$&$4/m'$&11.4.38&$4/m$&11.1.35&4&9.1.29&4\\
83.48&$P_c4/m$&d,$\,^1E\,^2E$&$4/m'$&11.4.38&$4/m$&11.1.35&4&9.1.29&4\\
\hline\end{longtable}

$\\$

\clearpage

\subsubsection{Exceptional Composite Band Coreps in the Type-IV Double MSGs}

$\\$
 
\begin{longtable}{|cc|c|cc|cc|cc|c|}
\caption{
Exceptional composite band coreps induced from site-symmetry coreps in the Type-IV double MSGs (Appendix~\ref{sec:type4}).  In order, the columns in this table list the number of the MSG in the BNS setting and the symbol of the MSG, the letter of the maximal Wyckoff position containing ${\bf q}$ and the double-valued corep of the site-symmetry group $G_{\bf q}$, the symbol of the MSG isomorphic to the site-symmetry group $G_{\bf q}$ in the Hermann-Mauguin notation of the~\href{http://www.cryst.ehu.es/cryst/mpoint.html}{MPOINT} tool on the BCS~\cite{BCSMag1,BCSMag2,BCSMag3,BCSMag4} and the number of the MSG isomorphic to $G_{\bf q}$ in the convention established by Litvin in Ref.~\onlinecite{MagneticBook}, the symbol and number of the MPG isomorphic to the reducing group $G_{{\bf q}'}$, the symbol and number of the MPG isomorphic to the intersection group $G_{{\bf q}_{0}}=G_{\bf q}\cap G_{{\bf q}'}$, and the dimension $d$ of the exceptional composite band corep.  See Appendix~\ref{sec:exceptions} for further information regarding exceptional composite band coreps.}
\\
\hline
\multicolumn{2}{|c|}{MSG}&Corep&\multicolumn{2}{|c}{$G_{\mathbf{q}}$}&\multicolumn{2}{|c}{$G_{\mathbf{q}'}$}&\multicolumn{2}{|c|}{$G_{\mathbf{q}_0}$}&d\\\hline
2.7&$P_S\bar{1}$&b,$\overline{A}\overline{A}$&$\bar{1}'$&2.3.5&$\bar{1}$&2.1.3&1&1.1.1&4\\
2.7&$P_S\bar{1}$&f,$\overline{A}\overline{A}$&$\bar{1}'$&2.3.5&$\bar{1}$&2.1.3&1&1.1.1&4\\
2.7&$P_S\bar{1}$&g,$\overline{A}\overline{A}$&$\bar{1}'$&2.3.5&$\bar{1}$&2.1.3&1&1.1.1&4\\
2.7&$P_S\bar{1}$&h,$\overline{A}\overline{A}$&$\bar{1}'$&2.3.5&$\bar{1}$&2.1.3&1&1.1.1&4\\
10.47&$P_a2/m$&d,$\,^1\overline{E}\,^2\overline{E}$&$2'/m$&5.3.14&$2/m$&5.1.12&m&4.1.9&4\\
10.47&$P_a2/m$&e,$\,^1\overline{E}\,^2\overline{E}$&$2'/m$&5.3.14&$2/m$&5.1.12&m&4.1.9&4\\
\hline\end{longtable}

\clearpage

\subsection{Maximum and Minimum Dimensions of the Single- and Double-Valued MEBRs}
\label{sec:EBCRdimension}

In this section we provide a complete tabulation of the maximum (M) and minimum (m) EBR dimension in each single and double SSG.  As discussed in Appendix~\ref{sec:mbandrep}, the EBRs are composed of the MEBRs of the Type-I MSGs and the PEBRs of the Type-II SGs previously computed for TQC~\cite{QuantumChemistry,Bandrep1,Bandrep2,Bandrep3,JenFragile1,BarryFragile}, as well as the MEBRs of the Type-III and Type-IV MSGs that were calculated for the present work.  For each EBR $\tilde{\rho}_{\bf q}^{G}$, the dimension of $\tilde{\rho}_{\bf q}^{G}$ is defined as $\chi_{\tilde{\rho}_{\bf q}}(E) \times n$, where $\tilde{\rho}_{\bf q}$ is the (co)rep of the site-symmetry group $G_{\bf q}$ from which $\tilde{\rho}_{\bf q}^{G}$ is induced [see Eq.~(\ref{eq:mainInductionMTQC}) and the surrounding text], and where $n$ is the multiplicity of the Wyckoff position indexed by ${\bf q}$ [see the text surrounding Eq.~(\ref{eq:finiteSitetoInfinite})].  For each single and double SSG, we have confirmed that the minimum EBR dimension $m$ is consistent with the minimum atomic insulator dimension previously calculated in Ref.~\onlinecite{AshvinMagnetic}.  We emphasize that the maximum EBR dimension M does not always coincide with the maximum band connectivity in an SSG, due to the possibility of decomposable [\emph{i.e.} disconnected or split] EBRs (see Refs.~\onlinecite{QuantumChemistry,Bandrep2,JenFragile1,BarryFragile,AndreiMaterials,LuisBasicBands,AshvinFragile} and the text surrounding Tables~\ref{tb:MEBRsSingle} and~\ref{tb:MEBRsDouble}).  To calculate the maximum band connectivity in the 1,651 SSGs, one must perform the intermediate tabulation of the \emph{basic bands}~\cite{LuisBasicBands}, which for each SSG are composed of the disconnected branches of the decomposable EBRs, the connected [indecomposable] EBRs, and the connected topological bands in the SSG (Appendix~\ref{sec:topologicalBands}).  The basic bands of the Type-II double SGs were previously tabulated in Ref.~\onlinecite{LuisBasicBands}; we leave the complete enumeration of the basic bands of the single SSGs and the double MSGs for future works.

\subsubsection{Maximum and Minimum Dimensions of the Single-Valued EBRs of the 1,651 Single SSGs}

\begin{scriptsize}

\begin{longtable}{|llrr|llrr|llrr|llrr|}

\caption{Maximum and minimum dimensions of the single-valued EBRs of the 1,651 single SSGs.  In order, the columns in this table list the symbol of the SSG, the number of the SSG in the BNS setting~\cite{BelovNotation}, the maximum EBR dimension in the SSG (M), and the minimum EBR dimension in the SSG (m).}
\\\hline
Symbol&BNS Number&M&m&Symbol&BNS Number&M&m&Symbol&BNS Number&M&m&Symbol&BNS Number&M&m\\\hline
$P1$&1.1&1&1&$P11'$&1.2&1&1&$P_S1$&1.3&2&2&$P\bar{1}$&2.4&1&1\\\hline
$P\bar{1}1'$&2.5&1&1&$P\bar{1}'$&2.6&1&1&$P_S\bar{1}$&2.7&2&2&$P2$&3.1&1&1\\\hline
$P21'$&3.2&1&1&$P2'$&3.3&1&1&$P_a2$&3.4&2&2&$P_b2$&3.5&2&2\\\hline
$P_C2$&3.6&2&2&$P2_1$&4.7&2&2&$P2_11'$&4.8&2&2&$P2_1'$&4.9&2&2\\\hline
$P_a2_1$&4.10&4&4&$P_b2_1$&4.11&2&2&$P_C2_1$&4.12&2&2&$C2$&5.13&1&1\\\hline
$C21'$&5.14&1&1&$C2'$&5.15&1&1&$C_c2$&5.16&2&2&$C_a2$&5.17&2&2\\
\hline
\end{longtable}

\end{scriptsize}

\clearpage

\subsubsection{Maximum and Minimum Dimensions of the Double-Valued EBRs of the 1,651 Double SSGs}

$\\$

\begin{scriptsize}

\begin{longtable}{|llrr|llrr|llrr|llrr|}

\caption{Maximum and minimum dimensions of the double-valued EBRs of the 1,651 double SSGs.  In order, the columns in this table list the symbol of the SSG, the number of the SSG in the BNS setting~\cite{BelovNotation}, the maximum EBR dimension in the SSG (M), and the minimum EBR dimension in the SSG (m).}
\\\hline
Symbol&BNS Number&M&m&Symbol&BNS Number&M&m&Symbol&BNS Number&M&m&Symbol&BNS Number&M&m\\\hline
$P1$&1.1&1&1&$P11'$&1.2&2&2&$P_S1$&1.3&2&2&$P\bar{1}$&2.4&1&1\\\hline
$P\bar{1}1'$&2.5&2&2&$P\bar{1}'$&2.6&2&2&$P_S\bar{1}$&2.7&2&2&$P2$&3.1&1&1\\\hline
$P21'$&3.2&2&2&$P2'$&3.3&1&1&$P_a2$&3.4&2&2&$P_b2$&3.5&2&2\\\hline
$P_C2$&3.6&2&2&$P2_1$&4.7&2&2&$P2_11'$&4.8&4&4&$P2_1'$&4.9&2&2\\\hline
$P_a2_1$&4.10&4&4&$P_b2_1$&4.11&2&2&$P_C2_1$&4.12&2&2&$C2$&5.13&1&1\\\hline
$C21'$&5.14&2&2&$C2'$&5.15&1&1&$C_c2$&5.16&2&2&$C_a2$&5.17&2&2\\
\hline
\end{longtable}

\end{scriptsize}

\clearpage

\subsection{Minimal SSG Dependencies for the Double SIs in the 1,651 Double SSGs}
\label{sec:minimalSSGTables}

In this section, we list the minimal double SSG $M$ with the minimal double SIs on which the double SIs in each double SSG $G$ are dependent (see Appendix~\ref{sec:minimalSIProcedure} for calculation details).

$\\$

\begin{longtable}{|c|c|c||c|c|c|||c|c|c||c|c|c|}
\caption{
Minimal SSG dependencies for the double SIs in the 1,651 double SSGs.  In order, the columns in this table list the symbol of the SSG $G$, the number of $G$ in the BNS setting~\cite{BelovNotation}, the type of $G$ (see Appendix~\ref{sec:MSGs}), the number of the minimal double SSG $M$ with the minimal double SIs on which the double SIs in $G$ are dependent (see Appendix~\ref{sec:minimalSIProcedure}), the symbol of $M$, and the type of $M$.  Entries for which $G=M$ are highlighted in bold text.  We have omitted double SSGs $G$ for which $|Z^{G}|=1$ [see Eq.~(\ref{eq:SIgroup}) and the surrounding text].}
\\
\hline
\multicolumn{2}{|c|}{Double SSG}&Type&\multicolumn{2}{c|}{Minimal SSG}&Type&\multicolumn{2}{c|}{Double SSG}&Type&\multicolumn{2}{c|}{Minimal SSG}&Type\\
\hline
\hline
\textbf{2.4}&$\mathbf{P\bar{1}}$&\textbf{I}&\textbf{2.4}&$\mathbf{P\bar{1}}$&\textbf{I}&\textbf{2.5}&$\mathbf{P\bar{1}1'}$&\textbf{II}&\textbf{2.5}&$\mathbf{P\bar{1}1'}$&\textbf{II}\\
\hline
2.7&$P_S\bar{1}$&IV&2.4&$P\bar{1}$&I&\textbf{3.1}&$\mathbf{P2}$&\textbf{I}&\textbf{3.1}&$\mathbf{P2}$&\textbf{I}\\
\hline
3.4&$P_a2$&IV&3.1&$P2$&I&\textbf{10.42}&$\mathbf{P2/m}$&\textbf{I}&\textbf{10.42}&$\mathbf{P2/m}$&\textbf{I}\\
\hline
10.43&$P2/m1'$&II&2.5&$P\bar{1}1'$&II&10.46&$P2'/m'$&III&2.4&$P\bar{1}$&I\\
\hline
10.47&$P_a2/m$&IV&10.42&$P2/m$&I&10.48&$P_b2/m$&IV&2.4&$P\bar{1}$&I\\
\hline
10.49&$P_C2/m$&IV&2.4&$P\bar{1}$&I&11.50&$P2_1/m$&I&2.4&$P\bar{1}$&I\\
\hline
\end{longtable}

\end{appendix}

\bibliographystyle{naturemag}
\bibliography{mEBR}

\end{document}